\newif\ifextended
\extendedtrue
\ifextended
\documentclass[acmsmall, screen, nonacm]{acmart}
\else
\documentclass[acmsmall]{acmart}
\fi

\AtEndPreamble{%
    \theoremstyle{acmdefinition}%
    \newtheorem{remark}[theorem]{Remark}}
\usepackage{mathpartir}
\usepackage{thm-restate}
\usepackage{resizegather}
\usepackage{graphicx}
\usepackage{hyperref}
\usepackage{tikz}
\usepackage{enumitem}

\newcommand{\eqStep}[0]{\hspace{2pt}\equiv\hspace{2pt}}
\newcommand{\smidge}[0]{\hspace{0.7pt}} % make [| . |] more balanced wrt to (| . |)

\newcommand{\SYSTEMdrule}[4][]{{\displaystyle\frac{\begin{array}{l}#2\end{array}}{#3}\quad\SYSTEMdrulename{#4}}}

\newcommand{\SYSTEMpremise}[1]{ #1 \\}
\newenvironment{SYSTEMdefnblock}[3][]{ \framebox{\mbox{#2}} \quad #3 \\[0pt]}{}

\newcommand{\SYSTEMnt}[1]{\mathit{#1}}
\newcommand{\SYSTEMmv}[1]{\mathit{#1}}

\newcommand{\SYSTEMsym}[1]{#1}

\newcommand{\SYSTEMdrulename}[1]{\textsc{#1}}

\IfFileExists{xcolor.sty}{
	\RequirePackage{xcolor}
}{
	\RequirePackage{color}
}

\usepackage{stmaryrd}
\definecolor{coeffectColor}{HTML}{0750D0}
\definecolor{effectColor}{HTML}{D64800}
\definecolor{GLcolor}{HTML}{7e4e99}
\definecolor{LGcolor}{HTML}{38721a}

\newcommand{\SYSTEMRenameRuleLinvar}[0]{\SYSTEMdrulename{Linvar}}
\newcommand{\SYSTEMdruleLinvar}[1]{\SYSTEMdrule[#1]{%
}{
   \SYSTEMmv{x}  :  \SYSTEMnt{A}    \vdash_{\textsc{l} }  \SYSTEMmv{x}  :  \SYSTEMnt{A} }{%
{\SYSTEMRenameRuleLinvar}{}%
}}

\newcommand{\SYSTEMRenameRuleLinabs}[0]{\SYSTEMdrulename{Linabs}}
\newcommand{\SYSTEMdruleLinabs}[1]{\SYSTEMdrule[#1]{%
\SYSTEMpremise{  \Gamma ,   \SYSTEMmv{x}  :  \SYSTEMnt{A}    \vdash_{\textsc{l} }  \SYSTEMnt{t}  :  \SYSTEMnt{B} }%
}{
 \Gamma  \vdash_{\textsc{l} }   \lambda  \SYSTEMmv{x}  .  \SYSTEMnt{t}   :   \SYSTEMnt{A}  \multimap  \SYSTEMnt{B}  }{%
{\SYSTEMRenameRuleLinabs}{}%
}}

\newcommand{\SYSTEMRenameRuleLinapp}[0]{\SYSTEMdrulename{Linapp}}
\newcommand{\SYSTEMdruleLinapp}[1]{\SYSTEMdrule[#1]{%
\SYSTEMpremise{ \begin{array}{cc}   \Gamma_{{\mathrm{1}}}  \vdash_{\textsc{l} }  \SYSTEMnt{t_{{\mathrm{1}}}}  :   \SYSTEMnt{A}  \multimap  \SYSTEMnt{B}    \; & \;   \Gamma_{{\mathrm{2}}}  \vdash_{\textsc{l} }  \SYSTEMnt{t_{{\mathrm{2}}}}  :  \SYSTEMnt{A}   \end{array} }%
}{
 \Gamma_{{\mathrm{1}}}  \SYSTEMsym{+}  \Gamma_{{\mathrm{2}}}  \vdash_{\textsc{l} }   \SYSTEMnt{t_{{\mathrm{1}}}} \,  \SYSTEMnt{t_{{\mathrm{2}}}}   :  \SYSTEMnt{B} }{%
{\SYSTEMRenameRuleLinapp}{}%
}}

\newcommand{\SYSTEMRenameRuleLinweak}[0]{\SYSTEMdrulename{Linweak}}
\newcommand{\SYSTEMdruleLinweak}[1]{\SYSTEMdrule[#1]{%
\SYSTEMpremise{ \begin{array}{cc}   \Gamma  \vdash_{\textsc{l} }  \SYSTEMnt{t}  :  \SYSTEMnt{A}   \; & \;   \mathrm{graded}( \Gamma' , \textcolor{coeffectColor}{ \SYSTEMsym{0} })   \end{array} }%
}{
  \Gamma  ,  \Gamma'   \vdash_{\textsc{l} }  \SYSTEMnt{t}  :  \SYSTEMnt{A} }{%
{\SYSTEMRenameRuleLinweak}{}%
}}

\newcommand{\SYSTEMRenameRuleLinder}[0]{\SYSTEMdrulename{Linder}}
\newcommand{\SYSTEMdruleLinder}[1]{\SYSTEMdrule[#1]{%
\SYSTEMpremise{  \Gamma ,   \SYSTEMmv{x}  :  \SYSTEMnt{A}    \vdash_{\textsc{l} }  \SYSTEMnt{t}  :  \SYSTEMnt{B} }%
}{
  \Gamma ,   \SYSTEMmv{x}  : \textcolor{coeffectColor}{[}  \SYSTEMnt{A} {\textcolor{coeffectColor}{]_{ \SYSTEMsym{1} } } }    \vdash_{\textsc{l} }  \SYSTEMnt{t}  :  \SYSTEMnt{B} }{%
{\SYSTEMRenameRuleLinder}{}%
}}

\newcommand{\SYSTEMRenameRuleLinpr}[0]{\SYSTEMdrulename{Linpr}}
\newcommand{\SYSTEMdruleLinpr}[1]{\SYSTEMdrule[#1]{%
\SYSTEMpremise{ \begin{array}{cc}   \Gamma  \vdash_{\textsc{l} }  \SYSTEMnt{t}  :  \SYSTEMnt{A}   \; & \;   \mathrm{graded}( \Gamma )   \end{array} }%
}{
  \textcolor{coeffectColor}{ \SYSTEMnt{r}  \cdot}  \Gamma   \vdash_{\textsc{l} }   \textcolor{coeffectColor}{[}  \SYSTEMnt{t}  \textcolor{coeffectColor}{]}   :   \textcolor{coeffectColor}{\square_{ \SYSTEMnt{r} } }  \SYSTEMnt{A}  }{%
{\SYSTEMRenameRuleLinpr}{}%
}}

\newcommand{\SYSTEMRenameRuleLinlet}[0]{\SYSTEMdrulename{Linlet}}
\newcommand{\SYSTEMdruleLinlet}[1]{\SYSTEMdrule[#1]{%
\SYSTEMpremise{ \begin{array}{cc}   \Gamma_{{\mathrm{1}}}  \vdash_{\textsc{l} }  \SYSTEMnt{t_{{\mathrm{1}}}}  :   \textcolor{coeffectColor}{\square_{ \SYSTEMnt{r} } }  \SYSTEMnt{A}    \; & \;    \Gamma_{{\mathrm{2}}} ,   \SYSTEMmv{x}  : \textcolor{coeffectColor}{[}  \SYSTEMnt{A} {\textcolor{coeffectColor}{]_{ \SYSTEMnt{r} } } }    \vdash_{\textsc{l} }  \SYSTEMnt{t_{{\mathrm{2}}}}  :  \SYSTEMnt{B}   \end{array} }%
}{
 \Gamma_{{\mathrm{1}}}  \SYSTEMsym{+}  \Gamma_{{\mathrm{2}}}  \vdash_{\textsc{l} }   \mathsf{let} \, \textcolor{coeffectColor}{[}  \SYSTEMmv{x}  \textcolor{coeffectColor}{]} =  \SYSTEMnt{t_{{\mathrm{1}}}}  \, \mathsf{in} \,  \SYSTEMnt{t_{{\mathrm{2}}}}   :  \SYSTEMnt{B} }{%
{\SYSTEMRenameRuleLinlet}{}%
}}

\newcommand{\SYSTEMRenameRuleLinapprox}[0]{\SYSTEMdrulename{Linapprox}}
\newcommand{\SYSTEMdruleLinapprox}[1]{\SYSTEMdrule[#1]{%
\SYSTEMpremise{ \begin{array}{cc}    \Gamma ,   \SYSTEMmv{x}  : \textcolor{coeffectColor}{[}  \SYSTEMnt{B} {\textcolor{coeffectColor}{]_{ \SYSTEMnt{r} } } }    \vdash_{\textsc{l} }  \SYSTEMnt{t}  :  \SYSTEMnt{A}   \; & \;   \SYSTEMnt{r}  \, \textcolor{coeffectColor}{\sqsubseteq} \,  \SYSTEMnt{s}   \end{array} }%
}{
  \Gamma ,   \SYSTEMmv{x}  : \textcolor{coeffectColor}{[}  \SYSTEMnt{B} {\textcolor{coeffectColor}{]_{ \SYSTEMnt{s} } } }    \vdash_{\textsc{l} }  \SYSTEMnt{t}  :  \SYSTEMnt{A} }{%
{\SYSTEMRenameRuleLinapprox}{}%
}}

\newcommand{\SYSTEMRenameRuleLinprodi}[0]{\SYSTEMdrulename{Linprodi}}
\newcommand{\SYSTEMdruleLinprodi}[1]{\SYSTEMdrule[#1]{%
\SYSTEMpremise{ \begin{array}{cc}   \Gamma_{{\mathrm{1}}}  \vdash_{\textsc{l} }  \SYSTEMnt{t_{{\mathrm{1}}}}  :  \SYSTEMnt{A}   \; & \;   \Gamma_{{\mathrm{2}}}  \vdash_{\textsc{l} }  \SYSTEMnt{t_{{\mathrm{2}}}}  :  \SYSTEMnt{B}   \end{array} }%
}{
 \Gamma_{{\mathrm{1}}}  \SYSTEMsym{+}  \Gamma_{{\mathrm{2}}}  \vdash_{\textsc{l} }   \langle  \SYSTEMnt{t_{{\mathrm{1}}}} ,  \SYSTEMnt{t_{{\mathrm{2}}}}  \rangle   :   \SYSTEMnt{A}  \otimes  \SYSTEMnt{B}  }{%
{\SYSTEMRenameRuleLinprodi}{}%
}}

\newcommand{\SYSTEMRenameRuleLinprode}[0]{\SYSTEMdrulename{Linprode}}
\newcommand{\SYSTEMdruleLinprode}[1]{\SYSTEMdrule[#1]{%
\SYSTEMpremise{ \begin{array}{cc}   \Gamma_{{\mathrm{1}}}  \vdash_{\textsc{l} }  \SYSTEMnt{t_{{\mathrm{1}}}}  :   \SYSTEMnt{A}  \otimes  \SYSTEMnt{B}    \; & \;     \Gamma_{{\mathrm{2}}} ,   \SYSTEMmv{x}  :  \SYSTEMnt{A}   ,   \SYSTEMmv{y}  :  \SYSTEMnt{B}    \vdash_{\textsc{l} }  \SYSTEMnt{t_{{\mathrm{2}}}}  :  \SYSTEMnt{C}   \end{array} }%
}{
 \Gamma_{{\mathrm{1}}}  \SYSTEMsym{+}  \Gamma_{{\mathrm{2}}}  \vdash_{\textsc{l} }   \mathsf{let} \, \langle  \SYSTEMmv{x} ,  \SYSTEMmv{y}  \rangle =  \SYSTEMnt{t_{{\mathrm{1}}}}  \, \mathsf{in} \,  \SYSTEMnt{t_{{\mathrm{2}}}}   :  \SYSTEMnt{C} }{%
{\SYSTEMRenameRuleLinprode}{}%
}}

\newcommand{\SYSTEMRenameRuleLinuniti}[0]{\SYSTEMdrulename{Linuniti}}
\newcommand{\SYSTEMdruleLinuniti}[1]{\SYSTEMdrule[#1]{%
}{
  \emptyset   \vdash_{\textsc{l} }   \langle \rangle   :   \mathrm{unit}  }{%
{\SYSTEMRenameRuleLinuniti}{}%
}}

\newcommand{\SYSTEMRenameRuleLinunite}[0]{\SYSTEMdrulename{Linunite}}
\newcommand{\SYSTEMdruleLinunite}[1]{\SYSTEMdrule[#1]{%
\SYSTEMpremise{ \begin{array}{cc}   \Gamma_{{\mathrm{1}}}  \vdash_{\textsc{l} }  \SYSTEMnt{t_{{\mathrm{1}}}}  :   \mathrm{unit}    \; & \;   \Gamma_{{\mathrm{2}}}  \vdash_{\textsc{l} }  \SYSTEMnt{t_{{\mathrm{2}}}}  :  \SYSTEMnt{A}   \end{array} }%
}{
 \Gamma_{{\mathrm{1}}}  \SYSTEMsym{+}  \Gamma_{{\mathrm{2}}}  \vdash_{\textsc{l} }   \mathsf{let} \, \langle \rangle =  \SYSTEMnt{t_{{\mathrm{1}}}}  \, \mathsf{in} \,  \SYSTEMnt{t_{{\mathrm{2}}}}   :  \SYSTEMnt{A} }{%
{\SYSTEMRenameRuleLinunite}{}%
}}

\newcommand{\SYSTEMRenameRuleLinpushprod}[0]{\SYSTEMdrulename{Linpushprod}}
\newcommand{\SYSTEMdruleLinpushprod}[1]{\SYSTEMdrule[#1]{%
\SYSTEMpremise{ \Gamma  \vdash_{\textsc{l} }  \SYSTEMnt{t}  :   \textcolor{coeffectColor}{\square_{ \SYSTEMnt{r} } }   (   \SYSTEMnt{A}  \otimes  \SYSTEMnt{B}   )   }%
}{
 \Gamma  \vdash_{\textsc{l} }   \textsf{push}_\otimes  \SYSTEMnt{t}   :     \textcolor{coeffectColor}{\square_{ \SYSTEMnt{r} } }  \SYSTEMnt{A}    \otimes    \textcolor{coeffectColor}{\square_{ \SYSTEMnt{r} } }  \SYSTEMnt{B}    }{%
{\SYSTEMRenameRuleLinpushprod}{}%
}}

\newcommand{\SYSTEMRenameRuleLinsumiOne}[0]{\SYSTEMdrulename{Linsumi1}}
\newcommand{\SYSTEMdruleLinsumiOne}[1]{\SYSTEMdrule[#1]{%
\SYSTEMpremise{ \Gamma  \vdash_{\textsc{l} }  \SYSTEMnt{t}  :  \SYSTEMnt{A} }%
}{
 \Gamma  \vdash_{\textsc{l} }   \mathsf{inj}_1 \,  \SYSTEMnt{t}   :   \SYSTEMnt{A}  \oplus  \SYSTEMnt{B}  }{%
{\SYSTEMRenameRuleLinsumiOne}{}%
}}

\newcommand{\SYSTEMRenameRuleLinsumiTwo}[0]{\SYSTEMdrulename{Linsumi2}}
\newcommand{\SYSTEMdruleLinsumiTwo}[1]{\SYSTEMdrule[#1]{%
\SYSTEMpremise{ \Gamma  \vdash_{\textsc{l} }  \SYSTEMnt{t}  :  \SYSTEMnt{B} }%
}{
 \Gamma  \vdash_{\textsc{l} }   \mathsf{inj}_2 \,  \SYSTEMnt{t}   :   \SYSTEMnt{A}  \oplus  \SYSTEMnt{B}  }{%
{\SYSTEMRenameRuleLinsumiTwo}{}%
}}

\newcommand{\SYSTEMRenameRuleLinsume}[0]{\SYSTEMdrulename{Linsume}}
\newcommand{\SYSTEMdruleLinsume}[1]{\SYSTEMdrule[#1]{%
\SYSTEMpremise{ \begin{array}{cc}    \begin{array}{cc}   \Gamma_{{\mathrm{1}}}  \vdash_{\textsc{l} }  \SYSTEMnt{t}  :   \SYSTEMnt{A}  \oplus  \SYSTEMnt{B}    \; & \;    \Gamma_{{\mathrm{2}}} ,   \SYSTEMmv{x}  :  \SYSTEMnt{A}    \vdash_{\textsc{l} }  \SYSTEMnt{t_{{\mathrm{1}}}}  :  \SYSTEMnt{C}   \end{array}    \; & \;    \Gamma_{{\mathrm{2}}} ,   \SYSTEMmv{y}  :  \SYSTEMnt{B}    \vdash_{\textsc{l} }  \SYSTEMnt{t_{{\mathrm{2}}}}  :  \SYSTEMnt{C}   \end{array} }%
}{
 \Gamma_{{\mathrm{1}}}  \SYSTEMsym{+}  \Gamma_{{\mathrm{2}}}  \vdash_{\textsc{l} }   \mathsf{case} \,  \SYSTEMnt{t}  \, \mathsf{of} \, \{ \mathsf{inj1} \,  \SYSTEMmv{x}  \rightarrow  \SYSTEMnt{t_{{\mathrm{1}}}}  ; \, \mathsf{inj2} \,  \SYSTEMmv{y}  \rightarrow  \SYSTEMnt{t_{{\mathrm{2}}}}  \}   :  \SYSTEMnt{C} }{%
{\SYSTEMRenameRuleLinsume}{}%
}}

\newcommand{\SYSTEMRenameRuleLinpushsum}[0]{\SYSTEMdrulename{Linpushsum}}
\newcommand{\SYSTEMdruleLinpushsum}[1]{\SYSTEMdrule[#1]{%
\SYSTEMpremise{ \begin{array}{cc}   \SYSTEMsym{1}  \, \textcolor{coeffectColor}{\sqsubseteq} \,  \SYSTEMnt{r}   \; & \;   \Gamma  \vdash_{\textsc{l} }  \SYSTEMnt{t}  :   \textcolor{coeffectColor}{\square_{ \SYSTEMnt{r} } }   (   \SYSTEMnt{A}  \oplus  \SYSTEMnt{B}   )     \end{array} }%
}{
 \Gamma  \vdash_{\textsc{l} }   \textsf{push}_\oplus  \SYSTEMnt{t}   :     \textcolor{coeffectColor}{\square_{ \SYSTEMnt{r} } }  \SYSTEMnt{A}    \oplus    \textcolor{coeffectColor}{\square_{ \SYSTEMnt{r} } }  \SYSTEMnt{B}    }{%
{\SYSTEMRenameRuleLinpushsum}{}%
}}

\newcommand{\SYSTEMRenameRuleLinpushunit}[0]{\SYSTEMdrulename{Linpushunit}}
\newcommand{\SYSTEMdruleLinpushunit}[1]{\SYSTEMdrule[#1]{%
\SYSTEMpremise{ \Gamma  \vdash_{\textsc{l} }  \SYSTEMnt{t}  :   \textcolor{coeffectColor}{\square_{ \SYSTEMnt{r} } }   \mathrm{unit}   }%
}{
 \Gamma  \vdash_{\textsc{l} }   \textsf{push}_{\mathrm{unit} }  \SYSTEMnt{t}   :   \mathrm{unit}  }{%
{\SYSTEMRenameRuleLinpushunit}{}%
}}

\newcommand{\SYSTEMRenameRuleGradvar}[0]{\SYSTEMdrulename{Gradvar}}
\newcommand{\SYSTEMdruleGradvar}[1]{\SYSTEMdrule[#1]{%
}{
   \SYSTEMmv{x}  :_{\textcolor{coeffectColor}{ \SYSTEMsym{1} } }  \SYSTEMnt{A}    \vdash_{\textsc{g} }  \SYSTEMmv{x}  :  \SYSTEMnt{A} }{%
{\SYSTEMRenameRuleGradvar}{}%
}}

\newcommand{\SYSTEMRenameRuleGradabs}[0]{\SYSTEMdrulename{Gradabs}}
\newcommand{\SYSTEMdruleGradabs}[1]{\SYSTEMdrule[#1]{%
\SYSTEMpremise{  \Delta ,   \SYSTEMmv{x}  :_{\textcolor{coeffectColor}{ \SYSTEMnt{r} } }  \SYSTEMnt{A}    \vdash_{\textsc{g} }  \SYSTEMnt{t}  :  \SYSTEMnt{B} }%
}{
 \Delta  \vdash_{\textsc{g} }   \lambda  \SYSTEMmv{x}  .  \SYSTEMnt{t}   :   \SYSTEMnt{A}  \xrightarrow{\textcolor{coeffectColor}{ \SYSTEMnt{r} } }  \SYSTEMnt{B}  }{%
{\SYSTEMRenameRuleGradabs}{}%
}}

\newcommand{\SYSTEMRenameRuleGradapp}[0]{\SYSTEMdrulename{Gradapp}}
\newcommand{\SYSTEMdruleGradapp}[1]{\SYSTEMdrule[#1]{%
\SYSTEMpremise{ \begin{array}{cc}   \Delta_{{\mathrm{1}}}  \vdash_{\textsc{g} }  \SYSTEMnt{t_{{\mathrm{1}}}}  :   \SYSTEMnt{A}  \xrightarrow{\textcolor{coeffectColor}{ \SYSTEMnt{r} } }  \SYSTEMnt{B}    \; & \;   \Delta_{{\mathrm{2}}}  \vdash_{\textsc{g} }  \SYSTEMnt{t_{{\mathrm{2}}}}  :  \SYSTEMnt{A}   \end{array} }%
}{
 \Delta_{{\mathrm{1}}}  \SYSTEMsym{+}   \textcolor{coeffectColor}{ \SYSTEMnt{r}  \cdot}  \Delta_{{\mathrm{2}}}   \vdash_{\textsc{g} }   \SYSTEMnt{t_{{\mathrm{1}}}} \,  \SYSTEMnt{t_{{\mathrm{2}}}}   :  \SYSTEMnt{B} }{%
{\SYSTEMRenameRuleGradapp}{}%
}}

\newcommand{\SYSTEMRenameRuleGradweak}[0]{\SYSTEMdrulename{Gradweak}}
\newcommand{\SYSTEMdruleGradweak}[1]{\SYSTEMdrule[#1]{%
\SYSTEMpremise{ \Delta  \vdash_{\textsc{g} }  \SYSTEMnt{t}  :  \SYSTEMnt{A} }%
}{
  \Delta  ,   \textcolor{coeffectColor}{ \SYSTEMsym{0}  \cdot}  \Delta'    \vdash_{\textsc{g} }  \SYSTEMnt{t}  :  \SYSTEMnt{A} }{%
{\SYSTEMRenameRuleGradweak}{}%
}}

\newcommand{\SYSTEMRenameRuleGradapprox}[0]{\SYSTEMdrulename{Gradapprox}}
\newcommand{\SYSTEMdruleGradapprox}[1]{\SYSTEMdrule[#1]{%
\SYSTEMpremise{ \begin{array}{cc}    \Delta ,   \SYSTEMmv{x}  :_{\textcolor{coeffectColor}{ \SYSTEMnt{r} } }  \SYSTEMnt{B}    \vdash_{\textsc{g} }  \SYSTEMnt{t}  :  \SYSTEMnt{A}   \; & \;   \SYSTEMnt{r}  \, \textcolor{coeffectColor}{\sqsubseteq} \,  \SYSTEMnt{s}   \end{array} }%
}{
  \Delta ,   \SYSTEMmv{x}  :_{\textcolor{coeffectColor}{ \SYSTEMnt{s} } }  \SYSTEMnt{B}    \vdash_{\textsc{g} }  \SYSTEMnt{t}  :  \SYSTEMnt{A} }{%
{\SYSTEMRenameRuleGradapprox}{}%
}}

\newcommand{\SYSTEMRenameRuleGradprodi}[0]{\SYSTEMdrulename{Gradprodi}}
\newcommand{\SYSTEMdruleGradprodi}[1]{\SYSTEMdrule[#1]{%
\SYSTEMpremise{ \begin{array}{cc}   \Delta_{{\mathrm{1}}}  \vdash_{\textsc{g} }  \SYSTEMnt{t_{{\mathrm{1}}}}  :  \SYSTEMnt{A}   \; & \;   \Delta_{{\mathrm{2}}}  \vdash_{\textsc{g} }  \SYSTEMnt{t_{{\mathrm{2}}}}  :  \SYSTEMnt{B}   \end{array} }%
}{
 \Delta_{{\mathrm{1}}}  \SYSTEMsym{+}  \Delta_{{\mathrm{2}}}  \vdash_{\textsc{g} }   \langle  \SYSTEMnt{t_{{\mathrm{1}}}} ,  \SYSTEMnt{t_{{\mathrm{2}}}}  \rangle   :   \SYSTEMnt{A}  \times  \SYSTEMnt{B}  }{%
{\SYSTEMRenameRuleGradprodi}{}%
}}

\newcommand{\SYSTEMRenameRuleGradprode}[0]{\SYSTEMdrulename{Gradprode}}
\newcommand{\SYSTEMdruleGradprode}[1]{\SYSTEMdrule[#1]{%
\SYSTEMpremise{ \begin{array}{cc}   \Delta_{{\mathrm{1}}}  \vdash_{\textsc{g} }  \SYSTEMnt{t_{{\mathrm{1}}}}  :   \SYSTEMnt{A}  \times  \SYSTEMnt{B}    \; & \;     \Delta_{{\mathrm{2}}} ,   \SYSTEMmv{x}  :_{\textcolor{coeffectColor}{ \SYSTEMnt{r} } }  \SYSTEMnt{A}   ,   \SYSTEMmv{y}  :_{\textcolor{coeffectColor}{ \SYSTEMnt{r} } }  \SYSTEMnt{B}    \vdash_{\textsc{g} }  \SYSTEMnt{t_{{\mathrm{2}}}}  :  \SYSTEMnt{C}   \end{array} }%
}{
   \textcolor{coeffectColor}{ \SYSTEMnt{r}  \cdot}  \Delta_{{\mathrm{1}}}    \SYSTEMsym{+}  \Delta_{{\mathrm{2}}}  \vdash_{\textsc{g} }   \mathsf{let} \, \langle  \SYSTEMmv{x} ,  \SYSTEMmv{y}  \rangle =  \SYSTEMnt{t_{{\mathrm{1}}}}  \, \mathsf{in} \,  \SYSTEMnt{t_{{\mathrm{2}}}}   :  \SYSTEMnt{C} }{%
{\SYSTEMRenameRuleGradprode}{}%
}}

\newcommand{\SYSTEMRenameRuleGraduniti}[0]{\SYSTEMdrulename{Graduniti}}
\newcommand{\SYSTEMdruleGraduniti}[1]{\SYSTEMdrule[#1]{%
}{
  \emptyset   \vdash_{\textsc{g} }   \langle \rangle   :   \mathrm{unit}  }{%
{\SYSTEMRenameRuleGraduniti}{}%
}}

\newcommand{\SYSTEMRenameRuleGradunite}[0]{\SYSTEMdrulename{Gradunite}}
\newcommand{\SYSTEMdruleGradunite}[1]{\SYSTEMdrule[#1]{%
\SYSTEMpremise{ \begin{array}{cc}   \Delta_{{\mathrm{1}}}  \vdash_{\textsc{g} }  \SYSTEMnt{t_{{\mathrm{1}}}}  :   \mathrm{unit}    \; & \;   \Delta_{{\mathrm{2}}}  \vdash_{\textsc{g} }  \SYSTEMnt{t_{{\mathrm{2}}}}  :  \SYSTEMnt{A}   \end{array} }%
}{
   \textcolor{coeffectColor}{ \SYSTEMnt{r}  \cdot}  \Delta_{{\mathrm{1}}}    \SYSTEMsym{+}  \Delta_{{\mathrm{2}}}  \vdash_{\textsc{g} }   \mathsf{let} \, \langle \rangle =  \SYSTEMnt{t_{{\mathrm{1}}}}  \, \mathsf{in} \,  \SYSTEMnt{t_{{\mathrm{2}}}}   :  \SYSTEMnt{A} }{%
{\SYSTEMRenameRuleGradunite}{}%
}}

\newcommand{\SYSTEMRenameRuleGradsumiOne}[0]{\SYSTEMdrulename{Gradsumi1}}
\newcommand{\SYSTEMdruleGradsumiOne}[1]{\SYSTEMdrule[#1]{%
\SYSTEMpremise{ \Delta  \vdash_{\textsc{g} }  \SYSTEMnt{t}  :  \SYSTEMnt{A} }%
}{
 \Delta  \vdash_{\textsc{g} }   \mathsf{inj}_1 \,  \SYSTEMnt{t}   :   \SYSTEMnt{A}  +  \SYSTEMnt{B}  }{%
{\SYSTEMRenameRuleGradsumiOne}{}%
}}

\newcommand{\SYSTEMRenameRuleGradsumiTwo}[0]{\SYSTEMdrulename{Gradsumi2}}
\newcommand{\SYSTEMdruleGradsumiTwo}[1]{\SYSTEMdrule[#1]{%
\SYSTEMpremise{ \Delta  \vdash_{\textsc{g} }  \SYSTEMnt{t}  :  \SYSTEMnt{B} }%
}{
 \Delta  \vdash_{\textsc{g} }   \mathsf{inj}_2 \,  \SYSTEMnt{t}   :   \SYSTEMnt{A}  +  \SYSTEMnt{B}  }{%
{\SYSTEMRenameRuleGradsumiTwo}{}%
}}

\newcommand{\SYSTEMRenameRuleGradsume}[0]{\SYSTEMdrulename{Gradsume}}
\newcommand{\SYSTEMdruleGradsume}[1]{\SYSTEMdrule[#1]{%
\SYSTEMpremise{      \Delta_{{\mathrm{1}}}  \vdash_{\textsc{g} }  \SYSTEMnt{t}  :   \SYSTEMnt{A}  +  \SYSTEMnt{B}    \quad    \Delta_{{\mathrm{2}}} ,   \SYSTEMmv{x}  :_{\textcolor{coeffectColor}{ \SYSTEMnt{r} } }  \SYSTEMnt{A}    \vdash_{\textsc{g} }  \SYSTEMnt{t_{{\mathrm{1}}}}  :  \SYSTEMnt{C}     \quad    \Delta_{{\mathrm{2}}} ,   \SYSTEMmv{y}  :_{\textcolor{coeffectColor}{ \SYSTEMnt{r} } }  \SYSTEMnt{B}    \vdash_{\textsc{g} }  \SYSTEMnt{t_{{\mathrm{2}}}}  :  \SYSTEMnt{C}     \quad   \SYSTEMsym{1}  \, \textcolor{coeffectColor}{\sqsubseteq} \,  \SYSTEMnt{r}  }%
}{
   \textcolor{coeffectColor}{ \SYSTEMnt{r}  \cdot}  \Delta_{{\mathrm{1}}}    \SYSTEMsym{+}  \Delta_{{\mathrm{2}}}  \vdash_{\textsc{g} }   \mathsf{case} \,  \SYSTEMnt{t}  \, \mathsf{of} \, \{ \mathsf{inj1} \,  \SYSTEMmv{x}  \rightarrow  \SYSTEMnt{t_{{\mathrm{1}}}}  ; \, \mathsf{inj2} \,  \SYSTEMmv{y}  \rightarrow  \SYSTEMnt{t_{{\mathrm{2}}}}  \}   :  \SYSTEMnt{C} }{%
{\SYSTEMRenameRuleGradsume}{}%
}}

\newcommand{\SYSTEMRenameRuleGradBoxpr}[0]{\SYSTEMdrulename{GradBoxpr}}
\newcommand{\SYSTEMdruleGradBoxpr}[1]{\SYSTEMdrule[#1]{%
\SYSTEMpremise{ \Delta  \vdash_{\textsc{g}_\square }  \SYSTEMnt{t}  :  \SYSTEMnt{A} }%
}{
  \textcolor{coeffectColor}{ \SYSTEMnt{r}  \cdot}  \Delta   \vdash_{\textsc{g}_\square }   \textcolor{coeffectColor}{[}  \SYSTEMnt{t}  \textcolor{coeffectColor}{]}   :   \textcolor{coeffectColor}{\square_{ \SYSTEMnt{r} } }  \SYSTEMnt{A}  }{%
{\SYSTEMRenameRuleGradBoxpr}{}%
}}

\newcommand{\SYSTEMRenameRuleGradBoxlet}[0]{\SYSTEMdrulename{GradBoxlet}}
\newcommand{\SYSTEMdruleGradBoxlet}[1]{\SYSTEMdrule[#1]{%
\SYSTEMpremise{ \begin{array}{cc}   \Delta_{{\mathrm{1}}}  \vdash_{\textsc{g}_\square }  \SYSTEMnt{t_{{\mathrm{1}}}}  :   \textcolor{coeffectColor}{\square_{ \SYSTEMnt{r} } }  \SYSTEMnt{A}    \; & \;    \Delta_{{\mathrm{2}}} ,   \SYSTEMmv{x}  :_{\textcolor{coeffectColor}{ \SYSTEMnt{r} } }  \SYSTEMnt{A}    \vdash_{\textsc{g}_\square }  \SYSTEMnt{t_{{\mathrm{2}}}}  :  \SYSTEMnt{B}   \end{array} }%
}{
 \Delta_{{\mathrm{1}}}  \SYSTEMsym{+}  \Delta_{{\mathrm{2}}}  \vdash_{\textsc{g}_\square }   \mathsf{let} \, \textcolor{coeffectColor}{[}  \SYSTEMmv{x}  \textcolor{coeffectColor}{]} =  \SYSTEMnt{t_{{\mathrm{1}}}}  \, \mathsf{in} \,  \SYSTEMnt{t_{{\mathrm{2}}}}   :  \SYSTEMnt{B} }{%
{\SYSTEMRenameRuleGradBoxlet}{}%
}}

\newcommand{\SYSTEMRenameRuleGradBoxletGen}[0]{\SYSTEMdrulename{GradBoxletGen}}
\newcommand{\SYSTEMdruleGradBoxletGen}[1]{\SYSTEMdrule[#1]{%
\SYSTEMpremise{ \begin{array}{cc}   \Delta_{{\mathrm{1}}}  \vdash_{\textsc{g}_\square }  \SYSTEMnt{t_{{\mathrm{1}}}}  :   \textcolor{coeffectColor}{\square_{ \SYSTEMnt{r} } }  \SYSTEMnt{A}    \; & \;    \Delta_{{\mathrm{2}}} ,   \SYSTEMmv{x}  :_{\textcolor{coeffectColor}{ \SYSTEMsym{(}   \SYSTEMnt{s}  \cdot  \SYSTEMnt{r}   \SYSTEMsym{)} } }  \SYSTEMnt{A}    \vdash_{\textsc{g}_\square }  \SYSTEMnt{t_{{\mathrm{2}}}}  :  \SYSTEMnt{B}   \end{array} }%
}{
  (   \textcolor{coeffectColor}{ \SYSTEMnt{s}  \cdot}  \Delta_{{\mathrm{1}}}   )   \SYSTEMsym{+}  \Delta_{{\mathrm{2}}}  \vdash_{\textsc{g}_\square }   \mathsf{let} \, \textcolor{coeffectColor}{[}  \SYSTEMmv{x}  \textcolor{coeffectColor}{]} =  \SYSTEMnt{t_{{\mathrm{1}}}}  \, \mathsf{in} \,  \SYSTEMnt{t_{{\mathrm{2}}}}   :  \SYSTEMnt{B} }{%
{\SYSTEMRenameRuleGradBoxletGen}{}%
}}

\newcommand{\SYSTEMRenameRuleLinEqbeta}[0]{\SYSTEMdrulename{LinEqbeta}}
\newcommand{\SYSTEMdruleLinEqbeta}[1]{\SYSTEMdrule[#1]{%
}{
  \SYSTEMsym{(}   \lambda  \SYSTEMmv{x}  .  \SYSTEMnt{t_{{\mathrm{2}}}}   \SYSTEMsym{)} \,  \SYSTEMnt{t_{{\mathrm{1}}}}   \,\equiv_{\textsc{l} }\,   [  \SYSTEMnt{t_{{\mathrm{1}}}}  /  \SYSTEMmv{x}  ]  \SYSTEMnt{t_{{\mathrm{2}}}}  }{%
{\SYSTEMRenameRuleLinEqbeta}{}%
}}

\newcommand{\SYSTEMRenameRuleLinEqeta}[0]{\SYSTEMdrulename{LinEqeta}}
\newcommand{\SYSTEMdruleLinEqeta}[1]{\SYSTEMdrule[#1]{%
\SYSTEMpremise{  \SYSTEMmv{x}  \,\#\,  \SYSTEMnt{t}  }%
}{
   \lambda  \SYSTEMmv{x}  .    \SYSTEMnt{t} \,  \SYSTEMmv{x}      \,\equiv_{\textsc{l} }\,  \SYSTEMnt{t} }{%
{\SYSTEMRenameRuleLinEqeta}{}%
}}

\newcommand{\SYSTEMRenameRuleLinEqbetaBox}[0]{\SYSTEMdrulename{LinEqbetaBox}}
\newcommand{\SYSTEMdruleLinEqbetaBox}[1]{\SYSTEMdrule[#1]{%
}{
  \mathsf{let} \, \textcolor{coeffectColor}{[}  \SYSTEMmv{x}  \textcolor{coeffectColor}{]} =   \textcolor{coeffectColor}{[}  \SYSTEMnt{t_{{\mathrm{1}}}}  \textcolor{coeffectColor}{]}   \, \mathsf{in} \,  \SYSTEMnt{t_{{\mathrm{2}}}}   \,\equiv_{\textsc{l} }\,   [  \SYSTEMnt{t_{{\mathrm{1}}}}  /  \SYSTEMmv{x}  ]  \SYSTEMnt{t_{{\mathrm{2}}}}  }{%
{\SYSTEMRenameRuleLinEqbetaBox}{}%
}}

\newcommand{\SYSTEMRenameRuleLinEqetaBox}[0]{\SYSTEMdrulename{LinEqetaBox}}
\newcommand{\SYSTEMdruleLinEqetaBox}[1]{\SYSTEMdrule[#1]{%
}{
  \mathsf{let} \, \textcolor{coeffectColor}{[}  \SYSTEMmv{x}  \textcolor{coeffectColor}{]} =  \SYSTEMnt{t}  \, \mathsf{in} \,   \textcolor{coeffectColor}{[}  \SYSTEMmv{x}  \textcolor{coeffectColor}{]}    \,\equiv_{\textsc{l} }\,  \SYSTEMnt{t} }{%
{\SYSTEMRenameRuleLinEqetaBox}{}%
}}

\newcommand{\SYSTEMRenameRuleLinEqletCommOne}[0]{\SYSTEMdrulename{LinEqletComm1}}
\newcommand{\SYSTEMdruleLinEqletCommOne}[1]{\SYSTEMdrule[#1]{%
\SYSTEMpremise{  \SYSTEMmv{x}  \,\#\,  \SYSTEMnt{t_{{\mathrm{0}}}}  }%
}{
  \SYSTEMnt{t_{{\mathrm{0}}}} \,  \SYSTEMsym{(}   \mathsf{let} \, \textcolor{coeffectColor}{[}  \SYSTEMmv{x}  \textcolor{coeffectColor}{]} =  \SYSTEMnt{t_{{\mathrm{1}}}}  \, \mathsf{in} \,  \SYSTEMnt{t_{{\mathrm{2}}}}   \SYSTEMsym{)}   \,\equiv_{\textsc{l} }\,   \mathsf{let} \, \textcolor{coeffectColor}{[}  \SYSTEMmv{x}  \textcolor{coeffectColor}{]} =  \SYSTEMnt{t_{{\mathrm{1}}}}  \, \mathsf{in} \,  \SYSTEMsym{(}   \SYSTEMnt{t_{{\mathrm{0}}}} \,  \SYSTEMnt{t_{{\mathrm{2}}}}   \SYSTEMsym{)}  }{%
{\SYSTEMRenameRuleLinEqletCommOne}{}%
}}

\newcommand{\SYSTEMRenameRuleLinEqletCommTwo}[0]{\SYSTEMdrulename{LinEqletComm2}}
\newcommand{\SYSTEMdruleLinEqletCommTwo}[1]{\SYSTEMdrule[#1]{%
\SYSTEMpremise{  \SYSTEMmv{x}  \,\#\,  \SYSTEMnt{t_{{\mathrm{0}}}}  }%
}{
  \SYSTEMsym{(}   \mathsf{let} \, \textcolor{coeffectColor}{[}  \SYSTEMmv{x}  \textcolor{coeffectColor}{]} =  \SYSTEMnt{t_{{\mathrm{1}}}}  \, \mathsf{in} \,  \SYSTEMnt{t_{{\mathrm{2}}}}   \SYSTEMsym{)} \,  \SYSTEMnt{t_{{\mathrm{0}}}}   \,\equiv_{\textsc{l} }\,   \mathsf{let} \, \textcolor{coeffectColor}{[}  \SYSTEMmv{x}  \textcolor{coeffectColor}{]} =  \SYSTEMnt{t_{{\mathrm{1}}}}  \, \mathsf{in} \,  \SYSTEMsym{(}   \SYSTEMnt{t_{{\mathrm{2}}}} \,  \SYSTEMnt{t_{{\mathrm{0}}}}   \SYSTEMsym{)}  }{%
{\SYSTEMRenameRuleLinEqletCommTwo}{}%
}}

\newcommand{\SYSTEMRenameRuleLinEqletCommBox}[0]{\SYSTEMdrulename{LinEqletCommBox}}
\newcommand{\SYSTEMdruleLinEqletCommBox}[1]{\SYSTEMdrule[#1]{%
}{
  \mathsf{let} \, \textcolor{coeffectColor}{[}  \SYSTEMmv{x}  \textcolor{coeffectColor}{]} =   \textcolor{coeffectColor}{[}  \SYSTEMnt{t_{{\mathrm{1}}}}  \textcolor{coeffectColor}{]}   \, \mathsf{in} \,   \textcolor{coeffectColor}{[}  \SYSTEMnt{t_{{\mathrm{2}}}}  \textcolor{coeffectColor}{]}    \,\equiv_{\textsc{l} }\,   \textcolor{coeffectColor}{[}   \mathsf{let} \, \textcolor{coeffectColor}{[}  \SYSTEMmv{x}  \textcolor{coeffectColor}{]} =   \textcolor{coeffectColor}{[}  \SYSTEMnt{t_{{\mathrm{1}}}}  \textcolor{coeffectColor}{]}   \, \mathsf{in} \,  \SYSTEMnt{t_{{\mathrm{2}}}}   \textcolor{coeffectColor}{]}  }{%
{\SYSTEMRenameRuleLinEqletCommBox}{}%
}}

\newcommand{\SYSTEMRenameRuleLinEqcongApp}[0]{\SYSTEMdrulename{LinEqcongApp}}
\newcommand{\SYSTEMdruleLinEqcongApp}[1]{\SYSTEMdrule[#1]{%
\SYSTEMpremise{ \begin{array}{cc}   \SYSTEMnt{t_{{\mathrm{1}}}}  \,\equiv_{\textsc{l} }\,  \SYSTEMnt{t'_{{\mathrm{1}}}}   \; & \;   \SYSTEMnt{t_{{\mathrm{2}}}}  \,\equiv_{\textsc{l} }\,  \SYSTEMnt{t'_{{\mathrm{2}}}}   \end{array} }%
}{
  \SYSTEMnt{t_{{\mathrm{1}}}} \,  \SYSTEMnt{t_{{\mathrm{2}}}}   \,\equiv_{\textsc{l} }\,   \SYSTEMnt{t'_{{\mathrm{1}}}} \,  \SYSTEMnt{t'_{{\mathrm{2}}}}  }{%
{\SYSTEMRenameRuleLinEqcongApp}{}%
}}

\newcommand{\SYSTEMRenameRuleLinEqcongAbs}[0]{\SYSTEMdrulename{LinEqcongAbs}}
\newcommand{\SYSTEMdruleLinEqcongAbs}[1]{\SYSTEMdrule[#1]{%
\SYSTEMpremise{ \SYSTEMnt{t}  \,\equiv_{\textsc{l} }\,  \SYSTEMnt{t'} }%
}{
  \lambda  \SYSTEMmv{x}  .  \SYSTEMnt{t}   \,\equiv_{\textsc{l} }\,   \lambda  \SYSTEMmv{x}  .  \SYSTEMnt{t'}  }{%
{\SYSTEMRenameRuleLinEqcongAbs}{}%
}}

\newcommand{\SYSTEMRenameRuleLinEqcongPr}[0]{\SYSTEMdrulename{LinEqcongPr}}
\newcommand{\SYSTEMdruleLinEqcongPr}[1]{\SYSTEMdrule[#1]{%
\SYSTEMpremise{ \SYSTEMnt{t}  \,\equiv_{\textsc{l} }\,  \SYSTEMnt{t'} }%
}{
  \textcolor{coeffectColor}{[}  \SYSTEMnt{t}  \textcolor{coeffectColor}{]}   \,\equiv_{\textsc{l} }\,   \textcolor{coeffectColor}{[}  \SYSTEMnt{t'}  \textcolor{coeffectColor}{]}  }{%
{\SYSTEMRenameRuleLinEqcongPr}{}%
}}

\newcommand{\SYSTEMRenameRuleLinEqcongLet}[0]{\SYSTEMdrulename{LinEqcongLet}}
\newcommand{\SYSTEMdruleLinEqcongLet}[1]{\SYSTEMdrule[#1]{%
\SYSTEMpremise{ \begin{array}{cc}   \SYSTEMnt{t_{{\mathrm{1}}}}  \,\equiv_{\textsc{l} }\,  \SYSTEMnt{t'_{{\mathrm{1}}}}   \; & \;   \SYSTEMnt{t_{{\mathrm{2}}}}  \,\equiv_{\textsc{l} }\,  \SYSTEMnt{t'_{{\mathrm{2}}}}   \end{array} }%
}{
 \SYSTEMsym{(}   \mathsf{let} \, \textcolor{coeffectColor}{[}  \SYSTEMmv{x}  \textcolor{coeffectColor}{]} =  \SYSTEMnt{t_{{\mathrm{1}}}}  \, \mathsf{in} \,  \SYSTEMnt{t_{{\mathrm{2}}}}   \SYSTEMsym{)}  \,\equiv_{\textsc{l} }\,  \SYSTEMsym{(}   \mathsf{let} \, \textcolor{coeffectColor}{[}  \SYSTEMmv{x}  \textcolor{coeffectColor}{]} =  \SYSTEMnt{t'_{{\mathrm{1}}}}  \, \mathsf{in} \,  \SYSTEMnt{t_{{\mathrm{2}}}}   \SYSTEMsym{)} }{%
{\SYSTEMRenameRuleLinEqcongLet}{}%
}}

\newcommand{\SYSTEMRenameRuleLinEqcongUnitE}[0]{\SYSTEMdrulename{LinEqcongUnitE}}
\newcommand{\SYSTEMdruleLinEqcongUnitE}[1]{\SYSTEMdrule[#1]{%
\SYSTEMpremise{ \begin{array}{cc}   \SYSTEMnt{t_{{\mathrm{1}}}}  \,\equiv_{\textsc{l} }\,  \SYSTEMnt{t'_{{\mathrm{1}}}}   \; & \;   \SYSTEMnt{t_{{\mathrm{2}}}}  \,\equiv_{\textsc{l} }\,  \SYSTEMnt{t'_{{\mathrm{2}}}}   \end{array} }%
}{
  \mathsf{let} \, \langle \rangle =  \SYSTEMnt{t_{{\mathrm{1}}}}  \, \mathsf{in} \,  \SYSTEMnt{t_{{\mathrm{2}}}}   \,\equiv_{\textsc{l} }\,   \mathsf{let} \, \langle \rangle =  \SYSTEMnt{t'_{{\mathrm{1}}}}  \, \mathsf{in} \,  \SYSTEMnt{t'_{{\mathrm{2}}}}  }{%
{\SYSTEMRenameRuleLinEqcongUnitE}{}%
}}

\newcommand{\SYSTEMRenameRuleLinEqbetaUnit}[0]{\SYSTEMdrulename{LinEqbetaUnit}}
\newcommand{\SYSTEMdruleLinEqbetaUnit}[1]{\SYSTEMdrule[#1]{%
}{
  \mathsf{let} \, \langle \rangle =   \langle \rangle   \, \mathsf{in} \,  \SYSTEMnt{t}   \,\equiv_{\textsc{l} }\,  \SYSTEMnt{t} }{%
{\SYSTEMRenameRuleLinEqbetaUnit}{}%
}}

\newcommand{\SYSTEMRenameRuleLinEqetaUnit}[0]{\SYSTEMdrulename{LinEqetaUnit}}
\newcommand{\SYSTEMdruleLinEqetaUnit}[1]{\SYSTEMdrule[#1]{%
}{
  \mathsf{let} \, \langle \rangle =  \SYSTEMnt{t}  \, \mathsf{in} \,   \langle \rangle    \,\equiv_{\textsc{l} }\,   \langle \rangle  }{%
{\SYSTEMRenameRuleLinEqetaUnit}{}%
}}

\newcommand{\SYSTEMRenameRuleLinEqcongProdE}[0]{\SYSTEMdrulename{LinEqcongProdE}}
\newcommand{\SYSTEMdruleLinEqcongProdE}[1]{\SYSTEMdrule[#1]{%
\SYSTEMpremise{ \begin{array}{cc}   \SYSTEMnt{t_{{\mathrm{1}}}}  \,\equiv_{\textsc{l} }\,  \SYSTEMnt{t_{{\mathrm{1}}}}   \; & \;   \SYSTEMnt{t_{{\mathrm{2}}}}  \,\equiv_{\textsc{l} }\,  \SYSTEMnt{t'_{{\mathrm{2}}}}   \end{array} }%
}{
 \SYSTEMsym{(}   \mathsf{let} \, \langle  \SYSTEMmv{x} ,  \SYSTEMmv{y}  \rangle =  \SYSTEMnt{t_{{\mathrm{1}}}}  \, \mathsf{in} \,  \SYSTEMnt{t_{{\mathrm{2}}}}   \SYSTEMsym{)}  \,\equiv_{\textsc{l} }\,  \SYSTEMsym{(}   \mathsf{let} \, \langle  \SYSTEMmv{x} ,  \SYSTEMmv{y}  \rangle =  \SYSTEMnt{t'_{{\mathrm{1}}}}  \, \mathsf{in} \,  \SYSTEMnt{t'_{{\mathrm{2}}}}   \SYSTEMsym{)} }{%
{\SYSTEMRenameRuleLinEqcongProdE}{}%
}}

\newcommand{\SYSTEMRenameRuleLinEqcongProdI}[0]{\SYSTEMdrulename{LinEqcongProdI}}
\newcommand{\SYSTEMdruleLinEqcongProdI}[1]{\SYSTEMdrule[#1]{%
\SYSTEMpremise{ \begin{array}{cc}   \SYSTEMnt{t_{{\mathrm{1}}}}  \,\equiv_{\textsc{l} }\,  \SYSTEMnt{t_{{\mathrm{1}}}}   \; & \;   \SYSTEMnt{t_{{\mathrm{2}}}}  \,\equiv_{\textsc{l} }\,  \SYSTEMnt{t'_{{\mathrm{2}}}}   \end{array} }%
}{
  \langle  \SYSTEMnt{t_{{\mathrm{1}}}} ,  \SYSTEMnt{t_{{\mathrm{2}}}}  \rangle   \,\equiv_{\textsc{l} }\,   \langle  \SYSTEMnt{t'_{{\mathrm{1}}}} ,  \SYSTEMnt{t'_{{\mathrm{2}}}}  \rangle  }{%
{\SYSTEMRenameRuleLinEqcongProdI}{}%
}}

\newcommand{\SYSTEMRenameRuleLinEqbetaProd}[0]{\SYSTEMdrulename{LinEqbetaProd}}
\newcommand{\SYSTEMdruleLinEqbetaProd}[1]{\SYSTEMdrule[#1]{%
}{
 \SYSTEMsym{(}   \mathsf{let} \, \langle  \SYSTEMmv{x} ,  \SYSTEMmv{y}  \rangle =   \langle  \SYSTEMnt{t_{{\mathrm{1}}}} ,  \SYSTEMnt{t_{{\mathrm{2}}}}  \rangle   \, \mathsf{in} \,  \SYSTEMnt{t}   \SYSTEMsym{)}  \,\equiv_{\textsc{l} }\,   [  \SYSTEMnt{t_{{\mathrm{1}}}}  /  \SYSTEMmv{x}  ]   [  \SYSTEMnt{t_{{\mathrm{2}}}}  /  \SYSTEMmv{y}  ]  \SYSTEMnt{t}   }{%
{\SYSTEMRenameRuleLinEqbetaProd}{}%
}}

\newcommand{\SYSTEMRenameRuleLinEqetaProd}[0]{\SYSTEMdrulename{LinEqetaProd}}
\newcommand{\SYSTEMdruleLinEqetaProd}[1]{\SYSTEMdrule[#1]{%
}{
  \mathsf{let} \, \langle  \SYSTEMmv{x} ,  \SYSTEMmv{y}  \rangle =  \SYSTEMnt{t}  \, \mathsf{in} \,   \langle  \SYSTEMmv{x} ,  \SYSTEMmv{y}  \rangle    \,\equiv_{\textsc{l} }\,  \SYSTEMnt{t} }{%
{\SYSTEMRenameRuleLinEqetaProd}{}%
}}

\newcommand{\SYSTEMRenameRuleLinEqcongSumE}[0]{\SYSTEMdrulename{LinEqcongSumE}}
\newcommand{\SYSTEMdruleLinEqcongSumE}[1]{\SYSTEMdrule[#1]{%
\SYSTEMpremise{ \begin{array}{cc}    \begin{array}{cc}   \SYSTEMnt{t}  \,\equiv_{\textsc{l} }\,  \SYSTEMnt{t'}   \; & \;   \SYSTEMnt{t_{{\mathrm{1}}}}  \,\equiv_{\textsc{l} }\,  \SYSTEMnt{t_{{\mathrm{1}}}}   \end{array}    \; & \;   \SYSTEMnt{t_{{\mathrm{2}}}}  \,\equiv_{\textsc{l} }\,  \SYSTEMnt{t'_{{\mathrm{2}}}}   \end{array} }%
}{
 \SYSTEMsym{(}   \mathsf{case} \,  \SYSTEMnt{t}  \, \mathsf{of} \, \{ \mathsf{inj1} \,  \SYSTEMmv{x}  \rightarrow  \SYSTEMnt{t_{{\mathrm{1}}}}  ; \, \mathsf{inj2} \,  \SYSTEMmv{y}  \rightarrow  \SYSTEMnt{t_{{\mathrm{2}}}}  \}   \SYSTEMsym{)}  \,\equiv_{\textsc{l} }\,  \SYSTEMsym{(}   \mathsf{case} \,  \SYSTEMnt{t'}  \, \mathsf{of} \, \{ \mathsf{inj1} \,  \SYSTEMmv{x}  \rightarrow  \SYSTEMnt{t'_{{\mathrm{1}}}}  ; \, \mathsf{inj2} \,  \SYSTEMmv{y}  \rightarrow  \SYSTEMnt{t'_{{\mathrm{2}}}}  \}   \SYSTEMsym{)} }{%
{\SYSTEMRenameRuleLinEqcongSumE}{}%
}}

\newcommand{\SYSTEMRenameRuleLinEqcongSumIOne}[0]{\SYSTEMdrulename{LinEqcongSumI1}}
\newcommand{\SYSTEMdruleLinEqcongSumIOne}[1]{\SYSTEMdrule[#1]{%
\SYSTEMpremise{ \SYSTEMnt{t}  \,\equiv_{\textsc{l} }\,  \SYSTEMnt{t'} }%
}{
  \mathsf{inj}_1 \,  \SYSTEMnt{t}   \,\equiv_{\textsc{l} }\,   \mathsf{inj}_1 \,  \SYSTEMnt{t'}  }{%
{\SYSTEMRenameRuleLinEqcongSumIOne}{}%
}}

\newcommand{\SYSTEMRenameRuleLinEqcongSumITwo}[0]{\SYSTEMdrulename{LinEqcongSumI2}}
\newcommand{\SYSTEMdruleLinEqcongSumITwo}[1]{\SYSTEMdrule[#1]{%
\SYSTEMpremise{ \SYSTEMnt{t}  \,\equiv_{\textsc{l} }\,  \SYSTEMnt{t'} }%
}{
  \mathsf{inj}_2 \,  \SYSTEMnt{t}   \,\equiv_{\textsc{l} }\,   \mathsf{inj}_2 \,  \SYSTEMnt{t'}  }{%
{\SYSTEMRenameRuleLinEqcongSumITwo}{}%
}}

\newcommand{\SYSTEMRenameRuleLinEqbetaSumOne}[0]{\SYSTEMdrulename{LinEqbetaSum1}}
\newcommand{\SYSTEMdruleLinEqbetaSumOne}[1]{\SYSTEMdrule[#1]{%
}{
 \SYSTEMsym{(}   \mathsf{case} \,   \mathsf{inj}_1 \,  \SYSTEMnt{t}   \, \mathsf{of} \, \{ \mathsf{inj1} \,  \SYSTEMmv{x}  \rightarrow  \SYSTEMnt{t_{{\mathrm{1}}}}  ; \, \mathsf{inj2} \,  \SYSTEMmv{y}  \rightarrow  \SYSTEMnt{t_{{\mathrm{2}}}}  \}   \SYSTEMsym{)}  \,\equiv_{\textsc{l} }\,   [  \SYSTEMnt{t}  /  \SYSTEMmv{x}  ]  \SYSTEMnt{t_{{\mathrm{1}}}}  }{%
{\SYSTEMRenameRuleLinEqbetaSumOne}{}%
}}

\newcommand{\SYSTEMRenameRuleLinEqbetaSumTwo}[0]{\SYSTEMdrulename{LinEqbetaSum2}}
\newcommand{\SYSTEMdruleLinEqbetaSumTwo}[1]{\SYSTEMdrule[#1]{%
}{
 \SYSTEMsym{(}   \mathsf{case} \,   \mathsf{inj}_2 \,  \SYSTEMnt{t}   \, \mathsf{of} \, \{ \mathsf{inj1} \,  \SYSTEMmv{x}  \rightarrow  \SYSTEMnt{t_{{\mathrm{1}}}}  ; \, \mathsf{inj2} \,  \SYSTEMmv{y}  \rightarrow  \SYSTEMnt{t_{{\mathrm{2}}}}  \}   \SYSTEMsym{)}  \,\equiv_{\textsc{l} }\,   [  \SYSTEMnt{t}  /  \SYSTEMmv{y}  ]  \SYSTEMnt{t_{{\mathrm{2}}}}  }{%
{\SYSTEMRenameRuleLinEqbetaSumTwo}{}%
}}

\newcommand{\SYSTEMRenameRuleLinEqetaSum}[0]{\SYSTEMdrulename{LinEqetaSum}}
\newcommand{\SYSTEMdruleLinEqetaSum}[1]{\SYSTEMdrule[#1]{%
}{
 \SYSTEMsym{(}   \mathsf{case} \,  \SYSTEMnt{t}  \, \mathsf{of} \, \{ \mathsf{inj1} \,  \SYSTEMmv{x}  \rightarrow   \mathsf{inj}_1 \,  \SYSTEMmv{x}   ; \, \mathsf{inj2} \,  \SYSTEMmv{y}  \rightarrow   \mathsf{inj}_2 \,  \SYSTEMmv{y}   \}   \SYSTEMsym{)}  \,\equiv_{\textsc{l} }\,  \SYSTEMnt{t} }{%
{\SYSTEMRenameRuleLinEqetaSum}{}%
}}

\newcommand{\SYSTEMRenameRuleLinEqpushProdCong}[0]{\SYSTEMdrulename{LinEqpushProdCong}}
\newcommand{\SYSTEMdruleLinEqpushProdCong}[1]{\SYSTEMdrule[#1]{%
\SYSTEMpremise{ \SYSTEMnt{t}  \,\equiv_{\textsc{l} }\,  \SYSTEMnt{t'} }%
}{
  \textsf{push}_\otimes  \SYSTEMnt{t}   \,\equiv_{\textsc{l} }\,   \textsf{push}_\otimes  \SYSTEMnt{t'}  }{%
{\SYSTEMRenameRuleLinEqpushProdCong}{}%
}}

\newcommand{\SYSTEMRenameRuleLinEqpushProdBeta}[0]{\SYSTEMdrulename{LinEqpushProdBeta}}
\newcommand{\SYSTEMdruleLinEqpushProdBeta}[1]{\SYSTEMdrule[#1]{%
}{
  \textsf{push}_\otimes   \textcolor{coeffectColor}{[}   \langle  \SYSTEMnt{t_{{\mathrm{1}}}} ,  \SYSTEMnt{t_{{\mathrm{2}}}}  \rangle   \textcolor{coeffectColor}{]}    \,\equiv_{\textsc{l} }\,   \langle   \textcolor{coeffectColor}{[}  \SYSTEMnt{t_{{\mathrm{1}}}}  \textcolor{coeffectColor}{]}  ,   \textcolor{coeffectColor}{[}  \SYSTEMnt{t_{{\mathrm{2}}}}  \textcolor{coeffectColor}{]}   \rangle  }{%
{\SYSTEMRenameRuleLinEqpushProdBeta}{}%
}}

\newcommand{\SYSTEMRenameRuleLinEqpushProdEta}[0]{\SYSTEMdrulename{LinEqpushProdEta}}
\newcommand{\SYSTEMdruleLinEqpushProdEta}[1]{\SYSTEMdrule[#1]{%
}{
  \mathsf{let} \, \langle  \SYSTEMmv{x'} ,  \SYSTEMmv{y'}  \rangle =   \textsf{push}_\otimes  \SYSTEMnt{t}   \, \mathsf{in} \,   \mathsf{let} \, \textcolor{coeffectColor}{[}  \SYSTEMmv{x}  \textcolor{coeffectColor}{]} =  \SYSTEMmv{x'}  \, \mathsf{in} \,   \mathsf{let} \, \textcolor{coeffectColor}{[}  \SYSTEMmv{y}  \textcolor{coeffectColor}{]} =  \SYSTEMmv{y'}  \, \mathsf{in} \,   \textcolor{coeffectColor}{[}   \langle  \SYSTEMmv{x} ,  \SYSTEMmv{y}  \rangle   \textcolor{coeffectColor}{]}      \,\equiv_{\textsc{l} }\,  \SYSTEMnt{t} }{%
{\SYSTEMRenameRuleLinEqpushProdEta}{}%
}}

\newcommand{\SYSTEMRenameRuleLinEqpushSumCong}[0]{\SYSTEMdrulename{LinEqpushSumCong}}
\newcommand{\SYSTEMdruleLinEqpushSumCong}[1]{\SYSTEMdrule[#1]{%
\SYSTEMpremise{ \SYSTEMnt{t}  \,\equiv_{\textsc{l} }\,  \SYSTEMnt{t'} }%
}{
  \textsf{push}_\oplus  \SYSTEMnt{t}   \,\equiv_{\textsc{l} }\,   \textsf{push}_\oplus  \SYSTEMnt{t'}  }{%
{\SYSTEMRenameRuleLinEqpushSumCong}{}%
}}

\newcommand{\SYSTEMRenameRuleLinEqpushSumBetaOne}[0]{\SYSTEMdrulename{LinEqpushSumBeta1}}
\newcommand{\SYSTEMdruleLinEqpushSumBetaOne}[1]{\SYSTEMdrule[#1]{%
}{
  \textsf{push}_\oplus   \textcolor{coeffectColor}{[}   \mathsf{inj}_1 \,  \SYSTEMnt{t}   \textcolor{coeffectColor}{]}    \,\equiv_{\textsc{l} }\,   \mathsf{inj}_1 \,   \textcolor{coeffectColor}{[}  \SYSTEMnt{t}  \textcolor{coeffectColor}{]}   }{%
{\SYSTEMRenameRuleLinEqpushSumBetaOne}{}%
}}

\newcommand{\SYSTEMRenameRuleLinEqpushSumBetaTwo}[0]{\SYSTEMdrulename{LinEqpushSumBeta2}}
\newcommand{\SYSTEMdruleLinEqpushSumBetaTwo}[1]{\SYSTEMdrule[#1]{%
}{
  \textsf{push}_\oplus   \textcolor{coeffectColor}{[}   \mathsf{inj}_2 \,  \SYSTEMnt{t}   \textcolor{coeffectColor}{]}    \,\equiv_{\textsc{l} }\,   \mathsf{inj}_2 \,   \textcolor{coeffectColor}{[}  \SYSTEMnt{t}  \textcolor{coeffectColor}{]}   }{%
{\SYSTEMRenameRuleLinEqpushSumBetaTwo}{}%
}}

\newcommand{\SYSTEMRenameRuleLinEqpushSumEta}[0]{\SYSTEMdrulename{LinEqpushSumEta}}
\newcommand{\SYSTEMdruleLinEqpushSumEta}[1]{\SYSTEMdrule[#1]{%
}{
  \mathsf{case} \,  \SYSTEMsym{(}   \textsf{push}_\oplus  \SYSTEMnt{t}   \SYSTEMsym{)}  \, \mathsf{of} \, \{ \mathsf{inj1} \,  \SYSTEMmv{x}  \rightarrow   \mathsf{let} \, \textcolor{coeffectColor}{[}  \SYSTEMmv{x'}  \textcolor{coeffectColor}{]} =  \SYSTEMmv{x}  \, \mathsf{in} \,   \textcolor{coeffectColor}{[}   \mathsf{inj}_1 \,  \SYSTEMmv{x'}   \textcolor{coeffectColor}{]}    ; \, \mathsf{inj2} \,  \SYSTEMmv{y}  \rightarrow   \mathsf{let} \, \textcolor{coeffectColor}{[}  \SYSTEMmv{y'}  \textcolor{coeffectColor}{]} =  \SYSTEMmv{y}  \, \mathsf{in} \,   \textcolor{coeffectColor}{[}   \mathsf{inj}_2 \,  \SYSTEMmv{y'}   \textcolor{coeffectColor}{]}    \}   \,\equiv_{\textsc{l} }\,  \SYSTEMnt{t} }{%
{\SYSTEMRenameRuleLinEqpushSumEta}{}%
}}

\newcommand{\SYSTEMRenameRuleLinEqpushUnitCong}[0]{\SYSTEMdrulename{LinEqpushUnitCong}}
\newcommand{\SYSTEMdruleLinEqpushUnitCong}[1]{\SYSTEMdrule[#1]{%
\SYSTEMpremise{ \SYSTEMnt{t}  \,\equiv_{\textsc{l} }\,  \SYSTEMnt{t'} }%
}{
  \textsf{push}_{\mathrm{unit} }  \SYSTEMnt{t}   \,\equiv_{\textsc{l} }\,   \textsf{push}_{\mathrm{unit} }  \SYSTEMnt{t'}  }{%
{\SYSTEMRenameRuleLinEqpushUnitCong}{}%
}}

\newcommand{\SYSTEMRenameRuleLinEqpushUnitBeta}[0]{\SYSTEMdrulename{LinEqpushUnitBeta}}
\newcommand{\SYSTEMdruleLinEqpushUnitBeta}[1]{\SYSTEMdrule[#1]{%
}{
  \textsf{push}_{\mathrm{unit} }   \textcolor{coeffectColor}{[}   \langle \rangle   \textcolor{coeffectColor}{]}    \,\equiv_{\textsc{l} }\,   \langle \rangle  }{%
{\SYSTEMRenameRuleLinEqpushUnitBeta}{}%
}}

\newcommand{\SYSTEMRenameRuleGradEqbeta}[0]{\SYSTEMdrulename{GradEqbeta}}
\newcommand{\SYSTEMdruleGradEqbeta}[1]{\SYSTEMdrule[#1]{%
}{
  \SYSTEMsym{(}   \lambda  \SYSTEMmv{x}  .  \SYSTEMnt{t_{{\mathrm{2}}}}   \SYSTEMsym{)} \,  \SYSTEMnt{t_{{\mathrm{1}}}}   \equiv_{\textsc{g} }   [  \SYSTEMnt{t_{{\mathrm{1}}}}  /  \SYSTEMmv{x}  ]  \SYSTEMnt{t_{{\mathrm{2}}}}  }{%
{\SYSTEMRenameRuleGradEqbeta}{}%
}}

\newcommand{\SYSTEMRenameRuleGradEqeta}[0]{\SYSTEMdrulename{GradEqeta}}
\newcommand{\SYSTEMdruleGradEqeta}[1]{\SYSTEMdrule[#1]{%
\SYSTEMpremise{  \SYSTEMmv{x}  \,\#\,  \SYSTEMnt{t}  }%
}{
   \lambda  \SYSTEMmv{x}  .    \SYSTEMnt{t} \,  \SYSTEMmv{x}      \equiv_{\textsc{g} }  \SYSTEMnt{t} }{%
{\SYSTEMRenameRuleGradEqeta}{}%
}}

\newcommand{\SYSTEMRenameRuleGradEqcongApp}[0]{\SYSTEMdrulename{GradEqcongApp}}
\newcommand{\SYSTEMdruleGradEqcongApp}[1]{\SYSTEMdrule[#1]{%
\SYSTEMpremise{ \begin{array}{cc}   \SYSTEMnt{t_{{\mathrm{1}}}}  \equiv_{\textsc{g} }  \SYSTEMnt{t'_{{\mathrm{1}}}}   \; & \;   \SYSTEMnt{t_{{\mathrm{2}}}}  \equiv_{\textsc{g} }  \SYSTEMnt{t'_{{\mathrm{2}}}}   \end{array} }%
}{
  \SYSTEMnt{t_{{\mathrm{1}}}} \,  \SYSTEMnt{t_{{\mathrm{2}}}}   \equiv_{\textsc{g} }   \SYSTEMnt{t'_{{\mathrm{1}}}} \,  \SYSTEMnt{t'_{{\mathrm{2}}}}  }{%
{\SYSTEMRenameRuleGradEqcongApp}{}%
}}

\newcommand{\SYSTEMRenameRuleGradEqcongAbs}[0]{\SYSTEMdrulename{GradEqcongAbs}}
\newcommand{\SYSTEMdruleGradEqcongAbs}[1]{\SYSTEMdrule[#1]{%
\SYSTEMpremise{ \SYSTEMnt{t}  \equiv_{\textsc{g} }  \SYSTEMnt{t'} }%
}{
  \lambda  \SYSTEMmv{x}  .  \SYSTEMnt{t}   \equiv_{\textsc{g} }   \lambda  \SYSTEMmv{x}  .  \SYSTEMnt{t'}  }{%
{\SYSTEMRenameRuleGradEqcongAbs}{}%
}}

\newcommand{\SYSTEMRenameRuleGradEqcongUnitE}[0]{\SYSTEMdrulename{GradEqcongUnitE}}
\newcommand{\SYSTEMdruleGradEqcongUnitE}[1]{\SYSTEMdrule[#1]{%
\SYSTEMpremise{ \begin{array}{cc}   \SYSTEMnt{t_{{\mathrm{1}}}}  \equiv_{\textsc{g} }  \SYSTEMnt{t'_{{\mathrm{1}}}}   \; & \;   \SYSTEMnt{t_{{\mathrm{2}}}}  \equiv_{\textsc{g} }  \SYSTEMnt{t'_{{\mathrm{2}}}}   \end{array} }%
}{
  \mathsf{let} \, \langle \rangle =  \SYSTEMnt{t_{{\mathrm{1}}}}  \, \mathsf{in} \,  \SYSTEMnt{t_{{\mathrm{2}}}}   \equiv_{\textsc{g} }   \mathsf{let} \, \langle \rangle =  \SYSTEMnt{t'_{{\mathrm{1}}}}  \, \mathsf{in} \,  \SYSTEMnt{t'_{{\mathrm{2}}}}  }{%
{\SYSTEMRenameRuleGradEqcongUnitE}{}%
}}

\newcommand{\SYSTEMRenameRuleGradEqbetaUnit}[0]{\SYSTEMdrulename{GradEqbetaUnit}}
\newcommand{\SYSTEMdruleGradEqbetaUnit}[1]{\SYSTEMdrule[#1]{%
}{
  \mathsf{let} \, \langle \rangle =   \langle \rangle   \, \mathsf{in} \,  \SYSTEMnt{t}   \equiv_{\textsc{g} }  \SYSTEMnt{t} }{%
{\SYSTEMRenameRuleGradEqbetaUnit}{}%
}}

\newcommand{\SYSTEMRenameRuleGradEqetaUnit}[0]{\SYSTEMdrulename{GradEqetaUnit}}
\newcommand{\SYSTEMdruleGradEqetaUnit}[1]{\SYSTEMdrule[#1]{%
}{
  \mathsf{let} \, \langle \rangle =  \SYSTEMnt{t}  \, \mathsf{in} \,   \langle \rangle    \equiv_{\textsc{g} }   \langle \rangle  }{%
{\SYSTEMRenameRuleGradEqetaUnit}{}%
}}

\newcommand{\SYSTEMRenameRuleGradEqcongProdE}[0]{\SYSTEMdrulename{GradEqcongProdE}}
\newcommand{\SYSTEMdruleGradEqcongProdE}[1]{\SYSTEMdrule[#1]{%
\SYSTEMpremise{ \begin{array}{cc}   \SYSTEMnt{t_{{\mathrm{1}}}}  \equiv_{\textsc{g} }  \SYSTEMnt{t_{{\mathrm{1}}}}   \; & \;   \SYSTEMnt{t_{{\mathrm{2}}}}  \equiv_{\textsc{g} }  \SYSTEMnt{t'_{{\mathrm{2}}}}   \end{array} }%
}{
 \SYSTEMsym{(}   \mathsf{let} \, \langle  \SYSTEMmv{x} ,  \SYSTEMmv{y}  \rangle =  \SYSTEMnt{t_{{\mathrm{1}}}}  \, \mathsf{in} \,  \SYSTEMnt{t_{{\mathrm{2}}}}   \SYSTEMsym{)}  \equiv_{\textsc{g} }  \SYSTEMsym{(}   \mathsf{let} \, \langle  \SYSTEMmv{x} ,  \SYSTEMmv{y}  \rangle =  \SYSTEMnt{t'_{{\mathrm{1}}}}  \, \mathsf{in} \,  \SYSTEMnt{t'_{{\mathrm{2}}}}   \SYSTEMsym{)} }{%
{\SYSTEMRenameRuleGradEqcongProdE}{}%
}}

\newcommand{\SYSTEMRenameRuleGradEqcongProdI}[0]{\SYSTEMdrulename{GradEqcongProdI}}
\newcommand{\SYSTEMdruleGradEqcongProdI}[1]{\SYSTEMdrule[#1]{%
\SYSTEMpremise{ \begin{array}{cc}   \SYSTEMnt{t_{{\mathrm{1}}}}  \equiv_{\textsc{g} }  \SYSTEMnt{t_{{\mathrm{1}}}}   \; & \;   \SYSTEMnt{t_{{\mathrm{2}}}}  \equiv_{\textsc{g} }  \SYSTEMnt{t'_{{\mathrm{2}}}}   \end{array} }%
}{
  \langle  \SYSTEMnt{t_{{\mathrm{1}}}} ,  \SYSTEMnt{t_{{\mathrm{2}}}}  \rangle   \equiv_{\textsc{g} }   \langle  \SYSTEMnt{t'_{{\mathrm{1}}}} ,  \SYSTEMnt{t'_{{\mathrm{2}}}}  \rangle  }{%
{\SYSTEMRenameRuleGradEqcongProdI}{}%
}}

\newcommand{\SYSTEMRenameRuleGradEqbetaProd}[0]{\SYSTEMdrulename{GradEqbetaProd}}
\newcommand{\SYSTEMdruleGradEqbetaProd}[1]{\SYSTEMdrule[#1]{%
}{
 \SYSTEMsym{(}   \mathsf{let} \, \langle  \SYSTEMmv{x} ,  \SYSTEMmv{y}  \rangle =   \langle  \SYSTEMnt{t_{{\mathrm{1}}}} ,  \SYSTEMnt{t_{{\mathrm{2}}}}  \rangle   \, \mathsf{in} \,  \SYSTEMnt{t}   \SYSTEMsym{)}  \equiv_{\textsc{g} }   [  \SYSTEMnt{t_{{\mathrm{1}}}}  /  \SYSTEMmv{x}  ]   [  \SYSTEMnt{t_{{\mathrm{2}}}}  /  \SYSTEMmv{y}  ]  \SYSTEMnt{t}   }{%
{\SYSTEMRenameRuleGradEqbetaProd}{}%
}}

\newcommand{\SYSTEMRenameRuleGradEqetaProd}[0]{\SYSTEMdrulename{GradEqetaProd}}
\newcommand{\SYSTEMdruleGradEqetaProd}[1]{\SYSTEMdrule[#1]{%
}{
  \mathsf{let} \, \langle  \SYSTEMmv{x} ,  \SYSTEMmv{y}  \rangle =  \SYSTEMnt{t}  \, \mathsf{in} \,   \langle  \SYSTEMmv{x} ,  \SYSTEMmv{y}  \rangle    \equiv_{\textsc{g} }  \SYSTEMnt{t} }{%
{\SYSTEMRenameRuleGradEqetaProd}{}%
}}

\newcommand{\SYSTEMRenameRuleGradEqcongSumE}[0]{\SYSTEMdrulename{GradEqcongSumE}}
\newcommand{\SYSTEMdruleGradEqcongSumE}[1]{\SYSTEMdrule[#1]{%
\SYSTEMpremise{ \begin{array}{cc}    \begin{array}{cc}   \SYSTEMnt{t}  \equiv_{\textsc{g} }  \SYSTEMnt{t'}   \; & \;   \SYSTEMnt{t_{{\mathrm{1}}}}  \equiv_{\textsc{g} }  \SYSTEMnt{t_{{\mathrm{1}}}}   \end{array}    \; & \;   \SYSTEMnt{t_{{\mathrm{2}}}}  \equiv_{\textsc{g} }  \SYSTEMnt{t'_{{\mathrm{2}}}}   \end{array} }%
}{
 \SYSTEMsym{(}   \mathsf{case} \,  \SYSTEMnt{t}  \, \mathsf{of} \, \{ \mathsf{inj1} \,  \SYSTEMmv{x}  \rightarrow  \SYSTEMnt{t_{{\mathrm{1}}}}  ; \, \mathsf{inj2} \,  \SYSTEMmv{y}  \rightarrow  \SYSTEMnt{t_{{\mathrm{2}}}}  \}   \SYSTEMsym{)}  \equiv_{\textsc{g} }  \SYSTEMsym{(}   \mathsf{case} \,  \SYSTEMnt{t'}  \, \mathsf{of} \, \{ \mathsf{inj1} \,  \SYSTEMmv{x}  \rightarrow  \SYSTEMnt{t'_{{\mathrm{1}}}}  ; \, \mathsf{inj2} \,  \SYSTEMmv{y}  \rightarrow  \SYSTEMnt{t'_{{\mathrm{2}}}}  \}   \SYSTEMsym{)} }{%
{\SYSTEMRenameRuleGradEqcongSumE}{}%
}}

\newcommand{\SYSTEMRenameRuleGradEqcongSumIOne}[0]{\SYSTEMdrulename{GradEqcongSumI1}}
\newcommand{\SYSTEMdruleGradEqcongSumIOne}[1]{\SYSTEMdrule[#1]{%
\SYSTEMpremise{ \SYSTEMnt{t}  \equiv_{\textsc{g} }  \SYSTEMnt{t'} }%
}{
  \mathsf{inj}_1 \,  \SYSTEMnt{t}   \equiv_{\textsc{g} }   \mathsf{inj}_1 \,  \SYSTEMnt{t'}  }{%
{\SYSTEMRenameRuleGradEqcongSumIOne}{}%
}}

\newcommand{\SYSTEMRenameRuleGradEqcongSumITwo}[0]{\SYSTEMdrulename{GradEqcongSumI2}}
\newcommand{\SYSTEMdruleGradEqcongSumITwo}[1]{\SYSTEMdrule[#1]{%
\SYSTEMpremise{ \SYSTEMnt{t}  \equiv_{\textsc{g} }  \SYSTEMnt{t'} }%
}{
  \mathsf{inj}_2 \,  \SYSTEMnt{t}   \equiv_{\textsc{g} }   \mathsf{inj}_2 \,  \SYSTEMnt{t'}  }{%
{\SYSTEMRenameRuleGradEqcongSumITwo}{}%
}}

\newcommand{\SYSTEMRenameRuleGradEqbetaSumOne}[0]{\SYSTEMdrulename{GradEqbetaSum1}}
\newcommand{\SYSTEMdruleGradEqbetaSumOne}[1]{\SYSTEMdrule[#1]{%
}{
 \SYSTEMsym{(}   \mathsf{case} \,   \mathsf{inj}_1 \,  \SYSTEMnt{t}   \, \mathsf{of} \, \{ \mathsf{inj1} \,  \SYSTEMmv{x}  \rightarrow  \SYSTEMnt{t_{{\mathrm{1}}}}  ; \, \mathsf{inj2} \,  \SYSTEMmv{y}  \rightarrow  \SYSTEMnt{t_{{\mathrm{2}}}}  \}   \SYSTEMsym{)}  \equiv_{\textsc{g} }   [  \SYSTEMnt{t}  /  \SYSTEMmv{x}  ]  \SYSTEMnt{t_{{\mathrm{1}}}}  }{%
{\SYSTEMRenameRuleGradEqbetaSumOne}{}%
}}

\newcommand{\SYSTEMRenameRuleGradEqbetaSumTwo}[0]{\SYSTEMdrulename{GradEqbetaSum2}}
\newcommand{\SYSTEMdruleGradEqbetaSumTwo}[1]{\SYSTEMdrule[#1]{%
}{
 \SYSTEMsym{(}   \mathsf{case} \,   \mathsf{inj}_2 \,  \SYSTEMnt{t}   \, \mathsf{of} \, \{ \mathsf{inj1} \,  \SYSTEMmv{x}  \rightarrow  \SYSTEMnt{t_{{\mathrm{1}}}}  ; \, \mathsf{inj2} \,  \SYSTEMmv{y}  \rightarrow  \SYSTEMnt{t_{{\mathrm{2}}}}  \}   \SYSTEMsym{)}  \equiv_{\textsc{g} }   [  \SYSTEMnt{t}  /  \SYSTEMmv{y}  ]  \SYSTEMnt{t_{{\mathrm{2}}}}  }{%
{\SYSTEMRenameRuleGradEqbetaSumTwo}{}%
}}

\newcommand{\SYSTEMRenameRuleGradEqetaSum}[0]{\SYSTEMdrulename{GradEqetaSum}}
\newcommand{\SYSTEMdruleGradEqetaSum}[1]{\SYSTEMdrule[#1]{%
}{
 \SYSTEMsym{(}   \mathsf{case} \,  \SYSTEMnt{t}  \, \mathsf{of} \, \{ \mathsf{inj1} \,  \SYSTEMmv{x}  \rightarrow   \mathsf{inj}_1 \,  \SYSTEMmv{x}   ; \, \mathsf{inj2} \,  \SYSTEMmv{y}  \rightarrow   \mathsf{inj}_2 \,  \SYSTEMmv{y}   \}   \SYSTEMsym{)}  \equiv_{\textsc{g} }  \SYSTEMnt{t} }{%
{\SYSTEMRenameRuleGradEqetaSum}{}%
}}

\newcommand{\SYSTEMRenameRuleGradMEqbetaBox}[0]{\SYSTEMdrulename{GradMEqbetaBox}}
\newcommand{\SYSTEMdruleGradMEqbetaBox}[1]{\SYSTEMdrule[#1]{%
}{
  \mathsf{let} \, \textcolor{coeffectColor}{[}  \SYSTEMmv{x}  \textcolor{coeffectColor}{]} =   \textcolor{coeffectColor}{[}  \SYSTEMnt{t_{{\mathrm{1}}}}  \textcolor{coeffectColor}{]}   \, \mathsf{in} \,  \SYSTEMnt{t_{{\mathrm{2}}}}   \,\equiv_{\textsc{g}_\Box}\,   [  \SYSTEMnt{t_{{\mathrm{1}}}}  /  \SYSTEMmv{x}  ]  \SYSTEMnt{t_{{\mathrm{2}}}}  }{%
{\SYSTEMRenameRuleGradMEqbetaBox}{}%
}}

\newcommand{\SYSTEMRenameRuleGradMEqetaBox}[0]{\SYSTEMdrulename{GradMEqetaBox}}
\newcommand{\SYSTEMdruleGradMEqetaBox}[1]{\SYSTEMdrule[#1]{%
}{
  \mathsf{let} \, \textcolor{coeffectColor}{[}  \SYSTEMmv{x}  \textcolor{coeffectColor}{]} =  \SYSTEMnt{t}  \, \mathsf{in} \,   \textcolor{coeffectColor}{[}  \SYSTEMmv{x}  \textcolor{coeffectColor}{]}    \,\equiv_{\textsc{g}_\Box}\,  \SYSTEMnt{t} }{%
{\SYSTEMRenameRuleGradMEqetaBox}{}%
}}

\newcommand{\SYSTEMRenameRuleGradMEqletCommOne}[0]{\SYSTEMdrulename{GradMEqletComm1}}
\newcommand{\SYSTEMdruleGradMEqletCommOne}[1]{\SYSTEMdrule[#1]{%
\SYSTEMpremise{  \SYSTEMmv{x}  \,\#\,  \SYSTEMnt{t_{{\mathrm{0}}}}  }%
}{
  \SYSTEMnt{t_{{\mathrm{0}}}} \,  \SYSTEMsym{(}   \mathsf{let} \, \textcolor{coeffectColor}{[}  \SYSTEMmv{x}  \textcolor{coeffectColor}{]} =  \SYSTEMnt{t_{{\mathrm{1}}}}  \, \mathsf{in} \,  \SYSTEMnt{t_{{\mathrm{2}}}}   \SYSTEMsym{)}   \,\equiv_{\textsc{g}_\Box}\,   \mathsf{let} \, \textcolor{coeffectColor}{[}  \SYSTEMmv{x}  \textcolor{coeffectColor}{]} =  \SYSTEMnt{t_{{\mathrm{1}}}}  \, \mathsf{in} \,  \SYSTEMsym{(}   \SYSTEMnt{t_{{\mathrm{0}}}} \,  \SYSTEMnt{t_{{\mathrm{2}}}}   \SYSTEMsym{)}  }{%
{\SYSTEMRenameRuleGradMEqletCommOne}{}%
}}

\newcommand{\SYSTEMRenameRuleGradMEqletCommTwo}[0]{\SYSTEMdrulename{GradMEqletComm2}}
\newcommand{\SYSTEMdruleGradMEqletCommTwo}[1]{\SYSTEMdrule[#1]{%
\SYSTEMpremise{  \SYSTEMmv{x}  \,\#\,  \SYSTEMnt{t_{{\mathrm{0}}}}  }%
}{
  \SYSTEMsym{(}   \mathsf{let} \, \textcolor{coeffectColor}{[}  \SYSTEMmv{x}  \textcolor{coeffectColor}{]} =  \SYSTEMnt{t_{{\mathrm{1}}}}  \, \mathsf{in} \,  \SYSTEMnt{t_{{\mathrm{2}}}}   \SYSTEMsym{)} \,  \SYSTEMnt{t_{{\mathrm{0}}}}   \,\equiv_{\textsc{g}_\Box}\,   \mathsf{let} \, \textcolor{coeffectColor}{[}  \SYSTEMmv{x}  \textcolor{coeffectColor}{]} =  \SYSTEMnt{t_{{\mathrm{1}}}}  \, \mathsf{in} \,  \SYSTEMsym{(}   \SYSTEMnt{t_{{\mathrm{2}}}} \,  \SYSTEMnt{t_{{\mathrm{0}}}}   \SYSTEMsym{)}  }{%
{\SYSTEMRenameRuleGradMEqletCommTwo}{}%
}}

\newcommand{\SYSTEMRenameRuleGradMEqletCommBox}[0]{\SYSTEMdrulename{GradMEqletCommBox}}
\newcommand{\SYSTEMdruleGradMEqletCommBox}[1]{\SYSTEMdrule[#1]{%
}{
  \mathsf{let} \, \textcolor{coeffectColor}{[}  \SYSTEMmv{x}  \textcolor{coeffectColor}{]} =   \textcolor{coeffectColor}{[}  \SYSTEMnt{t_{{\mathrm{1}}}}  \textcolor{coeffectColor}{]}   \, \mathsf{in} \,   \textcolor{coeffectColor}{[}  \SYSTEMnt{t_{{\mathrm{2}}}}  \textcolor{coeffectColor}{]}    \,\equiv_{\textsc{g}_\Box}\,   \textcolor{coeffectColor}{[}   \mathsf{let} \, \textcolor{coeffectColor}{[}  \SYSTEMmv{x}  \textcolor{coeffectColor}{]} =   \textcolor{coeffectColor}{[}  \SYSTEMnt{t_{{\mathrm{1}}}}  \textcolor{coeffectColor}{]}   \, \mathsf{in} \,  \SYSTEMnt{t_{{\mathrm{2}}}}   \textcolor{coeffectColor}{]}  }{%
{\SYSTEMRenameRuleGradMEqletCommBox}{}%
}}

\newcommand{\SYSTEMRenameRuleGradMEqcongPr}[0]{\SYSTEMdrulename{GradMEqcongPr}}
\newcommand{\SYSTEMdruleGradMEqcongPr}[1]{\SYSTEMdrule[#1]{%
\SYSTEMpremise{ \SYSTEMnt{t}  \,\equiv_{\textsc{g}_\Box}\,  \SYSTEMnt{t'} }%
}{
  \textcolor{coeffectColor}{[}  \SYSTEMnt{t}  \textcolor{coeffectColor}{]}   \,\equiv_{\textsc{g}_\Box}\,   \textcolor{coeffectColor}{[}  \SYSTEMnt{t'}  \textcolor{coeffectColor}{]}  }{%
{\SYSTEMRenameRuleGradMEqcongPr}{}%
}}

\newcommand{\SYSTEMRenameRuleGradMEqcongLet}[0]{\SYSTEMdrulename{GradMEqcongLet}}
\newcommand{\SYSTEMdruleGradMEqcongLet}[1]{\SYSTEMdrule[#1]{%
\SYSTEMpremise{ \begin{array}{cc}   \SYSTEMnt{t_{{\mathrm{1}}}}  \,\equiv_{\textsc{g}_\Box}\,  \SYSTEMnt{t'_{{\mathrm{1}}}}   \; & \;   \SYSTEMnt{t_{{\mathrm{2}}}}  \,\equiv_{\textsc{g}_\Box}\,  \SYSTEMnt{t'_{{\mathrm{2}}}}   \end{array} }%
}{
 \SYSTEMsym{(}   \mathsf{let} \, \textcolor{coeffectColor}{[}  \SYSTEMmv{x}  \textcolor{coeffectColor}{]} =  \SYSTEMnt{t_{{\mathrm{1}}}}  \, \mathsf{in} \,  \SYSTEMnt{t_{{\mathrm{2}}}}   \SYSTEMsym{)}  \,\equiv_{\textsc{g}_\Box}\,  \SYSTEMsym{(}   \mathsf{let} \, \textcolor{coeffectColor}{[}  \SYSTEMmv{x}  \textcolor{coeffectColor}{]} =  \SYSTEMnt{t_{{\mathrm{1}}}}  \, \mathsf{in} \,  \SYSTEMnt{t'_{{\mathrm{2}}}}   \SYSTEMsym{)} }{%
{\SYSTEMRenameRuleGradMEqcongLet}{}%
}}

\newcommand{\SYSTEMRenameRuleGradPolyEqbetaTy}[0]{\SYSTEMdrulename{GradPolyEqbetaTy}}
\newcommand{\SYSTEMdruleGradPolyEqbetaTy}[1]{\SYSTEMdrule[#1]{%
}{
  \SYSTEMsym{(}   \Lambda   \alpha   .  \SYSTEMnt{t}   \SYSTEMsym{)}  \text{@}  \SYSTEMnt{A}   \equiv_{\textsc{g}_\forall}   [  \SYSTEMnt{A}  /   \alpha   ]  \SYSTEMnt{t}  }{%
{\SYSTEMRenameRuleGradPolyEqbetaTy}{}%
}}

\newcommand{\SYSTEMRenameRuleGradPolyEqetaTy}[0]{\SYSTEMdrulename{GradPolyEqetaTy}}
\newcommand{\SYSTEMdruleGradPolyEqetaTy}[1]{\SYSTEMdrule[#1]{%
\SYSTEMpremise{  \SYSTEMmv{x}  \,\#\,  \SYSTEMnt{t}  }%
}{
   \Lambda   \alpha   .    \SYSTEMnt{t}  \text{@}   \alpha       \equiv_{\textsc{g}_\forall}  \SYSTEMnt{t} }{%
{\SYSTEMRenameRuleGradPolyEqetaTy}{}%
}}

\newcommand{\SYSTEMRenameRuleGradPolyEqtyAppCong}[0]{\SYSTEMdrulename{GradPolyEqtyAppCong}}
\newcommand{\SYSTEMdruleGradPolyEqtyAppCong}[1]{\SYSTEMdrule[#1]{%
\SYSTEMpremise{ \SYSTEMnt{t}  \equiv_{\textsc{g}_\forall}  \SYSTEMnt{t'} }%
}{
  \SYSTEMnt{t}  \text{@}  \SYSTEMnt{A}   \equiv_{\textsc{g}_\forall}   \SYSTEMnt{t'}  \text{@}  \SYSTEMnt{A}  }{%
{\SYSTEMRenameRuleGradPolyEqtyAppCong}{}%
}}

\newcommand{\SYSTEMRenameRuleGradPolyEqtyAbsCong}[0]{\SYSTEMdrulename{GradPolyEqtyAbsCong}}
\newcommand{\SYSTEMdruleGradPolyEqtyAbsCong}[1]{\SYSTEMdrule[#1]{%
\SYSTEMpremise{ \SYSTEMnt{t}  \equiv_{\textsc{g}_\forall}  \SYSTEMnt{t'} }%
}{
 \SYSTEMsym{(}   \Lambda   \alpha   .  \SYSTEMnt{t}   \SYSTEMsym{)}  \equiv_{\textsc{g}_\forall}  \SYSTEMsym{(}   \Lambda   \alpha   .  \SYSTEMnt{t'}   \SYSTEMsym{)} }{%
{\SYSTEMRenameRuleGradPolyEqtyAbsCong}{}%
}}

\newcommand{\SYSTEMRenameRuleGradPolytyAbs}[0]{\SYSTEMdrulename{GradPolytyAbs}}
\newcommand{\SYSTEMdruleGradPolytyAbs}[1]{\SYSTEMdrule[#1]{%
\SYSTEMpremise{ \begin{array}{cc}    \Delta  \vdash_{\textsc{g}_\forall }  \SYSTEMnt{t}  :  \SYSTEMnt{A}    \; & \;     \alpha   \,\#\,  \SYSTEMnt{t}    \end{array} }%
}{
 \Delta  \vdash_{\textsc{g}_\forall }   \Lambda   \alpha   .  \SYSTEMnt{t}   :   \forall   \alpha   .  \SYSTEMnt{A}  }{%
{\SYSTEMRenameRuleGradPolytyAbs}{}%
}}

\newcommand{\SYSTEMRenameRuleGradPolytyApp}[0]{\SYSTEMdrulename{GradPolytyApp}}
\newcommand{\SYSTEMdruleGradPolytyApp}[1]{\SYSTEMdrule[#1]{%
\SYSTEMpremise{ \Delta  \vdash_{\textsc{g}_\forall }  \SYSTEMnt{t}  :   \forall   \alpha   .  \SYSTEMnt{A}  }%
}{
 \Delta  \vdash_{\textsc{g}_\forall }   \SYSTEMnt{t}  \text{@}  \SYSTEMnt{B}   :   [  \SYSTEMnt{B}  /   \alpha   ]  \SYSTEMnt{A}  }{%
{\SYSTEMRenameRuleGradPolytyApp}{}%
}}

\newcommand{\SYSTEMRenameRuleSemLinbeta}[0]{\SYSTEMdrulename{SemLinbeta}}
\newcommand{\SYSTEMdruleSemLinbeta}[1]{\SYSTEMdrule[#1]{%
}{
  \SYSTEMsym{(}   \lambda  \SYSTEMmv{x}  .  \SYSTEMnt{t_{{\mathrm{2}}}}   \SYSTEMsym{)} \,  \SYSTEMnt{t_{{\mathrm{1}}}}   \rightsquigarrow_{\textsc{l} }   [  \SYSTEMnt{t_{{\mathrm{1}}}}  /  \SYSTEMmv{x}  ]  \SYSTEMnt{t_{{\mathrm{2}}}}  }{%
{\SYSTEMRenameRuleSemLinbeta}{}%
}}

\newcommand{\SYSTEMRenameRuleSemLincongAppL}[0]{\SYSTEMdrulename{SemLincongAppL}}
\newcommand{\SYSTEMdruleSemLincongAppL}[1]{\SYSTEMdrule[#1]{%
\SYSTEMpremise{ \SYSTEMnt{t_{{\mathrm{1}}}}  \rightsquigarrow_{\textsc{l} }  \SYSTEMnt{t'_{{\mathrm{1}}}} }%
}{
  \SYSTEMnt{t_{{\mathrm{1}}}} \,  \SYSTEMnt{t_{{\mathrm{2}}}}   \rightsquigarrow_{\textsc{l} }   \SYSTEMnt{t'_{{\mathrm{1}}}} \,  \SYSTEMnt{t_{{\mathrm{2}}}}  }{%
{\SYSTEMRenameRuleSemLincongAppL}{}%
}}

\newcommand{\SYSTEMRenameRuleSemLinbetaBox}[0]{\SYSTEMdrulename{SemLinbetaBox}}
\newcommand{\SYSTEMdruleSemLinbetaBox}[1]{\SYSTEMdrule[#1]{%
}{
  \mathsf{let} \, \textcolor{coeffectColor}{[}  \SYSTEMmv{x}  \textcolor{coeffectColor}{]} =   \textcolor{coeffectColor}{[}  \SYSTEMnt{t_{{\mathrm{1}}}}  \textcolor{coeffectColor}{]}   \, \mathsf{in} \,  \SYSTEMnt{t_{{\mathrm{2}}}}   \rightsquigarrow_{\textsc{l} }   [  \SYSTEMnt{t_{{\mathrm{1}}}}  /  \SYSTEMmv{x}  ]  \SYSTEMnt{t_{{\mathrm{2}}}}  }{%
{\SYSTEMRenameRuleSemLinbetaBox}{}%
}}

\newcommand{\SYSTEMRenameRuleSemLincongLetL}[0]{\SYSTEMdrulename{SemLincongLetL}}
\newcommand{\SYSTEMdruleSemLincongLetL}[1]{\SYSTEMdrule[#1]{%
\SYSTEMpremise{ \SYSTEMnt{t_{{\mathrm{1}}}}  \rightsquigarrow_{\textsc{l} }  \SYSTEMnt{t'_{{\mathrm{1}}}} }%
}{
  \mathsf{let} \, \textcolor{coeffectColor}{[}  \SYSTEMmv{x}  \textcolor{coeffectColor}{]} =  \SYSTEMnt{t_{{\mathrm{1}}}}  \, \mathsf{in} \,  \SYSTEMnt{t_{{\mathrm{2}}}}   \rightsquigarrow_{\textsc{l} }   \mathsf{let} \, \textcolor{coeffectColor}{[}  \SYSTEMmv{x}  \textcolor{coeffectColor}{]} =  \SYSTEMnt{t'_{{\mathrm{1}}}}  \, \mathsf{in} \,  \SYSTEMnt{t_{{\mathrm{2}}}}  }{%
{\SYSTEMRenameRuleSemLincongLetL}{}%
}}

\newcommand{\SYSTEMRenameRuleSemLincongCase}[0]{\SYSTEMdrulename{SemLincongCase}}
\newcommand{\SYSTEMdruleSemLincongCase}[1]{\SYSTEMdrule[#1]{%
\SYSTEMpremise{ \SYSTEMnt{t}  \rightsquigarrow_{\textsc{l} }  \SYSTEMnt{t'} }%
}{
  \mathsf{case} \,  \SYSTEMnt{t}  \, \mathsf{of} \, \{ \mathsf{inj1} \,  \SYSTEMmv{x}  \rightarrow  \SYSTEMnt{t_{{\mathrm{1}}}}  ; \, \mathsf{inj2} \,  \SYSTEMmv{y}  \rightarrow  \SYSTEMnt{t_{{\mathrm{2}}}}  \}   \rightsquigarrow_{\textsc{l} }   \mathsf{case} \,  \SYSTEMnt{t'}  \, \mathsf{of} \, \{ \mathsf{inj1} \,  \SYSTEMmv{x}  \rightarrow  \SYSTEMnt{t_{{\mathrm{1}}}}  ; \, \mathsf{inj2} \,  \SYSTEMmv{y}  \rightarrow  \SYSTEMnt{t_{{\mathrm{2}}}}  \}  }{%
{\SYSTEMRenameRuleSemLincongCase}{}%
}}

\newcommand{\SYSTEMRenameRuleSemLincaseInjOne}[0]{\SYSTEMdrulename{SemLincaseInj1}}
\newcommand{\SYSTEMdruleSemLincaseInjOne}[1]{\SYSTEMdrule[#1]{%
}{
  \mathsf{case} \,   \mathsf{inj}_1 \,  \SYSTEMnt{t}   \, \mathsf{of} \, \{ \mathsf{inj1} \,  \SYSTEMmv{x}  \rightarrow  \SYSTEMnt{t_{{\mathrm{1}}}}  ; \, \mathsf{inj2} \,  \SYSTEMmv{y}  \rightarrow  \SYSTEMnt{t_{{\mathrm{2}}}}  \}   \rightsquigarrow_{\textsc{l} }   [  \SYSTEMnt{t}  /  \SYSTEMmv{x}  ]  \SYSTEMnt{t_{{\mathrm{1}}}}  }{%
{\SYSTEMRenameRuleSemLincaseInjOne}{}%
}}

\newcommand{\SYSTEMRenameRuleSemLincaseInjTwo}[0]{\SYSTEMdrulename{SemLincaseInj2}}
\newcommand{\SYSTEMdruleSemLincaseInjTwo}[1]{\SYSTEMdrule[#1]{%
}{
  \mathsf{case} \,   \mathsf{inj}_2 \,  \SYSTEMnt{t}   \, \mathsf{of} \, \{ \mathsf{inj1} \,  \SYSTEMmv{x}  \rightarrow  \SYSTEMnt{t_{{\mathrm{1}}}}  ; \, \mathsf{inj2} \,  \SYSTEMmv{y}  \rightarrow  \SYSTEMnt{t_{{\mathrm{2}}}}  \}   \rightsquigarrow_{\textsc{l} }   [  \SYSTEMnt{t}  /  \SYSTEMmv{y}  ]  \SYSTEMnt{t_{{\mathrm{2}}}}  }{%
{\SYSTEMRenameRuleSemLincaseInjTwo}{}%
}}

\newcommand{\SYSTEMRenameRuleSemLinprodCong}[0]{\SYSTEMdrulename{SemLinprodCong}}
\newcommand{\SYSTEMdruleSemLinprodCong}[1]{\SYSTEMdrule[#1]{%
\SYSTEMpremise{ \SYSTEMnt{t_{{\mathrm{1}}}}  \rightsquigarrow_{\textsc{l} }  \SYSTEMnt{t'_{{\mathrm{1}}}} }%
}{
  \mathsf{let} \, \langle  \SYSTEMmv{x} ,  \SYSTEMmv{y}  \rangle =  \SYSTEMnt{t_{{\mathrm{1}}}}  \, \mathsf{in} \,  \SYSTEMnt{t_{{\mathrm{2}}}}   \rightsquigarrow_{\textsc{l} }   \mathsf{let} \, \langle  \SYSTEMmv{x} ,  \SYSTEMmv{y}  \rangle =  \SYSTEMnt{t'_{{\mathrm{1}}}}  \, \mathsf{in} \,  \SYSTEMnt{t_{{\mathrm{2}}}}  }{%
{\SYSTEMRenameRuleSemLinprodCong}{}%
}}

\newcommand{\SYSTEMRenameRuleSemLinprodBeta}[0]{\SYSTEMdrulename{SemLinprodBeta}}
\newcommand{\SYSTEMdruleSemLinprodBeta}[1]{\SYSTEMdrule[#1]{%
}{
  \mathsf{let} \, \langle  \SYSTEMmv{x} ,  \SYSTEMmv{y}  \rangle =   \langle  \SYSTEMnt{t_{{\mathrm{1}}}} ,  \SYSTEMnt{t_{{\mathrm{2}}}}  \rangle   \, \mathsf{in} \,  \SYSTEMnt{t_{{\mathrm{3}}}}   \rightsquigarrow_{\textsc{l} }   [  \SYSTEMnt{t_{{\mathrm{1}}}}  /  \SYSTEMmv{x}  ]   [  \SYSTEMnt{t_{{\mathrm{2}}}}  /  \SYSTEMmv{y}  ]  \SYSTEMnt{t_{{\mathrm{3}}}}   }{%
{\SYSTEMRenameRuleSemLinprodBeta}{}%
}}

\newcommand{\SYSTEMRenameRuleSemLinunitCong}[0]{\SYSTEMdrulename{SemLinunitCong}}
\newcommand{\SYSTEMdruleSemLinunitCong}[1]{\SYSTEMdrule[#1]{%
\SYSTEMpremise{ \SYSTEMnt{t_{{\mathrm{1}}}}  \rightsquigarrow_{\textsc{l} }  \SYSTEMnt{t'_{{\mathrm{1}}}} }%
}{
  \mathsf{let} \, \langle \rangle =  \SYSTEMnt{t_{{\mathrm{1}}}}  \, \mathsf{in} \,  \SYSTEMnt{t_{{\mathrm{2}}}}   \rightsquigarrow_{\textsc{l} }   \mathsf{let} \, \langle \rangle =  \SYSTEMnt{t'_{{\mathrm{1}}}}  \, \mathsf{in} \,  \SYSTEMnt{t_{{\mathrm{2}}}}  }{%
{\SYSTEMRenameRuleSemLinunitCong}{}%
}}

\newcommand{\SYSTEMRenameRuleSemLinunitBeta}[0]{\SYSTEMdrulename{SemLinunitBeta}}
\newcommand{\SYSTEMdruleSemLinunitBeta}[1]{\SYSTEMdrule[#1]{%
}{
  \mathsf{let} \, \langle \rangle =   \langle \rangle   \, \mathsf{in} \,  \SYSTEMnt{t_{{\mathrm{2}}}}   \rightsquigarrow_{\textsc{l} }  \SYSTEMnt{t_{{\mathrm{2}}}} }{%
{\SYSTEMRenameRuleSemLinunitBeta}{}%
}}

\newcommand{\SYSTEMRenameRuleSemLinpushProdCong}[0]{\SYSTEMdrulename{SemLinpushProdCong}}
\newcommand{\SYSTEMdruleSemLinpushProdCong}[1]{\SYSTEMdrule[#1]{%
\SYSTEMpremise{ \SYSTEMnt{t}  \rightsquigarrow_{\textsc{l} }  \SYSTEMnt{t'} }%
}{
  \textsf{push}_\otimes  \SYSTEMnt{t}   \rightsquigarrow_{\textsc{l} }   \textsf{push}_\otimes  \SYSTEMnt{t'}  }{%
{\SYSTEMRenameRuleSemLinpushProdCong}{}%
}}

\newcommand{\SYSTEMRenameRuleSemLinpushProdBoxCong}[0]{\SYSTEMdrulename{SemLinpushProdBoxCong}}
\newcommand{\SYSTEMdruleSemLinpushProdBoxCong}[1]{\SYSTEMdrule[#1]{%
\SYSTEMpremise{ \SYSTEMnt{t}  \rightsquigarrow_{\textsc{l} }  \SYSTEMnt{t'} }%
}{
  \textsf{push}_\otimes   \textcolor{coeffectColor}{[}  \SYSTEMnt{t}  \textcolor{coeffectColor}{]}    \rightsquigarrow_{\textsc{l} }   \textsf{push}_\otimes   \textcolor{coeffectColor}{[}  \SYSTEMnt{t'}  \textcolor{coeffectColor}{]}   }{%
{\SYSTEMRenameRuleSemLinpushProdBoxCong}{}%
}}

\newcommand{\SYSTEMRenameRuleSemLinpushProd}[0]{\SYSTEMdrulename{SemLinpushProd}}
\newcommand{\SYSTEMdruleSemLinpushProd}[1]{\SYSTEMdrule[#1]{%
}{
  \textsf{push}_\otimes   \textcolor{coeffectColor}{[}   \langle  \SYSTEMnt{t_{{\mathrm{1}}}} ,  \SYSTEMnt{t_{{\mathrm{2}}}}  \rangle   \textcolor{coeffectColor}{]}    \rightsquigarrow_{\textsc{l} }   \langle   \textcolor{coeffectColor}{[}  \SYSTEMnt{t_{{\mathrm{1}}}}  \textcolor{coeffectColor}{]}  ,   \textcolor{coeffectColor}{[}  \SYSTEMnt{t_{{\mathrm{2}}}}  \textcolor{coeffectColor}{]}   \rangle  }{%
{\SYSTEMRenameRuleSemLinpushProd}{}%
}}

\newcommand{\SYSTEMRenameRuleSemLinpushSumCong}[0]{\SYSTEMdrulename{SemLinpushSumCong}}
\newcommand{\SYSTEMdruleSemLinpushSumCong}[1]{\SYSTEMdrule[#1]{%
\SYSTEMpremise{ \SYSTEMnt{t}  \rightsquigarrow_{\textsc{l} }  \SYSTEMnt{t'} }%
}{
  \textsf{push}_\oplus  \SYSTEMnt{t}   \rightsquigarrow_{\textsc{l} }   \textsf{push}_\oplus  \SYSTEMnt{t'}  }{%
{\SYSTEMRenameRuleSemLinpushSumCong}{}%
}}

\newcommand{\SYSTEMRenameRuleSemLinpushSumBoxCong}[0]{\SYSTEMdrulename{SemLinpushSumBoxCong}}
\newcommand{\SYSTEMdruleSemLinpushSumBoxCong}[1]{\SYSTEMdrule[#1]{%
\SYSTEMpremise{ \SYSTEMnt{t}  \rightsquigarrow_{\textsc{l} }  \SYSTEMnt{t'} }%
}{
  \textsf{push}_\oplus   \textcolor{coeffectColor}{[}  \SYSTEMnt{t}  \textcolor{coeffectColor}{]}    \rightsquigarrow_{\textsc{l} }   \textsf{push}_\oplus   \textcolor{coeffectColor}{[}  \SYSTEMnt{t'}  \textcolor{coeffectColor}{]}   }{%
{\SYSTEMRenameRuleSemLinpushSumBoxCong}{}%
}}

\newcommand{\SYSTEMRenameRuleSemLinpushSumInjOne}[0]{\SYSTEMdrulename{SemLinpushSumInj1}}
\newcommand{\SYSTEMdruleSemLinpushSumInjOne}[1]{\SYSTEMdrule[#1]{%
}{
  \textsf{push}_\oplus   \textcolor{coeffectColor}{[}   \mathsf{inj}_1 \,  \SYSTEMnt{t}   \textcolor{coeffectColor}{]}    \rightsquigarrow_{\textsc{l} }   \mathsf{inj}_1 \,   \textcolor{coeffectColor}{[}  \SYSTEMnt{t}  \textcolor{coeffectColor}{]}   }{%
{\SYSTEMRenameRuleSemLinpushSumInjOne}{}%
}}

\newcommand{\SYSTEMRenameRuleSemLinpushSumInjTwo}[0]{\SYSTEMdrulename{SemLinpushSumInj2}}
\newcommand{\SYSTEMdruleSemLinpushSumInjTwo}[1]{\SYSTEMdrule[#1]{%
}{
  \textsf{push}_\oplus   \textcolor{coeffectColor}{[}   \mathsf{inj}_2 \,  \SYSTEMnt{t}   \textcolor{coeffectColor}{]}    \rightsquigarrow_{\textsc{l} }   \mathsf{inj}_2 \,   \textcolor{coeffectColor}{[}  \SYSTEMnt{t}  \textcolor{coeffectColor}{]}   }{%
{\SYSTEMRenameRuleSemLinpushSumInjTwo}{}%
}}

\newcommand{\SYSTEMRenameRuleSemLinpushSumInjGen}[0]{\SYSTEMdrulename{SemLinpushSumInjGen}}
\newcommand{\SYSTEMdruleSemLinpushSumInjGen}[1]{\SYSTEMdrule[#1]{%
}{
  \textsf{push}_\oplus   \textcolor{coeffectColor}{[}   \mathsf{inj}_i \,  \SYSTEMnt{t}   \textcolor{coeffectColor}{]}    \rightsquigarrow_{\textsc{l} }   \mathsf{inj}_i \,   \textcolor{coeffectColor}{[}  \SYSTEMnt{t}  \textcolor{coeffectColor}{]}   }{%
{\SYSTEMRenameRuleSemLinpushSumInjGen}{}%
}}

\newcommand{\SYSTEMRenameRuleSemLinpushUnitCong}[0]{\SYSTEMdrulename{SemLinpushUnitCong}}
\newcommand{\SYSTEMdruleSemLinpushUnitCong}[1]{\SYSTEMdrule[#1]{%
\SYSTEMpremise{ \SYSTEMnt{t}  \rightsquigarrow_{\textsc{l} }  \SYSTEMnt{t'} }%
}{
  \textsf{push}_{\mathrm{unit} }  \SYSTEMnt{t}   \rightsquigarrow_{\textsc{l} }   \textsf{push}_{\mathrm{unit} }  \SYSTEMnt{t'}  }{%
{\SYSTEMRenameRuleSemLinpushUnitCong}{}%
}}

\newcommand{\SYSTEMRenameRuleSemLinpushUnitBoxCong}[0]{\SYSTEMdrulename{SemLinpushUnitBoxCong}}
\newcommand{\SYSTEMdruleSemLinpushUnitBoxCong}[1]{\SYSTEMdrule[#1]{%
\SYSTEMpremise{ \SYSTEMnt{t}  \rightsquigarrow_{\textsc{l} }  \SYSTEMnt{t'} }%
}{
  \textsf{push}_{\mathrm{unit} }   \textcolor{coeffectColor}{[}  \SYSTEMnt{t}  \textcolor{coeffectColor}{]}    \rightsquigarrow_{\textsc{l} }   \textsf{push}_{\mathrm{unit} }   \textcolor{coeffectColor}{[}  \SYSTEMnt{t'}  \textcolor{coeffectColor}{]}   }{%
{\SYSTEMRenameRuleSemLinpushUnitBoxCong}{}%
}}

\newcommand{\SYSTEMRenameRuleSemLinpushUnit}[0]{\SYSTEMdrulename{SemLinpushUnit}}
\newcommand{\SYSTEMdruleSemLinpushUnit}[1]{\SYSTEMdrule[#1]{%
}{
  \textsf{push}_{\mathrm{unit} }   \textcolor{coeffectColor}{[}   \langle \rangle   \textcolor{coeffectColor}{]}    \rightsquigarrow_{\textsc{l} }   \langle \rangle  }{%
{\SYSTEMRenameRuleSemLinpushUnit}{}%
}}

\newcommand{\SYSTEMRenameRuleSemGrdModbetaBox}[0]{\SYSTEMdrulename{SemGrdModbetaBox}}
\newcommand{\SYSTEMdruleSemGrdModbetaBox}[1]{\SYSTEMdrule[#1]{%
}{
  \mathsf{let} \, \textcolor{coeffectColor}{[}  \SYSTEMmv{x}  \textcolor{coeffectColor}{]} =   \textcolor{coeffectColor}{[}  \SYSTEMnt{t_{{\mathrm{1}}}}  \textcolor{coeffectColor}{]}   \, \mathsf{in} \,  \SYSTEMnt{t_{{\mathrm{2}}}}   \rightsquigarrow_{\textsc{g}\Box}   [  \SYSTEMnt{t_{{\mathrm{1}}}}  /  \SYSTEMmv{x}  ]  \SYSTEMnt{t_{{\mathrm{2}}}}  }{%
{\SYSTEMRenameRuleSemGrdModbetaBox}{}%
}}

\newcommand{\SYSTEMRenameRuleSemGrdModcongLetL}[0]{\SYSTEMdrulename{SemGrdModcongLetL}}
\newcommand{\SYSTEMdruleSemGrdModcongLetL}[1]{\SYSTEMdrule[#1]{%
\SYSTEMpremise{ \SYSTEMnt{t_{{\mathrm{1}}}}  \rightsquigarrow_{\textsc{g}\Box}  \SYSTEMnt{t'_{{\mathrm{1}}}} }%
}{
  \mathsf{let} \, \textcolor{coeffectColor}{[}  \SYSTEMmv{x}  \textcolor{coeffectColor}{]} =  \SYSTEMnt{t_{{\mathrm{1}}}}  \, \mathsf{in} \,  \SYSTEMnt{t_{{\mathrm{2}}}}   \rightsquigarrow_{\textsc{g}\Box}   \mathsf{let} \, \textcolor{coeffectColor}{[}  \SYSTEMmv{x}  \textcolor{coeffectColor}{]} =  \SYSTEMnt{t'_{{\mathrm{1}}}}  \, \mathsf{in} \,  \SYSTEMnt{t_{{\mathrm{2}}}}  }{%
{\SYSTEMRenameRuleSemGrdModcongLetL}{}%
}}

\newcommand{\SYSTEMRenameRuleSemGrdbeta}[0]{\SYSTEMdrulename{SemGrdbeta}}
\newcommand{\SYSTEMdruleSemGrdbeta}[1]{\SYSTEMdrule[#1]{%
}{
  \SYSTEMsym{(}   \lambda  \SYSTEMmv{x}  .  \SYSTEMnt{t_{{\mathrm{2}}}}   \SYSTEMsym{)} \,  \SYSTEMnt{t_{{\mathrm{1}}}}   \rightsquigarrow_{\textsc{g} }   [  \SYSTEMnt{t_{{\mathrm{1}}}}  /  \SYSTEMmv{x}  ]  \SYSTEMnt{t_{{\mathrm{2}}}}  }{%
{\SYSTEMRenameRuleSemGrdbeta}{}%
}}

\newcommand{\SYSTEMRenameRuleSemGrdcongAppL}[0]{\SYSTEMdrulename{SemGrdcongAppL}}
\newcommand{\SYSTEMdruleSemGrdcongAppL}[1]{\SYSTEMdrule[#1]{%
\SYSTEMpremise{ \SYSTEMnt{t_{{\mathrm{1}}}}  \rightsquigarrow_{\textsc{g} }  \SYSTEMnt{t'_{{\mathrm{1}}}} }%
}{
  \SYSTEMnt{t_{{\mathrm{1}}}} \,  \SYSTEMnt{t_{{\mathrm{2}}}}   \rightsquigarrow_{\textsc{g} }   \SYSTEMnt{t'_{{\mathrm{1}}}} \,  \SYSTEMnt{t_{{\mathrm{2}}}}  }{%
{\SYSTEMRenameRuleSemGrdcongAppL}{}%
}}

\newcommand{\SYSTEMRenameRuleSemGrdcongAbs}[0]{\SYSTEMdrulename{SemGrdcongAbs}}

\newcommand{\SYSTEMRenameRuleSemGrdcongCase}[0]{\SYSTEMdrulename{SemGrdcongCase}}
\newcommand{\SYSTEMdruleSemGrdcongCase}[1]{\SYSTEMdrule[#1]{%
\SYSTEMpremise{ \SYSTEMnt{t}  \rightsquigarrow_{\textsc{g} }  \SYSTEMnt{t'} }%
}{
  \mathsf{case} \,  \SYSTEMnt{t}  \, \mathsf{of} \, \{ \mathsf{inj1} \,  \SYSTEMmv{x}  \rightarrow  \SYSTEMnt{t_{{\mathrm{1}}}}  ; \, \mathsf{inj2} \,  \SYSTEMmv{y}  \rightarrow  \SYSTEMnt{t_{{\mathrm{2}}}}  \}   \rightsquigarrow_{\textsc{g} }   \mathsf{case} \,  \SYSTEMnt{t'}  \, \mathsf{of} \, \{ \mathsf{inj1} \,  \SYSTEMmv{x}  \rightarrow  \SYSTEMnt{t_{{\mathrm{1}}}}  ; \, \mathsf{inj2} \,  \SYSTEMmv{y}  \rightarrow  \SYSTEMnt{t_{{\mathrm{2}}}}  \}  }{%
{\SYSTEMRenameRuleSemGrdcongCase}{}%
}}

\newcommand{\SYSTEMRenameRuleSemGrdcaseInjOne}[0]{\SYSTEMdrulename{SemGrdcaseInj1}}
\newcommand{\SYSTEMdruleSemGrdcaseInjOne}[1]{\SYSTEMdrule[#1]{%
}{
  \mathsf{case} \,   \mathsf{inj}_1 \,  \SYSTEMnt{t}   \, \mathsf{of} \, \{ \mathsf{inj1} \,  \SYSTEMmv{x}  \rightarrow  \SYSTEMnt{t_{{\mathrm{1}}}}  ; \, \mathsf{inj2} \,  \SYSTEMmv{y}  \rightarrow  \SYSTEMnt{t_{{\mathrm{2}}}}  \}   \rightsquigarrow_{\textsc{g} }   [  \SYSTEMnt{t}  /  \SYSTEMmv{x}  ]  \SYSTEMnt{t_{{\mathrm{1}}}}  }{%
{\SYSTEMRenameRuleSemGrdcaseInjOne}{}%
}}

\newcommand{\SYSTEMRenameRuleSemGrdcaseInjTwo}[0]{\SYSTEMdrulename{SemGrdcaseInj2}}
\newcommand{\SYSTEMdruleSemGrdcaseInjTwo}[1]{\SYSTEMdrule[#1]{%
}{
  \mathsf{case} \,   \mathsf{inj}_2 \,  \SYSTEMnt{t}   \, \mathsf{of} \, \{ \mathsf{inj1} \,  \SYSTEMmv{x}  \rightarrow  \SYSTEMnt{t_{{\mathrm{1}}}}  ; \, \mathsf{inj2} \,  \SYSTEMmv{y}  \rightarrow  \SYSTEMnt{t_{{\mathrm{2}}}}  \}   \rightsquigarrow_{\textsc{g} }   [  \SYSTEMnt{t}  /  \SYSTEMmv{y}  ]  \SYSTEMnt{t_{{\mathrm{2}}}}  }{%
{\SYSTEMRenameRuleSemGrdcaseInjTwo}{}%
}}

\newcommand{\SYSTEMRenameRuleSemGrdprodCong}[0]{\SYSTEMdrulename{SemGrdprodCong}}
\newcommand{\SYSTEMdruleSemGrdprodCong}[1]{\SYSTEMdrule[#1]{%
\SYSTEMpremise{ \SYSTEMnt{t_{{\mathrm{1}}}}  \rightsquigarrow_{\textsc{g} }  \SYSTEMnt{t'_{{\mathrm{1}}}} }%
}{
  \mathsf{let} \, \langle  \SYSTEMmv{x} ,  \SYSTEMmv{y}  \rangle =  \SYSTEMnt{t_{{\mathrm{1}}}}  \, \mathsf{in} \,  \SYSTEMnt{t_{{\mathrm{2}}}}   \rightsquigarrow_{\textsc{g} }   \mathsf{let} \, \langle  \SYSTEMmv{x} ,  \SYSTEMmv{y}  \rangle =  \SYSTEMnt{t'_{{\mathrm{1}}}}  \, \mathsf{in} \,  \SYSTEMnt{t_{{\mathrm{2}}}}  }{%
{\SYSTEMRenameRuleSemGrdprodCong}{}%
}}

\newcommand{\SYSTEMRenameRuleSemGrdprodBeta}[0]{\SYSTEMdrulename{SemGrdprodBeta}}
\newcommand{\SYSTEMdruleSemGrdprodBeta}[1]{\SYSTEMdrule[#1]{%
}{
  \mathsf{let} \, \langle  \SYSTEMmv{x} ,  \SYSTEMmv{y}  \rangle =   \langle  \SYSTEMnt{t_{{\mathrm{1}}}} ,  \SYSTEMnt{t_{{\mathrm{2}}}}  \rangle   \, \mathsf{in} \,  \SYSTEMnt{t_{{\mathrm{3}}}}   \rightsquigarrow_{\textsc{g} }   [  \SYSTEMnt{t_{{\mathrm{1}}}}  /  \SYSTEMmv{x}  ]   [  \SYSTEMnt{t_{{\mathrm{2}}}}  /  \SYSTEMmv{y}  ]  \SYSTEMnt{t_{{\mathrm{3}}}}   }{%
{\SYSTEMRenameRuleSemGrdprodBeta}{}%
}}

\newcommand{\SYSTEMRenameRuleSemGrdunitCong}[0]{\SYSTEMdrulename{SemGrdunitCong}}
\newcommand{\SYSTEMdruleSemGrdunitCong}[1]{\SYSTEMdrule[#1]{%
\SYSTEMpremise{ \SYSTEMnt{t_{{\mathrm{1}}}}  \rightsquigarrow_{\textsc{g} }  \SYSTEMnt{t'_{{\mathrm{1}}}} }%
}{
  \mathsf{let} \, \langle \rangle =  \SYSTEMnt{t_{{\mathrm{1}}}}  \, \mathsf{in} \,  \SYSTEMnt{t_{{\mathrm{2}}}}   \rightsquigarrow_{\textsc{g} }   \mathsf{let} \, \langle \rangle =  \SYSTEMnt{t'_{{\mathrm{1}}}}  \, \mathsf{in} \,  \SYSTEMnt{t_{{\mathrm{2}}}}  }{%
{\SYSTEMRenameRuleSemGrdunitCong}{}%
}}

\newcommand{\SYSTEMRenameRuleSemGrdunitBeta}[0]{\SYSTEMdrulename{SemGrdunitBeta}}
\newcommand{\SYSTEMdruleSemGrdunitBeta}[1]{\SYSTEMdrule[#1]{%
}{
  \mathsf{let} \, \langle \rangle =   \langle \rangle   \, \mathsf{in} \,  \SYSTEMnt{t_{{\mathrm{2}}}}   \rightsquigarrow_{\textsc{g} }  \SYSTEMnt{t_{{\mathrm{2}}}} }{%
{\SYSTEMRenameRuleSemGrdunitBeta}{}%
}}

\newcommand{\SYSTEMRenameRuleSemGrdPolycongTyApp}[0]{\SYSTEMdrulename{SemGrdPolycongTyApp}}
\newcommand{\SYSTEMdruleSemGrdPolycongTyApp}[1]{\SYSTEMdrule[#1]{%
\SYSTEMpremise{ \SYSTEMnt{t}  \rightsquigarrow_{\textsc{g}_\forall}  \SYSTEMnt{t'} }%
}{
  \SYSTEMnt{t}  \text{@}  \SYSTEMnt{A}   \rightsquigarrow_{\textsc{g}_\forall}   \SYSTEMnt{t'}  \text{@}  \SYSTEMnt{A}  }{%
{\SYSTEMRenameRuleSemGrdPolycongTyApp}{}%
}}

\newcommand{\SYSTEMRenameRuleSemGrdPolytyBeta}[0]{\SYSTEMdrulename{SemGrdPolytyBeta}}
\newcommand{\SYSTEMdruleSemGrdPolytyBeta}[1]{\SYSTEMdrule[#1]{%
}{
  \SYSTEMsym{(}   \Lambda   \alpha   .  \SYSTEMnt{t}   \SYSTEMsym{)}  \text{@}  \SYSTEMnt{B}   \rightsquigarrow_{\textsc{g}_\forall}   [  \SYSTEMnt{B}  /   \alpha   ]  \SYSTEMnt{t}  }{%
{\SYSTEMRenameRuleSemGrdPolytyBeta}{}%
}}

\newcommand{\SYSTEMRenameRuleTranslLtoGTmvar}[0]{\SYSTEMdrulename{TranslLtoGTmvar}}
\newcommand{\SYSTEMdruleTranslLtoGTmvar}[1]{\SYSTEMdrule[#1]{%
}{
 \textcolor{LGcolor}{\llparenthesis}  \SYSTEMmv{x}  \textcolor{LGcolor}{\rrparenthesis} \;\triangleq\;  \SYSTEMmv{x} }{%
{\SYSTEMRenameRuleTranslLtoGTmvar}{}%
}}

\newcommand{\SYSTEMRenameRuleTranslLtoGTmabs}[0]{\SYSTEMdrulename{TranslLtoGTmabs}}
\newcommand{\SYSTEMdruleTranslLtoGTmabs}[1]{\SYSTEMdrule[#1]{%
}{
 \textcolor{LGcolor}{\llparenthesis}   \lambda  \SYSTEMmv{x}  .  \SYSTEMnt{t}   \textcolor{LGcolor}{\rrparenthesis} \;\triangleq\;   \lambda  \SYSTEMmv{x}  .   \textcolor{LGcolor}{\llparenthesis} \smidge  \SYSTEMnt{t}  \smidge \textcolor{LGcolor}{\rrparenthesis}   }{%
{\SYSTEMRenameRuleTranslLtoGTmabs}{}%
}}

\newcommand{\SYSTEMRenameRuleTranslLtoGTmapp}[0]{\SYSTEMdrulename{TranslLtoGTmapp}}
\newcommand{\SYSTEMdruleTranslLtoGTmapp}[1]{\SYSTEMdrule[#1]{%
}{
 \textcolor{LGcolor}{\llparenthesis}   \SYSTEMnt{t_{{\mathrm{1}}}} \,  \SYSTEMnt{t_{{\mathrm{2}}}}   \textcolor{LGcolor}{\rrparenthesis} \;\triangleq\;    \textcolor{LGcolor}{\llparenthesis} \smidge  \SYSTEMnt{t_{{\mathrm{1}}}}  \smidge \textcolor{LGcolor}{\rrparenthesis}  \,   \textcolor{LGcolor}{\llparenthesis} \smidge  \SYSTEMnt{t_{{\mathrm{2}}}}  \smidge \textcolor{LGcolor}{\rrparenthesis}   }{%
{\SYSTEMRenameRuleTranslLtoGTmapp}{}%
}}

\newcommand{\SYSTEMRenameRuleTranslLtoGTmpr}[0]{\SYSTEMdrulename{TranslLtoGTmpr}}
\newcommand{\SYSTEMdruleTranslLtoGTmpr}[1]{\SYSTEMdrule[#1]{%
}{
 \textcolor{LGcolor}{\llparenthesis}   \textcolor{coeffectColor}{[}  \SYSTEMnt{t}  \textcolor{coeffectColor}{]}   \textcolor{LGcolor}{\rrparenthesis} \;\triangleq\;   \textcolor{coeffectColor}{[}   \textcolor{LGcolor}{\llparenthesis} \smidge  \SYSTEMnt{t}  \smidge \textcolor{LGcolor}{\rrparenthesis}   \textcolor{coeffectColor}{]}  }{%
{\SYSTEMRenameRuleTranslLtoGTmpr}{}%
}}

\newcommand{\SYSTEMRenameRuleTranslLtoGTmlet}[0]{\SYSTEMdrulename{TranslLtoGTmlet}}
\newcommand{\SYSTEMdruleTranslLtoGTmlet}[1]{\SYSTEMdrule[#1]{%
}{
 \textcolor{LGcolor}{\llparenthesis}   \mathsf{let} \, \textcolor{coeffectColor}{[}  \SYSTEMmv{x}  \textcolor{coeffectColor}{]} =  \SYSTEMnt{t_{{\mathrm{1}}}}  \, \mathsf{in} \,  \SYSTEMnt{t_{{\mathrm{2}}}}   \textcolor{LGcolor}{\rrparenthesis} \;\triangleq\;   \mathsf{let} \, \textcolor{coeffectColor}{[}  \SYSTEMmv{x}  \textcolor{coeffectColor}{]} =   \textcolor{LGcolor}{\llparenthesis} \smidge  \SYSTEMnt{t_{{\mathrm{1}}}}  \smidge \textcolor{LGcolor}{\rrparenthesis}   \, \mathsf{in} \,   \textcolor{LGcolor}{\llparenthesis} \smidge  \SYSTEMnt{t_{{\mathrm{2}}}}  \smidge \textcolor{LGcolor}{\rrparenthesis}   }{%
{\SYSTEMRenameRuleTranslLtoGTmlet}{}%
}}

\newcommand{\SYSTEMRenameRuleTranslLtoGTmprodi}[0]{\SYSTEMdrulename{TranslLtoGTmprodi}}
\newcommand{\SYSTEMdruleTranslLtoGTmprodi}[1]{\SYSTEMdrule[#1]{%
}{
 \textcolor{LGcolor}{\llparenthesis}   \langle  \SYSTEMnt{t_{{\mathrm{1}}}} ,  \SYSTEMnt{t_{{\mathrm{2}}}}  \rangle   \textcolor{LGcolor}{\rrparenthesis} \;\triangleq\;   \langle   \textcolor{LGcolor}{\llparenthesis} \smidge  \SYSTEMnt{t_{{\mathrm{1}}}}  \smidge \textcolor{LGcolor}{\rrparenthesis}  ,   \textcolor{LGcolor}{\llparenthesis} \smidge  \SYSTEMnt{t_{{\mathrm{2}}}}  \smidge \textcolor{LGcolor}{\rrparenthesis}   \rangle  }{%
{\SYSTEMRenameRuleTranslLtoGTmprodi}{}%
}}

\newcommand{\SYSTEMRenameRuleTranslLtoGTmprode}[0]{\SYSTEMdrulename{TranslLtoGTmprode}}
\newcommand{\SYSTEMdruleTranslLtoGTmprode}[1]{\SYSTEMdrule[#1]{%
}{
 \textcolor{LGcolor}{\llparenthesis}   \mathsf{let} \, \langle  \SYSTEMmv{x} ,  \SYSTEMmv{y}  \rangle =  \SYSTEMnt{t_{{\mathrm{1}}}}  \, \mathsf{in} \,  \SYSTEMnt{t_{{\mathrm{2}}}}   \textcolor{LGcolor}{\rrparenthesis} \;\triangleq\;   \mathsf{let} \, \langle  \SYSTEMmv{x} ,  \SYSTEMmv{y}  \rangle =   \textcolor{LGcolor}{\llparenthesis} \smidge  \SYSTEMnt{t_{{\mathrm{1}}}}  \smidge \textcolor{LGcolor}{\rrparenthesis}   \, \mathsf{in} \,   \textcolor{LGcolor}{\llparenthesis} \smidge  \SYSTEMnt{t_{{\mathrm{2}}}}  \smidge \textcolor{LGcolor}{\rrparenthesis}   }{%
{\SYSTEMRenameRuleTranslLtoGTmprode}{}%
}}

\newcommand{\SYSTEMRenameRuleTranslLtoGTmuniti}[0]{\SYSTEMdrulename{TranslLtoGTmuniti}}
\newcommand{\SYSTEMdruleTranslLtoGTmuniti}[1]{\SYSTEMdrule[#1]{%
}{
 \textcolor{LGcolor}{\llparenthesis}   \langle \rangle   \textcolor{LGcolor}{\rrparenthesis} \;\triangleq\;   \langle \rangle  }{%
{\SYSTEMRenameRuleTranslLtoGTmuniti}{}%
}}

\newcommand{\SYSTEMRenameRuleTranslLtoGTmunite}[0]{\SYSTEMdrulename{TranslLtoGTmunite}}
\newcommand{\SYSTEMdruleTranslLtoGTmunite}[1]{\SYSTEMdrule[#1]{%
}{
 \textcolor{LGcolor}{\llparenthesis}   \mathsf{let} \, \langle \rangle =  \SYSTEMnt{t_{{\mathrm{1}}}}  \, \mathsf{in} \,  \SYSTEMnt{t_{{\mathrm{2}}}}   \textcolor{LGcolor}{\rrparenthesis} \;\triangleq\;   \mathsf{let} \, \langle \rangle =   \textcolor{LGcolor}{\llparenthesis} \smidge  \SYSTEMnt{t_{{\mathrm{1}}}}  \smidge \textcolor{LGcolor}{\rrparenthesis}   \, \mathsf{in} \,   \textcolor{LGcolor}{\llparenthesis} \smidge  \SYSTEMnt{t_{{\mathrm{2}}}}  \smidge \textcolor{LGcolor}{\rrparenthesis}   }{%
{\SYSTEMRenameRuleTranslLtoGTmunite}{}%
}}

\newcommand{\SYSTEMRenameRuleTranslLtoGTmsumiOne}[0]{\SYSTEMdrulename{TranslLtoGTmsumi1}}
\newcommand{\SYSTEMdruleTranslLtoGTmsumiOne}[1]{\SYSTEMdrule[#1]{%
}{
 \textcolor{LGcolor}{\llparenthesis}   \mathsf{inj}_1 \,  \SYSTEMnt{t}   \textcolor{LGcolor}{\rrparenthesis} \;\triangleq\;   \mathsf{inj}_1 \,   \textcolor{LGcolor}{\llparenthesis} \smidge  \SYSTEMnt{t}  \smidge \textcolor{LGcolor}{\rrparenthesis}   }{%
{\SYSTEMRenameRuleTranslLtoGTmsumiOne}{}%
}}

\newcommand{\SYSTEMRenameRuleTranslLtoGTmsumiTwo}[0]{\SYSTEMdrulename{TranslLtoGTmsumi2}}
\newcommand{\SYSTEMdruleTranslLtoGTmsumiTwo}[1]{\SYSTEMdrule[#1]{%
}{
 \textcolor{LGcolor}{\llparenthesis}   \mathsf{inj}_2 \,  \SYSTEMnt{t}   \textcolor{LGcolor}{\rrparenthesis} \;\triangleq\;   \mathsf{inj}_2 \,   \textcolor{LGcolor}{\llparenthesis} \smidge  \SYSTEMnt{t}  \smidge \textcolor{LGcolor}{\rrparenthesis}   }{%
{\SYSTEMRenameRuleTranslLtoGTmsumiTwo}{}%
}}

\newcommand{\SYSTEMRenameRuleTranslLtoGTmsume}[0]{\SYSTEMdrulename{TranslLtoGTmsume}}
\newcommand{\SYSTEMdruleTranslLtoGTmsume}[1]{\SYSTEMdrule[#1]{%
}{
 \textcolor{LGcolor}{\llparenthesis}   \mathsf{case} \,  \SYSTEMnt{t}  \, \mathsf{of} \, \{ \mathsf{inj1} \,  \SYSTEMmv{x}  \rightarrow  \SYSTEMnt{t_{{\mathrm{1}}}}  ; \, \mathsf{inj2} \,  \SYSTEMmv{y}  \rightarrow  \SYSTEMnt{t_{{\mathrm{2}}}}  \}   \textcolor{LGcolor}{\rrparenthesis} \;\triangleq\;   \mathsf{case} \,   \textcolor{LGcolor}{\llparenthesis} \smidge  \SYSTEMnt{t}  \smidge \textcolor{LGcolor}{\rrparenthesis}   \, \mathsf{of} \, \{ \mathsf{inj1} \,  \SYSTEMmv{x}  \rightarrow   \textcolor{LGcolor}{\llparenthesis} \smidge  \SYSTEMnt{t_{{\mathrm{1}}}}  \smidge \textcolor{LGcolor}{\rrparenthesis}   ; \, \mathsf{inj2} \,  \SYSTEMmv{y}  \rightarrow   \textcolor{LGcolor}{\llparenthesis} \smidge  \SYSTEMnt{t_{{\mathrm{2}}}}  \smidge \textcolor{LGcolor}{\rrparenthesis}   \}  }{%
{\SYSTEMRenameRuleTranslLtoGTmsume}{}%
}}

\newcommand{\SYSTEMRenameRuleTranslLtoGTmpushProd}[0]{\SYSTEMdrulename{TranslLtoGTmpushProd}}
\newcommand{\SYSTEMdruleTranslLtoGTmpushProd}[1]{\SYSTEMdrule[#1]{%
\SYSTEMpremise{  \SYSTEMmv{x}  \,\#\,  \SYSTEMnt{t}  }%
}{
 \textcolor{LGcolor}{\llparenthesis}   \textsf{push}_\otimes  \SYSTEMnt{t}   \textcolor{LGcolor}{\rrparenthesis} \;\triangleq\;   \mathsf{let} \, \textcolor{coeffectColor}{[}  \SYSTEMmv{x}  \textcolor{coeffectColor}{]} =   \textcolor{LGcolor}{\llparenthesis} \smidge  \SYSTEMnt{t}  \smidge \textcolor{LGcolor}{\rrparenthesis}   \, \mathsf{in} \,   \mathsf{let} \, \langle  \SYSTEMmv{y} ,  \SYSTEMmv{z}  \rangle =  \SYSTEMmv{x}  \, \mathsf{in} \,   \langle   \textcolor{coeffectColor}{[}  \SYSTEMmv{y}  \textcolor{coeffectColor}{]}  ,   \textcolor{coeffectColor}{[}  \SYSTEMmv{z}  \textcolor{coeffectColor}{]}   \rangle    }{%
{\SYSTEMRenameRuleTranslLtoGTmpushProd}{}%
}}

\newcommand{\SYSTEMRenameRuleTranslLtoGTmpushSum}[0]{\SYSTEMdrulename{TranslLtoGTmpushSum}}
\newcommand{\SYSTEMdruleTranslLtoGTmpushSum}[1]{\SYSTEMdrule[#1]{%
\SYSTEMpremise{  \SYSTEMmv{x}  \,\#\,  \SYSTEMnt{t}  }%
}{
 \textcolor{LGcolor}{\llparenthesis}   \textsf{push}_\oplus  \SYSTEMnt{t}   \textcolor{LGcolor}{\rrparenthesis} \;\triangleq\;   \mathsf{let} \, \textcolor{coeffectColor}{[}  \SYSTEMmv{x}  \textcolor{coeffectColor}{]} =   \textcolor{LGcolor}{\llparenthesis} \smidge  \SYSTEMnt{t}  \smidge \textcolor{LGcolor}{\rrparenthesis}   \, \mathsf{in} \,   \mathsf{case} \,  \SYSTEMmv{x}  \, \mathsf{of} \, \{ \mathsf{inj1} \,  \SYSTEMmv{y}  \rightarrow   \mathsf{inj}_1 \,   \textcolor{coeffectColor}{[}  \SYSTEMmv{y}  \textcolor{coeffectColor}{]}    ; \, \mathsf{inj2} \,  \SYSTEMmv{z}  \rightarrow   \mathsf{inj}_2 \,   \textcolor{coeffectColor}{[}  \SYSTEMmv{z}  \textcolor{coeffectColor}{]}    \}   }{%
{\SYSTEMRenameRuleTranslLtoGTmpushSum}{}%
}}

\newcommand{\SYSTEMRenameRuleTranslLtoGTmpushUnit}[0]{\SYSTEMdrulename{TranslLtoGTmpushUnit}}
\newcommand{\SYSTEMdruleTranslLtoGTmpushUnit}[1]{\SYSTEMdrule[#1]{%
}{
 \textcolor{LGcolor}{\llparenthesis}   \textsf{push}_{\mathrm{unit} }  \SYSTEMnt{t}   \textcolor{LGcolor}{\rrparenthesis} \;\triangleq\;   \mathsf{let} \, \textcolor{coeffectColor}{[}  \SYSTEMmv{x}  \textcolor{coeffectColor}{]} =   \textcolor{LGcolor}{\llparenthesis} \smidge  \SYSTEMnt{t}  \smidge \textcolor{LGcolor}{\rrparenthesis}   \, \mathsf{in} \,   \mathsf{let} \, \langle \rangle =  \SYSTEMmv{x}  \, \mathsf{in} \,   \langle \rangle    }{%
{\SYSTEMRenameRuleTranslLtoGTmpushUnit}{}%
}}

\newcommand{\SYSTEMRenameRuleTranslLtoGAltTyfunTy}[0]{\SYSTEMdrulename{TranslLtoGAltTyfunTy}}
\newcommand{\SYSTEMdruleTranslLtoGAltTyfunTy}[1]{\SYSTEMdrule[#1]{%
}{
 \textcolor{LGcolor}{\llparenthesis}   \SYSTEMnt{A}  \multimap  \SYSTEMnt{B}   \textcolor{LGcolor}{\rrparenthesis}' \;\triangleq\;    \textcolor{LGcolor}{\llparenthesis} \smidge  \SYSTEMnt{A}  \smidge \textcolor{LGcolor}{\rrparenthesis}'   \xrightarrow{\textcolor{coeffectColor}{  \langle  \SYSTEMsym{1}  ,  \SYSTEMsym{1}  \rangle  } }   \textcolor{LGcolor}{\llparenthesis} \smidge  \SYSTEMnt{B}  \smidge \textcolor{LGcolor}{\rrparenthesis}'   }{%
{\SYSTEMRenameRuleTranslLtoGAltTyfunTy}{}%
}}

\newcommand{\SYSTEMRenameRuleTranslLtoGAltTyboxTy}[0]{\SYSTEMdrulename{TranslLtoGAltTyboxTy}}
\newcommand{\SYSTEMdruleTranslLtoGAltTyboxTy}[1]{\SYSTEMdrule[#1]{%
}{
 \textcolor{LGcolor}{\llparenthesis}   \textcolor{coeffectColor}{\square_{ \SYSTEMnt{r} } }  \SYSTEMnt{A}   \textcolor{LGcolor}{\rrparenthesis}' \;\triangleq\;   \textcolor{coeffectColor}{\square_{  \langle  \SYSTEMnt{r}  ,   \omega   \rangle  } }   \textcolor{LGcolor}{\llparenthesis} \smidge  \SYSTEMnt{A}  \smidge \textcolor{LGcolor}{\rrparenthesis}'   }{%
{\SYSTEMRenameRuleTranslLtoGAltTyboxTy}{}%
}}

\newcommand{\SYSTEMRenameRuleTranslLtoGAltTybaseTy}[0]{\SYSTEMdrulename{TranslLtoGAltTybaseTy}}
\newcommand{\SYSTEMdruleTranslLtoGAltTybaseTy}[1]{\SYSTEMdrule[#1]{%
}{
 \textcolor{LGcolor}{\llparenthesis}   \mathrm{K}   \textcolor{LGcolor}{\rrparenthesis}' \;\triangleq\;   \mathrm{K}  }{%
{\SYSTEMRenameRuleTranslLtoGAltTybaseTy}{}%
}}

\newcommand{\SYSTEMRenameRuleTranslLtoGAltGtxtgrad}[0]{\SYSTEMdrulename{TranslLtoGAltGtxtgrad}}
\newcommand{\SYSTEMdruleTranslLtoGAltGtxtgrad}[1]{\SYSTEMdrule[#1]{%
}{
 \textcolor{LGcolor}{\llparenthesis}   \Gamma ,   \SYSTEMmv{x}  : \textcolor{coeffectColor}{[}  \SYSTEMnt{A} {\textcolor{coeffectColor}{]_{ \SYSTEMnt{r} } } }    \textcolor{LGcolor}{\rrparenthesis}' \;\triangleq\;    \textcolor{LGcolor}{\llparenthesis}  \Gamma  \textcolor{LGcolor}{\rrparenthesis}'  ,   \SYSTEMmv{x}  :_{\textcolor{coeffectColor}{  \langle  \SYSTEMnt{r}  ,   \omega   \rangle  } }   \textcolor{LGcolor}{\llparenthesis} \smidge  \SYSTEMnt{A}  \smidge \textcolor{LGcolor}{\rrparenthesis}'    }{%
{\SYSTEMRenameRuleTranslLtoGAltGtxtgrad}{}%
}}

\newcommand{\SYSTEMRenameRuleTranslLtoGAltGtxtlin}[0]{\SYSTEMdrulename{TranslLtoGAltGtxtlin}}
\newcommand{\SYSTEMdruleTranslLtoGAltGtxtlin}[1]{\SYSTEMdrule[#1]{%
}{
 \textcolor{LGcolor}{\llparenthesis}   \Gamma ,   \SYSTEMmv{x}  :  \SYSTEMnt{A}    \textcolor{LGcolor}{\rrparenthesis}' \;\triangleq\;    \textcolor{LGcolor}{\llparenthesis}  \Gamma  \textcolor{LGcolor}{\rrparenthesis}'  ,   \SYSTEMmv{x}  :_{\textcolor{coeffectColor}{  \langle  \SYSTEMsym{1}  ,  \SYSTEMsym{1}  \rangle  } }   \textcolor{LGcolor}{\llparenthesis} \smidge  \SYSTEMnt{A}  \smidge \textcolor{LGcolor}{\rrparenthesis}'    }{%
{\SYSTEMRenameRuleTranslLtoGAltGtxtlin}{}%
}}

\newcommand{\SYSTEMRenameRuleTranslLtoGTybaseTy}[0]{\SYSTEMdrulename{TranslLtoGTybaseTy}}
\newcommand{\SYSTEMdruleTranslLtoGTybaseTy}[1]{\SYSTEMdrule[#1]{%
}{
 \textcolor{LGcolor}{\llparenthesis}   \mathrm{K}   \textcolor{LGcolor}{\rrparenthesis} \;\triangleq\;   \mathrm{K}  }{%
{\SYSTEMRenameRuleTranslLtoGTybaseTy}{}%
}}

\newcommand{\SYSTEMRenameRuleTranslLtoGTyfunTy}[0]{\SYSTEMdrulename{TranslLtoGTyfunTy}}
\newcommand{\SYSTEMdruleTranslLtoGTyfunTy}[1]{\SYSTEMdrule[#1]{%
}{
 \textcolor{LGcolor}{\llparenthesis}   \SYSTEMnt{A}  \multimap  \SYSTEMnt{B}   \textcolor{LGcolor}{\rrparenthesis} \;\triangleq\;    \textcolor{LGcolor}{\llparenthesis} \smidge  \SYSTEMnt{A}  \smidge \textcolor{LGcolor}{\rrparenthesis}   \xrightarrow{\textcolor{coeffectColor}{ \SYSTEMsym{1} } }   \textcolor{LGcolor}{\llparenthesis} \smidge  \SYSTEMnt{B}  \smidge \textcolor{LGcolor}{\rrparenthesis}   }{%
{\SYSTEMRenameRuleTranslLtoGTyfunTy}{}%
}}

\newcommand{\SYSTEMRenameRuleTranslLtoGTyboxTy}[0]{\SYSTEMdrulename{TranslLtoGTyboxTy}}
\newcommand{\SYSTEMdruleTranslLtoGTyboxTy}[1]{\SYSTEMdrule[#1]{%
}{
 \textcolor{LGcolor}{\llparenthesis}   \textcolor{coeffectColor}{\square_{ \SYSTEMnt{r} } }  \SYSTEMnt{A}   \textcolor{LGcolor}{\rrparenthesis} \;\triangleq\;   \textcolor{coeffectColor}{\square_{ \SYSTEMnt{r} } }   \textcolor{LGcolor}{\llparenthesis} \smidge  \SYSTEMnt{A}  \smidge \textcolor{LGcolor}{\rrparenthesis}   }{%
{\SYSTEMRenameRuleTranslLtoGTyboxTy}{}%
}}

\newcommand{\SYSTEMRenameRuleTranslLtoGTyprodTy}[0]{\SYSTEMdrulename{TranslLtoGTyprodTy}}
\newcommand{\SYSTEMdruleTranslLtoGTyprodTy}[1]{\SYSTEMdrule[#1]{%
}{
 \textcolor{LGcolor}{\llparenthesis}   \SYSTEMnt{A}  \otimes  \SYSTEMnt{B}   \textcolor{LGcolor}{\rrparenthesis} \;\triangleq\;    \textcolor{LGcolor}{\llparenthesis} \smidge  \SYSTEMnt{A}  \smidge \textcolor{LGcolor}{\rrparenthesis}   \times   \textcolor{LGcolor}{\llparenthesis} \smidge  \SYSTEMnt{B}  \smidge \textcolor{LGcolor}{\rrparenthesis}   }{%
{\SYSTEMRenameRuleTranslLtoGTyprodTy}{}%
}}

\newcommand{\SYSTEMRenameRuleTranslLtoGTysumTy}[0]{\SYSTEMdrulename{TranslLtoGTysumTy}}
\newcommand{\SYSTEMdruleTranslLtoGTysumTy}[1]{\SYSTEMdrule[#1]{%
}{
 \textcolor{LGcolor}{\llparenthesis}   \SYSTEMnt{A}  \oplus  \SYSTEMnt{B}   \textcolor{LGcolor}{\rrparenthesis} \;\triangleq\;    \textcolor{LGcolor}{\llparenthesis} \smidge  \SYSTEMnt{A}  \smidge \textcolor{LGcolor}{\rrparenthesis}   +   \textcolor{LGcolor}{\llparenthesis} \smidge  \SYSTEMnt{B}  \smidge \textcolor{LGcolor}{\rrparenthesis}   }{%
{\SYSTEMRenameRuleTranslLtoGTysumTy}{}%
}}

\newcommand{\SYSTEMRenameRuleTranslLtoGTyunitTy}[0]{\SYSTEMdrulename{TranslLtoGTyunitTy}}
\newcommand{\SYSTEMdruleTranslLtoGTyunitTy}[1]{\SYSTEMdrule[#1]{%
}{
 \textcolor{LGcolor}{\llparenthesis}   \mathrm{unit}   \textcolor{LGcolor}{\rrparenthesis} \;\triangleq\;   \mathrm{unit}  }{%
{\SYSTEMRenameRuleTranslLtoGTyunitTy}{}%
}}

\newcommand{\SYSTEMRenameRuleTranslLtoGTyboxTyPoly}[0]{\SYSTEMdrulename{TranslLtoGTyboxTyPoly}}
\newcommand{\SYSTEMdruleTranslLtoGTyboxTyPoly}[1]{\SYSTEMdrule[#1]{%
}{
 \textcolor{LGcolor}{\llparenthesis}   \textcolor{coeffectColor}{\square_{ \SYSTEMnt{r} } }  \SYSTEMnt{A}   \textcolor{LGcolor}{\rrparenthesis} \;\triangleq\;   \forall   \beta   .     (    \textcolor{LGcolor}{\llparenthesis} \smidge  \SYSTEMnt{A}  \smidge \textcolor{LGcolor}{\rrparenthesis}   \xrightarrow{\textcolor{coeffectColor}{ \SYSTEMnt{r} } }   \beta    )   \xrightarrow{\textcolor{coeffectColor}{ \SYSTEMsym{1} } }   \beta     }{%
{\SYSTEMRenameRuleTranslLtoGTyboxTyPoly}{}%
}}

\newcommand{\SYSTEMRenameRuleTranslLtoGCpsTmvar}[0]{\SYSTEMdrulename{TranslLtoGCpsTmvar}}
\newcommand{\SYSTEMdruleTranslLtoGCpsTmvar}[1]{\SYSTEMdrule[#1]{%
}{
 \textcolor{LGcolor}{\llparenthesis}  \SYSTEMmv{x}  \textcolor{LGcolor}{\rrparenthesis} \;\triangleq\;   \lambda  \SYSTEMmv{k}  .    \SYSTEMmv{x} \,  \SYSTEMmv{k}    }{%
{\SYSTEMRenameRuleTranslLtoGCpsTmvar}{}%
}}

\newcommand{\SYSTEMRenameRuleTranslLtoGCpsTmabs}[0]{\SYSTEMdrulename{TranslLtoGCpsTmabs}}
\newcommand{\SYSTEMdruleTranslLtoGCpsTmabs}[1]{\SYSTEMdrule[#1]{%
}{
 \textcolor{LGcolor}{\llparenthesis}   \lambda  \SYSTEMmv{x}  .  \SYSTEMnt{t}   \textcolor{LGcolor}{\rrparenthesis} \;\triangleq\;   \lambda  \SYSTEMmv{k}  .    \SYSTEMmv{k} \,  \SYSTEMsym{(}   \lambda  \SYSTEMmv{x}  .   \textcolor{LGcolor}{\llparenthesis} \smidge  \SYSTEMnt{t}  \smidge \textcolor{LGcolor}{\rrparenthesis}    \SYSTEMsym{)}    }{%
{\SYSTEMRenameRuleTranslLtoGCpsTmabs}{}%
}}

\newcommand{\SYSTEMRenameRuleTranslLtoGCpsTmapp}[0]{\SYSTEMdrulename{TranslLtoGCpsTmapp}}
\newcommand{\SYSTEMdruleTranslLtoGCpsTmapp}[1]{\SYSTEMdrule[#1]{%
}{
 \textcolor{LGcolor}{\llparenthesis}   \SYSTEMnt{t_{{\mathrm{1}}}} \,  \SYSTEMnt{t_{{\mathrm{2}}}}   \textcolor{LGcolor}{\rrparenthesis} \;\triangleq\;   \lambda  \SYSTEMmv{k}  .     \textcolor{LGcolor}{\llparenthesis} \smidge  \SYSTEMnt{t_{{\mathrm{1}}}}  \smidge \textcolor{LGcolor}{\rrparenthesis}  \,  \SYSTEMsym{(}   \lambda  \SYSTEMmv{f}  .      \SYSTEMmv{f} \,   \textcolor{LGcolor}{\llparenthesis} \smidge  \SYSTEMnt{t_{{\mathrm{2}}}}  \smidge \textcolor{LGcolor}{\rrparenthesis}    \,  \SYSTEMmv{k}     \SYSTEMsym{)}    }{%
{\SYSTEMRenameRuleTranslLtoGCpsTmapp}{}%
}}

\newcommand{\SYSTEMRenameRuleTranslLtoGCpsTmpr}[0]{\SYSTEMdrulename{TranslLtoGCpsTmpr}}
\newcommand{\SYSTEMdruleTranslLtoGCpsTmpr}[1]{\SYSTEMdrule[#1]{%
}{
 \textcolor{LGcolor}{\llparenthesis}   \textcolor{coeffectColor}{[}  \SYSTEMnt{t}  \textcolor{coeffectColor}{]}   \textcolor{LGcolor}{\rrparenthesis} \;\triangleq\;   \lambda  \SYSTEMmv{k}  .    \SYSTEMmv{k} \,   \textcolor{LGcolor}{\llparenthesis} \smidge  \SYSTEMnt{t}  \smidge \textcolor{LGcolor}{\rrparenthesis}     }{%
{\SYSTEMRenameRuleTranslLtoGCpsTmpr}{}%
}}

\newcommand{\SYSTEMRenameRuleTranslLtoGCpsTmlet}[0]{\SYSTEMdrulename{TranslLtoGCpsTmlet}}
\newcommand{\SYSTEMdruleTranslLtoGCpsTmlet}[1]{\SYSTEMdrule[#1]{%
}{
 \textcolor{LGcolor}{\llparenthesis}   \mathsf{let} \, \textcolor{coeffectColor}{[}  \SYSTEMmv{x}  \textcolor{coeffectColor}{]} =  \SYSTEMnt{t_{{\mathrm{1}}}}  \, \mathsf{in} \,  \SYSTEMnt{t_{{\mathrm{2}}}}   \textcolor{LGcolor}{\rrparenthesis} \;\triangleq\;   \lambda  \SYSTEMmv{k}  .     \textcolor{LGcolor}{\llparenthesis} \smidge  \SYSTEMnt{t_{{\mathrm{1}}}}  \smidge \textcolor{LGcolor}{\rrparenthesis}  \,  \SYSTEMsym{(}   \lambda  \SYSTEMmv{x}  .     \textcolor{LGcolor}{\llparenthesis} \smidge  \SYSTEMnt{t_{{\mathrm{2}}}}  \smidge \textcolor{LGcolor}{\rrparenthesis}  \,  \SYSTEMmv{k}     \SYSTEMsym{)}    }{%
{\SYSTEMRenameRuleTranslLtoGCpsTmlet}{}%
}}

\newcommand{\SYSTEMRenameRuleTranslLtoGCpsTybaseTy}[0]{\SYSTEMdrulename{TranslLtoGCpsTybaseTy}}
\newcommand{\SYSTEMdruleTranslLtoGCpsTybaseTy}[1]{\SYSTEMdrule[#1]{%
}{
 \textcolor{LGcolor}{\llparenthesis}   \mathrm{K}   \textcolor{LGcolor}{\rrparenthesis} \;\triangleq\;    (    \mathrm{K}   \xrightarrow{\textcolor{coeffectColor}{ \SYSTEMsym{1} } }   \mathrm{K}    )   \xrightarrow{\textcolor{coeffectColor}{ \SYSTEMsym{1} } }   \mathrm{K}   }{%
{\SYSTEMRenameRuleTranslLtoGCpsTybaseTy}{}%
}}

\newcommand{\SYSTEMRenameRuleTranslLtoGCpsTyfunTy}[0]{\SYSTEMdrulename{TranslLtoGCpsTyfunTy}}
\newcommand{\SYSTEMdruleTranslLtoGCpsTyfunTy}[1]{\SYSTEMdrule[#1]{%
}{
 \textcolor{LGcolor}{\llparenthesis}   \SYSTEMnt{A}  \multimap  \SYSTEMnt{B}   \textcolor{LGcolor}{\rrparenthesis} \;\triangleq\;    (    (    \textcolor{LGcolor}{\llparenthesis} \smidge  \SYSTEMnt{A}  \smidge \textcolor{LGcolor}{\rrparenthesis}   \xrightarrow{\textcolor{coeffectColor}{ \SYSTEMsym{1} } }   \textcolor{LGcolor}{\llparenthesis} \smidge  \SYSTEMnt{B}  \smidge \textcolor{LGcolor}{\rrparenthesis}    )   \xrightarrow{\textcolor{coeffectColor}{ \SYSTEMsym{1} } }   \mathrm{K}    )   \xrightarrow{\textcolor{coeffectColor}{ \SYSTEMsym{1} } }   \mathrm{K}   }{%
{\SYSTEMRenameRuleTranslLtoGCpsTyfunTy}{}%
}}

\newcommand{\SYSTEMRenameRuleTranslLtoGCpsTyboxTy}[0]{\SYSTEMdrulename{TranslLtoGCpsTyboxTy}}
\newcommand{\SYSTEMdruleTranslLtoGCpsTyboxTy}[1]{\SYSTEMdrule[#1]{%
}{
 \textcolor{LGcolor}{\llparenthesis}   \textcolor{coeffectColor}{\square_{ \SYSTEMnt{r} } }  \SYSTEMnt{A}   \textcolor{LGcolor}{\rrparenthesis} \;\triangleq\;    (    \textcolor{LGcolor}{\llparenthesis} \smidge  \SYSTEMnt{A}  \smidge \textcolor{LGcolor}{\rrparenthesis}   \xrightarrow{\textcolor{coeffectColor}{ \SYSTEMnt{r} } }   \mathrm{K}    )   \xrightarrow{\textcolor{coeffectColor}{ \SYSTEMsym{1} } }   \mathrm{K}   }{%
{\SYSTEMRenameRuleTranslLtoGCpsTyboxTy}{}%
}}

\newcommand{\SYSTEMRenameRuleTranslLtoGGtxtempty}[0]{\SYSTEMdrulename{TranslLtoGGtxtempty}}
\newcommand{\SYSTEMdruleTranslLtoGGtxtempty}[1]{\SYSTEMdrule[#1]{%
}{
 \textcolor{LGcolor}{\llparenthesis}   \emptyset   \textcolor{LGcolor}{\rrparenthesis} \;\triangleq\;   \emptyset  }{%
{\SYSTEMRenameRuleTranslLtoGGtxtempty}{}%
}}

\newcommand{\SYSTEMRenameRuleTranslLtoGGtxtgrad}[0]{\SYSTEMdrulename{TranslLtoGGtxtgrad}}
\newcommand{\SYSTEMdruleTranslLtoGGtxtgrad}[1]{\SYSTEMdrule[#1]{%
}{
 \textcolor{LGcolor}{\llparenthesis}   \Gamma ,   \SYSTEMmv{x}  : \textcolor{coeffectColor}{[}  \SYSTEMnt{A} {\textcolor{coeffectColor}{]_{ \SYSTEMnt{r} } } }    \textcolor{LGcolor}{\rrparenthesis} \;\triangleq\;    \textcolor{LGcolor}{\llparenthesis}  \Gamma  \textcolor{LGcolor}{\rrparenthesis}  ,   \SYSTEMmv{x}  :_{\textcolor{coeffectColor}{ \SYSTEMnt{r} } }   \textcolor{LGcolor}{\llparenthesis} \smidge  \SYSTEMnt{A}  \smidge \textcolor{LGcolor}{\rrparenthesis}    }{%
{\SYSTEMRenameRuleTranslLtoGGtxtgrad}{}%
}}

\newcommand{\SYSTEMRenameRuleTranslLtoGGtxtlin}[0]{\SYSTEMdrulename{TranslLtoGGtxtlin}}
\newcommand{\SYSTEMdruleTranslLtoGGtxtlin}[1]{\SYSTEMdrule[#1]{%
}{
 \textcolor{LGcolor}{\llparenthesis}   \Gamma ,   \SYSTEMmv{x}  :  \SYSTEMnt{A}    \textcolor{LGcolor}{\rrparenthesis} \;\triangleq\;    \textcolor{LGcolor}{\llparenthesis}  \Gamma  \textcolor{LGcolor}{\rrparenthesis}  ,   \SYSTEMmv{x}  :_{\textcolor{coeffectColor}{ \SYSTEMsym{1} } }   \textcolor{LGcolor}{\llparenthesis} \smidge  \SYSTEMnt{A}  \smidge \textcolor{LGcolor}{\rrparenthesis}    }{%
{\SYSTEMRenameRuleTranslLtoGGtxtlin}{}%
}}

\newcommand{\SYSTEMRenameRuleTranslGtoLTmvar}[0]{\SYSTEMdrulename{TranslGtoLTmvar}}
\newcommand{\SYSTEMdruleTranslGtoLTmvar}[1]{\SYSTEMdrule[#1]{%
}{
 \textcolor{GLcolor}{\llbracket} \smidge  \SYSTEMmv{x}  \smidge \textcolor{GLcolor}{\rrbracket} \;\triangleq\;  \SYSTEMmv{x} }{%
{\SYSTEMRenameRuleTranslGtoLTmvar}{}%
}}

\newcommand{\SYSTEMRenameRuleTranslGtoLTmabs}[0]{\SYSTEMdrulename{TranslGtoLTmabs}}
\newcommand{\SYSTEMdruleTranslGtoLTmabs}[1]{\SYSTEMdrule[#1]{%
\SYSTEMpremise{  \SYSTEMmv{y}  \,\#\,  \SYSTEMnt{t}  }%
}{
 \textcolor{GLcolor}{\llbracket} \smidge   \lambda  \SYSTEMmv{x}  .  \SYSTEMnt{t}   \smidge \textcolor{GLcolor}{\rrbracket} \;\triangleq\;   \lambda  \SYSTEMmv{y}  .   \mathsf{let} \, \textcolor{coeffectColor}{[}  \SYSTEMmv{x}  \textcolor{coeffectColor}{]} =  \SYSTEMmv{y}  \, \mathsf{in} \,   \textcolor{GLcolor}{\llbracket}  \SYSTEMnt{t}  \textcolor{GLcolor}{\rrbracket}    }{%
{\SYSTEMRenameRuleTranslGtoLTmabs}{}%
}}

\newcommand{\SYSTEMRenameRuleTranslGtoLTmapp}[0]{\SYSTEMdrulename{TranslGtoLTmapp}}
\newcommand{\SYSTEMdruleTranslGtoLTmapp}[1]{\SYSTEMdrule[#1]{%
}{
 \textcolor{GLcolor}{\llbracket} \smidge   \SYSTEMnt{t_{{\mathrm{1}}}} \,  \SYSTEMnt{t_{{\mathrm{2}}}}   \smidge \textcolor{GLcolor}{\rrbracket} \;\triangleq\;    \textcolor{GLcolor}{\llbracket}  \SYSTEMnt{t_{{\mathrm{1}}}}  \textcolor{GLcolor}{\rrbracket}  \,   \textcolor{coeffectColor}{[}   \textcolor{GLcolor}{\llbracket}  \SYSTEMnt{t_{{\mathrm{2}}}}  \textcolor{GLcolor}{\rrbracket}   \textcolor{coeffectColor}{]}   }{%
{\SYSTEMRenameRuleTranslGtoLTmapp}{}%
}}

\newcommand{\SYSTEMRenameRuleTranslGtoLTmpr}[0]{\SYSTEMdrulename{TranslGtoLTmpr}}
\newcommand{\SYSTEMdruleTranslGtoLTmpr}[1]{\SYSTEMdrule[#1]{%
}{
 \textcolor{GLcolor}{\llbracket} \smidge   \textcolor{coeffectColor}{[}  \SYSTEMnt{t}  \textcolor{coeffectColor}{]}   \smidge \textcolor{GLcolor}{\rrbracket} \;\triangleq\;   \textcolor{coeffectColor}{[}   \textcolor{GLcolor}{\llbracket}  \SYSTEMnt{t}  \textcolor{GLcolor}{\rrbracket}   \textcolor{coeffectColor}{]}  }{%
{\SYSTEMRenameRuleTranslGtoLTmpr}{}%
}}

\newcommand{\SYSTEMRenameRuleTranslGtoLTmlet}[0]{\SYSTEMdrulename{TranslGtoLTmlet}}
\newcommand{\SYSTEMdruleTranslGtoLTmlet}[1]{\SYSTEMdrule[#1]{%
}{
 \textcolor{GLcolor}{\llbracket} \smidge   \mathsf{let} \, \textcolor{coeffectColor}{[}  \SYSTEMmv{x}  \textcolor{coeffectColor}{]} =  \SYSTEMnt{t_{{\mathrm{1}}}}  \, \mathsf{in} \,  \SYSTEMnt{t_{{\mathrm{2}}}}   \smidge \textcolor{GLcolor}{\rrbracket} \;\triangleq\;   \mathsf{let} \, \textcolor{coeffectColor}{[}  \SYSTEMmv{x}  \textcolor{coeffectColor}{]} =   \textcolor{GLcolor}{\llbracket}  \SYSTEMnt{t_{{\mathrm{1}}}}  \textcolor{GLcolor}{\rrbracket}   \, \mathsf{in} \,   \textcolor{GLcolor}{\llbracket}  \SYSTEMnt{t_{{\mathrm{2}}}}  \textcolor{GLcolor}{\rrbracket}   }{%
{\SYSTEMRenameRuleTranslGtoLTmlet}{}%
}}

\newcommand{\SYSTEMRenameRuleTranslGtoLTmprodi}[0]{\SYSTEMdrulename{TranslGtoLTmprodi}}
\newcommand{\SYSTEMdruleTranslGtoLTmprodi}[1]{\SYSTEMdrule[#1]{%
}{
 \textcolor{GLcolor}{\llbracket} \smidge   \langle  \SYSTEMnt{t_{{\mathrm{1}}}} ,  \SYSTEMnt{t_{{\mathrm{2}}}}  \rangle   \smidge \textcolor{GLcolor}{\rrbracket} \;\triangleq\;   \langle   \textcolor{GLcolor}{\llbracket}  \SYSTEMnt{t_{{\mathrm{1}}}}  \textcolor{GLcolor}{\rrbracket}  ,   \textcolor{GLcolor}{\llbracket}  \SYSTEMnt{t_{{\mathrm{2}}}}  \textcolor{GLcolor}{\rrbracket}   \rangle  }{%
{\SYSTEMRenameRuleTranslGtoLTmprodi}{}%
}}

\newcommand{\SYSTEMRenameRuleTranslGtoLTmuniti}[0]{\SYSTEMdrulename{TranslGtoLTmuniti}}
\newcommand{\SYSTEMdruleTranslGtoLTmuniti}[1]{\SYSTEMdrule[#1]{%
}{
 \textcolor{GLcolor}{\llbracket} \smidge   \langle \rangle   \smidge \textcolor{GLcolor}{\rrbracket} \;\triangleq\;   \langle \rangle  }{%
{\SYSTEMRenameRuleTranslGtoLTmuniti}{}%
}}

\newcommand{\SYSTEMRenameRuleTranslGtoLTmunite}[0]{\SYSTEMdrulename{TranslGtoLTmunite}}
\newcommand{\SYSTEMdruleTranslGtoLTmunite}[1]{\SYSTEMdrule[#1]{%
}{
 \textcolor{GLcolor}{\llbracket} \smidge   \mathsf{let} \, \langle \rangle =  \SYSTEMnt{t_{{\mathrm{1}}}}  \, \mathsf{in} \,  \SYSTEMnt{t_{{\mathrm{2}}}}   \smidge \textcolor{GLcolor}{\rrbracket} \;\triangleq\;   \mathsf{let} \, \langle \rangle =   \textsf{push}_{\mathrm{unit} }   \textcolor{coeffectColor}{[}   \textcolor{GLcolor}{\llbracket}  \SYSTEMnt{t_{{\mathrm{1}}}}  \textcolor{GLcolor}{\rrbracket}   \textcolor{coeffectColor}{]}    \, \mathsf{in} \,   \textcolor{GLcolor}{\llbracket}  \SYSTEMnt{t_{{\mathrm{2}}}}  \textcolor{GLcolor}{\rrbracket}   }{%
{\SYSTEMRenameRuleTranslGtoLTmunite}{}%
}}

\newcommand{\SYSTEMRenameRuleTranslGtoLTmsumiOne}[0]{\SYSTEMdrulename{TranslGtoLTmsumi1}}
\newcommand{\SYSTEMdruleTranslGtoLTmsumiOne}[1]{\SYSTEMdrule[#1]{%
}{
 \textcolor{GLcolor}{\llbracket} \smidge   \mathsf{inj}_1 \,  \SYSTEMnt{t}   \smidge \textcolor{GLcolor}{\rrbracket} \;\triangleq\;   \mathsf{inj}_1 \,   \textcolor{GLcolor}{\llbracket}  \SYSTEMnt{t}  \textcolor{GLcolor}{\rrbracket}   }{%
{\SYSTEMRenameRuleTranslGtoLTmsumiOne}{}%
}}

\newcommand{\SYSTEMRenameRuleTranslGtoLTmsumiTwo}[0]{\SYSTEMdrulename{TranslGtoLTmsumi2}}
\newcommand{\SYSTEMdruleTranslGtoLTmsumiTwo}[1]{\SYSTEMdrule[#1]{%
}{
 \textcolor{GLcolor}{\llbracket} \smidge   \mathsf{inj}_2 \,  \SYSTEMnt{t}   \smidge \textcolor{GLcolor}{\rrbracket} \;\triangleq\;   \mathsf{inj}_2 \,   \textcolor{GLcolor}{\llbracket}  \SYSTEMnt{t}  \textcolor{GLcolor}{\rrbracket}   }{%
{\SYSTEMRenameRuleTranslGtoLTmsumiTwo}{}%
}}

\newcommand{\SYSTEMRenameRuleTranslGtoLTybaseTy}[0]{\SYSTEMdrulename{TranslGtoLTybaseTy}}
\newcommand{\SYSTEMdruleTranslGtoLTybaseTy}[1]{\SYSTEMdrule[#1]{%
}{
 \textcolor{GLcolor}{\llbracket} \smidge   \mathrm{K}   \smidge \textcolor{GLcolor}{\rrbracket} \;\triangleq\;   \mathrm{K}  }{%
{\SYSTEMRenameRuleTranslGtoLTybaseTy}{}%
}}

\newcommand{\SYSTEMRenameRuleTranslGtoLTyfunTy}[0]{\SYSTEMdrulename{TranslGtoLTyfunTy}}
\newcommand{\SYSTEMdruleTranslGtoLTyfunTy}[1]{\SYSTEMdrule[#1]{%
}{
 \textcolor{GLcolor}{\llbracket} \smidge   \SYSTEMnt{A}  \xrightarrow{\textcolor{coeffectColor}{ \SYSTEMnt{r} } }  \SYSTEMnt{B}   \smidge \textcolor{GLcolor}{\rrbracket} \;\triangleq\;     \textcolor{coeffectColor}{\square_{ \SYSTEMnt{r} } }   \textcolor{GLcolor}{\llbracket}  \SYSTEMnt{A}  \textcolor{GLcolor}{\rrbracket}     \multimap   \textcolor{GLcolor}{\llbracket}  \SYSTEMnt{B}  \textcolor{GLcolor}{\rrbracket}   }{%
{\SYSTEMRenameRuleTranslGtoLTyfunTy}{}%
}}

\newcommand{\SYSTEMRenameRuleTranslGtoLTyboxTy}[0]{\SYSTEMdrulename{TranslGtoLTyboxTy}}
\newcommand{\SYSTEMdruleTranslGtoLTyboxTy}[1]{\SYSTEMdrule[#1]{%
}{
 \textcolor{GLcolor}{\llbracket} \smidge   \textcolor{coeffectColor}{\square_{ \SYSTEMnt{r} } }  \SYSTEMnt{A}   \smidge \textcolor{GLcolor}{\rrbracket} \;\triangleq\;   \textcolor{coeffectColor}{\square_{ \SYSTEMnt{r} } }   \textcolor{GLcolor}{\llbracket}  \SYSTEMnt{A}  \textcolor{GLcolor}{\rrbracket}   }{%
{\SYSTEMRenameRuleTranslGtoLTyboxTy}{}%
}}

\newcommand{\SYSTEMRenameRuleTranslGtoLTyprodTy}[0]{\SYSTEMdrulename{TranslGtoLTyprodTy}}
\newcommand{\SYSTEMdruleTranslGtoLTyprodTy}[1]{\SYSTEMdrule[#1]{%
}{
 \textcolor{GLcolor}{\llbracket} \smidge   \SYSTEMnt{A}  \times  \SYSTEMnt{B}   \smidge \textcolor{GLcolor}{\rrbracket} \;\triangleq\;    \textcolor{GLcolor}{\llbracket}  \SYSTEMnt{A}  \textcolor{GLcolor}{\rrbracket}   \otimes   \textcolor{GLcolor}{\llbracket}  \SYSTEMnt{B}  \textcolor{GLcolor}{\rrbracket}   }{%
{\SYSTEMRenameRuleTranslGtoLTyprodTy}{}%
}}

\newcommand{\SYSTEMRenameRuleTranslGtoLTysumTy}[0]{\SYSTEMdrulename{TranslGtoLTysumTy}}
\newcommand{\SYSTEMdruleTranslGtoLTysumTy}[1]{\SYSTEMdrule[#1]{%
}{
 \textcolor{GLcolor}{\llbracket} \smidge   \SYSTEMnt{A}  +  \SYSTEMnt{B}   \smidge \textcolor{GLcolor}{\rrbracket} \;\triangleq\;    \textcolor{GLcolor}{\llbracket}  \SYSTEMnt{A}  \textcolor{GLcolor}{\rrbracket}   \oplus   \textcolor{GLcolor}{\llbracket}  \SYSTEMnt{B}  \textcolor{GLcolor}{\rrbracket}   }{%
{\SYSTEMRenameRuleTranslGtoLTysumTy}{}%
}}

\newcommand{\SYSTEMRenameRuleTranslGtoLTyunitTy}[0]{\SYSTEMdrulename{TranslGtoLTyunitTy}}
\newcommand{\SYSTEMdruleTranslGtoLTyunitTy}[1]{\SYSTEMdrule[#1]{%
}{
 \textcolor{GLcolor}{\llbracket} \smidge   \mathrm{unit}   \smidge \textcolor{GLcolor}{\rrbracket} \;\triangleq\;   \mathrm{unit}  }{%
{\SYSTEMRenameRuleTranslGtoLTyunitTy}{}%
}}

\newcommand{\SYSTEMRenameRuleTranslGtoLGtxtempty}[0]{\SYSTEMdrulename{TranslGtoLGtxtempty}}
\newcommand{\SYSTEMdruleTranslGtoLGtxtempty}[1]{\SYSTEMdrule[#1]{%
}{
 \textcolor{GLcolor}{\llbracket} \smidge   \emptyset   \smidge \textcolor{GLcolor}{\rrbracket} \;\triangleq\;   \emptyset  }{%
{\SYSTEMRenameRuleTranslGtoLGtxtempty}{}%
}}

\newcommand{\SYSTEMRenameRuleTranslGtoLGtxtgrad}[0]{\SYSTEMdrulename{TranslGtoLGtxtgrad}}
\newcommand{\SYSTEMdruleTranslGtoLGtxtgrad}[1]{\SYSTEMdrule[#1]{%
}{
 \textcolor{GLcolor}{\llbracket} \smidge   \Delta ,   \SYSTEMmv{x}  :_{\textcolor{coeffectColor}{ \SYSTEMnt{r} } }  \SYSTEMnt{A}    \smidge \textcolor{GLcolor}{\rrbracket} \;\triangleq\;    \textcolor{GLcolor}{\llbracket}  \Delta  \textcolor{GLcolor}{\rrbracket}  ,   \SYSTEMmv{x}  : \textcolor{coeffectColor}{[}   \textcolor{GLcolor}{\llbracket}  \SYSTEMnt{A}  \textcolor{GLcolor}{\rrbracket}  {\textcolor{coeffectColor}{]_{ \SYSTEMnt{r} } } }   }{%
{\SYSTEMRenameRuleTranslGtoLGtxtgrad}{}%
}}

\newcommand{\SYSTEMRenameRulePredicateGradednil}[0]{\SYSTEMdrulename{PredicateGradednil}}
\newcommand{\SYSTEMdrulePredicateGradednil}[1]{\SYSTEMdrule[#1]{%
}{
 \mathrm{graded}(  \emptyset  ) }{%
{\SYSTEMRenameRulePredicateGradednil}{}%
}}

\newcommand{\SYSTEMRenameRulePredicateGradedgrad}[0]{\SYSTEMdrulename{PredicateGradedgrad}}
\newcommand{\SYSTEMdrulePredicateGradedgrad}[1]{\SYSTEMdrule[#1]{%
\SYSTEMpremise{ \mathrm{graded}( \Gamma ) }%
}{
 \mathrm{graded}(  (   \Gamma ,   \SYSTEMmv{x}  : \textcolor{coeffectColor}{[}  \SYSTEMnt{A} {\textcolor{coeffectColor}{]_{ \SYSTEMnt{r} } } }    )  ) }{%
{\SYSTEMRenameRulePredicateGradedgrad}{}%
}}

\newcommand{\SYSTEMRenameRulePredicateGradedAtnil}[0]{\SYSTEMdrulename{PredicateGradedAtnil}}
\newcommand{\SYSTEMdrulePredicateGradedAtnil}[1]{\SYSTEMdrule[#1]{%
}{
 \mathrm{graded}(  \emptyset  , \textcolor{coeffectColor}{ \SYSTEMnt{r} }) }{%
{\SYSTEMRenameRulePredicateGradedAtnil}{}%
}}

\newcommand{\SYSTEMRenameRulePredicateGradedAtgrad}[0]{\SYSTEMdrulename{PredicateGradedAtgrad}}
\newcommand{\SYSTEMdrulePredicateGradedAtgrad}[1]{\SYSTEMdrule[#1]{%
\SYSTEMpremise{ \mathrm{graded}( \Gamma , \textcolor{coeffectColor}{ \SYSTEMnt{r} }) }%
}{
 \mathrm{graded}(  (   \Gamma ,   \SYSTEMmv{x}  : \textcolor{coeffectColor}{[}  \SYSTEMnt{A} {\textcolor{coeffectColor}{]_{ \SYSTEMnt{r} } } }    )  , \textcolor{coeffectColor}{ \SYSTEMnt{r} }) }{%
{\SYSTEMRenameRulePredicateGradedAtgrad}{}%
}}

\newcommand{\SYSTEMRenameRulesubstvarneq}[0]{\SYSTEMdrulename{substvarneq}}
\newcommand{\SYSTEMdrulesubstvarneq}[1]{\SYSTEMdrule[#1]{%
\SYSTEMpremise{ \SYSTEMmv{x}  \neq  \SYSTEMmv{y} }%
}{
 [  \SYSTEMnt{t}  /  \SYSTEMmv{x}  ]  \SYSTEMmv{y}  \;\triangleq\;  \SYSTEMmv{y} }{%
{\SYSTEMRenameRulesubstvarneq}{}%
}}

\newcommand{\SYSTEMRenameRulesubstvareq}[0]{\SYSTEMdrulename{substvareq}}
\newcommand{\SYSTEMdrulesubstvareq}[1]{\SYSTEMdrule[#1]{%
}{
 [  \SYSTEMnt{t}  /  \SYSTEMmv{x}  ]  \SYSTEMmv{x}  \;\triangleq\;  \SYSTEMnt{t} }{%
{\SYSTEMRenameRulesubstvareq}{}%
}}

\newcommand{\SYSTEMRenameRulesubstapp}[0]{\SYSTEMdrulename{substapp}}
\newcommand{\SYSTEMdrulesubstapp}[1]{\SYSTEMdrule[#1]{%
}{
 [  \SYSTEMnt{t}  /  \SYSTEMmv{x}  ]  \SYSTEMsym{(}   \SYSTEMnt{t_{{\mathrm{1}}}} \,  \SYSTEMnt{t_{{\mathrm{2}}}}   \SYSTEMsym{)}  \;\triangleq\;   \SYSTEMsym{(}   [  \SYSTEMnt{t}  /  \SYSTEMmv{x}  ]  \SYSTEMnt{t_{{\mathrm{1}}}}   \SYSTEMsym{)} \,  \SYSTEMsym{(}   [  \SYSTEMnt{t}  /  \SYSTEMmv{x}  ]  \SYSTEMnt{t_{{\mathrm{2}}}}   \SYSTEMsym{)}  }{%
{\SYSTEMRenameRulesubstapp}{}%
}}

\newcommand{\SYSTEMRenameRulesubstabs}[0]{\SYSTEMdrulename{substabs}}
\newcommand{\SYSTEMdrulesubstabs}[1]{\SYSTEMdrule[#1]{%
\SYSTEMpremise{  \SYSTEMmv{y'}  \,\#\,  \SYSTEMnt{t}  }%
}{
 [  \SYSTEMnt{t}  /  \SYSTEMmv{x}  ]   \lambda  \SYSTEMmv{y}  .  \SYSTEMnt{t'}   \;\triangleq\;   \lambda  \SYSTEMmv{y'}  .   [  \SYSTEMnt{t}  /  \SYSTEMmv{x}  ]   [  \SYSTEMmv{y'}  /  \SYSTEMmv{y}  ]  \SYSTEMnt{t'}    }{%
{\SYSTEMRenameRulesubstabs}{}%
}}

\newcommand{\SYSTEMRenameRulesubstpr}[0]{\SYSTEMdrulename{substpr}}
\newcommand{\SYSTEMdrulesubstpr}[1]{\SYSTEMdrule[#1]{%
}{
 [  \SYSTEMnt{t}  /  \SYSTEMmv{x}  ]   \textcolor{coeffectColor}{[}  \SYSTEMnt{t'}  \textcolor{coeffectColor}{]}   \;\triangleq\;   \textcolor{coeffectColor}{[}   [  \SYSTEMnt{t}  /  \SYSTEMmv{x}  ]  \SYSTEMnt{t'}   \textcolor{coeffectColor}{]}  }{%
{\SYSTEMRenameRulesubstpr}{}%
}}

\newcommand{\SYSTEMRenameRulesubstlet}[0]{\SYSTEMdrulename{substlet}}
\newcommand{\SYSTEMdrulesubstlet}[1]{\SYSTEMdrule[#1]{%
\SYSTEMpremise{  \SYSTEMmv{y'}  \,\#\,  \SYSTEMnt{t}  }%
}{
 [  \SYSTEMnt{t}  /  \SYSTEMmv{x}  ]   \mathsf{let} \, \textcolor{coeffectColor}{[}  \SYSTEMmv{y}  \textcolor{coeffectColor}{]} =  \SYSTEMnt{t_{{\mathrm{1}}}}  \, \mathsf{in} \,  \SYSTEMnt{t_{{\mathrm{2}}}}   \;\triangleq\;   \mathsf{let} \, \textcolor{coeffectColor}{[}  \SYSTEMmv{y'}  \textcolor{coeffectColor}{]} =   [  \SYSTEMnt{t}  /  \SYSTEMmv{x}  ]  \SYSTEMnt{t_{{\mathrm{1}}}}   \, \mathsf{in} \,   [  \SYSTEMnt{t}  /  \SYSTEMmv{x}  ]   [  \SYSTEMmv{y'}  /  \SYSTEMmv{y}  ]  \SYSTEMnt{t_{{\mathrm{2}}}}    }{%
{\SYSTEMRenameRulesubstlet}{}%
}}

\newcommand{\SYSTEMRenameRulesubstsumiOne}[0]{\SYSTEMdrulename{substsumi1}}
\newcommand{\SYSTEMdrulesubstsumiOne}[1]{\SYSTEMdrule[#1]{%
}{
 [  \SYSTEMnt{t}  /  \SYSTEMmv{x}  ]   \mathsf{inj}_1 \,  \SYSTEMnt{t'}   \;\triangleq\;   \mathsf{inj}_1 \,   [  \SYSTEMnt{t}  /  \SYSTEMmv{x}  ]  \SYSTEMnt{t'}   }{%
{\SYSTEMRenameRulesubstsumiOne}{}%
}}

\newcommand{\SYSTEMRenameRulesubstsumiTwo}[0]{\SYSTEMdrulename{substsumi2}}
\newcommand{\SYSTEMdrulesubstsumiTwo}[1]{\SYSTEMdrule[#1]{%
}{
 [  \SYSTEMnt{t}  /  \SYSTEMmv{x}  ]   \mathsf{inj}_2 \,  \SYSTEMnt{t'}   \;\triangleq\;   \mathsf{inj}_2 \,   [  \SYSTEMnt{t}  /  \SYSTEMmv{x}  ]  \SYSTEMnt{t'}   }{%
{\SYSTEMRenameRulesubstsumiTwo}{}%
}}

\newcommand{\SYSTEMRenameRulesubstsume}[0]{\SYSTEMdrulename{substsume}}
\newcommand{\SYSTEMdrulesubstsume}[1]{\SYSTEMdrule[#1]{%
\SYSTEMpremise{  \SYSTEMmv{y'} ,   \SYSTEMmv{z'}  \,\#\,  \SYSTEMnt{t}   }%
}{
 [  \SYSTEMnt{t}  /  \SYSTEMmv{x}  ]   \mathsf{case} \,  \SYSTEMnt{t'}  \, \mathsf{of} \, \{ \mathsf{inj1} \,  \SYSTEMmv{y}  \rightarrow  \SYSTEMnt{t_{{\mathrm{1}}}}  ; \, \mathsf{inj2} \,  \SYSTEMmv{z}  \rightarrow  \SYSTEMnt{t_{{\mathrm{2}}}}  \}   \;\triangleq\;   \mathsf{case} \,   [  \SYSTEMnt{t}  /  \SYSTEMmv{x}  ]  \SYSTEMnt{t'}   \, \mathsf{of} \, \{ \mathsf{inj1} \,  \SYSTEMmv{y'}  \rightarrow   [  \SYSTEMnt{t}  /  \SYSTEMmv{x}  ]   [  \SYSTEMmv{y'}  /  \SYSTEMmv{y}  ]  \SYSTEMnt{t_{{\mathrm{1}}}}    ; \, \mathsf{inj2} \,  \SYSTEMmv{z'}  \rightarrow   [  \SYSTEMnt{t}  /  \SYSTEMmv{x}  ]   [  \SYSTEMmv{z'}  /  \SYSTEMmv{z}  ]  \SYSTEMnt{t_{{\mathrm{2}}}}    \}  }{%
{\SYSTEMRenameRulesubstsume}{}%
}}

\newcommand{\SYSTEMRenameRulesubstprodi}[0]{\SYSTEMdrulename{substprodi}}
\newcommand{\SYSTEMdrulesubstprodi}[1]{\SYSTEMdrule[#1]{%
}{
 [  \SYSTEMnt{t}  /  \SYSTEMmv{x}  ]   \langle  \SYSTEMnt{t_{{\mathrm{1}}}} ,  \SYSTEMnt{t_{{\mathrm{2}}}}  \rangle   \;\triangleq\;   \langle   [  \SYSTEMnt{t}  /  \SYSTEMmv{x}  ]  \SYSTEMnt{t_{{\mathrm{1}}}}  ,   [  \SYSTEMnt{t}  /  \SYSTEMmv{x}  ]  \SYSTEMnt{t_{{\mathrm{2}}}}   \rangle  }{%
{\SYSTEMRenameRulesubstprodi}{}%
}}

\newcommand{\SYSTEMRenameRulesubstprode}[0]{\SYSTEMdrulename{substprode}}
\newcommand{\SYSTEMdrulesubstprode}[1]{\SYSTEMdrule[#1]{%
\SYSTEMpremise{  \SYSTEMmv{y'} ,   \SYSTEMmv{z'}  \,\#\,  \SYSTEMnt{t}   }%
}{
 [  \SYSTEMnt{t}  /  \SYSTEMmv{x}  ]   \mathsf{let} \, \langle  \SYSTEMmv{y} ,  \SYSTEMmv{z}  \rangle =  \SYSTEMnt{t_{{\mathrm{1}}}}  \, \mathsf{in} \,  \SYSTEMnt{t_{{\mathrm{2}}}}   \;\triangleq\;   \mathsf{let} \, \langle  \SYSTEMmv{y'} ,  \SYSTEMmv{z'}  \rangle =   [  \SYSTEMnt{t}  /  \SYSTEMmv{x}  ]  \SYSTEMnt{t_{{\mathrm{1}}}}   \, \mathsf{in} \,   [  \SYSTEMnt{t}  /  \SYSTEMmv{x}  ]   [  \SYSTEMmv{z'}  /  \SYSTEMmv{z}  ]   [  \SYSTEMmv{y'}  /  \SYSTEMmv{y}  ]  \SYSTEMnt{t_{{\mathrm{2}}}}     }{%
{\SYSTEMRenameRulesubstprode}{}%
}}

\newcommand{\SYSTEMRenameRulesubstuniti}[0]{\SYSTEMdrulename{substuniti}}
\newcommand{\SYSTEMdrulesubstuniti}[1]{\SYSTEMdrule[#1]{%
}{
 [  \SYSTEMnt{t}  /  \SYSTEMmv{x}  ]   \langle \rangle   \;\triangleq\;   \langle \rangle  }{%
{\SYSTEMRenameRulesubstuniti}{}%
}}

\newcommand{\SYSTEMRenameRulesubstunite}[0]{\SYSTEMdrulename{substunite}}
\newcommand{\SYSTEMdrulesubstunite}[1]{\SYSTEMdrule[#1]{%
}{
 [  \SYSTEMnt{t}  /  \SYSTEMmv{x}  ]   \mathsf{let} \, \langle \rangle =  \SYSTEMnt{t_{{\mathrm{1}}}}  \, \mathsf{in} \,  \SYSTEMnt{t_{{\mathrm{2}}}}   \;\triangleq\;   \mathsf{let} \, \langle \rangle =  \SYSTEMnt{t_{{\mathrm{1}}}}  \, \mathsf{in} \,   [  \SYSTEMnt{t}  /  \SYSTEMmv{x}  ]  \SYSTEMnt{t_{{\mathrm{2}}}}   }{%
{\SYSTEMRenameRulesubstunite}{}%
}}

\providecommand{\SYSTEMRenameRuleGradvar}{}
\providecommand{\SYSTEMRenameRuleGradabs}{}
\providecommand{\SYSTEMRenameRuleGradapp}{}
\providecommand{\SYSTEMRenameRuleGradweak}{}
\providecommand{\SYSTEMRenameRuleGradBoxpr}{}
\providecommand{\SYSTEMRenameRuleGradBoxlet}{}
\providecommand{\SYSTEMRenameRuleGradBoxletGen}{}
\providecommand{\SYSTEMRenameRuleGradapprox}{}
\providecommand{\SYSTEMRenameRuleGradprodi}{}
\providecommand{\SYSTEMRenameRuleGradprode}{}
\providecommand{\SYSTEMRenameRuleGraduniti}{}
\providecommand{\SYSTEMRenameRuleGradunite}{}
\providecommand{\SYSTEMRenameRuleGradsumiOne}{}
\providecommand{\SYSTEMRenameRuleGradsumiTwo}{}
\providecommand{\SYSTEMRenameRuleGradsume}{}
\providecommand{\SYSTEMRenameRuleGradPolytyAbs}{}
\providecommand{\SYSTEMRenameRuleGradPolytyApp}{}

\providecommand{\SYSTEMRenameRuleLinvar}{}
\providecommand{\SYSTEMRenameRuleLinabs}{}
\providecommand{\SYSTEMRenameRuleLinapp}{}
\providecommand{\SYSTEMRenameRuleLinweak}{}
\providecommand{\SYSTEMRenameRuleLinpr}{}
\providecommand{\SYSTEMRenameRuleLinlet}{}
\providecommand{\SYSTEMRenameRuleLinder}{}
\providecommand{\SYSTEMRenameRuleLinapprox}{}
\providecommand{\SYSTEMRenameRuleLinprodi}{}
\providecommand{\SYSTEMRenameRuleLinprode}{}
\providecommand{\SYSTEMRenameRuleLinuniti}{}
\providecommand{\SYSTEMRenameRuleLinunite}{}
\providecommand{\SYSTEMRenameRuleLinpushprod}{}
\providecommand{\SYSTEMRenameRuleLinsumiOne}{}
\providecommand{\SYSTEMRenameRuleLinsumiTwo}{}
\providecommand{\SYSTEMRenameRuleLinsume}{}
\providecommand{\SYSTEMRenameRuleLinpushsum}{}
\providecommand{\SYSTEMRenameRuleLinpushunit}{}

\providecommand{\SYSTEMRenameRuleSemGrdbeta}{}
\providecommand{\SYSTEMRenameRuleSemGrdcongAppL}{}
\providecommand{\SYSTEMRenameRuleSemGrdcongAbs}{}
\providecommand{\SYSTEMRenameRuleSemGrdcongCase}{}
\providecommand{\SYSTEMRenameRuleSemGrdcaseInjOne}{}
\providecommand{\SYSTEMRenameRuleSemGrdcaseInjTwo}{}
\providecommand{\SYSTEMRenameRuleSemGrdprodCong}{}
\providecommand{\SYSTEMRenameRuleSemGrdprodBeta}{}
\providecommand{\SYSTEMRenameRuleSemGrdunitCong}{}
\providecommand{\SYSTEMRenameRuleSemGrdunitBeta}{}

\providecommand{\SYSTEMRenameRuleSemLinbeta}{}
\providecommand{\SYSTEMRenameRuleSemLincongAppL}{}
\providecommand{\SYSTEMRenameRuleSemLinbetaBox}{}
\providecommand{\SYSTEMRenameRuleSemLincongLetL}{}
\providecommand{\SYSTEMRenameRuleSemLincongCase}{}
\providecommand{\SYSTEMRenameRuleSemLincaseInjOne}{}
\providecommand{\SYSTEMRenameRuleSemLincaseInjTwo}{}
\providecommand{\SYSTEMRenameRuleSemLinprodCong}{}
\providecommand{\SYSTEMRenameRuleSemLinprodBeta}{}
\providecommand{\SYSTEMRenameRuleSemLinunitCong}{}
\providecommand{\SYSTEMRenameRuleSemLinunitBeta}{}

\providecommand{\SYSTEMRenameRuleSemLinpushProdBoxCong}{}
\providecommand{\SYSTEMRenameRuleSemLinpushSumBoxCong}{}

\providecommand{\SYSTEMRenameRuleSemLinpushUnitCong}{}
\providecommand{\SYSTEMRenameRuleSemLinpushUnitBoxCong}{}
\providecommand{\SYSTEMRenameRuleSemLinpushUnit}{}
\providecommand{\SYSTEMRenameRuleSemLinpushProdCong}{}
\providecommand{\SYSTEMRenameRuleSemLinpushProd}{}
\providecommand{\SYSTEMRenameRuleSemLinpushSumCong}{}

\providecommand{\SYSTEMRenameRuleSemLinpushSumInjOne}{}
\providecommand{\SYSTEMRenameRuleSemLinpushSumInjTwo}{}
\providecommand{\SYSTEMRenameRuleSemLinpushSumInjGen}{}

\providecommand{\SYSTEMRenameRulePredicateGradednil}{}
\providecommand{\SYSTEMRenameRulePredicateGradedgrad}{}
\providecommand{\SYSTEMRenameRulePredicateGradedAtnil}{}
\providecommand{\SYSTEMRenameRulePredicateGradedAtgrad}{}

\providecommand{\SYSTEMRenameRuleSemGrdModbetaBox}{}
\providecommand{\SYSTEMRenameRuleSemGrdModcongLetL}{}

\renewcommand{\SYSTEMRenameRuleSemGrdModbetaBox}{$\Box\beta_{\textsf{G}}$}
\renewcommand{\SYSTEMRenameRuleSemGrdModcongLetL}{$\textsc{congLet}_{\textsf{G}}$}

\renewcommand{\SYSTEMRenameRuleGradvar}{\textsc{var$_{\textsf{G}}$}}
\renewcommand{\SYSTEMRenameRuleGradabs}{\textsc{abs$_{\textsf{G}}$}}
\renewcommand{\SYSTEMRenameRuleGradapp}{\textsc{app$_{\textsf{G}}$}}
\renewcommand{\SYSTEMRenameRuleGradweak}{\textsc{weak$_{\textsf{G}}$}}
\renewcommand{\SYSTEMRenameRuleGradBoxpr}{\textsc{pr$_{\textsf{G}}$}}
\renewcommand{\SYSTEMRenameRuleGradBoxlet}{\textsc{let$_{\textsf{G}}$}}
\renewcommand{\SYSTEMRenameRuleGradBoxletGen}{\textsc{let$_{\textsf{G}}^\prime$}}
\renewcommand{\SYSTEMRenameRuleGradapprox}{\textsc{approx$_{\textsf{G}}$}}
\renewcommand{\SYSTEMRenameRuleGradprodi}{\textsc{$\times$-i$_{\textsf{G}}$}}
\renewcommand{\SYSTEMRenameRuleGradprode}{\textsc{$\times$-e$_{\textsf{G}}$}}
\renewcommand{\SYSTEMRenameRuleGraduniti}{\textsc{$\mathsf{unit}$-i$_{\textsf{G}}$}}
\renewcommand{\SYSTEMRenameRuleGradunite}{\textsc{$\mathsf{unit}$-e$_{\textsf{G}}$}}
\renewcommand{\SYSTEMRenameRuleGradsumiOne}{\textsc{$+$-i$_{1\textsf{G}}$}}
\renewcommand{\SYSTEMRenameRuleGradsumiTwo}{\textsc{$+$-i$_{2\textsf{G}}$}}
\renewcommand{\SYSTEMRenameRuleGradsume}{\textsc{$+$-e$_{\textsf{G}}$}}
\renewcommand{\SYSTEMRenameRuleGradPolytyAbs}{\textsc{tyabs$_{\textsf{G}}$}}
\renewcommand{\SYSTEMRenameRuleGradPolytyApp}{\textsc{tyapp$_{\textsf{G}}$}}

\renewcommand{\SYSTEMRenameRuleLinvar}{\textsc{var$_{\textsf{L}}$}}
\renewcommand{\SYSTEMRenameRuleLinabs}{\textsc{abs$_{\textsf{L}}$}}
\renewcommand{\SYSTEMRenameRuleLinapp}{\textsc{app$_{\textsf{L}}$}}
\renewcommand{\SYSTEMRenameRuleLinweak}{\textsc{weak$_{\textsf{L}}$}}
\renewcommand{\SYSTEMRenameRuleLinpr}{\textsc{pr$_{\textsf{L}}$}}

\renewcommand{\SYSTEMRenameRuleLinlet}{\textsc{let$_{\textsf{L}}$}}
\renewcommand{\SYSTEMRenameRuleLinder}{\textsc{der$_{\textsf{L}}$}}
\renewcommand{\SYSTEMRenameRuleLinapprox}{\textsc{approx$_{\textsf{L}}$}}
\renewcommand{\SYSTEMRenameRuleLinprodi}{\textsc{$\otimes$-i$_{\textsf{L}}$}}
\renewcommand{\SYSTEMRenameRuleLinprode}{\textsc{$\otimes$-e$_{\textsf{L}}$}}
\renewcommand{\SYSTEMRenameRuleLinuniti}{\textsc{$\mathbf{1}$-i$_{\textsf{L}}$}}
\renewcommand{\SYSTEMRenameRuleLinunite}{\textsc{$\mathbf{1}$-e$_{\textsf{L}}$}}
\renewcommand{\SYSTEMRenameRuleLinpushprod}{push$_\otimes$}
\renewcommand{\SYSTEMRenameRuleLinsumiOne}{\textsc{$\oplus$-i$_{1\textsf{L}}$}}
\renewcommand{\SYSTEMRenameRuleLinsumiTwo}{\textsc{$\oplus$-i$_{2\textsf{L}}$}}
\renewcommand{\SYSTEMRenameRuleLinsume}{\textsc{$\oplus$-e$_{\textsf{L}}$}}
\renewcommand{\SYSTEMRenameRuleLinpushsum}{push$_\oplus$}
\renewcommand{\SYSTEMRenameRuleLinpushunit}{push$_\mathsf{unit}$}

\renewcommand{\SYSTEMRenameRuleSemGrdbeta}{$\beta_{\textsf{G}}$}
\renewcommand{\SYSTEMRenameRuleSemGrdcongAppL}{$\textsc{congApp}_{\textsf{G}}$}
\renewcommand{\SYSTEMRenameRuleSemGrdcongAbs}{$\textsc{congAbs}_{\textsf{G}}$}
\renewcommand{\SYSTEMRenameRuleSemGrdcongCase}{$\textsc{congCase}_\textsf{G}$}
\renewcommand{\SYSTEMRenameRuleSemGrdcaseInjOne}{$\textsc{caseInj1}_\textsf{G}$}
\renewcommand{\SYSTEMRenameRuleSemGrdcaseInjTwo}{$\textsc{caseInj2}_\textsf{G}$}
\renewcommand{\SYSTEMRenameRuleSemGrdprodCong}{$\textsc{congProd}_\textsf{G}$}
\renewcommand{\SYSTEMRenameRuleSemGrdprodBeta}{$\beta\textsc{Prod}_\textsf{G}$}
\renewcommand{\SYSTEMRenameRuleSemGrdunitCong}{$\textsc{congUnit}_\textsf{G}$}
\renewcommand{\SYSTEMRenameRuleSemGrdunitBeta}{$\beta\textsc{Unit}_\textsf{G}$}

\providecommand{\SYSTEMRenameRuleSemGrdPolycongTyApp}{}
\providecommand{\SYSTEMRenameRuleSemGrdPolytyBeta}{}

\renewcommand{\SYSTEMRenameRuleSemGrdPolycongTyApp}{$\textsc{cong}$\@}
\renewcommand{\SYSTEMRenameRuleSemGrdPolytyBeta}{$\beta_{\text{\@}}$}

\renewcommand{\SYSTEMRenameRuleSemLinbeta}{$\beta_{\textsf{L}}$}
\renewcommand{\SYSTEMRenameRuleSemLincongAppL}{$\textsc{congApp}_{\textsf{L}}$}
\renewcommand{\SYSTEMRenameRuleSemLinbetaBox}{$\Box\beta_{\textsf{L}}$}
\renewcommand{\SYSTEMRenameRuleSemLincongLetL}{$\textsc{congLet}_{\textsf{L}}$}
\renewcommand{\SYSTEMRenameRuleSemLincongCase}{$\textsc{congCase}_\textsf{L}$}
\renewcommand{\SYSTEMRenameRuleSemLincaseInjOne}{$\textsc{caseInj1}_\textsf{L}$}
\renewcommand{\SYSTEMRenameRuleSemLincaseInjTwo}{$\textsc{caseInj2}_\textsf{L}$}
\renewcommand{\SYSTEMRenameRuleSemLinprodCong}{$\textsc{congProd}_\textsf{L}$}
\renewcommand{\SYSTEMRenameRuleSemLinprodBeta}{$\beta_\textsf{L}$}
\renewcommand{\SYSTEMRenameRuleSemLinunitCong}{$\textsc{congUnit}_\textsf{L}$}
\renewcommand{\SYSTEMRenameRuleSemLinunitBeta}{$\beta\textsc{Unit}_\textsf{L}$}
\renewcommand{\SYSTEMRenameRuleSemLinpushProd}{$\beta\textsc{Push}_{\otimes \textsf{L}}$}
\renewcommand{\SYSTEMRenameRuleSemLinpushUnit}{$\beta\textsc{Push}_{\mathsf{unit} \textsf{L}}$}

\renewcommand{\SYSTEMRenameRuleSemLinpushProdBoxCong}{$\textsc{congPush}^\Box_{\otimes \textsf{L}}$}
\renewcommand{\SYSTEMRenameRuleSemLinpushSumBoxCong}{$\textsc{congPush}^\Box_{\oplus \textsf{L}}$}
\renewcommand{\SYSTEMRenameRuleSemLinpushUnitBoxCong}{$\textsc{congPush}^\Box_{\mathsf{unit} \textsf{L}}$}
\renewcommand{\SYSTEMRenameRuleSemLinpushUnitCong}{$\textsc{congPush}_{\mathsf{unit} \textsf{L}}$}
\renewcommand{\SYSTEMRenameRuleSemLinpushProdCong}{$\textsc{congPush}_{\otimes \textsf{L}}$}
\renewcommand{\SYSTEMRenameRuleSemLinpushSumCong}{$\textsc{congPush}_{\oplus \textsf{L}}$}
\renewcommand{\SYSTEMRenameRuleSemLinpushSumInjOne}{$\beta\textsc{push}_{\oplus1 \textsf{L}}$}
\renewcommand{\SYSTEMRenameRuleSemLinpushSumInjGen}{$\beta\textsc{push}_{\oplus_i \textsf{L}}$}
\renewcommand{\SYSTEMRenameRuleSemLinpushSumInjTwo}{$\beta\textsc{push}_{\oplus2 \textsf{L}}$}

\renewcommand{\SYSTEMRenameRulePredicateGradednil}{}%{$\textsc{g}_\textsc{nil}$}
\renewcommand{\SYSTEMRenameRulePredicateGradedgrad}{}%{$\textsc{g}_\textsc{ind}$}
\renewcommand{\SYSTEMRenameRulePredicateGradedAtnil}{}%{$\textsc{g}_\textsc{nil}^2$}
\renewcommand{\SYSTEMRenameRulePredicateGradedAtgrad}{}%{$\textsc{g}_\textsc{ind}^2$}

\renewcommand{\SYSTEMdrule}[4][]{{\displaystyle\frac{\begin{array}{c}#2\end{array}}{#3}\;\SYSTEMdrulename{#4}}}

\providecommand{\SYSTEMRenameRuleLinEqbeta}{}
\providecommand{\SYSTEMRenameRuleLinEqeta}{}
\providecommand{\SYSTEMRenameRuleLinEqbetaBox}{}
\providecommand{\SYSTEMRenameRuleLinEqetaBox}{}
\providecommand{\SYSTEMRenameRuleLinEqletCommBox}{}
\providecommand{\SYSTEMRenameRuleLinEqletCommOne}{}
\providecommand{\SYSTEMRenameRuleLinEqletCommTwo}{}
\providecommand{\SYSTEMRenameRuleLinEqcongApp}{}
\providecommand{\SYSTEMRenameRuleLinEqcongApp}{}
\providecommand{\SYSTEMRenameRuleLinEqcongAbs}{}
\providecommand{\SYSTEMRenameRuleLinEqcongPr}{}
\providecommand{\SYSTEMRenameRuleLinEqcongLet}{}
\providecommand{\SYSTEMRenameRuleLinEqcongLet}{}

\renewcommand{\SYSTEMRenameRuleLinEqbeta}{$\equiv_{\textsc{l}-\beta}$}
\renewcommand{\SYSTEMRenameRuleLinEqeta}{$\equiv_{\textsc{l}-\eta}$}
\renewcommand{\SYSTEMRenameRuleLinEqbetaBox}{$\equiv_{\textsc{l}-\Box\beta}$}
\renewcommand{\SYSTEMRenameRuleLinEqetaBox}{$\equiv_{\textsc{l}-\Box\eta}$}

\renewcommand{\SYSTEMRenameRuleLinEqletCommBox}{$\equiv_{\textsc{l}-\mathsf{let}\gamma\Box}$}
\renewcommand{\SYSTEMRenameRuleLinEqletCommOne}{$\equiv_{\textsc{l}-\mathsf{let}\gamma_1}$}
\renewcommand{\SYSTEMRenameRuleLinEqletCommTwo}{$\equiv_{\textsc{l}-\mathsf{let}\gamma_2}$}
\renewcommand{\SYSTEMRenameRuleLinEqcongApp}{$\equiv_{\textsc{l}-\mathsf{congApp}}$}
\renewcommand{\SYSTEMRenameRuleLinEqcongAbs}{$\equiv_{\textsc{l}-\mathsf{cong}\lambda}$}
\renewcommand{\SYSTEMRenameRuleLinEqcongPr}{$\equiv_{\textsc{l}-\mathsf{cong}\Box}$}
\renewcommand{\SYSTEMRenameRuleLinEqcongLet}{$\equiv_{\textsc{l}-\mathsf{congLet}}$}

\providecommand{\SYSTEMRenameRuleLinEqbetaUnit}{}
\providecommand{\SYSTEMRenameRuleLinEqetaUnit}{}
\providecommand{\SYSTEMRenameRuleLinEqbetaProd}{}
\providecommand{\SYSTEMRenameRuleLinEqetaProd}{}
\providecommand{\SYSTEMRenameRuleLinEqbetaSumOne}{}
\providecommand{\SYSTEMRenameRuleLinEqbetaSumTwo}{}
\providecommand{\SYSTEMRenameRuleLinEqetaSum}{}

\renewcommand{\SYSTEMRenameRuleLinEqbetaUnit}{$\equiv_{\textsc{L}-1\beta}$}
\renewcommand{\SYSTEMRenameRuleLinEqetaUnit}{$\equiv_{\textsc{L}-1\eta}$}
\renewcommand{\SYSTEMRenameRuleLinEqbetaProd}{$\equiv_{\textsc{L}-\otimes\beta}$}
\renewcommand{\SYSTEMRenameRuleLinEqetaProd}{$\equiv_{\textsc{L}-\otimes\eta}$}
\renewcommand{\SYSTEMRenameRuleLinEqbetaSumOne}{$\equiv_{\textsc{L}-\oplus\beta1}$}
\renewcommand{\SYSTEMRenameRuleLinEqbetaSumTwo}{$\equiv_{\textsc{L}-\oplus\beta2}$}
\renewcommand{\SYSTEMRenameRuleLinEqetaSum}{$\equiv_{\textsc{L}-\oplus\eta}$}

\providecommand{\SYSTEMRenameRuleLinEqcongUnitE}{}
\providecommand{\SYSTEMRenameRuleLinEqcongProdE}{}
\providecommand{\SYSTEMRenameRuleLinEqcongProdI}{}
\providecommand{\SYSTEMRenameRuleLinEqcongSumE}{}
\providecommand{\SYSTEMRenameRuleLinEqcongSumIOne}{}
\providecommand{\SYSTEMRenameRuleLinEqcongSumITwo}{}

\renewcommand{\SYSTEMRenameRuleLinEqcongUnitE}{$\equiv_{\textsc{L}_\mathsf{cong}1}$}
\renewcommand{\SYSTEMRenameRuleLinEqcongProdE}{$\equiv_{\textsc{L}_\mathsf{cong}\otimes{}E}$}
\renewcommand{\SYSTEMRenameRuleLinEqcongProdI}{$\equiv_{\textsc{L}_\mathsf{cong}\otimes{}I}$}
\renewcommand{\SYSTEMRenameRuleLinEqcongSumE}{$\equiv_{\textsc{L}_\mathsf{cong}\oplus{}E}$}
\renewcommand{\SYSTEMRenameRuleLinEqcongSumIOne}{$\equiv_{\textsc{L}_\mathsf{cong}\oplus{}I1}$}
\renewcommand{\SYSTEMRenameRuleLinEqcongSumITwo}{$\equiv_{\textsc{L}_\mathsf{cong}\oplus{}I2}$}

\providecommand{\SYSTEMRenameRuleLinEqpushUnitCong}{}
\providecommand{\SYSTEMRenameRuleLinEqpushUnitBeta}{}
\providecommand{\SYSTEMRenameRuleLinEqpushProdCong}{}
\providecommand{\SYSTEMRenameRuleLinEqpushProdEta}{}
\providecommand{\SYSTEMRenameRuleLinEqpushProdBeta}{}
\providecommand{\SYSTEMRenameRuleLinEqpushSumCong}{}
\providecommand{\SYSTEMRenameRuleLinEqpushSumBetaOne}{}
\providecommand{\SYSTEMRenameRuleLinEqpushSumBetaTwo}{}

\renewcommand{\SYSTEMRenameRuleLinEqpushUnitCong}{$\equiv_{\textsc{L}-\mathsf{cong}-\mathsf{push}\mathsf{unit}}$}
\renewcommand{\SYSTEMRenameRuleLinEqpushUnitBeta}{$\equiv_{\textsc{L}-\mathsf{push}\mathsf{unit}\beta}$}

\renewcommand{\SYSTEMRenameRuleLinEqpushProdCong}{$\equiv_{\textsc{L}-\mathsf{cong}-\mathsf{push}\otimes}$}
\renewcommand{\SYSTEMRenameRuleLinEqpushProdEta}{$\equiv_{\textsc{L}-\mathsf{push}\otimes\eta}$}
\renewcommand{\SYSTEMRenameRuleLinEqpushProdBeta}{$\equiv_{\textsc{L}-\mathsf{push}\otimes\beta}$}
\renewcommand{\SYSTEMRenameRuleLinEqpushSumCong}{$\equiv_{\textsc{L}-\mathsf{cong}-\mathsf{push}\oplus}$}
\renewcommand{\SYSTEMRenameRuleLinEqpushSumBetaOne}{$\equiv_{\textsc{L}-\mathsf{push}\oplus\beta1}$}
\renewcommand{\SYSTEMRenameRuleLinEqpushSumBetaTwo}{$\equiv_{\textsc{L}-\mathsf{push}\oplus\beta2}$}

\providecommand{\SYSTEMRenameRuleGradEqbeta}{}
\providecommand{\SYSTEMRenameRuleGradEqeta}{}
\providecommand{\SYSTEMRenameRuleGradEqcongApp}{}
\providecommand{\SYSTEMRenameRuleGradEqcongAbs}{}

\providecommand{\SYSTEMRenameRuleGradMEqbetaBox}{}
\providecommand{\SYSTEMRenameRuleGradMEqetaBox}{}
\providecommand{\SYSTEMRenameRuleGradMEqletCommBox}{}
\providecommand{\SYSTEMRenameRuleGradMEqletCommOne}{}
\providecommand{\SYSTEMRenameRuleGradMEqletCommTwo}{}
\providecommand{\SYSTEMRenameRuleGradMEqcongPr}{}
\providecommand{\SYSTEMRenameRuleGradMEqcongLet}{}

\renewcommand{\SYSTEMRenameRuleGradEqbeta}{$\equiv_{\textsc{g}-\beta}$}
\renewcommand{\SYSTEMRenameRuleGradEqeta}{$\equiv_{\textsc{g}-\eta}$}
\renewcommand{\SYSTEMRenameRuleGradEqcongApp}{$\equiv_{\textsc{g}-\mathsf{congApp}}$}
\renewcommand{\SYSTEMRenameRuleGradEqcongAbs}{$\equiv_{\textsc{g}-\mathsf{cong}\lambda}$}

\renewcommand{\SYSTEMRenameRuleGradMEqbetaBox}{$\equiv_{\textsc{g}-\Box\beta}$}
\renewcommand{\SYSTEMRenameRuleGradMEqetaBox}{$\equiv_{\textsc{g}-\Box\eta}$}
\renewcommand{\SYSTEMRenameRuleGradMEqletCommBox}{$\equiv_{\textsc{g}-\mathsf{let}\gamma\Box}$}
\renewcommand{\SYSTEMRenameRuleGradMEqletCommOne}{$\equiv_{\textsc{g}-\mathsf{let}\gamma_1}$}
\renewcommand{\SYSTEMRenameRuleGradMEqletCommTwo}{$\equiv_{\textsc{g}-\mathsf{let}\gamma_2}$}
\renewcommand{\SYSTEMRenameRuleGradMEqcongPr}{$\equiv_{\textsc{g}-\mathsf{cong}\Box}$}
\renewcommand{\SYSTEMRenameRuleGradMEqcongLet}{$\equiv_{\textsc{g}-\mathsf{congLet}}$}

\providecommand{\SYSTEMRenameRuleGradPolyEqbetaTy}{}
\providecommand{\SYSTEMRenameRuleGradPolyEqetaTy}{}
\providecommand{\SYSTEMRenameRuleGradPolyEqtyAppCong}{}
\providecommand{\SYSTEMRenameRuleGradPolyEqtyAbsCong}{}

\renewcommand{\SYSTEMRenameRuleGradPolyEqbetaTy}{$\equiv_{\textsc{g}-\forall\beta}$}
\renewcommand{\SYSTEMRenameRuleGradPolyEqetaTy}{$\equiv_{\textsc{g}-\forall\eta}$}
\renewcommand{\SYSTEMRenameRuleGradPolyEqtyAppCong}{$\equiv_{\textsc{g}-\mathsf{cong}@}$}
\renewcommand{\SYSTEMRenameRuleGradPolyEqtyAbsCong}{$\equiv_{\textsc{g}-\mathsf{cong}\forall}$}

\providecommand{\SYSTEMRenameRuleGradEqbetaUnit}{}
\providecommand{\SYSTEMRenameRuleGradEqetaUnit}{}
\providecommand{\SYSTEMRenameRuleGradEqbetaProd}{}
\providecommand{\SYSTEMRenameRuleGradEqetaProd}{}
\providecommand{\SYSTEMRenameRuleGradEqbetaSumOne}{}
\providecommand{\SYSTEMRenameRuleGradEqbetaSumTwo}{}
\providecommand{\SYSTEMRenameRuleGradEqetaSum}{}

\renewcommand{\SYSTEMRenameRuleGradEqbetaUnit}{$\equiv_{\textsc{G}-1\beta}$}
\renewcommand{\SYSTEMRenameRuleGradEqetaUnit}{$\equiv_{\textsc{G}-1\eta}$}
\renewcommand{\SYSTEMRenameRuleGradEqbetaProd}{$\equiv_{\textsc{G}-\otimes\beta}$}
\renewcommand{\SYSTEMRenameRuleGradEqetaProd}{$\equiv_{\textsc{G}-\otimes\eta}$}
\renewcommand{\SYSTEMRenameRuleGradEqbetaSumOne}{$\equiv_{\textsc{G}-\oplus\beta1}$}
\renewcommand{\SYSTEMRenameRuleGradEqbetaSumTwo}{$\equiv_{\textsc{G}-\oplus\beta2}$}
\renewcommand{\SYSTEMRenameRuleGradEqetaSum}{$\equiv_{\textsc{G}-\oplus\eta}$}

\providecommand{\SYSTEMRenameRuleGradEqcongUnitE}{}
\providecommand{\SYSTEMRenameRuleGradEqcongProdE}{}
\providecommand{\SYSTEMRenameRuleGradEqcongProdI}{}
\providecommand{\SYSTEMRenameRuleGradEqcongSumE}{}
\providecommand{\SYSTEMRenameRuleGradEqcongSumIOne}{}
\providecommand{\SYSTEMRenameRuleGradEqcongSumITwo}{}

\renewcommand{\SYSTEMRenameRuleGradEqcongUnitE}{$\equiv_{\textsc{G}_\mathsf{cong}1}$}
\renewcommand{\SYSTEMRenameRuleGradEqcongProdE}{$\equiv_{\textsc{G}_\mathsf{cong}\otimes{}E}$}
\renewcommand{\SYSTEMRenameRuleGradEqcongProdI}{$\equiv_{\textsc{G}_\mathsf{cong}\otimes{}I}$}
\renewcommand{\SYSTEMRenameRuleGradEqcongSumE}{$\equiv_{\textsc{G}_\mathsf{cong}\oplus{}E}$}
\renewcommand{\SYSTEMRenameRuleGradEqcongSumIOne}{$\equiv_{\textsc{G}_\mathsf{cong}\oplus{}I1}$}
\renewcommand{\SYSTEMRenameRuleGradEqcongSumITwo}{$\equiv_{\textsc{G}_\mathsf{cong}\oplus{}I2}$}

\usepackage{listings}
\DeclareFontShape{OT1}{cmtt}{bx}{n}{<5><6><7><8><9><10><10.95><12><14.4><17.28><20.74><24.88>cmttb10}{}
\lstdefinelanguage{Granule}{%
  mathescape=true,
  morecomment=[l]{--},
  moredelim=[s][\itshape]{`}{`},
  showspaces=false,
  aboveskip=0.5em,
  belowskip=0.5em,
  commentstyle=\itshape\color{black!60},
  basicstyle=\ttfamily\small,%\sffamily\small,%
  flexiblecolumns=true,
  columns=[l]flexible,
  columns=fullflexible,
  keepspaces=true,
  xleftmargin=2.5em,
  numbers=left,
  numberstyle=\tiny\color{gray},
  literate=%
  {<}{\textcolor{effectColor}{<}}1
  {>}{\textcolor{effectColor}{>}}1
  {[}{\textcolor{coeffectColor}{[}}1
  {]}{\textcolor{coeffectColor}{]}}1
  {\%}{\textcolor{coeffectColor}{\%}}1
  {[r' : R']}{[{\textcolor{coeffectColor}{r' : R'}}]}1
  {forall}{$\forall$}1
  {Inf}{$\infty$}1
  {->}{$\rightarrow$}1
  {-o}{$\multimap$}1
  {=>}{$\Rightarrow$}1
  {<-}{\textcolor{effectColor}{$\leftarrow$}}1
  {/\\}{$\sqcap$}1
  {\\/}{$\sqcup$}1
  {<=}{$\leqslant$}1
  {>=}{$\geqslant$}1
  {_1}{$\mathtt{_1}$}1
  {_2}{$\mathtt{_2}$}1
  {_3}{$\mathtt{_3}$}1
  {_4}{$\mathtt{_4}$}1
  {_L}{$\mathtt{_{L}}$}1
  {_LH}{$\mathtt{_{LH}}$}1
  {_Gr}{$\mathtt{_{Gr}}$}1
  {_p}{$\mathtt{_{p}}$}1
  {_q}{$\mathtt{_{q}}$}1
  {-o}{$\multimap$}1
  {\\times}{$\times$}1
  {--BLANK}{}1,
  keywordstyle = \bfseries,%\color{WildStrawberry!90!Black},
  keywords = {data, type, let, in, case, of, if, then, else, where,  import, Type, Semiring, language},
  keywordstyle = [4]\bfseries\color{effectColor},
  keywords     = [4]{pure}%{IO,
}

\lstdefinelanguage{GranuleErr}{%
  mathescape=true,
  mathescape=true,
  morecomment=[l]{--},
  moredelim=[s][\itshape]{`}{`},
  showspaces=false,
  aboveskip=0.5em,
  belowskip=0.5em,
  commentstyle=\itshape\color{black!60},
  basicstyle=\ttfamily\small\color{red!50!black},%\sffamily\small,%
  flexiblecolumns=true,
  columns=[l]flexible,
  columns=fullflexible,
  keepspaces=true,
  xleftmargin=2.5em,
}

\lstnewenvironment{granule}
{\lstset{language=Granule}}{}
\lstnewenvironment{granule-illtyped}
{\lstset{language=Granule}}{}
\lstnewenvironment{granule-misc}
{\lstset{language=Granule}}{}
\lstnewenvironment{granuleerr}
{\lstset{language=GranuleErr}}{}
\newcommand{\granin}[1]{\lstinline[language=Granule]{#1}}

\definecolor{multiplicity}{rgb}{0,0.3,0.08}

\lstdefinelanguage{Haskell}{%
  mathescape=true,
  morecomment=[l]{--},
  comment=[l]{\{-},
  moredelim=[s][\itshape]{`}{`},
  showspaces=false,
  aboveskip=0.5em,
  belowskip=0.5em,
  commentstyle=\itshape\color{black!60},
  basicstyle=\ttfamily\small,%\sffamily\small,%
  flexiblecolumns=true,
  columns=[l]flexible,
  columns=fullflexible,
  keepspaces=true,
  xleftmargin=2.5em,
  literate=%
   {->}{$\rightarrow$}1
   {\%r}{\textcolor{coeffectColor}{\%r}}1
   {0}{\textcolor{coeffectColor}{0}}1
   {\%'Many}{\textcolor{coeffectColor}{\%'Many}}1
   {\%Many}{\textcolor{coeffectColor}{\%Many}}1
   {\%One}{\textcolor{coeffectColor}{\%One}}1
   {\%1}{\textcolor{coeffectColor}{\%1}}1,
  keywordstyle = \bfseries,%\color{WildStrawberry!90!Black},
  keywords = {data, type, let, in, case, of, if, then, else, where,
  import, class, instance, LANGUAGE},
  numbers=left,
  numberstyle=\tiny\color{gray}
}

\lstnewenvironment{haskell}[1][]
{\lstset{language=Haskell, #1}}{}
\newcommand{\haskin}[1]{\lstinline[language=Haskell]{#1}}

\setlist[itemize]{itemsep=.6em}

\author{Vilem Liepelt}
\orcid{0000-0002-2198-1819}
\affiliation{%
    \institution{University of Kent}
    \department{School of Computing}
    \country{UK}
}
\email{mail@vilem.net}

\author{Danielle Marshall}
\orcid{0000-0002-4284-3757}
\affiliation{%
    \institution{Royal Holloway, University of London}
    \department{Department of Computer Science}
    \country{UK}
}
\email{Danielle.Marshall@rhul.ac.uk}

\author{Dominic Orchard}
\orcid{0000-0002-7058-7842}
\affiliation{%
    \institution{University of Kent}
    \department{School of Computing}
    \country{UK}
}
\affiliation{%
    \institution{University of Cambridge}
    \department{Department of Computer Science and Technology}
    \country{UK}
}
\email{D.A.Orchard@kent.ac.uk}

\setcopyright{cc}
\setcctype{by}
\ifextended
  \title{Same Coeffect, Different Base: Connecting Two Dominant Approaches to Graded Types {\small(with appendices)}}
\else
  \title{Same Coeffect, Different Base:\\Connecting Two Dominant Approaches to Graded Types}
  \usepackage{xr}
  \acmDOI{10.1145/3828697}
  \acmYear{2026}
  \acmJournal{PACMPL}
  \acmVolume{10}
  \acmNumber{ICFP}
  \acmArticle{299}
  \acmMonth{8}
  \acmSubmissionID{icfp26main-p85-p}
  \received{2026-02-19}
  \received[accepted]{2026-05-13}

\fi

\newcommand{\refappendix}[1]{%
  \ifextended
    \ref{#1}%
  \else
    \ref{A-#1}~\cite{extended-version}%
  \fi
}

\begin{document}

\begin{abstract}
\emph{Graded types} provide a way to augment a type system with fine-grained
information, e.g., to track side effects or context dependence and resource
use (called \emph{coeffects}). Graded types for coeffects have found their way into languages such as Haskell, Idris, and
Granule, enabling resourceful reasoning via coeffect analysis with varying levels of generality.
Two separate lineages of graded coeffect system have emerged in the last decade: those
in which coeffect annotations are pervasive, requiring annotations on function
types (which we call \emph{graded-base}) and those in which coeffects are added
by way of a graded modal type operator atop linear types (which we call
\emph{linear-base}). The latter has its origins in Girard's Linear Logic which
has been a rich humus for programming language research focused on resources,
whereas the graded-base approach emerged in the mid-2010s, seeing rapid adoption
in programming language theory and practice, e.g. in QTT and Linear Haskell. The
relationship between these two styles has however remained an open question. We answer this
question by giving translations between pairs of calculi of both lineages that
we prove type-, grade- and operational-semantics preserving. We show that the same notions of context
dependence can be expressed in either style, building a bridge
between the two lineages that enables transfer of results and ideas, while
helping language designers to make better informed choices.
\end{abstract}
\maketitle
\ifextended
\textit{\textbf{Extended Version:}
The official version of this paper (\href{https://doi.org/10.1145/3828697}{doi.org/10.1145/3828697}) appears in the proceedings of the
2026 ACM SIGPLAN International Conference on Functional Programming (ICFP 2026).
The appendices included herein provide proofs and definitions omitted due to space constraints.}
\fi
\iffalse
\begin{quotation}\noindent
\textit{One clearly needs more subtle logical tools taking into account that it costs something to make a
deduction, a guess, etc.; linear logic [...] could serve as a prototype for a more serious approach
to this subject.}

---Jean-Yves Girard, 1987 \cite{girard1987linear}
\end{quotation}
\fi
\section{Introduction}
The behaviour of useful programs often depends on execution context---whether through low-level factors like
hardware capabilities (e.g., the available I/O devices) or higher-level properties like information flow security.
Recent work has explored tracking such \emph{coeffects} via the type
system~\cite{DBLP:conf/icalp/PetricekOM13,petricek2014coeffects,brunel2014core,ghica2014bounded},
providing an analysis of context dependence, reified into the type system as an intrinsic part of the
language's meaning.

There exist in the literature two parallel threads on coeffect types systems: those which build atop
Linear Logic via the linear $\lambda$-calculus, which we call \emph{linear-base} systems, and those
which build atop the simply-typed $\lambda$-calculus or System F but with pervasive annotations on
all assumptions and function arrows, which we refer to as \emph{graded-base} systems (for reasons which will
become clear later). How these two approaches precisely relate is a long-standing open question that this paper
answers, giving type system and language designers a map of the graded type system landscape.

\paragraph{Generalising linearity to graded modal types}

\citet{girard1987linear} introduced Linear Logic, seeking
more fine-grained control over the structural rules
implicitly present in classical and intuitionistic logic. The relevance to computer science in
particular was very much at the forefront of the motivations for \citet{girard1987linear}: ``One of the main outputs of
Linear Logic seems to be in computer science''.

The more conscious use of structural properties in Linear Logic allows for reasoning about the cost
of computations and other properties, which are referred to under the umbrella term
\emph{resources}. As predicted by Girard himself, Linear Logic has become the mother to a profusion
of type systems for reasoning about various notions of resource~\cite{walker2005substructural}. Examples in the literature
include access to hardware devices \cite{DBLP:conf/icalp/PetricekOM13}, timing \cite{ghica2014bounded} and differential privacy
\cite{gaboardi2013linear}. These developments stand on the shoulders of a refinement of Linear
Logic, in the form of Bounded Linear Logic (BLL), due to \citet{girard1992bounded} in
the early 1990s. While Linear Logic delineates the world into `use strictly once' ($A$) and
`arbitrary usage' ($!A$), i.e. zero or more times, BLL indexes the \textit{of course} modality with a
polynomial bound, where $!_{n}A$ is usable up to $n$ times.
For example, the following type describes some higher-order function which
takes a function and composes it with itself, thus using it twice:
$!_2 (A \multimap A) \multimap (A \multimap A)$.

A significant evolution came in 2014, when \citet{brunel2014core} and \citet{ghica2014bounded}
independently generalised the natural number indices in BLL to
elements of a semiring $\mathcal{R}$ to capture more properties than
just variable (re)use, including privacy levels and hardware schedules.
This development then allowed other strands of research on \emph{coeffects} (general
context analysis) to be expressible in this linear-logic-derived setting.

The generalisation of BLL gave rise to a more general paradigm
of \emph{graded modal types}. \citet{DBLP:conf/icfp/GaboardiKOBU16} considered a calculus
with a semiring-graded modality $\Box_r A$ for coeffects alongside a
monoid-graded modality for effects $\Diamond_f A$, modelled
respectively by graded comonads and graded
monads.  \citet{DBLP:journals/pacmpl/OrchardLE19} crystallised the idea of
graded modal types as one in which ``graded modal types carry
information about semantic structure'' where the algebraic structure
of the indices reflects underlying program
structure. Others
have followed a similar approach~\cite{wood2020linear,DBLP:conf/esop/WoodA22,DBLP:journals/corr/abs-2112-14966,DBLP:conf/lopstr/HughesO20}.
In all these systems, Linear Logic has been at the core. We thus refer to
these as \emph{linear-base systems}.

\paragraph{Coeffect systems with pervasive grading and without inherent linearity}
In parallel, \citet{DBLP:conf/icalp/PetricekOM13,petricek2014coeffects} dualised type-and-effect
systems to \emph{coeffect systems}, capturing how programs depend on their
context.
Their approach built on top of the simply-typed
rather than the linear $\lambda$-calculus. Every assumption
has an associated coeffect (grade) and function arrows must also
be annotated with a coeffect describing how the parameter is used.
In the case of the `apply twice' higher-order function,
the analogous coeffect type is $(A \xrightarrow{1} A) \xrightarrow{2} (A
\xrightarrow {1} A)$. If the parameter function
is generalised from coeffect $1$ to
some arbitrary use $r$, then the type would be
$(A \xrightarrow{r} A) \xrightarrow{2} (A \xrightarrow {r * r} A)$.

Coeffect systems were noted by \citet{brunel2014core} to relate to a generalised
BLL system. Since then, many other graded
systems have taken a similar approach with no base notion
of linearity, but where every assumption has a grade and function arrows
are annotated, e.g., Linear Haskell (GHC/Haskell with the \texttt{LinearTypes}
extension)~\cite{DBLP:journals/pacmpl/BernardyBNJS18},
the general system of \citet{DBLP:journals/pacmpl/AbelB20},
Quantitative Type Theory, first due to \citet{McBride2016} then refined
by \citet{quantitative-type-theory} and used as the basis for Idris 2~\cite{DBLP:conf/ecoop/Brady21}, the \textsc{GraD} calculus
of \citet{grad}, Graded Modal Type Theory~\cite{grtt},
and various others~\cite{mycroftfest2024,DBLP:conf/esop/HughesO24,DBLP:journals/pacmpl/AbelDE23}. This style has even
been adapted to an imperative setting in the coeffectful Multi-Graded Featherweight Java~\cite{DBLP:journals/pacmpl/BianchiniDGZS22,DBLP:conf/ecoop/BianchiniDGZ23}. A similar type system structuring arose in a separate thread of work from \citet{wright1993usage} whose lineage has been recently been traced and summarised~\citep{wright_2026_20739966}. In these
approaches, grading is pervasive and there is no inherent notion of linearity as in Linear Logic
(although linearity can be achieved as an emergent phenomenon). We refer to these as
\emph{graded-base systems}.

So what is the relationship? Are graded coeffects with a linear basis
equally expressive as those without an inherent notion of linearity? This is the open question
that has become ever more glaring as these works have developed. We seek a more clear understanding of their
relationship and relative expressivity in order to inform future developments
and choices.

\begin{figure}
	\centering
	\includegraphics[width=0.6\textwidth]{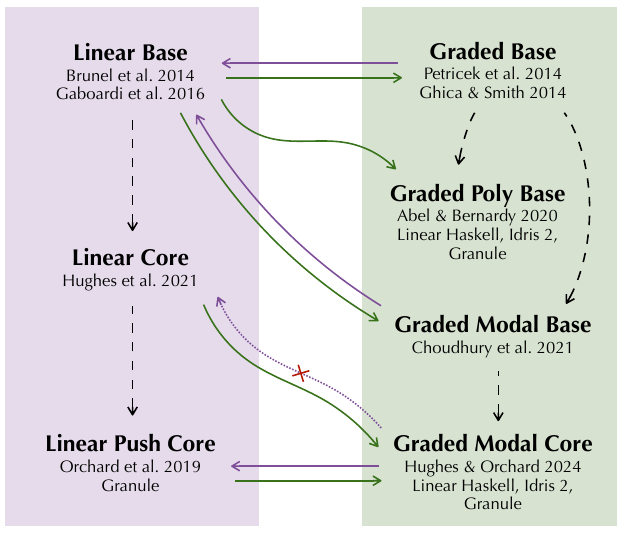}
	\caption{Relationship between graded calculi studied in this paper. For each calculus, a
   selection of representative papers and programming systems is given (that typically also extend the
   given calculus with orthogonal features, such as dependent types, which are not relevant to this paper). A dashed arrow denotes a
   subset relationship. A solid arrow denotes a semantics-preserving translation which we
   define in the present paper. A crossed-out arrow with a dotted shaft refers to a translation that
   is not possible.}
	\label{fig:relationship}
\end{figure}

\paragraph{Contributions and roadmap}

Our aim is to elucidate the connection between the two distinct strands of
coeffect reasoning that have emerged in the literature and in programming language
implementations, which we call \emph{linear-base} vs. \emph{graded-base} systems.
We show that, depending on the substrate, both systems can be given mutual translations to each other in a sound way.

Figure~\ref{fig:relationship}
gives an overview of these two strands, instantiated as several concrete calculi with differing
features which we summarise here with relevant section numbers:

Linear-base systems are shown on the left. Here \textbf{Linear~Base} (\ref{sec:linear-base}) represents a linear
$\lambda$-calculus extended with graded modal necessity $ \textcolor{coeffectColor}{\square_{ \SYSTEMnt{r} } }  \SYSTEMnt{A} $, corresponding to a
generalised \textit{of course} modality. \textbf{Linear~Core} (\ref{sec:linear-core}) augments this with products and sums
(multiplicative conjunction and additive disjunction). \textbf{Linear Push Core} (\ref{sec:linear-push-core}) further adds constructs for
allowing distributivity of graded modalities over products and sums, thereby losing the
correspondence to Linear Logic, as shown by \citet{hughes2021linear}. This system can model the
interaction between grades and products/sums in Granule as presented by
\citet{DBLP:journals/pacmpl/OrchardLE19}.

On the right then we have graded-base systems, for all of which \textbf{Graded~Base} (\ref{sec:graded-base}) is the
underlying calculus, corresponding to pervasively-graded simply-typed $\lambda$-calculus in the
style of \citet{petricek2014coeffects}, \citet{ghica2014bounded}, and Linear Haskell as presented by
\citet{DBLP:journals/pacmpl/BernardyBNJS18}. \textbf{Graded~Poly~Base} (\ref{sec:graded-poly-base}) augments this with
System-F-style type abstraction as in the work of \citet{DBLP:journals/pacmpl/AbelB20}, presenting a
substrate modelling more closely the actual implementation of Linear Haskell, as well as Idris 2
(where the dependent types have no interaction with grades)~\cite{DBLP:conf/ecoop/Brady21} and also
Granule with the \granin{GradedBase} language extension~\cite{DBLP:conf/esop/HughesO24}.
\textbf{Graded~Modal~Base} (\ref{sec:graded-modal-base}) instead augments Graded Base by adding a graded modal necessity akin to the
one in Linear Base. \textbf{Graded~Modal~Core} (\ref{sec:graded-core}) augments the latter system with products and sums and
thereby captures the interaction between grades and algebraic datatypes available e.g. in Granule with the
\granin{GradedBase} extension, and Linear Haskell and Idris 2 (where a modal operator can be
encoded via an algebraic datatype).

The paper is then structured as follows:

\begin{itemize}
\item
Section~\ref{sec:calculi} introduces Linear Base and Graded Base. Section \ref{sec:transl}
introduces syntax for translation, gives a \textbf{translation from Graded Base to Linear Base}
(\ref{subsec:transl-graded-base-to-linear-base}) with soundness results, and then
the converse \textbf{from Linear Base to Graded Base} via a CPS-like translation
(\ref{subsec:lin-to-grad-cps}).

\item Section~\ref{sec:calculi-with-modality} extends Graded Base to Graded Modal Base
(\ref{sec:graded-modal-base}), gives a \textbf{translation from Linear Base to Graded Modal Base}
(\ref{subsec:transl-lb2gmb}) with soundness results, gives a \textbf{translation from Graded Modal
Base to Linear Base} (\ref{subsec:transl-gmb2lb}) with soundness results, investigates the nature of
the relationship between both translations (\ref{subsec:mutual-relationship}), and gives a possible
generalisation of elimination for the graded modality that appears in the literature, along with tradeoffs
(\ref{sec:flattening}).

\item
Section~\ref{sec:poly} gives an alternative strategy for adding a graded modality to Graded Base,
via an encoding inspired by Linear Haskell, extends Graded Base to Graded Poly Base
(\ref{sec:graded-poly-base}), and gives a \textbf{translation from Linear Base to Graded Poly Base}
(\ref{subsec:transl-gpb2lb}) with soundness results.

\item
Section~\ref{sec:prods-sums} adds product and sum types, allowing us to study the interaction of
coeffects with data structures. It extends Linear Base to Linear Core (\ref{sec:linear-core}), extends
Graded Modal Base to Graded Modal Core (\ref{sec:graded-core}), gives a \textbf{translation from
Linear Core to Graded Modal Core} (\ref{sec:linear-core-to-graded}) with soundness results, shows
\textbf{there is no sound translation from Graded Modal Core to Linear Core}
(\ref{sec:no-trans-graded-core-to-linear-core}), extends Linear Core to Linear Push Core
(\ref{sec:linear-push-core}), and gives a \textbf{translation from Graded Modal
Core to Linear Push Core} with soundness results.

\item
Section~\ref{sec:discussion} asks: \textit{is one approach better than the other?}
(\ref{sec:better}), gives some simple example programs in languages that implement features from the
two lineages studied (\ref{sec:wild}, considers some related ideas (\ref{sec:comparison-with-effects}-\ref{sec:raos}), and concludes (\ref{sec:conclusion}).
\end{itemize}

\section{The Two Calculi: Linear and Graded Base}
\label{sec:calculi}
We present in turn two representative calculi for
the two dominant styles of graded coeffects:
Linear Base (Section~\ref{sec:linear-base}) and
Graded Base (Section~\ref{sec:graded-base}).

\subsection{Linear Base}
\label{sec:linear-base}

As discussed in the introduction, linear-base systems take
linear $\lambda$-calculus as their core. Such systems include
the core quantitative coeffect calculus of \citet{brunel2014core},
 the combined graded system for effects and coeffects of \citet{DBLP:conf/icfp/GaboardiKOBU16},
$\lambda\mathcal{R}$ of \citet{wood2020linear},
the core of Granule by \citet{DBLP:journals/pacmpl/OrchardLE19},
and associated work~\cite{DBLP:journals/corr/abs-2112-14966}.

The term syntax of our idealised system, Linear Base, is as follows:
\begin{align*}
    \SYSTEMnt{t} & ::= \SYSTEMmv{x} \mid  \SYSTEMnt{t_{{\mathrm{1}}}} \,  \SYSTEMnt{t_{{\mathrm{2}}}}  \mid  \lambda  \SYSTEMmv{x}  .  \SYSTEMnt{t}  \mid  \textcolor{coeffectColor}{[}  \SYSTEMnt{t}  \textcolor{coeffectColor}{]}  \mid  \mathsf{let} \, \textcolor{coeffectColor}{[}  \SYSTEMmv{x}  \textcolor{coeffectColor}{]} =  \SYSTEMnt{t_{{\mathrm{1}}}}  \, \mathsf{in} \,  \SYSTEMnt{t_{{\mathrm{2}}}} 
  \tag{terms}
\end{align*}
Terms comprise variables, applications, and abstraction as standard. Graded modal introduction
(formally \emph{promotion}, informally \emph{boxing}) is via $ \textcolor{coeffectColor}{[}  \SYSTEMnt{t}  \textcolor{coeffectColor}{]} $, which for now we can
think of as wrapping a term in a constructor. Graded modal elimination eliminates this constructor
via a \emph{let} binding, unwrapping the boxed value, which we refer to as \emph{unboxing}.

\noindent
Linear Base types include a linear function space, a
\emph{graded modality} which can be understood as generalising Linear Logic's exponential
modality $!$, and a base type $ \mathrm{K} $ so that types are well-founded:
\begin{align*}
  \SYSTEMnt{A} & ::=  \SYSTEMnt{A}  \multimap  \SYSTEMnt{B}  \mid  \textcolor{coeffectColor}{\square_{ \SYSTEMnt{r} } }  \SYSTEMnt{A}  \mid  \mathrm{K} 
  \tag{types}
\end{align*}
Here $\SYSTEMnt{r}$ is drawn from a pre-ordered
semiring $(\mathcal{R}, \cdot, 1, +, 0, \sqsubseteq)$
(where addition $+$ and multiplication $\cdot$ must
be monotonic wrt. to $\sqsubseteq$)
parameterising the calculus.

An intuition for $ \textcolor{coeffectColor}{\square_{ \SYSTEMnt{r} } }  \SYSTEMnt{A} $ is that it describes some capability $r$
with which terms of type $A$ contained inside can be used. For example,
for the natural numbers semiring $\mathbb{N}$ then $ \textcolor{coeffectColor}{\square_{ \SYSTEMsym{2} } }  \SYSTEMnt{A} $ denotes two copies
of an $A$, or in essence that its contents can be used twice.

Typing judgments are of the form $ \Gamma  \vdash_{\textsc{l} }  \SYSTEMnt{t}  :  \SYSTEMnt{A} $, relating a context $\Gamma$, term $\SYSTEMnt{t}$, and type $\SYSTEMnt{A}$. Contexts comprise linear assumptions
$ \SYSTEMmv{x}  :  \SYSTEMnt{A} $ and graded assumptions $ \SYSTEMmv{x}  : \textcolor{coeffectColor}{[}  \SYSTEMnt{A} {\textcolor{coeffectColor}{]_{ \SYSTEMnt{r} } } } $
with grade $\SYSTEMnt{r} \in \mathcal{R}$, of the form:
\begin{align*}
  \tag{contexts}
  \Gamma ::=  \emptyset  \mid  \Gamma ,   \SYSTEMmv{x}  :  \SYSTEMnt{A}   \mid  \Gamma ,   \SYSTEMmv{x}  : \textcolor{coeffectColor}{[}  \SYSTEMnt{A} {\textcolor{coeffectColor}{]_{ \SYSTEMnt{r} } } }  
\end{align*}
Contexts can be \textit{added} if they are disjoint in their linear assumptions and
using semiring addition to combine grades of any graded assumptions shared
between two contexts:
\begin{definition}[Context addition]
    For contexts $\Gamma_{{\mathrm{1}}}, \Gamma_{{\mathrm{2}}}$, then $\Gamma_{{\mathrm{1}}}  \SYSTEMsym{+}  \Gamma_{{\mathrm{2}}}$ is a partial operation
    computing the combined
    context providing \textit{contraction} on graded assumptions. Context addition is specified:
\begin{align*}
\Gamma + \emptyset = \Gamma \quad
 (   \Gamma_{{\mathrm{1}}} ,   \SYSTEMmv{x}  : \textcolor{coeffectColor}{[}  \SYSTEMnt{A} {\textcolor{coeffectColor}{]_{ \SYSTEMnt{r} } } }    )   \SYSTEMsym{+}   (   \Gamma_{{\mathrm{2}}} ,   \SYSTEMmv{x}  : \textcolor{coeffectColor}{[}  \SYSTEMnt{A} {\textcolor{coeffectColor}{]_{ \SYSTEMnt{s} } } }    )  & =   (  \Gamma_{{\mathrm{1}}}  \SYSTEMsym{+}  \Gamma_{{\mathrm{2}}}  )  ,   \SYSTEMmv{x}  : \textcolor{coeffectColor}{[}  \SYSTEMnt{A} {\textcolor{coeffectColor}{]_{  \SYSTEMnt{r}  \SYSTEMsym{+}  \SYSTEMnt{s}  } } }   \\
\emptyset + \Gamma = \Gamma \hspace{5.4em}  (   \Gamma_{{\mathrm{1}}} ,   \SYSTEMmv{x}  : \textcolor{coeffectColor}{[}  \SYSTEMnt{A} {\textcolor{coeffectColor}{]_{ \SYSTEMnt{r} } } }    )   \SYSTEMsym{+}  \Gamma_{{\mathrm{2}}} & =   (  \Gamma_{{\mathrm{1}}}  \SYSTEMsym{+}  \Gamma_{{\mathrm{2}}}  )  ,   \SYSTEMmv{x}  : \textcolor{coeffectColor}{[}  \SYSTEMnt{A} {\textcolor{coeffectColor}{]_{ \SYSTEMnt{r} } } }   & \mathrm{if}\ x \not\in \mathsf{dom}(\Gamma_{{\mathrm{2}}}) \\
 (   \Gamma_{{\mathrm{1}}} ,   \SYSTEMmv{x}  :  \SYSTEMnt{A}    )   \SYSTEMsym{+}  \Gamma_{{\mathrm{2}}} & =   (  \Gamma_{{\mathrm{1}}}  \SYSTEMsym{+}  \Gamma_{{\mathrm{2}}}  )  ,   \SYSTEMmv{x}  :  \SYSTEMnt{A}   & \mathrm{if}\ x \not\in \mathsf{dom}(\Gamma_{{\mathrm{2}}})
\end{align*}
That is, we union the disjoint parts of the contexts, and for the non-disjoint parts
of the contexts we merge with semiring addition. If the two contexts share linear assumptions
or if the two contexts share graded assumptions that do not agree on the type, then
the addition is undefined (in practical terms, this amounts to a linearity error).
Contexts are unordered, i.e., \emph{exchange} is permitted.
\end{definition}
We define two predicates for denoting contexts containing only graded assumptions
and those containing graded assumptions all with the same grade:
\begin{definition}[Graded contexts]
\label{dfn:graded-contexts}
  A context $\Gamma$ has property $ \mathrm{graded}( \Gamma ) $ if it contains only graded
  assumptions, and $ \mathrm{graded}( \Gamma , \textcolor{coeffectColor}{ \SYSTEMnt{r} }) $ if it contains only graded assumptions which are
  of grade $\SYSTEMnt{r}$. These predicates are defined inductively:
\begin{align*}
  \begin{array}{c}
\SYSTEMdrulePredicateGradednil{}
\quad
\SYSTEMdrulePredicateGradedgrad{}
\qquad
\SYSTEMdrulePredicateGradedAtnil{}
\quad
\SYSTEMdrulePredicateGradedAtgrad{}
\end{array}
\end{align*}
\end{definition}
\noindent
Lastly, contexts which contain only graded assumptions can be \emph{scaled} by
a grade:
\begin{definition}[Scalar multiplication]
  A context $\Gamma$ where $ \mathrm{graded}( \Gamma ) $ can be \emph{scaled}
  $ \textcolor{coeffectColor}{ \SYSTEMnt{r}  \cdot}  \Gamma $, defined:
  \begin{align*}
         \textcolor{coeffectColor}{ \SYSTEMnt{r}  \cdot}   \emptyset   \triangleq  \emptyset 
    \qquad
         \textcolor{coeffectColor}{ \SYSTEMnt{r}  \cdot}   (   \Gamma' ,   \SYSTEMmv{x}  : \textcolor{coeffectColor}{[}  \SYSTEMnt{A} {\textcolor{coeffectColor}{]_{ \SYSTEMnt{s} } } }    )   \triangleq   (   \textcolor{coeffectColor}{ \SYSTEMnt{r}  \cdot}  \Gamma'   )  ,   \SYSTEMmv{x}  : \textcolor{coeffectColor}{[}  \SYSTEMnt{A} {\textcolor{coeffectColor}{]_{   \SYSTEMnt{r}  \cdot  \SYSTEMnt{s}   } } }  
  \end{align*}
\end{definition}
\begin{figure}[t]
  \raggedright\noindent\framebox{$ \Gamma  \vdash_{\textsc{l} }  \SYSTEMnt{t}  :  \SYSTEMnt{A} $} \quad Typing rules
\begin{gather*}
\begin{align*}
  \begin{array}{c}
\SYSTEMdruleLinvar{}
\quad
\SYSTEMdruleLinabs{}
\quad
\SYSTEMdruleLinapp{}
\\[1.25em]
\SYSTEMdruleLinweak{}
\quad
\SYSTEMdruleLinder{}
\quad
\SYSTEMdruleLinpr{}
\\[1.25em]
\SYSTEMdruleLinlet{}
\qquad
\SYSTEMdruleLinapprox{}
\end{array}
\end{align*}
\end{gather*}
\caption{Linear Base typing.}
\label{fig:linear-base-types}
\end{figure}

% \newpage
Figure~\ref{fig:linear-base-types} defines typing rules. The \textit{variable}
($\SYSTEMRenameRuleLinvar{}$) and \textit{abstraction} ($\SYSTEMRenameRuleLinabs{}$) rules are standard for the
linear $\lambda$-calculus. The \textit{application} ($\SYSTEMRenameRuleLinapp{}$) rule is close to that of the
$\lambda$-calculus, but the separate contexts of the premises are added in the conclusion.
\textit{Weakening} ($\SYSTEMRenameRuleLinweak{}$) allows unused assumptions to be added, as long as they
are graded at the zero of the semiring. \textit{Dereliction} ($\SYSTEMRenameRuleLinder{}$) treats a linear
assumption as a graded assumption at the one element of the semiring, which connects the linear and
non-linear worlds. \textit{Promotion} ($\SYSTEMRenameRuleLinpr{}$) introduces grading to terms, making
use of scalar context multiplication to propagate the resource requirements to the assumptions. The
\textit{let} or \textit{unboxing} rule ($\SYSTEMRenameRuleLinlet{}$) eliminates an $\SYSTEMnt{r}$-graded
modality, making the unboxed value of $\SYSTEMnt{t_{{\mathrm{1}}}}$ available in the body of $\SYSTEMnt{t_{{\mathrm{2}}}}$ according to the capabilities
afforded by $\SYSTEMnt{r}$. \textit{Approximation} ($\SYSTEMRenameRuleLinapprox{}$) allows treating an $\SYSTEMnt{r}$-graded assumption as $\SYSTEMnt{s}$-graded if $ \SYSTEMnt{r}  \, \textcolor{coeffectColor}{\sqsubseteq} \,  \SYSTEMnt{s} $. Note the direction of the graded
inequality in the premise is according to \citet{petricek2014coeffects}. The opposite polarity is
also used in the literature, for example by \citet{brunel2014core}, without any semantic difference.

\begin{example}
\label{exm:nat-eq}
The natural numbers semiring $(\mathbb{N}, \ast, 1, +, 0, \equiv)$ whose ordering
is equality (i.e., a discrete ordering) captures a notion of exact usage.
For example, the following captures the typing of a linear function $f$ that
takes two inputs of type $\SYSTEMnt{A}$. The inputs are provided by first eliminating
an assumption $ \SYSTEMmv{x}  :   \textcolor{coeffectColor}{\square_{ \SYSTEMsym{2} } }  \SYSTEMnt{A}  $ into a graded assumption $ \SYSTEMmv{y}  : \textcolor{coeffectColor}{[}  \SYSTEMnt{A} {\textcolor{coeffectColor}{]_{ \SYSTEMsym{2} } } } $ which is then
used twice to apply $f$:
\begin{align*}
\inferrule*[right=\SYSTEMRenameRuleLinlet{}]
{
\ldots
   \inferrule*[right=\SYSTEMRenameRuleLinapp{}]
    {\inferrule*[right=\SYSTEMRenameRuleLinder{}]
      {\inferrule*[right=\SYSTEMRenameRuleLinapp{}]
        {\ldots}
        {    \SYSTEMmv{f}  :    \SYSTEMnt{A}  \multimap  \SYSTEMnt{A}   \multimap  \SYSTEMnt{B}    ,   \SYSTEMmv{y}  :  \SYSTEMnt{A}    \vdash_{\textsc{l} }   \SYSTEMmv{f} \,  \SYSTEMmv{y}   :   \SYSTEMnt{A}  \multimap  \SYSTEMnt{B}  }}
        {    \SYSTEMmv{f}  :    \SYSTEMnt{A}  \multimap  \SYSTEMnt{A}   \multimap  \SYSTEMnt{B}    ,   \SYSTEMmv{y}  : \textcolor{coeffectColor}{[}  \SYSTEMnt{A} {\textcolor{coeffectColor}{]_{ \SYSTEMsym{1} } } }    \vdash_{\textsc{l} }   \SYSTEMmv{f} \,  \SYSTEMmv{y}   :   \SYSTEMnt{A}  \multimap  \SYSTEMnt{B}  }
      \,
      \inferrule*[right=\SYSTEMRenameRuleLinder{}]
           {\inferrule*[right=\SYSTEMRenameRuleLinvar{}]
             {\quad}
             {   \SYSTEMmv{y}  :  \SYSTEMnt{A}    \vdash_{\textsc{l} }  \SYSTEMmv{y}  :  \SYSTEMnt{A} }
           }
           {   \SYSTEMmv{y}  : \textcolor{coeffectColor}{[}  \SYSTEMnt{A} {\textcolor{coeffectColor}{]_{ \SYSTEMsym{1} } } }    \vdash_{\textsc{l} }  \SYSTEMmv{y}  :  \SYSTEMnt{A} }
     \hspace{-2em}}
     {    \SYSTEMmv{f}  :    \SYSTEMnt{A}  \multimap  \SYSTEMnt{A}   \multimap  \SYSTEMnt{B}    ,   \SYSTEMmv{y}  : \textcolor{coeffectColor}{[}  \SYSTEMnt{A} {\textcolor{coeffectColor}{]_{ \SYSTEMsym{2} } } }    \vdash_{\textsc{l} }    \SYSTEMmv{f} \,  \SYSTEMmv{y}  \,  \SYSTEMmv{y}   :  \SYSTEMnt{B} \hspace{-6em}}
\hspace{-5em}}
{    \SYSTEMmv{f}  :    \SYSTEMnt{A}  \multimap  \SYSTEMnt{A}   \multimap  \SYSTEMnt{B}    ,   \SYSTEMmv{x}  :   \textcolor{coeffectColor}{\square_{ \SYSTEMsym{2} } }  \SYSTEMnt{A}     \vdash_{\textsc{l} }     \mathsf{let} \, \textcolor{coeffectColor}{[}  \SYSTEMmv{y}  \textcolor{coeffectColor}{]} =  \SYSTEMmv{x}  \, \mathsf{in} \,  \SYSTEMmv{f}  \,  \SYSTEMmv{y}  \,  \SYSTEMmv{y}   :  \SYSTEMnt{B} }
\end{align*}
\end{example}

\begin{example}[None-one-tons~\cite{McBride2016,quantitative-type-theory}, a.k.a. the Linearity Semiring]
  \label{exm:none-one-tons}
  Linearity can be captured with the semiring $\mathcal{R} = \{\SYSTEMsym{0}, \SYSTEMsym{1},  \omega \}$, with the required laws for $0$ and $1$, and $\SYSTEMsym{1}  \SYSTEMsym{+}  \SYSTEMsym{1} =  \omega $ and $ \omega   \SYSTEMsym{+}  \SYSTEMnt{r} =  \omega  = \SYSTEMnt{r}  \SYSTEMsym{+}   \omega $ and $  \omega   \cdot   \omega   =  \omega $, and ordering $ \SYSTEMsym{0}  \, \textcolor{coeffectColor}{\sqsubseteq} \,   \omega  $ and $ \SYSTEMsym{1}  \, \textcolor{coeffectColor}{\sqsubseteq} \,   \omega  $. This captures linear reasoning in a graded modal setting~\cite{DBLP:journals/pacmpl/AbelB20,DBLP:journals/pacmpl/AbelDE23},
  with $\SYSTEMsym{1}$ denoting linearity and $ \omega $ for
  non-linearity, and $\SYSTEMsym{0}$ providing a slightly finer-grained
  view when non-linearly discarding.
\end{example}

\begin{example}[Security~\cite{DBLP:journals/pacmpl/AbelB20,DBLP:conf/icfp/GaboardiKOBU16,mycroftfest2024}]
\label{exm:security-lattice}
Consider a semi-lattice of security levels $\mathcal{L} = \{\mathsf{Lo},
\mathsf{Hi}\}$ with ordering such that $\mathsf{Hi} \succeq \mathsf{Lo}$.
This forms a semiring with multiplication as join $\cdot = \vee$
with unit $1 = \mathsf{Lo}$, and addition as meet $+ = \wedge$
with unit $0 = \mathsf{Hi}$. We can type a function which maps
low-security to high-security values, using approximation where we set $\sqsubseteq \; = \; \succeq$:

\vspace{-1.5em}
\begin{align*}
\inferrule*[right=\SYSTEMRenameRuleLinapprox{}]
{
 \inferrule*[right=\SYSTEMRenameRuleLinpr{}]
 {
 \inferrule*[right=\SYSTEMRenameRuleLinder{}]
  {
    \inferrule*[right=\SYSTEMRenameRuleLinvar{}]
      {\quad}
      {   \SYSTEMmv{x}  :  \SYSTEMnt{A}    \vdash_{\textsc{l} }  \SYSTEMmv{x}  :  \SYSTEMnt{A} }
  }
  {   \SYSTEMmv{x}  : \textcolor{coeffectColor}{[}  \SYSTEMnt{A} {\textcolor{coeffectColor}{]_{  \mathsf{Lo}  } } }    \vdash_{\textsc{l} }  \SYSTEMmv{x}  :  \SYSTEMnt{A} }
 }
 {   \SYSTEMmv{x}  : \textcolor{coeffectColor}{[}  \SYSTEMnt{A} {\textcolor{coeffectColor}{]_{  \mathsf{Hi}  } } }    \vdash_{\textsc{l} }   \textcolor{coeffectColor}{[}  \SYSTEMmv{x}  \textcolor{coeffectColor}{]}   :   \textcolor{coeffectColor}{\square_{  \mathsf{Hi}  } }  \SYSTEMnt{A}   \qquad \mathsf{Hi} \succeq \mathsf{Lo}}
}
{   \SYSTEMmv{x}  : \textcolor{coeffectColor}{[}  \SYSTEMnt{A} {\textcolor{coeffectColor}{]_{  \mathsf{Lo}  } } }    \vdash_{\textsc{l} }   \textcolor{coeffectColor}{[}  \SYSTEMmv{x}  \textcolor{coeffectColor}{]}   :   \textcolor{coeffectColor}{\square_{  \mathsf{Hi}  } }  \SYSTEMnt{A}  }
\end{align*}
A function mapping from $ \mathsf{Hi} $ to $ \mathsf{Lo} $ values is derivable but cannot
use its input. This \emph{non-interference} property is proved by
\citet{DBLP:journals/pacmpl/AbelB20} and \citet{mycroftfest2024}.
\end{example}

\begin{example}[Product]
  \label{exm:product}
Given two pre-ordered semirings $\mathcal{R}$ and $\mathcal{S}$,
their product $\mathcal{R} \times \mathcal{S}$ is a semiring, with
zero element $\langle 0 , 0 \rangle \in (\mathcal{R} \times \mathcal{S})$,
one element $\langle 1 , 1 \rangle \in (\mathcal{R} \times \mathcal{S})$, and
operations and ordering acting pointwise. This is a standard
algebraic construction which has been used in graded systems to combine
analyses (e.g.~\cite{DBLP:journals/pacmpl/OrchardLE19}).

Section~\ref{subsec:linearity-preserving} uses this construction to
compose the linearity semiring (Example~\ref{exm:none-one-tons}) with some other
given semiring, thereby making explicit the linearity analysis inherent to Linear Base.

\citet{DBLP:journals/pacmpl/BianchiniDGZ23} present the \emph{smash product} construction
as a variation on the product construction which has some preconditions and filters
out the elements of the product that have either the left or the right $ = 0$,
instead providing a single canonical $\SYSTEMsym{0}$ element.
\end{example}

\begin{lemma}[Admissibility of substitution]
  There are two admissible substitutions:
  \begin{enumerate}
  \item (Linear) If $ \Gamma_{{\mathrm{1}}}  \vdash_{\textsc{l} }  \SYSTEMnt{t_{{\mathrm{1}}}}  :  \SYSTEMnt{A} $ and $  \Gamma_{{\mathrm{2}}} ,   \SYSTEMmv{x}  :  \SYSTEMnt{A}    \vdash_{\textsc{l} }  \SYSTEMnt{t_{{\mathrm{2}}}}  :  \SYSTEMnt{B} $
    then $ \Gamma_{{\mathrm{1}}}  \SYSTEMsym{+}  \Gamma_{{\mathrm{2}}}  \vdash_{\textsc{l} }   [  \SYSTEMnt{t_{{\mathrm{1}}}}  /  \SYSTEMmv{x}  ]  \SYSTEMnt{t_{{\mathrm{2}}}}   :  \SYSTEMnt{B} $;

  \item (Graded) If $ \Gamma_{{\mathrm{1}}}  \vdash_{\textsc{l} }  \SYSTEMnt{t_{{\mathrm{1}}}}  :  \SYSTEMnt{A} $ and $ \mathrm{graded}( \Gamma_{{\mathrm{1}}} ) $ and
    $  \Gamma_{{\mathrm{2}}} ,   \SYSTEMmv{x}  : \textcolor{coeffectColor}{[}  \SYSTEMnt{A} {\textcolor{coeffectColor}{]_{ \SYSTEMnt{r} } } }    \vdash_{\textsc{l} }  \SYSTEMnt{t_{{\mathrm{2}}}}  :  \SYSTEMnt{B} $
    then $  \textcolor{coeffectColor}{ \SYSTEMnt{r}  \cdot}  \Gamma_{{\mathrm{1}}}   \SYSTEMsym{+}  \Gamma_{{\mathrm{2}}}  \vdash_{\textsc{l} }   [  \SYSTEMnt{t_{{\mathrm{1}}}}  /  \SYSTEMmv{x}  ]  \SYSTEMnt{t_{{\mathrm{2}}}}   :  \SYSTEMnt{B} $.
  \end{enumerate}
  \label{lemma:linear-base-sub-admiss}
\end{lemma}

\begin{remark}
In the literature (e.g.~\citet{brunel2014core,DBLP:conf/icfp/GaboardiKOBU16,DBLP:journals/pacmpl/OrchardLE19}),
the graded context predicates (\ref{dfn:graded-contexts}) are often presented instead as partial identity functions on contexts:
written $[ \Gamma ]$, the function is identity when the context contains only
graded assumptions and is undefined otherwise, preventing a rule from
being applied (and similarly $[ \Gamma ]_r$ for contexts
containing only assumptions graded by $r$).
We chose to make the predicates explicit for clarity.
\end{remark}

\paragraph{Operational Semantics}

Figure~\ref{fig:linear-base-ops-and-eqns} (first group) gives the
operational semantics, which is that of the call-by-name
$\lambda$-calculus with an additional congruence rule
for graded modal elimination and $\beta$-reduction of graded modal
introduction followed by elimination. %Note the rule for
%evaluating inside of a graded modal introduction in keeping with
%the call-by-name style.

Note that the operational semantics we consider for each calculus in this paper represents the standard
models given in the literature in question, with progress and
preservation theorems, but these do not track specific resource
behaviour operationally (see discussion of resource-aware operational
models in Section~\ref{sec:raos}).

Appendix~\refappendix{app:definitions} gives the definition of syntactic
substitution which is standard.

\begin{figure}[t]
\raggedright\noindent\framebox{$ \SYSTEMnt{t}  \rightsquigarrow_{\textsc{l} }  \SYSTEMnt{t} $} \quad Operational semantics (Call-By-Name)
\begin{align*}
  \begin{array}{c}
  \SYSTEMdruleSemLinbeta{}
  \quad
  \SYSTEMdruleSemLincongAppL{}
  \\[1em]
  \SYSTEMdruleSemLinbetaBox{}
  \quad
  \SYSTEMdruleSemLincongLetL{} \\[1em]
  \end{array}
\end{align*}
\bigskip

\raggedright\noindent\framebox{$ \SYSTEMnt{t}  \,\equiv_{\textsc{l} }\,  \SYSTEMnt{t'} $}
\hspace{1em} Equational theory (congruence rules omitted).
\begin{gather*}
\begin{align*}
  \begin{array}{c}
\SYSTEMdruleLinEqbeta{} \;\;
\SYSTEMdruleLinEqeta{} \;\;
\SYSTEMdruleLinEqbetaBox{} \\[1em]
\SYSTEMdruleLinEqetaBox{} \quad
\SYSTEMdruleLinEqletCommBox{}  \\[1em]
\SYSTEMdruleLinEqletCommOne{} \, \SYSTEMdruleLinEqletCommTwo{}
  \end{array}
\end{align*}
\end{gather*}
\caption{Linear Base - Operational semantics and equations}
\label{fig:linear-base-ops-and-eqns}
\end{figure}

\paragraph{Equational theory}

Figure~\ref{fig:linear-base-ops-and-eqns} (second group) defines an equational theory, giving the key rules.
For brevity, we omit the congruence rules which apply to all
sub-term positions. Note that $ \rightsquigarrow_{\textsc{l} }  \subseteq  \equiv_{\textsc{l} } $.

\begin{definition}[Freshness]
  We make use of freshness here and throughout. It is defined as follows:
  \begin{align*}
    x_1, \ldots, x_n \# t \triangleq \{x_1, \ldots, x_n\} \cap \mathsf{fv}(t) = \emptyset
  \end{align*}
\end{definition}

\subsection{Graded Base}
\label{sec:graded-base}
The next calculus, Graded Base, is representative of the coeffect systems of
\citet{DBLP:conf/icalp/PetricekOM13,petricek2014coeffects},
the generalised bounded linear type system of \citet{ghica2014bounded},
the simply-typed core of work by \citet{DBLP:journals/pacmpl/AbelB20}, \citet{grad}, and \citet{DBLP:journals/pacmpl/AbelDE23},
the OO-based approach of \citet{DBLP:journals/pacmpl/BianchiniDGZS22},
and others.
The syntax of terms is that of the $\lambda$-calculus:
\begin{align*}
    \SYSTEMnt{t} & ::= \SYSTEMmv{x} \mid  \SYSTEMnt{t_{{\mathrm{1}}}} \,  \SYSTEMnt{t_{{\mathrm{2}}}}  \mid  \lambda  \SYSTEMmv{x}  .  \SYSTEMnt{t} 
  \tag{terms}
\end{align*}
We use $t$ to also range over Graded Base terms but disambiguate which
calculus terms belong to via the context of their later use (in translations).
Type comprise graded functions and a base type $ \mathrm{K}  $:
\begin{align*}
  \SYSTEMnt{A} & ::=  \SYSTEMnt{A}  \xrightarrow{\textcolor{coeffectColor}{ \SYSTEMnt{r} } }  \SYSTEMnt{B}  \mid  \mathrm{K} 
  \tag{types} \\
  \Delta & ::=  \emptyset  \mid  \Delta ,   \SYSTEMmv{x}  :_{\textcolor{coeffectColor}{ \SYSTEMnt{r} } }  \SYSTEMnt{A}  
  \tag{contexts}
\end{align*}
Instead of a linear function type and graded modality, Graded Base has just a \emph{graded function type}, where $r \in \mathcal{R}$. Contexts contain only graded assumptions which we instead denote as $ \SYSTEMmv{x}  :_{\textcolor{coeffectColor}{ \SYSTEMnt{r} } }  \SYSTEMnt{A} $.

Figure~\ref{fig:graded-base} (first group) gives the typing with judgments of the form $ \Delta  \vdash_{\textsc{g} }  \SYSTEMnt{t}  :  \SYSTEMnt{A} $. Note the singleton context graded at $\SYSTEMsym{1}$
in the \textit{variable} rule. The \textit{abstraction} rule introduces its binding into the context
graded at $\SYSTEMnt{r}$, per the graded function arrow. In the \textit{application} rule, the function
arrow propagates the function's resource requirements onto its argument $\SYSTEMnt{t_{{\mathrm{2}}}}$ via scalar
multiplication on $\Delta_{{\mathrm{2}}}$ (i.e., the context of $\SYSTEMnt{t_{{\mathrm{2}}}}$). Weakening scales by $0$ all the grades in an additional context $\Delta'$.
\begin{figure}[t]
\raggedright\noindent\framebox{$ \Delta  \vdash_{\textsc{g} }  \SYSTEMnt{t}  :  \SYSTEMnt{A} $} \quad Typing rules
\begin{align*}
\begin{array}{c}
\SYSTEMdruleGradvar{}
\quad
\SYSTEMdruleGradweak{}
\quad
\SYSTEMdruleGradapprox{}
\\[1.25em]
\SYSTEMdruleGradabs{}
\quad
\SYSTEMdruleGradapp{}
\end{array}
\end{align*}
\bigskip

\raggedright\noindent\framebox{$ \SYSTEMnt{t}  \rightsquigarrow_{\textsc{g} }  \SYSTEMnt{t'} $} \quad Operational semantics
\vspace{-1em}
\begin{align*}
  \begin{array}{c}
  \SYSTEMdruleSemGrdbeta{}
  \quad
  \SYSTEMdruleSemGrdcongAppL{}
  \end{array}
\end{align*}
\bigskip

\raggedright\noindent\framebox{$ \SYSTEMnt{t}  \equiv_{\textsc{g} }  \SYSTEMnt{t'} $} \quad Equational theory (congruence rules omitted)
\vspace{-1em}
\begin{align*}
  \begin{array}{c}
\SYSTEMdruleGradEqbeta{} \quad
\SYSTEMdruleGradEqeta{}
  \end{array}
\end{align*}
\caption{Graded Base system}
\label{fig:graded-base}
\end{figure}

Graded substitution is admissible, akin to the second clause
of Linear Base substitution (Lemma~\ref{lemma:linear-base-sub-admiss}):
\begin{lemma}[Admissibility of substitution]
   If $ \Delta_{{\mathrm{1}}}  \vdash_{\textsc{g} }  \SYSTEMnt{t_{{\mathrm{1}}}}  :  \SYSTEMnt{A} $ and
    $  \Delta_{{\mathrm{2}}} ,   \SYSTEMmv{x}  :_{\textcolor{coeffectColor}{ \SYSTEMnt{r} } }  \SYSTEMnt{A}    \vdash_{\textsc{g} }  \SYSTEMnt{t_{{\mathrm{2}}}}  :  \SYSTEMnt{B} $ then\\
    $  \textcolor{coeffectColor}{ \SYSTEMnt{r}  \cdot}  \Delta_{{\mathrm{1}}}   \SYSTEMsym{+}  \Delta_{{\mathrm{2}}}  \vdash_{\textsc{g} }   [  \SYSTEMnt{t_{{\mathrm{1}}}}  /  \SYSTEMmv{x}  ]  \SYSTEMnt{t_{{\mathrm{2}}}}   :  \SYSTEMnt{B} $.
\end{lemma}

\begin{example}
The simply-typed $\lambda$-calculus is recovered
by instantiating Graded Base with the single-point semiring $\mathcal{R} = \{\bullet\}$.
\end{example}
\begin{example}
The higher-order function $h =  \lambda  \SYSTEMmv{f}  .    \lambda  \SYSTEMmv{x}  .   \SYSTEMmv{f} \,  \SYSTEMsym{(}   \SYSTEMmv{f} \,  \SYSTEMmv{x}   \SYSTEMsym{)}    $ mentioned in the introduction, composing a function with
itself twice, has the following derivation for its body:
\begin{gather*}
\inferrule*[right=\SYSTEMRenameRuleGradapp{}]
{\inferrule*[right=\SYSTEMRenameRuleGradvar{}]{\;}{   \SYSTEMmv{f}  :_{\textcolor{coeffectColor}{ \SYSTEMsym{1} } }   (   \SYSTEMnt{A}  \xrightarrow{\textcolor{coeffectColor}{ \SYSTEMnt{r} } }  \SYSTEMnt{A}   )     \vdash_{\textsc{g} }  \SYSTEMmv{f}  :   \SYSTEMnt{A}  \xrightarrow{\textcolor{coeffectColor}{ \SYSTEMnt{r} } }  \SYSTEMnt{A}  }
 \inferrule*[right=\SYSTEMRenameRuleGradapp{}]
 {\inferrule*[right=\SYSTEMRenameRuleGradvar{}]{\;}{   \SYSTEMmv{f}  :_{\textcolor{coeffectColor}{ \SYSTEMsym{1} } }   (   \SYSTEMnt{A}  \xrightarrow{\textcolor{coeffectColor}{ \SYSTEMnt{r} } }  \SYSTEMnt{A}   )     \vdash_{\textsc{g} }  \SYSTEMmv{f}  :   \SYSTEMnt{A}  \xrightarrow{\textcolor{coeffectColor}{ \SYSTEMnt{r} } }  \SYSTEMnt{A}  }
  \inferrule*[right=\SYSTEMRenameRuleGradvar{}]{\;}{   \SYSTEMmv{x}  :_{\textcolor{coeffectColor}{ \SYSTEMsym{1} } }  \SYSTEMnt{A}    \vdash_{\textsc{g} }  \SYSTEMmv{x}  :  \SYSTEMnt{A} }}
  {    \SYSTEMmv{f}  :_{\textcolor{coeffectColor}{ \SYSTEMsym{1} } }   (   \SYSTEMnt{A}  \xrightarrow{\textcolor{coeffectColor}{ \SYSTEMnt{r} } }  \SYSTEMnt{A}   )    ,   \SYSTEMmv{x}  :_{\textcolor{coeffectColor}{ \SYSTEMnt{r} } }  \SYSTEMnt{A}    \vdash_{\textsc{g} }   \SYSTEMmv{f} \,  \SYSTEMmv{x}   :  \SYSTEMnt{A}  }
\hspace{-3em}}
{    \SYSTEMmv{f}  :_{\textcolor{coeffectColor}{ \SYSTEMsym{(}  \SYSTEMsym{1}  \SYSTEMsym{+}  \SYSTEMsym{1}  \SYSTEMsym{)} } }   (   \SYSTEMnt{A}  \xrightarrow{\textcolor{coeffectColor}{ \SYSTEMnt{r} } }  \SYSTEMnt{A}   )    ,   \SYSTEMmv{x}  :_{\textcolor{coeffectColor}{ \SYSTEMsym{(}   \SYSTEMnt{r}  \cdot  \SYSTEMnt{r}   \SYSTEMsym{)} } }  \SYSTEMnt{A}    \vdash_{\textsc{g} }   \SYSTEMmv{f} \,  \SYSTEMsym{(}   \SYSTEMmv{f} \,  \SYSTEMmv{x}   \SYSTEMsym{)}   :  \SYSTEMnt{A} }
\end{gather*}
\end{example}
All the semirings mentioned in Section~\ref{sec:linear-base} (Linear Base) can also be used with
Graded Base.

\paragraph{Operational Semantics and Equational theory}

Figure~\ref{fig:graded-base} (second group) gives the
operational semantics, which is that of the CBN
$\lambda$-calculus. Appendix~\refappendix{app:definitions} gives the definition of syntactic
substitution which is standard.

Figure~\ref{fig:graded-base} (third group) defines an equational theory
for the calculus, giving the key rules but omitting the congruence
rules which apply to all sub-term positions. Note $ \rightsquigarrow_{\textsc{g} }  \, \subseteq \,  \equiv_{\textsc{g} } $.

\subsection{Translation}
\label{sec:transl}
In the course of the paper, we define translations between several calculi of the Linear and Graded
strands, according to the overview in Figure~\ref{fig:relationship}. As for the term and type syntax
of the different calculi, we use the same interpretation brackets across different variations of
linear- and graded-base calculi respectively, making clear from the context which interpretation we are referring to.

The interpretation
$ \textcolor{GLcolor}{\llbracket}   -   \textcolor{GLcolor}{\rrbracket} $ takes graded-base terms to linear-base terms and $ \textcolor{LGcolor}{\llparenthesis} \smidge   -   \smidge \textcolor{LGcolor}{\rrparenthesis} $ takes linear-base
terms to graded-base terms. The following mnemonics have proven useful to the authors:
\begin{itemize}
  \item Given a graded term $t$, $ \textcolor{GLcolor}{\llbracket}  \SYSTEMnt{t}  \textcolor{GLcolor}{\rrbracket} $ is its translation, into a \underline{L}inear term,
    with the square \underline{L}ilac brackets mirroring the shape of the letter L.
  \item Given a linear term $t$, $ \textcolor{LGcolor}{\llparenthesis} \smidge  \SYSTEMnt{t}  \smidge \textcolor{LGcolor}{\rrparenthesis} $ is its translation into a \underline{G}raded term,
    with the rounded \underline{G}reen parentheses mirroring the shape of the letter G.
\end{itemize}
The same notation is used for type and context translations.

\subsubsection{Graded Base to Linear Base translation}
\label{subsec:transl-graded-base-to-linear-base}

\NewCommandCopy{\oldSYSTEMdrule}{\SYSTEMdrule}
\NewCommandCopy{\oldEquiv}{\equiv}
\NewCommandCopy{\oldTriangleq}{\triangleq}
\renewcommand{\triangleq}{& \oldTriangleq &}
\renewcommand{\equiv}{ \oldEquiv}
\renewcommand{\SYSTEMdrule}[4][]{#3 \qquad #2} % \quad\SYSTEMdrulename{#4}}}

% Our first defintion of Graded Base to Linear Base follows:
%
\begin{gather*}
{\setlength{\arraycolsep}{0.1em}
\begin{array}{lcl@{\hspace{2em}}r}
\SYSTEMdruleTranslGtoLTmvar{} \\
\SYSTEMdruleTranslGtoLTmabs{}
\SYSTEMdruleTranslGtoLTmapp{} & \text{(Term translation)} \\[0.75em]
\SYSTEMdruleTranslGtoLTybaseTy{} \\
\SYSTEMdruleTranslGtoLTyfunTy{} & \text{(Type translation)} \\[0.75em]
\SYSTEMdruleTranslGtoLGtxtempty{} \\
\SYSTEMdruleTranslGtoLGtxtgrad{} & \text{(Context translation)}
\end{array}}
\end{gather*}
\let\SYSTEMdrule\oldSYSTEMdrule
\let\equiv\oldEquiv
\let\triangleq\oldTriangleq
The key idea is to use the graded modality to capture the grading of assumptions and functions.
The graded function type translates to the Linear Base function type $     \multimap     $, with its left-hand side
under $ \textcolor{coeffectColor}{\square_{ \SYSTEMnt{r} } }     $.
This translation leads to the insight that Graded Base effectively
wraps a graded modality around function inputs, capturing the usage of the parameter in the body.
Since every term in Graded Base has an associated grading in the context,
every abstraction in the translation from Graded to Linear needs to do an unboxing (let) and every application needs to do a boxing (promotion).

Contexts are translated straightforwardly. The translation can never produce a linear assumption.

\begin{example} The judgment $    \SYSTEMmv{f}  :_{\textcolor{coeffectColor}{ \SYSTEMsym{(}  \SYSTEMsym{1}  \SYSTEMsym{+}  \SYSTEMsym{1}  \SYSTEMsym{)} } }   (   \SYSTEMnt{A}  \xrightarrow{\textcolor{coeffectColor}{ \SYSTEMnt{r} } }  \SYSTEMnt{A}   )    ,   \SYSTEMmv{x}  :_{\textcolor{coeffectColor}{ \SYSTEMsym{(}   \SYSTEMnt{r}  \cdot  \SYSTEMnt{r}   \SYSTEMsym{)} } }  \SYSTEMnt{A}    \vdash_{\textsc{g} }   \SYSTEMmv{f} \,  \SYSTEMsym{(}   \SYSTEMmv{f} \,  \SYSTEMmv{x}   \SYSTEMsym{)}   :  \SYSTEMnt{A} $
in Graded Base translates to $    \SYSTEMmv{f}  : \textcolor{coeffectColor}{[}    \textcolor{coeffectColor}{\square_{ \SYSTEMnt{r} } }   \textcolor{GLcolor}{\llbracket}  \SYSTEMnt{A}  \textcolor{GLcolor}{\rrbracket}    \multimap   \textcolor{GLcolor}{\llbracket}  \SYSTEMnt{A}  \textcolor{GLcolor}{\rrbracket}   {\textcolor{coeffectColor}{]_{ \SYSTEMsym{(}  \SYSTEMsym{1}  \SYSTEMsym{+}  \SYSTEMsym{1}  \SYSTEMsym{)} } } }   ,   \SYSTEMmv{x}  : \textcolor{coeffectColor}{[}   \textcolor{GLcolor}{\llbracket}  \SYSTEMnt{A}  \textcolor{GLcolor}{\rrbracket}  {\textcolor{coeffectColor}{]_{ \SYSTEMsym{(}   \SYSTEMnt{r}  \cdot  \SYSTEMnt{r}   \SYSTEMsym{)} } } }    \vdash_{\textsc{l} }   \SYSTEMmv{f} \,   \textcolor{coeffectColor}{[}   \SYSTEMmv{f} \,   \textcolor{coeffectColor}{[}  \SYSTEMmv{x}  \textcolor{coeffectColor}{]}    \textcolor{coeffectColor}{]}    :   \textcolor{GLcolor}{\llbracket}  \SYSTEMnt{A}  \textcolor{GLcolor}{\rrbracket}  $
in Linear Base. Note the right-hand sides of an application find themselves under a promotion.
\end{example}
\begin{remark}
We do not translate $ \SYSTEMnt{A}  \xrightarrow{\textcolor{coeffectColor}{ \SYSTEMsym{1} } }  \SYSTEMnt{B} $ to a linear function $  \textcolor{GLcolor}{\llbracket}  \SYSTEMnt{A}  \textcolor{GLcolor}{\rrbracket}   \multimap   \textcolor{GLcolor}{\llbracket}  \SYSTEMnt{B}  \textcolor{GLcolor}{\rrbracket}  $;
  this is because $ \SYSTEMnt{A}  \xrightarrow{\textcolor{coeffectColor}{ \SYSTEMsym{1} } }  \SYSTEMnt{B} $ does not necessarily mean that the function uses its argument linearly.
  Some choices of semiring mean that $1$ does not mean `used exactly once', the simplest example
  being the single-point (trivial) semiring in which $0 = 1$ and so $1$ would mean any usage.
\end{remark}
We now establish key soundness results. Firstly, that the translation is type
preserving, i.e., translation of terms,
contexts, and types is such that a well-typed derivation in Graded
Base is well-typed under its translation to Linear Base. This
requires various small lemmas that the translation is a
homomorphism between graded-base contexts and linear-base contexts
which we elide here. Secondly, the translation
preserves the operational semantics, and lastly it preserves the
equational theory. Appendix~\refappendix{app:proofs-grad-to-lin} provides the full proofs and supporting lemmas.
\begin{restatable}{thrm}{gradToLinTranslation}[Soundness of the Graded Base to Linear Base translation]
  \begin{itemize}[itemsep=0.3em]
  \item Type preservation: $ \Delta  \vdash_{\textsc{g} }  \SYSTEMnt{t}  :  \SYSTEMnt{A}  \implies   \textcolor{GLcolor}{\llbracket}  \Delta  \textcolor{GLcolor}{\rrbracket}   \vdash_{\textsc{l} }   \textcolor{GLcolor}{\llbracket}  \SYSTEMnt{t}  \textcolor{GLcolor}{\rrbracket}   :   \textcolor{GLcolor}{\llbracket}  \SYSTEMnt{A}  \textcolor{GLcolor}{\rrbracket}  $
  \item Operational correspondence: $ \SYSTEMnt{t}  \rightsquigarrow_{\textsc{g} }  \SYSTEMnt{t'}  \implies   \textcolor{GLcolor}{\llbracket}  \SYSTEMnt{t}  \textcolor{GLcolor}{\rrbracket}   \rightsquigarrow_{\textsc{l} }^\ast   \textcolor{GLcolor}{\llbracket}  \SYSTEMnt{t'}  \textcolor{GLcolor}{\rrbracket}  $
  \item Equation preservation: $ \SYSTEMnt{t}  \equiv_{\textsc{g} }  \SYSTEMnt{t'}  \implies   \textcolor{GLcolor}{\llbracket}  \SYSTEMnt{t}  \textcolor{GLcolor}{\rrbracket}   \,\equiv_{\textsc{l} }\,   \textcolor{GLcolor}{\llbracket}  \SYSTEMnt{t'}  \textcolor{GLcolor}{\rrbracket}  $
  \end{itemize}
  \label{thrm:gradToLinTranslation}
\end{restatable}

\subsubsection{Linear Base to Graded Base translation}
\label{subsec:lin-to-grad-cps}

A type-preserving translation from Linear Base to Graded Base is however more difficult
since we need to find some encoding of the graded modal type $ \textcolor{LGcolor}{\llparenthesis} \smidge   \textcolor{coeffectColor}{\square_{ \SYSTEMnt{r} } }  \SYSTEMnt{A}   \smidge \textcolor{LGcolor}{\rrparenthesis} $.
From the introduction rule of the graded modality (promotion), such a translation on
terms should incur a multiplication of the context by $r$ which in Graded Base is only
provided through application of a $ \SYSTEMnt{A}  \xrightarrow{\textcolor{coeffectColor}{ \SYSTEMnt{r} } }  \SYSTEMnt{B} $ function. Thus a translation must have such
a function in a negative position, suggesting a continuation-passing style (CPS)
or Church-encoding-like form.

Taking inspiration from~\citet{lindley2025encoding} on encoding products,
we extend Plotkin's CPS translation of the call-by-name $\lambda$-calculus~\cite{plotkin1975call}
with a CPS encoding of the graded modalities of Linear Base into Graded Base:
\renewcommand{\triangleq}{& \oldTriangleq &}
\renewcommand{\equiv}{& \oldEquiv &}
\renewcommand{\SYSTEMdrule}[4][]{#3 \qquad #2}
\begin{gather*}
{\setlength{\arraycolsep}{0.1em}
\begin{array}{lcl@{\hspace{2em}}r}
\SYSTEMdruleTranslLtoGCpsTyfunTy{} \\
\SYSTEMdruleTranslLtoGCpsTyboxTy{} \\
\SYSTEMdruleTranslLtoGCpsTybaseTy{} & \text{(Type translation)} \\[0.75em]
\SYSTEMdruleTranslLtoGCpsTmvar{} \\
\SYSTEMdruleTranslLtoGCpsTmabs{} \\
\SYSTEMdruleTranslLtoGCpsTmapp{} \\
\SYSTEMdruleTranslLtoGCpsTmpr{} \\
\SYSTEMdruleTranslLtoGCpsTmlet{} & \text{(Term translation)} \\[0.75em]
\SYSTEMdruleTranslLtoGGtxtempty{} \\
\SYSTEMdruleTranslLtoGGtxtlin{} \\
\SYSTEMdruleTranslLtoGGtxtgrad{} & \text{(Context translation)}
\end{array}}
\end{gather*}
\let\SYSTEMdrule\oldSYSTEMdrule
\let\equiv\oldEquiv
\let\triangleq\oldTriangleq
\noindent
The key point is to (recursively) encode $ \textcolor{coeffectColor}{\square_{ \SYSTEMnt{r} } }  \SYSTEMnt{A} $ as
 $  (   \SYSTEMnt{A}  \xrightarrow{\textcolor{coeffectColor}{ \SYSTEMnt{r} } }   \mathrm{K}    )   \xrightarrow{\textcolor{coeffectColor}{ \SYSTEMsym{1} } }   \mathrm{K}  $ so that in the case
of promotion $ \textcolor{coeffectColor}{[}  \SYSTEMnt{t}  \textcolor{coeffectColor}{]} $ we can use application to scale
the grades of $\SYSTEMnt{t}$. Note that linear assumptions get mapped
to graded assumptions with grade $1$, though this may not mean
`linear use' (see Section~\ref{subsec:linearity-preserving} for
a solution).

\begin{restatable}{thrm}{linToGradCPSTranslation}[Soundness of the Linear Base to Graded Base translation (proofs in Appendix~\refappendix{app:proofs-lin-to-grad-cps}).]
  \begin{itemize}[itemsep=0.2em]
  \item Type preservation: $ \Gamma  \vdash_{\textsc{l} }  \SYSTEMnt{t}  :  \SYSTEMnt{A}  \implies   \textcolor{LGcolor}{\llparenthesis}  \Gamma  \textcolor{LGcolor}{\rrparenthesis}   \vdash_{\textsc{g} }   \textcolor{LGcolor}{\llparenthesis} \smidge  \SYSTEMnt{t}  \smidge \textcolor{LGcolor}{\rrparenthesis}   :   \textcolor{LGcolor}{\llparenthesis} \smidge  \SYSTEMnt{A}  \smidge \textcolor{LGcolor}{\rrparenthesis}  $
  \item Operational correspondence: $ \SYSTEMnt{t}  \rightsquigarrow_{\textsc{l} }  \SYSTEMnt{t'}  \implies   \textcolor{LGcolor}{\llparenthesis} \smidge  \SYSTEMnt{t}  \smidge \textcolor{LGcolor}{\rrparenthesis}   \rightsquigarrow_{\beta\textsc{g} }^\ast   \textcolor{LGcolor}{\llparenthesis} \smidge  \SYSTEMnt{t'}  \smidge \textcolor{LGcolor}{\rrparenthesis}  $
  \item \vspace{-0.5em} Equation preservation: $\SYSTEMnt{t}  \equiv^{\beta_\to\beta\eta_\Box}_{\textsc{l} }  \SYSTEMnt{t'} \implies   \textcolor{LGcolor}{\llparenthesis} \smidge  \SYSTEMnt{t}  \smidge \textcolor{LGcolor}{\rrparenthesis}   \equiv_{\textsc{g} }   \textcolor{LGcolor}{\llparenthesis} \smidge  \SYSTEMnt{t'}  \smidge \textcolor{LGcolor}{\rrparenthesis}  $
  \end{itemize}
  \label{thrm:linToGradCPSTranslation}
\end{restatable}
\noindent
As we use Plotkin's CBN CPS translation, there are two notable implications:
Firstly, regarding operational correspondence, our source uses CBN but we need full
$\beta$-reduction (via the \SYSTEMRenameRuleSemGrdcongAbs{} rule) in the target to prove simulation.
Computational adequacy then follows from the $\lambda$-calculus standardisation theorem.
Secondly, regarding equation preservation: we
preserve only $\beta$-equality for functions and not $\eta$-equality---a well-known limitation of Plotkin's translation.
However, we preserve both $\beta$- and $\eta$-equality for the graded modality,
hence we label the equational theory here as $ \equiv^{\beta_\to\beta\eta_\Box}_{\textsc{l} } $.
Section~\ref{sec:poly} considers an alternate encoding in a polymorphic Graded Base setting.

\section{Extending Graded Base with a Graded Modality}
\label{sec:calculi-with-modality}
We now add a graded modality to
Graded Base thereby arriving at \textbf{Graded~Modal~Base}. This captures the type system
features of calculi like \textsc{GraD}~\cite{grad} and a monomorphic
subset of~\citet{DBLP:journals/pacmpl/AbelB20}, and permits a simple translation to Linear Base.

\subsection{Graded Modal Base}\label{sec:graded-modal-base}
Graded Modal Base can now share the syntax (and operational semantics) of Linear Base:
\begin{align*}
    \SYSTEMnt{t} & ::= \SYSTEMmv{x} \mid  \SYSTEMnt{t_{{\mathrm{1}}}} \,  \SYSTEMnt{t_{{\mathrm{2}}}}  \mid  \lambda  \SYSTEMmv{x}  .  \SYSTEMnt{t}  \mid  \textcolor{coeffectColor}{[}  \SYSTEMnt{t}  \textcolor{coeffectColor}{]}  \mid  \mathsf{let} \, \textcolor{coeffectColor}{[}  \SYSTEMmv{x}  \textcolor{coeffectColor}{]} =  \SYSTEMnt{t_{{\mathrm{1}}}}  \, \mathsf{in} \,  \SYSTEMnt{t_{{\mathrm{2}}}} 
    \tag{shared terms} \\
  \SYSTEMnt{A} & ::=  \SYSTEMnt{A}  \xrightarrow{\textcolor{coeffectColor}{ \SYSTEMnt{r} } }  \SYSTEMnt{B}  \mid  \textcolor{coeffectColor}{\square_{ \SYSTEMnt{r} } }  \SYSTEMnt{A}  \mid  \mathrm{K} 
    \tag{extended types}
\end{align*}
Typing judgments in Graded Modal Base are of the form $ \Delta  \vdash_{\textsc{g}_\square }  \SYSTEMnt{t}  :  \SYSTEMnt{A} $.
The typing rules are the same as Graded Base with addition of typing for
graded modal introduction and elimination:
\begin{align*}
\begin{array}{c}
\SYSTEMdruleGradBoxpr{}
\quad
\SYSTEMdruleGradBoxlet{}
\end{array}
\end{align*}
The equational theory extends to include $\beta\eta$-equality on
graded modalities (see Appendix~\refappendix{app:definitions-graded-modal-base}).

\subsection{Translation}

\subsubsection{Linear Base to Graded Modal Base translation}
\label{subsec:transl-lb2gmb}

We can now translate Linear Base to Graded \emph{Modal} Base as
a homomorphism on terms, rather than the CPS encoding of Section~\ref{sec:transl}:
\renewcommand{\triangleq}{& \oldTriangleq &}
\renewcommand{\equiv}{& \oldEquiv &}
\renewcommand{\SYSTEMdrule}[4][]{#3 \qquad #2}
\begin{gather*}
{\setlength{\arraycolsep}{0.1em}
\begin{array}{lcl@{\hspace{2em}}r}
\SYSTEMdruleTranslLtoGTmvar{} \\
\SYSTEMdruleTranslLtoGTmabs{} \\
\SYSTEMdruleTranslLtoGTmapp{} \\
\SYSTEMdruleTranslLtoGTmpr{} \\
\SYSTEMdruleTranslLtoGTmlet{} & \text{(Term translation)} \\[0.5em]
\SYSTEMdruleTranslLtoGTyfunTy{} \\
\SYSTEMdruleTranslLtoGTyboxTy{} \\
\SYSTEMdruleTranslLtoGTybaseTy{} & \text{(Type translation)} \\[0.5em]
\SYSTEMdruleTranslLtoGGtxtempty{} \\
\SYSTEMdruleTranslLtoGGtxtlin{} \\
\SYSTEMdruleTranslLtoGGtxtgrad{} & \text{(Context translation)}
\end{array}}
\end{gather*}
\let\SYSTEMdrule\oldSYSTEMdrule
\let\equiv\oldEquiv
\let\triangleq\oldTriangleq
For types, we translate $ \textcolor{coeffectColor}{\square_{ \SYSTEMnt{r} } }     $ in the straightforward way, directly
to the Graded Modal Base graded modality. To translate
the linear function space we grade the resulting function arrow
in Graded Modal Base at $1 \in \mathcal{R}$ and similarly for linear assumptions.

\begin{example}
  \label{exm:k-combinator}
  The following Linear Base judgment types a $K$-combinator-like expression:
  $$   \SYSTEMmv{x}  :  \SYSTEMnt{B}    \vdash_{\textsc{l} }   \lambda  \SYSTEMmv{y}  .   \mathsf{let} \, \textcolor{coeffectColor}{[}  \SYSTEMmv{z}  \textcolor{coeffectColor}{]} =  \SYSTEMmv{y}  \, \mathsf{in} \,  \SYSTEMmv{x}    :     \textcolor{coeffectColor}{\square_{ \SYSTEMsym{0} } }  \SYSTEMnt{A}    \multimap  \SYSTEMnt{B}  $$
  The parameter $y$ is used by eliminating the graded modality,
  binding to $z$ but then discarding the term. Its translation to Graded Modal Base results
  in an identical term but with translated types:
  $$   \SYSTEMmv{x}  :_{\textcolor{coeffectColor}{ \SYSTEMsym{1} } }   \textcolor{LGcolor}{\llparenthesis} \smidge  \SYSTEMnt{B}  \smidge \textcolor{LGcolor}{\rrparenthesis}     \vdash_{\textsc{g} }   \lambda  \SYSTEMmv{y}  .   \mathsf{let} \, \textcolor{coeffectColor}{[}  \SYSTEMmv{z}  \textcolor{coeffectColor}{]} =  \SYSTEMmv{y}  \, \mathsf{in} \,  \SYSTEMmv{x}    :     \textcolor{coeffectColor}{\square_{ \SYSTEMsym{0} } }   \textcolor{LGcolor}{\llparenthesis} \smidge  \SYSTEMnt{A}  \smidge \textcolor{LGcolor}{\rrparenthesis}     \xrightarrow{\textcolor{coeffectColor}{ \SYSTEMsym{1} } }   \textcolor{LGcolor}{\llparenthesis} \smidge  \SYSTEMnt{B}  \smidge \textcolor{LGcolor}{\rrparenthesis}   $$
\end{example}

\begin{restatable}{thrm}{linToGradTranslation}[Soundness of the Linear Base to Graded Modal Base translation]\label{thrm:linToGradTranslation}
  \begin{itemize}[itemsep=0.25em,topsep=0.25em]
    \item Type preservation: $ \Gamma  \vdash_{\textsc{l} }  \SYSTEMnt{t}  :  \SYSTEMnt{A}  \implies   \textcolor{LGcolor}{\llparenthesis}  \Gamma  \textcolor{LGcolor}{\rrparenthesis}   \vdash_{\textsc{g}_\square }   \textcolor{LGcolor}{\llparenthesis} \smidge  \SYSTEMnt{t}  \smidge \textcolor{LGcolor}{\rrparenthesis}   :   \textcolor{LGcolor}{\llparenthesis} \smidge  \SYSTEMnt{A}  \smidge \textcolor{LGcolor}{\rrparenthesis}  $
    \item Operational correspondence: $ \SYSTEMnt{t}  \rightsquigarrow_{\textsc{l} }  \SYSTEMnt{t'}  \implies   \textcolor{LGcolor}{\llparenthesis} \smidge  \SYSTEMnt{t}  \smidge \textcolor{LGcolor}{\rrparenthesis}   \rightsquigarrow_{\textsc{g}\Box }^\ast   \textcolor{LGcolor}{\llparenthesis} \smidge  \SYSTEMnt{t'}  \smidge \textcolor{LGcolor}{\rrparenthesis}  $
    \item Equation preservation: $ \SYSTEMnt{t}  \,\equiv_{\textsc{l} }\,  \SYSTEMnt{t'}  \implies   \textcolor{LGcolor}{\llparenthesis} \smidge  \SYSTEMnt{t}  \smidge \textcolor{LGcolor}{\rrparenthesis}   \,\equiv_{\textsc{g}_\Box}\,   \textcolor{LGcolor}{\llparenthesis} \smidge  \SYSTEMnt{t'}  \smidge \textcolor{LGcolor}{\rrparenthesis}  $
  \end{itemize}
\end{restatable}

\noindent
Appendix~\refappendix{app:proofs-lin-to-grad-mod} provides the proof along with its supporting lemmas.

\begin{remark}
  We might consider whether there is a dual to the above, a kind of completeness result,
i.e., that
given $ \Delta  \vdash_{\textsc{g} }  \SYSTEMnt{t}  :  \SYSTEMnt{A} $
there exists $\Gamma, \SYSTEMnt{t'}, \SYSTEMnt{A'}$
such that $ \Gamma  \vdash_{\textsc{l} }  \SYSTEMnt{t'}  :  \SYSTEMnt{A'} $ and $  \textcolor{LGcolor}{\llparenthesis}  \Gamma  \textcolor{LGcolor}{\rrparenthesis}   \vdash_{\textsc{g} }   \textcolor{LGcolor}{\llparenthesis} \smidge  \SYSTEMnt{t'}  \smidge \textcolor{LGcolor}{\rrparenthesis}   :   \textcolor{LGcolor}{\llparenthesis} \smidge  \SYSTEMnt{A'}  \smidge \textcolor{LGcolor}{\rrparenthesis}  $
and $\Delta \equiv  \textcolor{LGcolor}{\llparenthesis}  \Gamma  \textcolor{LGcolor}{\rrparenthesis} $
and $\SYSTEMnt{t} \equiv  \textcolor{LGcolor}{\llparenthesis} \smidge  \SYSTEMnt{t'}  \smidge \textcolor{LGcolor}{\rrparenthesis} $
and $\SYSTEMnt{A} \equiv  \textcolor{LGcolor}{\llparenthesis} \smidge  \SYSTEMnt{A'}  \smidge \textcolor{LGcolor}{\rrparenthesis} $.
However, an easy counterexample is the following judgment:
$$   \SYSTEMmv{x}  :_{\textcolor{coeffectColor}{ \SYSTEMsym{1} } }  \SYSTEMnt{B}    \vdash_{\textsc{g} }   \lambda  \SYSTEMmv{y}  .  \SYSTEMmv{x}   :   \SYSTEMnt{A}  \xrightarrow{\textcolor{coeffectColor}{ \SYSTEMsym{0} } }  \SYSTEMnt{B}  .$$
There is no type in Linear Base which translates to $ \SYSTEMnt{A}  \xrightarrow{\textcolor{coeffectColor}{ \SYSTEMsym{0} } }  \SYSTEMnt{B} $ as we always
translate the linear function type to $ \SYSTEMnt{A}  \xrightarrow{\textcolor{coeffectColor}{ \SYSTEMsym{1} } }  \SYSTEMnt{B} $. Indeed, the analogous
term in Linear Base was already shown in Example~\ref{exm:k-combinator}, where
we see that the result of the translation has type $   \textcolor{coeffectColor}{\square_{ \SYSTEMsym{0} } }   \textcolor{LGcolor}{\llparenthesis} \smidge  \SYSTEMnt{A}  \smidge \textcolor{LGcolor}{\rrparenthesis}     \xrightarrow{\textcolor{coeffectColor}{ \SYSTEMsym{1} } }   \textcolor{LGcolor}{\llparenthesis} \smidge  \SYSTEMnt{B}  \smidge \textcolor{LGcolor}{\rrparenthesis}  $.

Thus, the translation from Linear Base to Graded Modal Base is not surjective. Neither
is it injective: both the linear assumption $ \SYSTEMmv{x}  :  \SYSTEMnt{A} $ and the graded assumption
$ \SYSTEMmv{x}  : \textcolor{coeffectColor}{[}  \SYSTEMnt{A} {\textcolor{coeffectColor}{]_{ \SYSTEMsym{1} } } } $ are translated to $ \SYSTEMmv{x}  :_{\textcolor{coeffectColor}{ \SYSTEMsym{1} } }   \textcolor{LGcolor}{\llparenthesis} \smidge  \SYSTEMnt{A}  \smidge \textcolor{LGcolor}{\rrparenthesis}  $. For those semirings
in which $1$ does not capture linearity (e.g., the singleton semiring, any semiring
where $0 \leq 1$, and others), linear and non-linear
use is conflated by this translation.
\end{remark}

\subsubsection{A Linearity-Preserving Linear Base to Graded Modal Base Translation}
\label{subsec:linearity-preserving}

Instead, we can give an alternate translation which is injective
and \emph{linearity preserving}: keeping the translation of terms
as above, we can translate types and grades from Linear Base with semiring $\mathcal{R}$
into types and grades in Graded Modal Base with semiring $(\mathcal{R} \times \{\SYSTEMsym{0},
\SYSTEMsym{1},  \omega \})$, i.e., the product semiring (Example~\ref{exm:product})
of $\mathcal{R}$ with the linearity-capturing semiring (Example~\ref{exm:none-one-tons}).
\renewcommand{\triangleq}{& \oldTriangleq &}
\renewcommand{\equiv}{& \oldEquiv &}
\renewcommand{\SYSTEMdrule}[4][]{#3 \qquad #2}
\begin{gather*}
{\setlength{\arraycolsep}{0.1em}
\begin{array}{lcl@{\hspace{2em}}r}
\SYSTEMdruleTranslLtoGAltTyfunTy{} \\
\SYSTEMdruleTranslLtoGAltTyboxTy{} \\
\SYSTEMdruleTranslLtoGAltTybaseTy{} & \text{(Type translation)} \\[1em]
%\SYSTEMdruleTranslLtoGAltGtxtempty{} \\
\SYSTEMdruleTranslLtoGAltGtxtlin{} \\
\SYSTEMdruleTranslLtoGAltGtxtgrad{} & \text{(Context translation)}
\end{array}}
\end{gather*}
\let\SYSTEMdrule\oldSYSTEMdrule
\let\equiv\oldEquiv
\let\triangleq\oldTriangleq
%
%\newpage
Thus, the translation is now injective in the types since we translate a linear assumption $ \SYSTEMmv{x}  :  \SYSTEMnt{A} $ to
have grade $ \langle  \SYSTEMsym{1}  ,  \SYSTEMsym{1}  \rangle $ but a graded assumption $ \SYSTEMmv{x}  : \textcolor{coeffectColor}{[}  \SYSTEMnt{A} {\textcolor{coeffectColor}{]_{ \SYSTEMnt{r} } } } $ to $ \langle  \SYSTEMnt{r}  ,   \omega   \rangle $
which is a different grade even if $r = 1$. By taking the pairing with the linearity semiring,
we can capture the combination of linearity and grading that occurs in Linear Base, but
now in the Graded Modal Base setting.

  Operational correspondence and equation preservation hold as before in Theorem~\ref{thrm:linToGradTranslation} since the term translation is unchanged. We then re-restablish
  type preservation of the translation:
\begin{restatable}{thrm}{linToGradAltTranslation}[Soundness of the Linearity-Preserving Linear Base to Graded Modal Base translation]
  \begin{itemize}[itemsep=0.2em]
    \item Type preservation: $ \Gamma  \vdash_{\textsc{l} }  \SYSTEMnt{t}  :  \SYSTEMnt{A}  \implies   \textcolor{LGcolor}{\llparenthesis}  \Gamma  \textcolor{LGcolor}{\rrparenthesis}'   \vdash_{\textsc{g}_\square }   \textcolor{LGcolor}{\llparenthesis} \smidge  \SYSTEMnt{t}  \smidge \textcolor{LGcolor}{\rrparenthesis}   :   \textcolor{LGcolor}{\llparenthesis} \smidge  \SYSTEMnt{A}  \smidge \textcolor{LGcolor}{\rrparenthesis}'  $
  \end{itemize}
Appendix~\refappendix{app:proofs-lin-to-grad-mod-alt} provides the proof.
\end{restatable}

The key insight in this translation is that Linear Base essentially comprises two parallel
forms of analysis: a graded analysis and a latent linear analysis witnessed by the
corresponding two forms of typing assumption in Linear Base contexts.
Our translation here captures these two aspects systematically,
by taking the product of the same grading semiring and the linearity semiring.

A question may arise here:
\textit{What is the meaning of a grade such as $ \langle  \SYSTEMsym{2}  ,  \SYSTEMsym{1}  \rangle $,
where the first component looks non-linear in $\mathcal{R}$
but the second represents linearity?} Such a grade is not
in the image of our translation but we may still consider what it is
like to type terms in Graded (Modal) Base with $\mathcal{R} \times \{0, 1,
\omega\}$. First, let us decompose $ \langle  \SYSTEMsym{2}  ,  \SYSTEMsym{1}  \rangle $
into $ \langle  \SYSTEMsym{1}  \SYSTEMsym{+}  \SYSTEMsym{1}  ,  \SYSTEMsym{1}  \rangle $
for clarity, i.e., for any semiring $\mathcal{R}$ in the first component.
Such grades could be formed via approximation if
$\mathcal{R}$ permits $ \SYSTEMsym{1}  \, \textcolor{coeffectColor}{\sqsubseteq} \,  \SYSTEMsym{1}  \SYSTEMsym{+}  \SYSTEMsym{1} $ and thus $  \langle  \SYSTEMsym{1}  ,  \SYSTEMsym{1}  \rangle   \, \textcolor{coeffectColor}{\sqsubseteq} \,   \langle  \SYSTEMsym{1}  \SYSTEMsym{+}  \SYSTEMsym{1}  ,  \SYSTEMsym{1}  \rangle  $
is an over-approximation of grades, e.g., useful for when calculating the grade
across multiple program branches. Without such an
approximation, $ \langle  \SYSTEMsym{1}  \SYSTEMsym{+}  \SYSTEMsym{1}  ,  \SYSTEMsym{1}  \rangle $ cannot be derived from variable use
and contraction of variables alone since in $(\mathcal{R} \times \{\SYSTEMsym{0},
\SYSTEMsym{1},  \omega \})$ we have
$ \langle  \SYSTEMsym{1}  ,  \SYSTEMsym{1}  \rangle   \SYSTEMsym{+}   \langle  \SYSTEMsym{1}  ,  \SYSTEMsym{1}  \rangle  =  \langle  \SYSTEMsym{1}  \SYSTEMsym{+}  \SYSTEMsym{1}  ,   \omega   \rangle $.
Such a grade could only arise from an open term applying
a function, e.g.,
$$    \SYSTEMmv{f}  :_{\textcolor{coeffectColor}{ \SYSTEMsym{1} } }    \SYSTEMnt{A}  \xrightarrow{\textcolor{coeffectColor}{  \langle  \SYSTEMsym{1}  \SYSTEMsym{+}  \SYSTEMsym{1}  ,  \SYSTEMsym{1}  \rangle  } }  \SYSTEMnt{B}     ,   \SYSTEMmv{x}  :_{\textcolor{coeffectColor}{  \langle  \SYSTEMsym{1}  \SYSTEMsym{+}  \SYSTEMsym{1}  ,  \SYSTEMsym{1}  \rangle  } }  \SYSTEMnt{A}    \vdash_{\textsc{g} }   \SYSTEMmv{f} \,  \SYSTEMmv{x}   :  \SYSTEMnt{B} .$$
However there is no well-typed closed term that can be formed to substitute into
$f$ since we cannot form $ \langle  \SYSTEMsym{1}  \SYSTEMsym{+}  \SYSTEMsym{1}  ,  \SYSTEMsym{1}  \rangle $ without use of such an open term.
Consequently, for closed terms, we cannot form $ \langle  \SYSTEMsym{1}  \SYSTEMsym{+}  \SYSTEMsym{1}  ,  \SYSTEMsym{1}  \rangle $
unless it is permitted by approximation in $\mathcal{R}$.
In the approximation case, it is up to the choice (or design) of semiring
to permit approximation via the definition of $     \, \textcolor{coeffectColor}{\sqsubseteq} \,     $.

More generally, we may consider the advantage of pairing
semirings with the linearity semiring. Consider a semiring such as the security
lattice semiring (Example~\ref{exm:security-lattice}) where $1 =  \mathsf{Lo} $.
Since $ \mathsf{Lo}  +  \mathsf{Lo}  =  \mathsf{Lo} $ and $ \SYSTEMsym{0}  \, \textcolor{coeffectColor}{\sqsubseteq} \,   \mathsf{Lo}  $ then a $ \mathsf{Lo} $-graded variable
can represent both linear or non-linear usage. The pairing
$ \langle   \mathsf{Lo}   ,  \SYSTEMsym{1}  \rangle $ then distinguishes a linear low-security variable
from $ \langle   \mathsf{Lo}   ,   \omega   \rangle $ which instead represents a non-linear low-security variable.
Conversely, recall that $0 =  \mathsf{Hi} $,
giving the translation $  \textcolor{LGcolor}{\llparenthesis} \smidge   \textcolor{coeffectColor}{\square_{  \mathsf{Hi}  } }  \SYSTEMnt{A}   \smidge \textcolor{LGcolor}{\rrparenthesis}'   \equiv   \textcolor{coeffectColor}{\square_{  \langle   \mathsf{Hi}   ,   \omega   \rangle  } }  \SYSTEMnt{A}  $,
which allows unlimited usage, but limited to $ \mathsf{Hi} $ security contexts.

Our approach here is not necessarily the only sound and injective translation;
however, it seems particularly illuminating,
in that the product construction clearly demarcates and represents the two analyses present within Linear Base:
the explicit coeffect in the grades on the one hand and the latent linearity coeffect of the
system (not present in Graded Base) on the other.

% is it really a necessary construction? No there might be other choices
% Lin-completion is not satisfactory as it can give Lin <= 1 + 1
%

% Reviewer:
% The goal of translation seems to be to /explain/ one world in terms of another, but the graded world doesn't use semi-rings like this ... so is it an explanation or a proof tool that "just happens" to work?

% Our insight: that the Linear Base is really like having a pair of two semirings
% one for R and one for linearity, and this conceptually shows that there are two
% separate kinds of tracking going on , which we then naturally present as a product
% of two semirings in this translation to Graded Base

% Lin <= 1

% <1, 1>

%Does it matter that it seems to make no sense to "translate back" in the other direction?

\subsubsection{Graded Modal Base to Linear Base Translation}
\label{subsec:transl-gmb2lb}

\renewcommand{\triangleq}{& \oldTriangleq &}
\renewcommand{\equiv}{& \oldEquiv &}
\renewcommand{\SYSTEMdrule}[4][]{#3 \qquad #2}

We can now extend the previous translation from Graded Base to Linear Base
(Section~\ref{sec:transl}) to translate from Graded Modal Base to Linear Base:
\begin{gather*}
{\setlength{\arraycolsep}{0.1em}
\begin{array}{lcl@{\hspace{2em}}r}
\SYSTEMdruleTranslGtoLTmvar{} \\
\SYSTEMdruleTranslGtoLTmabs{}
\SYSTEMdruleTranslGtoLTmapp{} \\
\SYSTEMdruleTranslGtoLTmpr{} \\
\SYSTEMdruleTranslGtoLTmlet{} & \text{(Term translation)} \\[0.75em]
\SYSTEMdruleTranslGtoLTyfunTy{} \\
\SYSTEMdruleTranslGtoLTyboxTy{} \\
\SYSTEMdruleTranslGtoLTybaseTy{} & \text{(Type translation)} \\[0.75em]
\SYSTEMdruleTranslGtoLGtxtempty{} \\
\SYSTEMdruleTranslGtoLGtxtgrad{} & \text{(Context translation)}
\end{array}}
\end{gather*}
\let\SYSTEMdrule\oldSYSTEMdrule
\let\equiv\oldEquiv
\let\triangleq\oldTriangleq
The only new parts are the translation of the graded modality which maps directly
onto the graded modality in Linear Base. We then (re)establish soundness. Appendix~\refappendix{app:proofs-grad-mod-to-lin} provides the proofs.
\begin{restatable}{thrm}{gradModToLinTranslation}[Soundness of the Graded Modal Base to Linear Base translation]
\begin{itemize}[itemsep=0.2em]
  \item Type preservation: $ \Delta  \vdash_{\textsc{g}_\square }  \SYSTEMnt{t}  :  \SYSTEMnt{A}  \implies   \textcolor{GLcolor}{\llbracket}  \Delta  \textcolor{GLcolor}{\rrbracket}   \vdash_{\textsc{l} }   \textcolor{GLcolor}{\llbracket}  \SYSTEMnt{t}  \textcolor{GLcolor}{\rrbracket}   :   \textcolor{GLcolor}{\llbracket}  \SYSTEMnt{A}  \textcolor{GLcolor}{\rrbracket}  $
  \item Operational correspondence: $ \SYSTEMnt{t}  \rightsquigarrow_{\textsc{g}\Box}  \SYSTEMnt{t'}  \implies   \textcolor{GLcolor}{\llbracket}  \SYSTEMnt{t}  \textcolor{GLcolor}{\rrbracket}   \rightsquigarrow_{\textsc{l} }^\ast   \textcolor{GLcolor}{\llbracket}  \SYSTEMnt{t'}  \textcolor{GLcolor}{\rrbracket}  $
  \item Equation preservation: $ \SYSTEMnt{t}  \,\equiv_{\textsc{g}_\Box}\,  \SYSTEMnt{t'}  \implies   \textcolor{GLcolor}{\llbracket}  \SYSTEMnt{t}  \textcolor{GLcolor}{\rrbracket}   \,\equiv_{\textsc{l} }\,   \textcolor{GLcolor}{\llbracket}  \SYSTEMnt{t'}  \textcolor{GLcolor}{\rrbracket}  $.
 \end{itemize}
  \label{ref:gradModToLinTranslation}
\end{restatable}

\subsubsection{Mutual relationship}
\label{subsec:mutual-relationship}

Given the pair of translations here, we might wonder: what is the relationship between
the translations? Are they mutually inverse? By considering just the translation of function
types we can quickly see that the translations are not inverse:
\begin{align*}
 \textcolor{GLcolor}{\llbracket}   \textcolor{LGcolor}{\llparenthesis} \smidge   \SYSTEMnt{A}  \multimap  \SYSTEMnt{B}   \smidge \textcolor{LGcolor}{\rrparenthesis}   \textcolor{GLcolor}{\rrbracket}  & =   \textcolor{GLcolor}{\llbracket}   \SYSTEMnt{A}  \xrightarrow{\textcolor{coeffectColor}{ \SYSTEMsym{1} } }  \SYSTEMnt{B}   \textcolor{GLcolor}{\rrbracket}   \equiv    \textcolor{coeffectColor}{\square_{ \SYSTEMsym{1} } }    \textcolor{GLcolor}{\llbracket}   \textcolor{LGcolor}{\llparenthesis} \smidge  \SYSTEMnt{A}  \smidge \textcolor{LGcolor}{\rrparenthesis}   \textcolor{GLcolor}{\rrbracket}     \multimap    \textcolor{GLcolor}{\llbracket}   \textcolor{LGcolor}{\llparenthesis} \smidge  \SYSTEMnt{B}  \smidge \textcolor{LGcolor}{\rrparenthesis}   \textcolor{GLcolor}{\rrbracket}    
\\
 \textcolor{LGcolor}{\llparenthesis} \smidge   \textcolor{GLcolor}{\llbracket}   \SYSTEMnt{A}  \xrightarrow{\textcolor{coeffectColor}{ \SYSTEMnt{r} } }  \SYSTEMnt{B}   \textcolor{GLcolor}{\rrbracket}   \smidge \textcolor{LGcolor}{\rrparenthesis}  & =   \textcolor{LGcolor}{\llparenthesis} \smidge    \textcolor{coeffectColor}{\square_{ \SYSTEMnt{r} } }  \SYSTEMnt{A}   \multimap  \SYSTEMnt{B}   \smidge \textcolor{LGcolor}{\rrparenthesis}   \equiv    \textcolor{coeffectColor}{\square_{ \SYSTEMnt{r} } }   \textcolor{LGcolor}{\llparenthesis} \smidge   \textcolor{GLcolor}{\llbracket}  \SYSTEMnt{A}  \textcolor{GLcolor}{\rrbracket}   \smidge \textcolor{LGcolor}{\rrparenthesis}    \xrightarrow{\textcolor{coeffectColor}{ \SYSTEMsym{1} } }   \textcolor{LGcolor}{\llparenthesis} \smidge   \textcolor{GLcolor}{\llbracket}  \SYSTEMnt{B}  \textcolor{GLcolor}{\rrbracket}   \smidge \textcolor{LGcolor}{\rrparenthesis}   
\end{align*}
Similarly, given $   \SYSTEMmv{x}  :  \SYSTEMnt{A}    \vdash_{\textsc{l} }  \SYSTEMmv{x}  :  \SYSTEMnt{A} $ in Linear Base
then its translation to Graded Modal Base is $   \SYSTEMmv{x}  :_{\textcolor{coeffectColor}{ \SYSTEMsym{1} } }  \SYSTEMnt{A}    \vdash_{\textsc{g}_\square }  \SYSTEMmv{x}  :  \SYSTEMnt{A} $
which is translated back to Linear Base as $   \SYSTEMmv{x}  : \textcolor{coeffectColor}{[}  \SYSTEMnt{A} {\textcolor{coeffectColor}{]_{ \SYSTEMsym{1} } } }    \vdash_{\textsc{l} }  \SYSTEMmv{x}  :  \SYSTEMnt{A} $.

However, by composing operational correspondence theorems, then round-tripping the translations
yields a term that corresponds operationally to the original, in both directions:
\begin{corollary}[Observational isomorphism]
If $ \SYSTEMnt{t}  \rightsquigarrow_{\textsc{l} }  \SYSTEMnt{t'} $ then $  \textcolor{GLcolor}{\llbracket}   \textcolor{LGcolor}{\llparenthesis} \smidge  \SYSTEMnt{t}  \smidge \textcolor{LGcolor}{\rrparenthesis}   \textcolor{GLcolor}{\rrbracket}   \rightsquigarrow_{\textsc{l} }^\ast   \textcolor{GLcolor}{\llbracket}   \textcolor{LGcolor}{\llparenthesis} \smidge  \SYSTEMnt{t'}  \smidge \textcolor{LGcolor}{\rrparenthesis}   \textcolor{GLcolor}{\rrbracket}  $
and if $ \SYSTEMnt{t}  \rightsquigarrow_{\textsc{g}\Box}  \SYSTEMnt{t'} $ then $  \textcolor{LGcolor}{\llparenthesis} \smidge   \textcolor{GLcolor}{\llbracket}  \SYSTEMnt{t}  \textcolor{GLcolor}{\rrbracket}   \smidge \textcolor{LGcolor}{\rrparenthesis}   \rightsquigarrow_{\textsc{g}\Box }^\ast   \textcolor{LGcolor}{\llparenthesis} \smidge   \textcolor{GLcolor}{\llbracket}  \SYSTEMnt{t'}  \textcolor{GLcolor}{\rrbracket}   \smidge \textcolor{LGcolor}{\rrparenthesis}  $.
\end{corollary}
We might also wonder if there is an adjoint relationship between the translations. If $ \textcolor{GLcolor}{\llbracket}   -   \textcolor{GLcolor}{\rrbracket}  \dashv  \textcolor{LGcolor}{\llparenthesis} \smidge   -   \smidge \textcolor{LGcolor}{\rrparenthesis} $ then we would expect to be able to construct a mapping in the category of types (whose morphisms are derivations of judgments with source as inputs and the target as the result):
\[
\eta_A : A \rightarrow  \textcolor{GLcolor}{\llbracket}   \textcolor{LGcolor}{\llparenthesis} \smidge  \SYSTEMnt{A}  \smidge \textcolor{LGcolor}{\rrparenthesis}   \textcolor{GLcolor}{\rrbracket} 
\]
However, there is no such mapping for $ \SYSTEMnt{A}  \multimap  \SYSTEMnt{B} $ to $  \textcolor{GLcolor}{\llbracket}   \textcolor{LGcolor}{\llparenthesis} \smidge   \SYSTEMnt{A}  \multimap  \SYSTEMnt{B}   \smidge \textcolor{LGcolor}{\rrparenthesis}   \textcolor{GLcolor}{\rrbracket}   \equiv    \textcolor{coeffectColor}{\square_{ \SYSTEMsym{1} } }   \textcolor{GLcolor}{\llbracket}   \textcolor{LGcolor}{\llparenthesis} \smidge  \SYSTEMnt{A}  \smidge \textcolor{LGcolor}{\rrparenthesis}   \textcolor{GLcolor}{\rrbracket}    \multimap   \textcolor{GLcolor}{\llbracket}   \textcolor{LGcolor}{\llparenthesis} \smidge  \SYSTEMnt{B}  \smidge \textcolor{LGcolor}{\rrparenthesis}   \textcolor{GLcolor}{\rrbracket}   $ in general since this is undefined for a higher-order type.

Consider $  (    \mathrm{K}   \multimap   \mathrm{K}    )   \multimap   \mathrm{K}  $ for the base type $ \mathrm{K} $ whose translation between
the systems is the identity, and thus $ \textcolor{GLcolor}{\llbracket}   \textcolor{LGcolor}{\llparenthesis} \smidge   \mathrm{K}   \smidge \textcolor{LGcolor}{\rrparenthesis}   \textcolor{GLcolor}{\rrbracket}  =  \mathrm{K} $. Then $\eta$ would need to provide
a derivation of the judgment:
\begin{align*}
  (    \mathrm{K}   \multimap   \mathrm{K}    )   \multimap   \mathrm{K}   \vdash   \textcolor{coeffectColor}{\square_{ \SYSTEMsym{1} } }   (    \textcolor{coeffectColor}{\square_{ \SYSTEMsym{1} } }   \mathrm{K}    \multimap   \mathrm{K}    )    \multimap   \mathrm{K}  
\end{align*}
Such a derivation would require us to map a value of type $ \textcolor{coeffectColor}{\square_{ \SYSTEMsym{1} } }   (    \textcolor{coeffectColor}{\square_{ \SYSTEMsym{1} } }   \mathrm{K}    \multimap   \mathrm{K}    )  $
to one of type $  \mathrm{K}   \multimap   \mathrm{K}  $. We can certainly derive $ \textcolor{coeffectColor}{\square_{ \SYSTEMsym{1} } }   (    \textcolor{coeffectColor}{\square_{ \SYSTEMsym{1} } }   \mathrm{K}    \multimap   \mathrm{K}    )   \vdash   \textcolor{coeffectColor}{\square_{ \SYSTEMsym{1} } }   \mathrm{K}    \multimap   \mathrm{K}  $ by elimination
and dereliction, but $  \textcolor{coeffectColor}{\square_{ \SYSTEMsym{1} } }   \mathrm{K}    \multimap   \mathrm{K}   \not\vdash   \mathrm{K}   \multimap   \mathrm{K}  $. To see why, consider
the single-point semiring $0 = 1$ and thus a function typed $  \textcolor{coeffectColor}{\square_{ \SYSTEMsym{1} } }   \mathrm{K}    \multimap   \mathrm{K}  $ may not use its argument
at all, whereas $  \mathrm{K}   \multimap   \mathrm{K}  $ must use its argument exactly once.

Given the above negative result, we could then consider the opposite adjoint relationship, i.e., perhaps
 $ \textcolor{LGcolor}{\llparenthesis} \smidge   -   \smidge \textcolor{LGcolor}{\rrparenthesis}  \dashv  \textcolor{GLcolor}{\llbracket}   -   \textcolor{GLcolor}{\rrbracket} $? In which case, we would require the counit
 operation:
\[
\epsilon_A :  \textcolor{GLcolor}{\llbracket}   \textcolor{LGcolor}{\llparenthesis} \smidge  \SYSTEMnt{A}  \smidge \textcolor{LGcolor}{\rrparenthesis}   \textcolor{GLcolor}{\rrbracket}  \to A
\]
However, there is no such mapping, even in a first-order case: for
$  \textcolor{GLcolor}{\llbracket}   \textcolor{LGcolor}{\llparenthesis} \smidge    \mathrm{K}   \multimap   \mathrm{K}    \smidge \textcolor{LGcolor}{\rrparenthesis}   \textcolor{GLcolor}{\rrbracket}   \equiv    \textcolor{coeffectColor}{\square_{ \SYSTEMsym{1} } }   \textcolor{GLcolor}{\llbracket}   \textcolor{LGcolor}{\llparenthesis} \smidge   \mathrm{K}   \smidge \textcolor{LGcolor}{\rrparenthesis}   \textcolor{GLcolor}{\rrbracket}    \multimap   \textcolor{GLcolor}{\llbracket}   \textcolor{LGcolor}{\llparenthesis} \smidge   \mathrm{K}   \smidge \textcolor{LGcolor}{\rrparenthesis}   \textcolor{GLcolor}{\rrbracket}   $ we need to
map to $  \mathrm{K}   \multimap   \mathrm{K}  $ but as described above, no such derivation is possible.

Thus the translations are neither inverses nor adjoint. We leave it as future work to
consider whether more complex invertible translations are possible.

\subsection{Generalised $\Box$-elimination}
\label{sec:flattening}
Elimination of graded modalities in both linear- and graded-base systems looks almost identical,
eliminating a graded modality at grade $r$ by a kind of
`cut' (substitution) against a variable with matching grade $r$:
\begin{gather*}
\SYSTEMdruleLinlet{} \quad \SYSTEMdruleGradBoxlet{}
\end{gather*}
Some graded-base systems in the literature however
have a more general elimination rule~\cite{DBLP:journals/pacmpl/AbelB20,grtt},
of the form:
\begin{align*}
\SYSTEMdruleGradBoxletGen{}
\end{align*}
The generalised rule allows an additional scaling of the resources
for $\SYSTEMnt{t_{{\mathrm{1}}}}$ due to additional use of $x$ in $\SYSTEMnt{t_{{\mathrm{2}}}}$. Such
a rule has not appeared in any Linear Base system, but could
easily be added to both systems to provide equivalent power
across both flavours of coeffect types.

This generalised typing enables the
derivation of a function $  \textcolor{coeffectColor}{\square_{ \SYSTEMnt{r} } }   (   \textcolor{coeffectColor}{\square_{ \SYSTEMnt{s} } }  \SYSTEMnt{A}   )    \multimap   \textcolor{coeffectColor}{\square_{ \SYSTEMsym{(}   \SYSTEMnt{r}  \cdot  \SYSTEMnt{s}   \SYSTEMsym{)} } }  \SYSTEMnt{A}  $,
akin to the multiplication (join) of a graded monad (the dual to comultiplication
of a graded comonad structure):
\begin{gather*}
  \vspace{-1em}
  \inferrule*[right=\SYSTEMRenameRuleGradBoxletGen{}]
             {\inferrule*[right=\SYSTEMRenameRuleGradvar{}]
                {\;}{   \SYSTEMmv{x}  :_{\textcolor{coeffectColor}{ \SYSTEMsym{1} } }    \textcolor{coeffectColor}{\square_{ \SYSTEMnt{r} } }   (   \textcolor{coeffectColor}{\square_{ \SYSTEMnt{s} } }  \SYSTEMnt{A}   )       \vdash_{\textsc{g}_\square }  \SYSTEMmv{x}  :   \textcolor{coeffectColor}{\square_{ \SYSTEMnt{r} } }   (   \textcolor{coeffectColor}{\square_{ \SYSTEMnt{s} } }  \SYSTEMnt{A}   )   }
               \inferrule*[right=\SYSTEMRenameRuleGradBoxletGen{}]
                          {\inferrule*[right=\SYSTEMRenameRuleGradvar{}]{\;}
                            {   \SYSTEMmv{y}  :_{\textcolor{coeffectColor}{ \SYSTEMsym{1} } }    \textcolor{coeffectColor}{\square_{ \SYSTEMnt{s} } }  \SYSTEMnt{A}      \vdash_{\textsc{g}_\square }  \SYSTEMmv{y}  :   \textcolor{coeffectColor}{\square_{ \SYSTEMnt{s} } }  \SYSTEMnt{A}  }
                            \inferrule*[right=\SYSTEMRenameRuleGradBoxpr{}]
                                       {\inferrule*[right=\SYSTEMRenameRuleGradvar{}]{\;}
                                         {   \SYSTEMmv{z}  :_{\textcolor{coeffectColor}{ \SYSTEMsym{1} } }  \SYSTEMnt{A}    \vdash_{\textsc{g}_\square }  \SYSTEMmv{z}  :  \SYSTEMnt{A} }}
                                       {   \SYSTEMmv{z}  :_{\textcolor{coeffectColor}{   \SYSTEMnt{r}  \cdot  \SYSTEMnt{s}   } }  \SYSTEMnt{A}    \vdash_{\textsc{g}_\square }   \textcolor{coeffectColor}{[}  \SYSTEMmv{z}  \textcolor{coeffectColor}{]}   :   \textcolor{coeffectColor}{\square_{   \SYSTEMnt{r}  \cdot  \SYSTEMnt{s}   } }  \SYSTEMnt{A}  }
                          \hspace{-2.4em}}
                          {   \SYSTEMmv{y}  :_{\textcolor{coeffectColor}{ \SYSTEMnt{r} } }    \textcolor{coeffectColor}{\square_{ \SYSTEMnt{s} } }  \SYSTEMnt{A}      \vdash_{\textsc{g}_\square }   \mathsf{let} \, \textcolor{coeffectColor}{[}  \SYSTEMmv{z}  \textcolor{coeffectColor}{]} =  \SYSTEMmv{y}  \, \mathsf{in} \,   \textcolor{coeffectColor}{[}  \SYSTEMmv{z}  \textcolor{coeffectColor}{]}    :   \textcolor{coeffectColor}{\square_{   \SYSTEMnt{r}  \cdot  \SYSTEMnt{s}   } }  \SYSTEMnt{A}  \hspace{-3em}}
             \hspace{-3em}}
             {    \SYSTEMmv{x}  :_{\textcolor{coeffectColor}{ \SYSTEMsym{1} } }    \textcolor{coeffectColor}{\square_{ \SYSTEMnt{r} } }   (   \textcolor{coeffectColor}{\square_{ \SYSTEMnt{s} } }  \SYSTEMnt{A}   )       \vdash_{\textsc{g}_\square }   \mathsf{let} \, \textcolor{coeffectColor}{[}  \SYSTEMmv{y}  \textcolor{coeffectColor}{]} =  \SYSTEMmv{x}  \, \mathsf{in} \,   \mathsf{let} \, \textcolor{coeffectColor}{[}  \SYSTEMmv{z}  \textcolor{coeffectColor}{]} =  \SYSTEMmv{y}  \, \mathsf{in} \,   \textcolor{coeffectColor}{[}  \SYSTEMmv{z}  \textcolor{coeffectColor}{]}     :   \textcolor{coeffectColor}{\square_{   \SYSTEMnt{r}  \cdot  \SYSTEMnt{s}   } }  \SYSTEMnt{A}   }
\end{gather*}
Indeed, the generalised $\Box$-elimination can be modelled in a
categorical semantics by a graded comonad along with an additional
graded multiplication operation $\mu_{r,s,A} :   \textcolor{coeffectColor}{\square_{ \SYSTEMnt{r} } }   (   \textcolor{coeffectColor}{\square_{ \SYSTEMnt{s} } }  \SYSTEMnt{A}   )    \multimap   \textcolor{coeffectColor}{\square_{ \SYSTEMsym{(}   \SYSTEMnt{r}  \cdot  \SYSTEMnt{s}   \SYSTEMsym{)} } }  \SYSTEMnt{A}  $ (as one might find in a graded monad structure).
However, not all graded comonads modelling $ \textcolor{coeffectColor}{\square_{ \SYSTEMnt{r} } }  \SYSTEMnt{A} $
admit such an operation and so having the generalised
$\Box$-elimination may not be desirable in general graded comonadic calculi.
We thus omitted this generalised rule from the main part of our results, considering
the simpler $\SYSTEMRenameRuleLinlet{}$ and $\SYSTEMRenameRuleGradBoxlet{}$ rules.

\section{Deriving a Graded Modality with a Polymorphic Graded Base}
\label{sec:poly}
\citet{DBLP:journals/pacmpl/BernardyBNJS18} develop an extension to GHC/Haskell that
provides a notion of linear typing by using a graded type system.  The
core calculus is akin to that of Graded Base, but specialised to
the $\{0, 1, \omega\}$ semiring (Example~\ref{exm:none-one-tons}). Their calculus has no graded
modality, but the implementation as part of GHC/Haskell can
encode
the \textit{unrestricted} modality
\texttt{Ur} via an algebraic datatype:\footnote{\url{https://hackage.haskell.org/package/linear-base-0.4.0/docs/Data-Unrestricted-Linear.html\#t:Ur}}
\begin{haskell}
data Ur a where
  Ur :: a -> Ur a
\end{haskell}
where the arrow \haskin{->} denotes an $\omega$-graded arrow for the two-point semiring
$\{1, \omega\}$ ($0$ is not available in the implementation)
and thus this constructor is essentially of type $A
\xrightarrow{\omega} \mathsf{Ur}\ A$ in the Graded Base syntax. \citet{DBLP:journals/corr/abs-2112-14966} give a generalisation of this definition to a graded modality:
\begin{haskell}
data Box r a where
   Box :: a %r -> Box r a
\end{haskell}
This suggests an alternate Graded Base calculus in relation to Linear Base.
We previously showed (Section~\ref{subsec:lin-to-grad-cps}) that a type-preserving translation of Linear Base terms into
Graded Base requires a CPS encoding. By changing the substrate of Graded Base to System F, we have
\textbf{Graded~Poly~Base}, unlocking the ability for a direct-style
polymorphic Church encoding of this \texttt{Box} data type:
\begin{align}
\label{eq:derived-box}
   \textcolor{coeffectColor}{\square_{ \SYSTEMnt{r} } }  \SYSTEMnt{A}  =  \forall   \beta   .     (   \SYSTEMnt{A}  \xrightarrow{\textcolor{coeffectColor}{ \SYSTEMnt{r} } }   \beta    )   \xrightarrow{\textcolor{coeffectColor}{ \SYSTEMsym{1} } }   \beta    
\end{align}
This enables a less intrusive translation, without the need to CPS translate all terms.

\subsection{Graded Poly Base}\label{sec:graded-poly-base}
We thus extend Graded Base with polymorphic typing, the syntax and
typing of which is standard:
\begin{align*}
t & ::= \ldots \mid  \SYSTEMnt{t}  \text{@}  \SYSTEMnt{A}  \mid  \Lambda  \alpha  .  \SYSTEMnt{t} 
\tag{Terms} \\
\SYSTEMnt{A} & ::= \ldots \mid \alpha \mid  \forall  \alpha  .  \SYSTEMnt{B}  \tag{Types}
\end{align*}
where $ \alpha ,  \beta $ range over type variables, with
a corresponding standard substitution at the type-level
written $ [  \SYSTEMnt{A}  /  \alpha  ]  \SYSTEMnt{B} $.
Typing judgments are of the form $ \Delta  \vdash_{\textsc{g}_\forall }  \SYSTEMnt{t}  :  \SYSTEMnt{A} $ extending
the previous typing with:
\begin{align*}
\SYSTEMdruleGradPolytyAbs{} \quad
\SYSTEMdruleGradPolytyApp{}
\end{align*}
The operational semantics has the standard $\beta$ rule for type abstraction/application
and congruence:
\begin{align*}
\SYSTEMdruleSemGrdPolytyBeta{} \quad
\SYSTEMdruleSemGrdPolycongTyApp{}
\end{align*}
Introduction and elimination for the derived $\Box_r$ (from~\eqref{eq:derived-box})
can therefore be derived respectively as follows (with their typing derivations shown):
%
%\begin{align*}
%  \dfrac{ \Delta_{{\mathrm{1}}}  \vdash_{\textsc{g} }  \SYSTEMnt{t}  :  \SYSTEMnt{A} }
%        {  \textcolor{coeffectColor}{ \SYSTEMnt{r}  \cdot}  \Delta_{{\mathrm{1}}}   \vdash_{\textsc{g} }    \Lambda   \beta   .   \lambda  \SYSTEMmv{f}  .  \SYSTEMmv{f}   \,  \SYSTEMnt{t}   :    \forall   \beta   .   (   \SYSTEMnt{A}  \xrightarrow{\textcolor{coeffectColor}{ \SYSTEMnt{r} } }   \beta    )    \xrightarrow{\textcolor{coeffectColor}{ \SYSTEMsym{1} } }   \beta   }
%        \Box_i
%  \qquad
%  \dfrac{ \Delta_{{\mathrm{1}}}  \vdash_{\textsc{g} }  \SYSTEMnt{t_{{\mathrm{1}}}}  :    \forall   \beta   .   (   \SYSTEMnt{A}  \xrightarrow{\textcolor{coeffectColor}{ \SYSTEMnt{r} } }   \beta    )    \xrightarrow{\textcolor{coeffectColor}{ \SYSTEMsym{1} } }   \beta     \quad
%           \Delta_{{\mathrm{2}}} ,   \SYSTEMmv{x}  :_{\textcolor{coeffectColor}{ \SYSTEMnt{r} } }  \SYSTEMnt{A}    \vdash_{\textsc{g} }  \SYSTEMnt{t_{{\mathrm{2}}}}  :  \SYSTEMnt{B} }
%        { \Delta_{{\mathrm{1}}}  \SYSTEMsym{+}  \Delta_{{\mathrm{2}}}  \vdash_{\textsc{g} }   \SYSTEMsym{(}   \SYSTEMnt{t_{{\mathrm{1}}}}  \text{@}  \SYSTEMnt{B}   \SYSTEMsym{)} \,  \SYSTEMsym{(}   \lambda  \SYSTEMmv{x}  .  \SYSTEMnt{t_{{\mathrm{2}}}}   \SYSTEMsym{)}   :  \SYSTEMnt{B} }
%        \Box_e
%\end{align*}

\vspace{-1.5em}
\begin{align}
\inferrule*[Right=\SYSTEMRenameRuleGradPolytyAbs{}]
{
\inferrule*[Right=\SYSTEMRenameRuleGradabs{}]
{
\inferrule*[Right=\SYSTEMRenameRuleGradapp{}]
{  \Delta  \vdash_{\textsc{g}_\forall }  \SYSTEMnt{t}  :  \SYSTEMnt{A}  \quad
  \inferrule*[Right=\SYSTEMRenameRuleGradvar{}]
    {\quad}{   \SYSTEMmv{f}  :_{\textcolor{coeffectColor}{ \SYSTEMsym{1} } }    \SYSTEMnt{A}  \xrightarrow{\textcolor{coeffectColor}{ \SYSTEMnt{r} } }   \beta'       \vdash_{\textsc{g}_\forall }  \SYSTEMmv{f}  :    \SYSTEMnt{A}  \xrightarrow{\textcolor{coeffectColor}{ \SYSTEMnt{r} } }   \beta'     }}
{    \textcolor{coeffectColor}{ \SYSTEMnt{r}  \cdot}  \Delta  ,   \SYSTEMmv{f}  :_{\textcolor{coeffectColor}{ \SYSTEMsym{1} } }    \SYSTEMnt{A}  \xrightarrow{\textcolor{coeffectColor}{ \SYSTEMnt{r} } }   \beta'       \vdash_{\textsc{g}_\forall }   \SYSTEMmv{f} \,  \SYSTEMnt{t}   :   \beta'   }
}
{  \textcolor{coeffectColor}{ \SYSTEMnt{r}  \cdot}  \Delta   \vdash_{\textsc{g}_\forall }    \lambda  \SYSTEMmv{f}  .  \SYSTEMmv{f}  \,  \SYSTEMnt{t}   :    (   \SYSTEMnt{A}  \xrightarrow{\textcolor{coeffectColor}{ \SYSTEMnt{r} } }   \beta'    )   \xrightarrow{\textcolor{coeffectColor}{ \SYSTEMsym{1} } }   \beta'    }
}
{
  \textcolor{coeffectColor}{ \SYSTEMnt{r}  \cdot}  \Delta   \vdash_{\textsc{g}_\forall }    \Lambda   \beta   .   \lambda  \SYSTEMmv{f}  .  \SYSTEMmv{f}   \,  \SYSTEMnt{t}   :    \forall   \beta   .   (   \SYSTEMnt{A}  \xrightarrow{\textcolor{coeffectColor}{ \SYSTEMnt{r} } }   \beta    )    \xrightarrow{\textcolor{coeffectColor}{ \SYSTEMsym{1} } }   \beta   
}\qquad
\tag{$\Box_i$}
\end{align}
\begin{align}
\inferrule*[Right=\SYSTEMRenameRuleGradapp{}]
{
  \inferrule*[Right=\SYSTEMRenameRuleGradPolytyApp{}]
     { \Delta_{{\mathrm{1}}}  \vdash_{\textsc{g}_\forall }  \SYSTEMnt{t_{{\mathrm{1}}}}  :    \forall   \beta   .   (   \SYSTEMnt{A}  \xrightarrow{\textcolor{coeffectColor}{ \SYSTEMnt{r} } }   \beta    )    \xrightarrow{\textcolor{coeffectColor}{ \SYSTEMsym{1} } }   \beta   }
     { \Delta_{{\mathrm{1}}}  \vdash_{\textsc{g}_\forall }   \SYSTEMnt{t_{{\mathrm{1}}}}  \text{@}  \SYSTEMnt{B}   :    (   \SYSTEMnt{A}  \xrightarrow{\textcolor{coeffectColor}{ \SYSTEMnt{r} } }  \SYSTEMnt{B}   )   \xrightarrow{\textcolor{coeffectColor}{ \SYSTEMsym{1} } }  \SYSTEMnt{B}  }
  \\ \quad
  \inferrule*[Right=\SYSTEMRenameRuleGradabs{}]
    {  \Delta_{{\mathrm{2}}} ,   \SYSTEMmv{x}  :_{\textcolor{coeffectColor}{ \SYSTEMnt{r} } }  \SYSTEMnt{A}    \vdash_{\textsc{g}_\forall }  \SYSTEMnt{t_{{\mathrm{2}}}}  :  \SYSTEMnt{B} }
    { \Delta_{{\mathrm{2}}}  \vdash_{\textsc{g}_\forall }   \lambda  \SYSTEMmv{x}  .  \SYSTEMnt{t_{{\mathrm{2}}}}   :   \SYSTEMnt{A}  \xrightarrow{\textcolor{coeffectColor}{ \SYSTEMnt{r} } }  \SYSTEMnt{B}  }
}
{ \Delta_{{\mathrm{1}}}  \SYSTEMsym{+}  \Delta_{{\mathrm{2}}}  \vdash_{\textsc{g}_\forall }   \SYSTEMsym{(}   \SYSTEMnt{t_{{\mathrm{1}}}}  \text{@}  \SYSTEMnt{B}   \SYSTEMsym{)} \,  \SYSTEMsym{(}   \lambda  \SYSTEMmv{x}  .  \SYSTEMnt{t_{{\mathrm{2}}}}   \SYSTEMsym{)}   :  \SYSTEMnt{B} }
\quad
\tag{$\Box_e$}
\end{align}
\subsubsection{Linear Base to Graded Poly Base Translation}
\label{subsec:transl-gpb2lb}
We thus adapt the translation of types to:
\renewcommand{\triangleq}{& \oldTriangleq &}
\renewcommand{\equiv}{& \oldEquiv &}
\renewcommand{\SYSTEMdrule}[4][]{#3 \qquad #2}
\begin{gather*}
{\setlength{\arraycolsep}{0.1em}
\begin{array}{lcl@{\hspace{2em}}r}
\SYSTEMdruleTranslLtoGTyboxTyPoly{} & \text{(Type translation)}
\end{array}}
\end{gather*}
\let\SYSTEMdrule\oldSYSTEMdrule
\let\equiv\oldEquiv
\let\triangleq\oldTriangleq
We might expect that we can interpret the syntax of terms then directly
as in the above derivations, i.e.,
$ \textcolor{LGcolor}{\llparenthesis} \smidge   \textcolor{coeffectColor}{[}  \SYSTEMnt{t}  \textcolor{coeffectColor}{]}   \smidge \textcolor{LGcolor}{\rrparenthesis}  \triangleq   \Lambda   \beta   .    \lambda  \SYSTEMmv{f}  .    \SYSTEMmv{f} \,   \textcolor{LGcolor}{\llparenthesis} \smidge  \SYSTEMnt{t}  \smidge \textcolor{LGcolor}{\rrparenthesis}       $
and
$ \textcolor{LGcolor}{\llparenthesis} \smidge   \mathsf{let} \, \textcolor{coeffectColor}{[}  \SYSTEMmv{x}  \textcolor{coeffectColor}{]} =  \SYSTEMnt{t_{{\mathrm{1}}}}  \, \mathsf{in} \,  \SYSTEMnt{t_{{\mathrm{2}}}}   \smidge \textcolor{LGcolor}{\rrparenthesis}  \triangleq  \SYSTEMsym{(}    \textcolor{LGcolor}{\llparenthesis} \smidge  \SYSTEMnt{t_{{\mathrm{1}}}}  \smidge \textcolor{LGcolor}{\rrparenthesis}   \text{@}  \SYSTEMnt{B}   \SYSTEMsym{)} \,  \SYSTEMsym{(}   \lambda  \SYSTEMmv{x}  .   \textcolor{LGcolor}{\llparenthesis} \smidge  \SYSTEMnt{t_{{\mathrm{2}}}}  \smidge \textcolor{LGcolor}{\rrparenthesis}    \SYSTEMsym{)} $.
However, note the presence of $B$ on the right-hand side of $\Box$-elimination,
which has now appeared out of nowhere. Instead the translation must
ingest the entire \emph{typing derivation}, not merely the syntax of terms and types
separately. We can however translate the types separately.

Thus, we define the translation as mapping from typing derivations to terms,
where we denote derivation trees by $\Pi$:
\begin{align*}
\tag{Term translation}
\tikz[anchor=base, baseline]{
    \draw[line width=1pt,   LGcolor] plot[smooth, tension=.9] coordinates {(0,-0.4) (-0.1,-0.2) (-0.14,.25) (-0.1,0.7) (0,0.9)};
    \draw[line width=1pt, LGcolor] plot[smooth, tension=1] coordinates {(-.11,-0.15) (-0.13,.25) (-.11,0.65)};
    \draw[thick, LGcolor, cap=round] plot[smooth, tension=1] coordinates {(0,-0.405) (0,0.905)};
}
\dfrac
{\dfrac{\Pi}{ \Gamma  \vdash_{\textsc{l} }  \SYSTEMnt{t}  :  \SYSTEMnt{A} } \quad  \mathrm{graded}( \Gamma ) }
{  \textcolor{coeffectColor}{ \SYSTEMnt{r}  \cdot}  \Gamma   \vdash_{\textsc{l} }   \textcolor{coeffectColor}{[}  \SYSTEMnt{t}  \textcolor{coeffectColor}{]}   :   \textcolor{coeffectColor}{\square_{ \SYSTEMnt{r} } }  \SYSTEMnt{A}  }\SYSTEMRenameRuleLinpr
\tikz[anchor=base, baseline]{
    \draw[line width=1pt,   LGcolor] plot[smooth, tension=.9] coordinates {(0,-0.4) (0.1,-0.2) (0.14,.25) (0.1,0.7) (0,0.9)};
    \draw[line width=1pt, LGcolor] plot[smooth, tension=1] coordinates {(.11,-0.15) (0.13,.25) (.11,0.65)};
    \draw[thick, LGcolor, cap=round] plot[smooth, tension=1] coordinates {(0,-0.405) (0,0.905)};
}
& \;\triangleq\;
  \Lambda   \beta   .   \lambda  \SYSTEMmv{f}  .  \SYSTEMmv{f}   \,   \textcolor{LGcolor}{\llparenthesis}  \Pi  \textcolor{LGcolor}{\rrparenthesis}   \\[0.5em] %: (A r ->  tyb)
\tikz[anchor=base, baseline]{
    \draw[line width=1pt,   LGcolor] plot[smooth, tension=.9] coordinates {(0,-0.4) (-0.1,-0.2) (-0.14,.25) (-0.1,0.7) (0,0.9)};
    \draw[line width=1pt, LGcolor] plot[smooth, tension=1] coordinates {(-.11,-0.15) (-0.13,.25) (-.11,0.65)};
    \draw[thick, LGcolor, cap=round] plot[smooth, tension=1] coordinates {(0,-0.405) (0,0.905)};
}
\dfrac
{\dfrac{\Pi_1}
            { \Gamma_{{\mathrm{1}}}  \vdash_{\textsc{l} }  \SYSTEMnt{t_{{\mathrm{1}}}}  :   \textcolor{coeffectColor}{\square_{ \SYSTEMnt{r} } }  \SYSTEMnt{A}  }
 \quad
 \dfrac{\Pi_2}
            {  \Gamma_{{\mathrm{2}}} ,   \SYSTEMmv{x}  : \textcolor{coeffectColor}{[}  \SYSTEMnt{A} {\textcolor{coeffectColor}{]_{ \SYSTEMnt{r} } } }    \vdash_{\textsc{l} }  \SYSTEMnt{t_{{\mathrm{2}}}}  :  \SYSTEMnt{B} }
}
{ \Gamma_{{\mathrm{1}}}  \SYSTEMsym{+}  \Gamma_{{\mathrm{2}}}  \vdash_{\textsc{l} }   \mathsf{let} \, \textcolor{coeffectColor}{[}  \SYSTEMmv{x}  \textcolor{coeffectColor}{]} =  \SYSTEMnt{t_{{\mathrm{1}}}}  \, \mathsf{in} \,  \SYSTEMnt{t_{{\mathrm{2}}}}   :  \SYSTEMnt{B} }\SYSTEMRenameRuleLinlet
\tikz[anchor=base, baseline]{
    \draw[line width=1pt,   LGcolor] plot[smooth, tension=.9] coordinates {(0,-0.4) (0.1,-0.2) (0.14,.25) (0.1,0.7) (0,0.9)};
    \draw[line width=1pt, LGcolor] plot[smooth, tension=1] coordinates {(.11,-0.15) (0.13,.25) (.11,0.65)};
    \draw[thick, LGcolor, cap=round] plot[smooth, tension=1] coordinates {(0,-0.405) (0,0.905)};
}
& \;\triangleq\;
 \SYSTEMsym{(}    \textcolor{LGcolor}{\llparenthesis}  \Pi_{{\mathrm{1}}}  \textcolor{LGcolor}{\rrparenthesis}   \text{@}   \textcolor{LGcolor}{\llparenthesis} \smidge  \SYSTEMnt{B}  \smidge \textcolor{LGcolor}{\rrparenthesis}    \SYSTEMsym{)} \,  \SYSTEMsym{(}   \lambda  \SYSTEMmv{x}  .   \textcolor{LGcolor}{\llparenthesis}  \Pi_{{\mathrm{2}}}  \textcolor{LGcolor}{\rrparenthesis}    \SYSTEMsym{)} 
\end{align*}
The rest of the translation follows the structure of the previous term
translation.

In the following, we write $\Pi :  \Gamma  \vdash_{\textsc{l} }  \SYSTEMnt{t}  :  \SYSTEMnt{A} $
to mean that $\Pi$ is the derivation of a judgment $ \Gamma  \vdash_{\textsc{l} }  \SYSTEMnt{t}  :  \SYSTEMnt{A} $.

\begin{restatable}{thrm}{linToGradTranslationPoly}[Soundness of the Linear Base to Graded Poly Base translation]
 \begin{itemize}[itemsep=0.5em,topsep=0.5em,leftmargin=1em]
  \item Type preservation: $\Pi :  \Gamma  \vdash_{\textsc{l} }  \SYSTEMnt{t}  :  \SYSTEMnt{A}  \implies   \textcolor{LGcolor}{\llparenthesis}  \Gamma  \textcolor{LGcolor}{\rrparenthesis}   \vdash_{\textsc{g}_\forall }   \textcolor{LGcolor}{\llparenthesis}  \Pi  \textcolor{LGcolor}{\rrparenthesis}   :   \textcolor{LGcolor}{\llparenthesis} \smidge  \SYSTEMnt{A}  \smidge \textcolor{LGcolor}{\rrparenthesis}  $
  \item Operational correspondence:  $(\Pi :  \Gamma  \vdash_{\textsc{l} }  \SYSTEMnt{t}  :  \SYSTEMnt{A} ) \wedge (\Pi' :  \Gamma  \vdash_{\textsc{l} }  \SYSTEMnt{t'}  :  \SYSTEMnt{A} ) \wedge ( \SYSTEMnt{t}  \rightsquigarrow_{\textsc{l} }  \SYSTEMnt{t'} ) \implies   \textcolor{LGcolor}{\llparenthesis}  \Pi  \textcolor{LGcolor}{\rrparenthesis}   \rightsquigarrow_{\textsc{g}\forall }^\ast   \textcolor{LGcolor}{\llparenthesis}  \Pi'  \textcolor{LGcolor}{\rrparenthesis}  $
  \item Equation preservation: $(\Pi :  \Gamma  \vdash_{\textsc{l} }  \SYSTEMnt{t}  :  \SYSTEMnt{A} ) \wedge (\Pi' :  \Gamma  \vdash_{\textsc{l} }  \SYSTEMnt{t'}  :  \SYSTEMnt{A} ) \wedge (\SYSTEMnt{t}  \equiv^{-\eta\Box}_{\textsc{l} }  \SYSTEMnt{t'}) \implies   \textcolor{LGcolor}{\llparenthesis}  \Pi  \textcolor{LGcolor}{\rrparenthesis}   \equiv_{\textsc{g}_\forall}   \textcolor{LGcolor}{\llparenthesis}  \Pi'  \textcolor{LGcolor}{\rrparenthesis}  $
\end{itemize}
Appendix~\refappendix{app:lin-to-grad-poly} provides the proofs.

Note the operational correspondence
needs to have typing derivations of the source term and target term.
Furthermore, the translation does not preserve $\eta$-equality for the graded modality; only the $\beta$-equality rules, commuting conversions, and congruence rules are preserved;
we thus denote the subset of the preserved equational theory here as $ \equiv^{-\eta\Box}_{\textsc{l} } $.
This is because the translation of a term $t$ of type $ \textcolor{coeffectColor}{\square_{ \SYSTEMnt{r} } }  \SYSTEMnt{A} $
may not be of the form $  \Lambda   \beta   .   \lambda  \SYSTEMmv{f}  .  \SYSTEMmv{f}   \,  \SYSTEMnt{t'} $ for some $t'$ (e.g., it could be just a free variable), which is required for $\eta$-equality to apply.
Instead, parametricity reasoning (based on the polymorphic typing) would provide that $t$ is \emph{observationally equivalent} to $  \Lambda   \beta   .   \lambda  \SYSTEMmv{f}  .  \SYSTEMmv{f}   \,  \SYSTEMnt{t'} $ for some $t'$,
and thus the translation preserves $\eta$-equality up to observational equivalence, though not up to syntactic equality;
we leave the details for further work.
\label{thrm:linToGradTranslationPoly}
\end{restatable}

\begin{remark}
(\haskin{Ur} is not the $!$-modality)
\citet{DBLP:journals/pacmpl/BernardyBNJS18} suggest \haskin{Ur} as
an encoding of Linear Logic's $!$-modality, although \citet{hughes2021linear} later show that this
encoding in Linear Haskell and similarly in Idris 2, as well as the graded modality in Granule, do not exactly correspond to Linear Logic's $!$-modality. Once (multiplicative) products
are included, Linear Logic does not admit distributing $!$ over products, i.e.,
$!(A \otimes B) \multimap \,! A\, \otimes \,! B$ is not derivable. However the aforementioned systems permit its derivation due to the interaction between graded modalities and products
(and similarly sums). We consider calculi with products and sums next
as this takes us closer towards practical programming and it allows us to uncover
other expressivity differences in the literature.
\end{remark}

\section{Extending Linear Base and Graded Modal Base with Products and Sums}
\label{sec:prods-sums}
Thus far the calculi studied have only covered the implicational
fragments of the two approaches, i.e., just function types and
variables.  Adding product and sum types to the calculi brings the
theory closer to a useful basis for understanding actual
programming languages such as the linear-types extension to Haskell (graded-base) or Granule
(linear-base).
We thus now add products and sums to Linear Base and Graded Modal Base
following standard approaches in the literature for both flavours
of linear-base and graded-base systems.  This raises interesting questions
that delineate the expressive power of linear- vs. graded-base
calculi. From linear-base to graded-base, a straightforward
homomorphic translation is possible. However, for the opposite
direction we will see a difference in expressivity that needs handling
by extending the linear-base system.

\paragraph{Syntax} We extend Linear Base and Graded Modal Base (which already share the same
syntax) to \textbf{Linear~Core} and \textbf{Graded~Modal~Core} respectively, with the same extended syntax:
\begin{align*}
  \SYSTEMnt{t} & ::= ... \mid  \langle  \SYSTEMnt{t_{{\mathrm{1}}}} ,  \SYSTEMnt{t_{{\mathrm{2}}}}  \rangle  \mid  \mathsf{let} \, \langle  \SYSTEMmv{x} ,  \SYSTEMmv{y}  \rangle =  \SYSTEMnt{t_{{\mathrm{1}}}}  \, \mathsf{in} \,  \SYSTEMnt{t_{{\mathrm{2}}}}  \mid  \langle \rangle  \mid  \mathsf{let} \, \langle \rangle =  \SYSTEMnt{t_{{\mathrm{1}}}}  \, \mathsf{in} \,  \SYSTEMnt{t_{{\mathrm{2}}}}  \\
         \mid & \; \mathsf{inj}_1 \,  \SYSTEMnt{t}  \mid  \mathsf{inj}_2 \,  \SYSTEMnt{t}  \mid  \mathsf{case} \,  \SYSTEMnt{t}  \, \mathsf{of} \, \{ \mathsf{inj1} \,  \SYSTEMmv{x}  \rightarrow  \SYSTEMnt{t_{{\mathrm{1}}}}  ; \, \mathsf{inj2} \,  \SYSTEMmv{y}  \rightarrow  \SYSTEMnt{t_{{\mathrm{2}}}}  \} 
  \tag{terms}
\end{align*}
These additional term formers provide product introduction and elimination, unit introduction
and elimination, and sum (coproduct) introductions and elimination respectively.

The operational semantics are the standard call-by-name operational semantics for products and sums given in Figure~\ref{fig:linear-core} for Linear Core, with the same
added to operational semantics of Graded Modal Core. We elide the $\beta\eta$-equational theory which is standard (see Appendices~\refappendix{app:definitions-linear-core} and
~\refappendix{app:definitions-graded-core}).

\begin{figure}[t]
\raggedright\noindent\framebox{$ \SYSTEMnt{t}  \rightsquigarrow_{\textsc{l} }  \SYSTEMnt{t} $} \quad Standard operational semantics (Call-By-Name) for products, units, and sums
\begin{gather*}
  \begin{array}{c}
  \SYSTEMdruleSemLinprodCong{}
  \;
  \SYSTEMdruleSemLinprodBeta{}
  \\[1.5em]
  \SYSTEMdruleSemLinunitCong{}
  \quad
  \SYSTEMdruleSemLinunitBeta{}
  \\[1.5em]
  \SYSTEMdruleSemLincongCase{}
  \\[1.5em]
  \SYSTEMdruleSemLincaseInjOne{}
  \\[1.5em]
  \SYSTEMdruleSemLincaseInjTwo{}
  \end{array}
\end{gather*}
\bigskip

\raggedright\noindent\framebox{$ \Gamma  \vdash_{\textsc{l} }  \SYSTEMnt{t}  :  \SYSTEMnt{A} $} \quad Typing rules
\begin{gather*}
\begin{align*}
\begin{array}{c}
\SYSTEMdruleLinprodi{}
\quad
\SYSTEMdruleLinprode{}
\\[1.5em]
\SYSTEMdruleLinuniti{}
\quad
\SYSTEMdruleLinunite{}
\quad
\SYSTEMdruleLinsumiOne{}
\\[1.5em]
\SYSTEMdruleLinsumiTwo{}
\quad
\SYSTEMdruleLinsume{}
\end{array}
\end{align*}
\end{gather*}
\caption{Linear Core semantics and typing}
\label{fig:linear-core}
\end{figure}

% The additional syntax for types differs only in
%the symbol used for the type constructors of sums and products.
% ($     \otimes     $ vs $     \times     $ and $     \oplus     $ vs $     +     $).
% \newpage
To type the new syntax, we extend the type system of Linear Base to Linear Core
(Section~\ref{sec:linear-core}), and extend Graded Modal Base to
Graded Modal Core (Section~\ref{sec:graded-core}).
Section~\ref{sec:linear-core-to-graded} then gives the
translation from Linear Core to Graded Modal Core which is
straightforwardly homomorphic.
Section~\ref{sec:no-trans-graded-core-to-linear-core} then
shows that the opposite direction is not possible, requiring an
extension of Linear Core to \textbf{Linear~Push~Core} in
Section~\ref{sec:linear-push-core}, at the cost of weakening the correspondence between our $\Box$-modality
and Linear Logic's $!$-modality.  We then give the translation
from Graded Modal Core to Linear Push Core
(Section~\ref{sec:trans-graded-core-to-linear-push-core}) and the translation
from Linear Push Core to Graded Modal Core
(Section~\ref{sec:trans-linear-push-core-to-graded-core}).

\subsection{Linear Core}
\label{sec:linear-core}

Linear Base augmented with products and sums becomes what we call \textbf{Linear~Core}.
Linear Core features the following new type constructors for products,
units, and sums:
\begin{align*}
  \SYSTEMnt{A} & ::= ... \mid  \SYSTEMnt{A}  \otimes  \SYSTEMnt{B}  \mid  \mathrm{unit}  \mid  \SYSTEMnt{A}  \oplus  \SYSTEMnt{B} 
  \tag{types}
\end{align*}

\paragraph{Typing} Linear Core extends Linear Base typing (given in Figure~\ref{fig:linear-base-types})
with the judgments for multiplicative products and their unit, along with additive disjunction, in Figure~\ref{fig:linear-core}, which
are standard from Linear Logic~\cite{girard1987linear}. This is the same typing as in linear-base systems such as Granule~\cite{DBLP:journals/pacmpl/OrchardLE19,DBLP:journals/corr/abs-2112-14966}, the synthesis calculus of \citet{hughes2021linear}, and that of \citet{brunel2014core} (although
their system has a less general natural numbers data type instead of products and sums, but the same insights apply).

As in Linear Base, any terms comprising multiple sub-terms have their contexts
combined using context addition $+$ which adds together the grades of
graded assumptions using semiring addition.

\subsection{Graded Modal Core}
\label{sec:graded-core}

Graded Modal Base augmented with products and sums becomes \textbf{Graded Modal Core}.
Graded Modal Core features the following new type constructors for products, units, and sums:
\begin{align*}
  \SYSTEMnt{A} & ::= ... \mid  \SYSTEMnt{A}  \times  \SYSTEMnt{B}  \mid  \mathrm{unit}  \mid  \SYSTEMnt{A}  +  \SYSTEMnt{B} 
  \tag{types}
\end{align*}

\paragraph{Typing}

Graded Modal Core extends Graded Modal Base typing (given in Section~\ref{sec:graded-modal-base}) by the judgments
given in Figure~\ref{fig:graded-base-types-products-sums}. The typing is akin to that shown
by \citet{DBLP:journals/pacmpl/AbelB20}, in Linear Haskell~\cite{DBLP:journals/pacmpl/BernardyBNJS18}, and by \citet{grad} (although they restrict their
products initially to $r = 1$).

In the elimination of the products we may use each variable
according to grade $\SYSTEMnt{r}$ and thus the context $\Delta_{{\mathrm{1}}}$ for the eliminated product
is multiplied by $\SYSTEMnt{r}$ in the conclusion. Thus, the usage of the components is propagated
to demands on the dependencies for the product. The elimination of sums
has a similar idea, propagating demands on the variables $x$ and $y$ through to the
dependencies of the eliminated sum term $t$. There is an additional side condition here too
 ($ \SYSTEMsym{1}  \, \textcolor{coeffectColor}{\sqsubseteq} \,  \SYSTEMnt{r} $) accounting for usage incurred by the inspection of the sum
constructor, and thus some information gained. This side condition is key to applications of
graded coeffect types for information-flow control~(\ref{exm:security-lattice}) \cite{DBLP:journals/pacmpl/AbelB20,mycroftfest2024}:
if the grade here is $r = \mathsf{Hi}$, representing
a high-security value, then the side condition fails as $1 = \mathsf{Lo}$ and
thus $\mathsf{Lo} \not\succeq \mathsf{Hi}$. This prevents pattern matching on a high-security
value as a control-flow attack to leak information. Lastly note the elimination of
the unit type similarly allows (an arbitrary) scaling of the scrutinee's context, in line with
presentations in the literature (e.g.,~\cite{DBLP:journals/pacmpl/AbelB20}).
\begin{figure}[t]
\raggedright\noindent\framebox{$ \Delta  \vdash_{\textsc{g} }  \SYSTEMnt{t}  :  \SYSTEMnt{A} $} \quad Typing rules
\begin{gather*}
\begin{align*}
\begin{array}{c}
\SYSTEMdruleGradprodi{}
\quad
\SYSTEMdruleGradprode{}
\\[1.25em]
\SYSTEMdruleGraduniti{}
\quad
\SYSTEMdruleGradunite{}
\quad
\SYSTEMdruleGradsumiOne{}
\\[1.25em]
\SYSTEMdruleGradsumiTwo{}
\;
\SYSTEMdruleGradsume{}
\end{array}
\end{align*}
\end{gather*}
\caption{Graded Modal Core additional typing rules.}
\label{fig:graded-base-types-products-sums}
\end{figure}

\subsection{Translation from Linear Core to Graded Modal Core}
\label{sec:linear-core-to-graded}

We extend the translation from Section \ref{subsec:transl-lb2gmb} via a homomorphism
on the additional syntax:
%
% TEMPORARILY REDEFINE SYSTEM DRULE FOR THE TRANSLATION WHICH IS REALLY
% JUST A FUNCTIOn
\renewcommand{\triangleq}{& \oldTriangleq &}
\renewcommand{\equiv}{& \oldEquiv &}
\renewcommand{\SYSTEMdrule}[4][]{#3 \qquad #2}
\begin{gather*}
{\setlength{\arraycolsep}{0.1em}
\begin{array}{lcl@{\hspace{0em}}r}
\SYSTEMdruleTranslLtoGTmprodi{} \\
\SYSTEMdruleTranslLtoGTmprode{} \\
\SYSTEMdruleTranslLtoGTmuniti{} \\
\SYSTEMdruleTranslLtoGTmunite{} \\
\SYSTEMdruleTranslLtoGTmsumiOne{} \\
\SYSTEMdruleTranslLtoGTmsumiTwo{} & \text{(Term translation)} \\
\SYSTEMdruleTranslLtoGTmsume{}  \\[0.5em]
\SYSTEMdruleTranslLtoGTyprodTy{} \\
\SYSTEMdruleTranslLtoGTysumTy{} \\
\SYSTEMdruleTranslLtoGTyunitTy{} & \text{(Type translation)}
\end{array}}
\end{gather*}
\let\SYSTEMdrule\oldSYSTEMdrule
\let\equiv\oldEquiv
\let\triangleq\oldTriangleq
We can then re-establish the two key results about the translation (proofs in
Appendix~\refappendix{app:proofs-lin-core-to-grad-core}):
\begin{restatable}{thrm}{linCanToGradTranslationExt}[Soundness of the Linear Core to Graded Modal Core translation]
 \begin{itemize}[itemsep=0.2em]
  \item Type preservation: $ \Gamma  \vdash_{\textsc{l} }  \SYSTEMnt{t}  :  \SYSTEMnt{A}  \implies   \textcolor{LGcolor}{\llparenthesis}  \Gamma  \textcolor{LGcolor}{\rrparenthesis}   \vdash_{\textsc{g} }   \textcolor{LGcolor}{\llparenthesis} \smidge  \SYSTEMnt{t}  \smidge \textcolor{LGcolor}{\rrparenthesis}   :   \textcolor{LGcolor}{\llparenthesis} \smidge  \SYSTEMnt{A}  \smidge \textcolor{LGcolor}{\rrparenthesis}  $;
  \item Operational correspondence: $ \SYSTEMnt{t}  \rightsquigarrow_{\textsc{l} }  \SYSTEMnt{t'}  \implies   \textcolor{LGcolor}{\llparenthesis} \smidge  \SYSTEMnt{t}  \smidge \textcolor{LGcolor}{\rrparenthesis}   \rightsquigarrow_{\textsc{g} }   \textcolor{LGcolor}{\llparenthesis} \smidge  \SYSTEMnt{t'}  \smidge \textcolor{LGcolor}{\rrparenthesis}  $;
  \item Equation preservation: $ \SYSTEMnt{t}  \,\equiv_{\textsc{l} }\,  \SYSTEMnt{t'}  \implies   \textcolor{LGcolor}{\llparenthesis} \smidge  \SYSTEMnt{t}  \smidge \textcolor{LGcolor}{\rrparenthesis}   \equiv_{\textsc{g} }   \textcolor{LGcolor}{\llparenthesis} \smidge  \SYSTEMnt{t'}  \smidge \textcolor{LGcolor}{\rrparenthesis}  $.
\end{itemize}
\end{restatable}
A notable point of the type preservation proof is that when translating sum elimination from Linear Core to
Graded Modal Core, we pick the grade $r = 1$ which satisfies the side condition that $ \SYSTEMsym{1}  \, \textcolor{coeffectColor}{\sqsubseteq} \,  \SYSTEMsym{1} $
trivially by reflexivity of the pre-order.

\subsection{No Type-Sound Translation from Graded Modal Core to Linear Core}
\label{sec:no-trans-graded-core-to-linear-core}

We now attempt a similar homomorphic translation ($  \textcolor{GLcolor}{\llbracket}   -   \textcolor{GLcolor}{\rrbracket}  _\circ $) back from Graded Modal Core
to Linear Core, focusing on just products and sums:
\renewcommand{\equiv}{ \oldEquiv}
\renewcommand{\SYSTEMdrule}[4][]{{\displaystyle{#3}}}
\begin{align}
 \tag{Term translation} \\[-1.5em]
\notag & \begin{array}{l}
  \textcolor{GLcolor}{\llbracket}   \langle  \SYSTEMnt{t_{{\mathrm{1}}}} ,  \SYSTEMnt{t_{{\mathrm{2}}}}  \rangle   \textcolor{GLcolor}{\rrbracket}  _\circ  \triangleq  \langle    \textcolor{GLcolor}{\llbracket}  \SYSTEMnt{t_{{\mathrm{1}}}}  \textcolor{GLcolor}{\rrbracket}  _\circ  ,    \textcolor{GLcolor}{\llbracket}  \SYSTEMnt{t_{{\mathrm{2}}}}  \textcolor{GLcolor}{\rrbracket}  _\circ   \rangle  \\
  \textcolor{GLcolor}{\llbracket}   \mathsf{let} \, \langle  \SYSTEMmv{x} ,  \SYSTEMmv{y}  \rangle =  \SYSTEMnt{t_{{\mathrm{1}}}}  \, \mathsf{in} \,  \SYSTEMnt{t_{{\mathrm{2}}}}   \textcolor{GLcolor}{\rrbracket}  _\circ  \triangleq   \mathsf{let} \, \langle  \SYSTEMmv{x} ,  \SYSTEMmv{y}  \rangle =    \textcolor{GLcolor}{\llbracket}  \SYSTEMnt{t_{{\mathrm{1}}}}  \textcolor{GLcolor}{\rrbracket}  _\circ   \, \mathsf{in} \,   \textcolor{GLcolor}{\llbracket}  \SYSTEMnt{t_{{\mathrm{2}}}}  \textcolor{GLcolor}{\rrbracket}   _\circ \\
  \textcolor{GLcolor}{\llbracket}   \mathsf{inj}_i \,  \SYSTEMnt{t}   \textcolor{GLcolor}{\rrbracket}  _\circ  \triangleq   \mathsf{inj}_i \,   \textcolor{GLcolor}{\llbracket}  \SYSTEMnt{t}  \textcolor{GLcolor}{\rrbracket}   _\circ \\
  \textcolor{GLcolor}{\llbracket}   \mathsf{inj}_2 \,  \SYSTEMnt{t}   \textcolor{GLcolor}{\rrbracket}  _\circ  \triangleq   \mathsf{inj}_2 \,   \textcolor{GLcolor}{\llbracket}  \SYSTEMnt{t}  \textcolor{GLcolor}{\rrbracket}   _\circ \\
  \textcolor{GLcolor}{\llbracket}   \mathsf{case} \,  \SYSTEMnt{t}  \, \mathsf{of} \, \{ \mathsf{inj1} \,  \SYSTEMmv{x}  \rightarrow  \SYSTEMnt{t_{{\mathrm{1}}}}  ; \, \mathsf{inj2} \,  \SYSTEMmv{y}  \rightarrow  \SYSTEMnt{t_{{\mathrm{2}}}}  \}   \textcolor{GLcolor}{\rrbracket}  _\circ  \triangleq  \mathsf{case} \,    \textcolor{GLcolor}{\llbracket}  \SYSTEMnt{t}  \textcolor{GLcolor}{\rrbracket}  _\circ   \, \mathsf{of} \, \{ \mathsf{inj1} \,  \SYSTEMmv{x}  \rightarrow    \textcolor{GLcolor}{\llbracket}  \SYSTEMnt{t_{{\mathrm{1}}}}  \textcolor{GLcolor}{\rrbracket}  _\circ   ; \, \mathsf{inj2} \,  \SYSTEMmv{y}  \rightarrow    \textcolor{GLcolor}{\llbracket}  \SYSTEMnt{t_{{\mathrm{2}}}}  \textcolor{GLcolor}{\rrbracket}  _\circ   \} 
\end{array} \\
\tag{Type translation}
& \begin{array}{l}
  \textcolor{GLcolor}{\llbracket}   \SYSTEMnt{A}  \times  \SYSTEMnt{B}   \textcolor{GLcolor}{\rrbracket}  _\circ  \triangleq     \textcolor{GLcolor}{\llbracket}  \SYSTEMnt{A}  \textcolor{GLcolor}{\rrbracket}  _\circ   \otimes   \textcolor{GLcolor}{\llbracket}  \SYSTEMnt{B}  \textcolor{GLcolor}{\rrbracket}   _\circ  \\
  \textcolor{GLcolor}{\llbracket}   \SYSTEMnt{A}  +  \SYSTEMnt{B}   \textcolor{GLcolor}{\rrbracket}  _\circ  \triangleq     \textcolor{GLcolor}{\llbracket}  \SYSTEMnt{A}  \textcolor{GLcolor}{\rrbracket}  _\circ   \oplus   \textcolor{GLcolor}{\llbracket}  \SYSTEMnt{B}  \textcolor{GLcolor}{\rrbracket}   _\circ 
\end{array}
\end{align}
\let\SYSTEMdrule\oldSYSTEMdrule
\let\equiv\oldEquiv
However, this translation is unsound: consider the following Graded Modal Core terms, $f$ and $g$,
which manifest a distributivity of grading over products and sums:
\begin{align*}
\begin{array}{rl}
f &:   (   \SYSTEMnt{A}  \times  \SYSTEMnt{B}   )   \xrightarrow{\textcolor{coeffectColor}{ \SYSTEMnt{r} } }   (    \textcolor{coeffectColor}{\square_{ \SYSTEMnt{r} } }  \SYSTEMnt{A}   \times   \textcolor{coeffectColor}{\square_{ \SYSTEMnt{r} } }  \SYSTEMnt{B}    )   \\
f &=  \lambda  \SYSTEMmv{p}  .  \SYSTEMsym{(}   \mathsf{let} \, \langle  \SYSTEMmv{x} ,  \SYSTEMmv{y}  \rangle =  \SYSTEMmv{p}  \, \mathsf{in} \,   \langle   \textcolor{coeffectColor}{[}  \SYSTEMmv{x}  \textcolor{coeffectColor}{]}  ,   \textcolor{coeffectColor}{[}  \SYSTEMmv{y}  \textcolor{coeffectColor}{]}   \rangle    \SYSTEMsym{)} 
\end{array}
\begin{array}{rl}
g &:   (   \SYSTEMnt{A}  +  \SYSTEMnt{B}   )   \xrightarrow{\textcolor{coeffectColor}{ \SYSTEMnt{r} } }   (    \textcolor{coeffectColor}{\square_{ \SYSTEMnt{r} } }  \SYSTEMnt{A}   +   \textcolor{coeffectColor}{\square_{ \SYSTEMnt{r} } }  \SYSTEMnt{B}    )   \quad (\text{where }  \SYSTEMsym{1}  \, \textcolor{coeffectColor}{\sqsubseteq} \,  \SYSTEMnt{r} )\\
g &=  \lambda  \SYSTEMmv{e}  .   \mathsf{case} \,  \SYSTEMmv{e}  \, \mathsf{of} \, \{ \mathsf{inj1} \,  \SYSTEMmv{x}  \rightarrow   \mathsf{inj}_1 \,   \textcolor{coeffectColor}{[}  \SYSTEMmv{x}  \textcolor{coeffectColor}{]}    ; \, \mathsf{inj2} \,  \SYSTEMmv{y}  \rightarrow   \mathsf{inj}_2 \,   \textcolor{coeffectColor}{[}  \SYSTEMmv{y}  \textcolor{coeffectColor}{]}    \}  
\end{array}
\end{align*}
Thus $f$ is a function that distributes the resource requirements on a pair onto its components, and similarly, $g$ distributes the resource requirements on a sum onto the components of its injections.

Applying the above translation to $f$ and $g$ yields the following ill-typed terms
(to avoid clutter, we consider $  \textcolor{GLcolor}{\llbracket}  \SYSTEMnt{A}  \textcolor{GLcolor}{\rrbracket}  _\circ  $ interchangeable with $ \SYSTEMnt{A} $):
\begin{align*}
  \textcolor{GLcolor}{\llbracket}  \SYSTEMmv{f}  \textcolor{GLcolor}{\rrbracket}  _\circ  &:   \textcolor{coeffectColor}{\square_{ \SYSTEMnt{r} } }   (   \SYSTEMnt{A}  \otimes  \SYSTEMnt{B}   )    \multimap   (    \textcolor{coeffectColor}{\square_{ \SYSTEMnt{r} } }  \SYSTEMnt{A}   \otimes   \textcolor{coeffectColor}{\square_{ \SYSTEMnt{r} } }  \SYSTEMnt{B}    )   \tag{ill-typed} \\
\notag
  \textcolor{GLcolor}{\llbracket}  \SYSTEMmv{f}  \textcolor{GLcolor}{\rrbracket}  _\circ  &=  \lambda  \SYSTEMmv{b}  .  \SYSTEMsym{(}   \mathsf{let} \, \textcolor{coeffectColor}{[}  \SYSTEMmv{p}  \textcolor{coeffectColor}{]} =  \SYSTEMmv{b}  \, \mathsf{in} \,   \mathsf{let} \, \langle  \SYSTEMmv{x} ,  \SYSTEMmv{y}  \rangle =  \SYSTEMmv{p}  \, \mathsf{in} \,   \langle   \textcolor{coeffectColor}{[}  \SYSTEMmv{x}  \textcolor{coeffectColor}{]}  ,   \textcolor{coeffectColor}{[}  \SYSTEMmv{y}  \textcolor{coeffectColor}{]}   \rangle     \SYSTEMsym{)}  \\[.25em]
\notag
  \textcolor{GLcolor}{\llbracket}  \SYSTEMmv{g}  \textcolor{GLcolor}{\rrbracket}  _\circ  &:   \textcolor{coeffectColor}{\square_{ \SYSTEMnt{r} } }   (   \SYSTEMnt{A}  \oplus  \SYSTEMnt{B}   )    \multimap   (    \textcolor{coeffectColor}{\square_{ \SYSTEMnt{r} } }  \SYSTEMnt{A}   \oplus   \textcolor{coeffectColor}{\square_{ \SYSTEMnt{r} } }  \SYSTEMnt{B}    )   \\
\notag
  \textcolor{GLcolor}{\llbracket}  \SYSTEMmv{g}  \textcolor{GLcolor}{\rrbracket}  _\circ  &=  \lambda  \SYSTEMmv{b}  .  \SYSTEMsym{(}   \mathsf{let} \, \textcolor{coeffectColor}{[}  \SYSTEMmv{e}  \textcolor{coeffectColor}{]} =  \SYSTEMmv{b}  \, \mathsf{in} \,   \mathsf{case} \,  \SYSTEMmv{e}  \, \mathsf{of} \, \{ \mathsf{inj1} \,  \SYSTEMmv{x}  \rightarrow   \mathsf{inj}_1 \,   \textcolor{coeffectColor}{[}  \SYSTEMmv{x}  \textcolor{coeffectColor}{]}    ; \, \mathsf{inj2} \,  \SYSTEMmv{y}  \rightarrow   \mathsf{inj}_2 \,   \textcolor{coeffectColor}{[}  \SYSTEMmv{y}  \textcolor{coeffectColor}{]}    \}    \SYSTEMsym{)} 
\end{align*}
To see why these are ill-typed, the following gives a partial derivation for the body of
$  \textcolor{GLcolor}{\llbracket}  \SYSTEMmv{f}  \textcolor{GLcolor}{\rrbracket}  _\circ $:

\begin{gather*}
    \inferrule*[right=\SYSTEMRenameRuleLinlet{}]
      {
       % \inferrule*[Right=\SYSTEMRenameRuleLinvar{}, vdots=5.5em, rightskip=12em]
      %    { }
       %   {   \SYSTEMmv{b}  :   \textcolor{coeffectColor}{\square_{ \SYSTEMnt{r} } }   (   \SYSTEMnt{A}  \otimes  \SYSTEMnt{B}   )      \vdash_{\textsc{l} }  \SYSTEMmv{b}  :   \textcolor{coeffectColor}{\square_{ \SYSTEMnt{r} } }   (   \SYSTEMnt{A}  \otimes  \SYSTEMnt{B}   )   }
        \inferrule*[Right=\SYSTEMRenameRuleLinprode{}]
          {
            \inferrule*[Right=\SYSTEMRenameRuleLinder{}*]
              {\inferrule*[Right=\SYSTEMRenameRuleLinvar{}]{ }{   \SYSTEMmv{p}  :   \SYSTEMnt{A}  \otimes  \SYSTEMnt{B}     \vdash_{\textsc{l} }  \SYSTEMmv{p}  :   \SYSTEMnt{A}  \otimes  \SYSTEMnt{B}  } \hspace{5em} r = 1 }
              {   \SYSTEMmv{p}  : \textcolor{coeffectColor}{[}   \SYSTEMnt{A}  \otimes  \SYSTEMnt{B}  {\textcolor{coeffectColor}{]_{ \SYSTEMnt{r} } } }    \vdash_{\textsc{l} }  \SYSTEMmv{p}  :   \SYSTEMnt{A}  \otimes  \SYSTEMnt{B}  }
            \hspace{4em}
            \inferrule*[Right=\SYSTEMRenameRuleLinprodi{}]
              {
                \inferrule*[Right=?]
                  { }
                  {    \SYSTEMmv{x}  :  \SYSTEMnt{A}    \vdash_{\textsc{l} }   \textcolor{coeffectColor}{[}  \SYSTEMmv{x}  \textcolor{coeffectColor}{]}   :   \textcolor{coeffectColor}{\square_{ \SYSTEMnt{r} } }  \SYSTEMnt{A}   }
                \hspace{4em}
                \inferrule*[Right=?]
                  { }
                  {    \SYSTEMmv{y}  :  \SYSTEMnt{B}    \vdash_{\textsc{l} }   \textcolor{coeffectColor}{[}  \SYSTEMmv{y}  \textcolor{coeffectColor}{]}   :   \textcolor{coeffectColor}{\square_{ \SYSTEMnt{r} } }  \SYSTEMnt{B}   }
              }
              {    \SYSTEMmv{x}  :   \textcolor{coeffectColor}{\square_{ \SYSTEMnt{r} } }  \SYSTEMnt{A}     \SYSTEMsym{+}    \SYSTEMmv{y}  :   \textcolor{coeffectColor}{\square_{ \SYSTEMnt{r} } }  \SYSTEMnt{B}     \vdash_{\textsc{l} }   \langle   \textcolor{coeffectColor}{[}  \SYSTEMmv{x}  \textcolor{coeffectColor}{]}  ,   \textcolor{coeffectColor}{[}  \SYSTEMmv{y}  \textcolor{coeffectColor}{]}   \rangle   :    \textcolor{coeffectColor}{\square_{ \SYSTEMnt{r} } }  \SYSTEMnt{A}   \otimes   \textcolor{coeffectColor}{\square_{ \SYSTEMnt{r} } }  \SYSTEMnt{B}   }
          }
          {   \SYSTEMmv{p}  : \textcolor{coeffectColor}{[}   \SYSTEMnt{A}  \otimes  \SYSTEMnt{B}  {\textcolor{coeffectColor}{]_{ \SYSTEMnt{r} } } }    \vdash_{\textsc{l} }   \mathsf{let} \, \langle  \SYSTEMmv{x} ,  \SYSTEMmv{y}  \rangle =  \SYSTEMmv{p}  \, \mathsf{in} \,   \langle   \textcolor{coeffectColor}{[}  \SYSTEMmv{x}  \textcolor{coeffectColor}{]}  ,   \textcolor{coeffectColor}{[}  \SYSTEMmv{y}  \textcolor{coeffectColor}{]}   \rangle    :    \textcolor{coeffectColor}{\square_{ \SYSTEMnt{r} } }  \SYSTEMnt{A}   \otimes   \textcolor{coeffectColor}{\square_{ \SYSTEMnt{r} } }  \SYSTEMnt{B}    }
      }
      {
        %\inferrule*[Right=\SYSTEMRenameRuleLinabs{}]
          {   \SYSTEMmv{b}  :   \textcolor{coeffectColor}{\square_{ \SYSTEMnt{r} } }   (   \SYSTEMnt{A}  \otimes  \SYSTEMnt{B}   )      \vdash_{\textsc{l} }   \mathsf{let} \, \textcolor{coeffectColor}{[}  \SYSTEMmv{p}  \textcolor{coeffectColor}{]} =  \SYSTEMmv{b}  \, \mathsf{in} \,   \mathsf{let} \, \langle  \SYSTEMmv{x} ,  \SYSTEMmv{y}  \rangle =  \SYSTEMmv{p}  \, \mathsf{in} \,   \langle   \textcolor{coeffectColor}{[}  \SYSTEMmv{x}  \textcolor{coeffectColor}{]}  ,   \textcolor{coeffectColor}{[}  \SYSTEMmv{y}  \textcolor{coeffectColor}{]}   \rangle     :    \textcolor{coeffectColor}{\square_{ \SYSTEMnt{r} } }  \SYSTEMnt{A}   \otimes   \textcolor{coeffectColor}{\square_{ \SYSTEMnt{r} } }  \SYSTEMnt{B}   }
        %  {  \emptyset   \vdash_{\textsc{l} }   \lambda  \SYSTEMmv{b}  .  \SYSTEMsym{(}   \mathsf{let} \, \textcolor{coeffectColor}{[}  \SYSTEMmv{p}  \textcolor{coeffectColor}{]} =  \SYSTEMmv{b}  \, \mathsf{in} \,   \mathsf{let} \, \langle  \SYSTEMmv{x} ,  \SYSTEMmv{y}  \rangle =  \SYSTEMmv{p}  \, \mathsf{in} \,   \langle   \textcolor{coeffectColor}{[}  \SYSTEMmv{x}  \textcolor{coeffectColor}{]}  ,   \textcolor{coeffectColor}{[}  \SYSTEMmv{y}  \textcolor{coeffectColor}{]}   \rangle     \SYSTEMsym{)}   :     \textcolor{coeffectColor}{\square_{ \SYSTEMnt{r} } }   (   \SYSTEMnt{A}  \otimes  \SYSTEMnt{B}   )    \multimap   \textcolor{coeffectColor}{\square_{ \SYSTEMnt{r} } }  \SYSTEMnt{A}    \otimes   \textcolor{coeffectColor}{\square_{ \SYSTEMnt{r} } }  \SYSTEMnt{B}   }
      }
\end{gather*}
At the points marked with question marks (?), we cannot apply promotion (\SYSTEMRenameRuleLinpr{}) to
$x$ and $y$, because the \textit{graded} predicate in the
antecedent does not hold; they are linear variables. Further, to apply \SYSTEMRenameRuleLinder{} at
the point marked with an asterisk (*), we would have to restrict to $r = 1$.
A similar attempted derivation for $  \textcolor{GLcolor}{\llbracket}  \SYSTEMmv{g}  \textcolor{GLcolor}{\rrbracket}  _\circ $, commuting the graded modality
and sum types, encounters the same issues.

One may wonder if any sound translation is possible at all. \citet{hughes2021linear} have explored
precisely this point and show that linear-base and graded-base systems exhibit a fundamental
difference in expressivity. Linear Core is akin to Linear Logic, in that it does
not admit derivation of the distributive law $!(A \otimes B) \multimap \, !A \otimes !B$ as the
semantics of multiplicative tensor $\otimes$ is that each component must be used equally; such a
distributive law would allow $A$ and $B$ to be decoupled and used differently. On the other hand,
Graded Base and the graded-base systems in the literature cited above admit such a distributive law
as seen with $f$. Thus Graded Base cannot model Linear Logic's $!$-modality exactly through \emph{any}
choice of semiring for the graded modality.

\citet{hughes2021linear} propose extensions to graded systems to bring back the restrictions of
Linear Logic. Here we explore a different route, namely we extend Linear Core by further constructs
which admit the augmented expressivity of Graded Modal Core, giving the system \textbf{Linear~Push~Core}.

%We could `drag back' the expressivity of Graded Base by requiring that
%$r = 1$ in the product and sum eliminations, however even that is
%still more general.

\subsection{Linear Push Core}
\label{sec:linear-push-core}
We extend Linear Core to \textbf{Linear~Push~Core} with new syntax, heaving it to a level of
expressivity on par with Graded Modal Core, leaving behind the restrictions of Linear Logic
discussed in \ref{sec:no-trans-graded-core-to-linear-core}.

Since the essential missing piece is the distributive laws between the graded modality and
products and sums, we complete the syntax of Linear Push Core with the following two constructs:
\begin{align*}
  \SYSTEMnt{t} & ::= ... \mid  \textsf{push}_\otimes  \SYSTEMnt{t}  \mid  \textsf{push}_\oplus  \SYSTEMnt{t}  \mid  \textsf{push}_{\mathrm{unit} }  \SYSTEMnt{t} 
  \tag{terms}
\end{align*}
The syntax is non-standard but isomorphic to other approaches; Granule achieves the same
expressivity implicitly via pattern matching~\cite{DBLP:journals/pacmpl/OrchardLE19}, but
we prefer to introduce explicit syntax here. These two constructs provide exactly
the distributive behaviour captured in product and sum elimination in Graded Modal Core.
Figure~\ref{fig:linear-core-push} gives the typing and operational model.

\begin{figure}[t]
\raggedright\noindent\framebox{$ \Gamma  \vdash_{\textsc{l} }  \SYSTEMnt{t}  :  \SYSTEMnt{A} $} \quad Typing rules
\begin{gather*}
\begin{align*}
\begin{array}{c}
\SYSTEMdruleLinpushprod{}
\,
\SYSTEMdruleLinpushsum{}
\,
\SYSTEMdruleLinpushunit{}
\end{array}
\end{align*}
\end{gather*}
\bigskip

\raggedright\noindent\framebox{$ \SYSTEMnt{t}  \rightsquigarrow_{\textsc{l} }  \SYSTEMnt{t'} $} \quad Operational semantics for push constructors
\begin{gather*}
\begin{align*}
\begin{array}{c}
    \SYSTEMdruleSemLinpushProd{}
   \;\;
   \SYSTEMdruleSemLinpushUnit{}
   \;\;
   \SYSTEMdruleSemLinpushSumInjGen{}
  \\[1em]
  \SYSTEMdruleSemLinpushProdCong{}
  \;
  \SYSTEMdruleSemLinpushSumCong{}
  \;
  \SYSTEMdruleSemLinpushUnitCong{}
  \\[1em]
  \SYSTEMdruleSemLinpushProdBoxCong{}
  \;
  \SYSTEMdruleSemLinpushSumBoxCong{}
  \;
  \SYSTEMdruleSemLinpushUnitBoxCong{}
  \end{array}
\end{align*}
\end{gather*}
\caption{Linear Push Core}
\label{fig:linear-core-push}
\end{figure}

\subsection{Translation from Graded Modal Core to Linear Push Core}
\label{sec:trans-graded-core-to-linear-push-core}
We define the translation
from Graded Modal Core into Linear Push Core as follows
(fixing the unsound translation to Linear Core from Section~\ref{sec:no-trans-graded-core-to-linear-core}
in the product and sum elimination):
\renewcommand{\triangleq}{& \oldTriangleq &}
\renewcommand{\equiv}{& \oldEquiv &}
\renewcommand{\SYSTEMdrule}[4][]{#3 \qquad #2}
\begin{gather*}
{\setlength{\arraycolsep}{0.1em}
\begin{array}{lcl@{\hspace{-4em}}r}
\SYSTEMdruleTranslGtoLTmprodi{} \\
 \textcolor{GLcolor}{\llbracket}   \mathsf{let} \, \langle  \SYSTEMmv{x} ,  \SYSTEMmv{y}  \rangle =  \SYSTEMnt{t_{{\mathrm{1}}}}  \, \mathsf{in} \,  \SYSTEMnt{t_{{\mathrm{2}}}}   \textcolor{GLcolor}{\rrbracket}  \triangleq  \mathsf{let} \, \langle  \SYSTEMmv{x'} ,  \SYSTEMmv{y'}  \rangle =   \textsf{push}_\otimes   \textcolor{coeffectColor}{[}   \textcolor{GLcolor}{\llbracket}  \SYSTEMnt{t_{{\mathrm{1}}}}  \textcolor{GLcolor}{\rrbracket}   \textcolor{coeffectColor}{]}    \, \mathsf{in} \,   \mathsf{let} \, \textcolor{coeffectColor}{[}  \SYSTEMmv{x}  \textcolor{coeffectColor}{]} =  \SYSTEMmv{x'}  \, \mathsf{in} \,   \mathsf{let} \, \textcolor{coeffectColor}{[}  \SYSTEMmv{y}  \textcolor{coeffectColor}{]} =  \SYSTEMmv{y'}  \, \mathsf{in} \,   \textcolor{GLcolor}{\llbracket}  \SYSTEMnt{t_{{\mathrm{2}}}}  \textcolor{GLcolor}{\rrbracket}     \\
\SYSTEMdruleTranslGtoLTmsumiOne{} \\
\SYSTEMdruleTranslGtoLTmsumiTwo{} \\
  \textcolor{GLcolor}{\llbracket}   \mathsf{case} \,  \SYSTEMnt{t}  \, \mathsf{of} \, \{ \mathsf{inj1} \,  \SYSTEMmv{x}  \rightarrow  \SYSTEMnt{t_{{\mathrm{1}}}}  ; \, \mathsf{inj2} \,  \SYSTEMmv{y}  \rightarrow  \SYSTEMnt{t_{{\mathrm{2}}}}  \}   \textcolor{GLcolor}{\rrbracket}  \\
\multicolumn{3}{l}{\hspace{0.2em}\;\quad \oldTriangleq \;  \mathsf{case} \,   \textsf{push}_\oplus   \textcolor{coeffectColor}{[}   \textcolor{GLcolor}{\llbracket}  \SYSTEMnt{t}  \textcolor{GLcolor}{\rrbracket}   \textcolor{coeffectColor}{]}    \, \mathsf{of} \, \{ \mathsf{inj1} \,  \SYSTEMmv{x'}  \rightarrow   \mathsf{let} \, \textcolor{coeffectColor}{[}  \SYSTEMmv{x}  \textcolor{coeffectColor}{]} =  \SYSTEMmv{x'}  \, \mathsf{in} \,   \textcolor{GLcolor}{\llbracket}  \SYSTEMnt{t_{{\mathrm{1}}}}  \textcolor{GLcolor}{\rrbracket}    ; \, \mathsf{inj2} \,  \SYSTEMmv{y'}  \rightarrow   \mathsf{let} \, \textcolor{coeffectColor}{[}  \SYSTEMmv{y}  \textcolor{coeffectColor}{]} =  \SYSTEMmv{y'}  \, \mathsf{in} \,   \textcolor{GLcolor}{\llbracket}  \SYSTEMnt{t_{{\mathrm{2}}}}  \textcolor{GLcolor}{\rrbracket}    \}  }\\
\SYSTEMdruleTranslGtoLTmuniti{} \\
\SYSTEMdruleTranslGtoLTmunite{} & \text{(Term translation)} \\[0.5em]
\SYSTEMdruleTranslGtoLTyprodTy{} \\
\SYSTEMdruleTranslGtoLTysumTy{} \\
\SYSTEMdruleTranslGtoLTyunitTy{} & \text{(Type translation)}
\end{array}}
\end{gather*}
\let\SYSTEMdrule\oldSYSTEMdrule
\let\equiv\oldEquiv
\let\triangleq\oldTriangleq
Translation of a product elimination on $\SYSTEMnt{t_{{\mathrm{1}}}}$ first promotes
the product to a graded modality which is then `pushed into' each component of the product via $\mathsf{push}_\otimes$. The resulting graded modal components $x'$ and $y'$ are then eliminated in the scope of $\SYSTEMnt{t_{{\mathrm{2}}}}$. A similar idea is applied for sum elimination. Thus this translation
gets at the heart of what is happening inside Graded Modal Core (and why it differs from
Linear Core): its eliminators distribute the coeffect analysis inside products and sums,
which we now capture explicitly here in Linear Push Core.

We now re-establish the core results (proofs for which are
in Appendix~\refappendix{app:proofs-grad-core-to-lin-core}):
\begin{restatable}{thrm}{gradToLinTranslationExt}[Soundness of the Graded Modal Core to Linear Push Core translation]
 \begin{itemize}[itemsep=0.2em]
  \item Type preservation: $ \Delta  \vdash_{\textsc{g} }  \SYSTEMnt{t}  :  \SYSTEMnt{A}  \implies   \textcolor{GLcolor}{\llbracket}  \Delta  \textcolor{GLcolor}{\rrbracket}   \vdash_{\textsc{l} }   \textcolor{GLcolor}{\llbracket}  \SYSTEMnt{t}  \textcolor{GLcolor}{\rrbracket}   :   \textcolor{GLcolor}{\llbracket}  \SYSTEMnt{A}  \textcolor{GLcolor}{\rrbracket}  $;
  \item Operational correspondence: $ \SYSTEMnt{t}  \rightsquigarrow_{\textsc{g} }  \SYSTEMnt{t'}  \implies   \textcolor{GLcolor}{\llbracket}  \SYSTEMnt{t}  \textcolor{GLcolor}{\rrbracket}   \rightsquigarrow_{\textsc{l} }^\ast   \textcolor{GLcolor}{\llbracket}  \SYSTEMnt{t'}  \textcolor{GLcolor}{\rrbracket}  $.
  \item Equation preservation: $(\SYSTEMnt{t}  \equiv^{-\eta\otimes\oplus\mathsf{unit} }_{\textsc{g} }  \SYSTEMnt{t'}) \implies   \textcolor{GLcolor}{\llbracket}  \SYSTEMnt{t}  \textcolor{GLcolor}{\rrbracket}   \,\equiv_{\textsc{l} }\,   \textcolor{GLcolor}{\llbracket}  \SYSTEMnt{t'}  \textcolor{GLcolor}{\rrbracket}  $.
\end{itemize}
Our translation does not, in general, preserve $\eta$-equalities for
products, sums, and unit (it can only preserve $\eta$ on these types
for closed terms); i.e., our translation is not extensional due to
additional `push'-structure for handling nesting of graded modalities
and products, sums, and units.
\end{restatable}
Given that we have changed Linear Core to Linear Push Core by adding two new constructs,
we now re-establish that we can translate Linear Push Core back into Graded Modal Core.

\subsection{Translation from Linear Push Core to Graded Modal Core}
\label{sec:trans-linear-push-core-to-graded-core}
We extend the Linear Core to Graded Modal Core translation with the following
additional cases to handle the extra constructs of Linear Push Core:
\renewcommand{\equiv}{ \oldEquiv}
\renewcommand{\SYSTEMdrule}[4][]{{\displaystyle{#3}}}
\begin{align}
\tag{Term translation} \\[-1.5em]
\notag & \begin{array}{l}
\SYSTEMdruleTranslLtoGTmpushProd{} \\
\SYSTEMdruleTranslLtoGTmpushSum{} \\
\SYSTEMdruleTranslLtoGTmpushUnit{}
\end{array}
\end{align}
\let\SYSTEMdrule\oldSYSTEMdrule
\let\equiv\oldEquiv
Note that (for $ \textsf{push}_\otimes     $ and $ \textsf{push}_\oplus     $) these are essentially the programs $f$ and $g$ shown in Section~\ref{sec:no-trans-graded-core-to-linear-core}
which are the derivation of $\textsf{push}$ combinators in Graded Modal Core.

We thus (re)establish the key results (proofs in Appendix~\refappendix{app:proofs-lin-core-to-grad-core}):

\begin{restatable}{thrm}{linToGradTranslationExt}[Soundness of the Linear Push Core to Graded Modal Core translation]
 \begin{itemize}[itemsep=0.2em]
  \item Type preservation: $ \Gamma  \vdash_{\textsc{l} }  \SYSTEMnt{t}  :  \SYSTEMnt{A}  \implies   \textcolor{LGcolor}{\llparenthesis}  \Gamma  \textcolor{LGcolor}{\rrparenthesis}   \vdash_{\textsc{g} }   \textcolor{LGcolor}{\llparenthesis} \smidge  \SYSTEMnt{t}  \smidge \textcolor{LGcolor}{\rrparenthesis}   :   \textcolor{LGcolor}{\llparenthesis} \smidge  \SYSTEMnt{A}  \smidge \textcolor{LGcolor}{\rrparenthesis}  $;
  \item Operational correspondence: $ \SYSTEMnt{t}  \rightsquigarrow_{\textsc{l} }  \SYSTEMnt{t'}  \implies   \textcolor{LGcolor}{\llparenthesis} \smidge  \SYSTEMnt{t}  \smidge \textcolor{LGcolor}{\rrparenthesis}   \rightsquigarrow_{\textsc{g} }^\ast   \textcolor{LGcolor}{\llparenthesis} \smidge  \SYSTEMnt{t'}  \smidge \textcolor{LGcolor}{\rrparenthesis}  $;
  \item Equation preservation: $ \SYSTEMnt{t}  \,\equiv_{\textsc{l} }\,  \SYSTEMnt{t'}  \implies   \textcolor{LGcolor}{\llparenthesis} \smidge  \SYSTEMnt{t}  \smidge \textcolor{LGcolor}{\rrparenthesis}   \equiv_{\textsc{g} }   \textcolor{LGcolor}{\llparenthesis} \smidge  \SYSTEMnt{t'}  \smidge \textcolor{LGcolor}{\rrparenthesis}  $.
\end{itemize}
\end{restatable}
Note that whilst the operational correspondence from Linear Base to Graded Modal Base mapped a single
reduction $ \SYSTEMnt{t}  \rightsquigarrow_{\textsc{l} }  \SYSTEMnt{t'} $ to a single graded-base reduction $  \textcolor{LGcolor}{\llparenthesis} \smidge  \SYSTEMnt{t}  \smidge \textcolor{LGcolor}{\rrparenthesis}   \rightsquigarrow_{\textsc{g} }   \textcolor{LGcolor}{\llparenthesis} \smidge  \SYSTEMnt{t'}  \smidge \textcolor{LGcolor}{\rrparenthesis}  $ this
is now not possible for Linear Push Core. Instead, a single reduction in Linear Push Core for
the $\mathsf{push}$ combinators corresponds to several reductions in Graded Modal Core.

\begin{remark}
Graded Modal Core extends Graded Modal Base rather than Graded Base.
If Graded Modal Core did not have a graded modality it could not give a precise type to projection on products:
\begin{align*}
  \emptyset   \vdash_{\textsc{g} }  \SYSTEMsym{(}   \lambda  \SYSTEMmv{z}  .   \mathsf{let} \, \langle  \SYSTEMmv{x} ,  \SYSTEMmv{y}  \rangle =  \SYSTEMmv{z}  \, \mathsf{in} \,  \SYSTEMmv{x}    \SYSTEMsym{)}  :    (   \SYSTEMnt{A}  \times  \SYSTEMnt{B}   )   \xrightarrow{\textcolor{coeffectColor}{ \SYSTEMnt{r} } }  \SYSTEMnt{A}  
\end{align*}
where $ \SYSTEMsym{1}  \, \textcolor{coeffectColor}{\sqsubseteq} \,  \SYSTEMnt{r} $ due to the use of $x$ and $ \SYSTEMsym{0}  \, \textcolor{coeffectColor}{\sqsubseteq} \,  \SYSTEMnt{r} $ due to the
absence of $y$.  For the exact-usage natural numbers semiring this
would not be typeable as there is no such $r$ satisfying both
constraints. For a semiring of intervals denoting lower-bound and
upper-bound usage~\cite{DBLP:journals/pacmpl/OrchardLE19}, the
constraints would be satisfied by $r = [0,1]$.

We instead opted to include the graded modality in Graded Modal Core as in the work of \citet{DBLP:journals/pacmpl/AbelB20}, \citet{grtt}, and others~\cite{DBLP:conf/esop/HughesO24,mycroftfest2024}.
Thus we can specify for this example a more fine-grained type via the graded modality,
where the $\SYSTEMnt{B}$ component of the product is unused:
\[
    \emptyset   \vdash_{\textsc{g} }   \lambda  \SYSTEMmv{z}  .   \mathsf{let} \, \langle  \SYSTEMmv{x} ,  \SYSTEMmv{y}  \rangle =  \SYSTEMmv{z}  \, \mathsf{in} \,   \mathsf{let} \, \textcolor{coeffectColor}{[}  \SYSTEMmv{y'}  \textcolor{coeffectColor}{]} =  \SYSTEMmv{y}  \, \mathsf{in} \,  \SYSTEMmv{x}     :    (   \SYSTEMnt{A}  \times   \textcolor{coeffectColor}{\square_{ \SYSTEMsym{0} } }  \SYSTEMnt{B}    )   \xrightarrow{\textcolor{coeffectColor}{ \SYSTEMsym{1} } }  \SYSTEMnt{A}  
\]
Without the inclusion of graded modalities, Graded Modal Core is less expressive than Linear Core.

Note that some dependently-typed graded-base systems such as \textsc{GraD}~\cite{grad}
and QTT~\cite{quantitative-type-theory} omit the graded modality but include a graded
dependent $\Sigma$ type which can be used to encode the graded modality.
\end{remark}

\section{Discussion and Conclusion}
\label{sec:discussion}
\subsection{Is one approach better than the other?}
\label{sec:better}
A natural question at this point from the language designer's perspective is: which approach to
linearity should I use in my language? We do not try to offer an authoritative answer, but instead give
some observations and points to consider.

Clearly, graded-base style approaches afford a simpler system for the end-user with less syntactic overhead. The
pervasive grading obviates the need for most promotions and \textsf{let}-unboxings.
But while having explicit \textsf{push} syntax is an additional burden to the user of a
linear-base system, in practice this is not necessary. For example, Granule handles this implicitly
via pattern matching~\cite{DBLP:journals/pacmpl/OrchardLE19, hughes2021linear}.

On the flip-side, it may be desirable to disallow \textsf{push} for reasoning about certain
domains, where a coeffect should not distribute over the components of a pair. An example for this
might be for reasoning about interdependent threads that need to run in parallel. Given what we
have shown in this paper, linear base systems would seem better equipped to offer this more restrictive
modality, however \citet{hughes2021linear} have shown that it is possible to recover the non-distributive behaviour of the $\Box$-modality with respect to products in a graded-base setting.
They argue that any useful language implementation would reasonably have to offer both flavours of
the modality.

What about the burden for the language designer? Empirically, the authors have found that proofs for
graded-base systems are significantly shorter, due, in part, to the homogeneous treatment of
assumptions, i.e. the lack of split between linear and non-linear free variables. Especially
when retrofitting linear or coeffect types to an existing language, only Graded Base offers a
backwards-compatible way and in such settings this is the only practical option (see, e.g.,
the addition of graded types to Haskell for linearity~\cite{DBLP:journals/pacmpl/BernardyBNJS18}).

\subsection{Graded Type Systems in Practice}
\label{sec:wild}
The \haskin{LinearTypes} language extension available since GHC 9.0.1~\cite{GHCLinearTypes}
enables Linear Haskell in the context of an
industrial-strength programming language. As discussed in this paper, Linear Haskell is a graded-base system, which is currently semiring-monomorphic: in the theory
of \citet{DBLP:journals/pacmpl/BernardyBNJS18} the \textit{none-one-tons} semiring is used,
$\{0, 1, \omega\}$. However, as of the time of writing,
the implementation in GHC offers no $0$, and
thus the grades range over $\{1, \omega\}$, written in Haskell as \haskin{One} and \haskin{Many}, respectively.
Consider the following Linear Haskell program:
\begin{haskell}
{-# LANGUAGE LinearTypes #-}
{-# LANGUAGE DataKinds #-}
import GHC.Types
const :: a %One -> b %Many -> a
const x y = x
\end{haskell}
By default arrows are \haskin{Many} graded, thus the above can also be written
as:
\begin{haskell}[firstnumber=last]
const' :: a %One -> b -> a
const' x y = x
\end{haskell}
Notably, Linear Haskell supports \textit{multiplicity polymorphism}, however the multiplicities
are inhabited by a single semiring.
Future work would be a \haskin{GradedTypes} extension to generalise Haskell to be polymorphic over
semirings, a feature supported by Granule, where \citet{DBLP:journals/pacmpl/OrchardLE19} use an
SMT-solver to discharge constraints over the grades. It is not yet clear how this could be
practically integrated with GHC, perhaps as a compiler plugin akin to Liquid Haskell \cite{Vazou2020LiquidHaskell}.

Idris 2 also implements a graded-base system, based on Quantitative
Type Theory (due initially to \citet{McBride2016} and later adapted
slightly by \citet{quantitative-type-theory}), also providing
grades in the \textit{none-one-tons} semiring~\cite{DBLP:conf/ecoop/Brady21,DBLP:journals/darts/Brady21}. For example, \haskin{const} can be rendered in Idris 2:
\begin{haskell}
const : {0 a : Type} -> {0 b : Type} -> a -> (0 x : b) -> a
const x y = x
\end{haskell}

Granule is a research language serving as a test bed for ideas
in graded types~\cite{DBLP:journals/pacmpl/OrchardLE19}. In our classification,
Granule is a linear-base system by default.

Semiring-graded modalities written here as $ \textcolor{coeffectColor}{\square_{ \SYSTEMnt{r} } }  \SYSTEMnt{A} $ in the
core calculi are written as `\granin{a [r]}' in Granule.
While Granule uses \granin{->}  for the function arrow (\texttt{->} in ASCII), as in Haskell or ML, it is really a linear arrow.
Our running example looks as follows in Granule:
\begin{granule}
const : forall {a : Type, b : Type, s : Semiring} . a -> b [0 : s] -> a
const x y = let [_] = y in x
\end{granule}
Granule also includes a language extension that changes its semantics to Graded
Base and whose syntax for graded arrows is borrowed from Linear Haskell, but supporting
semiring polymorphism:
\begin{granule}
language GradedBase
const' : forall {a : Type, b : Type, s : Semiring} . a -> b % (0 : s) -> a
const' x y = x
\end{granule}
%
%The graded-base approach has been explored as a target
%for program synthesis~\cite{DBLP:conf/esop/HughesO24}.

\subsection{Comparison with Effects}
\label{sec:comparison-with-effects}
Here we have considered the relationship between Linear Base
with linear function spaces and linear assumptions as well
as graded, and Graded Base where grades are pervasive. This
situation is somewhat analogous to the relationship between
\emph{effect systems} and \emph{monads}.

The type-and-effect systems of Gifford and
Lucassen~[\citeyear{gifford1986integrating,lucassen1988polymorphic}]
provide a technique for augmenting a type system with information
about the side effects a program may perform whilst computing
a result. Judgments and function arrows carry effect information:
\begin{align*}
\Gamma \vdash M : \tau, F \qquad\quad
\Gamma' \vdash \lambda x . M' : \sigma \xrightarrow{F'} \tau, \emptyset
\end{align*}
i.e., the program $M$ computes a result of type $\tau$ and may perform side effects described by
$F$. Traditionally $F$ is a set of effect `labels', describing
operations that have been performed.  This was later generalised to
semilattices and ordered
monoids~\cite{nielson1999type,DBLP:conf/birthday/MycroftOP16},
enabling other kinds of analysis. On the right,
the function term encapsulates side effects $F'$ and is itself pure,
marked with $\emptyset$. This is analogous to Graded Base, where coeffect grades annotate function types
(instead representing usage of the parameter).

Monads on the other hand provide a semantic technique for giving a
denotational model of various kinds of
effect~\cite{DBLP:journals/iandc/Moggi91,DBLP:conf/lics/Moggi89}.
The \emph{monadic meta language} provides a type constructor $T$
where $T \tau$ captures the type of computations performing some side
effects in order to compute a $\tau$ value~\cite{DBLP:journals/iandc/Moggi91}.
Thus, terms of the monadic meta language have types
of the form:
\begin{align*}
\Gamma \vdash M : T \tau \qquad\quad
\Gamma \vdash \lambda x . M' : \sigma \rightarrow T \tau
\end{align*}
On the right is an example of a function that encapsulates
side effects in its result, but the function itself does not need
wrapping in the monadic $T$ constructor since abstraction is pure
(side-effect free). This is analogous to Linear Base if we had
only a single comonadic modality.

\citet{wadler2003marriage} observed that the structure of effect systems
and the structure of the monadic meta language are the same.
They thus married the two by annotating the monadic constructor
with effect information, i.e.,
\begin{align*}
\Gamma \vdash M : T_F \tau \qquad\quad
\Gamma' \vdash \lambda x . M' : T_\emptyset (\sigma \rightarrow T_{F'} \tau)
\end{align*}
Later, this indexed structure was captured generally by the notion of
\emph{graded monads}~\cite{DBLP:conf/popl/Katsumata14,DBLP:journals/corr/OrchardPM14}. Graded
monads generalise monads to a family of constructors indexed by the
elements of a monoid, and whose operations are stratified and
accounted for by the monoidal structure. Graded monads thus unify
the semantics of effects with a syntactic analysis in the types.

Linear Base with graded modality $ \textcolor{coeffectColor}{\square_{ \SYSTEMnt{r} } }  \SYSTEMnt{A} $ is thus a dual
notion to the graded monadic calculus, unifying the semantics of
coeffects with a static analysis in the types, but with linearity
as its base typing system. We could thus see Linear Base as a \emph{graded comonadic meta language}.
In contrast, Graded Base is analogous to effect systems. Indeed the early
work on \emph{coeffect systems} (captured here as Graded Base) has no graded modality and was motivated
by dualising effect systems~\cite{DBLP:conf/icalp/PetricekOM13,petricek2014coeffects}.
As seen, without this graded modality, Graded Base cannot localise information
to sub-parts of complex types as discussed in Section~\ref{sec:prods-sums}.

\subsection{Adjoint Relationship}

\citet{DBLP:journals/corr/abs-2401-17199} follow
the Mixed Linear and Non-Linear logic approach of \citet{DBLP:conf/csl/Benton94}
to provide a pair of graded calculi with an adjunction mediating between them.
One calculus is a linear-base system with linear functions
and a mixed context of linear and graded assumptions. The other calculus
is a graded-base system but with no graded function arrow. A modality
$\mathsf{Lin}$ embeds formulae from the linear-base system into the graded-base. Conversely, a graded modality
$\mathsf{Grd}_r$ embeds formulae from the graded-base system into
the linear-base system, capturing their grade $r$ and incurring a promotion (scaling of
the graded context). From $\mathsf{Lin}$ and
$\mathsf{Grd}_r$ a graded comonadic modality is derived following the adjoint resolution of graded
comonads~\cite{DBLP:conf/fossacs/FujiiKM16}. %They then study the categorical semantics for this system.

The work seems related but on a more semantic level. Whilst we showed
that our translation is not adjoint in Section~\ref{sec:transl}, our two calculi have
function types on both sides which seemed to be the main blocker to forming an adjunction.
Further work is to ascertain whether our translation could
be restricted to the systems considered by \citet{DBLP:journals/corr/abs-2401-17199} and then used
as a model of the adjunction by using our
translations as the semantics of the modalities operating on
syntactic categories of terms. Another avenue
would be to explore whether generalising function spaces on the graded-base
side would undermine their adjunction construction in general.

\subsection{Filling Out the Design Space: Cartesian Base}
\label{sec:cartesian-base}

A central idea in this work has been to consider the `base' notion of usage in
graded systems: linear-base systems have linear propositions, with a linear
axiom rule, and linear function type; graded-base systems have graded propositions,
with a graded axiom rule, and a graded function type. This framing suggest
an unexplored point in the design space: what about a graded system with
a cartesian base, i.e., with a cartesian axiom rule and standard cartesian
function (type), with contexts comprising both cartesian and graded
assumptions?

Crucially we must then consider how the cartesian part interacts with
the graded part. On the left, we have the rules for dereliction and
application in Linear Base, on the right have the potential rules
for a Cartesian Base. Since cartesian propositions can be used arbitrarily
we now require an element $\omega \in \mathcal{R}$ of the semiring which
behaves in an absorbing fashion for both addition $+$ and multiplication $\cdot$.
Dereliction in Cartesian Base can then mark a cartesian assumption
with $\omega$. Application must also scale the argument context $\Gamma_2$
by $\omega$ since the function arrow is cartesian and so may use
its argument arbitrarily:

\begin{minipage}{0.5\linewidth}
\begin{align*}
\begin{array}{c}
\text{(Linear Base)} \\[0.5em]
\dfrac{\Gamma, x : A \vdash_{\textsc{l}} t : B}
      {\Gamma, x : [A]_1 \vdash_{\textsc{l}} t : B} (der)
\\[1.5em]
\dfrac{\Gamma_1 \vdash_{\textsc{l}} t_1 : A \multimap B \quad \Gamma_2 \vdash_{\textsc{l}} t_2 : A}
      {\Gamma_1 + \Gamma_2 \vdash_{\textsc{l}} t_1\ t_2 : B}(app)
\end{array}
\\
\end{align*}
\end{minipage}
\begin{minipage}{0.5\linewidth}
\begin{align*}
\begin{array}{c}
\text{(Cartesian Base)} \\[0.5em]
\dfrac{\Gamma, x : A \vdash t : B}
      {\Gamma, x : [A]_\omega \vdash t : B}(der)
\\[1.5em]
\dfrac{\Gamma_1 \vdash t_1 : A \rightarrow B \quad \Gamma_2 \vdash t_2 : A}
      {\Gamma_1 + \omega \ast \Gamma_2 \vdash t_1\ t_2 : B}(app)
\end{array}
\\
\end{align*}
\end{minipage}

\noindent
The above seems a reasonable approach to building a
Cartesian Base system. However, this system
provides little benefit as the grading quickly degenerates to
being always $\omega$ in derivations. Consider below the derivation of
the `copy' operation on an axiom to produce a pair:

\begin{align*}
\begin{array}{cc}
\dfrac{
\dfrac{x : A \vdash_{\textsc{l}} x : A}
      {x : [A]_1 \vdash_{\textsc{l}} x : A}
\quad
\dfrac{x : A \vdash_{\textsc{l}} x : A}
      {x : [A]_1 \vdash_{\textsc{l}} x : A}
}
{x : [A]_{1 + 1} \vdash_{\textsc{l}} (x, x) : A \otimes A}
&
\dfrac{
\dfrac{x : A \vdash x : A}
      {x : [A]_\omega \vdash x : A}
\quad
\dfrac{x : A \vdash x : A}
      {x : [A]_\omega \vdash x : A}
}
{x : [A]_\omega \vdash (x, x) : A \otimes A}
\end{array}
\end{align*}
In Linear Base, there is a precise account of the non-linearity
incurred: the term uses $x$ with graded $1+1$. However, in
Cartesian Base we can only say $\omega$ due to the dereliction rule.
In the end, the Cartesian Base system
is ultimately no more informative as a graded system than
standard intuitionistic, cartesian systems.

\subsection{Resource-Aware Operational Semantics}
\label{sec:raos}

The operational semantics presented here leverage standard $\lambda$-calculus CBN operational semantics,
which are typical for the literature studied.
More recent work has developed \emph{resource-aware} operational
semantics~\cite{grad, DBLP:journals/pacmpl/BianchiniDGZ23, DBLP:journals/corr/abs-2311-11795, DBLP:conf/ecoop/BianchiniDGZ23}. These approaches are based on an abstract machine
approach with a variable environment, rather than a substitution-based model.
Conceptually, a resource-aware semantics keeps track not just of variable mappings in the evaluation environment,
but more broadly tracks resource supplies, which are consumed by variable use,
thereby making the connection between static and dynamic resource guarantees:
A \emph{resource-sound} type system only accepts \emph{well-resourced} programs,
whose evaluation under the resource-aware operational semantics respects the static resource bounds,
meaning (1) evaluation does not get stuck due to depletion (\emph{resource progress});
and (2) every evaluation step leads to a program with comparable resource requirements
under the (transitive) approximation relation ($\sqsubseteq$) of the resource algebra (\emph{resource conservation}).

Most recently, \citet{DBLP:journals/corr/abs-2507-13792} extend these guarantees to the converse property,
ensuring both upper and lower resource bounds, thereby also preventing unused resources (\emph{wastefulness}).

Extending the various flavours of linear- and graded-base calculi and their translations,
which form the substrate of this paper, to such resource-aware operational semantics goes
beyond the scope of what we are able to investigate here. Instead, we leave this as future work.

\subsection{Conclusion}
\label{sec:conclusion}

Our goal here has been to serve as a bridge between two different approaches
in the literature for constructing graded type systems that capture coeffects (context dependence).
We answer the long open question about the relationship between
the linear-base and graded-base styles and give a detailed analysis and metatheoretic results. Overall we see that we can translate between
the two flavours: with Linear Base and Graded Base having mutual embeddings
(but requiring a global CPS encoding for the Linear Base into Graded Base direction);
Linear Base and Graded Modal Base having (more straightforward) mutual embeddings;
and Linear Push Core and Graded Modal Core having mutual embeddings: these are the points
where the two styles of calculi can be inter-translated (see Figure~\ref{fig:relationship}).

Now that graded types for tracking various kinds of program property are moving
from the laboratory to the field, with Haskell~\cite{DBLP:journals/pacmpl/BernardyBNJS18}, Idris~\cite{DBLP:conf/ecoop/Brady21}, Granule~\cite{DBLP:journals/pacmpl/OrchardLE19}, and lately OCaml with its moding system~\cite{DBLP:journals/pacmpl/LorenzenWDEL24,DBLP:journals/pacmpl/GeorgesPEWDECPD25},
our hope is that this study can serve as a design guide for
further theoretical work and for language designers wishing to integrate graded types
in some way, drawing from a wider arsenal of techniques and metatheoretic results.

\paragraph{Acknowledgments}

We thank the anonymous reviewers of this work for their comments and feedback.
During its early stages, this work was partly supported by EPSRC grant EP/T013516/1.
Orchard also received support through Schmidt Sciences, LLC.

\bibliography{references}

\ifextended
\newpage
\appendix
\section{Collected and Additional Definitions}
\label{app:definitions}
\paragraph{Syntactic substitution}
\begin{gather*}
  \begin{align*}
\begin{array}{l}
\SYSTEMdrulesubstvarneq{}\\[2em]
\SYSTEMdrulesubstvareq{}\\[2em]
\SYSTEMdrulesubstapp{}\\[2em]
\SYSTEMdrulesubstabs{}\\[2em]
\SYSTEMdrulesubstpr{}\\[2em]
\SYSTEMdrulesubstlet{}\\[2em]
\SYSTEMdrulesubstsumiOne{}\\[2em]
\SYSTEMdrulesubstsumiTwo{}\\[2em]
\SYSTEMdrulesubstsume{}\\[2em]
\SYSTEMdrulesubstprodi{}\\[2em]
\SYSTEMdrulesubstprode{}\\[2em]
\SYSTEMdrulesubstuniti{}\\[2em]
\SYSTEMdrulesubstunite{}
\end{array}
\end{align*}
\end{gather*}

\begin{definition}[Freshness]
  \begin{align*}
    x_1, \ldots, x_n \# t \triangleq \{x_1, \ldots, x_n\} \cap \mathsf{fv}(t) = \emptyset
  \end{align*}
\end{definition}

\subsection{Linear Base}

\begin{align}
    \SYSTEMnt{t} & ::= \SYSTEMmv{x} \mid  \SYSTEMnt{t_{{\mathrm{1}}}} \,  \SYSTEMnt{t_{{\mathrm{2}}}}  \mid  \lambda  \SYSTEMmv{x}  .  \SYSTEMnt{t}  \mid  \textcolor{coeffectColor}{[}  \SYSTEMnt{t}  \textcolor{coeffectColor}{]}  \mid  \mathsf{let} \, \textcolor{coeffectColor}{[}  \SYSTEMmv{x}  \textcolor{coeffectColor}{]} =  \SYSTEMnt{t_{{\mathrm{1}}}}  \, \mathsf{in} \,  \SYSTEMnt{t_{{\mathrm{2}}}} 
    \tag{terms} \\
  \SYSTEMnt{A} & ::=  \SYSTEMnt{A}  \multimap  \SYSTEMnt{B}  \mid  \textcolor{coeffectColor}{\square_{ \SYSTEMnt{r} } }  \SYSTEMnt{A}  \mid  \mathrm{K} 
  \tag{types} \\
    \tag{contexts}
  \Gamma & ::=  \emptyset  \mid  \Gamma ,   \SYSTEMmv{x}  :  \SYSTEMnt{A}   \mid  \Gamma ,   \SYSTEMmv{x}  : \textcolor{coeffectColor}{[}  \SYSTEMnt{A} {\textcolor{coeffectColor}{]_{ \SYSTEMnt{r} } } }  
\end{align}

\paragraph{Typing}

\begin{gather*}
\begin{align*}
  \begin{array}{c}
\SYSTEMdruleLinvar{}
\quad
\SYSTEMdruleLinabs{}
\quad
\SYSTEMdruleLinapp{}
\\[2em]
\SYSTEMdruleLinweak{}
\quad
\SYSTEMdruleLinder{}
\quad
\SYSTEMdruleLinpr{}
\\[2em]
\SYSTEMdruleLinlet{}
\qquad
\SYSTEMdruleLinapprox{}
\\
\end{array}
\end{align*}
\end{gather*}

\paragraph{Operational Semantics}

\begin{gather*}
\begin{align*}
  \begin{array}{c}
  \SYSTEMdruleSemLinbeta{}
  \quad
  \SYSTEMdruleSemLincongAppL{}
  \\[2em]
  \SYSTEMdruleSemLinbetaBox{}
  \quad
  \SYSTEMdruleSemLincongLetL{} \\[2em]
  \end{array}
\end{align*}
\end{gather*}

\paragraph{Equational Theory}

  \begin{gather*}
\begin{align*}
  \begin{array}{c}
\SYSTEMdruleLinEqbeta{} \quad
\SYSTEMdruleLinEqeta{} \\[2em]
\SYSTEMdruleLinEqbetaBox{} \quad
\SYSTEMdruleLinEqetaBox{} \\[2em]
\SYSTEMdruleLinEqletCommBox{} \\[2em]
\SYSTEMdruleLinEqletCommOne{} \\[2em]
\SYSTEMdruleLinEqletCommTwo{} \\[2em]
\SYSTEMdruleLinEqcongApp{} \quad
\SYSTEMdruleLinEqcongAbs{} \\[2em]
\SYSTEMdruleLinEqcongPr{} \quad
\SYSTEMdruleLinEqcongLet{}
  \end{array}
\end{align*}
\end{gather*}

\subsection{Linear Core}
\label{app:definitions-linear-core}

\begin{align}
  \notag    \SYSTEMnt{t} & ::= \SYSTEMmv{x} \mid  \SYSTEMnt{t_{{\mathrm{1}}}} \,  \SYSTEMnt{t_{{\mathrm{2}}}}  \mid  \lambda  \SYSTEMmv{x}  .  \SYSTEMnt{t}  \mid  \textcolor{coeffectColor}{[}  \SYSTEMnt{t}  \textcolor{coeffectColor}{]}  \mid  \mathsf{let} \, \textcolor{coeffectColor}{[}  \SYSTEMmv{x}  \textcolor{coeffectColor}{]} =  \SYSTEMnt{t_{{\mathrm{1}}}}  \, \mathsf{in} \,  \SYSTEMnt{t_{{\mathrm{2}}}}  \mid  \langle  \SYSTEMnt{t_{{\mathrm{1}}}} ,  \SYSTEMnt{t_{{\mathrm{2}}}}  \rangle  \mid  \mathsf{let} \, \langle  \SYSTEMmv{x} ,  \SYSTEMmv{y}  \rangle =  \SYSTEMnt{t_{{\mathrm{1}}}}  \, \mathsf{in} \,  \SYSTEMnt{t_{{\mathrm{2}}}}  \mid  \langle \rangle  \mid  \mathsf{let} \, \langle \rangle =  \SYSTEMnt{t_{{\mathrm{1}}}}  \, \mathsf{in} \,  \SYSTEMnt{t_{{\mathrm{2}}}}  \\
         \mid & \; \mathsf{inj}_1 \,  \SYSTEMnt{t}  \mid  \mathsf{inj}_2 \,  \SYSTEMnt{t}  \mid  \mathsf{case} \,  \SYSTEMnt{t}  \, \mathsf{of} \, \{ \mathsf{inj1} \,  \SYSTEMmv{x}  \rightarrow  \SYSTEMnt{t_{{\mathrm{1}}}}  ; \, \mathsf{inj2} \,  \SYSTEMmv{y}  \rightarrow  \SYSTEMnt{t_{{\mathrm{2}}}}  \} 
    \tag{terms} \\
  \SYSTEMnt{A} & ::=  \SYSTEMnt{A}  \multimap  \SYSTEMnt{B}  \mid  \textcolor{coeffectColor}{\square_{ \SYSTEMnt{r} } }  \SYSTEMnt{A}  \mid  \SYSTEMnt{A}  \otimes  \SYSTEMnt{B}  \mid  \mathrm{unit}  \mid  \SYSTEMnt{A}  \oplus  \SYSTEMnt{B} 
  \tag{types} \\
    \tag{contexts}
  \Gamma & ::=  \emptyset  \mid  \Gamma ,   \SYSTEMmv{x}  :  \SYSTEMnt{A}   \mid  \Gamma ,   \SYSTEMmv{x}  : \textcolor{coeffectColor}{[}  \SYSTEMnt{A} {\textcolor{coeffectColor}{]_{ \SYSTEMnt{r} } } }  
\end{align}

\paragraph{Typing}

Linear Base typing plus the following:

\begin{gather*}
\begin{align*}
  \begin{array}{c}
    \SYSTEMdruleLinprodi{}
\quad
\SYSTEMdruleLinprode{}
\\[2em]
\SYSTEMdruleLinuniti{}
\quad
\SYSTEMdruleLinunite{}
\quad
\SYSTEMdruleLinsumiOne{}
\\[2em]
\SYSTEMdruleLinsumiTwo{}
\quad
\SYSTEMdruleLinsume{}
\end{array}
\end{align*}
\end{gather*}

\paragraph{Operational Semantics}

Linear Base operational semantics plus the following:

\begin{gather*}
\begin{align*}
  \begin{array}{c}
    \SYSTEMdruleSemLinprodCong{}
    \;\;
    \SYSTEMdruleSemLinprodBeta{}
    \\[2em]
    \SYSTEMdruleSemLinunitCong{}
    \;\;
    \SYSTEMdruleSemLinunitBeta{}
    \\[2em]
    \SYSTEMdruleSemLincongCase{}
    \\[2em]
    \;\;
    \SYSTEMdruleSemLincaseInjOne{}
    \;\;
    \SYSTEMdruleSemLincaseInjTwo{}
  \end{array}
\end{align*}
\end{gather*}

\paragraph{Equational Theory}

Linear Base equational theory plus the following:

  \begin{gather*}
\begin{align*}
  \begin{array}{c}
     \SYSTEMdruleLinEqbetaUnit{} \quad
\SYSTEMdruleLinEqetaUnit{} \\[2em]
\SYSTEMdruleLinEqbetaProd{}
\SYSTEMdruleLinEqetaProd{} \\[2em]
\SYSTEMdruleLinEqbetaSumOne{} \quad
\SYSTEMdruleLinEqbetaSumTwo{} \\[2em]
\SYSTEMdruleLinEqetaSum{} \\[2em]
    \SYSTEMdruleLinEqcongUnitE{}
\SYSTEMdruleLinEqcongProdE{} \\[2em]
\SYSTEMdruleLinEqcongProdI{}
\SYSTEMdruleLinEqcongSumIOne{}
\SYSTEMdruleLinEqcongSumITwo{} \\[2em]
\SYSTEMdruleLinEqcongSumE{}
  \end{array}
\end{align*}
\end{gather*}

\subsection{Linear Push Core}
\label{app:definitions-linear-push-core}

\begin{align}
  \notag    \SYSTEMnt{t} & ::= \SYSTEMmv{x} \mid  \SYSTEMnt{t_{{\mathrm{1}}}} \,  \SYSTEMnt{t_{{\mathrm{2}}}}  \mid  \lambda  \SYSTEMmv{x}  .  \SYSTEMnt{t}  \mid  \textcolor{coeffectColor}{[}  \SYSTEMnt{t}  \textcolor{coeffectColor}{]}  \mid  \mathsf{let} \, \textcolor{coeffectColor}{[}  \SYSTEMmv{x}  \textcolor{coeffectColor}{]} =  \SYSTEMnt{t_{{\mathrm{1}}}}  \, \mathsf{in} \,  \SYSTEMnt{t_{{\mathrm{2}}}}  \mid  \langle  \SYSTEMnt{t_{{\mathrm{1}}}} ,  \SYSTEMnt{t_{{\mathrm{2}}}}  \rangle  \mid  \mathsf{let} \, \langle  \SYSTEMmv{x} ,  \SYSTEMmv{y}  \rangle =  \SYSTEMnt{t_{{\mathrm{1}}}}  \, \mathsf{in} \,  \SYSTEMnt{t_{{\mathrm{2}}}}  \mid  \langle \rangle  \mid  \mathsf{let} \, \langle \rangle =  \SYSTEMnt{t_{{\mathrm{1}}}}  \, \mathsf{in} \,  \SYSTEMnt{t_{{\mathrm{2}}}}  \\
  \notag  \mid & \; \mathsf{inj}_1 \,  \SYSTEMnt{t}  \mid  \mathsf{inj}_2 \,  \SYSTEMnt{t}  \mid  \mathsf{case} \,  \SYSTEMnt{t}  \, \mathsf{of} \, \{ \mathsf{inj1} \,  \SYSTEMmv{x}  \rightarrow  \SYSTEMnt{t_{{\mathrm{1}}}}  ; \, \mathsf{inj2} \,  \SYSTEMmv{y}  \rightarrow  \SYSTEMnt{t_{{\mathrm{2}}}}  \}  \\
  \mid & \;  \textsf{push}_\otimes  \SYSTEMnt{t}  \mid  \textsf{push}_\oplus  \SYSTEMnt{t}  \mid  \textsf{push}_{\mathrm{unit} }  \SYSTEMnt{t} 
    \tag{terms} \\
  \SYSTEMnt{A} & ::=  \SYSTEMnt{A}  \multimap  \SYSTEMnt{B}  \mid  \textcolor{coeffectColor}{\square_{ \SYSTEMnt{r} } }  \SYSTEMnt{A}  \mid  \SYSTEMnt{A}  \otimes  \SYSTEMnt{B}  \mid  \mathrm{unit}  \mid  \SYSTEMnt{A}  \oplus  \SYSTEMnt{B} 
  \tag{types} \\
    \tag{contexts}
  \Gamma & ::=  \emptyset  \mid  \Gamma ,   \SYSTEMmv{x}  :  \SYSTEMnt{A}   \mid  \Gamma ,   \SYSTEMmv{x}  : \textcolor{coeffectColor}{[}  \SYSTEMnt{A} {\textcolor{coeffectColor}{]_{ \SYSTEMnt{r} } } }  
\end{align}

\paragraph{Typing}

Linear Core typing plus the following:

\begin{gather*}
\begin{align*}
  \begin{array}{c}
    \SYSTEMdruleLinpushprod{}
\quad
\SYSTEMdruleLinpushsum{} \\[1.5em]
\,
\SYSTEMdruleLinpushunit{} \\
\end{array}
\end{align*}
\end{gather*}

\paragraph{Operational Semantics}

Linear Core operational semantics plus the following:

\begin{gather*}
\begin{align*}
  \begin{array}{c}
    \SYSTEMdruleSemLinpushProd{}
    \;\;
    \SYSTEMdruleSemLinpushSumInjOne{}
    \;\;
    \SYSTEMdruleSemLinpushSumInjTwo{}
    \\[1.5em]
    \SYSTEMdruleSemLinpushProdCong{}
    \quad
    \SYSTEMdruleSemLinpushSumCong{}
    \\[1.5em]
    \SYSTEMdruleSemLinpushProdBoxCong{}
    \quad
    \SYSTEMdruleSemLinpushSumBoxCong{}
    \\[1.5em]
    \SYSTEMdruleSemLinpushUnitCong{}
    \quad
    \SYSTEMdruleSemLinpushUnitBoxCong{}
    \quad
    \SYSTEMdruleSemLinpushUnit{}
  \end{array}
\end{align*}
\end{gather*}

\paragraph{Equational Theory}

Linear Core equational theory plus the following:

  \begin{gather*}
\begin{align*}
  \begin{array}{c}
\SYSTEMdruleLinEqpushProdCong{} \quad
\SYSTEMdruleLinEqpushProdBeta{} \\[2em]
\SYSTEMdruleLinEqpushProdEta{} \\[2em]
\SYSTEMdruleLinEqpushSumCong{} \\[2em]
\SYSTEMdruleLinEqpushSumBetaOne{} \quad
\SYSTEMdruleLinEqpushSumBetaTwo{} \\[2em]
\SYSTEMdruleLinEqpushSumEta{} \\[2em]
\SYSTEMdruleLinEqpushUnitCong{} \quad
\SYSTEMdruleLinEqpushUnitBeta{}
  \end{array}
\end{align*}
\end{gather*}

\subsection{Graded Base}

\begin{align*}
    \SYSTEMnt{t} & ::= \SYSTEMmv{x} \mid  \SYSTEMnt{t_{{\mathrm{1}}}} \,  \SYSTEMnt{t_{{\mathrm{2}}}}  \mid  \lambda  \SYSTEMmv{x}  .  \SYSTEMnt{t} 
    \tag{terms} \\
      \SYSTEMnt{A} & ::=  \SYSTEMnt{A}  \xrightarrow{\textcolor{coeffectColor}{ \SYSTEMnt{r} } }  \SYSTEMnt{B}  \mid  \mathrm{K} 
  \tag{types} \\
  \Delta & ::=  \emptyset  \mid  \Delta ,   \SYSTEMmv{x}  :_{\textcolor{coeffectColor}{ \SYSTEMnt{r} } }  \SYSTEMnt{A}  
  \tag{contexts}
\end{align*}

\paragraph{Typing}

\begin{gather*}
\begin{align*}
\begin{array}{c}
\SYSTEMdruleGradvar{}
\;
\SYSTEMdruleGradweak{}
\;
\SYSTEMdruleGradapprox{}
\\[1em]
\SYSTEMdruleGradabs{}
\quad
\SYSTEMdruleGradapp{} \\
\end{array}
\end{align*}
\end{gather*}

\paragraph{Operational Semantics}

\begin{gather*}
\begin{align*}
  \begin{array}{c}
  \SYSTEMdruleSemGrdbeta{}
  \quad
  \SYSTEMdruleSemGrdcongAppL{} \\
  \end{array}
\end{align*}
\end{gather*}

\paragraph{Equational Theory}

  \begin{gather*}
\begin{align*}
  \begin{array}{c}
\SYSTEMdruleGradEqbeta{} \quad
\SYSTEMdruleGradEqeta{} \\[2em]
\SYSTEMdruleGradEqcongApp{} \quad
\SYSTEMdruleGradEqcongAbs{} \\
  \end{array}
\end{align*}
\end{gather*}

\subsection{Graded Modal Base}
\label{app:definitions-graded-modal-base}

\begin{align*}
      \SYSTEMnt{t} & ::= \SYSTEMmv{x} \mid  \SYSTEMnt{t_{{\mathrm{1}}}} \,  \SYSTEMnt{t_{{\mathrm{2}}}}  \mid  \lambda  \SYSTEMmv{x}  .  \SYSTEMnt{t}  \mid  \textcolor{coeffectColor}{[}  \SYSTEMnt{t}  \textcolor{coeffectColor}{]}  \mid  \mathsf{let} \, \textcolor{coeffectColor}{[}  \SYSTEMmv{x}  \textcolor{coeffectColor}{]} =  \SYSTEMnt{t_{{\mathrm{1}}}}  \, \mathsf{in} \,  \SYSTEMnt{t_{{\mathrm{2}}}} 
    \tag{terms} \\
  \SYSTEMnt{A} & ::=  \SYSTEMnt{A}  \xrightarrow{\textcolor{coeffectColor}{ \SYSTEMnt{r} } }  \SYSTEMnt{B}  \mid  \textcolor{coeffectColor}{\square_{ \SYSTEMnt{r} } }  \SYSTEMnt{A}  \mid  \mathrm{K} 
    \tag{types} \\
  \Delta & ::=  \emptyset  \mid  \Delta ,   \SYSTEMmv{x}  :_{\textcolor{coeffectColor}{ \SYSTEMnt{r} } }  \SYSTEMnt{A}  
  \tag{contexts}
\end{align*}

\paragraph{Typing}

Graded Base typing plus the following:

\begin{gather*}
\begin{align*}
  \begin{array}{c}\
\SYSTEMdruleGradBoxpr{}
\quad
\SYSTEMdruleGradBoxlet{}
\end{array}
\end{align*}
\end{gather*}

\paragraph{Operational Semantics}

Graded Base operational semantics plus the following:

\begin{gather*}
\begin{align*}
  \begin{array}{c}
  \SYSTEMdruleSemGrdModbetaBox{}
  \quad
  \SYSTEMdruleSemGrdModcongLetL{} \\
  \end{array}
\end{align*}
\end{gather*}

\paragraph{Equational Theory}

Graded Base equational theory plus the following:

  \begin{gather*}
\begin{align*}
  \begin{array}{c}
\SYSTEMdruleGradMEqbetaBox{} \quad
\SYSTEMdruleGradMEqetaBox{} \\[2em]
\SYSTEMdruleGradMEqletCommBox{} \\[2em]
\SYSTEMdruleGradMEqletCommOne{} \\[2em]
\SYSTEMdruleGradMEqletCommTwo{} \quad
\SYSTEMdruleGradMEqcongPr{} \\[2em]
\SYSTEMdruleGradMEqcongLet{} \\
  \end{array}
\end{align*}
\end{gather*}

\subsection{Graded Poly Base}
\label{app:definitions-graded-poly-base}

\begin{align*}
      \SYSTEMnt{t} & ::= \SYSTEMmv{x} \mid  \SYSTEMnt{t_{{\mathrm{1}}}} \,  \SYSTEMnt{t_{{\mathrm{2}}}}  \mid  \lambda  \SYSTEMmv{x}  .  \SYSTEMnt{t}  \mid  \SYSTEMnt{t}  \text{@}  \SYSTEMnt{A}  \mid  \Lambda  \alpha  .  \SYSTEMnt{t} 
    \tag{terms} \\
  \SYSTEMnt{A} & ::=  \SYSTEMnt{A}  \xrightarrow{\textcolor{coeffectColor}{ \SYSTEMnt{r} } }  \SYSTEMnt{B}  \mid  \textcolor{coeffectColor}{\square_{ \SYSTEMnt{r} } }  \SYSTEMnt{A}  \mid \alpha \mid  \forall  \alpha  .  \SYSTEMnt{B}  \mid  \mathrm{K} 
    \tag{types} \\
  \Delta & ::=  \emptyset  \mid  \Delta ,   \SYSTEMmv{x}  :_{\textcolor{coeffectColor}{ \SYSTEMnt{r} } }  \SYSTEMnt{A}  
  \tag{contexts}
\end{align*}

\paragraph{Typing}

Graded Base typing plus the following:

\begin{gather*}
\begin{align*}
  \begin{array}{c}
    \SYSTEMdruleGradPolytyAbs{} \quad
\SYSTEMdruleGradPolytyApp{} \\
\end{array}
\end{align*}
\end{gather*}

\paragraph{Operational Semantics}

Graded Base operational semantics plus the following:

\begin{gather*}
\begin{align*}
  \begin{array}{c}
\SYSTEMdruleSemGrdPolytyBeta{} \quad
\SYSTEMdruleSemGrdPolycongTyApp{} \\
  \end{array}
\end{align*}
\end{gather*}

\paragraph{Equational Theory}

Graded Base equational theory the following:

  \begin{gather*}
\begin{align*}
  \begin{array}{c}
\SYSTEMdruleGradPolyEqbetaTy{} \quad
\SYSTEMdruleGradPolyEqetaTy{} \\[2em]
\SYSTEMdruleGradPolyEqtyAppCong{} \quad
\SYSTEMdruleGradPolyEqtyAbsCong{} \\
  \end{array}
\end{align*}
\end{gather*}

\subsection{Graded Modal Core}
\label{app:definitions-graded-core}

\begin{align*}
\notag  \SYSTEMnt{t} ::= & \SYSTEMmv{x} \mid  \SYSTEMnt{t_{{\mathrm{1}}}} \,  \SYSTEMnt{t_{{\mathrm{2}}}}  \mid  \lambda  \SYSTEMmv{x}  .  \SYSTEMnt{t}  \mid  \textcolor{coeffectColor}{[}  \SYSTEMnt{t}  \textcolor{coeffectColor}{]}  \mid  \mathsf{let} \, \textcolor{coeffectColor}{[}  \SYSTEMmv{x}  \textcolor{coeffectColor}{]} =  \SYSTEMnt{t_{{\mathrm{1}}}}  \, \mathsf{in} \,  \SYSTEMnt{t_{{\mathrm{2}}}}  \\
\notag   \mid & \;  \langle  \SYSTEMnt{t_{{\mathrm{1}}}} ,  \SYSTEMnt{t_{{\mathrm{2}}}}  \rangle  \mid  \mathsf{let} \, \langle  \SYSTEMmv{x} ,  \SYSTEMmv{y}  \rangle =  \SYSTEMnt{t_{{\mathrm{1}}}}  \, \mathsf{in} \,  \SYSTEMnt{t_{{\mathrm{2}}}}  \mid  \langle \rangle  \mid  \mathsf{let} \, \langle \rangle =  \SYSTEMnt{t_{{\mathrm{1}}}}  \, \mathsf{in} \,  \SYSTEMnt{t_{{\mathrm{2}}}}  \\
         \mid & \; \mathsf{inj}_1 \,  \SYSTEMnt{t}  \mid  \mathsf{inj}_2 \,  \SYSTEMnt{t}  \mid  \mathsf{case} \,  \SYSTEMnt{t}  \, \mathsf{of} \, \{ \mathsf{inj1} \,  \SYSTEMmv{x}  \rightarrow  \SYSTEMnt{t_{{\mathrm{1}}}}  ; \, \mathsf{inj2} \,  \SYSTEMmv{y}  \rightarrow  \SYSTEMnt{t_{{\mathrm{2}}}}  \} 
    \tag{terms} \\
  \SYSTEMnt{A} & ::=  \SYSTEMnt{A}  \xrightarrow{\textcolor{coeffectColor}{ \SYSTEMnt{r} } }  \SYSTEMnt{B}  \mid  \textcolor{coeffectColor}{\square_{ \SYSTEMnt{r} } }  \SYSTEMnt{A}  \mid  \SYSTEMnt{A}  \times  \SYSTEMnt{B}  \mid  \mathrm{unit}  \mid  \SYSTEMnt{A}  +  \SYSTEMnt{B} 
    \tag{types} \\
  \Delta & ::=  \emptyset  \mid  \Delta ,   \SYSTEMmv{x}  :_{\textcolor{coeffectColor}{ \SYSTEMnt{r} } }  \SYSTEMnt{A}  
  \tag{contexts}
\end{align*}

\paragraph{Typing}

Graded Modal Base typing plus the following:

\begin{gather*}
\begin{align*}
  \begin{array}{c}
    \SYSTEMdruleGradprodi{}
\quad
\SYSTEMdruleGradprode{}
\\[2em]
\SYSTEMdruleGraduniti{}
\quad
\SYSTEMdruleGradunite{}
\quad
\SYSTEMdruleGradsumiOne{}
\\[2em]
\SYSTEMdruleGradsumiTwo{}
\quad
\SYSTEMdruleGradsume{} \\
\end{array}
\end{align*}
\end{gather*}

\paragraph{Operational Semantics}

Graded Modal Base operational semantics plus the following:

\begin{gather*}
\begin{align*}
  \begin{array}{c}
      \SYSTEMdruleSemGrdprodCong{}
  \quad
  \SYSTEMdruleSemGrdprodBeta{}
  \\[1.25em]
  \SYSTEMdruleSemGrdunitCong{}
  \quad
  \SYSTEMdruleSemGrdunitBeta{}
  \\[1.25em]
  \SYSTEMdruleSemGrdcongCase{}
  \\[1.25em]
  \SYSTEMdruleSemGrdcaseInjOne{}
  \\[1.25em]
  \SYSTEMdruleSemGrdcaseInjTwo{} \\[2em]
  \end{array}
\end{align*}
\end{gather*}

\paragraph{Equational Theory}

Graded Modal Base equational theory plus the following:
  \begin{gather*}
\begin{align*}
  \begin{array}{c}
 \SYSTEMdruleGradEqbetaUnit{} \quad
\SYSTEMdruleGradEqetaUnit{} \\[2em]
\SYSTEMdruleGradEqbetaProd{}
\SYSTEMdruleGradEqetaProd{} \\[2em]
\SYSTEMdruleGradEqbetaSumOne{} \quad
\SYSTEMdruleGradEqbetaSumTwo{} \\[2em]
\SYSTEMdruleGradEqetaSum{} \\[2em]
    \SYSTEMdruleGradEqcongUnitE{} \quad
\SYSTEMdruleGradEqcongProdE{} \\[2em]
\SYSTEMdruleGradEqcongProdI{} \quad
\SYSTEMdruleGradEqcongSumIOne{} \quad
\SYSTEMdruleGradEqcongSumITwo{} \\[2em]
\SYSTEMdruleGradEqcongSumE{}
  \end{array}
\end{align*}
\end{gather*}

\section{Proofs}
\label{app:proofs}
\subsection{Proof of Soundness for Graded Base to Linear Base}
\label{app:proofs-grad-to-lin}

\ifextended\gradToLinTranslation*\fi

\noindent
Type preservation is in Appendix~\ref{app:proofs-grad-to-lin-typ},
operational correspondence in Appendix~\ref{app:proofs-grad-to-lin-ops},
and equation preservation in Appendix~\ref{app:proofs-grad-to-lin-eqs}.

\subsubsection{Type preservation}
\label{app:proofs-grad-to-lin-typ}

\begin{lemma}[The interpretation of a zeroed Graded Base context gives a Linear Base context
that is graded at zero]
\label{lemma:zero-interp-lemma-grad-to-lin}
For all graded-base contexts $\Delta$, the predicate $ \mathrm{graded}(  \textcolor{GLcolor}{\llbracket}   \textcolor{coeffectColor}{ \SYSTEMsym{0}  \cdot}  \Delta   \textcolor{GLcolor}{\rrbracket}  , \textcolor{coeffectColor}{ \SYSTEMsym{0} }) $ holds.
\end{lemma}

\begin{proof} By induction on the structure of $\Delta$.
\begin{itemize}
  \item (empty context)
  For $ \emptyset $, we prove the goal $ \mathrm{graded}(  \textcolor{GLcolor}{\llbracket}   \textcolor{coeffectColor}{ \SYSTEMsym{0}  \cdot}   \emptyset    \textcolor{GLcolor}{\rrbracket}  , \textcolor{coeffectColor}{ \SYSTEMsym{0} }) $ syntactically: % todo: phrasing

    \begin{align*}
  \begin{array}{rll}
 \textit{(defn. graded)} & &  \mathrm{graded}(  \emptyset  , \textcolor{coeffectColor}{ \SYSTEMsym{0} })  \\
 \textit{(defn. interepretation)}& \implies &  \mathrm{graded}(  \textcolor{GLcolor}{\llbracket}   \emptyset   \textcolor{GLcolor}{\rrbracket}  , \textcolor{coeffectColor}{ \SYSTEMsym{0} })  \\
 \textit{(defn. multiplication)}  & \implies &  \mathrm{graded}(  \textcolor{GLcolor}{\llbracket}   \textcolor{coeffectColor}{ \SYSTEMsym{0}  \cdot}   \emptyset    \textcolor{GLcolor}{\rrbracket}  , \textcolor{coeffectColor}{ \SYSTEMsym{0} }) 
 \end{array}
 \end{align*}

  \item (graded context extension)
  For $ \Delta ,   \SYSTEMmv{x}  :_{\textcolor{coeffectColor}{ \SYSTEMnt{r} } }  \SYSTEMnt{A}  $,

  By induction, we have that $ \mathrm{graded}(  \textcolor{GLcolor}{\llbracket}   \textcolor{coeffectColor}{ \SYSTEMsym{0}  \cdot}  \Delta   \textcolor{GLcolor}{\rrbracket}  , \textcolor{coeffectColor}{ \SYSTEMsym{0} }) $, from which we conclude
  with the following steps:
  \begin{align*}
  \begin{array}{rll}
\textit{(i.h.)}\;\;  & &  \mathrm{graded}(  \textcolor{GLcolor}{\llbracket}   \textcolor{coeffectColor}{ \SYSTEMsym{0}  \cdot}  \Delta   \textcolor{GLcolor}{\rrbracket}  , \textcolor{coeffectColor}{ \SYSTEMsym{0} })  \\
\textit{(defn. graded)} \;\;  & \implies & \mathrm{graded}(   \textcolor{GLcolor}{\llbracket}   \textcolor{coeffectColor}{ \SYSTEMsym{0}  \cdot}  \Delta   \textcolor{GLcolor}{\rrbracket}  ,   \SYSTEMmv{x}  : \textcolor{coeffectColor}{[}   \textcolor{GLcolor}{\llbracket}  \SYSTEMnt{A}  \textcolor{GLcolor}{\rrbracket}  {\textcolor{coeffectColor}{]_{ \SYSTEMsym{0} } } }   , \textcolor{coeffectColor}{ \SYSTEMsym{0} })  \\
\textit{(defn. interpretation)} \;\;  & \implies &    \mathrm{graded}(  \textcolor{GLcolor}{\llbracket}    (   \textcolor{coeffectColor}{ \SYSTEMsym{0}  \cdot}  \Delta   )  ,   \SYSTEMmv{x}  :_{\textcolor{coeffectColor}{ \SYSTEMsym{0} } }  \SYSTEMnt{A}    \textcolor{GLcolor}{\rrbracket}  , \textcolor{coeffectColor}{ \SYSTEMsym{0} })  \\
\textit{(defn. multiplication)} \;\;  &  \implies &  \mathrm{graded}(  \textcolor{GLcolor}{\llbracket}   \textcolor{coeffectColor}{ \SYSTEMsym{0}  \cdot}   (   \Delta ,   \SYSTEMmv{x}  :_{\textcolor{coeffectColor}{ \SYSTEMnt{r} } }  \SYSTEMnt{A}    )    \textcolor{GLcolor}{\rrbracket}  , \textcolor{coeffectColor}{ \SYSTEMsym{0} }) 
\end{array}
  \end{align*}
\end{itemize}
\end{proof}

\begin{lemma}[Interpretation of a graded context is graded in Linear Base]
\label{lemma:interp-grad-ctx-is-graded}
For all contexts $\Delta$, $ \mathrm{graded}(  \textcolor{GLcolor}{\llbracket}  \Delta  \textcolor{GLcolor}{\rrbracket}  ) $.
\end{lemma}

\begin{proof} By induction over the structure of $\Delta$.
\begin{itemize}
  \item (empty context)
      \begin{align*}
  \begin{array}{rll}
 \textit{(defn. graded)} & &  \mathrm{graded}(  \emptyset  )  \\
 \textit{(defn. interpretation)} & \implies &  \mathrm{graded}(  \textcolor{GLcolor}{\llbracket}   \emptyset   \textcolor{GLcolor}{\rrbracket}  ) 
 \end{array}
 \end{align*}

  \item (graded context extension)
        \begin{align*}
  \begin{array}{rll}
\textit{(i.h.)} & &  \mathrm{graded}(  \textcolor{GLcolor}{\llbracket}  \Delta  \textcolor{GLcolor}{\rrbracket}  )  \\
\textit{(defn. graded)} & \implies &  \mathrm{graded}(   \textcolor{GLcolor}{\llbracket}  \Delta  \textcolor{GLcolor}{\rrbracket}  ,   \SYSTEMmv{x}  : \textcolor{coeffectColor}{[}   \textcolor{GLcolor}{\llbracket}  \SYSTEMnt{A}  \textcolor{GLcolor}{\rrbracket}  {\textcolor{coeffectColor}{]_{ \SYSTEMnt{r} } } }   )  \\
\textit{(defn. interpretation)} & \implies &  \mathrm{graded}(  \textcolor{GLcolor}{\llbracket}   \Delta ,   \SYSTEMmv{x}  :_{\textcolor{coeffectColor}{ \SYSTEMnt{r} } }  \SYSTEMnt{A}    \textcolor{GLcolor}{\rrbracket}  ) 
\end{array}
\end{align*}

\end{itemize}
\end{proof}

\begin{proof}
  By induction on the typing derivations of Graded Base.
\begin{itemize}
\item (var)
$$
\SYSTEMdruleGradvar{}
$$
Therefore we construct the goal typing:
$$
\inferrule*[Right=\SYSTEMRenameRuleLinder{}]
 {\inferrule*[Right=\SYSTEMRenameRuleLinvar{}]
    { }
    {   \SYSTEMmv{x}  :   \textcolor{GLcolor}{\llbracket}  \SYSTEMnt{A}  \textcolor{GLcolor}{\rrbracket}     \vdash_{\textsc{l} }  \SYSTEMmv{x}  :   \textcolor{GLcolor}{\llbracket}  \SYSTEMnt{A}  \textcolor{GLcolor}{\rrbracket}  }
 }
 {   \SYSTEMmv{x}  : \textcolor{coeffectColor}{[}   \textcolor{GLcolor}{\llbracket}  \SYSTEMnt{A}  \textcolor{GLcolor}{\rrbracket}  {\textcolor{coeffectColor}{]_{ \SYSTEMsym{1} } } }    \vdash_{\textsc{l} }  \SYSTEMmv{x}  :   \textcolor{GLcolor}{\llbracket}  \SYSTEMnt{A}  \textcolor{GLcolor}{\rrbracket}  }
$$

\item (abs)
$$
\SYSTEMdruleGradabs{}
$$
By induction on the premise we have $   \textcolor{GLcolor}{\llbracket}  \Delta  \textcolor{GLcolor}{\rrbracket}  ,   \SYSTEMmv{x}  : \textcolor{coeffectColor}{[}   \textcolor{GLcolor}{\llbracket}  \SYSTEMnt{A}  \textcolor{GLcolor}{\rrbracket}  {\textcolor{coeffectColor}{]_{ \SYSTEMnt{r} } } }    \vdash_{\textsc{l} }   \textcolor{GLcolor}{\llbracket}  \SYSTEMnt{t}  \textcolor{GLcolor}{\rrbracket}   :   \textcolor{GLcolor}{\llbracket}  \SYSTEMnt{B}  \textcolor{GLcolor}{\rrbracket}  $.

Then we construct, with $y \# t$:
$$
\inferrule*[Right=\SYSTEMRenameRuleLinabs{}]
    {\inferrule*[Right=\SYSTEMRenameRuleLinlet{}]
      {\inferrule*[Right=\SYSTEMRenameRuleLinvar{}]
         { }{   \SYSTEMmv{y}  :   \textcolor{coeffectColor}{\square_{ \SYSTEMnt{r} } }   \textcolor{GLcolor}{\llbracket}  \SYSTEMnt{A}  \textcolor{GLcolor}{\rrbracket}      \vdash_{\textsc{l} }  \SYSTEMmv{y}  :   \textcolor{coeffectColor}{\square_{ \SYSTEMnt{r} } }   \textcolor{GLcolor}{\llbracket}  \SYSTEMnt{A}  \textcolor{GLcolor}{\rrbracket}   }
       \\
        \inferrule*{ih}{   \textcolor{GLcolor}{\llbracket}  \Delta  \textcolor{GLcolor}{\rrbracket}  ,   \SYSTEMmv{x}  : \textcolor{coeffectColor}{[}   \textcolor{GLcolor}{\llbracket}  \SYSTEMnt{A}  \textcolor{GLcolor}{\rrbracket}  {\textcolor{coeffectColor}{]_{ \SYSTEMnt{r} } } }    \vdash_{\textsc{l} }   \textcolor{GLcolor}{\llbracket}  \SYSTEMnt{t}  \textcolor{GLcolor}{\rrbracket}   :   \textcolor{GLcolor}{\llbracket}  \SYSTEMnt{B}  \textcolor{GLcolor}{\rrbracket}  }
      }
      {   \textcolor{GLcolor}{\llbracket}  \Delta  \textcolor{GLcolor}{\rrbracket}  ,   \SYSTEMmv{y}  :   \textcolor{coeffectColor}{\square_{ \SYSTEMnt{r} } }   \textcolor{GLcolor}{\llbracket}  \SYSTEMnt{A}  \textcolor{GLcolor}{\rrbracket}      \vdash_{\textsc{l} }   \mathsf{let} \, \textcolor{coeffectColor}{[}  \SYSTEMmv{x}  \textcolor{coeffectColor}{]} =  \SYSTEMmv{y}  \, \mathsf{in} \,   \textcolor{GLcolor}{\llbracket}  \SYSTEMnt{t}  \textcolor{GLcolor}{\rrbracket}    :   \textcolor{GLcolor}{\llbracket}  \SYSTEMnt{B}  \textcolor{GLcolor}{\rrbracket}  }
     }
    {  \textcolor{GLcolor}{\llbracket}  \Delta  \textcolor{GLcolor}{\rrbracket}   \vdash_{\textsc{l} }   \lambda  \SYSTEMmv{y}  .   \mathsf{let} \, \textcolor{coeffectColor}{[}  \SYSTEMmv{x}  \textcolor{coeffectColor}{]} =  \SYSTEMmv{y}  \, \mathsf{in} \,   \textcolor{GLcolor}{\llbracket}  \SYSTEMnt{t}  \textcolor{GLcolor}{\rrbracket}     :    \textcolor{coeffectColor}{\square_{ \SYSTEMnt{r} } }   \textcolor{GLcolor}{\llbracket}  \SYSTEMnt{A}  \textcolor{GLcolor}{\rrbracket}    \multimap   \textcolor{GLcolor}{\llbracket}  \SYSTEMnt{B}  \textcolor{GLcolor}{\rrbracket}   }
$$
which satisfies the goal.

\item (app)
$$
\SYSTEMdruleGradapp{}
$$
By induction we have: $  \textcolor{GLcolor}{\llbracket}  \Delta_{{\mathrm{1}}}  \textcolor{GLcolor}{\rrbracket}   \vdash_{\textsc{l} }   \textcolor{GLcolor}{\llbracket}  \SYSTEMnt{t_{{\mathrm{1}}}}  \textcolor{GLcolor}{\rrbracket}   :    \textcolor{coeffectColor}{\square_{ \SYSTEMnt{r} } }   \textcolor{GLcolor}{\llbracket}  \SYSTEMnt{A}  \textcolor{GLcolor}{\rrbracket}    \multimap   \textcolor{GLcolor}{\llbracket}  \SYSTEMnt{B}  \textcolor{GLcolor}{\rrbracket}   $
and $  \textcolor{GLcolor}{\llbracket}  \Delta_{{\mathrm{2}}}  \textcolor{GLcolor}{\rrbracket}   \vdash_{\textsc{l} }   \textcolor{GLcolor}{\llbracket}  \SYSTEMnt{t_{{\mathrm{2}}}}  \textcolor{GLcolor}{\rrbracket}   :   \textcolor{GLcolor}{\llbracket}  \SYSTEMnt{A}  \textcolor{GLcolor}{\rrbracket}  $.

Then we construct:
$$
\inferrule*[Right=\SYSTEMRenameRuleLinapp{}]
    {  \textcolor{GLcolor}{\llbracket}  \Delta_{{\mathrm{1}}}  \textcolor{GLcolor}{\rrbracket}   \vdash_{\textsc{l} }   \textcolor{GLcolor}{\llbracket}  \SYSTEMnt{t_{{\mathrm{1}}}}  \textcolor{GLcolor}{\rrbracket}   :    \textcolor{coeffectColor}{\square_{ \SYSTEMnt{r} } }   \textcolor{GLcolor}{\llbracket}  \SYSTEMnt{A}  \textcolor{GLcolor}{\rrbracket}    \multimap   \textcolor{GLcolor}{\llbracket}  \SYSTEMnt{B}  \textcolor{GLcolor}{\rrbracket}   
    \\\inferrule*[Right=\SYSTEMRenameRuleLinpr{}]
        {  \textcolor{GLcolor}{\llbracket}  \Delta_{{\mathrm{2}}}  \textcolor{GLcolor}{\rrbracket}   \vdash_{\textsc{l} }   \textcolor{GLcolor}{\llbracket}  \SYSTEMnt{t_{{\mathrm{2}}}}  \textcolor{GLcolor}{\rrbracket}   :   \textcolor{GLcolor}{\llbracket}  \SYSTEMnt{A}  \textcolor{GLcolor}{\rrbracket}  }
        {  \textcolor{coeffectColor}{ \SYSTEMnt{r}  \cdot}   \textcolor{GLcolor}{\llbracket}  \Delta_{{\mathrm{2}}}  \textcolor{GLcolor}{\rrbracket}    \vdash_{\textsc{l} }   \textcolor{coeffectColor}{[}   \textcolor{GLcolor}{\llbracket}  \SYSTEMnt{t_{{\mathrm{2}}}}  \textcolor{GLcolor}{\rrbracket}   \textcolor{coeffectColor}{]}   :   \textcolor{coeffectColor}{\square_{ \SYSTEMnt{r} } }   \textcolor{GLcolor}{\llbracket}  \SYSTEMnt{A}  \textcolor{GLcolor}{\rrbracket}   }}
    {  \textcolor{GLcolor}{\llbracket}  \Delta_{{\mathrm{1}}}  \textcolor{GLcolor}{\rrbracket}   \SYSTEMsym{+}   \textcolor{coeffectColor}{ \SYSTEMnt{r}  \cdot}   \textcolor{GLcolor}{\llbracket}  \Delta_{{\mathrm{2}}}  \textcolor{GLcolor}{\rrbracket}    \vdash_{\textsc{l} }    \textcolor{GLcolor}{\llbracket}  \SYSTEMnt{t_{{\mathrm{1}}}}  \textcolor{GLcolor}{\rrbracket}  \,   \textcolor{coeffectColor}{[}   \textcolor{GLcolor}{\llbracket}  \SYSTEMnt{t_{{\mathrm{2}}}}  \textcolor{GLcolor}{\rrbracket}   \textcolor{coeffectColor}{]}    :   \textcolor{GLcolor}{\llbracket}  \SYSTEMnt{B}  \textcolor{GLcolor}{\rrbracket}  }
$$
    Since translation of contexts is a homomorphism then:
    $ \textcolor{GLcolor}{\llbracket}  \Delta_{{\mathrm{1}}}  \SYSTEMsym{+}   \textcolor{coeffectColor}{ \SYSTEMnt{r}  \cdot}  \Delta_{{\mathrm{2}}}   \textcolor{GLcolor}{\rrbracket}  \equiv  \textcolor{GLcolor}{\llbracket}  \Delta_{{\mathrm{1}}}  \textcolor{GLcolor}{\rrbracket}   \SYSTEMsym{+}   \textcolor{coeffectColor}{ \SYSTEMnt{r}  \cdot}   \textcolor{GLcolor}{\llbracket}  \Delta_{{\mathrm{2}}}  \textcolor{GLcolor}{\rrbracket}  $

\item (weak)
$$
\SYSTEMdruleGradweak{}
$$
By induction on the premises we have $  \textcolor{GLcolor}{\llbracket}  \Delta  \textcolor{GLcolor}{\rrbracket}   \vdash_{\textsc{l} }   \textcolor{GLcolor}{\llbracket}  \SYSTEMnt{t}  \textcolor{GLcolor}{\rrbracket}   :   \textcolor{GLcolor}{\llbracket}  \SYSTEMnt{A}  \textcolor{GLcolor}{\rrbracket}  $.

Therefore we construct:
$$
\inferrule*[Right=\SYSTEMRenameRuleLinweak{}]
    {  \textcolor{GLcolor}{\llbracket}  \Delta  \textcolor{GLcolor}{\rrbracket}   \vdash_{\textsc{l} }   \textcolor{GLcolor}{\llbracket}  \SYSTEMnt{t}  \textcolor{GLcolor}{\rrbracket}   :   \textcolor{GLcolor}{\llbracket}  \SYSTEMnt{A}  \textcolor{GLcolor}{\rrbracket}   \quad  \mathrm{graded}(  \textcolor{GLcolor}{\llbracket}   \textcolor{coeffectColor}{ \SYSTEMsym{0}  \cdot}  \Delta'   \textcolor{GLcolor}{\rrbracket}  , \textcolor{coeffectColor}{ \SYSTEMsym{0} }) }
    {   \textcolor{GLcolor}{\llbracket}  \Delta  \textcolor{GLcolor}{\rrbracket}   ,    \textcolor{GLcolor}{\llbracket}   \textcolor{coeffectColor}{ \SYSTEMsym{0}  \cdot}  \Delta'   \textcolor{GLcolor}{\rrbracket}     \vdash_{\textsc{l} }   \textcolor{GLcolor}{\llbracket}  \SYSTEMnt{t}  \textcolor{GLcolor}{\rrbracket}   :   \textcolor{GLcolor}{\llbracket}  \SYSTEMnt{A}  \textcolor{GLcolor}{\rrbracket}  }
$$
where the second premise is provided by Lemma~\ref{lemma:zero-interp-lemma-grad-to-lin}.
\item (approx)
$$
\SYSTEMdruleGradapprox{}
$$
% $$
% \SYSTEMdruleLinapprox{}
% $$
By induction on the premise we have\\
$   \textcolor{GLcolor}{\llbracket}  \Delta  \textcolor{GLcolor}{\rrbracket}  ,   \SYSTEMmv{x}  : \textcolor{coeffectColor}{[}   \textcolor{GLcolor}{\llbracket}  \SYSTEMnt{B}  \textcolor{GLcolor}{\rrbracket}  {\textcolor{coeffectColor}{]_{ \SYSTEMnt{s} } } }    \vdash_{\textsc{l} }   \textcolor{GLcolor}{\llbracket}  \SYSTEMnt{t}  \textcolor{GLcolor}{\rrbracket}   :   \textcolor{GLcolor}{\llbracket}  \SYSTEMnt{A}  \textcolor{GLcolor}{\rrbracket}  $.

Therefore we can construct:
$$
\inferrule*[Right=\SYSTEMRenameRuleLinapprox]
{ \begin{array}{cc}     \textcolor{GLcolor}{\llbracket}  \Delta  \textcolor{GLcolor}{\rrbracket}  ,   \SYSTEMmv{x}  : \textcolor{coeffectColor}{[}   \textcolor{GLcolor}{\llbracket}  \SYSTEMnt{B}  \textcolor{GLcolor}{\rrbracket}  {\textcolor{coeffectColor}{]_{ \SYSTEMnt{s} } } }    \vdash_{\textsc{l} }   \textcolor{GLcolor}{\llbracket}  \SYSTEMnt{t}  \textcolor{GLcolor}{\rrbracket}   :   \textcolor{GLcolor}{\llbracket}  \SYSTEMnt{A}  \textcolor{GLcolor}{\rrbracket}    \; & \;   \SYSTEMnt{s}  \, \textcolor{coeffectColor}{\sqsubseteq} \,  \SYSTEMnt{r}   \end{array} }
{   \textcolor{GLcolor}{\llbracket}  \Delta  \textcolor{GLcolor}{\rrbracket}  ,   \SYSTEMmv{x}  : \textcolor{coeffectColor}{[}   \textcolor{GLcolor}{\llbracket}  \SYSTEMnt{B}  \textcolor{GLcolor}{\rrbracket}  {\textcolor{coeffectColor}{]_{ \SYSTEMnt{s} } } }    \vdash_{\textsc{l} }   \textcolor{GLcolor}{\llbracket}  \SYSTEMnt{t}  \textcolor{GLcolor}{\rrbracket}   :   \textcolor{GLcolor}{\llbracket}  \SYSTEMnt{A}  \textcolor{GLcolor}{\rrbracket}   }
$$
\end{itemize}
\end{proof}

\subsubsection{Operational correspondence}
\label{app:proofs-grad-to-lin-ops}

\begin{lemma}[Interpretation preserves substitution]
\label{lemma:interp-grd-to-lin-preserves-subst}
For all Graded Base terms $\SYSTEMnt{t}, \SYSTEMnt{t'}$ then
$ \textcolor{GLcolor}{\llbracket}   [  \SYSTEMnt{t}  /  \SYSTEMmv{x}  ]  \SYSTEMnt{t'}   \textcolor{GLcolor}{\rrbracket}  \equiv  [   \textcolor{GLcolor}{\llbracket}  \SYSTEMnt{t}  \textcolor{GLcolor}{\rrbracket}   /  \SYSTEMmv{x}  ]   \textcolor{GLcolor}{\llbracket}  \SYSTEMnt{t'}  \textcolor{GLcolor}{\rrbracket}  $.
\end{lemma}

\begin{proof}
By induction on the receiving term $\SYSTEMnt{t'}$:
\begin{itemize}
\item (var) $\SYSTEMnt{t'} \equiv \SYSTEMmv{y}$:

\begin{itemize}
\item $\SYSTEMmv{x} \equiv \SYSTEMmv{y}$ then we refine the goal:
\begin{align*}
\begin{array}{rcrlll}
\textit{(goal)} & &  \textcolor{GLcolor}{\llbracket}   [  \SYSTEMnt{t}  /  \SYSTEMmv{x}  ]  \SYSTEMmv{x}   \textcolor{GLcolor}{\rrbracket}  & \eqStep &  [   \textcolor{GLcolor}{\llbracket}  \SYSTEMnt{t}  \textcolor{GLcolor}{\rrbracket}   /  \SYSTEMmv{x}  ]   \textcolor{GLcolor}{\llbracket}  \SYSTEMmv{x}  \textcolor{GLcolor}{\rrbracket}   \\
\textit{(defn. subst [lhs])} & \Rightarrow &  \textcolor{GLcolor}{\llbracket}  \SYSTEMnt{t}  \textcolor{GLcolor}{\rrbracket}  & \eqStep &  [   \textcolor{GLcolor}{\llbracket}  \SYSTEMnt{t}  \textcolor{GLcolor}{\rrbracket}   /  \SYSTEMmv{x}  ]   \textcolor{GLcolor}{\llbracket}  \SYSTEMmv{x}  \textcolor{GLcolor}{\rrbracket}   \\
\textit{(defn. interp [rhs])} & \Rightarrow &  \textcolor{GLcolor}{\llbracket}  \SYSTEMnt{t}  \textcolor{GLcolor}{\rrbracket}  & \eqStep &  [   \textcolor{GLcolor}{\llbracket}  \SYSTEMnt{t}  \textcolor{GLcolor}{\rrbracket}   /  \SYSTEMmv{x}  ]  \SYSTEMmv{x}  \\
\textit{(defn. subst [rhs])} & \Rightarrow &  \textcolor{GLcolor}{\llbracket}  \SYSTEMnt{t}  \textcolor{GLcolor}{\rrbracket}  & \eqStep &  \textcolor{GLcolor}{\llbracket}  \SYSTEMnt{t}  \textcolor{GLcolor}{\rrbracket}  \\
\textit{(reflexivity)} & \Rightarrow & \top
\end{array}
\end{align*}

\item $\SYSTEMmv{x} \not\equiv \SYSTEMmv{y}$ then we refine the goal:
\begin{align*}
\begin{array}{rcrlll}
\textit{(goal)} & &  \textcolor{GLcolor}{\llbracket}   [  \SYSTEMnt{t}  /  \SYSTEMmv{x}  ]  \SYSTEMmv{y}   \textcolor{GLcolor}{\rrbracket}  & \eqStep &  [   \textcolor{GLcolor}{\llbracket}  \SYSTEMnt{t}  \textcolor{GLcolor}{\rrbracket}   /  \SYSTEMmv{x}  ]   \textcolor{GLcolor}{\llbracket}  \SYSTEMmv{y}  \textcolor{GLcolor}{\rrbracket}   \\
\textit{(defn. subst [lhs])} & \Rightarrow & \SYSTEMmv{y} & \eqStep &  [   \textcolor{GLcolor}{\llbracket}  \SYSTEMnt{t}  \textcolor{GLcolor}{\rrbracket}   /  \SYSTEMmv{x}  ]   \textcolor{GLcolor}{\llbracket}  \SYSTEMmv{y}  \textcolor{GLcolor}{\rrbracket}   \\
\textit{(defn. interp [rhs])} & \Rightarrow & \SYSTEMmv{y} & \eqStep &  [   \textcolor{GLcolor}{\llbracket}  \SYSTEMnt{t}  \textcolor{GLcolor}{\rrbracket}   /  \SYSTEMmv{x}  ]  \SYSTEMmv{y}  \\
\textit{(defn. subst [rhs])} & \Rightarrow & \SYSTEMmv{y} & \eqStep & \SYSTEMmv{y} \\
\textit{(reflexivity)} & \Rightarrow & \top
\end{array}
\end{align*}
\end{itemize}
% END (var)
\item (abs) $\SYSTEMnt{t'} \equiv  \lambda  \SYSTEMmv{y}  .  \SYSTEMnt{t_{{\mathrm{1}}}} $ (with $ \SYSTEMmv{y}  \,\#\,  \SYSTEMnt{t} $);
\begin{gather*}
\begin{align*}
\begin{array}{rcrlll}
\textit{(goal)} & &  \textcolor{GLcolor}{\llbracket}   [  \SYSTEMnt{t}  /  \SYSTEMmv{x}  ]  \SYSTEMsym{(}   \lambda  \SYSTEMmv{y}  .  \SYSTEMnt{t_{{\mathrm{1}}}}   \SYSTEMsym{)}   \textcolor{GLcolor}{\rrbracket}  & \eqStep &  [   \textcolor{GLcolor}{\llbracket}  \SYSTEMnt{t}  \textcolor{GLcolor}{\rrbracket}   /  \SYSTEMmv{x}  ]   \textcolor{GLcolor}{\llbracket}  \SYSTEMsym{(}   \lambda  \SYSTEMmv{y}  .  \SYSTEMnt{t_{{\mathrm{1}}}}   \SYSTEMsym{)}  \textcolor{GLcolor}{\rrbracket}   \\
\textit{(defn. subst [lhs])} & \Rightarrow &  \textcolor{GLcolor}{\llbracket}   \lambda  \SYSTEMmv{y}  .   [  \SYSTEMnt{t}  /  \SYSTEMmv{x}  ]  \SYSTEMnt{t_{{\mathrm{1}}}}    \textcolor{GLcolor}{\rrbracket}  & \eqStep &  [   \textcolor{GLcolor}{\llbracket}  \SYSTEMnt{t}  \textcolor{GLcolor}{\rrbracket}   /  \SYSTEMmv{x}  ]   \textcolor{GLcolor}{\llbracket}  \SYSTEMsym{(}   \lambda  \SYSTEMmv{y}  .  \SYSTEMnt{t_{{\mathrm{1}}}}   \SYSTEMsym{)}  \textcolor{GLcolor}{\rrbracket}   \\
\textit{(defn. interp [rhs])} & \Rightarrow &  \textcolor{GLcolor}{\llbracket}   \lambda  \SYSTEMmv{y}  .   [  \SYSTEMnt{t}  /  \SYSTEMmv{x}  ]  \SYSTEMnt{t_{{\mathrm{1}}}}    \textcolor{GLcolor}{\rrbracket}  & \eqStep &  [   \textcolor{GLcolor}{\llbracket}  \SYSTEMnt{t}  \textcolor{GLcolor}{\rrbracket}   /  \SYSTEMmv{x}  ]   \lambda  \SYSTEMmv{y'}  .   \mathsf{let} \, \textcolor{coeffectColor}{[}  \SYSTEMmv{y}  \textcolor{coeffectColor}{]} =  \SYSTEMmv{y'}  \, \mathsf{in} \,   \textcolor{GLcolor}{\llbracket}  \SYSTEMnt{t_{{\mathrm{1}}}}  \textcolor{GLcolor}{\rrbracket}     \\
\textit{(defn. subst [rhs])} & \Rightarrow &  \textcolor{GLcolor}{\llbracket}   \lambda  \SYSTEMmv{y}  .   [  \SYSTEMnt{t}  /  \SYSTEMmv{x}  ]  \SYSTEMnt{t_{{\mathrm{1}}}}    \textcolor{GLcolor}{\rrbracket}  & \eqStep &  \lambda  \SYSTEMmv{y'}  .   \mathsf{let} \, \textcolor{coeffectColor}{[}  \SYSTEMmv{y}  \textcolor{coeffectColor}{]} =  \SYSTEMmv{y'}  \, \mathsf{in} \,   [   \textcolor{GLcolor}{\llbracket}  \SYSTEMnt{t}  \textcolor{GLcolor}{\rrbracket}   /  \SYSTEMmv{x}  ]   \textcolor{GLcolor}{\llbracket}  \SYSTEMnt{t_{{\mathrm{1}}}}  \textcolor{GLcolor}{\rrbracket}     \\
\textit{(defn. interp [lhs])} & \Rightarrow &  \lambda  \SYSTEMmv{y'}  .   \mathsf{let} \, \textcolor{coeffectColor}{[}  \SYSTEMmv{y}  \textcolor{coeffectColor}{]} =  \SYSTEMmv{y'}  \, \mathsf{in} \,   \textcolor{GLcolor}{\llbracket}   [  \SYSTEMnt{t}  /  \SYSTEMmv{x}  ]  \SYSTEMnt{t_{{\mathrm{1}}}}   \textcolor{GLcolor}{\rrbracket}    & \eqStep &  \lambda  \SYSTEMmv{y'}  .   \mathsf{let} \, \textcolor{coeffectColor}{[}  \SYSTEMmv{y}  \textcolor{coeffectColor}{]} =  \SYSTEMmv{y'}  \, \mathsf{in} \,   [   \textcolor{GLcolor}{\llbracket}  \SYSTEMnt{t}  \textcolor{GLcolor}{\rrbracket}   /  \SYSTEMmv{x}  ]   \textcolor{GLcolor}{\llbracket}  \SYSTEMnt{t_{{\mathrm{1}}}}  \textcolor{GLcolor}{\rrbracket}     \\
\textit{(induction on $t1$)} & \Rightarrow & \top &
\end{array}
\end{align*}
\end{gather*}
% END (abs)
\item (app) $\SYSTEMnt{t'} \equiv  \SYSTEMnt{t_{{\mathrm{1}}}} \,  \SYSTEMnt{t_{{\mathrm{2}}}} $;
\begin{gather*}
\begin{align*}
\begin{array}{rcrlll}
\textit{(goal)} & &  \textcolor{GLcolor}{\llbracket}   [  \SYSTEMnt{t}  /  \SYSTEMmv{x}  ]  \SYSTEMsym{(}   \SYSTEMnt{t_{{\mathrm{1}}}} \,  \SYSTEMnt{t_{{\mathrm{2}}}}   \SYSTEMsym{)}   \textcolor{GLcolor}{\rrbracket}  & \eqStep &  [   \textcolor{GLcolor}{\llbracket}  \SYSTEMnt{t}  \textcolor{GLcolor}{\rrbracket}   /  \SYSTEMmv{x}  ]   \textcolor{GLcolor}{\llbracket}   \SYSTEMnt{t_{{\mathrm{1}}}} \,  \SYSTEMnt{t_{{\mathrm{2}}}}   \textcolor{GLcolor}{\rrbracket}   \\
\textit{(defn. subst [lhs])} & \Rightarrow &  \textcolor{GLcolor}{\llbracket}     [  \SYSTEMnt{t}  /  \SYSTEMmv{x}  ]  \SYSTEMnt{t_{{\mathrm{1}}}}   \,    [  \SYSTEMnt{t}  /  \SYSTEMmv{x}  ]  \SYSTEMnt{t_{{\mathrm{2}}}}     \textcolor{GLcolor}{\rrbracket}  & \eqStep &  [   \textcolor{GLcolor}{\llbracket}  \SYSTEMnt{t}  \textcolor{GLcolor}{\rrbracket}   /  \SYSTEMmv{x}  ]   \textcolor{GLcolor}{\llbracket}   \SYSTEMnt{t_{{\mathrm{1}}}} \,  \SYSTEMnt{t_{{\mathrm{2}}}}   \textcolor{GLcolor}{\rrbracket}   \\
\textit{(defn. interp [rhs])} & \Rightarrow &  \textcolor{GLcolor}{\llbracket}     [  \SYSTEMnt{t}  /  \SYSTEMmv{x}  ]  \SYSTEMnt{t_{{\mathrm{1}}}}   \,    [  \SYSTEMnt{t}  /  \SYSTEMmv{x}  ]  \SYSTEMnt{t_{{\mathrm{2}}}}     \textcolor{GLcolor}{\rrbracket}  & \eqStep &  [   \textcolor{GLcolor}{\llbracket}  \SYSTEMnt{t}  \textcolor{GLcolor}{\rrbracket}   /  \SYSTEMmv{x}  ]  \SYSTEMsym{(}    \textcolor{GLcolor}{\llbracket}  \SYSTEMnt{t_{{\mathrm{1}}}}  \textcolor{GLcolor}{\rrbracket}  \,   \textcolor{coeffectColor}{[}   \textcolor{GLcolor}{\llbracket}  \SYSTEMnt{t_{{\mathrm{2}}}}  \textcolor{GLcolor}{\rrbracket}   \textcolor{coeffectColor}{]}    \SYSTEMsym{)}  \\
\textit{(defn. subst [rhs])} & \Rightarrow &  \textcolor{GLcolor}{\llbracket}     [  \SYSTEMnt{t}  /  \SYSTEMmv{x}  ]  \SYSTEMnt{t_{{\mathrm{1}}}}   \,    [  \SYSTEMnt{t}  /  \SYSTEMmv{x}  ]  \SYSTEMnt{t_{{\mathrm{2}}}}     \textcolor{GLcolor}{\rrbracket}  & \eqStep &  \SYSTEMsym{(}   [   \textcolor{GLcolor}{\llbracket}  \SYSTEMnt{t}  \textcolor{GLcolor}{\rrbracket}   /  \SYSTEMmv{x}  ]   \textcolor{GLcolor}{\llbracket}  \SYSTEMnt{t_{{\mathrm{1}}}}  \textcolor{GLcolor}{\rrbracket}    \SYSTEMsym{)} \,  \SYSTEMsym{(}   [   \textcolor{GLcolor}{\llbracket}  \SYSTEMnt{t}  \textcolor{GLcolor}{\rrbracket}   /  \SYSTEMmv{x}  ]   \textcolor{coeffectColor}{[}   \textcolor{GLcolor}{\llbracket}  \SYSTEMnt{t_{{\mathrm{2}}}}  \textcolor{GLcolor}{\rrbracket}   \textcolor{coeffectColor}{]}    \SYSTEMsym{)}  \\
\textit{(defn. interp [lhs])} & \Rightarrow &   \textcolor{GLcolor}{\llbracket}   [  \SYSTEMnt{t}  /  \SYSTEMmv{x}  ]  \SYSTEMnt{t_{{\mathrm{1}}}}   \textcolor{GLcolor}{\rrbracket}  \,   \textcolor{coeffectColor}{[}   \textcolor{GLcolor}{\llbracket}   [  \SYSTEMnt{t}  /  \SYSTEMmv{x}  ]  \SYSTEMnt{t_{{\mathrm{2}}}}   \textcolor{GLcolor}{\rrbracket}   \textcolor{coeffectColor}{]}   & \eqStep &  \SYSTEMsym{(}   [   \textcolor{GLcolor}{\llbracket}  \SYSTEMnt{t}  \textcolor{GLcolor}{\rrbracket}   /  \SYSTEMmv{x}  ]   \textcolor{GLcolor}{\llbracket}  \SYSTEMnt{t_{{\mathrm{1}}}}  \textcolor{GLcolor}{\rrbracket}    \SYSTEMsym{)} \,  \SYSTEMsym{(}   [   \textcolor{GLcolor}{\llbracket}  \SYSTEMnt{t}  \textcolor{GLcolor}{\rrbracket}   /  \SYSTEMmv{x}  ]   \textcolor{coeffectColor}{[}   \textcolor{GLcolor}{\llbracket}  \SYSTEMnt{t_{{\mathrm{2}}}}  \textcolor{GLcolor}{\rrbracket}   \textcolor{coeffectColor}{]}    \SYSTEMsym{)}  \\
\textit{(defn. subst [rhs])} & \Rightarrow &   \textcolor{GLcolor}{\llbracket}   [  \SYSTEMnt{t}  /  \SYSTEMmv{x}  ]  \SYSTEMnt{t_{{\mathrm{1}}}}   \textcolor{GLcolor}{\rrbracket}  \,   \textcolor{coeffectColor}{[}   \textcolor{GLcolor}{\llbracket}   [  \SYSTEMnt{t}  /  \SYSTEMmv{x}  ]  \SYSTEMnt{t_{{\mathrm{2}}}}   \textcolor{GLcolor}{\rrbracket}   \textcolor{coeffectColor}{]}   & \eqStep &  \SYSTEMsym{(}   [   \textcolor{GLcolor}{\llbracket}  \SYSTEMnt{t}  \textcolor{GLcolor}{\rrbracket}   /  \SYSTEMmv{x}  ]   \textcolor{GLcolor}{\llbracket}  \SYSTEMnt{t_{{\mathrm{1}}}}  \textcolor{GLcolor}{\rrbracket}    \SYSTEMsym{)} \,   \textcolor{coeffectColor}{[}   [   \textcolor{GLcolor}{\llbracket}  \SYSTEMnt{t}  \textcolor{GLcolor}{\rrbracket}   /  \SYSTEMmv{x}  ]   \textcolor{GLcolor}{\llbracket}  \SYSTEMnt{t_{{\mathrm{2}}}}  \textcolor{GLcolor}{\rrbracket}    \textcolor{coeffectColor}{]}   \\
\textit{(induction on $\SYSTEMnt{t_{{\mathrm{1}}}}$ and $\SYSTEMnt{t_{{\mathrm{2}}}}$)} & \Rightarrow & \top
\end{array}
\end{align*}
\end{gather*}
\end{itemize}
% END (app)
\end{proof}

The operational correspondence now follows via the following proof:

\begin{proof}
For convenience, we expand the statement to $ \SYSTEMnt{t}  \rightsquigarrow_{\textsc{g} }  \SYSTEMnt{t'} $
  then $\exists t'' .   \textcolor{GLcolor}{\llbracket}  \SYSTEMnt{t}  \textcolor{GLcolor}{\rrbracket}   \rightsquigarrow_{\textsc{l} }^\ast  \SYSTEMnt{t''} $
   $\, \wedge \,  \textcolor{GLcolor}{\llbracket}  \SYSTEMnt{t'}  \textcolor{GLcolor}{\rrbracket}  \equiv \SYSTEMnt{t''}$.

By induction on reductions:
\begin{itemize}
\item (beta)
\[
\SYSTEMdruleSemGrdbeta{}
\]
The interpretation of the reducing term is:
\[
  \textcolor{GLcolor}{\llbracket}   \SYSTEMsym{(}   \lambda  \SYSTEMmv{x}  .  \SYSTEMnt{t_{{\mathrm{2}}}}   \SYSTEMsym{)} \,  \SYSTEMnt{t_{{\mathrm{1}}}}   \textcolor{GLcolor}{\rrbracket}   \equiv   \SYSTEMsym{(}   \lambda  \SYSTEMmv{y}  .   \mathsf{let} \, \textcolor{coeffectColor}{[}  \SYSTEMmv{x}  \textcolor{coeffectColor}{]} =  \SYSTEMmv{y}  \, \mathsf{in} \,   \textcolor{GLcolor}{\llbracket}  \SYSTEMnt{t_{{\mathrm{2}}}}  \textcolor{GLcolor}{\rrbracket}     \SYSTEMsym{)} \,   \textcolor{coeffectColor}{[}   \textcolor{GLcolor}{\llbracket}  \SYSTEMnt{t_{{\mathrm{1}}}}  \textcolor{GLcolor}{\rrbracket}   \textcolor{coeffectColor}{]}   
\]
where $y \# t$.
Then we construct the following reduction sequence in Linear Base:
\begin{align*}
& \inferrule*[right=\SYSTEMRenameRuleSemLinbeta{}]
{ }
{  \SYSTEMsym{(}   \lambda  \SYSTEMmv{y}  .   \mathsf{let} \, \textcolor{coeffectColor}{[}  \SYSTEMmv{x}  \textcolor{coeffectColor}{]} =  \SYSTEMmv{y}  \, \mathsf{in} \,   \textcolor{GLcolor}{\llbracket}  \SYSTEMnt{t_{{\mathrm{2}}}}  \textcolor{GLcolor}{\rrbracket}     \SYSTEMsym{)} \,   \textcolor{coeffectColor}{[}   \textcolor{GLcolor}{\llbracket}  \SYSTEMnt{t_{{\mathrm{1}}}}  \textcolor{GLcolor}{\rrbracket}   \textcolor{coeffectColor}{]}    \rightsquigarrow_{\textsc{l} }   \mathsf{let} \, \textcolor{coeffectColor}{[}  \SYSTEMmv{x}  \textcolor{coeffectColor}{]} =   \textcolor{coeffectColor}{[}   \textcolor{GLcolor}{\llbracket}  \SYSTEMnt{t_{{\mathrm{1}}}}  \textcolor{GLcolor}{\rrbracket}   \textcolor{coeffectColor}{]}   \, \mathsf{in} \,   \textcolor{GLcolor}{\llbracket}  \SYSTEMnt{t_{{\mathrm{2}}}}  \textcolor{GLcolor}{\rrbracket}   }
\\
& \inferrule*[right=\SYSTEMRenameRuleSemLinbetaBox{}]
{ }
{  \mathsf{let} \, \textcolor{coeffectColor}{[}  \SYSTEMmv{x}  \textcolor{coeffectColor}{]} =   \textcolor{coeffectColor}{[}   \textcolor{GLcolor}{\llbracket}  \SYSTEMnt{t_{{\mathrm{1}}}}  \textcolor{GLcolor}{\rrbracket}   \textcolor{coeffectColor}{]}   \, \mathsf{in} \,   \textcolor{GLcolor}{\llbracket}  \SYSTEMnt{t_{{\mathrm{2}}}}  \textcolor{GLcolor}{\rrbracket}    \rightsquigarrow_{\textsc{l} }   [   \textcolor{GLcolor}{\llbracket}  \SYSTEMnt{t_{{\mathrm{1}}}}  \textcolor{GLcolor}{\rrbracket}   /  \SYSTEMmv{x}  ]   \textcolor{GLcolor}{\llbracket}  \SYSTEMnt{t_{{\mathrm{2}}}}  \textcolor{GLcolor}{\rrbracket}   }
\end{align*}
And since $ \textcolor{GLcolor}{\llbracket}   [  \SYSTEMnt{t_{{\mathrm{1}}}}  /  \SYSTEMmv{x}  ]  \SYSTEMnt{t_{{\mathrm{2}}}}   \textcolor{GLcolor}{\rrbracket}  \equiv
 [   \textcolor{GLcolor}{\llbracket}  \SYSTEMnt{t_{{\mathrm{1}}}}  \textcolor{GLcolor}{\rrbracket}   /  \SYSTEMmv{x}  ]   \textcolor{GLcolor}{\llbracket}  \SYSTEMnt{t_{{\mathrm{2}}}}  \textcolor{GLcolor}{\rrbracket}  $
by Lemma~\ref{lemma:interp-grd-to-lin-preserves-subst}
then the goal is satisfied.

\item (congAppL)
\[
\SYSTEMdruleSemGrdcongAppL{}
\]
The interpretation of the reducing term is:
\[
  \textcolor{GLcolor}{\llbracket}   \SYSTEMnt{t_{{\mathrm{1}}}} \,  \SYSTEMnt{t_{{\mathrm{2}}}}   \textcolor{GLcolor}{\rrbracket}   \equiv    \textcolor{GLcolor}{\llbracket}  \SYSTEMnt{t_{{\mathrm{1}}}}  \textcolor{GLcolor}{\rrbracket}  \,   \textcolor{coeffectColor}{[}   \textcolor{GLcolor}{\llbracket}  \SYSTEMnt{t_{{\mathrm{2}}}}  \textcolor{GLcolor}{\rrbracket}   \textcolor{coeffectColor}{]}   
\]
By induction on $ \SYSTEMnt{t_{{\mathrm{1}}}}  \rightsquigarrow_{\textsc{g} }  \SYSTEMnt{t'_{{\mathrm{1}}}} $, we have
$\exists \SYSTEMnt{t''_{{\mathrm{1}}}} .   \textcolor{GLcolor}{\llbracket}  \SYSTEMnt{t_{{\mathrm{1}}}}  \textcolor{GLcolor}{\rrbracket}   \rightsquigarrow_{\textsc{l} }^\ast  \SYSTEMnt{t''_{{\mathrm{1}}}}  \wedge
 \SYSTEMnt{t''_{{\mathrm{1}}}}  \equiv   \textcolor{GLcolor}{\llbracket}  \SYSTEMnt{t'_{{\mathrm{1}}}}  \textcolor{GLcolor}{\rrbracket}  $. Therefore, we construct
a chain of reductions in Linear Base:
\[
\inferrule*[right=\SYSTEMRenameRuleSemLincongAppL{}]
{  \textcolor{GLcolor}{\llbracket}  \SYSTEMnt{t_{{\mathrm{1}}}}  \textcolor{GLcolor}{\rrbracket}   \rightsquigarrow_{\textsc{l} }  \SYSTEMnt{t''_{{\mathrm{1}}\,\SYSTEMmv{i}}} }
{   \textcolor{GLcolor}{\llbracket}  \SYSTEMnt{t_{{\mathrm{1}}}}  \textcolor{GLcolor}{\rrbracket}  \,   \textcolor{coeffectColor}{[}   \textcolor{GLcolor}{\llbracket}  \SYSTEMnt{t_{{\mathrm{2}}}}  \textcolor{GLcolor}{\rrbracket}   \textcolor{coeffectColor}{]}    \rightsquigarrow_{\textsc{l} }   \SYSTEMnt{t''_{{\mathrm{1}}\,\SYSTEMmv{i}}} \,   \textcolor{coeffectColor}{[}   \textcolor{GLcolor}{\llbracket}  \SYSTEMnt{t_{{\mathrm{2}}}}  \textcolor{GLcolor}{\rrbracket}   \textcolor{coeffectColor}{]}   }
\;\;
\cdots
\;\;
\inferrule*[right=\SYSTEMRenameRuleSemLincongAppL{}]
{  \SYSTEMnt{t''_{{\mathrm{1}}\,\SYSTEMmv{m}}} \,  \SYSTEMnt{t_{{\mathrm{1}}}}   \rightsquigarrow_{\textsc{l} }  \SYSTEMnt{t''_{{\mathrm{1}}}} }
{  \SYSTEMnt{t''_{{\mathrm{1}}\,\SYSTEMmv{m}}} \,   \textcolor{coeffectColor}{[}   \textcolor{GLcolor}{\llbracket}  \SYSTEMnt{t_{{\mathrm{2}}}}  \textcolor{GLcolor}{\rrbracket}   \textcolor{coeffectColor}{]}    \rightsquigarrow_{\textsc{l} }   \SYSTEMnt{t''_{{\mathrm{1}}}} \,   \textcolor{coeffectColor}{[}   \textcolor{GLcolor}{\llbracket}  \SYSTEMnt{t_{{\mathrm{2}}}}  \textcolor{GLcolor}{\rrbracket}   \textcolor{coeffectColor}{]}   }
\]
where $m$ is the length of the reduction sequence from
the induction.

satisfying the goal, since $ \SYSTEMnt{t''_{{\mathrm{1}}}}  \equiv   \textcolor{GLcolor}{\llbracket}  \SYSTEMnt{t'_{{\mathrm{1}}}}  \textcolor{GLcolor}{\rrbracket}  $
and therefore
$  \SYSTEMnt{t''_{{\mathrm{1}}}} \,   \textcolor{coeffectColor}{[}   \textcolor{GLcolor}{\llbracket}  \SYSTEMnt{t_{{\mathrm{2}}}}  \textcolor{GLcolor}{\rrbracket}   \textcolor{coeffectColor}{]}    \equiv    \textcolor{GLcolor}{\llbracket}  \SYSTEMnt{t'_{{\mathrm{1}}}}  \textcolor{GLcolor}{\rrbracket}  \,   \textcolor{coeffectColor}{[}   \textcolor{GLcolor}{\llbracket}  \SYSTEMnt{t_{{\mathrm{2}}}}  \textcolor{GLcolor}{\rrbracket}   \textcolor{coeffectColor}{]}   $.

\end{itemize}
\end{proof}

\subsubsection{Equation preservation}
\label{app:proofs-grad-to-lin-eqs}

%\gradToLinEq*

\begin{proof}
  By induction on the definition of the equational theory.
  \begin{itemize}
  \item \[\SYSTEMdruleGradEqbeta{}\]
    Follows by Theorem~\ref{thrm:gradToLinTranslation} (preservation of operational
    semantics), since this includes the $\beta$ case as a reduction.

  \item \[\SYSTEMdruleGradEqeta{}\]

    Preservation then follows by:
    \begin{align*}
      \begin{array}{rll}
                                  &  \textcolor{GLcolor}{\llbracket}    \lambda  \SYSTEMmv{x}  .  \SYSTEMnt{t}  \,  \SYSTEMmv{x}   \textcolor{GLcolor}{\rrbracket}  &  \\
   \textit{\{defn. translation\}} & =   \lambda  \SYSTEMmv{y}  .   \mathsf{let} \, \textcolor{coeffectColor}{[}  \SYSTEMmv{x}  \textcolor{coeffectColor}{]} =  \SYSTEMmv{y}  \, \mathsf{in} \,   \textcolor{GLcolor}{\llbracket}  \SYSTEMnt{t}  \textcolor{GLcolor}{\rrbracket}    \,   \textcolor{coeffectColor}{[}  \SYSTEMmv{x}  \textcolor{coeffectColor}{]}   \\
   \textit{\{\SYSTEMRenameRuleLinEqcongAbs{}+\SYSTEMRenameRuleLinEqletCommOne\}}
                                  & \equiv_L   \lambda  \SYSTEMmv{y}  .   \textcolor{GLcolor}{\llbracket}  \SYSTEMnt{t}  \textcolor{GLcolor}{\rrbracket}   \,  \SYSTEMsym{(}   \mathsf{let} \, \textcolor{coeffectColor}{[}  \SYSTEMmv{x}  \textcolor{coeffectColor}{]} =  \SYSTEMmv{y}  \, \mathsf{in} \,   \textcolor{coeffectColor}{[}  \SYSTEMmv{x}  \textcolor{coeffectColor}{]}    \SYSTEMsym{)}  \\
   \textit{\{\SYSTEMRenameRuleLinEqcongAbs{}+\SYSTEMRenameRuleLinEqbetaBox\}}
                                  & \equiv_L   \lambda  \SYSTEMmv{y}  .   \textcolor{GLcolor}{\llbracket}  \SYSTEMnt{t}  \textcolor{GLcolor}{\rrbracket}   \,  \SYSTEMmv{y}  \\
   \textit{\{\SYSTEMRenameRuleLinEqeta\}}
                                  & \equiv_L  \textcolor{GLcolor}{\llbracket}  \SYSTEMnt{t}  \textcolor{GLcolor}{\rrbracket} 
      \end{array}
    \end{align*}
  \item The remaining rules are congruences and all follow by induction with
    the translation.

  \end{itemize}

\end{proof}

\subsection{Proof of Soundness for Graded Base to Linear Base}
\label{app:proofs-lin-to-grad-cps}

\ifextended\linToGradCPSTranslation*\fi

\noindent
Type preservation is in Appendix~\ref{app:proofs-lin-to-grad-cps-typ},
operational correspondence in Appendix~\ref{app:proofs-lin-to-grad-cps-ops},
and equation preservation in Appendix~\ref{app:proofs-lin-to-grad-cps-eqns}.

\subsubsection{Type preservation}
\label{app:proofs-lin-to-grad-cps-typ}

\begin{lemma}[CPS form]
  \label{lemm:cps-form}
For all types $\SYSTEMnt{A}$ then
$\exists \SYSTEMnt{B}.\  \textcolor{LGcolor}{\llparenthesis} \smidge  \SYSTEMnt{A}  \smidge \textcolor{LGcolor}{\rrparenthesis}  =  \SYSTEMnt{B}  \xrightarrow{\textcolor{coeffectColor}{ \SYSTEMsym{1} } }   \mathrm{K}  $.
\end{lemma}

\begin{proof}
  By cases on the CPS translation
\end{proof}

\begin{proof}
By induction on the Linear Base typing:
\begin{itemize}
\item (var)
$$
\SYSTEMdruleLinvar{}
$$
By lemma~\ref{lemm:cps-form}, let $ \textcolor{LGcolor}{\llparenthesis} \smidge  \SYSTEMnt{A}  \smidge \textcolor{LGcolor}{\rrparenthesis}  =  \SYSTEMnt{B}  \xrightarrow{\textcolor{coeffectColor}{ \SYSTEMsym{1} } }   \mathrm{K}  $
for some $B$.

Therefore we construct the goal typing:
$$
\inferrule*[Right=\SYSTEMRenameRuleGradabs{}]
{
  \inferrule*[Right=\SYSTEMRenameRuleGradapp{}]
   {   \SYSTEMmv{x}  :_{\textcolor{coeffectColor}{ \SYSTEMsym{1} } }   (   \SYSTEMnt{B}  \xrightarrow{\textcolor{coeffectColor}{ \SYSTEMsym{1} } }   \mathrm{K}    )     \vdash_{\textsc{g} }  \SYSTEMmv{x}  :   (   \SYSTEMnt{B}  \xrightarrow{\textcolor{coeffectColor}{ \SYSTEMsym{1} } }   \mathrm{K}    )   \quad
       \SYSTEMmv{k}  :_{\textcolor{coeffectColor}{ \SYSTEMsym{1} } }  \SYSTEMnt{B}    \vdash_{\textsc{g} }  \SYSTEMmv{k}  :  \SYSTEMnt{B} }
   {    \SYSTEMmv{x}  :_{\textcolor{coeffectColor}{ \SYSTEMsym{1} } }   (   \SYSTEMnt{B}  \xrightarrow{\textcolor{coeffectColor}{ \SYSTEMsym{1} } }   \mathrm{K}    )    ,   \SYSTEMmv{k}  :_{\textcolor{coeffectColor}{ \SYSTEMsym{1} } }  \SYSTEMnt{B}    \vdash_{\textsc{g} }   \SYSTEMmv{x} \,  \SYSTEMmv{k}   :   \mathrm{K}  }
}
{    \SYSTEMmv{x}  :_{\textcolor{coeffectColor}{ \SYSTEMsym{1} } }   (   \SYSTEMnt{B}  \xrightarrow{\textcolor{coeffectColor}{ \SYSTEMsym{1} } }   \mathrm{K}    )     \vdash_{\textsc{g} }    \lambda  \SYSTEMmv{k}  .  \SYSTEMmv{x}  \,  \SYSTEMmv{k}   :   (   \SYSTEMnt{B}  \xrightarrow{\textcolor{coeffectColor}{ \SYSTEMsym{1} } }   \mathrm{K}    )   }
$$
\end{itemize}

\item (app)
$$
\SYSTEMdruleLinapp{}
$$
By induction we have that:
\begin{align*}
  &   \textcolor{LGcolor}{\llparenthesis}  \Gamma_{{\mathrm{1}}}  \textcolor{LGcolor}{\rrparenthesis}   \vdash_{\textsc{g} }   \textcolor{LGcolor}{\llparenthesis} \smidge  \SYSTEMnt{t_{{\mathrm{1}}}}  \smidge \textcolor{LGcolor}{\rrparenthesis}   :    (    (    \textcolor{LGcolor}{\llparenthesis} \smidge  \SYSTEMnt{A}  \smidge \textcolor{LGcolor}{\rrparenthesis}   \xrightarrow{\textcolor{coeffectColor}{ \SYSTEMsym{1} } }   \textcolor{LGcolor}{\llparenthesis} \smidge  \SYSTEMnt{B}  \smidge \textcolor{LGcolor}{\rrparenthesis}    )   \xrightarrow{\textcolor{coeffectColor}{ \SYSTEMsym{1} } }   \mathrm{K}    )   \xrightarrow{\textcolor{coeffectColor}{ \SYSTEMsym{1} } }   \mathrm{K}    \\
  &   \textcolor{LGcolor}{\llparenthesis}  \Gamma_{{\mathrm{2}}}  \textcolor{LGcolor}{\rrparenthesis}   \vdash_{\textsc{g} }   \textcolor{LGcolor}{\llparenthesis} \smidge  \SYSTEMnt{t_{{\mathrm{2}}}}  \smidge \textcolor{LGcolor}{\rrparenthesis}   :   \textcolor{LGcolor}{\llparenthesis} \smidge  \SYSTEMnt{A}  \smidge \textcolor{LGcolor}{\rrparenthesis}  
\end{align*}
and by lemma~\ref{lemm:cps-form}, $ \textcolor{LGcolor}{\llparenthesis} \smidge  \SYSTEMnt{B}  \smidge \textcolor{LGcolor}{\rrparenthesis}  =  \SYSTEMnt{C}  \xrightarrow{\textcolor{coeffectColor}{ \SYSTEMsym{1} } }   \mathrm{K}  $ for some
$C$.
\begin{gather*}
\begin{align*}
  \inferrule*[right=\SYSTEMRenameRuleGradabs{}]
{
  \inferrule*[right=\SYSTEMRenameRuleGradapp{}]
  { (ih1) :   (    (    \textcolor{LGcolor}{\llparenthesis} \smidge  \SYSTEMnt{A}  \smidge \textcolor{LGcolor}{\rrparenthesis}   \xrightarrow{\textcolor{coeffectColor}{ \SYSTEMsym{1} } }   \textcolor{LGcolor}{\llparenthesis} \smidge  \SYSTEMnt{B}  \smidge \textcolor{LGcolor}{\rrparenthesis}    )   \xrightarrow{\textcolor{coeffectColor}{ \SYSTEMsym{1} } }   \mathrm{K}    )   \xrightarrow{\textcolor{coeffectColor}{ \SYSTEMsym{1} } }   \mathrm{K}   \quad
  \inferrule*[right=\SYSTEMRenameRuleLinabs{}]
    {
      \inferrule*[right=\SYSTEMRenameRuleGradapp{}]
      {
        \inferrule*[right=\SYSTEMRenameRuleGradapp{}]
        {   \SYSTEMmv{f}  :    \textcolor{LGcolor}{\llparenthesis} \smidge  \SYSTEMnt{A}  \smidge \textcolor{LGcolor}{\rrparenthesis}   \xrightarrow{\textcolor{coeffectColor}{ \SYSTEMsym{1} } }   \textcolor{LGcolor}{\llparenthesis} \smidge  \SYSTEMnt{B}  \smidge \textcolor{LGcolor}{\rrparenthesis}      \vdash_{\textsc{l} }  \SYSTEMmv{f}  :    \textcolor{LGcolor}{\llparenthesis} \smidge  \SYSTEMnt{A}  \smidge \textcolor{LGcolor}{\rrparenthesis}   \xrightarrow{\textcolor{coeffectColor}{ \SYSTEMsym{1} } }   \textcolor{LGcolor}{\llparenthesis} \smidge  \SYSTEMnt{B}  \smidge \textcolor{LGcolor}{\rrparenthesis}    \quad   \textcolor{LGcolor}{\llparenthesis}  \Gamma_{{\mathrm{2}}}  \textcolor{LGcolor}{\rrparenthesis}   \vdash_{\textsc{g} }   \textcolor{LGcolor}{\llparenthesis} \smidge  \SYSTEMnt{t_{{\mathrm{2}}}}  \smidge \textcolor{LGcolor}{\rrparenthesis}   :   \textcolor{LGcolor}{\llparenthesis} \smidge  \SYSTEMnt{A}  \smidge \textcolor{LGcolor}{\rrparenthesis}  }
        {   \textcolor{LGcolor}{\llparenthesis}  \Gamma_{{\mathrm{2}}}  \textcolor{LGcolor}{\rrparenthesis}  ,   \SYSTEMmv{f}  :_{\textcolor{coeffectColor}{ \SYSTEMsym{1} } }     \textcolor{LGcolor}{\llparenthesis} \smidge  \SYSTEMnt{A}  \smidge \textcolor{LGcolor}{\rrparenthesis}   \xrightarrow{\textcolor{coeffectColor}{ \SYSTEMsym{1} } }   \textcolor{LGcolor}{\llparenthesis} \smidge  \SYSTEMnt{B}  \smidge \textcolor{LGcolor}{\rrparenthesis}       \vdash_{\textsc{g} }   \SYSTEMmv{f} \,   \textcolor{LGcolor}{\llparenthesis} \smidge  \SYSTEMnt{t_{{\mathrm{2}}}}  \smidge \textcolor{LGcolor}{\rrparenthesis}    :   \textcolor{LGcolor}{\llparenthesis} \smidge  \SYSTEMnt{B}  \smidge \textcolor{LGcolor}{\rrparenthesis}  }
      }
      {    \textcolor{LGcolor}{\llparenthesis}  \Gamma_{{\mathrm{2}}}  \textcolor{LGcolor}{\rrparenthesis}  ,   \SYSTEMmv{k}  :_{\textcolor{coeffectColor}{ \SYSTEMsym{1} } }  \SYSTEMnt{C}   ,   \SYSTEMmv{f}  :_{\textcolor{coeffectColor}{ \SYSTEMsym{1} } }     \textcolor{LGcolor}{\llparenthesis} \smidge  \SYSTEMnt{A}  \smidge \textcolor{LGcolor}{\rrparenthesis}   \xrightarrow{\textcolor{coeffectColor}{ \SYSTEMsym{1} } }   \textcolor{LGcolor}{\llparenthesis} \smidge  \SYSTEMnt{B}  \smidge \textcolor{LGcolor}{\rrparenthesis}       \vdash_{\textsc{g} }     \SYSTEMmv{f} \,   \textcolor{LGcolor}{\llparenthesis} \smidge  \SYSTEMnt{t_{{\mathrm{2}}}}  \smidge \textcolor{LGcolor}{\rrparenthesis}    \,  \SYSTEMmv{k}   :   \mathrm{K}  }
    }
    {   \textcolor{LGcolor}{\llparenthesis}  \Gamma_{{\mathrm{2}}}  \textcolor{LGcolor}{\rrparenthesis}  ,   \SYSTEMmv{k}  :_{\textcolor{coeffectColor}{ \SYSTEMsym{1} } }  \SYSTEMnt{C}    \vdash_{\textsc{g} }   \lambda  \SYSTEMmv{f}  .      \SYSTEMmv{f} \,   \textcolor{LGcolor}{\llparenthesis} \smidge  \SYSTEMnt{t_{{\mathrm{2}}}}  \smidge \textcolor{LGcolor}{\rrparenthesis}    \,  \SYSTEMmv{k}     :    (    \textcolor{LGcolor}{\llparenthesis} \smidge  \SYSTEMnt{A}  \smidge \textcolor{LGcolor}{\rrparenthesis}   \xrightarrow{\textcolor{coeffectColor}{ \SYSTEMsym{1} } }   \textcolor{LGcolor}{\llparenthesis} \smidge  \SYSTEMnt{B}  \smidge \textcolor{LGcolor}{\rrparenthesis}    )   \xrightarrow{\textcolor{coeffectColor}{ \SYSTEMsym{1} } }   \mathrm{K}    }
  }
  {      \textcolor{LGcolor}{\llparenthesis}  \Gamma_{{\mathrm{1}}}  \textcolor{LGcolor}{\rrparenthesis}   ,   \textcolor{LGcolor}{\llparenthesis}  \Gamma_{{\mathrm{2}}}  \textcolor{LGcolor}{\rrparenthesis}    ,   \SYSTEMmv{k}  :_{\textcolor{coeffectColor}{ \SYSTEMsym{1} } }  \SYSTEMnt{C}    \vdash_{\textsc{g} }     \textcolor{LGcolor}{\llparenthesis} \smidge  \SYSTEMnt{t_{{\mathrm{1}}}}  \smidge \textcolor{LGcolor}{\rrparenthesis}  \,  \SYSTEMsym{(}   \lambda  \SYSTEMmv{f}  .      \SYSTEMmv{f} \,   \textcolor{LGcolor}{\llparenthesis} \smidge  \SYSTEMnt{t_{{\mathrm{2}}}}  \smidge \textcolor{LGcolor}{\rrparenthesis}    \,  \SYSTEMmv{k}     \SYSTEMsym{)}    :   \mathrm{K}   }
}
{   \textcolor{LGcolor}{\llparenthesis}  \Gamma_{{\mathrm{1}}}  \textcolor{LGcolor}{\rrparenthesis}   ,   \textcolor{LGcolor}{\llparenthesis}  \Gamma_{{\mathrm{2}}}  \textcolor{LGcolor}{\rrparenthesis}    \vdash_{\textsc{g} }   \lambda  \SYSTEMmv{k}  .     \textcolor{LGcolor}{\llparenthesis} \smidge  \SYSTEMnt{t_{{\mathrm{1}}}}  \smidge \textcolor{LGcolor}{\rrparenthesis}  \,  \SYSTEMsym{(}   \lambda  \SYSTEMmv{f}  .      \SYSTEMmv{f} \,   \textcolor{LGcolor}{\llparenthesis} \smidge  \SYSTEMnt{t_{{\mathrm{2}}}}  \smidge \textcolor{LGcolor}{\rrparenthesis}    \,  \SYSTEMmv{k}     \SYSTEMsym{)}     :   \SYSTEMnt{C}  \xrightarrow{\textcolor{coeffectColor}{ \SYSTEMsym{1} } }   \mathrm{K}   }
\end{align*}
\end{gather*}

\item (abs)
$$
\SYSTEMdruleLinabs{}
$$
By induction we have that:
\begin{align*}
   \textcolor{LGcolor}{\llparenthesis}  \Gamma  \textcolor{LGcolor}{\rrparenthesis}  ,   \SYSTEMmv{x}  :_{\textcolor{coeffectColor}{ \SYSTEMsym{1} } }   \textcolor{LGcolor}{\llparenthesis} \smidge  \SYSTEMnt{A}  \smidge \textcolor{LGcolor}{\rrparenthesis}     \vdash_{\textsc{g} }   \textcolor{LGcolor}{\llparenthesis} \smidge  \SYSTEMnt{t}  \smidge \textcolor{LGcolor}{\rrparenthesis}   :   \textcolor{LGcolor}{\llparenthesis} \smidge  \SYSTEMnt{B}  \smidge \textcolor{LGcolor}{\rrparenthesis}  
\end{align*}
Then we construct:
\begin{gather*}
\begin{align*}
  \inferrule*[right=\SYSTEMRenameRuleGradabs{}]
{
  \inferrule*[right=\SYSTEMRenameRuleGradapp{}]
  {
    \inferrule*[right=\SYSTEMRenameRuleGradvar{}]
    {~}
    {   \SYSTEMmv{k}  :_{\textcolor{coeffectColor}{ \SYSTEMsym{1} } }     (    \textcolor{LGcolor}{\llparenthesis} \smidge  \SYSTEMnt{A}  \smidge \textcolor{LGcolor}{\rrparenthesis}   \xrightarrow{\textcolor{coeffectColor}{ \SYSTEMsym{1} } }   \textcolor{LGcolor}{\llparenthesis} \smidge  \SYSTEMnt{B}  \smidge \textcolor{LGcolor}{\rrparenthesis}    )   \xrightarrow{\textcolor{coeffectColor}{ \SYSTEMsym{1} } }   \mathrm{K}       \vdash_{\textsc{g} }  \SYSTEMmv{k}  :    (    \textcolor{LGcolor}{\llparenthesis} \smidge  \SYSTEMnt{A}  \smidge \textcolor{LGcolor}{\rrparenthesis}   \xrightarrow{\textcolor{coeffectColor}{ \SYSTEMsym{1} } }   \textcolor{LGcolor}{\llparenthesis} \smidge  \SYSTEMnt{B}  \smidge \textcolor{LGcolor}{\rrparenthesis}    )   \xrightarrow{\textcolor{coeffectColor}{ \SYSTEMsym{1} } }   \mathrm{K}   }
    \quad
    \inferrule*[right=\SYSTEMRenameRuleGradabs{}]
    {
      (ih) :    \textcolor{LGcolor}{\llparenthesis}  \Gamma  \textcolor{LGcolor}{\rrparenthesis}  ,   \SYSTEMmv{x}  :_{\textcolor{coeffectColor}{ \SYSTEMsym{1} } }   \textcolor{LGcolor}{\llparenthesis} \smidge  \SYSTEMnt{A}  \smidge \textcolor{LGcolor}{\rrparenthesis}     \vdash_{\textsc{g} }   \textcolor{LGcolor}{\llparenthesis} \smidge  \SYSTEMnt{t}  \smidge \textcolor{LGcolor}{\rrparenthesis}   :   \textcolor{LGcolor}{\llparenthesis} \smidge  \SYSTEMnt{B}  \smidge \textcolor{LGcolor}{\rrparenthesis}  
    }
    {  \textcolor{LGcolor}{\llparenthesis}  \Gamma  \textcolor{LGcolor}{\rrparenthesis}   \vdash_{\textsc{g} }   \lambda  \SYSTEMmv{x}  .   \textcolor{LGcolor}{\llparenthesis} \smidge  \SYSTEMnt{t}  \smidge \textcolor{LGcolor}{\rrparenthesis}    :    \textcolor{LGcolor}{\llparenthesis} \smidge  \SYSTEMnt{A}  \smidge \textcolor{LGcolor}{\rrparenthesis}   \xrightarrow{\textcolor{coeffectColor}{ \SYSTEMsym{1} } }   \textcolor{LGcolor}{\llparenthesis} \smidge  \SYSTEMnt{B}  \smidge \textcolor{LGcolor}{\rrparenthesis}   }
  }
  {     \textcolor{LGcolor}{\llparenthesis}  \Gamma  \textcolor{LGcolor}{\rrparenthesis}   ,   \SYSTEMmv{k}  :_{\textcolor{coeffectColor}{ \SYSTEMsym{1} } }     (    \textcolor{LGcolor}{\llparenthesis} \smidge  \SYSTEMnt{A}  \smidge \textcolor{LGcolor}{\rrparenthesis}   \xrightarrow{\textcolor{coeffectColor}{ \SYSTEMsym{1} } }   \textcolor{LGcolor}{\llparenthesis} \smidge  \SYSTEMnt{B}  \smidge \textcolor{LGcolor}{\rrparenthesis}    )   \xrightarrow{\textcolor{coeffectColor}{ \SYSTEMsym{1} } }   \mathrm{K}       \vdash_{\textsc{g} }    \SYSTEMmv{k} \,  \SYSTEMsym{(}   \lambda  \SYSTEMmv{x}  .   \textcolor{LGcolor}{\llparenthesis} \smidge  \SYSTEMnt{t}  \smidge \textcolor{LGcolor}{\rrparenthesis}    \SYSTEMsym{)}    :   \mathrm{K}   }
}
{  \textcolor{LGcolor}{\llparenthesis}  \Gamma  \textcolor{LGcolor}{\rrparenthesis}   \vdash_{\textsc{g} }   \lambda  \SYSTEMmv{k}  .    \SYSTEMmv{k} \,  \SYSTEMsym{(}   \lambda  \SYSTEMmv{x}  .   \textcolor{LGcolor}{\llparenthesis} \smidge  \SYSTEMnt{t}  \smidge \textcolor{LGcolor}{\rrparenthesis}    \SYSTEMsym{)}     :    (    (    \textcolor{LGcolor}{\llparenthesis} \smidge  \SYSTEMnt{A}  \smidge \textcolor{LGcolor}{\rrparenthesis}   \xrightarrow{\textcolor{coeffectColor}{ \SYSTEMsym{1} } }   \textcolor{LGcolor}{\llparenthesis} \smidge  \SYSTEMnt{B}  \smidge \textcolor{LGcolor}{\rrparenthesis}    )   \xrightarrow{\textcolor{coeffectColor}{ \SYSTEMsym{1} } }   \mathrm{K}    )   \xrightarrow{\textcolor{coeffectColor}{ \SYSTEMsym{1} } }   \mathrm{K}   }
\end{align*}
\end{gather*}

\item (pr)
$$
\SYSTEMdruleLinpr{}
$$
By induction we have that:
\begin{align*}
  \textcolor{LGcolor}{\llparenthesis}  \Gamma  \textcolor{LGcolor}{\rrparenthesis}   \vdash_{\textsc{g} }   \textcolor{LGcolor}{\llparenthesis} \smidge  \SYSTEMnt{t}  \smidge \textcolor{LGcolor}{\rrparenthesis}   :   \textcolor{LGcolor}{\llparenthesis} \smidge  \SYSTEMnt{A}  \smidge \textcolor{LGcolor}{\rrparenthesis}  
\end{align*}
Then we construct:
\begin{align*}
  \inferrule*[right=\SYSTEMRenameRuleGradabs{}]
{
  \inferrule*[right=\SYSTEMRenameRuleGradapp{}]
  {
    \inferrule*[right=\SYSTEMRenameRuleGradvar{}]
    {~}
    {   \SYSTEMmv{k}  :_{\textcolor{coeffectColor}{ \SYSTEMsym{1} } }   (    \textcolor{LGcolor}{\llparenthesis} \smidge  \SYSTEMnt{A}  \smidge \textcolor{LGcolor}{\rrparenthesis}   \xrightarrow{\textcolor{coeffectColor}{ \SYSTEMnt{r} } }   \mathrm{K}    )     \vdash_{\textsc{g} }  \SYSTEMmv{k}  :    \textcolor{LGcolor}{\llparenthesis} \smidge  \SYSTEMnt{A}  \smidge \textcolor{LGcolor}{\rrparenthesis}   \xrightarrow{\textcolor{coeffectColor}{ \SYSTEMnt{r} } }   \mathrm{K}   }
    \quad
    (ih) :   \textcolor{LGcolor}{\llparenthesis}  \Gamma  \textcolor{LGcolor}{\rrparenthesis}   \vdash_{\textsc{g} }   \textcolor{LGcolor}{\llparenthesis} \smidge  \SYSTEMnt{t}  \smidge \textcolor{LGcolor}{\rrparenthesis}   :   \textcolor{LGcolor}{\llparenthesis} \smidge  \SYSTEMnt{A}  \smidge \textcolor{LGcolor}{\rrparenthesis}  
  }
  {     \textcolor{coeffectColor}{ \SYSTEMnt{r}  \cdot}   \textcolor{LGcolor}{\llparenthesis}  \Gamma  \textcolor{LGcolor}{\rrparenthesis}    ,   \SYSTEMmv{k}  :_{\textcolor{coeffectColor}{ \SYSTEMsym{1} } }   (    \textcolor{LGcolor}{\llparenthesis} \smidge  \SYSTEMnt{A}  \smidge \textcolor{LGcolor}{\rrparenthesis}   \xrightarrow{\textcolor{coeffectColor}{ \SYSTEMnt{r} } }   \mathrm{K}    )     \vdash_{\textsc{g} }    \SYSTEMmv{k} \,   \textcolor{LGcolor}{\llparenthesis} \smidge  \SYSTEMnt{t}  \smidge \textcolor{LGcolor}{\rrparenthesis}     :   \mathrm{K}   }
}
{  \textcolor{coeffectColor}{ \SYSTEMnt{r}  \cdot}   \textcolor{LGcolor}{\llparenthesis}  \Gamma  \textcolor{LGcolor}{\rrparenthesis}    \vdash_{\textsc{g} }   \lambda  \SYSTEMmv{k}  .    \SYSTEMmv{k} \,   \textcolor{LGcolor}{\llparenthesis} \smidge  \SYSTEMnt{t}  \smidge \textcolor{LGcolor}{\rrparenthesis}      :    (    \textcolor{LGcolor}{\llparenthesis} \smidge  \SYSTEMnt{A}  \smidge \textcolor{LGcolor}{\rrparenthesis}   \xrightarrow{\textcolor{coeffectColor}{ \SYSTEMnt{r} } }   \mathrm{K}    )   \xrightarrow{\textcolor{coeffectColor}{ \SYSTEMsym{1} } }   \mathrm{K}   }
\end{align*}

\item (let)
$$
\SYSTEMdruleLinlet{}
$$
By induction we have that:
\begin{align*}
&   \textcolor{LGcolor}{\llparenthesis}  \Gamma_{{\mathrm{1}}}  \textcolor{LGcolor}{\rrparenthesis}   \vdash_{\textsc{g} }   \textcolor{LGcolor}{\llparenthesis} \smidge  \SYSTEMnt{t_{{\mathrm{1}}}}  \smidge \textcolor{LGcolor}{\rrparenthesis}   :   \textcolor{LGcolor}{\llparenthesis} \smidge   \textcolor{coeffectColor}{\square_{ \SYSTEMnt{r} } }  \SYSTEMnt{A}   \smidge \textcolor{LGcolor}{\rrparenthesis}   \\
&    \textcolor{LGcolor}{\llparenthesis}  \Gamma_{{\mathrm{2}}}  \textcolor{LGcolor}{\rrparenthesis}  ,   \SYSTEMmv{x}  :_{\textcolor{coeffectColor}{ \SYSTEMnt{r} } }   \textcolor{LGcolor}{\llparenthesis} \smidge  \SYSTEMnt{A}  \smidge \textcolor{LGcolor}{\rrparenthesis}     \vdash_{\textsc{g} }   \textcolor{LGcolor}{\llparenthesis} \smidge  \SYSTEMnt{t_{{\mathrm{2}}}}  \smidge \textcolor{LGcolor}{\rrparenthesis}   :   \textcolor{LGcolor}{\llparenthesis} \smidge  \SYSTEMnt{B}  \smidge \textcolor{LGcolor}{\rrparenthesis}   \\
\end{align*}
and by lemma~\ref{lemm:cps-form}, $ \textcolor{LGcolor}{\llparenthesis} \smidge  \SYSTEMnt{B}  \smidge \textcolor{LGcolor}{\rrparenthesis}  =  \SYSTEMnt{C}  \xrightarrow{\textcolor{coeffectColor}{ \SYSTEMsym{1} } }   \mathrm{K}  $ for some
$C$, then we construct:
\begin{gather*}
\begin{align*}
  \inferrule*[right=\SYSTEMRenameRuleGradabs{}]
{
  \inferrule*[right=\SYSTEMRenameRuleGradapp{}]
  {
    (ih1) :   \textcolor{LGcolor}{\llparenthesis}  \Gamma_{{\mathrm{1}}}  \textcolor{LGcolor}{\rrparenthesis}   \vdash_{\textsc{g} }   \textcolor{LGcolor}{\llparenthesis} \smidge  \SYSTEMnt{t_{{\mathrm{1}}}}  \smidge \textcolor{LGcolor}{\rrparenthesis}   :    (    \textcolor{LGcolor}{\llparenthesis} \smidge  \SYSTEMnt{A}  \smidge \textcolor{LGcolor}{\rrparenthesis}   \xrightarrow{\textcolor{coeffectColor}{ \SYSTEMnt{r} } }   \mathrm{K}    )   \xrightarrow{\textcolor{coeffectColor}{ \SYSTEMsym{1} } }   \mathrm{K}   
    \quad
    \inferrule*[right=\SYSTEMRenameRuleGradabs{}]
    {
      \inferrule*[right=\SYSTEMRenameRuleGradapp{}]
      {
        (ih2) :    \textcolor{LGcolor}{\llparenthesis}  \Gamma_{{\mathrm{2}}}  \textcolor{LGcolor}{\rrparenthesis}  ,   \SYSTEMmv{x}  :_{\textcolor{coeffectColor}{ \SYSTEMnt{r} } }   \textcolor{LGcolor}{\llparenthesis} \smidge  \SYSTEMnt{A}  \smidge \textcolor{LGcolor}{\rrparenthesis}     \vdash_{\textsc{g} }   \textcolor{LGcolor}{\llparenthesis} \smidge  \SYSTEMnt{t_{{\mathrm{2}}}}  \smidge \textcolor{LGcolor}{\rrparenthesis}   :   \textcolor{LGcolor}{\llparenthesis} \smidge  \SYSTEMnt{B}  \smidge \textcolor{LGcolor}{\rrparenthesis}  
        \quad
        \inferrule*[right=\SYSTEMRenameRuleGradvar{}]
        {~}
        {   \SYSTEMmv{k}  :_{\textcolor{coeffectColor}{ \SYSTEMsym{1} } }  \SYSTEMnt{C}    \vdash_{\textsc{g} }  \SYSTEMmv{k}  :  \SYSTEMnt{C} }
      }
      {    \textcolor{LGcolor}{\llparenthesis}  \Gamma_{{\mathrm{2}}}  \textcolor{LGcolor}{\rrparenthesis}  ,   \SYSTEMmv{x}  :_{\textcolor{coeffectColor}{ \SYSTEMnt{r} } }   \textcolor{LGcolor}{\llparenthesis} \smidge  \SYSTEMnt{A}  \smidge \textcolor{LGcolor}{\rrparenthesis}    ,   \SYSTEMmv{k}  :_{\textcolor{coeffectColor}{ \SYSTEMsym{1} } }  \SYSTEMnt{C}    \vdash_{\textsc{g} }     \textcolor{LGcolor}{\llparenthesis} \smidge  \SYSTEMnt{t_{{\mathrm{2}}}}  \smidge \textcolor{LGcolor}{\rrparenthesis}  \,  \SYSTEMmv{k}    :   \mathrm{K}  }
    }
    {   \textcolor{LGcolor}{\llparenthesis}  \Gamma_{{\mathrm{2}}}  \textcolor{LGcolor}{\rrparenthesis}  ,   \SYSTEMmv{k}  :_{\textcolor{coeffectColor}{ \SYSTEMsym{1} } }  \SYSTEMnt{C}    \vdash_{\textsc{g} }   \lambda  \SYSTEMmv{x}  .     \textcolor{LGcolor}{\llparenthesis} \smidge  \SYSTEMnt{t_{{\mathrm{2}}}}  \smidge \textcolor{LGcolor}{\rrparenthesis}  \,  \SYSTEMmv{k}     :    \textcolor{LGcolor}{\llparenthesis} \smidge  \SYSTEMnt{A}  \smidge \textcolor{LGcolor}{\rrparenthesis}   \xrightarrow{\textcolor{coeffectColor}{ \SYSTEMnt{r} } }   \mathrm{K}   }
  }
  {      \textcolor{LGcolor}{\llparenthesis}  \Gamma_{{\mathrm{1}}}  \textcolor{LGcolor}{\rrparenthesis}   ,   \textcolor{LGcolor}{\llparenthesis}  \Gamma_{{\mathrm{2}}}  \textcolor{LGcolor}{\rrparenthesis}    ,   \SYSTEMmv{k}  :_{\textcolor{coeffectColor}{ \SYSTEMsym{1} } }  \SYSTEMnt{C}    \vdash_{\textsc{g} }     \textcolor{LGcolor}{\llparenthesis} \smidge  \SYSTEMnt{t_{{\mathrm{1}}}}  \smidge \textcolor{LGcolor}{\rrparenthesis}  \,  \SYSTEMsym{(}   \lambda  \SYSTEMmv{x}  .     \textcolor{LGcolor}{\llparenthesis} \smidge  \SYSTEMnt{t_{{\mathrm{2}}}}  \smidge \textcolor{LGcolor}{\rrparenthesis}  \,  \SYSTEMmv{k}     \SYSTEMsym{)}    :   \mathrm{K}   }
}
{   \textcolor{LGcolor}{\llparenthesis}  \Gamma_{{\mathrm{1}}}  \textcolor{LGcolor}{\rrparenthesis}   ,   \textcolor{LGcolor}{\llparenthesis}  \Gamma_{{\mathrm{2}}}  \textcolor{LGcolor}{\rrparenthesis}    \vdash_{\textsc{g} }   \lambda  \SYSTEMmv{k}  .     \textcolor{LGcolor}{\llparenthesis} \smidge  \SYSTEMnt{t_{{\mathrm{1}}}}  \smidge \textcolor{LGcolor}{\rrparenthesis}  \,  \SYSTEMsym{(}   \lambda  \SYSTEMmv{x}  .     \textcolor{LGcolor}{\llparenthesis} \smidge  \SYSTEMnt{t_{{\mathrm{2}}}}  \smidge \textcolor{LGcolor}{\rrparenthesis}  \,  \SYSTEMmv{k}     \SYSTEMsym{)}     :   \SYSTEMnt{C}  \xrightarrow{\textcolor{coeffectColor}{ \SYSTEMsym{1} } }   \mathrm{K}   }
\end{align*}
\end{gather*}
\end{proof}

\subsubsection{Operational correspondence}
\label{app:proofs-lin-to-grad-cps-ops}

\begin{lemma}[Interpretation preserves substitution]
\label{lemma:interp-lin-to-grd-cps-preserves-subst}
For all Linear Base terms $\SYSTEMnt{t}, \SYSTEMnt{t'}$ then
$ \textcolor{LGcolor}{\llparenthesis} \smidge   [  \SYSTEMnt{t}  /  \SYSTEMmv{x}  ]  \SYSTEMnt{t'}   \smidge \textcolor{LGcolor}{\rrparenthesis}  \equiv  [   \textcolor{LGcolor}{\llparenthesis} \smidge  \SYSTEMnt{t}  \smidge \textcolor{LGcolor}{\rrparenthesis}   /  \SYSTEMmv{x}  ]   \textcolor{LGcolor}{\llparenthesis} \smidge  \SYSTEMnt{t'}  \smidge \textcolor{LGcolor}{\rrparenthesis}  $.
\end{lemma}

\begin{proof}
By induction on the receiving term $\SYSTEMnt{t'}$:
\begin{itemize}
\item (var) $\SYSTEMnt{t'} \equiv \SYSTEMmv{y}$:

\begin{itemize}
\item $\SYSTEMmv{x} \equiv \SYSTEMmv{y}$ then we refine the goal:
\begin{align*}
\begin{array}{rcrlll}
\textit{(goal)} & &  \textcolor{LGcolor}{\llparenthesis} \smidge   [  \SYSTEMnt{t}  /  \SYSTEMmv{x}  ]  \SYSTEMmv{x}   \smidge \textcolor{LGcolor}{\rrparenthesis}  & \eqStep &  [   \textcolor{LGcolor}{\llparenthesis} \smidge  \SYSTEMnt{t}  \smidge \textcolor{LGcolor}{\rrparenthesis}   /  \SYSTEMmv{x}  ]   \textcolor{LGcolor}{\llparenthesis} \smidge  \SYSTEMmv{x}  \smidge \textcolor{LGcolor}{\rrparenthesis}   \\
\textit{(defn. subst [lhs])} & \Rightarrow &  \textcolor{LGcolor}{\llparenthesis} \smidge  \SYSTEMnt{t}  \smidge \textcolor{LGcolor}{\rrparenthesis}  & \eqStep &  [   \textcolor{LGcolor}{\llparenthesis} \smidge  \SYSTEMnt{t}  \smidge \textcolor{LGcolor}{\rrparenthesis}   /  \SYSTEMmv{x}  ]   \textcolor{LGcolor}{\llparenthesis} \smidge  \SYSTEMmv{x}  \smidge \textcolor{LGcolor}{\rrparenthesis}   \\
\textit{(defn. interp [rhs])} & \Rightarrow &  \textcolor{LGcolor}{\llparenthesis} \smidge  \SYSTEMnt{t}  \smidge \textcolor{LGcolor}{\rrparenthesis}  & \eqStep &  [   \textcolor{LGcolor}{\llparenthesis} \smidge  \SYSTEMnt{t}  \smidge \textcolor{LGcolor}{\rrparenthesis}   /  \SYSTEMmv{x}  ]   \lambda  \SYSTEMmv{k}  .    \SYSTEMmv{x} \,  \SYSTEMmv{k}     \\
\textit{($\equiv_\eta [rhs]$)} & \Rightarrow &  \textcolor{LGcolor}{\llparenthesis} \smidge  \SYSTEMnt{t}  \smidge \textcolor{LGcolor}{\rrparenthesis}  & \eqStep &  [   \textcolor{LGcolor}{\llparenthesis} \smidge  \SYSTEMnt{t}  \smidge \textcolor{LGcolor}{\rrparenthesis}   /  \SYSTEMmv{x}  ]  \SYSTEMmv{x}  \\
\textit{(defn. subst [rhs])} & \Rightarrow &  \textcolor{LGcolor}{\llparenthesis} \smidge  \SYSTEMnt{t}  \smidge \textcolor{LGcolor}{\rrparenthesis}  & \eqStep &  \textcolor{LGcolor}{\llparenthesis} \smidge  \SYSTEMnt{t}  \smidge \textcolor{LGcolor}{\rrparenthesis}  \\
\textit{(reflexivity)} & \Rightarrow & \top
\end{array}
\end{align*}

\item $\SYSTEMmv{x} \not\equiv \SYSTEMmv{y}$ then we refine the goal:
\begin{align*}
\begin{array}{rcrlll}
\textit{(goal)} & &  \textcolor{LGcolor}{\llparenthesis} \smidge   [  \SYSTEMnt{t}  /  \SYSTEMmv{x}  ]  \SYSTEMmv{y}   \smidge \textcolor{LGcolor}{\rrparenthesis}  & \eqStep &  [   \textcolor{LGcolor}{\llparenthesis} \smidge  \SYSTEMnt{t}  \smidge \textcolor{LGcolor}{\rrparenthesis}   /  \SYSTEMmv{x}  ]   \textcolor{LGcolor}{\llparenthesis} \smidge  \SYSTEMmv{y}  \smidge \textcolor{LGcolor}{\rrparenthesis}   \\
\textit{(defn. subst [lhs])} & \Rightarrow &  \textcolor{LGcolor}{\llparenthesis} \smidge  \SYSTEMmv{y}  \smidge \textcolor{LGcolor}{\rrparenthesis}  & \eqStep &  [   \textcolor{LGcolor}{\llparenthesis} \smidge  \SYSTEMnt{t}  \smidge \textcolor{LGcolor}{\rrparenthesis}   /  \SYSTEMmv{x}  ]   \textcolor{LGcolor}{\llparenthesis} \smidge  \SYSTEMmv{y}  \smidge \textcolor{LGcolor}{\rrparenthesis}   \\
\textit{(defn. interp [lhs])} & \Rightarrow &  \lambda  \SYSTEMmv{k}  .    \SYSTEMmv{y} \,  \SYSTEMmv{k}    & \eqStep &  [   \textcolor{LGcolor}{\llparenthesis} \smidge  \SYSTEMnt{t}  \smidge \textcolor{LGcolor}{\rrparenthesis}   /  \SYSTEMmv{x}  ]   \textcolor{LGcolor}{\llparenthesis} \smidge  \SYSTEMmv{y}  \smidge \textcolor{LGcolor}{\rrparenthesis}   \\
\textit{(defn. interp [rhs])} & \Rightarrow &  \lambda  \SYSTEMmv{k}  .    \SYSTEMmv{y} \,  \SYSTEMmv{k}    & \eqStep &  [   \textcolor{LGcolor}{\llparenthesis} \smidge  \SYSTEMnt{t}  \smidge \textcolor{LGcolor}{\rrparenthesis}   /  \SYSTEMmv{x}  ]   \lambda  \SYSTEMmv{k}  .    \SYSTEMmv{y} \,  \SYSTEMmv{k}     \\
\textit{(defn. subst [rhs])} & \Rightarrow &  \lambda  \SYSTEMmv{k}  .    \SYSTEMmv{y} \,  \SYSTEMmv{k}    & \eqStep &  \lambda  \SYSTEMmv{k}  .    \SYSTEMmv{y} \,  \SYSTEMmv{k}    \\
\textit{(reflexivity)} & \Rightarrow & \top
\end{array}
\end{align*}

\end{itemize}

\item (abs) $\SYSTEMnt{t'} \equiv  \lambda  \SYSTEMmv{y}  .  \SYSTEMnt{t_{{\mathrm{1}}}} $ (with $ \SYSTEMmv{k} ,   \SYSTEMmv{y}  \,\#\,  \SYSTEMnt{t}  $);
\begin{gather*}
\begin{align*}
\begin{array}{rcrlll}
\textit{(goal)} & &  \textcolor{LGcolor}{\llparenthesis} \smidge   [  \SYSTEMnt{t}  /  \SYSTEMmv{x}  ]  \SYSTEMsym{(}   \lambda  \SYSTEMmv{y}  .  \SYSTEMnt{t_{{\mathrm{1}}}}   \SYSTEMsym{)}   \smidge \textcolor{LGcolor}{\rrparenthesis}  & \eqStep &  [   \textcolor{LGcolor}{\llparenthesis} \smidge  \SYSTEMnt{t}  \smidge \textcolor{LGcolor}{\rrparenthesis}   /  \SYSTEMmv{x}  ]   \textcolor{LGcolor}{\llparenthesis} \smidge  \SYSTEMsym{(}   \lambda  \SYSTEMmv{y}  .  \SYSTEMnt{t_{{\mathrm{1}}}}   \SYSTEMsym{)}  \smidge \textcolor{LGcolor}{\rrparenthesis}   \\
\textit{(defn. subst [lhs])} & \Rightarrow &  \textcolor{LGcolor}{\llparenthesis} \smidge   \lambda  \SYSTEMmv{y}  .   [  \SYSTEMnt{t}  /  \SYSTEMmv{x}  ]  \SYSTEMnt{t_{{\mathrm{1}}}}    \smidge \textcolor{LGcolor}{\rrparenthesis}  & \eqStep &  [   \textcolor{LGcolor}{\llparenthesis} \smidge  \SYSTEMnt{t}  \smidge \textcolor{LGcolor}{\rrparenthesis}   /  \SYSTEMmv{x}  ]   \textcolor{LGcolor}{\llparenthesis} \smidge  \SYSTEMsym{(}   \lambda  \SYSTEMmv{y}  .  \SYSTEMnt{t_{{\mathrm{1}}}}   \SYSTEMsym{)}  \smidge \textcolor{LGcolor}{\rrparenthesis}   \\
\textit{(defn. interp [rhs])} & \Rightarrow &  \textcolor{LGcolor}{\llparenthesis} \smidge   \lambda  \SYSTEMmv{y}  .   [  \SYSTEMnt{t}  /  \SYSTEMmv{x}  ]  \SYSTEMnt{t_{{\mathrm{1}}}}    \smidge \textcolor{LGcolor}{\rrparenthesis}  & \eqStep &  [   \textcolor{LGcolor}{\llparenthesis} \smidge  \SYSTEMnt{t}  \smidge \textcolor{LGcolor}{\rrparenthesis}   /  \SYSTEMmv{x}  ]   \lambda  \SYSTEMmv{k}  .    \SYSTEMmv{k} \,  \SYSTEMsym{(}   \lambda  \SYSTEMmv{y}  .   \textcolor{LGcolor}{\llparenthesis} \smidge  \SYSTEMnt{t_{{\mathrm{1}}}}  \smidge \textcolor{LGcolor}{\rrparenthesis}    \SYSTEMsym{)}     \\
\textit{(defn. subst [rhs])} & \Rightarrow &  \textcolor{LGcolor}{\llparenthesis} \smidge   \lambda  \SYSTEMmv{y}  .   [  \SYSTEMnt{t}  /  \SYSTEMmv{x}  ]  \SYSTEMnt{t_{{\mathrm{1}}}}    \smidge \textcolor{LGcolor}{\rrparenthesis}  & \eqStep &  \lambda  \SYSTEMmv{k}  .    \SYSTEMmv{k} \,  \SYSTEMsym{(}   \lambda  \SYSTEMmv{y}  .   [   \textcolor{LGcolor}{\llparenthesis} \smidge  \SYSTEMnt{t}  \smidge \textcolor{LGcolor}{\rrparenthesis}   /  \SYSTEMmv{x}  ]   \textcolor{LGcolor}{\llparenthesis} \smidge  \SYSTEMnt{t_{{\mathrm{1}}}}  \smidge \textcolor{LGcolor}{\rrparenthesis}     \SYSTEMsym{)}    \\
\textit{(defn. interp [lhs])} & \Rightarrow &  \lambda  \SYSTEMmv{k}  .    \SYSTEMmv{k} \,  \SYSTEMsym{(}   \lambda  \SYSTEMmv{y}  .   \textcolor{LGcolor}{\llparenthesis} \smidge   [  \SYSTEMnt{t}  /  \SYSTEMmv{x}  ]  \SYSTEMnt{t_{{\mathrm{1}}}}   \smidge \textcolor{LGcolor}{\rrparenthesis}    \SYSTEMsym{)}    & \eqStep &  \lambda  \SYSTEMmv{k}  .    \SYSTEMmv{k} \,  \SYSTEMsym{(}   \lambda  \SYSTEMmv{y}  .   [   \textcolor{LGcolor}{\llparenthesis} \smidge  \SYSTEMnt{t}  \smidge \textcolor{LGcolor}{\rrparenthesis}   /  \SYSTEMmv{x}  ]   \textcolor{LGcolor}{\llparenthesis} \smidge  \SYSTEMnt{t_{{\mathrm{1}}}}  \smidge \textcolor{LGcolor}{\rrparenthesis}     \SYSTEMsym{)}    \\
\textit{(induction on $t1$)} & \Rightarrow & \top
\end{array}
\end{align*}
\end{gather*}

\item (app) $\SYSTEMnt{t'} \equiv  \SYSTEMnt{t_{{\mathrm{1}}}} \,  \SYSTEMnt{t_{{\mathrm{2}}}} $ (with $ \SYSTEMmv{f} ,   \SYSTEMmv{k}  \,\#\,  \SYSTEMnt{t}  $);
\begin{gather*}
\begin{align*}
\begin{array}{rcrlll}
\textit{(goal)} & &  \textcolor{LGcolor}{\llparenthesis} \smidge   [  \SYSTEMnt{t}  /  \SYSTEMmv{x}  ]  \SYSTEMsym{(}   \SYSTEMnt{t_{{\mathrm{1}}}} \,  \SYSTEMnt{t_{{\mathrm{2}}}}   \SYSTEMsym{)}   \smidge \textcolor{LGcolor}{\rrparenthesis}  & \eqStep &  [   \textcolor{LGcolor}{\llparenthesis} \smidge  \SYSTEMnt{t}  \smidge \textcolor{LGcolor}{\rrparenthesis}   /  \SYSTEMmv{x}  ]   \textcolor{LGcolor}{\llparenthesis} \smidge   \SYSTEMnt{t_{{\mathrm{1}}}} \,  \SYSTEMnt{t_{{\mathrm{2}}}}   \smidge \textcolor{LGcolor}{\rrparenthesis}   \\
\textit{(defn. subst [lhs])} & \Rightarrow &  \textcolor{LGcolor}{\llparenthesis} \smidge     [  \SYSTEMnt{t}  /  \SYSTEMmv{x}  ]  \SYSTEMnt{t_{{\mathrm{1}}}}   \,    [  \SYSTEMnt{t}  /  \SYSTEMmv{x}  ]  \SYSTEMnt{t_{{\mathrm{2}}}}     \smidge \textcolor{LGcolor}{\rrparenthesis}  & \eqStep &  [   \textcolor{LGcolor}{\llparenthesis} \smidge  \SYSTEMnt{t}  \smidge \textcolor{LGcolor}{\rrparenthesis}   /  \SYSTEMmv{x}  ]   \textcolor{LGcolor}{\llparenthesis} \smidge   \SYSTEMnt{t_{{\mathrm{1}}}} \,  \SYSTEMnt{t_{{\mathrm{2}}}}   \smidge \textcolor{LGcolor}{\rrparenthesis}   \\
\textit{(defn. interp [rhs])} & \Rightarrow &  \textcolor{LGcolor}{\llparenthesis} \smidge     [  \SYSTEMnt{t}  /  \SYSTEMmv{x}  ]  \SYSTEMnt{t_{{\mathrm{1}}}}   \,    [  \SYSTEMnt{t}  /  \SYSTEMmv{x}  ]  \SYSTEMnt{t_{{\mathrm{2}}}}     \smidge \textcolor{LGcolor}{\rrparenthesis}  & \eqStep &  [   \textcolor{LGcolor}{\llparenthesis} \smidge  \SYSTEMnt{t}  \smidge \textcolor{LGcolor}{\rrparenthesis}   /  \SYSTEMmv{x}  ]  \SYSTEMsym{(}   \lambda  \SYSTEMmv{k}  .     \textcolor{LGcolor}{\llparenthesis} \smidge  \SYSTEMnt{t_{{\mathrm{1}}}}  \smidge \textcolor{LGcolor}{\rrparenthesis}  \,  \SYSTEMsym{(}   \lambda  \SYSTEMmv{f}  .      \SYSTEMmv{f} \,   \textcolor{LGcolor}{\llparenthesis} \smidge  \SYSTEMnt{t_{{\mathrm{2}}}}  \smidge \textcolor{LGcolor}{\rrparenthesis}    \,  \SYSTEMmv{k}     \SYSTEMsym{)}     \SYSTEMsym{)}  \\
\textit{(defn. subst [rhs])} & \Rightarrow &  \textcolor{LGcolor}{\llparenthesis} \smidge     [  \SYSTEMnt{t}  /  \SYSTEMmv{x}  ]  \SYSTEMnt{t_{{\mathrm{1}}}}   \,    [  \SYSTEMnt{t}  /  \SYSTEMmv{x}  ]  \SYSTEMnt{t_{{\mathrm{2}}}}     \smidge \textcolor{LGcolor}{\rrparenthesis}  & \eqStep &  \lambda  \SYSTEMmv{k}  .    \SYSTEMsym{(}   [   \textcolor{LGcolor}{\llparenthesis} \smidge  \SYSTEMnt{t}  \smidge \textcolor{LGcolor}{\rrparenthesis}   /  \SYSTEMmv{x}  ]   \textcolor{LGcolor}{\llparenthesis} \smidge  \SYSTEMnt{t_{{\mathrm{1}}}}  \smidge \textcolor{LGcolor}{\rrparenthesis}    \SYSTEMsym{)} \,  \SYSTEMsym{(}   \lambda  \SYSTEMmv{f}  .      \SYSTEMmv{f} \,  \SYSTEMsym{(}   [   \textcolor{LGcolor}{\llparenthesis} \smidge  \SYSTEMnt{t}  \smidge \textcolor{LGcolor}{\rrparenthesis}   /  \SYSTEMmv{x}  ]   \textcolor{LGcolor}{\llparenthesis} \smidge  \SYSTEMnt{t_{{\mathrm{2}}}}  \smidge \textcolor{LGcolor}{\rrparenthesis}    \SYSTEMsym{)}   \,  \SYSTEMmv{k}     \SYSTEMsym{)}    \\
\textit{(defn. interp [lhs])} & \Rightarrow &  \lambda  \SYSTEMmv{k}  .     \textcolor{LGcolor}{\llparenthesis} \smidge   [  \SYSTEMnt{t}  /  \SYSTEMmv{x}  ]  \SYSTEMnt{t_{{\mathrm{1}}}}   \smidge \textcolor{LGcolor}{\rrparenthesis}  \,  \SYSTEMsym{(}   \lambda  \SYSTEMmv{f}  .      \SYSTEMmv{f} \,   \textcolor{LGcolor}{\llparenthesis} \smidge   [  \SYSTEMnt{t}  /  \SYSTEMmv{x}  ]  \SYSTEMnt{t_{{\mathrm{2}}}}   \smidge \textcolor{LGcolor}{\rrparenthesis}    \,  \SYSTEMmv{k}     \SYSTEMsym{)}    & \eqStep &  \lambda  \SYSTEMmv{k}  .    \SYSTEMsym{(}   [   \textcolor{LGcolor}{\llparenthesis} \smidge  \SYSTEMnt{t}  \smidge \textcolor{LGcolor}{\rrparenthesis}   /  \SYSTEMmv{x}  ]   \textcolor{LGcolor}{\llparenthesis} \smidge  \SYSTEMnt{t_{{\mathrm{1}}}}  \smidge \textcolor{LGcolor}{\rrparenthesis}    \SYSTEMsym{)} \,  \SYSTEMsym{(}   \lambda  \SYSTEMmv{f}  .      \SYSTEMmv{f} \,  \SYSTEMsym{(}   [   \textcolor{LGcolor}{\llparenthesis} \smidge  \SYSTEMnt{t}  \smidge \textcolor{LGcolor}{\rrparenthesis}   /  \SYSTEMmv{x}  ]   \textcolor{LGcolor}{\llparenthesis} \smidge  \SYSTEMnt{t_{{\mathrm{2}}}}  \smidge \textcolor{LGcolor}{\rrparenthesis}    \SYSTEMsym{)}   \,  \SYSTEMmv{k}     \SYSTEMsym{)}    \\
\textit{(induction on $\SYSTEMnt{t_{{\mathrm{1}}}}$ and $\SYSTEMnt{t_{{\mathrm{2}}}}$)} & \Rightarrow & \top
\end{array}
\end{align*}
\end{gather*}
\item (pr) $\SYSTEMnt{t'} \equiv  \textcolor{coeffectColor}{[}  \SYSTEMnt{t_{{\mathrm{1}}}}  \textcolor{coeffectColor}{]} $ (with $ \SYSTEMmv{k}  \,\#\,  \SYSTEMnt{t} $);
\begin{align*}
\begin{array}{rcrlll}
\textit{(goal)} & &  \textcolor{LGcolor}{\llparenthesis} \smidge   [  \SYSTEMnt{t}  /  \SYSTEMmv{x}  ]   \textcolor{coeffectColor}{[}  \SYSTEMnt{t_{{\mathrm{1}}}}  \textcolor{coeffectColor}{]}    \smidge \textcolor{LGcolor}{\rrparenthesis}  & \eqStep &  [   \textcolor{LGcolor}{\llparenthesis} \smidge  \SYSTEMnt{t}  \smidge \textcolor{LGcolor}{\rrparenthesis}   /  \SYSTEMmv{x}  ]   \textcolor{LGcolor}{\llparenthesis} \smidge   \textcolor{coeffectColor}{[}  \SYSTEMnt{t_{{\mathrm{1}}}}  \textcolor{coeffectColor}{]}   \smidge \textcolor{LGcolor}{\rrparenthesis}   \\
\textit{(defn. subst [lhs])} & \Rightarrow &  \textcolor{LGcolor}{\llparenthesis} \smidge   \textcolor{coeffectColor}{[}   [  \SYSTEMnt{t}  /  \SYSTEMmv{x}  ]  \SYSTEMnt{t_{{\mathrm{1}}}}   \textcolor{coeffectColor}{]}   \smidge \textcolor{LGcolor}{\rrparenthesis}  & \eqStep &  [   \textcolor{LGcolor}{\llparenthesis} \smidge  \SYSTEMnt{t}  \smidge \textcolor{LGcolor}{\rrparenthesis}   /  \SYSTEMmv{x}  ]   \textcolor{LGcolor}{\llparenthesis} \smidge   \textcolor{coeffectColor}{[}  \SYSTEMnt{t_{{\mathrm{1}}}}  \textcolor{coeffectColor}{]}   \smidge \textcolor{LGcolor}{\rrparenthesis}   \\
\textit{(defn. interp [rhs])} & \Rightarrow &  \textcolor{LGcolor}{\llparenthesis} \smidge   \textcolor{coeffectColor}{[}   [  \SYSTEMnt{t}  /  \SYSTEMmv{x}  ]  \SYSTEMnt{t_{{\mathrm{1}}}}   \textcolor{coeffectColor}{]}   \smidge \textcolor{LGcolor}{\rrparenthesis}  & \eqStep &  [   \textcolor{LGcolor}{\llparenthesis} \smidge  \SYSTEMnt{t}  \smidge \textcolor{LGcolor}{\rrparenthesis}   /  \SYSTEMmv{x}  ]  \SYSTEMsym{(}   \lambda  \SYSTEMmv{k}  .    \SYSTEMmv{k} \,   \textcolor{LGcolor}{\llparenthesis} \smidge  \SYSTEMnt{t_{{\mathrm{1}}}}  \smidge \textcolor{LGcolor}{\rrparenthesis}      \SYSTEMsym{)}  \\
\textit{(defn. subst [rhs])} & \Rightarrow &  \textcolor{LGcolor}{\llparenthesis} \smidge   \textcolor{coeffectColor}{[}   [  \SYSTEMnt{t}  /  \SYSTEMmv{x}  ]  \SYSTEMnt{t_{{\mathrm{1}}}}   \textcolor{coeffectColor}{]}   \smidge \textcolor{LGcolor}{\rrparenthesis}  & \eqStep &  \lambda  \SYSTEMmv{k}  .    \SYSTEMmv{k} \,  \SYSTEMsym{(}   [   \textcolor{LGcolor}{\llparenthesis} \smidge  \SYSTEMnt{t}  \smidge \textcolor{LGcolor}{\rrparenthesis}   /  \SYSTEMmv{x}  ]   \textcolor{LGcolor}{\llparenthesis} \smidge  \SYSTEMnt{t_{{\mathrm{1}}}}  \smidge \textcolor{LGcolor}{\rrparenthesis}    \SYSTEMsym{)}    \\
\textit{(defn. interp [lhs])} & \Rightarrow &  \lambda  \SYSTEMmv{k}  .    \SYSTEMmv{k} \,   \textcolor{LGcolor}{\llparenthesis} \smidge   [  \SYSTEMnt{t}  /  \SYSTEMmv{x}  ]  \SYSTEMnt{t_{{\mathrm{1}}}}   \smidge \textcolor{LGcolor}{\rrparenthesis}     & \eqStep &  \lambda  \SYSTEMmv{k}  .    \SYSTEMmv{k} \,  \SYSTEMsym{(}   [   \textcolor{LGcolor}{\llparenthesis} \smidge  \SYSTEMnt{t}  \smidge \textcolor{LGcolor}{\rrparenthesis}   /  \SYSTEMmv{x}  ]   \textcolor{LGcolor}{\llparenthesis} \smidge  \SYSTEMnt{t_{{\mathrm{1}}}}  \smidge \textcolor{LGcolor}{\rrparenthesis}    \SYSTEMsym{)}    \\
\textit{(induction on $t1$)} & \Rightarrow & \top
\end{array}
\end{align*}
\item (let) $\SYSTEMnt{t'} \equiv  \mathsf{let} \, \textcolor{coeffectColor}{[}  \SYSTEMmv{y}  \textcolor{coeffectColor}{]} =  \SYSTEMnt{t_{{\mathrm{1}}}}  \, \mathsf{in} \,  \SYSTEMnt{t_{{\mathrm{2}}}} $ (with $ \SYSTEMmv{k} ,   \SYSTEMmv{y}  \,\#\,  \SYSTEMnt{t}  $);
\begin{gather*}
\begin{align*}
\begin{array}{rcrlll}
\textit{(goal)} & &  \textcolor{LGcolor}{\llparenthesis} \smidge   [  \SYSTEMnt{t}  /  \SYSTEMmv{x}  ]  \SYSTEMsym{(}   \mathsf{let} \, \textcolor{coeffectColor}{[}  \SYSTEMmv{y}  \textcolor{coeffectColor}{]} =  \SYSTEMnt{t_{{\mathrm{1}}}}  \, \mathsf{in} \,  \SYSTEMnt{t_{{\mathrm{2}}}}   \SYSTEMsym{)}   \smidge \textcolor{LGcolor}{\rrparenthesis}  & \eqStep &  [   \textcolor{LGcolor}{\llparenthesis} \smidge  \SYSTEMnt{t}  \smidge \textcolor{LGcolor}{\rrparenthesis}   /  \SYSTEMmv{x}  ]   \textcolor{LGcolor}{\llparenthesis} \smidge   \mathsf{let} \, \textcolor{coeffectColor}{[}  \SYSTEMmv{y}  \textcolor{coeffectColor}{]} =  \SYSTEMnt{t_{{\mathrm{1}}}}  \, \mathsf{in} \,  \SYSTEMnt{t_{{\mathrm{2}}}}   \smidge \textcolor{LGcolor}{\rrparenthesis}   \\
\textit{(defn. subst [lhs])} & \Rightarrow &  \textcolor{LGcolor}{\llparenthesis} \smidge   \mathsf{let} \, \textcolor{coeffectColor}{[}  \SYSTEMmv{y}  \textcolor{coeffectColor}{]} =   [  \SYSTEMnt{t}  /  \SYSTEMmv{x}  ]  \SYSTEMnt{t_{{\mathrm{1}}}}   \, \mathsf{in} \,    [  \SYSTEMnt{t}  /  \SYSTEMmv{x}  ]  \SYSTEMnt{t_{{\mathrm{2}}}}     \smidge \textcolor{LGcolor}{\rrparenthesis}  & \eqStep &  [   \textcolor{LGcolor}{\llparenthesis} \smidge  \SYSTEMnt{t}  \smidge \textcolor{LGcolor}{\rrparenthesis}   /  \SYSTEMmv{x}  ]   \textcolor{LGcolor}{\llparenthesis} \smidge   \mathsf{let} \, \textcolor{coeffectColor}{[}  \SYSTEMmv{y}  \textcolor{coeffectColor}{]} =  \SYSTEMnt{t_{{\mathrm{1}}}}  \, \mathsf{in} \,  \SYSTEMnt{t_{{\mathrm{2}}}}   \smidge \textcolor{LGcolor}{\rrparenthesis}   \\
\textit{(defn. interp [rhs])} & \Rightarrow &  \textcolor{LGcolor}{\llparenthesis} \smidge   \mathsf{let} \, \textcolor{coeffectColor}{[}  \SYSTEMmv{y}  \textcolor{coeffectColor}{]} =   [  \SYSTEMnt{t}  /  \SYSTEMmv{x}  ]  \SYSTEMnt{t_{{\mathrm{1}}}}   \, \mathsf{in} \,    [  \SYSTEMnt{t}  /  \SYSTEMmv{x}  ]  \SYSTEMnt{t_{{\mathrm{2}}}}     \smidge \textcolor{LGcolor}{\rrparenthesis}  & \eqStep &  [   \textcolor{LGcolor}{\llparenthesis} \smidge  \SYSTEMnt{t}  \smidge \textcolor{LGcolor}{\rrparenthesis}   /  \SYSTEMmv{x}  ]  \SYSTEMsym{(}   \lambda  \SYSTEMmv{k}  .     \textcolor{LGcolor}{\llparenthesis} \smidge  \SYSTEMnt{t_{{\mathrm{1}}}}  \smidge \textcolor{LGcolor}{\rrparenthesis}  \,  \SYSTEMsym{(}   \lambda  \SYSTEMmv{y}  .     \textcolor{LGcolor}{\llparenthesis} \smidge  \SYSTEMnt{t_{{\mathrm{2}}}}  \smidge \textcolor{LGcolor}{\rrparenthesis}  \,  \SYSTEMmv{k}     \SYSTEMsym{)}     \SYSTEMsym{)}  \\
\textit{(defn. subst [rhs])} & \Rightarrow &  \textcolor{LGcolor}{\llparenthesis} \smidge   \mathsf{let} \, \textcolor{coeffectColor}{[}  \SYSTEMmv{y}  \textcolor{coeffectColor}{]} =   [  \SYSTEMnt{t}  /  \SYSTEMmv{x}  ]  \SYSTEMnt{t_{{\mathrm{1}}}}   \, \mathsf{in} \,    [  \SYSTEMnt{t}  /  \SYSTEMmv{x}  ]  \SYSTEMnt{t_{{\mathrm{2}}}}     \smidge \textcolor{LGcolor}{\rrparenthesis}  & \eqStep &  \lambda  \SYSTEMmv{k}  .    \SYSTEMsym{(}   [   \textcolor{LGcolor}{\llparenthesis} \smidge  \SYSTEMnt{t}  \smidge \textcolor{LGcolor}{\rrparenthesis}   /  \SYSTEMmv{x}  ]   \textcolor{LGcolor}{\llparenthesis} \smidge  \SYSTEMnt{t_{{\mathrm{1}}}}  \smidge \textcolor{LGcolor}{\rrparenthesis}    \SYSTEMsym{)} \,  \SYSTEMsym{(}   \lambda  \SYSTEMmv{y}  .    \SYSTEMsym{(}   [   \textcolor{LGcolor}{\llparenthesis} \smidge  \SYSTEMnt{t}  \smidge \textcolor{LGcolor}{\rrparenthesis}   /  \SYSTEMmv{x}  ]   \textcolor{LGcolor}{\llparenthesis} \smidge  \SYSTEMnt{t_{{\mathrm{2}}}}  \smidge \textcolor{LGcolor}{\rrparenthesis}    \SYSTEMsym{)} \,  \SYSTEMmv{k}     \SYSTEMsym{)}    \\
\textit{(defn. interp [lhs])} & \Rightarrow &  \lambda  \SYSTEMmv{k}  .     \textcolor{LGcolor}{\llparenthesis} \smidge   [  \SYSTEMnt{t}  /  \SYSTEMmv{x}  ]  \SYSTEMnt{t_{{\mathrm{1}}}}   \smidge \textcolor{LGcolor}{\rrparenthesis}  \,  \SYSTEMsym{(}   \lambda  \SYSTEMmv{y}  .     \textcolor{LGcolor}{\llparenthesis} \smidge   [  \SYSTEMnt{t}  /  \SYSTEMmv{x}  ]  \SYSTEMnt{t_{{\mathrm{2}}}}   \smidge \textcolor{LGcolor}{\rrparenthesis}  \,  \SYSTEMmv{k}     \SYSTEMsym{)}    & \eqStep &  \lambda  \SYSTEMmv{k}  .    \SYSTEMsym{(}   [   \textcolor{LGcolor}{\llparenthesis} \smidge  \SYSTEMnt{t}  \smidge \textcolor{LGcolor}{\rrparenthesis}   /  \SYSTEMmv{x}  ]   \textcolor{LGcolor}{\llparenthesis} \smidge  \SYSTEMnt{t_{{\mathrm{1}}}}  \smidge \textcolor{LGcolor}{\rrparenthesis}    \SYSTEMsym{)} \,  \SYSTEMsym{(}   \lambda  \SYSTEMmv{y}  .    \SYSTEMsym{(}   [   \textcolor{LGcolor}{\llparenthesis} \smidge  \SYSTEMnt{t}  \smidge \textcolor{LGcolor}{\rrparenthesis}   /  \SYSTEMmv{x}  ]   \textcolor{LGcolor}{\llparenthesis} \smidge  \SYSTEMnt{t_{{\mathrm{2}}}}  \smidge \textcolor{LGcolor}{\rrparenthesis}    \SYSTEMsym{)} \,  \SYSTEMmv{k}     \SYSTEMsym{)}    \\
\textit{(induction on $\SYSTEMnt{t_{{\mathrm{1}}}}$ and $\SYSTEMnt{t_{{\mathrm{2}}}}$)} & \Rightarrow & \top
\end{array}
\end{align*}
\end{gather*}
\end{itemize}
\end{proof}

We now prove the operational correspondence as follows:

\begin{proof}
We consider the theorem in an expanded form: for every $ \SYSTEMnt{t}  \rightsquigarrow_{\textsc{l} }  \SYSTEMnt{t'} $
then $\exists t'' .   \textcolor{LGcolor}{\llparenthesis} \smidge  \SYSTEMnt{t}  \smidge \textcolor{LGcolor}{\rrparenthesis}   \rightsquigarrow_{\beta\textsc{g} }^\ast  \SYSTEMnt{t''}  \wedge  \textcolor{LGcolor}{\llparenthesis} \smidge  \SYSTEMnt{t'}  \smidge \textcolor{LGcolor}{\rrparenthesis}  \equiv \SYSTEMnt{t''}$.

The proof then follows by induction on reductions:
\begin{itemize}
\item (beta)
\[
\SYSTEMdruleSemLinbeta{}
\]
The interpretation of the reducing term is:
\begin{align*}
 \textcolor{LGcolor}{\llparenthesis} \smidge   \SYSTEMsym{(}   \lambda  \SYSTEMmv{x}  .  \SYSTEMnt{t_{{\mathrm{2}}}}   \SYSTEMsym{)} \,  \SYSTEMnt{t_{{\mathrm{1}}}}   \smidge \textcolor{LGcolor}{\rrparenthesis}  \;\equiv\;\;&  \lambda  \SYSTEMmv{k}  .     \textcolor{LGcolor}{\llparenthesis} \smidge  \SYSTEMsym{(}   \lambda  \SYSTEMmv{x}  .  \SYSTEMnt{t_{{\mathrm{2}}}}   \SYSTEMsym{)}  \smidge \textcolor{LGcolor}{\rrparenthesis}  \,  \SYSTEMsym{(}   \lambda  \SYSTEMmv{f}  .      \SYSTEMmv{f} \,   \textcolor{LGcolor}{\llparenthesis} \smidge  \SYSTEMnt{t_{{\mathrm{1}}}}  \smidge \textcolor{LGcolor}{\rrparenthesis}    \,  \SYSTEMmv{k}     \SYSTEMsym{)}    \\
                             \equiv\;\;&  \lambda  \SYSTEMmv{k}  .    \SYSTEMsym{(}   \lambda  \SYSTEMmv{k'}  .    \SYSTEMmv{k'} \,  \SYSTEMsym{(}   \lambda  \SYSTEMmv{x}  .   \textcolor{LGcolor}{\llparenthesis} \smidge  \SYSTEMnt{t_{{\mathrm{2}}}}  \smidge \textcolor{LGcolor}{\rrparenthesis}    \SYSTEMsym{)}     \SYSTEMsym{)} \,  \SYSTEMsym{(}   \lambda  \SYSTEMmv{f}  .      \SYSTEMmv{f} \,   \textcolor{LGcolor}{\llparenthesis} \smidge  \SYSTEMnt{t_{{\mathrm{1}}}}  \smidge \textcolor{LGcolor}{\rrparenthesis}    \,  \SYSTEMmv{k}     \SYSTEMsym{)}   
\end{align*}
Then we construct the reduction sequence in Graded Base:
\begin{align*}
&\inferrule*[right=\SYSTEMRenameRuleSemGrdcongAbs{}]
{
  \inferrule*[right=\SYSTEMRenameRuleSemGrdbeta{}]
  { }
  {  \SYSTEMsym{(}   \lambda  \SYSTEMmv{k'}  .    \SYSTEMmv{k'} \,  \SYSTEMsym{(}   \lambda  \SYSTEMmv{x}  .   \textcolor{LGcolor}{\llparenthesis} \smidge  \SYSTEMnt{t_{{\mathrm{2}}}}  \smidge \textcolor{LGcolor}{\rrparenthesis}    \SYSTEMsym{)}     \SYSTEMsym{)} \,  \SYSTEMsym{(}   \lambda  \SYSTEMmv{f}  .      \SYSTEMmv{f} \,   \textcolor{LGcolor}{\llparenthesis} \smidge  \SYSTEMnt{t_{{\mathrm{1}}}}  \smidge \textcolor{LGcolor}{\rrparenthesis}    \,  \SYSTEMmv{k}     \SYSTEMsym{)}   \rightsquigarrow_{\textsc{g} }    \SYSTEMsym{(}   \lambda  \SYSTEMmv{f}  .      \SYSTEMmv{f} \,   \textcolor{LGcolor}{\llparenthesis} \smidge  \SYSTEMnt{t_{{\mathrm{1}}}}  \smidge \textcolor{LGcolor}{\rrparenthesis}    \,  \SYSTEMmv{k}     \SYSTEMsym{)} \,  \SYSTEMsym{(}   \lambda  \SYSTEMmv{x}  .   \textcolor{LGcolor}{\llparenthesis} \smidge  \SYSTEMnt{t_{{\mathrm{2}}}}  \smidge \textcolor{LGcolor}{\rrparenthesis}    \SYSTEMsym{)}   }
}
{  \lambda  \SYSTEMmv{k}  .    \SYSTEMsym{(}   \lambda  \SYSTEMmv{k'}  .    \SYSTEMmv{k'} \,  \SYSTEMsym{(}   \lambda  \SYSTEMmv{x}  .   \textcolor{LGcolor}{\llparenthesis} \smidge  \SYSTEMnt{t_{{\mathrm{2}}}}  \smidge \textcolor{LGcolor}{\rrparenthesis}    \SYSTEMsym{)}     \SYSTEMsym{)} \,  \SYSTEMsym{(}   \lambda  \SYSTEMmv{f}  .      \SYSTEMmv{f} \,   \textcolor{LGcolor}{\llparenthesis} \smidge  \SYSTEMnt{t_{{\mathrm{1}}}}  \smidge \textcolor{LGcolor}{\rrparenthesis}    \,  \SYSTEMmv{k}     \SYSTEMsym{)}     \rightsquigarrow_{\textsc{g} }   \lambda  \SYSTEMmv{k}  .    \SYSTEMsym{(}   \lambda  \SYSTEMmv{f}  .      \SYSTEMmv{f} \,   \textcolor{LGcolor}{\llparenthesis} \smidge  \SYSTEMnt{t_{{\mathrm{1}}}}  \smidge \textcolor{LGcolor}{\rrparenthesis}    \,  \SYSTEMmv{k}     \SYSTEMsym{)} \,  \SYSTEMsym{(}   \lambda  \SYSTEMmv{x}  .   \textcolor{LGcolor}{\llparenthesis} \smidge  \SYSTEMnt{t_{{\mathrm{2}}}}  \smidge \textcolor{LGcolor}{\rrparenthesis}    \SYSTEMsym{)}    }
\end{align*}
\begin{align*}
&\inferrule*[right=\SYSTEMRenameRuleSemGrdcongAbs{}]
{
  \inferrule*[right=\SYSTEMRenameRuleSemGrdbeta{}]
  { }
  {  \SYSTEMsym{(}   \lambda  \SYSTEMmv{f}  .      \SYSTEMmv{f} \,   \textcolor{LGcolor}{\llparenthesis} \smidge  \SYSTEMnt{t_{{\mathrm{1}}}}  \smidge \textcolor{LGcolor}{\rrparenthesis}    \,  \SYSTEMmv{k}     \SYSTEMsym{)} \,  \SYSTEMsym{(}   \lambda  \SYSTEMmv{x}  .   \textcolor{LGcolor}{\llparenthesis} \smidge  \SYSTEMnt{t_{{\mathrm{2}}}}  \smidge \textcolor{LGcolor}{\rrparenthesis}    \SYSTEMsym{)}   \rightsquigarrow_{\textsc{g} }  \SYSTEMsym{(}   \lambda  \SYSTEMmv{f}  .      \SYSTEMsym{(}   \lambda  \SYSTEMmv{x}  .   \textcolor{LGcolor}{\llparenthesis} \smidge  \SYSTEMnt{t_{{\mathrm{2}}}}  \smidge \textcolor{LGcolor}{\rrparenthesis}    \SYSTEMsym{)} \,   \textcolor{LGcolor}{\llparenthesis} \smidge  \SYSTEMnt{t_{{\mathrm{1}}}}  \smidge \textcolor{LGcolor}{\rrparenthesis}    \,  \SYSTEMmv{k}     \SYSTEMsym{)} }
}
{  \lambda  \SYSTEMmv{k}  .    \SYSTEMsym{(}   \lambda  \SYSTEMmv{f}  .      \SYSTEMmv{f} \,   \textcolor{LGcolor}{\llparenthesis} \smidge  \SYSTEMnt{t_{{\mathrm{1}}}}  \smidge \textcolor{LGcolor}{\rrparenthesis}    \,  \SYSTEMmv{k}     \SYSTEMsym{)} \,  \SYSTEMsym{(}   \lambda  \SYSTEMmv{x}  .   \textcolor{LGcolor}{\llparenthesis} \smidge  \SYSTEMnt{t_{{\mathrm{2}}}}  \smidge \textcolor{LGcolor}{\rrparenthesis}    \SYSTEMsym{)}     \rightsquigarrow_{\textsc{g} }   \lambda  \SYSTEMmv{k}  .   \SYSTEMsym{(}   \lambda  \SYSTEMmv{f}  .      \SYSTEMsym{(}   \lambda  \SYSTEMmv{x}  .   \textcolor{LGcolor}{\llparenthesis} \smidge  \SYSTEMnt{t_{{\mathrm{2}}}}  \smidge \textcolor{LGcolor}{\rrparenthesis}    \SYSTEMsym{)} \,   \textcolor{LGcolor}{\llparenthesis} \smidge  \SYSTEMnt{t_{{\mathrm{1}}}}  \smidge \textcolor{LGcolor}{\rrparenthesis}    \,  \SYSTEMmv{k}     \SYSTEMsym{)}   }
\end{align*}
\begin{align*}
&\inferrule*[right=\SYSTEMRenameRuleSemGrdcongAbs{}]
{
  \inferrule*[right=\SYSTEMRenameRuleSemGrdbeta{}]
  { }
  {    \SYSTEMsym{(}   \lambda  \SYSTEMmv{x}  .   \textcolor{LGcolor}{\llparenthesis} \smidge  \SYSTEMnt{t_{{\mathrm{2}}}}  \smidge \textcolor{LGcolor}{\rrparenthesis}    \SYSTEMsym{)} \,   \textcolor{LGcolor}{\llparenthesis} \smidge  \SYSTEMnt{t_{{\mathrm{1}}}}  \smidge \textcolor{LGcolor}{\rrparenthesis}    \,  \SYSTEMmv{k}   \rightsquigarrow_{\textsc{g} }   \SYSTEMsym{(}   [   \textcolor{LGcolor}{\llparenthesis} \smidge  \SYSTEMnt{t_{{\mathrm{1}}}}  \smidge \textcolor{LGcolor}{\rrparenthesis}   /  \SYSTEMmv{x}  ]   \textcolor{LGcolor}{\llparenthesis} \smidge  \SYSTEMnt{t_{{\mathrm{2}}}}  \smidge \textcolor{LGcolor}{\rrparenthesis}    \SYSTEMsym{)} \,  \SYSTEMmv{k}  }
}
{  \lambda  \SYSTEMmv{k}  .   \SYSTEMsym{(}   \lambda  \SYSTEMmv{f}  .      \SYSTEMsym{(}   \lambda  \SYSTEMmv{x}  .   \textcolor{LGcolor}{\llparenthesis} \smidge  \SYSTEMnt{t_{{\mathrm{2}}}}  \smidge \textcolor{LGcolor}{\rrparenthesis}    \SYSTEMsym{)} \,   \textcolor{LGcolor}{\llparenthesis} \smidge  \SYSTEMnt{t_{{\mathrm{1}}}}  \smidge \textcolor{LGcolor}{\rrparenthesis}    \,  \SYSTEMmv{k}     \SYSTEMsym{)}    \rightsquigarrow_{\textsc{g} }   \lambda  \SYSTEMmv{k}  .    \SYSTEMsym{(}   [   \textcolor{LGcolor}{\llparenthesis} \smidge  \SYSTEMnt{t_{{\mathrm{1}}}}  \smidge \textcolor{LGcolor}{\rrparenthesis}   /  \SYSTEMmv{x}  ]   \textcolor{LGcolor}{\llparenthesis} \smidge  \SYSTEMnt{t_{{\mathrm{2}}}}  \smidge \textcolor{LGcolor}{\rrparenthesis}    \SYSTEMsym{)} \,  \SYSTEMmv{k}    }
\end{align*}
By \SYSTEMRenameRuleGradEqeta{} and Lemma~\ref{lemma:interp-lin-to-grd-cps-preserves-subst} then interpreting
the linear-base reduced term is equal to the graded-base
reduced term: $ \textcolor{LGcolor}{\llparenthesis} \smidge   [  \SYSTEMnt{t_{{\mathrm{1}}}}  /  \SYSTEMmv{x}  ]  \SYSTEMnt{t_{{\mathrm{2}}}}   \smidge \textcolor{LGcolor}{\rrparenthesis}  \equiv  \lambda  \SYSTEMmv{k}  .    \SYSTEMsym{(}   [   \textcolor{LGcolor}{\llparenthesis} \smidge  \SYSTEMnt{t_{{\mathrm{1}}}}  \smidge \textcolor{LGcolor}{\rrparenthesis}   /  \SYSTEMmv{x}  ]   \textcolor{LGcolor}{\llparenthesis} \smidge  \SYSTEMnt{t_{{\mathrm{2}}}}  \smidge \textcolor{LGcolor}{\rrparenthesis}    \SYSTEMsym{)} \,  \SYSTEMmv{k}   $.

\item (congAppL)
\[
\SYSTEMdruleSemLincongAppL{}
\]
The interpretation of the reducing term is:
\[
 \textcolor{LGcolor}{\llparenthesis} \smidge   \SYSTEMnt{t_{{\mathrm{1}}}} \,  \SYSTEMnt{t_{{\mathrm{2}}}}   \smidge \textcolor{LGcolor}{\rrparenthesis}  \equiv  \lambda  \SYSTEMmv{k}  .     \textcolor{LGcolor}{\llparenthesis} \smidge  \SYSTEMnt{t_{{\mathrm{1}}}}  \smidge \textcolor{LGcolor}{\rrparenthesis}  \,  \SYSTEMsym{(}   \lambda  \SYSTEMmv{f}  .      \SYSTEMmv{f} \,   \textcolor{LGcolor}{\llparenthesis} \smidge  \SYSTEMnt{t_{{\mathrm{2}}}}  \smidge \textcolor{LGcolor}{\rrparenthesis}    \,  \SYSTEMmv{k}     \SYSTEMsym{)}   
\]
By induction on $ \SYSTEMnt{t_{{\mathrm{1}}}}  \rightsquigarrow_{\textsc{l} }  \SYSTEMnt{t'_{{\mathrm{1}}}} $, we have
$\exists \SYSTEMnt{t''_{{\mathrm{1}}}} .   \textcolor{LGcolor}{\llparenthesis} \smidge  \SYSTEMnt{t_{{\mathrm{1}}}}  \smidge \textcolor{LGcolor}{\rrparenthesis}   \rightsquigarrow_{\textsc{g} }^\ast  \SYSTEMnt{t''_{{\mathrm{1}}}}  \wedge
\SYSTEMnt{t''_{{\mathrm{1}}}} \equiv  \textcolor{LGcolor}{\llparenthesis} \smidge  \SYSTEMnt{t'_{{\mathrm{1}}}}  \smidge \textcolor{LGcolor}{\rrparenthesis} $. Therefore, we construct
a chain of reductions in Graded Base:
\begin{gather*}
\begin{align*}
\inferrule*[right=\SYSTEMRenameRuleSemGrdcongAbs{}]
{
  \inferrule*[right=\SYSTEMRenameRuleSemLincongAppL{}]
  {  \textcolor{LGcolor}{\llparenthesis} \smidge  \SYSTEMnt{t_{{\mathrm{1}}}}  \smidge \textcolor{LGcolor}{\rrparenthesis}   \rightsquigarrow_{\textsc{l} }  \SYSTEMnt{t''_{{\mathrm{1}}\,\SYSTEMmv{i}}} }
  {   \textcolor{LGcolor}{\llparenthesis} \smidge  \SYSTEMnt{t_{{\mathrm{1}}}}  \smidge \textcolor{LGcolor}{\rrparenthesis}  \,  \SYSTEMsym{(}   \lambda  \SYSTEMmv{f}  .      \SYSTEMmv{f} \,   \textcolor{LGcolor}{\llparenthesis} \smidge  \SYSTEMnt{t_{{\mathrm{2}}}}  \smidge \textcolor{LGcolor}{\rrparenthesis}    \,  \SYSTEMmv{k}     \SYSTEMsym{)}   \rightsquigarrow_{\textsc{l} }   \SYSTEMnt{t''_{{\mathrm{1}}\,\SYSTEMmv{i}}} \,  \SYSTEMsym{(}   \lambda  \SYSTEMmv{f}  .      \SYSTEMmv{f} \,   \textcolor{LGcolor}{\llparenthesis} \smidge  \SYSTEMnt{t_{{\mathrm{2}}}}  \smidge \textcolor{LGcolor}{\rrparenthesis}    \,  \SYSTEMmv{k}     \SYSTEMsym{)}  }
}
{  \lambda  \SYSTEMmv{k}  .     \textcolor{LGcolor}{\llparenthesis} \smidge  \SYSTEMnt{t_{{\mathrm{1}}}}  \smidge \textcolor{LGcolor}{\rrparenthesis}  \,  \SYSTEMsym{(}   \lambda  \SYSTEMmv{f}  .      \SYSTEMmv{f} \,   \textcolor{LGcolor}{\llparenthesis} \smidge  \SYSTEMnt{t_{{\mathrm{2}}}}  \smidge \textcolor{LGcolor}{\rrparenthesis}    \,  \SYSTEMmv{k}     \SYSTEMsym{)}     \rightsquigarrow_{\textsc{l} }   \lambda  \SYSTEMmv{k}  .    \SYSTEMnt{t''_{{\mathrm{1}}\,\SYSTEMmv{i}}} \,  \SYSTEMsym{(}   \lambda  \SYSTEMmv{f}  .      \SYSTEMmv{f} \,   \textcolor{LGcolor}{\llparenthesis} \smidge  \SYSTEMnt{t_{{\mathrm{2}}}}  \smidge \textcolor{LGcolor}{\rrparenthesis}    \,  \SYSTEMmv{k}     \SYSTEMsym{)}    }
\;\;
\cdots
\;\;
\inferrule*[right=\SYSTEMRenameRuleSemGrdcongAbs{}]
{
  \inferrule*[right=\SYSTEMRenameRuleSemLincongAppL{}]
  { \SYSTEMnt{t''_{{\mathrm{1}}\,\SYSTEMmv{m}}}  \rightsquigarrow_{\textsc{l} }  \SYSTEMnt{t''_{{\mathrm{1}}}} }
  {  \SYSTEMnt{t''_{{\mathrm{1}}\,\SYSTEMmv{m}}} \,  \SYSTEMsym{(}   \lambda  \SYSTEMmv{f}  .      \SYSTEMmv{f} \,   \textcolor{LGcolor}{\llparenthesis} \smidge  \SYSTEMnt{t_{{\mathrm{2}}}}  \smidge \textcolor{LGcolor}{\rrparenthesis}    \,  \SYSTEMmv{k}     \SYSTEMsym{)}   \rightsquigarrow_{\textsc{l} }   \SYSTEMnt{t''_{{\mathrm{1}}}} \,  \SYSTEMsym{(}   \lambda  \SYSTEMmv{f}  .      \SYSTEMmv{f} \,   \textcolor{LGcolor}{\llparenthesis} \smidge  \SYSTEMnt{t_{{\mathrm{2}}}}  \smidge \textcolor{LGcolor}{\rrparenthesis}    \,  \SYSTEMmv{k}     \SYSTEMsym{)}  }
}
{  \lambda  \SYSTEMmv{k}  .    \SYSTEMnt{t''_{{\mathrm{1}}\,\SYSTEMmv{m}}} \,  \SYSTEMsym{(}   \lambda  \SYSTEMmv{f}  .      \SYSTEMmv{f} \,   \textcolor{LGcolor}{\llparenthesis} \smidge  \SYSTEMnt{t_{{\mathrm{2}}}}  \smidge \textcolor{LGcolor}{\rrparenthesis}    \,  \SYSTEMmv{k}     \SYSTEMsym{)}     \rightsquigarrow_{\textsc{l} }   \lambda  \SYSTEMmv{k}  .    \SYSTEMnt{t''_{{\mathrm{1}}}} \,  \SYSTEMsym{(}   \lambda  \SYSTEMmv{f}  .      \SYSTEMmv{f} \,   \textcolor{LGcolor}{\llparenthesis} \smidge  \SYSTEMnt{t_{{\mathrm{2}}}}  \smidge \textcolor{LGcolor}{\rrparenthesis}    \,  \SYSTEMmv{k}     \SYSTEMsym{)}    }
\end{align*}
\end{gather*}
where $m$ is the length of the reduction sequence from
the induction.

satisfying the goal, since $ \SYSTEMnt{t''_{{\mathrm{1}}}}  \equiv   \textcolor{LGcolor}{\llparenthesis} \smidge  \SYSTEMnt{t'_{{\mathrm{1}}}}  \smidge \textcolor{LGcolor}{\rrparenthesis}  $
and therefore
$  \lambda  \SYSTEMmv{k}  .    \SYSTEMnt{t''_{{\mathrm{1}}}} \,  \SYSTEMsym{(}   \lambda  \SYSTEMmv{f}  .      \SYSTEMmv{f} \,   \textcolor{LGcolor}{\llparenthesis} \smidge  \SYSTEMnt{t_{{\mathrm{2}}}}  \smidge \textcolor{LGcolor}{\rrparenthesis}    \,  \SYSTEMmv{k}     \SYSTEMsym{)}     \equiv   \lambda  \SYSTEMmv{k}  .     \textcolor{LGcolor}{\llparenthesis} \smidge  \SYSTEMnt{t'_{{\mathrm{1}}}}  \smidge \textcolor{LGcolor}{\rrparenthesis}  \,  \SYSTEMsym{(}   \lambda  \SYSTEMmv{f}  .      \SYSTEMmv{f} \,   \textcolor{LGcolor}{\llparenthesis} \smidge  \SYSTEMnt{t_{{\mathrm{2}}}}  \smidge \textcolor{LGcolor}{\rrparenthesis}    \,  \SYSTEMmv{k}     \SYSTEMsym{)}    $.
\item (betaBox)
\[
\SYSTEMdruleSemLinbetaBox{}
\]
The interpretation of the reducing term is:
\[
 \textcolor{LGcolor}{\llparenthesis} \smidge   \mathsf{let} \, \textcolor{coeffectColor}{[}  \SYSTEMmv{x}  \textcolor{coeffectColor}{]} =   \textcolor{coeffectColor}{[}  \SYSTEMnt{t_{{\mathrm{1}}}}  \textcolor{coeffectColor}{]}   \, \mathsf{in} \,  \SYSTEMnt{t_{{\mathrm{2}}}}   \smidge \textcolor{LGcolor}{\rrparenthesis}  \equiv  \lambda  \SYSTEMmv{k}  .    \SYSTEMsym{(}   \lambda  \SYSTEMmv{k'}  .    \SYSTEMmv{k'} \,   \textcolor{LGcolor}{\llparenthesis} \smidge  \SYSTEMnt{t_{{\mathrm{1}}}}  \smidge \textcolor{LGcolor}{\rrparenthesis}      \SYSTEMsym{)} \,  \SYSTEMsym{(}   \lambda  \SYSTEMmv{x}  .     \textcolor{LGcolor}{\llparenthesis} \smidge  \SYSTEMnt{t_{{\mathrm{2}}}}  \smidge \textcolor{LGcolor}{\rrparenthesis}  \,  \SYSTEMmv{k}     \SYSTEMsym{)}   
\]
Then we construct the reduction in Graded Base:
\begin{align*}
&\inferrule*[right=\SYSTEMRenameRuleSemGrdcongAbs{}]
{
  \inferrule*[right=\SYSTEMRenameRuleSemGrdbeta{}]
  { }
  {  \SYSTEMsym{(}   \lambda  \SYSTEMmv{k'}  .    \SYSTEMmv{k'} \,   \textcolor{LGcolor}{\llparenthesis} \smidge  \SYSTEMnt{t_{{\mathrm{1}}}}  \smidge \textcolor{LGcolor}{\rrparenthesis}      \SYSTEMsym{)} \,  \SYSTEMsym{(}   \lambda  \SYSTEMmv{x}  .     \textcolor{LGcolor}{\llparenthesis} \smidge  \SYSTEMnt{t_{{\mathrm{2}}}}  \smidge \textcolor{LGcolor}{\rrparenthesis}  \,  \SYSTEMmv{k}     \SYSTEMsym{)}   \rightsquigarrow_{\textsc{g} }   \SYSTEMsym{(}   \lambda  \SYSTEMmv{x}  .     \textcolor{LGcolor}{\llparenthesis} \smidge  \SYSTEMnt{t_{{\mathrm{2}}}}  \smidge \textcolor{LGcolor}{\rrparenthesis}  \,  \SYSTEMmv{k}     \SYSTEMsym{)} \,   \textcolor{LGcolor}{\llparenthesis} \smidge  \SYSTEMnt{t_{{\mathrm{1}}}}  \smidge \textcolor{LGcolor}{\rrparenthesis}   }
}
{  \lambda  \SYSTEMmv{k}  .    \SYSTEMsym{(}   \lambda  \SYSTEMmv{k'}  .    \SYSTEMmv{k'} \,   \textcolor{LGcolor}{\llparenthesis} \smidge  \SYSTEMnt{t_{{\mathrm{1}}}}  \smidge \textcolor{LGcolor}{\rrparenthesis}      \SYSTEMsym{)} \,  \SYSTEMsym{(}   \lambda  \SYSTEMmv{x}  .     \textcolor{LGcolor}{\llparenthesis} \smidge  \SYSTEMnt{t_{{\mathrm{2}}}}  \smidge \textcolor{LGcolor}{\rrparenthesis}  \,  \SYSTEMmv{k}     \SYSTEMsym{)}     \rightsquigarrow_{\textsc{g} }   \lambda  \SYSTEMmv{k}  .    \SYSTEMsym{(}   \lambda  \SYSTEMmv{x}  .     \textcolor{LGcolor}{\llparenthesis} \smidge  \SYSTEMnt{t_{{\mathrm{2}}}}  \smidge \textcolor{LGcolor}{\rrparenthesis}  \,  \SYSTEMmv{k}     \SYSTEMsym{)} \,   \textcolor{LGcolor}{\llparenthesis} \smidge  \SYSTEMnt{t_{{\mathrm{1}}}}  \smidge \textcolor{LGcolor}{\rrparenthesis}     }
\end{align*}
\begin{align*}
&\inferrule*[right=\SYSTEMRenameRuleSemGrdcongAbs{}]
{
  \inferrule*[right=\SYSTEMRenameRuleSemGrdbeta{}]
  { }
  {  \SYSTEMsym{(}   \lambda  \SYSTEMmv{x}  .     \textcolor{LGcolor}{\llparenthesis} \smidge  \SYSTEMnt{t_{{\mathrm{2}}}}  \smidge \textcolor{LGcolor}{\rrparenthesis}  \,  \SYSTEMmv{k}     \SYSTEMsym{)} \,   \textcolor{LGcolor}{\llparenthesis} \smidge  \SYSTEMnt{t_{{\mathrm{1}}}}  \smidge \textcolor{LGcolor}{\rrparenthesis}    \rightsquigarrow_{\textsc{g} }   [   \textcolor{LGcolor}{\llparenthesis} \smidge  \SYSTEMnt{t_{{\mathrm{1}}}}  \smidge \textcolor{LGcolor}{\rrparenthesis}   /  \SYSTEMmv{x}  ]  \SYSTEMsym{(}    \textcolor{LGcolor}{\llparenthesis} \smidge  \SYSTEMnt{t_{{\mathrm{2}}}}  \smidge \textcolor{LGcolor}{\rrparenthesis}  \,  \SYSTEMmv{k}   \SYSTEMsym{)}  }
}
{  \lambda  \SYSTEMmv{k}  .    \SYSTEMsym{(}   \lambda  \SYSTEMmv{x}  .     \textcolor{LGcolor}{\llparenthesis} \smidge  \SYSTEMnt{t_{{\mathrm{2}}}}  \smidge \textcolor{LGcolor}{\rrparenthesis}  \,  \SYSTEMmv{k}     \SYSTEMsym{)} \,   \textcolor{LGcolor}{\llparenthesis} \smidge  \SYSTEMnt{t_{{\mathrm{1}}}}  \smidge \textcolor{LGcolor}{\rrparenthesis}      \rightsquigarrow_{\textsc{g} }   \lambda  \SYSTEMmv{k}  .    [   \textcolor{LGcolor}{\llparenthesis} \smidge  \SYSTEMnt{t_{{\mathrm{1}}}}  \smidge \textcolor{LGcolor}{\rrparenthesis}   /  \SYSTEMmv{x}  ]  \SYSTEMsym{(}    \textcolor{LGcolor}{\llparenthesis} \smidge  \SYSTEMnt{t_{{\mathrm{2}}}}  \smidge \textcolor{LGcolor}{\rrparenthesis}  \,  \SYSTEMmv{k}   \SYSTEMsym{)}    }
\end{align*}
By \SYSTEMRenameRuleGradEqeta{} and Lemma~\ref{lemma:interp-lin-to-grd-cps-preserves-subst} then interpreting
the linear-base reduced term is equal to the graded-base
reduced term: $ \textcolor{LGcolor}{\llparenthesis} \smidge   [  \SYSTEMnt{t_{{\mathrm{1}}}}  /  \SYSTEMmv{x}  ]  \SYSTEMnt{t_{{\mathrm{2}}}}   \smidge \textcolor{LGcolor}{\rrparenthesis}  \equiv  \lambda  \SYSTEMmv{k}  .    [   \textcolor{LGcolor}{\llparenthesis} \smidge  \SYSTEMnt{t_{{\mathrm{1}}}}  \smidge \textcolor{LGcolor}{\rrparenthesis}   /  \SYSTEMmv{x}  ]  \SYSTEMsym{(}    \textcolor{LGcolor}{\llparenthesis} \smidge  \SYSTEMnt{t_{{\mathrm{2}}}}  \smidge \textcolor{LGcolor}{\rrparenthesis}  \,  \SYSTEMmv{k}   \SYSTEMsym{)}   $.

\item (congLetL)
\[
\SYSTEMdruleSemLincongLetL{}
\]
By induction, analogously to the case for (congAppL).
\end{itemize}
\end{proof}

\subsubsection{Equation preservation}
\label{app:proofs-lin-to-grad-cps-eqns}

\begin{itemize}
  \item \[\SYSTEMdruleLinEqbeta{}\]

  \begin{align*}
                   & =  \textcolor{LGcolor}{\llparenthesis} \smidge   \SYSTEMsym{(}   \lambda  \SYSTEMmv{x}  .  \SYSTEMnt{t_{{\mathrm{2}}}}   \SYSTEMsym{)} \,  \SYSTEMnt{t_{{\mathrm{1}}}}   \smidge \textcolor{LGcolor}{\rrparenthesis}  \\
\{\textit{defn.}\} & =   \lambda  \SYSTEMmv{k}  .  \SYSTEMsym{(}    \lambda  \SYSTEMmv{k}  .  \SYSTEMmv{k}  \,  \SYSTEMsym{(}   \lambda  \SYSTEMmv{x}  .   \textcolor{LGcolor}{\llparenthesis} \smidge  \SYSTEMnt{t_{{\mathrm{2}}}}  \smidge \textcolor{LGcolor}{\rrparenthesis}    \SYSTEMsym{)}   \SYSTEMsym{)}  \,  \SYSTEMsym{(}     \lambda  \SYSTEMmv{f}  .  \SYSTEMmv{f}  \,   \textcolor{LGcolor}{\llparenthesis} \smidge  \SYSTEMnt{t_{{\mathrm{1}}}}  \smidge \textcolor{LGcolor}{\rrparenthesis}   \,  \SYSTEMmv{k}   \SYSTEMsym{)}  \\
\{\text{\SYSTEMRenameRuleGradEqbeta{}}\} & =   \lambda  \SYSTEMmv{k}  .  \SYSTEMsym{(}     \lambda  \SYSTEMmv{f}  .  \SYSTEMmv{f}  \,   \textcolor{LGcolor}{\llparenthesis} \smidge  \SYSTEMnt{t_{{\mathrm{1}}}}  \smidge \textcolor{LGcolor}{\rrparenthesis}   \,  \SYSTEMmv{k}   \SYSTEMsym{)}  \,  \SYSTEMsym{(}   \lambda  \SYSTEMmv{x}  .   \textcolor{LGcolor}{\llparenthesis} \smidge  \SYSTEMnt{t_{{\mathrm{2}}}}  \smidge \textcolor{LGcolor}{\rrparenthesis}    \SYSTEMsym{)}  \\
\{\text{\SYSTEMRenameRuleGradEqbeta{}}\} & =  \lambda  \SYSTEMmv{k}  .  \SYSTEMsym{(}    \SYSTEMsym{(}   \lambda  \SYSTEMmv{x}  .   \textcolor{LGcolor}{\llparenthesis} \smidge  \SYSTEMnt{t_{{\mathrm{2}}}}  \smidge \textcolor{LGcolor}{\rrparenthesis}    \SYSTEMsym{)} \,   \textcolor{LGcolor}{\llparenthesis} \smidge  \SYSTEMnt{t_{{\mathrm{1}}}}  \smidge \textcolor{LGcolor}{\rrparenthesis}   \,  \SYSTEMmv{k}   \SYSTEMsym{)}  \\
\{\text{\SYSTEMRenameRuleGradEqbeta{}}\} & =   \lambda  \SYSTEMmv{k}  .  \SYSTEMsym{(}   [   \textcolor{LGcolor}{\llparenthesis} \smidge  \SYSTEMnt{t_{{\mathrm{1}}}}  \smidge \textcolor{LGcolor}{\rrparenthesis}   /  \SYSTEMmv{x}  ]   \textcolor{LGcolor}{\llparenthesis} \smidge  \SYSTEMnt{t_{{\mathrm{2}}}}  \smidge \textcolor{LGcolor}{\rrparenthesis}    \SYSTEMsym{)}  \,  \SYSTEMmv{k}  \\
\{\text{\SYSTEMRenameRuleGradEqeta{}}\} & =  [   \textcolor{LGcolor}{\llparenthesis} \smidge  \SYSTEMnt{t_{{\mathrm{1}}}}  \smidge \textcolor{LGcolor}{\rrparenthesis}   /  \SYSTEMmv{x}  ]   \textcolor{LGcolor}{\llparenthesis} \smidge  \SYSTEMnt{t_{{\mathrm{2}}}}  \smidge \textcolor{LGcolor}{\rrparenthesis}   \\
\{\textit{defn.}\} & =  \textcolor{LGcolor}{\llparenthesis} \smidge   [  \SYSTEMnt{t_{{\mathrm{1}}}}  /  \SYSTEMmv{x}  ]  \SYSTEMnt{t_{{\mathrm{2}}}}   \smidge \textcolor{LGcolor}{\rrparenthesis} 
\end{align*}

% Keeping this here as notes.
% \item \[\SYSTEMdruleLinEqeta{}\]
% % we can't preserve eta equality--this is actually well known for Plotkin's CPS
%   \begin{align*}
%                    & =  \textcolor{LGcolor}{\llparenthesis} \smidge   \lambda  \SYSTEMmv{x}  .    \SYSTEMnt{t} \,  \SYSTEMmv{x}     \smidge \textcolor{LGcolor}{\rrparenthesis}  \\
% \{\textit{defn.}\} & =   \lambda  \SYSTEMmv{k}  .  \SYSTEMmv{k}  \,  \SYSTEMsym{(}   \lambda  \SYSTEMmv{x}  .   \textcolor{LGcolor}{\llparenthesis} \smidge   \SYSTEMnt{t} \,  \SYSTEMmv{x}   \smidge \textcolor{LGcolor}{\rrparenthesis}    \SYSTEMsym{)}  \\
% \{\textit{defn.}\} & =   \lambda  \SYSTEMmv{k}  .  \SYSTEMmv{k}  \,  \SYSTEMsym{(}    \lambda  \SYSTEMmv{x}  .   \lambda  \SYSTEMmv{k'}  .   \textcolor{LGcolor}{\llparenthesis} \smidge  \SYSTEMnt{t}  \smidge \textcolor{LGcolor}{\rrparenthesis}    \,  \SYSTEMsym{(}     \lambda  \SYSTEMmv{f}  .  \SYSTEMmv{f}  \,   \textcolor{LGcolor}{\llparenthesis} \smidge  \SYSTEMmv{x}  \smidge \textcolor{LGcolor}{\rrparenthesis}   \,  \SYSTEMmv{k'}   \SYSTEMsym{)}   \SYSTEMsym{)}  \\
% \{\textit{defn.}\} & =   \lambda  \SYSTEMmv{k}  .  \SYSTEMmv{k}  \,  \SYSTEMsym{(}   \lambda  \SYSTEMmv{x}  .  \SYSTEMsym{(}    \lambda  \SYSTEMmv{k'}  .   \textcolor{LGcolor}{\llparenthesis} \smidge  \SYSTEMnt{t}  \smidge \textcolor{LGcolor}{\rrparenthesis}   \,  \SYSTEMsym{(}     \lambda  \SYSTEMmv{f}  .  \SYSTEMmv{f}  \,  \SYSTEMmv{x}  \,  \SYSTEMmv{k'}   \SYSTEMsym{)}   \SYSTEMsym{)}   \SYSTEMsym{)} 
% %% STUCK
% \end{align*}
%

  \item \[\SYSTEMdruleLinEqbetaBox{}\]

  \begin{align*}
    & =  \textcolor{LGcolor}{\llparenthesis} \smidge   \mathsf{let} \, \textcolor{coeffectColor}{[}  \SYSTEMmv{x}  \textcolor{coeffectColor}{]} =   \textcolor{coeffectColor}{[}  \SYSTEMnt{t_{{\mathrm{1}}}}  \textcolor{coeffectColor}{]}   \, \mathsf{in} \,  \SYSTEMnt{t_{{\mathrm{2}}}}   \smidge \textcolor{LGcolor}{\rrparenthesis}  \\
\{\textit{defn.}\} & =   \lambda  \SYSTEMmv{k}  .   \textcolor{LGcolor}{\llparenthesis} \smidge   \textcolor{coeffectColor}{[}  \SYSTEMnt{t_{{\mathrm{1}}}}  \textcolor{coeffectColor}{]}   \smidge \textcolor{LGcolor}{\rrparenthesis}   \,  \SYSTEMsym{(}    \lambda  \SYSTEMmv{x}  .   \textcolor{LGcolor}{\llparenthesis} \smidge  \SYSTEMnt{t_{{\mathrm{2}}}}  \smidge \textcolor{LGcolor}{\rrparenthesis}   \,  \SYSTEMmv{k}   \SYSTEMsym{)}  \\
\{\textit{defn.}\} & =   \lambda  \SYSTEMmv{k}  .  \SYSTEMsym{(}    \lambda  \SYSTEMmv{k}  .  \SYSTEMmv{k}  \,   \textcolor{LGcolor}{\llparenthesis} \smidge  \SYSTEMnt{t_{{\mathrm{1}}}}  \smidge \textcolor{LGcolor}{\rrparenthesis}    \SYSTEMsym{)}  \,  \SYSTEMsym{(}    \lambda  \SYSTEMmv{x}  .   \textcolor{LGcolor}{\llparenthesis} \smidge  \SYSTEMnt{t_{{\mathrm{2}}}}  \smidge \textcolor{LGcolor}{\rrparenthesis}   \,  \SYSTEMmv{k}   \SYSTEMsym{)}  \\
\{\text{\SYSTEMRenameRuleGradEqbeta{}}\} & =   \lambda  \SYSTEMmv{k}  .  \SYSTEMsym{(}    \lambda  \SYSTEMmv{x}  .   \textcolor{LGcolor}{\llparenthesis} \smidge  \SYSTEMnt{t_{{\mathrm{2}}}}  \smidge \textcolor{LGcolor}{\rrparenthesis}   \,  \SYSTEMmv{k}   \SYSTEMsym{)}  \,   \textcolor{LGcolor}{\llparenthesis} \smidge  \SYSTEMnt{t_{{\mathrm{1}}}}  \smidge \textcolor{LGcolor}{\rrparenthesis}   \\
\{\text{\SYSTEMRenameRuleGradEqbeta{}}\} & =   \lambda  \SYSTEMmv{k}  .   [   \textcolor{LGcolor}{\llparenthesis} \smidge  \SYSTEMnt{t_{{\mathrm{1}}}}  \smidge \textcolor{LGcolor}{\rrparenthesis}   /  \SYSTEMmv{x}  ]   \textcolor{LGcolor}{\llparenthesis} \smidge  \SYSTEMnt{t_{{\mathrm{2}}}}  \smidge \textcolor{LGcolor}{\rrparenthesis}    \,  \SYSTEMmv{k}  \\
\{\text{\SYSTEMRenameRuleGradEqeta{}}\} & =  [   \textcolor{LGcolor}{\llparenthesis} \smidge  \SYSTEMnt{t_{{\mathrm{1}}}}  \smidge \textcolor{LGcolor}{\rrparenthesis}   /  \SYSTEMmv{x}  ]   \textcolor{LGcolor}{\llparenthesis} \smidge  \SYSTEMnt{t_{{\mathrm{2}}}}  \smidge \textcolor{LGcolor}{\rrparenthesis}   \\
\{\textit{defn.}\} & =  \textcolor{LGcolor}{\llparenthesis} \smidge   [  \SYSTEMnt{t_{{\mathrm{1}}}}  /  \SYSTEMmv{x}  ]  \SYSTEMnt{t_{{\mathrm{2}}}}   \smidge \textcolor{LGcolor}{\rrparenthesis} 
  \end{align*}

\item \[\SYSTEMdruleLinEqetaBox{}\]

\begin{align*}
    & =  \textcolor{LGcolor}{\llparenthesis} \smidge   \mathsf{let} \, \textcolor{coeffectColor}{[}  \SYSTEMmv{x}  \textcolor{coeffectColor}{]} =  \SYSTEMnt{t}  \, \mathsf{in} \,   \textcolor{coeffectColor}{[}  \SYSTEMmv{x}  \textcolor{coeffectColor}{]}    \smidge \textcolor{LGcolor}{\rrparenthesis}  \\
\{\textit{defn.}\} & =   \lambda  \SYSTEMmv{k}  .   \textcolor{LGcolor}{\llparenthesis} \smidge  \SYSTEMnt{t}  \smidge \textcolor{LGcolor}{\rrparenthesis}   \,  \SYSTEMsym{(}    \lambda  \SYSTEMmv{x}  .   \textcolor{LGcolor}{\llparenthesis} \smidge   \textcolor{coeffectColor}{[}  \SYSTEMmv{x}  \textcolor{coeffectColor}{]}   \smidge \textcolor{LGcolor}{\rrparenthesis}   \,  \SYSTEMmv{k}   \SYSTEMsym{)}  \\
\{\textit{defn.}\} & =   \lambda  \SYSTEMmv{k}  .   \textcolor{LGcolor}{\llparenthesis} \smidge  \SYSTEMnt{t}  \smidge \textcolor{LGcolor}{\rrparenthesis}   \,  \SYSTEMsym{(}    \lambda  \SYSTEMmv{x}  .  \SYSTEMsym{(}    \lambda  \SYSTEMmv{k}  .  \SYSTEMmv{k}  \,  \SYSTEMmv{x}   \SYSTEMsym{)}  \,  \SYSTEMmv{k}   \SYSTEMsym{)}  \\
\{\textit{\SYSTEMRenameRuleGradEqbeta{}}\} & =   \lambda  \SYSTEMmv{k}  .   \textcolor{LGcolor}{\llparenthesis} \smidge  \SYSTEMnt{t}  \smidge \textcolor{LGcolor}{\rrparenthesis}   \,  \SYSTEMsym{(}    \lambda  \SYSTEMmv{x}  .  \SYSTEMmv{k}  \,  \SYSTEMmv{x}   \SYSTEMsym{)}  \\
\{\textit{\SYSTEMRenameRuleGradEqeta{}}\} & =   \lambda  \SYSTEMmv{k}  .   \textcolor{LGcolor}{\llparenthesis} \smidge  \SYSTEMnt{t}  \smidge \textcolor{LGcolor}{\rrparenthesis}   \,  \SYSTEMmv{k}  \\
\{\textit{\SYSTEMRenameRuleGradEqeta{}}\} & =  \textcolor{LGcolor}{\llparenthesis} \smidge  \SYSTEMnt{t}  \smidge \textcolor{LGcolor}{\rrparenthesis} 
\end{align*}

\item All congruence rules follow straightforwardly by induction.

\end{itemize}

\subsection{Proof of Soundness for Linear Base to Graded Modal Base}
\label{app:proofs-lin-to-grad-mod}

\ifextended\linToGradTranslation*\fi

\subsubsection{Type preservation}
\label{app:proofs-lin-to-grad-mod-typ}

\begin{lemma}
\label{lemma:zero-interp-lemma-lin-to-grad}
For all linear-base contexts $\Gamma$, if $ \mathrm{graded}( \Gamma , \textcolor{coeffectColor}{ \SYSTEMsym{0} }) $ then
$ \textcolor{LGcolor}{\llparenthesis}  \Gamma  \textcolor{LGcolor}{\rrparenthesis}  \equiv  \textcolor{coeffectColor}{ \SYSTEMsym{0}  \cdot}   \textcolor{LGcolor}{\llparenthesis}  \Gamma  \textcolor{LGcolor}{\rrparenthesis}  $.
\end{lemma}

\begin{proof} By induction on the structure of $\Gamma$.
\begin{itemize}
  \item (empty context)
  For $ \emptyset $, $ \textcolor{LGcolor}{\llparenthesis}   \emptyset   \textcolor{LGcolor}{\rrparenthesis} $ is the graded context $ \emptyset $. We have
  $ \emptyset  \equiv  \textcolor{coeffectColor}{ \SYSTEMsym{0}  \cdot}   \emptyset  $, thus the goal is trivially satisfied.
  \item (linear context extension)
  For $ \Gamma ,   \SYSTEMmv{x}  :  \SYSTEMnt{A}  $, the predicate  $ \mathrm{graded}(  (   \Gamma ,   \SYSTEMmv{x}  :  \SYSTEMnt{A}    )  , \textcolor{coeffectColor}{ \SYSTEMsym{0} }) $ by \textit{ex falso quodlibet}, the goal holds trivially.
  \item (graded context extension)
  For $ \Gamma ,   \SYSTEMmv{x}  : \textcolor{coeffectColor}{[}  \SYSTEMnt{A} {\textcolor{coeffectColor}{]_{ \SYSTEMnt{r} } } }  $, given  $ \mathrm{graded}(  (   \Gamma ,   \SYSTEMmv{x}  : \textcolor{coeffectColor}{[}  \SYSTEMnt{A} {\textcolor{coeffectColor}{]_{ \SYSTEMnt{r} } } }    )  , \textcolor{coeffectColor}{ \SYSTEMsym{0} }) $, unification implies $\SYSTEMnt{r} \equiv \SYSTEMsym{0}$, given this and the definition of $\mathsf{graded}$, we have $ \mathrm{graded}( \Gamma , \textcolor{coeffectColor}{ \SYSTEMsym{0} }) $. Applying this with induction on $\Gamma$ we get $ \textcolor{LGcolor}{\llparenthesis}  \Gamma  \textcolor{LGcolor}{\rrparenthesis}  \equiv  \textcolor{coeffectColor}{ \SYSTEMsym{0}  \cdot}   \textcolor{LGcolor}{\llparenthesis}  \Gamma  \textcolor{LGcolor}{\rrparenthesis}  $ from which we conclude with the following reasoning steps:
  \begin{align*}
    \textcolor{LGcolor}{\llparenthesis}  \Gamma  \textcolor{LGcolor}{\rrparenthesis}  ,   \SYSTEMmv{x}  :_{\textcolor{coeffectColor}{ \SYSTEMsym{0} } }  \SYSTEMnt{A}   &\equiv    \textcolor{coeffectColor}{ \SYSTEMsym{0}  \cdot}   \textcolor{LGcolor}{\llparenthesis}  \Gamma  \textcolor{LGcolor}{\rrparenthesis}    ,   \SYSTEMmv{x}  :_{\textcolor{coeffectColor}{ \SYSTEMsym{0} } }  \SYSTEMnt{A}  \\
    \textcolor{LGcolor}{\llparenthesis}  \Gamma  \textcolor{LGcolor}{\rrparenthesis}  ,   \SYSTEMmv{x}  :_{\textcolor{coeffectColor}{ \SYSTEMsym{0} } }  \SYSTEMnt{A}   &\equiv  \textcolor{coeffectColor}{ \SYSTEMsym{0}  \cdot}   (    \textcolor{LGcolor}{\llparenthesis}  \Gamma  \textcolor{LGcolor}{\rrparenthesis}  ,   \SYSTEMmv{x}  :_{\textcolor{coeffectColor}{ \SYSTEMsym{0} } }  \SYSTEMnt{A}    )  \\
   \textcolor{LGcolor}{\llparenthesis}   \Gamma ,   \SYSTEMmv{x}  : \textcolor{coeffectColor}{[}  \SYSTEMnt{A} {\textcolor{coeffectColor}{]_{ \SYSTEMsym{0} } } }    \textcolor{LGcolor}{\rrparenthesis}  &\equiv  \textcolor{coeffectColor}{ \SYSTEMsym{0}  \cdot}   \textcolor{LGcolor}{\llparenthesis}   \Gamma ,   \SYSTEMmv{x}  : \textcolor{coeffectColor}{[}  \SYSTEMnt{A} {\textcolor{coeffectColor}{]_{ \SYSTEMsym{0} } } }    \textcolor{LGcolor}{\rrparenthesis}  
  \end{align*}

\end{itemize}
\end{proof}

Type preservation then follows:

\begin{proof}
By induction on the Linear Base typing:
\begin{itemize}
\item (var)
$$
\SYSTEMdruleLinvar{}
$$
Therefore we construct the goal typing:
$$
\SYSTEMdruleGradvar{}
$$
%%%%%%%%%%%%%%%%%%%%%%%%%%%%%%%%%%%%%%%%%%%%%%%%%%%%%%%%%%%%%%%%%%%%%%%%%%%%%%%%
\item (abs)
$$
\SYSTEMdruleLinabs{}
$$
By induction on the premise we have
${   \textcolor{LGcolor}{\llparenthesis}  \Gamma  \textcolor{LGcolor}{\rrparenthesis}  ,   \SYSTEMmv{x}  :_{\textcolor{coeffectColor}{ \SYSTEMsym{1} } }   \textcolor{LGcolor}{\llparenthesis} \smidge  \SYSTEMnt{A}  \smidge \textcolor{LGcolor}{\rrparenthesis}     \vdash_{\textsc{g} }   \textcolor{LGcolor}{\llparenthesis} \smidge  \SYSTEMnt{t}  \smidge \textcolor{LGcolor}{\rrparenthesis}   :   \textcolor{LGcolor}{\llparenthesis} \smidge  \SYSTEMnt{B}  \smidge \textcolor{LGcolor}{\rrparenthesis}  }$.

Therefore we construct the goal typing:
$$\inferrule*[Right=\SYSTEMRenameRuleGradabs{}]
    {   \textcolor{LGcolor}{\llparenthesis}  \Gamma  \textcolor{LGcolor}{\rrparenthesis}  ,   \SYSTEMmv{x}  :_{\textcolor{coeffectColor}{ \SYSTEMsym{1} } }   \textcolor{LGcolor}{\llparenthesis} \smidge  \SYSTEMnt{A}  \smidge \textcolor{LGcolor}{\rrparenthesis}     \vdash_{\textsc{g} }   \textcolor{LGcolor}{\llparenthesis} \smidge  \SYSTEMnt{t}  \smidge \textcolor{LGcolor}{\rrparenthesis}   :   \textcolor{LGcolor}{\llparenthesis} \smidge  \SYSTEMnt{B}  \smidge \textcolor{LGcolor}{\rrparenthesis}  }
    {  \textcolor{LGcolor}{\llparenthesis}  \Gamma  \textcolor{LGcolor}{\rrparenthesis}   \vdash_{\textsc{g} }   \lambda  \SYSTEMmv{x}  .   \textcolor{LGcolor}{\llparenthesis} \smidge  \SYSTEMnt{t}  \smidge \textcolor{LGcolor}{\rrparenthesis}    :    \textcolor{LGcolor}{\llparenthesis} \smidge  \SYSTEMnt{A}  \smidge \textcolor{LGcolor}{\rrparenthesis}   \xrightarrow{\textcolor{coeffectColor}{ \SYSTEMsym{1} } }   \textcolor{LGcolor}{\llparenthesis} \smidge  \SYSTEMnt{B}  \smidge \textcolor{LGcolor}{\rrparenthesis}   }
$$
%%%%%%%%%%%%%%%%%%%%%%%%%%%%%%%%%%%%%%%%%%%%%%%%%%%%%%%%%%%%%%%%%%%%%%%%%%%%%%%%
\item (app)
$$
\SYSTEMdruleLinapp{}
$$
By induction on the premises we have
$  \textcolor{LGcolor}{\llparenthesis}  \Gamma_{{\mathrm{1}}}  \textcolor{LGcolor}{\rrparenthesis}   \vdash_{\textsc{g} }   \textcolor{LGcolor}{\llparenthesis} \smidge  \SYSTEMnt{t_{{\mathrm{1}}}}  \smidge \textcolor{LGcolor}{\rrparenthesis}   :    \textcolor{LGcolor}{\llparenthesis} \smidge  \SYSTEMnt{A}  \smidge \textcolor{LGcolor}{\rrparenthesis}   \xrightarrow{\textcolor{coeffectColor}{ \SYSTEMsym{1} } }   \textcolor{LGcolor}{\llparenthesis} \smidge  \SYSTEMnt{B}  \smidge \textcolor{LGcolor}{\rrparenthesis}   $
and
$  \textcolor{LGcolor}{\llparenthesis}  \Gamma_{{\mathrm{2}}}  \textcolor{LGcolor}{\rrparenthesis}   \vdash_{\textsc{g} }   \textcolor{LGcolor}{\llparenthesis} \smidge  \SYSTEMnt{t_{{\mathrm{2}}}}  \smidge \textcolor{LGcolor}{\rrparenthesis}   :   \textcolor{LGcolor}{\llparenthesis} \smidge  \SYSTEMnt{A}  \smidge \textcolor{LGcolor}{\rrparenthesis}  $.

Therefore we construct:
$$
\inferrule*[Right=\SYSTEMRenameRuleGradapp{}]
    {  \textcolor{LGcolor}{\llparenthesis}  \Gamma_{{\mathrm{1}}}  \textcolor{LGcolor}{\rrparenthesis}   \vdash_{\textsc{g} }   \textcolor{LGcolor}{\llparenthesis} \smidge  \SYSTEMnt{t_{{\mathrm{1}}}}  \smidge \textcolor{LGcolor}{\rrparenthesis}   :    \textcolor{LGcolor}{\llparenthesis} \smidge  \SYSTEMnt{A}  \smidge \textcolor{LGcolor}{\rrparenthesis}   \xrightarrow{\textcolor{coeffectColor}{ \SYSTEMsym{1} } }   \textcolor{LGcolor}{\llparenthesis} \smidge  \SYSTEMnt{B}  \smidge \textcolor{LGcolor}{\rrparenthesis}   
    \\  \textcolor{LGcolor}{\llparenthesis}  \Gamma_{{\mathrm{2}}}  \textcolor{LGcolor}{\rrparenthesis}   \vdash_{\textsc{g} }   \textcolor{LGcolor}{\llparenthesis} \smidge  \SYSTEMnt{t_{{\mathrm{2}}}}  \smidge \textcolor{LGcolor}{\rrparenthesis}   :   \textcolor{LGcolor}{\llparenthesis} \smidge  \SYSTEMnt{A}  \smidge \textcolor{LGcolor}{\rrparenthesis}  }
    {  \textcolor{LGcolor}{\llparenthesis}  \Gamma_{{\mathrm{1}}}  \textcolor{LGcolor}{\rrparenthesis}   \SYSTEMsym{+}   \textcolor{coeffectColor}{ \SYSTEMsym{1}  \cdot}   \textcolor{LGcolor}{\llparenthesis}  \Gamma_{{\mathrm{2}}}  \textcolor{LGcolor}{\rrparenthesis}    \vdash_{\textsc{g} }    \textcolor{LGcolor}{\llparenthesis} \smidge  \SYSTEMnt{t_{{\mathrm{1}}}}  \smidge \textcolor{LGcolor}{\rrparenthesis}  \,   \textcolor{LGcolor}{\llparenthesis} \smidge  \SYSTEMnt{t_{{\mathrm{2}}}}  \smidge \textcolor{LGcolor}{\rrparenthesis}    :   \textcolor{LGcolor}{\llparenthesis} \smidge  \SYSTEMnt{B}  \smidge \textcolor{LGcolor}{\rrparenthesis}  }
$$

which matches the goal type by:
$$
\begin{array}{rll}
&& \textcolor{LGcolor}{\llparenthesis}  \Gamma_{{\mathrm{1}}}  \textcolor{LGcolor}{\rrparenthesis}   \SYSTEMsym{+}   \textcolor{coeffectColor}{ \SYSTEMsym{1}  \cdot}   \textcolor{LGcolor}{\llparenthesis}  \Gamma_{{\mathrm{2}}}  \textcolor{LGcolor}{\rrparenthesis}  \\
\text{unit property}&\equiv& \textcolor{LGcolor}{\llparenthesis}  \Gamma_{{\mathrm{1}}}  \textcolor{LGcolor}{\rrparenthesis}   \SYSTEMsym{+}   \textcolor{LGcolor}{\llparenthesis}  \Gamma_{{\mathrm{2}}}  \textcolor{LGcolor}{\rrparenthesis} \\
\text{homomorphism}&\equiv& \textcolor{LGcolor}{\llparenthesis}  \Gamma_{{\mathrm{1}}}  \SYSTEMsym{+}  \Gamma_{{\mathrm{2}}}  \textcolor{LGcolor}{\rrparenthesis} 
\end{array}
$$
%%%%%%%%%%%%%%%%%%%%%%%%%%%%%%%%%%%%%%%%%%%%%%%%%%%%%%%%%%%%%%%%%%%%%%%%%%%%%%%%
\item (weak)
$$
\SYSTEMdruleLinweak{}
$$

By induction on the premise we have
$  \textcolor{LGcolor}{\llparenthesis}  \Gamma  \textcolor{LGcolor}{\rrparenthesis}   \vdash_{\textsc{g} }   \textcolor{LGcolor}{\llparenthesis} \smidge  \SYSTEMnt{t}  \smidge \textcolor{LGcolor}{\rrparenthesis}   :   \textcolor{LGcolor}{\llparenthesis} \smidge  \SYSTEMnt{A}  \smidge \textcolor{LGcolor}{\rrparenthesis}  $.

Therefore we construct:
$$
\inferrule*[Right=\SYSTEMRenameRuleGradweak{}]
    {  \textcolor{LGcolor}{\llparenthesis}  \Gamma  \textcolor{LGcolor}{\rrparenthesis}   \vdash_{\textsc{g} }   \textcolor{LGcolor}{\llparenthesis} \smidge  \SYSTEMnt{t}  \smidge \textcolor{LGcolor}{\rrparenthesis}   :   \textcolor{LGcolor}{\llparenthesis} \smidge  \SYSTEMnt{A}  \smidge \textcolor{LGcolor}{\rrparenthesis}  }
    {   \textcolor{LGcolor}{\llparenthesis}  \Gamma  \textcolor{LGcolor}{\rrparenthesis}   ,   \textcolor{coeffectColor}{ \SYSTEMsym{0}  \cdot}   \textcolor{LGcolor}{\llparenthesis}  \Gamma'  \textcolor{LGcolor}{\rrparenthesis}     \vdash_{\textsc{g} }   \textcolor{LGcolor}{\llparenthesis} \smidge  \SYSTEMnt{t}  \smidge \textcolor{LGcolor}{\rrparenthesis}   :   \textcolor{LGcolor}{\llparenthesis} \smidge  \SYSTEMnt{A}  \smidge \textcolor{LGcolor}{\rrparenthesis}  }
$$
which satisfies the goal by Lemma~\ref{lemma:zero-interp-lemma-lin-to-grad}.

%%%%%%%%%%%%%%%%%%%%%%%%%%%%%%%%%%%%%%%%%%%%%%%%%%%%%%%%%%%%%%%%%%%%%%%%%%%%%%%%
\item (der)
$$
\SYSTEMdruleLinder{}
$$
By induction on the premise, we have $   \textcolor{LGcolor}{\llparenthesis}  \Gamma  \textcolor{LGcolor}{\rrparenthesis}  ,   \SYSTEMmv{x}  :_{\textcolor{coeffectColor}{ \SYSTEMsym{1} } }   \textcolor{LGcolor}{\llparenthesis} \smidge  \SYSTEMnt{A}  \smidge \textcolor{LGcolor}{\rrparenthesis}     \vdash_{\textsc{g} }   \textcolor{LGcolor}{\llparenthesis} \smidge  \SYSTEMnt{t}  \smidge \textcolor{LGcolor}{\rrparenthesis}   :   \textcolor{LGcolor}{\llparenthesis} \smidge  \SYSTEMnt{B}  \smidge \textcolor{LGcolor}{\rrparenthesis}  $
which gives the goal.

%%%%%%%%%%%%%%%%%%%%%%%%%%%%%%%%%%%%%%%%%%%%%%%%%%%%%%%%%%%%%%%%%%%%%%%%%%%%%%%%
\item (pr)
$$
\SYSTEMdruleLinpr{}
$$
By induction on the premise we have $  \textcolor{LGcolor}{\llparenthesis}  \Gamma  \textcolor{LGcolor}{\rrparenthesis}   \vdash_{\textsc{g} }   \textcolor{LGcolor}{\llparenthesis} \smidge  \SYSTEMnt{t}  \smidge \textcolor{LGcolor}{\rrparenthesis}   :   \textcolor{LGcolor}{\llparenthesis} \smidge  \SYSTEMnt{A}  \smidge \textcolor{LGcolor}{\rrparenthesis}  $.

Therefore we can construct:
$$
\inferrule*[Right=\SYSTEMRenameRuleGradBoxpr{}]
{   \textcolor{LGcolor}{\llparenthesis}  \Gamma  \textcolor{LGcolor}{\rrparenthesis}   \vdash_{\textsc{g} }   \textcolor{LGcolor}{\llparenthesis} \smidge  \SYSTEMnt{t}  \smidge \textcolor{LGcolor}{\rrparenthesis}   :   \textcolor{LGcolor}{\llparenthesis} \smidge  \SYSTEMnt{A}  \smidge \textcolor{LGcolor}{\rrparenthesis}   }
{   \textcolor{coeffectColor}{ \SYSTEMnt{r}  \cdot}   \textcolor{LGcolor}{\llparenthesis}  \Gamma  \textcolor{LGcolor}{\rrparenthesis}    \vdash_{\textsc{g} }   \textcolor{LGcolor}{\llparenthesis} \smidge  \SYSTEMnt{t}  \smidge \textcolor{LGcolor}{\rrparenthesis}   :   \textcolor{coeffectColor}{\square_{ \SYSTEMnt{r} } }   \textcolor{LGcolor}{\llparenthesis} \smidge  \SYSTEMnt{A}  \smidge \textcolor{LGcolor}{\rrparenthesis}    }
$$
which satisfies the goal since $ \textcolor{LGcolor}{\llparenthesis}   \textcolor{coeffectColor}{ \SYSTEMnt{r}  \cdot}  \Gamma   \textcolor{LGcolor}{\rrparenthesis}  \equiv  \textcolor{coeffectColor}{ \SYSTEMnt{r}  \cdot}   \textcolor{LGcolor}{\llparenthesis}  \Gamma  \textcolor{LGcolor}{\rrparenthesis}  $ (homomorphism).

\item (let)
$$
\SYSTEMdruleLinlet{}
$$
By induction on both premises we have
$  \textcolor{LGcolor}{\llparenthesis}  \Gamma_{{\mathrm{1}}}  \textcolor{LGcolor}{\rrparenthesis}   \vdash_{\textsc{g} }   \textcolor{LGcolor}{\llparenthesis} \smidge  \SYSTEMnt{t_{{\mathrm{1}}}}  \smidge \textcolor{LGcolor}{\rrparenthesis}   :   \textcolor{coeffectColor}{\square_{ \SYSTEMnt{r} } }   \textcolor{LGcolor}{\llparenthesis} \smidge  \SYSTEMnt{A}  \smidge \textcolor{LGcolor}{\rrparenthesis}   $
and
$   \textcolor{LGcolor}{\llparenthesis}  \Gamma_{{\mathrm{2}}}  \textcolor{LGcolor}{\rrparenthesis}  ,   \SYSTEMmv{x}  :_{\textcolor{coeffectColor}{ \SYSTEMnt{r} } }   \textcolor{LGcolor}{\llparenthesis} \smidge  \SYSTEMnt{A}  \smidge \textcolor{LGcolor}{\rrparenthesis}     \vdash_{\textsc{g} }   \textcolor{LGcolor}{\llparenthesis} \smidge  \SYSTEMnt{t_{{\mathrm{2}}}}  \smidge \textcolor{LGcolor}{\rrparenthesis}   :   \textcolor{LGcolor}{\llparenthesis} \smidge  \SYSTEMnt{B}  \smidge \textcolor{LGcolor}{\rrparenthesis}  $.

Therefore we can construct:
$$
\inferrule*[Right=\SYSTEMRenameRuleGradBoxlet{}]
{  \textcolor{LGcolor}{\llparenthesis}  \Gamma_{{\mathrm{1}}}  \textcolor{LGcolor}{\rrparenthesis}   \vdash_{\textsc{g} }   \textcolor{LGcolor}{\llparenthesis} \smidge  \SYSTEMnt{t_{{\mathrm{1}}}}  \smidge \textcolor{LGcolor}{\rrparenthesis}   :   \textcolor{coeffectColor}{\square_{ \SYSTEMnt{r} } }   \textcolor{LGcolor}{\llparenthesis} \smidge  \SYSTEMnt{A}  \smidge \textcolor{LGcolor}{\rrparenthesis}    \\
    \textcolor{LGcolor}{\llparenthesis}  \Gamma_{{\mathrm{2}}}  \textcolor{LGcolor}{\rrparenthesis}  ,   \SYSTEMmv{x}  :_{\textcolor{coeffectColor}{ \SYSTEMnt{r} } }   \textcolor{LGcolor}{\llparenthesis} \smidge  \SYSTEMnt{A}  \smidge \textcolor{LGcolor}{\rrparenthesis}     \vdash_{\textsc{g} }   \textcolor{LGcolor}{\llparenthesis} \smidge  \SYSTEMnt{t_{{\mathrm{2}}}}  \smidge \textcolor{LGcolor}{\rrparenthesis}   :   \textcolor{LGcolor}{\llparenthesis} \smidge  \SYSTEMnt{B}  \smidge \textcolor{LGcolor}{\rrparenthesis}  }
{  \textcolor{LGcolor}{\llparenthesis}  \Gamma_{{\mathrm{1}}}  \textcolor{LGcolor}{\rrparenthesis}   \SYSTEMsym{+}   \textcolor{LGcolor}{\llparenthesis}  \Gamma_{{\mathrm{2}}}  \textcolor{LGcolor}{\rrparenthesis}   \vdash_{\textsc{g} }   \mathsf{let} \, \textcolor{coeffectColor}{[}  \SYSTEMmv{x}  \textcolor{coeffectColor}{]} =   \textcolor{LGcolor}{\llparenthesis} \smidge  \SYSTEMnt{t_{{\mathrm{1}}}}  \smidge \textcolor{LGcolor}{\rrparenthesis}   \, \mathsf{in} \,   \textcolor{LGcolor}{\llparenthesis} \smidge  \SYSTEMnt{t_{{\mathrm{2}}}}  \smidge \textcolor{LGcolor}{\rrparenthesis}    :   \textcolor{LGcolor}{\llparenthesis} \smidge  \SYSTEMnt{B}  \smidge \textcolor{LGcolor}{\rrparenthesis}   }
$$

\item (approx)
$$
\SYSTEMdruleLinapprox{}
$$
By induction on the premise we have
$   \textcolor{LGcolor}{\llparenthesis}  \Gamma  \textcolor{LGcolor}{\rrparenthesis}  ,   \SYSTEMmv{x}  :_{\textcolor{coeffectColor}{ \SYSTEMnt{s} } }   \textcolor{LGcolor}{\llparenthesis} \smidge  \SYSTEMnt{B}  \smidge \textcolor{LGcolor}{\rrparenthesis}     \vdash_{\textsc{g} }   \textcolor{LGcolor}{\llparenthesis} \smidge  \SYSTEMnt{t}  \smidge \textcolor{LGcolor}{\rrparenthesis}   :   \textcolor{LGcolor}{\llparenthesis} \smidge  \SYSTEMnt{A}  \smidge \textcolor{LGcolor}{\rrparenthesis}  $.

Therefore we can construct:
$$
\inferrule*[Right=\SYSTEMRenameRuleGradBoxlet{}]
{ \begin{array}{cc}     \textcolor{LGcolor}{\llparenthesis}  \Gamma  \textcolor{LGcolor}{\rrparenthesis}  ,   \SYSTEMmv{x}  :_{\textcolor{coeffectColor}{ \SYSTEMnt{r} } }   \textcolor{LGcolor}{\llparenthesis} \smidge  \SYSTEMnt{B}  \smidge \textcolor{LGcolor}{\rrparenthesis}     \vdash_{\textsc{g} }   \textcolor{LGcolor}{\llparenthesis} \smidge  \SYSTEMnt{t}  \smidge \textcolor{LGcolor}{\rrparenthesis}   :   \textcolor{LGcolor}{\llparenthesis} \smidge  \SYSTEMnt{A}  \smidge \textcolor{LGcolor}{\rrparenthesis}    \; & \;   \SYSTEMnt{r}  \, \textcolor{coeffectColor}{\sqsubseteq} \,  \SYSTEMnt{s}   \end{array} }
{   \textcolor{LGcolor}{\llparenthesis}  \Gamma  \textcolor{LGcolor}{\rrparenthesis}  ,   \SYSTEMmv{x}  :_{\textcolor{coeffectColor}{ \SYSTEMnt{s} } }   \textcolor{LGcolor}{\llparenthesis} \smidge  \SYSTEMnt{B}  \smidge \textcolor{LGcolor}{\rrparenthesis}     \vdash_{\textsc{g} }   \textcolor{LGcolor}{\llparenthesis} \smidge  \SYSTEMnt{t}  \smidge \textcolor{LGcolor}{\rrparenthesis}   :   \textcolor{LGcolor}{\llparenthesis} \smidge  \SYSTEMnt{A}  \smidge \textcolor{LGcolor}{\rrparenthesis}  }
$$

\end{itemize}
\end{proof}

\subsubsection{Operational correspondence}
\label{app:proofs-lin-to-grad-mod-ops}

\begin{lemma}[Interpretation preserves substitution]
\label{lemma:interp-lin-to-grd-mod-preserves-subst}
For all Linear Base terms $\SYSTEMnt{t}, \SYSTEMnt{t'}$ then
$ \textcolor{LGcolor}{\llparenthesis} \smidge   [  \SYSTEMnt{t}  /  \SYSTEMmv{x}  ]  \SYSTEMnt{t'}   \smidge \textcolor{LGcolor}{\rrparenthesis}  \equiv  [   \textcolor{LGcolor}{\llparenthesis} \smidge  \SYSTEMnt{t}  \smidge \textcolor{LGcolor}{\rrparenthesis}   /  \SYSTEMmv{x}  ]   \textcolor{LGcolor}{\llparenthesis} \smidge  \SYSTEMnt{t'}  \smidge \textcolor{LGcolor}{\rrparenthesis}  $.
\end{lemma}

\begin{proof}
By induction on the receiving term $\SYSTEMnt{t'}$:
\begin{itemize}
\item (var) $\SYSTEMnt{t'} \equiv \SYSTEMmv{y}$:

\begin{itemize}
\item $\SYSTEMmv{x} \equiv \SYSTEMmv{y}$ then we refine the goal:
\begin{align*}
\begin{array}{rcrlll}
\textit{(goal)} & &  \textcolor{LGcolor}{\llparenthesis} \smidge   [  \SYSTEMnt{t}  /  \SYSTEMmv{x}  ]  \SYSTEMmv{x}   \smidge \textcolor{LGcolor}{\rrparenthesis}  & \eqStep &  [   \textcolor{LGcolor}{\llparenthesis} \smidge  \SYSTEMnt{t}  \smidge \textcolor{LGcolor}{\rrparenthesis}   /  \SYSTEMmv{x}  ]   \textcolor{LGcolor}{\llparenthesis} \smidge  \SYSTEMmv{x}  \smidge \textcolor{LGcolor}{\rrparenthesis}   \\
\textit{(defn. subst [lhs])} & \Rightarrow &  \textcolor{LGcolor}{\llparenthesis} \smidge  \SYSTEMnt{t}  \smidge \textcolor{LGcolor}{\rrparenthesis}  & \eqStep &  [   \textcolor{LGcolor}{\llparenthesis} \smidge  \SYSTEMnt{t}  \smidge \textcolor{LGcolor}{\rrparenthesis}   /  \SYSTEMmv{x}  ]   \textcolor{LGcolor}{\llparenthesis} \smidge  \SYSTEMmv{x}  \smidge \textcolor{LGcolor}{\rrparenthesis}   \\
\textit{(defn. interp [rhs])} & \Rightarrow &  \textcolor{LGcolor}{\llparenthesis} \smidge  \SYSTEMnt{t}  \smidge \textcolor{LGcolor}{\rrparenthesis}  & \eqStep &  [   \textcolor{LGcolor}{\llparenthesis} \smidge  \SYSTEMnt{t}  \smidge \textcolor{LGcolor}{\rrparenthesis}   /  \SYSTEMmv{x}  ]  \SYSTEMmv{x}  \\
\textit{(defn. subst [rhs])} & \Rightarrow &  \textcolor{LGcolor}{\llparenthesis} \smidge  \SYSTEMnt{t}  \smidge \textcolor{LGcolor}{\rrparenthesis}  & \eqStep &  \textcolor{LGcolor}{\llparenthesis} \smidge  \SYSTEMnt{t}  \smidge \textcolor{LGcolor}{\rrparenthesis}  \\
\textit{(reflexivity)} & \Rightarrow & \top
\end{array}
\end{align*}

\item $\SYSTEMmv{x} \not\equiv \SYSTEMmv{y}$ then we refine the goal:
\begin{align*}
\begin{array}{rcrlll}
\textit{(goal)} & &  \textcolor{LGcolor}{\llparenthesis} \smidge   [  \SYSTEMnt{t}  /  \SYSTEMmv{x}  ]  \SYSTEMmv{y}   \smidge \textcolor{LGcolor}{\rrparenthesis}  & \eqStep &  [   \textcolor{LGcolor}{\llparenthesis} \smidge  \SYSTEMnt{t}  \smidge \textcolor{LGcolor}{\rrparenthesis}   /  \SYSTEMmv{x}  ]   \textcolor{LGcolor}{\llparenthesis} \smidge  \SYSTEMmv{y}  \smidge \textcolor{LGcolor}{\rrparenthesis}   \\
\textit{(defn. subst [lhs])} & \Rightarrow &  \textcolor{LGcolor}{\llparenthesis} \smidge  \SYSTEMmv{y}  \smidge \textcolor{LGcolor}{\rrparenthesis}  & \eqStep &  [   \textcolor{LGcolor}{\llparenthesis} \smidge  \SYSTEMnt{t}  \smidge \textcolor{LGcolor}{\rrparenthesis}   /  \SYSTEMmv{x}  ]   \textcolor{LGcolor}{\llparenthesis} \smidge  \SYSTEMmv{y}  \smidge \textcolor{LGcolor}{\rrparenthesis}   \\
\textit{(defn. interp [lhs])} & \Rightarrow & \SYSTEMmv{y} & \eqStep &  [   \textcolor{LGcolor}{\llparenthesis} \smidge  \SYSTEMnt{t}  \smidge \textcolor{LGcolor}{\rrparenthesis}   /  \SYSTEMmv{x}  ]   \textcolor{LGcolor}{\llparenthesis} \smidge  \SYSTEMmv{y}  \smidge \textcolor{LGcolor}{\rrparenthesis}   \\
\textit{(defn. interp [rhs])} & \Rightarrow & \SYSTEMmv{y} & \eqStep &  [   \textcolor{LGcolor}{\llparenthesis} \smidge  \SYSTEMnt{t}  \smidge \textcolor{LGcolor}{\rrparenthesis}   /  \SYSTEMmv{x}  ]  \SYSTEMmv{y}  \\
\textit{(defn. subst [rhs])} & \Rightarrow & \SYSTEMmv{y} & \eqStep & \SYSTEMmv{y} \\
\textit{(reflexivity)} & \Rightarrow & \top
\end{array}
\end{align*}

\end{itemize}

\item (abs) $\SYSTEMnt{t'} \equiv  \lambda  \SYSTEMmv{y}  .  \SYSTEMnt{t_{{\mathrm{1}}}} $ (with $ \SYSTEMmv{y}  \,\#\,  \SYSTEMnt{t} $);
\begin{gather*}
\begin{align*}
\begin{array}{rcrlll}
\textit{(goal)} & &  \textcolor{LGcolor}{\llparenthesis} \smidge   [  \SYSTEMnt{t}  /  \SYSTEMmv{x}  ]  \SYSTEMsym{(}   \lambda  \SYSTEMmv{y}  .  \SYSTEMnt{t_{{\mathrm{1}}}}   \SYSTEMsym{)}   \smidge \textcolor{LGcolor}{\rrparenthesis}  & \eqStep &  [   \textcolor{LGcolor}{\llparenthesis} \smidge  \SYSTEMnt{t}  \smidge \textcolor{LGcolor}{\rrparenthesis}   /  \SYSTEMmv{x}  ]   \textcolor{LGcolor}{\llparenthesis} \smidge  \SYSTEMsym{(}   \lambda  \SYSTEMmv{y}  .  \SYSTEMnt{t_{{\mathrm{1}}}}   \SYSTEMsym{)}  \smidge \textcolor{LGcolor}{\rrparenthesis}   \\
\textit{(defn. subst [lhs])} & \Rightarrow &  \textcolor{LGcolor}{\llparenthesis} \smidge   \lambda  \SYSTEMmv{y}  .   [  \SYSTEMnt{t}  /  \SYSTEMmv{x}  ]  \SYSTEMnt{t_{{\mathrm{1}}}}    \smidge \textcolor{LGcolor}{\rrparenthesis}  & \eqStep &  [   \textcolor{LGcolor}{\llparenthesis} \smidge  \SYSTEMnt{t}  \smidge \textcolor{LGcolor}{\rrparenthesis}   /  \SYSTEMmv{x}  ]   \textcolor{LGcolor}{\llparenthesis} \smidge  \SYSTEMsym{(}   \lambda  \SYSTEMmv{y}  .  \SYSTEMnt{t_{{\mathrm{1}}}}   \SYSTEMsym{)}  \smidge \textcolor{LGcolor}{\rrparenthesis}   \\
\textit{(defn. interp [rhs])} & \Rightarrow &  \textcolor{LGcolor}{\llparenthesis} \smidge   \lambda  \SYSTEMmv{y}  .   [  \SYSTEMnt{t}  /  \SYSTEMmv{x}  ]  \SYSTEMnt{t_{{\mathrm{1}}}}    \smidge \textcolor{LGcolor}{\rrparenthesis}  & \eqStep &  [   \textcolor{LGcolor}{\llparenthesis} \smidge  \SYSTEMnt{t}  \smidge \textcolor{LGcolor}{\rrparenthesis}   /  \SYSTEMmv{x}  ]   \lambda  \SYSTEMmv{y}  .   \textcolor{LGcolor}{\llparenthesis} \smidge  \SYSTEMnt{t_{{\mathrm{1}}}}  \smidge \textcolor{LGcolor}{\rrparenthesis}    \\
\textit{(defn. subst [rhs])} & \Rightarrow &  \textcolor{LGcolor}{\llparenthesis} \smidge   \lambda  \SYSTEMmv{y}  .   [  \SYSTEMnt{t}  /  \SYSTEMmv{x}  ]  \SYSTEMnt{t_{{\mathrm{1}}}}    \smidge \textcolor{LGcolor}{\rrparenthesis}  & \eqStep &  \lambda  \SYSTEMmv{y}  .   [   \textcolor{LGcolor}{\llparenthesis} \smidge  \SYSTEMnt{t}  \smidge \textcolor{LGcolor}{\rrparenthesis}   /  \SYSTEMmv{x}  ]   \textcolor{LGcolor}{\llparenthesis} \smidge  \SYSTEMnt{t_{{\mathrm{1}}}}  \smidge \textcolor{LGcolor}{\rrparenthesis}    \\
\textit{(defn. interp [lhs])} & \Rightarrow &  \lambda  \SYSTEMmv{y}  .   \textcolor{LGcolor}{\llparenthesis} \smidge   [  \SYSTEMnt{t}  /  \SYSTEMmv{x}  ]  \SYSTEMnt{t_{{\mathrm{1}}}}   \smidge \textcolor{LGcolor}{\rrparenthesis}   & \eqStep &  \lambda  \SYSTEMmv{y}  .   [   \textcolor{LGcolor}{\llparenthesis} \smidge  \SYSTEMnt{t}  \smidge \textcolor{LGcolor}{\rrparenthesis}   /  \SYSTEMmv{x}  ]   \textcolor{LGcolor}{\llparenthesis} \smidge  \SYSTEMnt{t_{{\mathrm{1}}}}  \smidge \textcolor{LGcolor}{\rrparenthesis}    \\
\textit{(induction on $t1$)} & \Rightarrow & \top
\end{array}
\end{align*}
\end{gather*}

\item (app) $\SYSTEMnt{t'} \equiv  \SYSTEMnt{t_{{\mathrm{1}}}} \,  \SYSTEMnt{t_{{\mathrm{2}}}} $;
\begin{gather*}
\begin{align*}
\begin{array}{rcrlll}
\textit{(goal)} & &  \textcolor{LGcolor}{\llparenthesis} \smidge   [  \SYSTEMnt{t}  /  \SYSTEMmv{x}  ]  \SYSTEMsym{(}   \SYSTEMnt{t_{{\mathrm{1}}}} \,  \SYSTEMnt{t_{{\mathrm{2}}}}   \SYSTEMsym{)}   \smidge \textcolor{LGcolor}{\rrparenthesis}  & \eqStep &  [   \textcolor{LGcolor}{\llparenthesis} \smidge  \SYSTEMnt{t}  \smidge \textcolor{LGcolor}{\rrparenthesis}   /  \SYSTEMmv{x}  ]   \textcolor{LGcolor}{\llparenthesis} \smidge   \SYSTEMnt{t_{{\mathrm{1}}}} \,  \SYSTEMnt{t_{{\mathrm{2}}}}   \smidge \textcolor{LGcolor}{\rrparenthesis}   \\
\textit{(defn. subst [lhs])} & \Rightarrow &  \textcolor{LGcolor}{\llparenthesis} \smidge     [  \SYSTEMnt{t}  /  \SYSTEMmv{x}  ]  \SYSTEMnt{t_{{\mathrm{1}}}}   \,    [  \SYSTEMnt{t}  /  \SYSTEMmv{x}  ]  \SYSTEMnt{t_{{\mathrm{2}}}}     \smidge \textcolor{LGcolor}{\rrparenthesis}  & \eqStep &  [   \textcolor{LGcolor}{\llparenthesis} \smidge  \SYSTEMnt{t}  \smidge \textcolor{LGcolor}{\rrparenthesis}   /  \SYSTEMmv{x}  ]   \textcolor{LGcolor}{\llparenthesis} \smidge   \SYSTEMnt{t_{{\mathrm{1}}}} \,  \SYSTEMnt{t_{{\mathrm{2}}}}   \smidge \textcolor{LGcolor}{\rrparenthesis}   \\
\textit{(defn. interp [rhs])} & \Rightarrow &  \textcolor{LGcolor}{\llparenthesis} \smidge     [  \SYSTEMnt{t}  /  \SYSTEMmv{x}  ]  \SYSTEMnt{t_{{\mathrm{1}}}}   \,    [  \SYSTEMnt{t}  /  \SYSTEMmv{x}  ]  \SYSTEMnt{t_{{\mathrm{2}}}}     \smidge \textcolor{LGcolor}{\rrparenthesis}  & \eqStep &  [   \textcolor{LGcolor}{\llparenthesis} \smidge  \SYSTEMnt{t}  \smidge \textcolor{LGcolor}{\rrparenthesis}   /  \SYSTEMmv{x}  ]  \SYSTEMsym{(}    \textcolor{LGcolor}{\llparenthesis} \smidge  \SYSTEMnt{t_{{\mathrm{1}}}}  \smidge \textcolor{LGcolor}{\rrparenthesis}  \,   \textcolor{LGcolor}{\llparenthesis} \smidge  \SYSTEMnt{t_{{\mathrm{2}}}}  \smidge \textcolor{LGcolor}{\rrparenthesis}    \SYSTEMsym{)}  \\
\textit{(defn. subst [rhs])} & \Rightarrow &  \textcolor{LGcolor}{\llparenthesis} \smidge     [  \SYSTEMnt{t}  /  \SYSTEMmv{x}  ]  \SYSTEMnt{t_{{\mathrm{1}}}}   \,    [  \SYSTEMnt{t}  /  \SYSTEMmv{x}  ]  \SYSTEMnt{t_{{\mathrm{2}}}}     \smidge \textcolor{LGcolor}{\rrparenthesis}  & \eqStep &  \SYSTEMsym{(}   [   \textcolor{LGcolor}{\llparenthesis} \smidge  \SYSTEMnt{t}  \smidge \textcolor{LGcolor}{\rrparenthesis}   /  \SYSTEMmv{x}  ]   \textcolor{LGcolor}{\llparenthesis} \smidge  \SYSTEMnt{t_{{\mathrm{1}}}}  \smidge \textcolor{LGcolor}{\rrparenthesis}    \SYSTEMsym{)} \,  \SYSTEMsym{(}   [   \textcolor{LGcolor}{\llparenthesis} \smidge  \SYSTEMnt{t}  \smidge \textcolor{LGcolor}{\rrparenthesis}   /  \SYSTEMmv{x}  ]   \textcolor{LGcolor}{\llparenthesis} \smidge  \SYSTEMnt{t_{{\mathrm{2}}}}  \smidge \textcolor{LGcolor}{\rrparenthesis}    \SYSTEMsym{)}  \\
\textit{(defn. interp [lhs])} & \Rightarrow &   \textcolor{LGcolor}{\llparenthesis} \smidge   [  \SYSTEMnt{t}  /  \SYSTEMmv{x}  ]  \SYSTEMnt{t_{{\mathrm{1}}}}   \smidge \textcolor{LGcolor}{\rrparenthesis}  \,   \textcolor{LGcolor}{\llparenthesis} \smidge   [  \SYSTEMnt{t}  /  \SYSTEMmv{x}  ]  \SYSTEMnt{t_{{\mathrm{2}}}}   \smidge \textcolor{LGcolor}{\rrparenthesis}   & \eqStep &  \SYSTEMsym{(}   [   \textcolor{LGcolor}{\llparenthesis} \smidge  \SYSTEMnt{t}  \smidge \textcolor{LGcolor}{\rrparenthesis}   /  \SYSTEMmv{x}  ]   \textcolor{LGcolor}{\llparenthesis} \smidge  \SYSTEMnt{t_{{\mathrm{1}}}}  \smidge \textcolor{LGcolor}{\rrparenthesis}    \SYSTEMsym{)} \,  \SYSTEMsym{(}   [   \textcolor{LGcolor}{\llparenthesis} \smidge  \SYSTEMnt{t}  \smidge \textcolor{LGcolor}{\rrparenthesis}   /  \SYSTEMmv{x}  ]   \textcolor{LGcolor}{\llparenthesis} \smidge  \SYSTEMnt{t_{{\mathrm{2}}}}  \smidge \textcolor{LGcolor}{\rrparenthesis}    \SYSTEMsym{)}  \\
\textit{(induction on $\SYSTEMnt{t_{{\mathrm{1}}}}$ and $\SYSTEMnt{t_{{\mathrm{2}}}}$)} & \Rightarrow & \top
\end{array}
\end{align*}
\end{gather*}
\item (pr) $\SYSTEMnt{t'} \equiv  \textcolor{coeffectColor}{[}  \SYSTEMnt{t_{{\mathrm{1}}}}  \textcolor{coeffectColor}{]} $;
\begin{align*}
\begin{array}{rcrlll}
\textit{(goal)} & &  \textcolor{LGcolor}{\llparenthesis} \smidge   [  \SYSTEMnt{t}  /  \SYSTEMmv{x}  ]   \textcolor{coeffectColor}{[}  \SYSTEMnt{t_{{\mathrm{1}}}}  \textcolor{coeffectColor}{]}    \smidge \textcolor{LGcolor}{\rrparenthesis}  & \eqStep &  [   \textcolor{LGcolor}{\llparenthesis} \smidge  \SYSTEMnt{t}  \smidge \textcolor{LGcolor}{\rrparenthesis}   /  \SYSTEMmv{x}  ]   \textcolor{LGcolor}{\llparenthesis} \smidge   \textcolor{coeffectColor}{[}  \SYSTEMnt{t_{{\mathrm{1}}}}  \textcolor{coeffectColor}{]}   \smidge \textcolor{LGcolor}{\rrparenthesis}   \\
\textit{(defn. subst [lhs])} & \Rightarrow &  \textcolor{LGcolor}{\llparenthesis} \smidge   \textcolor{coeffectColor}{[}   [  \SYSTEMnt{t}  /  \SYSTEMmv{x}  ]  \SYSTEMnt{t_{{\mathrm{1}}}}   \textcolor{coeffectColor}{]}   \smidge \textcolor{LGcolor}{\rrparenthesis}  & \eqStep &  [   \textcolor{LGcolor}{\llparenthesis} \smidge  \SYSTEMnt{t}  \smidge \textcolor{LGcolor}{\rrparenthesis}   /  \SYSTEMmv{x}  ]   \textcolor{LGcolor}{\llparenthesis} \smidge   \textcolor{coeffectColor}{[}  \SYSTEMnt{t_{{\mathrm{1}}}}  \textcolor{coeffectColor}{]}   \smidge \textcolor{LGcolor}{\rrparenthesis}   \\
\textit{(defn. interp [rhs])} & \Rightarrow &  \textcolor{LGcolor}{\llparenthesis} \smidge   \textcolor{coeffectColor}{[}   [  \SYSTEMnt{t}  /  \SYSTEMmv{x}  ]  \SYSTEMnt{t_{{\mathrm{1}}}}   \textcolor{coeffectColor}{]}   \smidge \textcolor{LGcolor}{\rrparenthesis}  & \eqStep &  [   \textcolor{LGcolor}{\llparenthesis} \smidge  \SYSTEMnt{t}  \smidge \textcolor{LGcolor}{\rrparenthesis}   /  \SYSTEMmv{x}  ]   \textcolor{coeffectColor}{[}   \textcolor{LGcolor}{\llparenthesis} \smidge  \SYSTEMnt{t_{{\mathrm{1}}}}  \smidge \textcolor{LGcolor}{\rrparenthesis}   \textcolor{coeffectColor}{]}   \\
\textit{(defn. subst [rhs])} & \Rightarrow &  \textcolor{LGcolor}{\llparenthesis} \smidge   \textcolor{coeffectColor}{[}   [  \SYSTEMnt{t}  /  \SYSTEMmv{x}  ]  \SYSTEMnt{t_{{\mathrm{1}}}}   \textcolor{coeffectColor}{]}   \smidge \textcolor{LGcolor}{\rrparenthesis}  & \eqStep &  \textcolor{coeffectColor}{[}   [   \textcolor{LGcolor}{\llparenthesis} \smidge  \SYSTEMnt{t}  \smidge \textcolor{LGcolor}{\rrparenthesis}   /  \SYSTEMmv{x}  ]   \textcolor{LGcolor}{\llparenthesis} \smidge  \SYSTEMnt{t_{{\mathrm{1}}}}  \smidge \textcolor{LGcolor}{\rrparenthesis}    \textcolor{coeffectColor}{]}  \\
\textit{(defn. interp [lhs])} & \Rightarrow &  \textcolor{coeffectColor}{[}   \textcolor{LGcolor}{\llparenthesis} \smidge   [  \SYSTEMnt{t}  /  \SYSTEMmv{x}  ]  \SYSTEMnt{t_{{\mathrm{1}}}}   \smidge \textcolor{LGcolor}{\rrparenthesis}   \textcolor{coeffectColor}{]}  & \eqStep &  \textcolor{coeffectColor}{[}   [   \textcolor{LGcolor}{\llparenthesis} \smidge  \SYSTEMnt{t}  \smidge \textcolor{LGcolor}{\rrparenthesis}   /  \SYSTEMmv{x}  ]   \textcolor{LGcolor}{\llparenthesis} \smidge  \SYSTEMnt{t_{{\mathrm{1}}}}  \smidge \textcolor{LGcolor}{\rrparenthesis}    \textcolor{coeffectColor}{]}  \\
\textit{(induction on $t1$)} & \Rightarrow & \top
\end{array}
\end{align*}
\item (let) $\SYSTEMnt{t'} \equiv  \mathsf{let} \, \textcolor{coeffectColor}{[}  \SYSTEMmv{y}  \textcolor{coeffectColor}{]} =  \SYSTEMnt{t_{{\mathrm{1}}}}  \, \mathsf{in} \,  \SYSTEMnt{t_{{\mathrm{2}}}} $ (with $ \SYSTEMmv{y}  \,\#\,  \SYSTEMnt{t} $);
\begin{gather*}
\begin{align*}
\begin{array}{rcrlll}
\textit{(goal)} & &  \textcolor{LGcolor}{\llparenthesis} \smidge   [  \SYSTEMnt{t}  /  \SYSTEMmv{x}  ]  \SYSTEMsym{(}   \mathsf{let} \, \textcolor{coeffectColor}{[}  \SYSTEMmv{y}  \textcolor{coeffectColor}{]} =  \SYSTEMnt{t_{{\mathrm{1}}}}  \, \mathsf{in} \,  \SYSTEMnt{t_{{\mathrm{2}}}}   \SYSTEMsym{)}   \smidge \textcolor{LGcolor}{\rrparenthesis}  & \eqStep &  [   \textcolor{LGcolor}{\llparenthesis} \smidge  \SYSTEMnt{t}  \smidge \textcolor{LGcolor}{\rrparenthesis}   /  \SYSTEMmv{x}  ]   \textcolor{LGcolor}{\llparenthesis} \smidge   \mathsf{let} \, \textcolor{coeffectColor}{[}  \SYSTEMmv{y}  \textcolor{coeffectColor}{]} =  \SYSTEMnt{t_{{\mathrm{1}}}}  \, \mathsf{in} \,  \SYSTEMnt{t_{{\mathrm{2}}}}   \smidge \textcolor{LGcolor}{\rrparenthesis}   \\
\textit{(defn. subst [lhs])} & \Rightarrow &  \textcolor{LGcolor}{\llparenthesis} \smidge   \mathsf{let} \, \textcolor{coeffectColor}{[}  \SYSTEMmv{y}  \textcolor{coeffectColor}{]} =   [  \SYSTEMnt{t}  /  \SYSTEMmv{x}  ]  \SYSTEMnt{t_{{\mathrm{1}}}}   \, \mathsf{in} \,    [  \SYSTEMnt{t}  /  \SYSTEMmv{x}  ]  \SYSTEMnt{t_{{\mathrm{2}}}}     \smidge \textcolor{LGcolor}{\rrparenthesis}  & \eqStep &  [   \textcolor{LGcolor}{\llparenthesis} \smidge  \SYSTEMnt{t}  \smidge \textcolor{LGcolor}{\rrparenthesis}   /  \SYSTEMmv{x}  ]   \textcolor{LGcolor}{\llparenthesis} \smidge   \mathsf{let} \, \textcolor{coeffectColor}{[}  \SYSTEMmv{y}  \textcolor{coeffectColor}{]} =  \SYSTEMnt{t_{{\mathrm{1}}}}  \, \mathsf{in} \,  \SYSTEMnt{t_{{\mathrm{2}}}}   \smidge \textcolor{LGcolor}{\rrparenthesis}   \\
\textit{(defn. interp [rhs])} & \Rightarrow &  \textcolor{LGcolor}{\llparenthesis} \smidge   \mathsf{let} \, \textcolor{coeffectColor}{[}  \SYSTEMmv{y}  \textcolor{coeffectColor}{]} =   [  \SYSTEMnt{t}  /  \SYSTEMmv{x}  ]  \SYSTEMnt{t_{{\mathrm{1}}}}   \, \mathsf{in} \,    [  \SYSTEMnt{t}  /  \SYSTEMmv{x}  ]  \SYSTEMnt{t_{{\mathrm{2}}}}     \smidge \textcolor{LGcolor}{\rrparenthesis}  & \eqStep &  [   \textcolor{LGcolor}{\llparenthesis} \smidge  \SYSTEMnt{t}  \smidge \textcolor{LGcolor}{\rrparenthesis}   /  \SYSTEMmv{x}  ]  \SYSTEMsym{(}   \mathsf{let} \, \textcolor{coeffectColor}{[}  \SYSTEMmv{y}  \textcolor{coeffectColor}{]} =   \textcolor{LGcolor}{\llparenthesis} \smidge  \SYSTEMnt{t_{{\mathrm{1}}}}  \smidge \textcolor{LGcolor}{\rrparenthesis}   \, \mathsf{in} \,   \textcolor{LGcolor}{\llparenthesis} \smidge  \SYSTEMnt{t_{{\mathrm{2}}}}  \smidge \textcolor{LGcolor}{\rrparenthesis}    \SYSTEMsym{)}  \\
\textit{(defn. subst [rhs])} & \Rightarrow &  \textcolor{LGcolor}{\llparenthesis} \smidge   \mathsf{let} \, \textcolor{coeffectColor}{[}  \SYSTEMmv{y}  \textcolor{coeffectColor}{]} =   [  \SYSTEMnt{t}  /  \SYSTEMmv{x}  ]  \SYSTEMnt{t_{{\mathrm{1}}}}   \, \mathsf{in} \,    [  \SYSTEMnt{t}  /  \SYSTEMmv{x}  ]  \SYSTEMnt{t_{{\mathrm{2}}}}     \smidge \textcolor{LGcolor}{\rrparenthesis}  & \eqStep &  \mathsf{let} \, \textcolor{coeffectColor}{[}  \SYSTEMmv{y}  \textcolor{coeffectColor}{]} =    [   \textcolor{LGcolor}{\llparenthesis} \smidge  \SYSTEMnt{t}  \smidge \textcolor{LGcolor}{\rrparenthesis}   /  \SYSTEMmv{x}  ]   \textcolor{LGcolor}{\llparenthesis} \smidge  \SYSTEMnt{t_{{\mathrm{1}}}}  \smidge \textcolor{LGcolor}{\rrparenthesis}     \, \mathsf{in} \,    [   \textcolor{LGcolor}{\llparenthesis} \smidge  \SYSTEMnt{t}  \smidge \textcolor{LGcolor}{\rrparenthesis}   /  \SYSTEMmv{x}  ]   \textcolor{LGcolor}{\llparenthesis} \smidge  \SYSTEMnt{t_{{\mathrm{2}}}}  \smidge \textcolor{LGcolor}{\rrparenthesis}     \\
\textit{(defn. interp [lhs])} & \Rightarrow &  \mathsf{let} \, \textcolor{coeffectColor}{[}  \SYSTEMmv{y}  \textcolor{coeffectColor}{]} =   \textcolor{LGcolor}{\llparenthesis} \smidge   [  \SYSTEMnt{t}  /  \SYSTEMmv{x}  ]  \SYSTEMnt{t_{{\mathrm{1}}}}   \smidge \textcolor{LGcolor}{\rrparenthesis}   \, \mathsf{in} \,   \textcolor{LGcolor}{\llparenthesis} \smidge   [  \SYSTEMnt{t}  /  \SYSTEMmv{x}  ]  \SYSTEMnt{t_{{\mathrm{2}}}}   \smidge \textcolor{LGcolor}{\rrparenthesis}   & \eqStep &  \mathsf{let} \, \textcolor{coeffectColor}{[}  \SYSTEMmv{y}  \textcolor{coeffectColor}{]} =    [   \textcolor{LGcolor}{\llparenthesis} \smidge  \SYSTEMnt{t}  \smidge \textcolor{LGcolor}{\rrparenthesis}   /  \SYSTEMmv{x}  ]   \textcolor{LGcolor}{\llparenthesis} \smidge  \SYSTEMnt{t_{{\mathrm{1}}}}  \smidge \textcolor{LGcolor}{\rrparenthesis}     \, \mathsf{in} \,    [   \textcolor{LGcolor}{\llparenthesis} \smidge  \SYSTEMnt{t}  \smidge \textcolor{LGcolor}{\rrparenthesis}   /  \SYSTEMmv{x}  ]   \textcolor{LGcolor}{\llparenthesis} \smidge  \SYSTEMnt{t_{{\mathrm{2}}}}  \smidge \textcolor{LGcolor}{\rrparenthesis}     \\
\textit{(induction on $\SYSTEMnt{t_{{\mathrm{1}}}}$ and $\SYSTEMnt{t_{{\mathrm{2}}}}$)} & \Rightarrow & \top
\end{array}
\end{align*}
\end{gather*}
\end{itemize}
\end{proof}

We now prove the operational correspondence as follows:

\begin{proof}
We consider the theorem in an expanded form: for every $ \SYSTEMnt{t}  \rightsquigarrow_{\textsc{l} }  \SYSTEMnt{t'} $
then $\exists t'' .   \textcolor{LGcolor}{\llparenthesis} \smidge  \SYSTEMnt{t}  \smidge \textcolor{LGcolor}{\rrparenthesis}   \rightsquigarrow_{\textsc{g} }^\ast  \SYSTEMnt{t''}  \wedge  \textcolor{LGcolor}{\llparenthesis} \smidge  \SYSTEMnt{t'}  \smidge \textcolor{LGcolor}{\rrparenthesis}  \equiv \SYSTEMnt{t''}$.

The proof then follows by induction on reductions:
\begin{itemize}
\item (beta)
\[
\SYSTEMdruleSemLinbeta{}
\]
The interpretation of the reducing term is:
\[
 \textcolor{LGcolor}{\llparenthesis} \smidge   \SYSTEMsym{(}   \lambda  \SYSTEMmv{x}  .  \SYSTEMnt{t_{{\mathrm{2}}}}   \SYSTEMsym{)} \,  \SYSTEMnt{t_{{\mathrm{1}}}}   \smidge \textcolor{LGcolor}{\rrparenthesis}  \equiv  \SYSTEMsym{(}   \lambda  \SYSTEMmv{x}  .   \textcolor{LGcolor}{\llparenthesis} \smidge  \SYSTEMnt{t_{{\mathrm{2}}}}  \smidge \textcolor{LGcolor}{\rrparenthesis}    \SYSTEMsym{)} \,   \textcolor{LGcolor}{\llparenthesis} \smidge  \SYSTEMnt{t_{{\mathrm{1}}}}  \smidge \textcolor{LGcolor}{\rrparenthesis}  
\]
Then we construct the reduction in Graded Base:
\begin{align*}
\inferrule*[right=\SYSTEMRenameRuleSemGrdbeta{}]
{ }
{  \SYSTEMsym{(}   \lambda  \SYSTEMmv{x}  .   \textcolor{LGcolor}{\llparenthesis} \smidge  \SYSTEMnt{t_{{\mathrm{2}}}}  \smidge \textcolor{LGcolor}{\rrparenthesis}    \SYSTEMsym{)} \,   \textcolor{LGcolor}{\llparenthesis} \smidge  \SYSTEMnt{t_{{\mathrm{1}}}}  \smidge \textcolor{LGcolor}{\rrparenthesis}    \rightsquigarrow_{\textsc{g} }   [   \textcolor{LGcolor}{\llparenthesis} \smidge  \SYSTEMnt{t_{{\mathrm{1}}}}  \smidge \textcolor{LGcolor}{\rrparenthesis}   /  \SYSTEMmv{x}  ]   \textcolor{LGcolor}{\llparenthesis} \smidge  \SYSTEMnt{t_{{\mathrm{2}}}}  \smidge \textcolor{LGcolor}{\rrparenthesis}   }
\end{align*}
By Lemma~\ref{lemma:interp-lin-to-grd-mod-preserves-subst} then interpreting
the linear-base reduced term is equal to the graded-base
reduced term: $ \textcolor{LGcolor}{\llparenthesis} \smidge   [  \SYSTEMnt{t_{{\mathrm{1}}}}  /  \SYSTEMmv{x}  ]  \SYSTEMnt{t_{{\mathrm{2}}}}   \smidge \textcolor{LGcolor}{\rrparenthesis}  \equiv  [   \textcolor{LGcolor}{\llparenthesis} \smidge  \SYSTEMnt{t_{{\mathrm{1}}}}  \smidge \textcolor{LGcolor}{\rrparenthesis}   /  \SYSTEMmv{x}  ]   \textcolor{LGcolor}{\llparenthesis} \smidge  \SYSTEMnt{t_{{\mathrm{2}}}}  \smidge \textcolor{LGcolor}{\rrparenthesis}  $.

\item (congAppL)
\[
\SYSTEMdruleSemLincongAppL{}
\]
The interpretation of the reducing term is:
\[
 \textcolor{LGcolor}{\llparenthesis} \smidge   \SYSTEMnt{t_{{\mathrm{1}}}} \,  \SYSTEMnt{t_{{\mathrm{2}}}}   \smidge \textcolor{LGcolor}{\rrparenthesis}  \equiv   \textcolor{LGcolor}{\llparenthesis} \smidge  \SYSTEMnt{t_{{\mathrm{1}}}}  \smidge \textcolor{LGcolor}{\rrparenthesis}  \,   \textcolor{LGcolor}{\llparenthesis} \smidge  \SYSTEMnt{t_{{\mathrm{2}}}}  \smidge \textcolor{LGcolor}{\rrparenthesis}  
\]
By induction on $ \SYSTEMnt{t_{{\mathrm{1}}}}  \rightsquigarrow_{\textsc{l} }  \SYSTEMnt{t'_{{\mathrm{1}}}} $, we have
$\exists \SYSTEMnt{t''_{{\mathrm{1}}}} .   \textcolor{LGcolor}{\llparenthesis} \smidge  \SYSTEMnt{t_{{\mathrm{1}}}}  \smidge \textcolor{LGcolor}{\rrparenthesis}   \rightsquigarrow_{\textsc{g} }^\ast  \SYSTEMnt{t''_{{\mathrm{1}}}}  \wedge
\SYSTEMnt{t''_{{\mathrm{1}}}} \equiv  \textcolor{LGcolor}{\llparenthesis} \smidge  \SYSTEMnt{t'_{{\mathrm{1}}}}  \smidge \textcolor{LGcolor}{\rrparenthesis} $. Therefore, we construct
a chain of reductions in Graded Base:
\[
\inferrule*[right=\SYSTEMRenameRuleSemLincongAppL{}]
{  \textcolor{LGcolor}{\llparenthesis} \smidge  \SYSTEMnt{t_{{\mathrm{1}}}}  \smidge \textcolor{LGcolor}{\rrparenthesis}   \rightsquigarrow_{\textsc{l} }  \SYSTEMnt{t''_{{\mathrm{1}}\,\SYSTEMmv{i}}} }
{   \textcolor{LGcolor}{\llparenthesis} \smidge  \SYSTEMnt{t_{{\mathrm{1}}}}  \smidge \textcolor{LGcolor}{\rrparenthesis}  \,   \textcolor{LGcolor}{\llparenthesis} \smidge  \SYSTEMnt{t_{{\mathrm{2}}}}  \smidge \textcolor{LGcolor}{\rrparenthesis}    \rightsquigarrow_{\textsc{l} }   \SYSTEMnt{t''_{{\mathrm{1}}\,\SYSTEMmv{i}}} \,   \textcolor{LGcolor}{\llparenthesis} \smidge  \SYSTEMnt{t_{{\mathrm{2}}}}  \smidge \textcolor{LGcolor}{\rrparenthesis}   }
\;\;
\cdots
\;\;
\inferrule*[right=\SYSTEMRenameRuleSemLincongAppL{}]
{  \SYSTEMnt{t''_{{\mathrm{1}}\,\SYSTEMmv{m}}} \,  \SYSTEMnt{t_{{\mathrm{1}}}}   \rightsquigarrow_{\textsc{l} }  \SYSTEMnt{t''_{{\mathrm{1}}}} }
{  \SYSTEMnt{t''_{{\mathrm{1}}\,\SYSTEMmv{m}}} \,   \textcolor{LGcolor}{\llparenthesis} \smidge  \SYSTEMnt{t_{{\mathrm{2}}}}  \smidge \textcolor{LGcolor}{\rrparenthesis}    \rightsquigarrow_{\textsc{l} }   \SYSTEMnt{t''_{{\mathrm{1}}}} \,   \textcolor{LGcolor}{\llparenthesis} \smidge  \SYSTEMnt{t_{{\mathrm{2}}}}  \smidge \textcolor{LGcolor}{\rrparenthesis}   }
\]
where $m$ is the length of the reduction sequence from
the induction.

satisfying the goal, since $ \SYSTEMnt{t''_{{\mathrm{1}}}}  \equiv   \textcolor{LGcolor}{\llparenthesis} \smidge  \SYSTEMnt{t'_{{\mathrm{1}}}}  \smidge \textcolor{LGcolor}{\rrparenthesis}  $
and therefore
$  \SYSTEMnt{t''_{{\mathrm{1}}}} \,   \textcolor{LGcolor}{\llparenthesis} \smidge  \SYSTEMnt{t_{{\mathrm{2}}}}  \smidge \textcolor{LGcolor}{\rrparenthesis}    \equiv    \textcolor{LGcolor}{\llparenthesis} \smidge  \SYSTEMnt{t'_{{\mathrm{1}}}}  \smidge \textcolor{LGcolor}{\rrparenthesis}  \,   \textcolor{LGcolor}{\llparenthesis} \smidge  \SYSTEMnt{t_{{\mathrm{2}}}}  \smidge \textcolor{LGcolor}{\rrparenthesis}   $.
\item (betaBox)
\[
\SYSTEMdruleSemLinbetaBox{}
\]
The interpretation of the reducing term is:
\[
 \textcolor{LGcolor}{\llparenthesis} \smidge   \mathsf{let} \, \textcolor{coeffectColor}{[}  \SYSTEMmv{x}  \textcolor{coeffectColor}{]} =   \textcolor{coeffectColor}{[}  \SYSTEMnt{t_{{\mathrm{1}}}}  \textcolor{coeffectColor}{]}   \, \mathsf{in} \,  \SYSTEMnt{t_{{\mathrm{2}}}}   \smidge \textcolor{LGcolor}{\rrparenthesis}  \equiv  \mathsf{let} \, \textcolor{coeffectColor}{[}  \SYSTEMmv{x}  \textcolor{coeffectColor}{]} =   \textcolor{coeffectColor}{[}   \textcolor{LGcolor}{\llparenthesis} \smidge  \SYSTEMnt{t_{{\mathrm{1}}}}  \smidge \textcolor{LGcolor}{\rrparenthesis}   \textcolor{coeffectColor}{]}   \, \mathsf{in} \,   \textcolor{LGcolor}{\llparenthesis} \smidge  \SYSTEMnt{t_{{\mathrm{2}}}}  \smidge \textcolor{LGcolor}{\rrparenthesis}  
\]
Then we construct the reduction in Graded Base:
\begin{align*}
\inferrule*[right=\SYSTEMRenameRuleSemGrdModbetaBox{}]
{ }
{  \mathsf{let} \, \textcolor{coeffectColor}{[}  \SYSTEMmv{x}  \textcolor{coeffectColor}{]} =   \textcolor{coeffectColor}{[}   \textcolor{LGcolor}{\llparenthesis} \smidge  \SYSTEMnt{t_{{\mathrm{1}}}}  \smidge \textcolor{LGcolor}{\rrparenthesis}   \textcolor{coeffectColor}{]}   \, \mathsf{in} \,   \textcolor{LGcolor}{\llparenthesis} \smidge  \SYSTEMnt{t_{{\mathrm{2}}}}  \smidge \textcolor{LGcolor}{\rrparenthesis}    \rightsquigarrow_{\textsc{g} }   [   \textcolor{LGcolor}{\llparenthesis} \smidge  \SYSTEMnt{t_{{\mathrm{1}}}}  \smidge \textcolor{LGcolor}{\rrparenthesis}   /  \SYSTEMmv{x}  ]   \textcolor{LGcolor}{\llparenthesis} \smidge  \SYSTEMnt{t_{{\mathrm{2}}}}  \smidge \textcolor{LGcolor}{\rrparenthesis}   }
\end{align*}
By Lemma~\ref{lemma:interp-lin-to-grd-mod-preserves-subst} then interpreting
the linear-base reduced term is equal to the graded-base
reduced term: $ \textcolor{LGcolor}{\llparenthesis} \smidge   [  \SYSTEMnt{t_{{\mathrm{1}}}}  /  \SYSTEMmv{x}  ]  \SYSTEMnt{t_{{\mathrm{2}}}}   \smidge \textcolor{LGcolor}{\rrparenthesis}  \equiv  [   \textcolor{LGcolor}{\llparenthesis} \smidge  \SYSTEMnt{t_{{\mathrm{1}}}}  \smidge \textcolor{LGcolor}{\rrparenthesis}   /  \SYSTEMmv{x}  ]   \textcolor{LGcolor}{\llparenthesis} \smidge  \SYSTEMnt{t_{{\mathrm{2}}}}  \smidge \textcolor{LGcolor}{\rrparenthesis}  $.

\item (congLetL)
\[
\SYSTEMdruleSemLincongLetL{}
\]
By induction, analogously to the case for (congAppL).

\end{itemize}
\end{proof}

\subsubsection{Equation correspondence}
\label{app:proofs-lin-to-grad-mod-eqs}

\begin{proof}
  \begin{itemize}
    \item \[\SYSTEMdruleLinEqeta{}\]
    \begin{align*}
      \begin{array}{rll}
                                  &  \textcolor{LGcolor}{\llparenthesis} \smidge    \lambda  \SYSTEMmv{x}  .  \SYSTEMnt{t}  \,  \SYSTEMmv{x}   \smidge \textcolor{LGcolor}{\rrparenthesis}  &  \\
   \textit{\{defn. translation\}} & =   \lambda  \SYSTEMmv{x}  .   \textcolor{LGcolor}{\llparenthesis} \smidge  \SYSTEMnt{t}  \smidge \textcolor{LGcolor}{\rrparenthesis}   \,  \SYSTEMmv{x}  \; (\textit{$x \# t$}) &  \\
   \textit{\{\SYSTEMRenameRuleGradEqeta{}\}}  & =  \textcolor{LGcolor}{\llparenthesis} \smidge  \SYSTEMnt{t}  \smidge \textcolor{LGcolor}{\rrparenthesis} 
      \end{array}
    \end{align*}

  \item \[\SYSTEMdruleLinEqbeta{}\]
    Follows from the preservation of operational semantics (see above), Theorem~\ref{ref:gradModToLinTranslation}.

  \item \[\SYSTEMdruleLinEqetaBox{}\]
    \begin{align*}
      \begin{array}{rll}
                                  &  \textcolor{LGcolor}{\llparenthesis} \smidge   \mathsf{let} \, \textcolor{coeffectColor}{[}  \SYSTEMmv{x}  \textcolor{coeffectColor}{]} =  \SYSTEMnt{t}  \, \mathsf{in} \,   \textcolor{coeffectColor}{[}  \SYSTEMmv{x}  \textcolor{coeffectColor}{]}    \smidge \textcolor{LGcolor}{\rrparenthesis}  &  \\
   \textit{\{defn. translation\}} & =  \mathsf{let} \, \textcolor{coeffectColor}{[}  \SYSTEMmv{x}  \textcolor{coeffectColor}{]} =   \textcolor{LGcolor}{\llparenthesis} \smidge  \SYSTEMnt{t}  \smidge \textcolor{LGcolor}{\rrparenthesis}   \, \mathsf{in} \,   \textcolor{coeffectColor}{[}  \SYSTEMmv{x}  \textcolor{coeffectColor}{]}   \\
   \textit{\{\SYSTEMRenameRuleGradMEqetaBox{}\}}  & =  \textcolor{LGcolor}{\llparenthesis} \smidge  \SYSTEMnt{t}  \smidge \textcolor{LGcolor}{\rrparenthesis} 
      \end{array}
    \end{align*}

    \item \[\SYSTEMdruleLinEqbetaBox{}\]
    Follows from the preservation of operational semantics (see above), Theorem~\ref{ref:gradModToLinTranslation}.

  \item \[\SYSTEMdruleLinEqletCommBox{}\]
        \begin{align*}
      \begin{array}{rll}
                                  &  \textcolor{LGcolor}{\llparenthesis} \smidge   \mathsf{let} \, \textcolor{coeffectColor}{[}  \SYSTEMmv{x}  \textcolor{coeffectColor}{]} =   \textcolor{coeffectColor}{[}  \SYSTEMnt{t_{{\mathrm{1}}}}  \textcolor{coeffectColor}{]}   \, \mathsf{in} \,   \textcolor{coeffectColor}{[}  \SYSTEMnt{t_{{\mathrm{2}}}}  \textcolor{coeffectColor}{]}    \smidge \textcolor{LGcolor}{\rrparenthesis}  &  \\
   \textit{\{defn. translation\}} & =  \mathsf{let} \, \textcolor{coeffectColor}{[}  \SYSTEMmv{x}  \textcolor{coeffectColor}{]} =   \textcolor{coeffectColor}{[}   \textcolor{LGcolor}{\llparenthesis} \smidge  \SYSTEMnt{t_{{\mathrm{1}}}}  \smidge \textcolor{LGcolor}{\rrparenthesis}   \textcolor{coeffectColor}{]}   \, \mathsf{in} \,   \textcolor{coeffectColor}{[}   \textcolor{LGcolor}{\llparenthesis} \smidge  \SYSTEMnt{t_{{\mathrm{2}}}}  \smidge \textcolor{LGcolor}{\rrparenthesis}   \textcolor{coeffectColor}{]}   \\
   \textit{\{\SYSTEMRenameRuleGradMEqletCommBox{}\}}  & =  \textcolor{coeffectColor}{[}   \mathsf{let} \, \textcolor{coeffectColor}{[}  \SYSTEMmv{x}  \textcolor{coeffectColor}{]} =   \textcolor{coeffectColor}{[}   \textcolor{LGcolor}{\llparenthesis} \smidge  \SYSTEMnt{t_{{\mathrm{1}}}}  \smidge \textcolor{LGcolor}{\rrparenthesis}   \textcolor{coeffectColor}{]}   \, \mathsf{in} \,   \textcolor{LGcolor}{\llparenthesis} \smidge  \SYSTEMnt{t_{{\mathrm{2}}}}  \smidge \textcolor{LGcolor}{\rrparenthesis}    \textcolor{coeffectColor}{]} 
      \end{array}
        \end{align*}

  \item \[\SYSTEMdruleLinEqletCommOne{}\]
        \begin{align*}
      \begin{array}{rll}
                                  &  \textcolor{LGcolor}{\llparenthesis} \smidge   \SYSTEMnt{t_{{\mathrm{0}}}} \,  \SYSTEMsym{(}   \mathsf{let} \, \textcolor{coeffectColor}{[}  \SYSTEMmv{x}  \textcolor{coeffectColor}{]} =  \SYSTEMnt{t_{{\mathrm{1}}}}  \, \mathsf{in} \,  \SYSTEMnt{t_{{\mathrm{2}}}}   \SYSTEMsym{)}   \smidge \textcolor{LGcolor}{\rrparenthesis}  &  \\
   \textit{\{defn. translation\}} & =   \textcolor{LGcolor}{\llparenthesis} \smidge  \SYSTEMnt{t_{{\mathrm{0}}}}  \smidge \textcolor{LGcolor}{\rrparenthesis}  \,  \SYSTEMsym{(}   \mathsf{let} \, \textcolor{coeffectColor}{[}  \SYSTEMmv{x}  \textcolor{coeffectColor}{]} =   \textcolor{LGcolor}{\llparenthesis} \smidge  \SYSTEMnt{t_{{\mathrm{1}}}}  \smidge \textcolor{LGcolor}{\rrparenthesis}   \, \mathsf{in} \,   \textcolor{LGcolor}{\llparenthesis} \smidge  \SYSTEMnt{t_{{\mathrm{2}}}}  \smidge \textcolor{LGcolor}{\rrparenthesis}    \SYSTEMsym{)}  \\
   \textit{\{\SYSTEMRenameRuleGradMEqletCommOne{}\}}  & =   \mathsf{let} \, \textcolor{coeffectColor}{[}  \SYSTEMmv{x}  \textcolor{coeffectColor}{]} =   \textcolor{LGcolor}{\llparenthesis} \smidge  \SYSTEMnt{t_{{\mathrm{1}}}}  \smidge \textcolor{LGcolor}{\rrparenthesis}   \, \mathsf{in} \,   \textcolor{LGcolor}{\llparenthesis} \smidge  \SYSTEMnt{t_{{\mathrm{0}}}}  \smidge \textcolor{LGcolor}{\rrparenthesis}   \,   \textcolor{LGcolor}{\llparenthesis} \smidge  \SYSTEMnt{t_{{\mathrm{2}}}}  \smidge \textcolor{LGcolor}{\rrparenthesis}  
      \end{array}
        \end{align*}

  \item \[\SYSTEMdruleLinEqletCommTwo{}\]
        \begin{align*}
      \begin{array}{rll}
                                  &  \textcolor{LGcolor}{\llparenthesis} \smidge   \SYSTEMsym{(}   \mathsf{let} \, \textcolor{coeffectColor}{[}  \SYSTEMmv{x}  \textcolor{coeffectColor}{]} =  \SYSTEMnt{t_{{\mathrm{1}}}}  \, \mathsf{in} \,  \SYSTEMnt{t_{{\mathrm{2}}}}   \SYSTEMsym{)} \,  \SYSTEMnt{t_{{\mathrm{0}}}}   \smidge \textcolor{LGcolor}{\rrparenthesis}  &  \\
   \textit{\{defn. translation\}} & =  \SYSTEMsym{(}   \mathsf{let} \, \textcolor{coeffectColor}{[}  \SYSTEMmv{x}  \textcolor{coeffectColor}{]} =   \textcolor{LGcolor}{\llparenthesis} \smidge  \SYSTEMnt{t_{{\mathrm{1}}}}  \smidge \textcolor{LGcolor}{\rrparenthesis}   \, \mathsf{in} \,   \textcolor{LGcolor}{\llparenthesis} \smidge  \SYSTEMnt{t_{{\mathrm{2}}}}  \smidge \textcolor{LGcolor}{\rrparenthesis}    \SYSTEMsym{)} \,   \textcolor{LGcolor}{\llparenthesis} \smidge  \SYSTEMnt{t_{{\mathrm{0}}}}  \smidge \textcolor{LGcolor}{\rrparenthesis}   \\
   \textit{\{\SYSTEMRenameRuleGradMEqletCommTwo{}\}}  & =   \mathsf{let} \, \textcolor{coeffectColor}{[}  \SYSTEMmv{x}  \textcolor{coeffectColor}{]} =   \textcolor{LGcolor}{\llparenthesis} \smidge  \SYSTEMnt{t_{{\mathrm{1}}}}  \smidge \textcolor{LGcolor}{\rrparenthesis}   \, \mathsf{in} \,   \textcolor{LGcolor}{\llparenthesis} \smidge  \SYSTEMnt{t_{{\mathrm{2}}}}  \smidge \textcolor{LGcolor}{\rrparenthesis}   \,   \textcolor{LGcolor}{\llparenthesis} \smidge  \SYSTEMnt{t_{{\mathrm{0}}}}  \smidge \textcolor{LGcolor}{\rrparenthesis}  
      \end{array}
        \end{align*}

   \item All the congruence rules are simply by induction.
  \end{itemize}
\end{proof}

%%%%%%%%%%%%%%%%%%%%%%%%%%%%%%%%%%%%%%%%%

\subsection{Proof of Soundness for Linearity-Preserving translation of Linear Base to Graded Modal Base}
\label{app:proofs-lin-to-grad-mod-alt}

\ifextended\linToGradAltTranslation*\fi

\subsubsection{Type preservation}
\label{app:proofs-lin-to-grad-mod-alt-typ}

\begin{proof}
By induction on the Linear Base typing:
\begin{itemize}
\item (var)
$$
\SYSTEMdruleLinvar{}
$$
Therefore we construct the goal typing since the multiplicative unit is
$ \langle  \SYSTEMsym{1}  ,  \SYSTEMsym{1}  \rangle $ now:
$$
\inferrule*[Right=\SYSTEMRenameRuleGradvar{}]
{}
{   \SYSTEMmv{x}  :_{\textcolor{coeffectColor}{  \langle  \SYSTEMsym{1}  ,  \SYSTEMsym{1}  \rangle  } }  \SYSTEMnt{A}    \vdash_{\textsc{g} }  \SYSTEMmv{x}  :  \SYSTEMnt{A} }
$$
%%%%%%%%%%%%%%%%%%%%%%%%%%%%%%%%%%%%%%%%%%%%%%%%%%%%%%%%%%%%%%%%%%%%%%%%%%%%%%%%
\item (abs)
$$
\SYSTEMdruleLinabs{}
$$
By induction on the premise we have
${   \textcolor{LGcolor}{\llparenthesis}  \Gamma  \textcolor{LGcolor}{\rrparenthesis}'  ,   \SYSTEMmv{x}  :_{\textcolor{coeffectColor}{  \langle  \SYSTEMsym{1}  ,  \SYSTEMsym{1}  \rangle  } }   \textcolor{LGcolor}{\llparenthesis} \smidge  \SYSTEMnt{A}  \smidge \textcolor{LGcolor}{\rrparenthesis}'     \vdash_{\textsc{g} }   \textcolor{LGcolor}{\llparenthesis} \smidge  \SYSTEMnt{t}  \smidge \textcolor{LGcolor}{\rrparenthesis}   :   \textcolor{LGcolor}{\llparenthesis} \smidge  \SYSTEMnt{B}  \smidge \textcolor{LGcolor}{\rrparenthesis}'  }$.

Therefore we construct the goal typing:
$$\inferrule*[Right=\SYSTEMRenameRuleGradabs{}]
    {   \textcolor{LGcolor}{\llparenthesis}  \Gamma  \textcolor{LGcolor}{\rrparenthesis}'  ,   \SYSTEMmv{x}  :_{\textcolor{coeffectColor}{  \langle  \SYSTEMsym{1}  ,  \SYSTEMsym{1}  \rangle  } }   \textcolor{LGcolor}{\llparenthesis} \smidge  \SYSTEMnt{A}  \smidge \textcolor{LGcolor}{\rrparenthesis}'     \vdash_{\textsc{g} }   \textcolor{LGcolor}{\llparenthesis} \smidge  \SYSTEMnt{t}  \smidge \textcolor{LGcolor}{\rrparenthesis}   :   \textcolor{LGcolor}{\llparenthesis} \smidge  \SYSTEMnt{B}  \smidge \textcolor{LGcolor}{\rrparenthesis}'  }
    {  \textcolor{LGcolor}{\llparenthesis}  \Gamma  \textcolor{LGcolor}{\rrparenthesis}'   \vdash_{\textsc{g} }   \lambda  \SYSTEMmv{x}  .   \textcolor{LGcolor}{\llparenthesis} \smidge  \SYSTEMnt{t}  \smidge \textcolor{LGcolor}{\rrparenthesis}    :    \textcolor{LGcolor}{\llparenthesis} \smidge  \SYSTEMnt{A}  \smidge \textcolor{LGcolor}{\rrparenthesis}'   \xrightarrow{\textcolor{coeffectColor}{  \langle  \SYSTEMsym{1}  ,  \SYSTEMsym{1}  \rangle  } }   \textcolor{LGcolor}{\llparenthesis} \smidge  \SYSTEMnt{B}  \smidge \textcolor{LGcolor}{\rrparenthesis}'   }
$$
%%%%%%%%%%%%%%%%%%%%%%%%%%%%%%%%%%%%%%%%%%%%%%%%%%%%%%%%%%%%%%%%%%%%%%%%%%%%%%%%
\item (app)
$$
\SYSTEMdruleLinapp{}
$$
By induction on the premises we have
$  \textcolor{LGcolor}{\llparenthesis}  \Gamma_{{\mathrm{1}}}  \textcolor{LGcolor}{\rrparenthesis}'   \vdash_{\textsc{g} }   \textcolor{LGcolor}{\llparenthesis} \smidge  \SYSTEMnt{t_{{\mathrm{1}}}}  \smidge \textcolor{LGcolor}{\rrparenthesis}   :    \textcolor{LGcolor}{\llparenthesis} \smidge  \SYSTEMnt{A}  \smidge \textcolor{LGcolor}{\rrparenthesis}'   \xrightarrow{\textcolor{coeffectColor}{  \langle  \SYSTEMsym{1}  ,  \SYSTEMsym{1}  \rangle  } }   \textcolor{LGcolor}{\llparenthesis} \smidge  \SYSTEMnt{B}  \smidge \textcolor{LGcolor}{\rrparenthesis}'   $
and
$  \textcolor{LGcolor}{\llparenthesis}  \Gamma_{{\mathrm{2}}}  \textcolor{LGcolor}{\rrparenthesis}'   \vdash_{\textsc{g} }   \textcolor{LGcolor}{\llparenthesis} \smidge  \SYSTEMnt{t_{{\mathrm{2}}}}  \smidge \textcolor{LGcolor}{\rrparenthesis}   :   \textcolor{LGcolor}{\llparenthesis} \smidge  \SYSTEMnt{A}  \smidge \textcolor{LGcolor}{\rrparenthesis}'  $.

Therefore we construct:
$$
\inferrule*[Right=\SYSTEMRenameRuleGradapp{}]
    {  \textcolor{LGcolor}{\llparenthesis}  \Gamma_{{\mathrm{1}}}  \textcolor{LGcolor}{\rrparenthesis}'   \vdash_{\textsc{g} }   \textcolor{LGcolor}{\llparenthesis} \smidge  \SYSTEMnt{t_{{\mathrm{1}}}}  \smidge \textcolor{LGcolor}{\rrparenthesis}   :    \textcolor{LGcolor}{\llparenthesis} \smidge  \SYSTEMnt{A}  \smidge \textcolor{LGcolor}{\rrparenthesis}'   \xrightarrow{\textcolor{coeffectColor}{  \langle  \SYSTEMsym{1}  ,  \SYSTEMsym{1}  \rangle  } }   \textcolor{LGcolor}{\llparenthesis} \smidge  \SYSTEMnt{B}  \smidge \textcolor{LGcolor}{\rrparenthesis}'   
    \\  \textcolor{LGcolor}{\llparenthesis}  \Gamma_{{\mathrm{2}}}  \textcolor{LGcolor}{\rrparenthesis}'   \vdash_{\textsc{g} }   \textcolor{LGcolor}{\llparenthesis} \smidge  \SYSTEMnt{t_{{\mathrm{2}}}}  \smidge \textcolor{LGcolor}{\rrparenthesis}   :   \textcolor{LGcolor}{\llparenthesis} \smidge  \SYSTEMnt{A}  \smidge \textcolor{LGcolor}{\rrparenthesis}'  }
    {  \textcolor{LGcolor}{\llparenthesis}  \Gamma_{{\mathrm{1}}}  \textcolor{LGcolor}{\rrparenthesis}'   \SYSTEMsym{+}   \textcolor{coeffectColor}{  \langle  \SYSTEMsym{1}  ,  \SYSTEMsym{1}  \rangle   \cdot}   \textcolor{LGcolor}{\llparenthesis}  \Gamma_{{\mathrm{2}}}  \textcolor{LGcolor}{\rrparenthesis}'    \vdash_{\textsc{g} }    \textcolor{LGcolor}{\llparenthesis} \smidge  \SYSTEMnt{t_{{\mathrm{1}}}}  \smidge \textcolor{LGcolor}{\rrparenthesis}  \,   \textcolor{LGcolor}{\llparenthesis} \smidge  \SYSTEMnt{t_{{\mathrm{2}}}}  \smidge \textcolor{LGcolor}{\rrparenthesis}    :   \textcolor{LGcolor}{\llparenthesis} \smidge  \SYSTEMnt{B}  \smidge \textcolor{LGcolor}{\rrparenthesis}'  }
$$

which matches the goal type by:
$$
\begin{array}{rll}
&& \textcolor{LGcolor}{\llparenthesis}  \Gamma_{{\mathrm{1}}}  \textcolor{LGcolor}{\rrparenthesis}'   \SYSTEMsym{+}   \textcolor{coeffectColor}{  \langle  \SYSTEMsym{1}  ,  \SYSTEMsym{1}  \rangle   \cdot}   \textcolor{LGcolor}{\llparenthesis}  \Gamma_{{\mathrm{2}}}  \textcolor{LGcolor}{\rrparenthesis}'  \\
\text{unit property}&\equiv& \textcolor{LGcolor}{\llparenthesis}  \Gamma_{{\mathrm{1}}}  \textcolor{LGcolor}{\rrparenthesis}'   \SYSTEMsym{+}   \textcolor{LGcolor}{\llparenthesis}  \Gamma_{{\mathrm{2}}}  \textcolor{LGcolor}{\rrparenthesis}' \\
\text{homomorphism}&\equiv& \textcolor{LGcolor}{\llparenthesis}  \Gamma_{{\mathrm{1}}}  \SYSTEMsym{+}  \Gamma_{{\mathrm{2}}}  \textcolor{LGcolor}{\rrparenthesis}' 
\end{array}
$$
%%%%%%%%%%%%%%%%%%%%%%%%%%%%%%%%%%%%%%%%%%%%%%%%%%%%%%%%%%%%%%%%%%%%%%%%%%%%%%%%
\item (weak)
$$
\SYSTEMdruleLinweak{}
$$

By induction on the premise we have
$  \textcolor{LGcolor}{\llparenthesis}  \Gamma  \textcolor{LGcolor}{\rrparenthesis}'   \vdash_{\textsc{g} }   \textcolor{LGcolor}{\llparenthesis} \smidge  \SYSTEMnt{t}  \smidge \textcolor{LGcolor}{\rrparenthesis}   :   \textcolor{LGcolor}{\llparenthesis} \smidge  \SYSTEMnt{A}  \smidge \textcolor{LGcolor}{\rrparenthesis}'  $.

Therefore we construct:
$$
\inferrule*[Right=\SYSTEMRenameRuleGradweak{}]
    {  \textcolor{LGcolor}{\llparenthesis}  \Gamma  \textcolor{LGcolor}{\rrparenthesis}'   \vdash_{\textsc{g} }   \textcolor{LGcolor}{\llparenthesis} \smidge  \SYSTEMnt{t}  \smidge \textcolor{LGcolor}{\rrparenthesis}   :   \textcolor{LGcolor}{\llparenthesis} \smidge  \SYSTEMnt{A}  \smidge \textcolor{LGcolor}{\rrparenthesis}'  }
    {   \textcolor{LGcolor}{\llparenthesis}  \Gamma  \textcolor{LGcolor}{\rrparenthesis}'   ,   \textcolor{coeffectColor}{  \langle  \SYSTEMsym{0}  ,  \SYSTEMsym{0}  \rangle   \cdot}   \textcolor{LGcolor}{\llparenthesis}  \Gamma'  \textcolor{LGcolor}{\rrparenthesis}'     \vdash_{\textsc{g} }   \textcolor{LGcolor}{\llparenthesis} \smidge  \SYSTEMnt{t}  \smidge \textcolor{LGcolor}{\rrparenthesis}   :   \textcolor{LGcolor}{\llparenthesis} \smidge  \SYSTEMnt{A}  \smidge \textcolor{LGcolor}{\rrparenthesis}'  }
$$
which satisfies the goal by Lemma~\ref{lemma:zero-interp-lemma-lin-to-grad}.

%%%%%%%%%%%%%%%%%%%%%%%%%%%%%%%%%%%%%%%%%%%%%%%%%%%%%%%%%%%%%%%%%%%%%%%%%%%%%%%%
\item (der)
$$
\SYSTEMdruleLinder{}
$$
By induction on the premise, we have $   \textcolor{LGcolor}{\llparenthesis}  \Gamma  \textcolor{LGcolor}{\rrparenthesis}'  ,   \SYSTEMmv{x}  :_{\textcolor{coeffectColor}{  \langle  \SYSTEMsym{1}  ,  \SYSTEMsym{1}  \rangle  } }   \textcolor{LGcolor}{\llparenthesis} \smidge  \SYSTEMnt{A}  \smidge \textcolor{LGcolor}{\rrparenthesis}'     \vdash_{\textsc{g} }   \textcolor{LGcolor}{\llparenthesis} \smidge  \SYSTEMnt{t}  \smidge \textcolor{LGcolor}{\rrparenthesis}   :   \textcolor{LGcolor}{\llparenthesis} \smidge  \SYSTEMnt{B}  \smidge \textcolor{LGcolor}{\rrparenthesis}'  $
which we then use with approximation:
$$
\inferrule*[Right=\SYSTEMRenameRuleGradapprox{}]
    {   \textcolor{LGcolor}{\llparenthesis}  \Gamma  \textcolor{LGcolor}{\rrparenthesis}'  ,   \SYSTEMmv{x}  :_{\textcolor{coeffectColor}{  \langle  \SYSTEMsym{1}  ,  \SYSTEMsym{1}  \rangle  } }   \textcolor{LGcolor}{\llparenthesis} \smidge  \SYSTEMnt{A}  \smidge \textcolor{LGcolor}{\rrparenthesis}'     \vdash_{\textsc{g} }   \textcolor{LGcolor}{\llparenthesis} \smidge  \SYSTEMnt{t}  \smidge \textcolor{LGcolor}{\rrparenthesis}   :   \textcolor{LGcolor}{\llparenthesis} \smidge  \SYSTEMnt{B}  \smidge \textcolor{LGcolor}{\rrparenthesis}'   \quad
      \langle  \SYSTEMsym{1}  ,  \SYSTEMsym{1}  \rangle   \, \textcolor{coeffectColor}{\sqsubseteq} \,   \langle  \SYSTEMsym{1}  ,   \omega   \rangle   }
    {   \textcolor{LGcolor}{\llparenthesis}  \Gamma  \textcolor{LGcolor}{\rrparenthesis}'  ,   \SYSTEMmv{x}  :_{\textcolor{coeffectColor}{  \langle  \SYSTEMsym{1}  ,   \omega   \rangle  } }   \textcolor{LGcolor}{\llparenthesis} \smidge  \SYSTEMnt{A}  \smidge \textcolor{LGcolor}{\rrparenthesis}'     \vdash_{\textsc{g} }   \textcolor{LGcolor}{\llparenthesis} \smidge  \SYSTEMnt{t}  \smidge \textcolor{LGcolor}{\rrparenthesis}   :   \textcolor{LGcolor}{\llparenthesis} \smidge  \SYSTEMnt{B}  \smidge \textcolor{LGcolor}{\rrparenthesis}'  }
$$
which matches the goal.

%%%%%%%%%%%%%%%%%%%%%%%%%%%%%%%%%%%%%%%%%%%%%%%%%%%%%%%%%%%%%%%%%%%%%%%%%%%%%%%%
\item (pr)
$$
\SYSTEMdruleLinpr{}
$$
By induction on the premise we have $  \textcolor{LGcolor}{\llparenthesis}  \Gamma  \textcolor{LGcolor}{\rrparenthesis}'   \vdash_{\textsc{g} }   \textcolor{LGcolor}{\llparenthesis} \smidge  \SYSTEMnt{t}  \smidge \textcolor{LGcolor}{\rrparenthesis}   :   \textcolor{LGcolor}{\llparenthesis} \smidge  \SYSTEMnt{A}  \smidge \textcolor{LGcolor}{\rrparenthesis}'  $.

Therefore we can construct:
$$
\inferrule*[Right=\SYSTEMRenameRuleGradBoxpr{}]
{   \textcolor{LGcolor}{\llparenthesis}  \Gamma  \textcolor{LGcolor}{\rrparenthesis}'   \vdash_{\textsc{g} }   \textcolor{LGcolor}{\llparenthesis} \smidge  \SYSTEMnt{t}  \smidge \textcolor{LGcolor}{\rrparenthesis}   :   \textcolor{LGcolor}{\llparenthesis} \smidge  \SYSTEMnt{A}  \smidge \textcolor{LGcolor}{\rrparenthesis}'   }
{   \textcolor{coeffectColor}{  \langle  \SYSTEMnt{r}  ,   \omega   \rangle   \cdot}   \textcolor{LGcolor}{\llparenthesis}  \Gamma  \textcolor{LGcolor}{\rrparenthesis}'    \vdash_{\textsc{g} }   \textcolor{LGcolor}{\llparenthesis} \smidge  \SYSTEMnt{t}  \smidge \textcolor{LGcolor}{\rrparenthesis}   :   \textcolor{coeffectColor}{\square_{  \langle  \SYSTEMnt{r}  ,   \omega   \rangle  } }   \textcolor{LGcolor}{\llparenthesis} \smidge  \SYSTEMnt{A}  \smidge \textcolor{LGcolor}{\rrparenthesis}'    }
$$
which satisfies the goal since $ \textcolor{LGcolor}{\llparenthesis}   \textcolor{coeffectColor}{ \SYSTEMnt{r}  \cdot}  \Gamma   \textcolor{LGcolor}{\rrparenthesis}'  \equiv  \textcolor{coeffectColor}{  \langle  \SYSTEMnt{r}  ,   \omega   \rangle   \cdot}   \textcolor{LGcolor}{\llparenthesis}  \Gamma  \textcolor{LGcolor}{\rrparenthesis}'  $ (homomorphism).

\item (let)
$$
\SYSTEMdruleLinlet{}
$$
By induction on both premises we have
$  \textcolor{LGcolor}{\llparenthesis}  \Gamma_{{\mathrm{1}}}  \textcolor{LGcolor}{\rrparenthesis}'   \vdash_{\textsc{g} }   \textcolor{LGcolor}{\llparenthesis} \smidge  \SYSTEMnt{t_{{\mathrm{1}}}}  \smidge \textcolor{LGcolor}{\rrparenthesis}   :   \textcolor{coeffectColor}{\square_{  \langle  \SYSTEMnt{r}  ,   \omega   \rangle  } }   \textcolor{LGcolor}{\llparenthesis} \smidge  \SYSTEMnt{A}  \smidge \textcolor{LGcolor}{\rrparenthesis}   $
and
$   \textcolor{LGcolor}{\llparenthesis}  \Gamma_{{\mathrm{2}}}  \textcolor{LGcolor}{\rrparenthesis}'  ,   \SYSTEMmv{x}  :_{\textcolor{coeffectColor}{  \langle  \SYSTEMnt{r}  ,   \omega   \rangle  } }   \textcolor{LGcolor}{\llparenthesis} \smidge  \SYSTEMnt{A}  \smidge \textcolor{LGcolor}{\rrparenthesis}'     \vdash_{\textsc{g} }   \textcolor{LGcolor}{\llparenthesis} \smidge  \SYSTEMnt{t_{{\mathrm{2}}}}  \smidge \textcolor{LGcolor}{\rrparenthesis}   :   \textcolor{LGcolor}{\llparenthesis} \smidge  \SYSTEMnt{B}  \smidge \textcolor{LGcolor}{\rrparenthesis}'  $.

Therefore we can construct:
$$
\inferrule*[Right=\SYSTEMRenameRuleGradBoxlet{}]
{
  \textcolor{LGcolor}{\llparenthesis}  \Gamma_{{\mathrm{1}}}  \textcolor{LGcolor}{\rrparenthesis}'   \vdash_{\textsc{g} }   \textcolor{LGcolor}{\llparenthesis} \smidge  \SYSTEMnt{t_{{\mathrm{1}}}}  \smidge \textcolor{LGcolor}{\rrparenthesis}   :   \textcolor{coeffectColor}{\square_{  \langle  \SYSTEMnt{r}  ,   \omega   \rangle  } }   \textcolor{LGcolor}{\llparenthesis} \smidge  \SYSTEMnt{A}  \smidge \textcolor{LGcolor}{\rrparenthesis}'    \quad
    \textcolor{LGcolor}{\llparenthesis}  \Gamma_{{\mathrm{2}}}  \textcolor{LGcolor}{\rrparenthesis}'  ,   \SYSTEMmv{x}  :_{\textcolor{coeffectColor}{  \langle  \SYSTEMnt{r}  ,   \omega   \rangle  } }   \textcolor{LGcolor}{\llparenthesis} \smidge  \SYSTEMnt{A}  \smidge \textcolor{LGcolor}{\rrparenthesis}'     \vdash_{\textsc{g} }   \textcolor{LGcolor}{\llparenthesis} \smidge  \SYSTEMnt{t_{{\mathrm{2}}}}  \smidge \textcolor{LGcolor}{\rrparenthesis}   :   \textcolor{LGcolor}{\llparenthesis} \smidge  \SYSTEMnt{B}  \smidge \textcolor{LGcolor}{\rrparenthesis}'  }
{  \textcolor{LGcolor}{\llparenthesis}  \Gamma_{{\mathrm{1}}}  \textcolor{LGcolor}{\rrparenthesis}'   \SYSTEMsym{+}   \textcolor{LGcolor}{\llparenthesis}  \Gamma_{{\mathrm{2}}}  \textcolor{LGcolor}{\rrparenthesis}'   \vdash_{\textsc{g} }   \mathsf{let} \, \textcolor{coeffectColor}{[}  \SYSTEMmv{x}  \textcolor{coeffectColor}{]} =   \textcolor{LGcolor}{\llparenthesis} \smidge  \SYSTEMnt{t_{{\mathrm{1}}}}  \smidge \textcolor{LGcolor}{\rrparenthesis}   \, \mathsf{in} \,   \textcolor{LGcolor}{\llparenthesis} \smidge  \SYSTEMnt{t_{{\mathrm{2}}}}  \smidge \textcolor{LGcolor}{\rrparenthesis}    :   \textcolor{LGcolor}{\llparenthesis} \smidge  \SYSTEMnt{B}  \smidge \textcolor{LGcolor}{\rrparenthesis}'   }
$$
\item (approx)
$$
\SYSTEMdruleLinapprox{}
$$
By induction on the premise we have
$   \textcolor{LGcolor}{\llparenthesis}  \Gamma  \textcolor{LGcolor}{\rrparenthesis}'  ,   \SYSTEMmv{x}  :_{\textcolor{coeffectColor}{  \langle  \SYSTEMnt{s}  ,   \omega   \rangle  } }   \textcolor{LGcolor}{\llparenthesis} \smidge  \SYSTEMnt{B}  \smidge \textcolor{LGcolor}{\rrparenthesis}'     \vdash_{\textsc{g} }   \textcolor{LGcolor}{\llparenthesis} \smidge  \SYSTEMnt{t}  \smidge \textcolor{LGcolor}{\rrparenthesis}   :   \textcolor{LGcolor}{\llparenthesis} \smidge  \SYSTEMnt{A}  \smidge \textcolor{LGcolor}{\rrparenthesis}'  $.

Therefore we can construct:
$$
\inferrule*[Right=\SYSTEMRenameRuleGradBoxlet{}]
{ \begin{array}{cc}     \textcolor{LGcolor}{\llparenthesis}  \Gamma  \textcolor{LGcolor}{\rrparenthesis}'  ,   \SYSTEMmv{x}  :_{\textcolor{coeffectColor}{  \langle  \SYSTEMnt{r}  ,   \omega   \rangle  } }   \textcolor{LGcolor}{\llparenthesis} \smidge  \SYSTEMnt{B}  \smidge \textcolor{LGcolor}{\rrparenthesis}'     \vdash_{\textsc{g} }   \textcolor{LGcolor}{\llparenthesis} \smidge  \SYSTEMnt{t}  \smidge \textcolor{LGcolor}{\rrparenthesis}   :   \textcolor{LGcolor}{\llparenthesis} \smidge  \SYSTEMnt{A}  \smidge \textcolor{LGcolor}{\rrparenthesis}'    \; & \;    \langle  \SYSTEMnt{r}  ,   \omega   \rangle   \, \textcolor{coeffectColor}{\sqsubseteq} \,   \langle  \SYSTEMnt{s}  ,   \omega   \rangle    \end{array} }
{   \textcolor{LGcolor}{\llparenthesis}  \Gamma  \textcolor{LGcolor}{\rrparenthesis}'  ,   \SYSTEMmv{x}  :_{\textcolor{coeffectColor}{  \langle  \SYSTEMnt{s}  ,   \omega   \rangle  } }   \textcolor{LGcolor}{\llparenthesis} \smidge  \SYSTEMnt{B}  \smidge \textcolor{LGcolor}{\rrparenthesis}'     \vdash_{\textsc{g} }   \textcolor{LGcolor}{\llparenthesis} \smidge  \SYSTEMnt{t}  \smidge \textcolor{LGcolor}{\rrparenthesis}   :   \textcolor{LGcolor}{\llparenthesis} \smidge  \SYSTEMnt{A}  \smidge \textcolor{LGcolor}{\rrparenthesis}'  }
$$

\end{itemize}
\end{proof}

\subsection{Proof of Soundness for Graded Modal Base to Linear Base Translation}
\label{app:proofs-grad-mod-to-lin}

\ifextended\gradModToLinTranslation*\fi

Type preservation is in Appendix~\ref{app:proofs-grad-mod-to-lin-typ},
operational correspondence in Appendix~\ref{app:proofs-grad-mod-to-lin-ops},
and equation preservation in Appendix~\ref{app:proofs-grad-mod-to-lin-eqs}.

\subsubsection{Type preservation}
\label{app:proofs-grad-mod-to-lin-typ}

\begin{proof}
  \begin{itemize}
  \item (pr)
    $$
    \SYSTEMdruleGradBoxpr{}
    $$
  By induction on the premise we have: $  \textcolor{GLcolor}{\llbracket}  \Delta  \textcolor{GLcolor}{\rrbracket}   \vdash_{\textsc{l} }   \textcolor{GLcolor}{\llbracket}  \SYSTEMnt{t}  \textcolor{GLcolor}{\rrbracket}   :   \textcolor{GLcolor}{\llbracket}  \SYSTEMnt{A}  \textcolor{GLcolor}{\rrbracket}  $. $ \mathrm{graded}(  \textcolor{GLcolor}{\llbracket}  \Delta  \textcolor{GLcolor}{\rrbracket}  ) $ follows from lemma~\eqref{lemma:interp-grad-ctx-is-graded}.

Therefore we construct:
$$
\inferrule*[Right=\SYSTEMRenameRuleLinpr{}]
    {  \textcolor{GLcolor}{\llbracket}  \Delta  \textcolor{GLcolor}{\rrbracket}   \vdash_{\textsc{l} }   \textcolor{GLcolor}{\llbracket}  \SYSTEMnt{t}  \textcolor{GLcolor}{\rrbracket}   :   \textcolor{GLcolor}{\llbracket}  \SYSTEMnt{A}  \textcolor{GLcolor}{\rrbracket}   \quad  \mathrm{graded}(  \textcolor{GLcolor}{\llbracket}  \Delta  \textcolor{GLcolor}{\rrbracket}  ) }
    {  \textcolor{coeffectColor}{ \SYSTEMnt{r}  \cdot}   \textcolor{GLcolor}{\llbracket}  \Delta  \textcolor{GLcolor}{\rrbracket}    \vdash_{\textsc{l} }   \textcolor{coeffectColor}{[}   \textcolor{GLcolor}{\llbracket}  \SYSTEMnt{t}  \textcolor{GLcolor}{\rrbracket}   \textcolor{coeffectColor}{]}   :   \textcolor{coeffectColor}{\square_{ \SYSTEMnt{r} } }   \textcolor{GLcolor}{\llbracket}  \SYSTEMnt{A}  \textcolor{GLcolor}{\rrbracket}   }
$$
\item (let)
$$
\SYSTEMdruleGradBoxlet{}
$$
By induction on the premises we have\\
$  \textcolor{GLcolor}{\llbracket}  \Delta_{{\mathrm{1}}}  \textcolor{GLcolor}{\rrbracket}   \vdash_{\textsc{l} }   \textcolor{GLcolor}{\llbracket}  \SYSTEMnt{t_{{\mathrm{1}}}}  \textcolor{GLcolor}{\rrbracket}   :   \textcolor{GLcolor}{\llbracket}   \textcolor{coeffectColor}{\square_{ \SYSTEMnt{r} } }  \SYSTEMnt{A}   \textcolor{GLcolor}{\rrbracket}  $ and $   \textcolor{GLcolor}{\llbracket}  \Delta_{{\mathrm{2}}}  \textcolor{GLcolor}{\rrbracket}  ,   \SYSTEMmv{x}  : \textcolor{coeffectColor}{[}   \textcolor{GLcolor}{\llbracket}  \SYSTEMnt{A}  \textcolor{GLcolor}{\rrbracket}  {\textcolor{coeffectColor}{]_{ \SYSTEMnt{r} } } }    \vdash_{\textsc{l} }   \textcolor{GLcolor}{\llbracket}  \SYSTEMnt{t_{{\mathrm{2}}}}  \textcolor{GLcolor}{\rrbracket}   :   \textcolor{GLcolor}{\llbracket}  \SYSTEMnt{B}  \textcolor{GLcolor}{\rrbracket}  $.

Therefore we can construct:
$$
\inferrule*[Right=\SYSTEMRenameRuleLinlet{}]
{  \textcolor{GLcolor}{\llbracket}  \Delta_{{\mathrm{1}}}  \textcolor{GLcolor}{\rrbracket}   \vdash_{\textsc{l} }   \textcolor{GLcolor}{\llbracket}  \SYSTEMnt{t_{{\mathrm{1}}}}  \textcolor{GLcolor}{\rrbracket}   :   \textcolor{coeffectColor}{\square_{ \SYSTEMnt{r} } }   \textcolor{GLcolor}{\llbracket}  \SYSTEMnt{A}  \textcolor{GLcolor}{\rrbracket}    \\
    \textcolor{GLcolor}{\llbracket}  \Delta_{{\mathrm{2}}}  \textcolor{GLcolor}{\rrbracket}  ,   \SYSTEMmv{x}  : \textcolor{coeffectColor}{[}   \textcolor{GLcolor}{\llbracket}  \SYSTEMnt{A}  \textcolor{GLcolor}{\rrbracket}  {\textcolor{coeffectColor}{]_{ \SYSTEMnt{r} } } }    \vdash_{\textsc{l} }   \textcolor{GLcolor}{\llbracket}  \SYSTEMnt{t_{{\mathrm{2}}}}  \textcolor{GLcolor}{\rrbracket}   :   \textcolor{GLcolor}{\llbracket}  \SYSTEMnt{B}  \textcolor{GLcolor}{\rrbracket}  }
{  \textcolor{GLcolor}{\llbracket}  \Delta_{{\mathrm{1}}}  \textcolor{GLcolor}{\rrbracket}   \SYSTEMsym{+}   \textcolor{GLcolor}{\llbracket}  \Delta_{{\mathrm{2}}}  \textcolor{GLcolor}{\rrbracket}   \vdash_{\textsc{l} }   \mathsf{let} \, \textcolor{coeffectColor}{[}  \SYSTEMmv{x}  \textcolor{coeffectColor}{]} =   \textcolor{GLcolor}{\llbracket}  \SYSTEMnt{t_{{\mathrm{1}}}}  \textcolor{GLcolor}{\rrbracket}   \, \mathsf{in} \,   \textcolor{GLcolor}{\llbracket}  \SYSTEMnt{t_{{\mathrm{2}}}}  \textcolor{GLcolor}{\rrbracket}    :   \textcolor{GLcolor}{\llbracket}  \SYSTEMnt{B}  \textcolor{GLcolor}{\rrbracket}   }
$$
  \end{itemize}
\end{proof}

\subsubsection{Operational correspondence}
\label{app:proofs-grad-mod-to-lin-ops}

\begin{lemma}[Interpretation preserves substitution]
\label{lemma:interp-grd-modal-to-lin-preserves-subst}
For all graded modal base terms $\SYSTEMnt{t}, \SYSTEMnt{t'}$ then
$ \textcolor{GLcolor}{\llbracket}   [  \SYSTEMnt{t}  /  \SYSTEMmv{x}  ]  \SYSTEMnt{t'}   \textcolor{GLcolor}{\rrbracket}  \equiv  [   \textcolor{GLcolor}{\llbracket}  \SYSTEMnt{t}  \textcolor{GLcolor}{\rrbracket}   /  \SYSTEMmv{x}  ]   \textcolor{GLcolor}{\llbracket}  \SYSTEMnt{t'}  \textcolor{GLcolor}{\rrbracket}  $.
\end{lemma}

\begin{proof}
  Extending the previous proof of Lemma~\ref{lemma:interp-grd-to-lin-preserves-subst}.

  \begin{itemize}
\item (pr) $\SYSTEMnt{t'} \equiv  \textcolor{coeffectColor}{[}  \SYSTEMnt{t_{{\mathrm{1}}}}  \textcolor{coeffectColor}{]} $;
\begin{gather*}
\begin{align*}
\begin{array}{rcrlll}
\textit{(goal)} & &  \textcolor{GLcolor}{\llbracket}   [  \SYSTEMnt{t}  /  \SYSTEMmv{x}  ]   \textcolor{coeffectColor}{[}  \SYSTEMnt{t_{{\mathrm{1}}}}  \textcolor{coeffectColor}{]}    \textcolor{GLcolor}{\rrbracket}  & \eqStep &  [   \textcolor{GLcolor}{\llbracket}  \SYSTEMnt{t}  \textcolor{GLcolor}{\rrbracket}   /  \SYSTEMmv{x}  ]   \textcolor{GLcolor}{\llbracket}   \textcolor{coeffectColor}{[}  \SYSTEMnt{t_{{\mathrm{1}}}}  \textcolor{coeffectColor}{]}   \textcolor{GLcolor}{\rrbracket}   \\
\textit{(defn. subst [lhs])} & \Rightarrow &  \textcolor{GLcolor}{\llbracket}   \textcolor{coeffectColor}{[}   [  \SYSTEMnt{t}  /  \SYSTEMmv{x}  ]  \SYSTEMnt{t_{{\mathrm{1}}}}   \textcolor{coeffectColor}{]}   \textcolor{GLcolor}{\rrbracket}  & \eqStep &  [   \textcolor{GLcolor}{\llbracket}  \SYSTEMnt{t}  \textcolor{GLcolor}{\rrbracket}   /  \SYSTEMmv{x}  ]   \textcolor{GLcolor}{\llbracket}   \textcolor{coeffectColor}{[}  \SYSTEMnt{t_{{\mathrm{1}}}}  \textcolor{coeffectColor}{]}   \textcolor{GLcolor}{\rrbracket}   \\
\textit{(defn. interp [rhs])} & \Rightarrow &  \textcolor{GLcolor}{\llbracket}   \textcolor{coeffectColor}{[}   [  \SYSTEMnt{t}  /  \SYSTEMmv{x}  ]  \SYSTEMnt{t_{{\mathrm{1}}}}   \textcolor{coeffectColor}{]}   \textcolor{GLcolor}{\rrbracket}  & \eqStep &  [   \textcolor{GLcolor}{\llbracket}  \SYSTEMnt{t}  \textcolor{GLcolor}{\rrbracket}   /  \SYSTEMmv{x}  ]   \textcolor{coeffectColor}{[}   \textcolor{GLcolor}{\llbracket}  \SYSTEMnt{t_{{\mathrm{1}}}}  \textcolor{GLcolor}{\rrbracket}   \textcolor{coeffectColor}{]}   \\
\textit{(defn. subst [rhs])} & \Rightarrow &  \textcolor{GLcolor}{\llbracket}   \textcolor{coeffectColor}{[}   [  \SYSTEMnt{t}  /  \SYSTEMmv{x}  ]  \SYSTEMnt{t_{{\mathrm{1}}}}   \textcolor{coeffectColor}{]}   \textcolor{GLcolor}{\rrbracket}  & \eqStep &  \textcolor{coeffectColor}{[}   [   \textcolor{GLcolor}{\llbracket}  \SYSTEMnt{t}  \textcolor{GLcolor}{\rrbracket}   /  \SYSTEMmv{x}  ]   \textcolor{GLcolor}{\llbracket}  \SYSTEMnt{t_{{\mathrm{1}}}}  \textcolor{GLcolor}{\rrbracket}    \textcolor{coeffectColor}{]}  \\
\textit{(defn. interp [lhs])} & \Rightarrow &  \textcolor{coeffectColor}{[}   \textcolor{GLcolor}{\llbracket}   [  \SYSTEMnt{t}  /  \SYSTEMmv{x}  ]  \SYSTEMnt{t_{{\mathrm{1}}}}   \textcolor{GLcolor}{\rrbracket}   \textcolor{coeffectColor}{]}  & \eqStep &  \textcolor{coeffectColor}{[}   [   \textcolor{GLcolor}{\llbracket}  \SYSTEMnt{t}  \textcolor{GLcolor}{\rrbracket}   /  \SYSTEMmv{x}  ]   \textcolor{GLcolor}{\llbracket}  \SYSTEMnt{t_{{\mathrm{1}}}}  \textcolor{GLcolor}{\rrbracket}    \textcolor{coeffectColor}{]}  \\
\textit{(induction on $t1$)} & \Rightarrow & \top
\end{array}
\end{align*}
\end{gather*}
% END (pr)
\item (let) $\SYSTEMnt{t'} \equiv  \mathsf{let} \, \textcolor{coeffectColor}{[}  \SYSTEMmv{y}  \textcolor{coeffectColor}{]} =  \SYSTEMnt{t_{{\mathrm{1}}}}  \, \mathsf{in} \,  \SYSTEMnt{t_{{\mathrm{2}}}} $ (with $ \SYSTEMmv{y}  \,\#\,  \SYSTEMnt{t} $);
\begin{gather*}
\begin{align*}
\begin{array}{rcrlll}
\textit{(goal)}               &             &  \textcolor{GLcolor}{\llbracket}   [  \SYSTEMnt{t}  /  \SYSTEMmv{x}  ]  \SYSTEMsym{(}   \mathsf{let} \, \textcolor{coeffectColor}{[}  \SYSTEMmv{y}  \textcolor{coeffectColor}{]} =  \SYSTEMnt{t_{{\mathrm{1}}}}  \, \mathsf{in} \,  \SYSTEMnt{t_{{\mathrm{2}}}}   \SYSTEMsym{)}   \textcolor{GLcolor}{\rrbracket}  & \eqStep &  [   \textcolor{GLcolor}{\llbracket}  \SYSTEMnt{t}  \textcolor{GLcolor}{\rrbracket}   /  \SYSTEMmv{x}  ]   \textcolor{GLcolor}{\llbracket}   \mathsf{let} \, \textcolor{coeffectColor}{[}  \SYSTEMmv{y}  \textcolor{coeffectColor}{]} =  \SYSTEMnt{t_{{\mathrm{1}}}}  \, \mathsf{in} \,  \SYSTEMnt{t_{{\mathrm{2}}}}   \textcolor{GLcolor}{\rrbracket}   \\
\textit{(defn. subst [lhs])}  & \Rightarrow &  \textcolor{GLcolor}{\llbracket}   \mathsf{let} \, \textcolor{coeffectColor}{[}  \SYSTEMmv{y}  \textcolor{coeffectColor}{]} =   [  \SYSTEMnt{t}  /  \SYSTEMmv{x}  ]  \SYSTEMnt{t_{{\mathrm{1}}}}   \, \mathsf{in} \,    [  \SYSTEMnt{t}  /  \SYSTEMmv{x}  ]  \SYSTEMnt{t_{{\mathrm{2}}}}     \textcolor{GLcolor}{\rrbracket}  & \eqStep &  [   \textcolor{GLcolor}{\llbracket}  \SYSTEMnt{t}  \textcolor{GLcolor}{\rrbracket}   /  \SYSTEMmv{x}  ]   \textcolor{GLcolor}{\llbracket}   \mathsf{let} \, \textcolor{coeffectColor}{[}  \SYSTEMmv{y}  \textcolor{coeffectColor}{]} =  \SYSTEMnt{t_{{\mathrm{1}}}}  \, \mathsf{in} \,  \SYSTEMnt{t_{{\mathrm{2}}}}   \textcolor{GLcolor}{\rrbracket}   \\
\textit{(defn. interp [rhs])} & \Rightarrow &  \textcolor{GLcolor}{\llbracket}   \mathsf{let} \, \textcolor{coeffectColor}{[}  \SYSTEMmv{y}  \textcolor{coeffectColor}{]} =   [  \SYSTEMnt{t}  /  \SYSTEMmv{x}  ]  \SYSTEMnt{t_{{\mathrm{1}}}}   \, \mathsf{in} \,    [  \SYSTEMnt{t}  /  \SYSTEMmv{x}  ]  \SYSTEMnt{t_{{\mathrm{2}}}}     \textcolor{GLcolor}{\rrbracket}  & \eqStep &  [   \textcolor{GLcolor}{\llbracket}  \SYSTEMnt{t}  \textcolor{GLcolor}{\rrbracket}   /  \SYSTEMmv{x}  ]  \SYSTEMsym{(}   \mathsf{let} \, \textcolor{coeffectColor}{[}  \SYSTEMmv{y}  \textcolor{coeffectColor}{]} =   \textcolor{GLcolor}{\llbracket}  \SYSTEMnt{t_{{\mathrm{1}}}}  \textcolor{GLcolor}{\rrbracket}   \, \mathsf{in} \,   \textcolor{GLcolor}{\llbracket}  \SYSTEMnt{t_{{\mathrm{2}}}}  \textcolor{GLcolor}{\rrbracket}    \SYSTEMsym{)}  \\
\textit{(defn. subst [rhs])}  & \Rightarrow &  \textcolor{GLcolor}{\llbracket}   \mathsf{let} \, \textcolor{coeffectColor}{[}  \SYSTEMmv{y}  \textcolor{coeffectColor}{]} =   [  \SYSTEMnt{t}  /  \SYSTEMmv{x}  ]  \SYSTEMnt{t_{{\mathrm{1}}}}   \, \mathsf{in} \,    [  \SYSTEMnt{t}  /  \SYSTEMmv{x}  ]  \SYSTEMnt{t_{{\mathrm{2}}}}     \textcolor{GLcolor}{\rrbracket}  & \eqStep &  \mathsf{let} \, \textcolor{coeffectColor}{[}  \SYSTEMmv{y}  \textcolor{coeffectColor}{]} =    [   \textcolor{GLcolor}{\llbracket}  \SYSTEMnt{t}  \textcolor{GLcolor}{\rrbracket}   /  \SYSTEMmv{x}  ]   \textcolor{GLcolor}{\llbracket}  \SYSTEMnt{t_{{\mathrm{1}}}}  \textcolor{GLcolor}{\rrbracket}     \, \mathsf{in} \,    [   \textcolor{GLcolor}{\llbracket}  \SYSTEMnt{t}  \textcolor{GLcolor}{\rrbracket}   /  \SYSTEMmv{x}  ]   \textcolor{GLcolor}{\llbracket}  \SYSTEMnt{t_{{\mathrm{2}}}}  \textcolor{GLcolor}{\rrbracket}     \\
\textit{(defn. interp [lhs])} & \Rightarrow &  \mathsf{let} \, \textcolor{coeffectColor}{[}  \SYSTEMmv{y}  \textcolor{coeffectColor}{]} =   \textcolor{GLcolor}{\llbracket}   [  \SYSTEMnt{t}  /  \SYSTEMmv{x}  ]  \SYSTEMnt{t_{{\mathrm{1}}}}   \textcolor{GLcolor}{\rrbracket}   \, \mathsf{in} \,   \textcolor{GLcolor}{\llbracket}   [  \SYSTEMnt{t}  /  \SYSTEMmv{x}  ]  \SYSTEMnt{t_{{\mathrm{2}}}}   \textcolor{GLcolor}{\rrbracket}   & \eqStep &  \mathsf{let} \, \textcolor{coeffectColor}{[}  \SYSTEMmv{y}  \textcolor{coeffectColor}{]} =    [   \textcolor{GLcolor}{\llbracket}  \SYSTEMnt{t}  \textcolor{GLcolor}{\rrbracket}   /  \SYSTEMmv{x}  ]   \textcolor{GLcolor}{\llbracket}  \SYSTEMnt{t_{{\mathrm{1}}}}  \textcolor{GLcolor}{\rrbracket}     \, \mathsf{in} \,    [   \textcolor{GLcolor}{\llbracket}  \SYSTEMnt{t}  \textcolor{GLcolor}{\rrbracket}   /  \SYSTEMmv{x}  ]   \textcolor{GLcolor}{\llbracket}  \SYSTEMnt{t_{{\mathrm{2}}}}  \textcolor{GLcolor}{\rrbracket}     \\
\textit{(induction on $\SYSTEMnt{t_{{\mathrm{1}}}}$ and $\SYSTEMnt{t_{{\mathrm{2}}}}$)} & \Rightarrow & \top
\end{array}
\end{align*}
\end{gather*}
  \end{itemize}
  \end{proof}

The operational correspondence then follows via the following proof:

\begin{proof}
  We extend the proof for operational correspondence of Graded Base to Linear
  Base (Appendix~\ref{app:proofs-grad-to-lin-ops}) with the requisite additional
  cases for the Graded Modal Base extension.

    \begin{itemize}
\item (betaBox)
\[
\SYSTEMdruleSemGrdModbetaBox{}
\]
The interpretation of the reducing term is:
\[
  \textcolor{GLcolor}{\llbracket}   \mathsf{let} \, \textcolor{coeffectColor}{[}  \SYSTEMmv{x}  \textcolor{coeffectColor}{]} =   \textcolor{coeffectColor}{[}  \SYSTEMnt{t_{{\mathrm{1}}}}  \textcolor{coeffectColor}{]}   \, \mathsf{in} \,  \SYSTEMnt{t_{{\mathrm{2}}}}   \textcolor{GLcolor}{\rrbracket}   \equiv   \mathsf{let} \, \textcolor{coeffectColor}{[}  \SYSTEMmv{x}  \textcolor{coeffectColor}{]} =   \textcolor{coeffectColor}{[}   \textcolor{GLcolor}{\llbracket}  \SYSTEMnt{t_{{\mathrm{1}}}}  \textcolor{GLcolor}{\rrbracket}   \textcolor{coeffectColor}{]}   \, \mathsf{in} \,   \textcolor{GLcolor}{\llbracket}  \SYSTEMnt{t_{{\mathrm{2}}}}  \textcolor{GLcolor}{\rrbracket}   
\]
Then we construct the reduction in Linear Base:
\begin{align*}
\inferrule*[right=\SYSTEMRenameRuleSemLinbetaBox{}]
{ }
{  \mathsf{let} \, \textcolor{coeffectColor}{[}  \SYSTEMmv{x}  \textcolor{coeffectColor}{]} =   \textcolor{coeffectColor}{[}   \textcolor{GLcolor}{\llbracket}  \SYSTEMnt{t_{{\mathrm{1}}}}  \textcolor{GLcolor}{\rrbracket}   \textcolor{coeffectColor}{]}   \, \mathsf{in} \,   \textcolor{GLcolor}{\llbracket}  \SYSTEMnt{t_{{\mathrm{2}}}}  \textcolor{GLcolor}{\rrbracket}    \rightsquigarrow_{\textsc{l} }   [   \textcolor{GLcolor}{\llbracket}  \SYSTEMnt{t_{{\mathrm{1}}}}  \textcolor{GLcolor}{\rrbracket}   /  \SYSTEMmv{x}  ]   \textcolor{GLcolor}{\llbracket}  \SYSTEMnt{t_{{\mathrm{2}}}}  \textcolor{GLcolor}{\rrbracket}   }
\end{align*}
By Lemma~\ref{lemma:interp-grd-to-lin-preserves-subst} then interpreting
the linear-base reduced term is equal to the graded-base
reduced term: $  \textcolor{GLcolor}{\llbracket}   [  \SYSTEMnt{t_{{\mathrm{1}}}}  /  \SYSTEMmv{x}  ]  \SYSTEMnt{t_{{\mathrm{2}}}}   \textcolor{GLcolor}{\rrbracket}   \equiv   [   \textcolor{GLcolor}{\llbracket}  \SYSTEMnt{t_{{\mathrm{1}}}}  \textcolor{GLcolor}{\rrbracket}   /  \SYSTEMmv{x}  ]   \textcolor{GLcolor}{\llbracket}  \SYSTEMnt{t_{{\mathrm{2}}}}  \textcolor{GLcolor}{\rrbracket}   $.

\item (congLetL)
\[
\SYSTEMdruleSemGrdModcongLetL{}
\]
By induction, analogously to the case for (congAppL).

\end{itemize}
\end{proof}

\subsubsection{Equation preservation}
\label{app:proofs-grad-mod-to-lin-eqs}

\begin{proof}
  This proof extends that for preservation of equations by
  the translation of Graded Base to Linear Base (Appendix~\ref{app:proofs-grad-to-lin-eqs}).

  \begin{itemize}
      \item \[\SYSTEMdruleGradMEqbetaBox{} \]
    Follows by Theorem~\ref{ref:gradModToLinTranslation} (preservation of operational
    semantics), since this includes the $\Box\beta$ case as a reduction.

  \item \[\SYSTEMdruleGradMEqetaBox{} \]

    Preservation then follows by:
    \begin{align*}
      \begin{array}{rl}
        &  \textcolor{GLcolor}{\llbracket}  \SYSTEMsym{(}   \mathsf{let} \, \textcolor{coeffectColor}{[}  \SYSTEMmv{x}  \textcolor{coeffectColor}{]} =  \SYSTEMnt{t}  \, \mathsf{in} \,   \textcolor{coeffectColor}{[}  \SYSTEMmv{x}  \textcolor{coeffectColor}{]}    \SYSTEMsym{)}  \textcolor{GLcolor}{\rrbracket}  \\
    \textit{\{defn. translation\}} & =  \mathsf{let} \, \textcolor{coeffectColor}{[}  \SYSTEMmv{x}  \textcolor{coeffectColor}{]} =   \textcolor{GLcolor}{\llbracket}  \SYSTEMnt{t}  \textcolor{GLcolor}{\rrbracket}   \, \mathsf{in} \,   \textcolor{coeffectColor}{[}  \SYSTEMmv{x}  \textcolor{coeffectColor}{]}   \\
    \textit{\{\SYSTEMRenameRuleLinEqetaBox\}}
                                  & \equiv_L  \textcolor{GLcolor}{\llbracket}  \SYSTEMnt{t}  \textcolor{GLcolor}{\rrbracket} 
      \end{array}
      \end{align*}

  \item \[\SYSTEMdruleGradMEqletCommBox{}\]

    \begin{align*}
      \begin{array}{rl}
        &  \textcolor{GLcolor}{\llbracket}   \mathsf{let} \, \textcolor{coeffectColor}{[}  \SYSTEMmv{x}  \textcolor{coeffectColor}{]} =   \textcolor{coeffectColor}{[}  \SYSTEMnt{t_{{\mathrm{1}}}}  \textcolor{coeffectColor}{]}   \, \mathsf{in} \,   \textcolor{coeffectColor}{[}  \SYSTEMnt{t_{{\mathrm{2}}}}  \textcolor{coeffectColor}{]}    \textcolor{GLcolor}{\rrbracket}  \\
        \textit{\{defn. translation\}} & \equiv_L  \mathsf{let} \, \textcolor{coeffectColor}{[}  \SYSTEMmv{x}  \textcolor{coeffectColor}{]} =   \textcolor{coeffectColor}{[}   \textcolor{GLcolor}{\llbracket}  \SYSTEMnt{t_{{\mathrm{1}}}}  \textcolor{GLcolor}{\rrbracket}   \textcolor{coeffectColor}{]}   \, \mathsf{in} \,   \textcolor{coeffectColor}{[}   \textcolor{GLcolor}{\llbracket}  \SYSTEMnt{t_{{\mathrm{2}}}}  \textcolor{GLcolor}{\rrbracket}   \textcolor{coeffectColor}{]}   \\
        \textit{\{\SYSTEMRenameRuleLinEqletCommBox{}}\} & \equiv_L  \textcolor{coeffectColor}{[}   \mathsf{let} \, \textcolor{coeffectColor}{[}  \SYSTEMmv{x}  \textcolor{coeffectColor}{]} =   \textcolor{GLcolor}{\llbracket}  \SYSTEMnt{t_{{\mathrm{1}}}}  \textcolor{GLcolor}{\rrbracket}   \, \mathsf{in} \,   \textcolor{GLcolor}{\llbracket}  \SYSTEMnt{t_{{\mathrm{2}}}}  \textcolor{GLcolor}{\rrbracket}    \textcolor{coeffectColor}{]}  \\
        \textit{\{defn. translation\}} & \equiv_L  \textcolor{GLcolor}{\llbracket}   \textcolor{coeffectColor}{[}   \mathsf{let} \, \textcolor{coeffectColor}{[}  \SYSTEMmv{x}  \textcolor{coeffectColor}{]} =   \textcolor{coeffectColor}{[}  \SYSTEMnt{t_{{\mathrm{1}}}}  \textcolor{coeffectColor}{]}   \, \mathsf{in} \,  \SYSTEMnt{t_{{\mathrm{2}}}}   \textcolor{coeffectColor}{]}   \textcolor{GLcolor}{\rrbracket}  \\
      \end{array}
    \end{align*}

  \item \[\SYSTEMdruleGradMEqletCommOne{}\]

    Preservation then follows by:
    \begin{align*}
      \begin{array}{rl}
        &  \textcolor{GLcolor}{\llbracket}   \SYSTEMnt{t_{{\mathrm{0}}}} \,  \SYSTEMsym{(}   \mathsf{let} \, \textcolor{coeffectColor}{[}  \SYSTEMmv{x}  \textcolor{coeffectColor}{]} =  \SYSTEMnt{t_{{\mathrm{1}}}}  \, \mathsf{in} \,  \SYSTEMnt{t_{{\mathrm{2}}}}   \SYSTEMsym{)}   \textcolor{GLcolor}{\rrbracket}  \\
        \textit{\{defn. translation\}} & =   \textcolor{GLcolor}{\llbracket}  \SYSTEMnt{t_{{\mathrm{0}}}}  \textcolor{GLcolor}{\rrbracket}  \,   \textcolor{coeffectColor}{[}   \mathsf{let} \, \textcolor{coeffectColor}{[}  \SYSTEMmv{x}  \textcolor{coeffectColor}{]} =   \textcolor{GLcolor}{\llbracket}  \SYSTEMnt{t_{{\mathrm{1}}}}  \textcolor{GLcolor}{\rrbracket}   \, \mathsf{in} \,   \textcolor{GLcolor}{\llbracket}  \SYSTEMnt{t_{{\mathrm{2}}}}  \textcolor{GLcolor}{\rrbracket}    \textcolor{coeffectColor}{]}   \\
   \textit{\{\SYSTEMRenameRuleLinEqcongApp{}+\SYSTEMRenameRuleLinEqbeta\}}
        & \equiv_L   \textcolor{GLcolor}{\llbracket}  \SYSTEMnt{t_{{\mathrm{0}}}}  \textcolor{GLcolor}{\rrbracket}  \,  \SYSTEMsym{(}   \SYSTEMsym{(}   \lambda  \SYSTEMmv{z}  .   \textcolor{coeffectColor}{[}  \SYSTEMmv{z}  \textcolor{coeffectColor}{]}    \SYSTEMsym{)} \,  \SYSTEMsym{(}   \mathsf{let} \, \textcolor{coeffectColor}{[}  \SYSTEMmv{x}  \textcolor{coeffectColor}{]} =   \textcolor{GLcolor}{\llbracket}  \SYSTEMnt{t_{{\mathrm{1}}}}  \textcolor{GLcolor}{\rrbracket}   \, \mathsf{in} \,   \textcolor{GLcolor}{\llbracket}  \SYSTEMnt{t_{{\mathrm{2}}}}  \textcolor{GLcolor}{\rrbracket}    \SYSTEMsym{)}   \SYSTEMsym{)}  \\
   \textit{\{\SYSTEMRenameRuleLinEqcongApp{}+\SYSTEMRenameRuleLinEqletCommOne\}}
        & \equiv_L   \textcolor{GLcolor}{\llbracket}  \SYSTEMnt{t_{{\mathrm{0}}}}  \textcolor{GLcolor}{\rrbracket}  \,  \SYSTEMsym{(}    \mathsf{let} \, \textcolor{coeffectColor}{[}  \SYSTEMmv{x}  \textcolor{coeffectColor}{]} =   \textcolor{GLcolor}{\llbracket}  \SYSTEMnt{t_{{\mathrm{1}}}}  \textcolor{GLcolor}{\rrbracket}   \, \mathsf{in} \,  \SYSTEMsym{(}   \lambda  \SYSTEMmv{z}  .   \textcolor{coeffectColor}{[}  \SYSTEMmv{z}  \textcolor{coeffectColor}{]}    \SYSTEMsym{)}  \,   \textcolor{GLcolor}{\llbracket}  \SYSTEMnt{t_{{\mathrm{2}}}}  \textcolor{GLcolor}{\rrbracket}    \SYSTEMsym{)}  \\
   \textit{\{\SYSTEMRenameRuleLinEqcongApp{}+\SYSTEMRenameRuleLinEqcongLet{}+\SYSTEMRenameRuleLinEqbeta\}}
        & \equiv_L   \textcolor{GLcolor}{\llbracket}  \SYSTEMnt{t_{{\mathrm{0}}}}  \textcolor{GLcolor}{\rrbracket}  \,  \SYSTEMsym{(}   \mathsf{let} \, \textcolor{coeffectColor}{[}  \SYSTEMmv{x}  \textcolor{coeffectColor}{]} =   \textcolor{GLcolor}{\llbracket}  \SYSTEMnt{t_{{\mathrm{1}}}}  \textcolor{GLcolor}{\rrbracket}   \, \mathsf{in} \,   \textcolor{coeffectColor}{[}    \textcolor{GLcolor}{\llbracket}  \SYSTEMnt{t_{{\mathrm{2}}}}  \textcolor{GLcolor}{\rrbracket}    \textcolor{coeffectColor}{]}    \SYSTEMsym{)}  \\
   \textit{\{\SYSTEMRenameRuleLinEqletCommOne{}\}}
        & \equiv_L   \mathsf{let} \, \textcolor{coeffectColor}{[}  \SYSTEMmv{x}  \textcolor{coeffectColor}{]} =   \textcolor{GLcolor}{\llbracket}  \SYSTEMnt{t_{{\mathrm{1}}}}  \textcolor{GLcolor}{\rrbracket}   \, \mathsf{in} \,   \textcolor{GLcolor}{\llbracket}  \SYSTEMnt{t_{{\mathrm{0}}}}  \textcolor{GLcolor}{\rrbracket}   \,   \textcolor{coeffectColor}{[}    \textcolor{GLcolor}{\llbracket}  \SYSTEMnt{t_{{\mathrm{2}}}}  \textcolor{GLcolor}{\rrbracket}    \textcolor{coeffectColor}{]}   \\
   \textit{\{defn. translation}\}
        & =  \textcolor{GLcolor}{\llbracket}    \mathsf{let} \, \textcolor{coeffectColor}{[}  \SYSTEMmv{x}  \textcolor{coeffectColor}{]} =  \SYSTEMnt{t_{{\mathrm{1}}}}  \, \mathsf{in} \,  \SYSTEMnt{t_{{\mathrm{0}}}}  \,  \SYSTEMnt{t_{{\mathrm{2}}}}   \textcolor{GLcolor}{\rrbracket} 
      \end{array}
      \end{align*}

  \item \[\SYSTEMdruleGradMEqletCommTwo{}\]

    Preservation then follows by:
    \begin{align*}
      \begin{array}{rl}
        &  \textcolor{GLcolor}{\llbracket}   \SYSTEMsym{(}   \mathsf{let} \, \textcolor{coeffectColor}{[}  \SYSTEMmv{x}  \textcolor{coeffectColor}{]} =  \SYSTEMnt{t_{{\mathrm{1}}}}  \, \mathsf{in} \,  \SYSTEMnt{t_{{\mathrm{2}}}}   \SYSTEMsym{)} \,  \SYSTEMnt{t_{{\mathrm{0}}}}   \textcolor{GLcolor}{\rrbracket}  \\
        \textit{\{defn. translation\}} & =   \textcolor{coeffectColor}{[}   \mathsf{let} \, \textcolor{coeffectColor}{[}  \SYSTEMmv{x}  \textcolor{coeffectColor}{]} =   \textcolor{GLcolor}{\llbracket}  \SYSTEMnt{t_{{\mathrm{1}}}}  \textcolor{GLcolor}{\rrbracket}   \, \mathsf{in} \,   \textcolor{GLcolor}{\llbracket}  \SYSTEMnt{t_{{\mathrm{2}}}}  \textcolor{GLcolor}{\rrbracket}    \textcolor{coeffectColor}{]}  \,   \textcolor{GLcolor}{\llbracket}  \SYSTEMnt{t_{{\mathrm{0}}}}  \textcolor{GLcolor}{\rrbracket}   \\
   \textit{\{\SYSTEMRenameRuleLinEqcongApp{}+\SYSTEMRenameRuleLinEqbeta\}}
        & \equiv_L  \SYSTEMsym{(}   \SYSTEMsym{(}   \lambda  \SYSTEMmv{z}  .   \textcolor{coeffectColor}{[}  \SYSTEMmv{z}  \textcolor{coeffectColor}{]}    \SYSTEMsym{)} \,  \SYSTEMsym{(}   \mathsf{let} \, \textcolor{coeffectColor}{[}  \SYSTEMmv{x}  \textcolor{coeffectColor}{]} =   \textcolor{GLcolor}{\llbracket}  \SYSTEMnt{t_{{\mathrm{1}}}}  \textcolor{GLcolor}{\rrbracket}   \, \mathsf{in} \,   \textcolor{GLcolor}{\llbracket}  \SYSTEMnt{t_{{\mathrm{2}}}}  \textcolor{GLcolor}{\rrbracket}    \SYSTEMsym{)}   \SYSTEMsym{)} \,   \textcolor{GLcolor}{\llbracket}  \SYSTEMnt{t_{{\mathrm{0}}}}  \textcolor{GLcolor}{\rrbracket}   \\
   \textit{\{\SYSTEMRenameRuleLinEqcongApp{}+\SYSTEMRenameRuleLinEqletCommOne\}}
        & \equiv_L  \SYSTEMsym{(}    \mathsf{let} \, \textcolor{coeffectColor}{[}  \SYSTEMmv{x}  \textcolor{coeffectColor}{]} =   \textcolor{GLcolor}{\llbracket}  \SYSTEMnt{t_{{\mathrm{1}}}}  \textcolor{GLcolor}{\rrbracket}   \, \mathsf{in} \,  \SYSTEMsym{(}   \lambda  \SYSTEMmv{z}  .   \textcolor{coeffectColor}{[}  \SYSTEMmv{z}  \textcolor{coeffectColor}{]}    \SYSTEMsym{)}  \,   \textcolor{GLcolor}{\llbracket}  \SYSTEMnt{t_{{\mathrm{2}}}}  \textcolor{GLcolor}{\rrbracket}    \SYSTEMsym{)} \,   \textcolor{GLcolor}{\llbracket}  \SYSTEMnt{t_{{\mathrm{0}}}}  \textcolor{GLcolor}{\rrbracket}   \\
   \textit{\{\SYSTEMRenameRuleLinEqcongApp{}+\SYSTEMRenameRuleLinEqcongLet{}+\SYSTEMRenameRuleLinEqbeta\}}
        & \equiv_L  \SYSTEMsym{(}   \mathsf{let} \, \textcolor{coeffectColor}{[}  \SYSTEMmv{x}  \textcolor{coeffectColor}{]} =   \textcolor{GLcolor}{\llbracket}  \SYSTEMnt{t_{{\mathrm{1}}}}  \textcolor{GLcolor}{\rrbracket}   \, \mathsf{in} \,   \textcolor{coeffectColor}{[}    \textcolor{GLcolor}{\llbracket}  \SYSTEMnt{t_{{\mathrm{2}}}}  \textcolor{GLcolor}{\rrbracket}    \textcolor{coeffectColor}{]}    \SYSTEMsym{)} \,   \textcolor{GLcolor}{\llbracket}  \SYSTEMnt{t_{{\mathrm{0}}}}  \textcolor{GLcolor}{\rrbracket}   \\
   \textit{\{\SYSTEMRenameRuleLinEqletCommTwo{}\}}
        & \equiv_L   \mathsf{let} \, \textcolor{coeffectColor}{[}  \SYSTEMmv{x}  \textcolor{coeffectColor}{]} =   \textcolor{GLcolor}{\llbracket}  \SYSTEMnt{t_{{\mathrm{1}}}}  \textcolor{GLcolor}{\rrbracket}   \, \mathsf{in} \,   \textcolor{GLcolor}{\llbracket}  \SYSTEMnt{t_{{\mathrm{2}}}}  \textcolor{GLcolor}{\rrbracket}   \,   \textcolor{coeffectColor}{[}    \textcolor{GLcolor}{\llbracket}  \SYSTEMnt{t_{{\mathrm{0}}}}  \textcolor{GLcolor}{\rrbracket}    \textcolor{coeffectColor}{]}   \\
   \textit{\{defn. translation}\}
        & =  \textcolor{GLcolor}{\llbracket}    \mathsf{let} \, \textcolor{coeffectColor}{[}  \SYSTEMmv{x}  \textcolor{coeffectColor}{]} =  \SYSTEMnt{t_{{\mathrm{1}}}}  \, \mathsf{in} \,  \SYSTEMnt{t_{{\mathrm{2}}}}  \,  \SYSTEMnt{t_{{\mathrm{0}}}}   \textcolor{GLcolor}{\rrbracket} 
      \end{array}
      \end{align*}

    \item The remaining rules are congruences and all follow by induction with
    the translation.
  \end{itemize}
\end{proof}

\subsection{Proof of Soundness for Linear Base to Graded Poly Base translation}
\label{app:lin-to-grad-poly}

\ifextended\linToGradTranslationPoly*\fi

\subsubsection{Type preservation}
\label{app:proofs-lin-to-grad-poly-typ}

\begin{proof}
By induction on the Linear Base typing, considering just the two
cases that are new here:
\begin{itemize}
\item (pr)
$$
\SYSTEMdruleLinpr{}
$$
By induction on the premise we have $  \textcolor{LGcolor}{\llparenthesis}  \Gamma  \textcolor{LGcolor}{\rrparenthesis}   \vdash_{\textsc{g} }   \textcolor{LGcolor}{\llparenthesis}  \Pi  \textcolor{LGcolor}{\rrparenthesis}   :   \textcolor{LGcolor}{\llparenthesis} \smidge  \SYSTEMnt{A}  \smidge \textcolor{LGcolor}{\rrparenthesis}  $
(i.e., assuming the derivation of the premise is $\Pi$).

Therefore we can construct:
$$
\inferrule*[Right=\SYSTEMRenameRuleGradPolytyAbs{}]
{
\inferrule*[Right=\SYSTEMRenameRuleGradabs{}]
{
\inferrule*[Right=\SYSTEMRenameRuleGradapp{}]
{   \textcolor{LGcolor}{\llparenthesis}  \Gamma  \textcolor{LGcolor}{\rrparenthesis}   \vdash_{\textsc{g} }   \textcolor{LGcolor}{\llparenthesis}  \Pi  \textcolor{LGcolor}{\rrparenthesis}   :   \textcolor{LGcolor}{\llparenthesis} \smidge  \SYSTEMnt{A}  \smidge \textcolor{LGcolor}{\rrparenthesis}   \quad
  \inferrule*[Right=\SYSTEMRenameRuleGradvar{}]
    {\quad}{   \SYSTEMmv{f}  :_{\textcolor{coeffectColor}{ \SYSTEMsym{1} } }     \textcolor{LGcolor}{\llparenthesis} \smidge  \SYSTEMnt{A}  \smidge \textcolor{LGcolor}{\rrparenthesis}   \xrightarrow{\textcolor{coeffectColor}{ \SYSTEMnt{r} } }   \beta'       \vdash_{\textsc{g} }  \SYSTEMmv{f}  :     \textcolor{LGcolor}{\llparenthesis} \smidge  \SYSTEMnt{A}  \smidge \textcolor{LGcolor}{\rrparenthesis}   \xrightarrow{\textcolor{coeffectColor}{ \SYSTEMnt{r} } }   \beta'     }}
{    \textcolor{coeffectColor}{ \SYSTEMnt{r}  \cdot}   \textcolor{LGcolor}{\llparenthesis}  \Gamma  \textcolor{LGcolor}{\rrparenthesis}   ,   \SYSTEMmv{f}  :_{\textcolor{coeffectColor}{ \SYSTEMsym{1} } }     \textcolor{LGcolor}{\llparenthesis} \smidge  \SYSTEMnt{A}  \smidge \textcolor{LGcolor}{\rrparenthesis}   \xrightarrow{\textcolor{coeffectColor}{ \SYSTEMnt{r} } }   \beta'       \vdash_{\textsc{g} }   \SYSTEMmv{f} \,   \textcolor{LGcolor}{\llparenthesis}  \Pi  \textcolor{LGcolor}{\rrparenthesis}    :   \beta'   }
}
{  \textcolor{coeffectColor}{ \SYSTEMnt{r}  \cdot}   \textcolor{LGcolor}{\llparenthesis}  \Gamma  \textcolor{LGcolor}{\rrparenthesis}    \vdash_{\textsc{g} }    \lambda  \SYSTEMmv{f}  .  \SYSTEMmv{f}  \,   \textcolor{LGcolor}{\llparenthesis}  \Pi  \textcolor{LGcolor}{\rrparenthesis}    :    (    \textcolor{LGcolor}{\llparenthesis} \smidge  \SYSTEMnt{A}  \smidge \textcolor{LGcolor}{\rrparenthesis}   \xrightarrow{\textcolor{coeffectColor}{ \SYSTEMnt{r} } }   \beta'    )   \xrightarrow{\textcolor{coeffectColor}{ \SYSTEMsym{1} } }   \beta'    }
}
{
  \textcolor{coeffectColor}{ \SYSTEMnt{r}  \cdot}   \textcolor{LGcolor}{\llparenthesis}  \Gamma  \textcolor{LGcolor}{\rrparenthesis}    \vdash_{\textsc{g} }    \Lambda   \beta   .   \lambda  \SYSTEMmv{f}  .  \SYSTEMmv{f}   \,   \textcolor{LGcolor}{\llparenthesis}  \Pi  \textcolor{LGcolor}{\rrparenthesis}    :    \forall   \beta   .   (    \textcolor{LGcolor}{\llparenthesis} \smidge  \SYSTEMnt{A}  \smidge \textcolor{LGcolor}{\rrparenthesis}   \xrightarrow{\textcolor{coeffectColor}{ \SYSTEMnt{r} } }   \beta    )    \xrightarrow{\textcolor{coeffectColor}{ \SYSTEMsym{1} } }   \beta   
}
$$
\item (let)
$$
\SYSTEMdruleLinlet{}
$$
By induction on both premises (whose derivations we denote $\Pi_{{\mathrm{1}}}$ and
$\Pi_{{\mathrm{2}}}$), we have:
$  \textcolor{LGcolor}{\llparenthesis}  \Gamma_{{\mathrm{1}}}  \textcolor{LGcolor}{\rrparenthesis}   \vdash_{\textsc{g} }   \textcolor{LGcolor}{\llparenthesis}  \Pi_{{\mathrm{1}}}  \textcolor{LGcolor}{\rrparenthesis}   :   \textcolor{coeffectColor}{\square_{ \SYSTEMnt{r} } }   \textcolor{LGcolor}{\llparenthesis} \smidge  \SYSTEMnt{A}  \smidge \textcolor{LGcolor}{\rrparenthesis}   $
and
$   \textcolor{LGcolor}{\llparenthesis}  \Gamma_{{\mathrm{2}}}  \textcolor{LGcolor}{\rrparenthesis}  ,   \SYSTEMmv{x}  :_{\textcolor{coeffectColor}{ \SYSTEMnt{r} } }   \textcolor{LGcolor}{\llparenthesis} \smidge  \SYSTEMnt{A}  \smidge \textcolor{LGcolor}{\rrparenthesis}     \vdash_{\textsc{g} }   \textcolor{LGcolor}{\llparenthesis}  \Pi_{{\mathrm{2}}}  \textcolor{LGcolor}{\rrparenthesis}   :   \textcolor{LGcolor}{\llparenthesis} \smidge  \SYSTEMnt{B}  \smidge \textcolor{LGcolor}{\rrparenthesis}  $.

Therefore we can construct:
\begin{gather*}
\inferrule*[Right=\SYSTEMRenameRuleGradapp{}]
{
  \inferrule*[Right=\SYSTEMRenameRuleGradPolytyApp{}]
     {  \textcolor{LGcolor}{\llparenthesis}  \Gamma_{{\mathrm{1}}}  \textcolor{LGcolor}{\rrparenthesis}   \vdash_{\textsc{g} }   \textcolor{LGcolor}{\llparenthesis}  \Pi_{{\mathrm{1}}}  \textcolor{LGcolor}{\rrparenthesis}   :    \forall   \beta   .   (    \textcolor{LGcolor}{\llparenthesis} \smidge  \SYSTEMnt{A}  \smidge \textcolor{LGcolor}{\rrparenthesis}   \xrightarrow{\textcolor{coeffectColor}{ \SYSTEMnt{r} } }   \beta    )    \xrightarrow{\textcolor{coeffectColor}{ \SYSTEMsym{1} } }   \beta   }
     {  \textcolor{LGcolor}{\llparenthesis}  \Gamma_{{\mathrm{1}}}  \textcolor{LGcolor}{\rrparenthesis}   \vdash_{\textsc{g} }    \textcolor{LGcolor}{\llparenthesis}  \Pi_{{\mathrm{1}}}  \textcolor{LGcolor}{\rrparenthesis}   \text{@}   \textcolor{LGcolor}{\llparenthesis} \smidge  \SYSTEMnt{B}  \smidge \textcolor{LGcolor}{\rrparenthesis}    :    (    \textcolor{LGcolor}{\llparenthesis} \smidge  \SYSTEMnt{A}  \smidge \textcolor{LGcolor}{\rrparenthesis}   \xrightarrow{\textcolor{coeffectColor}{ \SYSTEMnt{r} } }   \textcolor{LGcolor}{\llparenthesis} \smidge  \SYSTEMnt{B}  \smidge \textcolor{LGcolor}{\rrparenthesis}    )   \xrightarrow{\textcolor{coeffectColor}{ \SYSTEMsym{1} } }   \textcolor{LGcolor}{\llparenthesis} \smidge  \SYSTEMnt{B}  \smidge \textcolor{LGcolor}{\rrparenthesis}   }
  \\ \quad
  \inferrule*[Right=\SYSTEMRenameRuleGradabs{}]
    {   \textcolor{LGcolor}{\llparenthesis}  \Gamma_{{\mathrm{2}}}  \textcolor{LGcolor}{\rrparenthesis}  ,   \SYSTEMmv{x}  :_{\textcolor{coeffectColor}{ \SYSTEMnt{r} } }  \SYSTEMnt{A}    \vdash_{\textsc{g} }   \textcolor{LGcolor}{\llparenthesis}  \Pi_{{\mathrm{2}}}  \textcolor{LGcolor}{\rrparenthesis}   :   \textcolor{LGcolor}{\llparenthesis} \smidge  \SYSTEMnt{B}  \smidge \textcolor{LGcolor}{\rrparenthesis}  }
    {  \textcolor{LGcolor}{\llparenthesis}  \Gamma_{{\mathrm{2}}}  \textcolor{LGcolor}{\rrparenthesis}   \vdash_{\textsc{g} }   \lambda  \SYSTEMmv{x}  .   \textcolor{LGcolor}{\llparenthesis}  \Pi_{{\mathrm{2}}}  \textcolor{LGcolor}{\rrparenthesis}    :    \textcolor{LGcolor}{\llparenthesis} \smidge  \SYSTEMnt{A}  \smidge \textcolor{LGcolor}{\rrparenthesis}   \xrightarrow{\textcolor{coeffectColor}{ \SYSTEMnt{r} } }   \textcolor{LGcolor}{\llparenthesis} \smidge  \SYSTEMnt{B}  \smidge \textcolor{LGcolor}{\rrparenthesis}   }
}
{  \textcolor{LGcolor}{\llparenthesis}  \Gamma_{{\mathrm{1}}}  \textcolor{LGcolor}{\rrparenthesis}   \SYSTEMsym{+}   \textcolor{LGcolor}{\llparenthesis}  \Gamma_{{\mathrm{2}}}  \textcolor{LGcolor}{\rrparenthesis}   \vdash_{\textsc{g} }   \SYSTEMsym{(}    \textcolor{LGcolor}{\llparenthesis}  \Pi_{{\mathrm{1}}}  \textcolor{LGcolor}{\rrparenthesis}   \text{@}   \textcolor{LGcolor}{\llparenthesis} \smidge  \SYSTEMnt{B}  \smidge \textcolor{LGcolor}{\rrparenthesis}    \SYSTEMsym{)} \,  \SYSTEMsym{(}   \lambda  \SYSTEMmv{x}  .   \textcolor{LGcolor}{\llparenthesis}  \Pi_{{\mathrm{2}}}  \textcolor{LGcolor}{\rrparenthesis}    \SYSTEMsym{)}   :   \textcolor{LGcolor}{\llparenthesis} \smidge  \SYSTEMnt{B}  \smidge \textcolor{LGcolor}{\rrparenthesis}  }
\end{gather*}
\end{itemize}
The remaining cases follow as before, following the structure of the derivations.
\end{proof}

\subsubsection{Operational correspondence}
\label{app:proofs-lin-to-grad-poly-ops}

Some intermediate lemmas are needed.

\begin{restatable}{lemm}{substGradPoly}[Interpretation preserves substitution of derivations]
\label{lemma:substGradPoly}
For all terms $\SYSTEMnt{t_{{\mathrm{1}}}}$ and $\SYSTEMnt{t_{{\mathrm{2}}}}$ and derivations $\Pi_{{\mathrm{1}}} :  \Gamma_{{\mathrm{1}}}  \vdash_{\textsc{l} }  \SYSTEMnt{t_{{\mathrm{1}}}}  :   \textcolor{coeffectColor}{\square_{ \SYSTEMnt{r} } }  \SYSTEMnt{A}  $
and $\Pi_{{\mathrm{2}}} :   \Gamma_{{\mathrm{2}}} ,   \SYSTEMmv{x}  : \textcolor{coeffectColor}{[}  \SYSTEMnt{A} {\textcolor{coeffectColor}{]_{ \SYSTEMnt{r} } } }    \vdash_{\textsc{l} }  \SYSTEMnt{t_{{\mathrm{2}}}}  :  \SYSTEMnt{B} $ then
there exists $\Pi'$ such that $\Pi' :  \Gamma_{{\mathrm{1}}}  \SYSTEMsym{+}  \Gamma_{{\mathrm{2}}}  \vdash_{\textsc{l} }   [  \SYSTEMnt{t_{{\mathrm{1}}}}  /  \SYSTEMmv{x}  ]  \SYSTEMnt{t_{{\mathrm{2}}}}   :  \SYSTEMnt{B} $
and
$ [   \textcolor{LGcolor}{\llparenthesis}  \Pi_{{\mathrm{1}}}  \textcolor{LGcolor}{\rrparenthesis}   /  \SYSTEMmv{x}  ]   \textcolor{LGcolor}{\llparenthesis}  \Pi_{{\mathrm{2}}}  \textcolor{LGcolor}{\rrparenthesis}   =  \textcolor{LGcolor}{\llparenthesis}  \Pi'  \textcolor{LGcolor}{\rrparenthesis} $.
\end{restatable}

\begin{proof}
By induction on $\Pi_{{\mathrm{2}}}$, generalising the admissibility of substitution.
\end{proof}

\begin{restatable}{lemm}{polySubst}[Type variables in the interpretation are irrelevant]
\label{lemma:substIrrelevant}
Given $\Pi' :  \Gamma  \vdash_{\textsc{l} }  \SYSTEMnt{t}  :  \SYSTEMnt{A} $
then for $\Pi :   \textcolor{coeffectColor}{ \SYSTEMnt{r}  \cdot}  \Gamma   \vdash_{\textsc{l} }   \textcolor{coeffectColor}{[}  \SYSTEMnt{t}  \textcolor{coeffectColor}{]}   :   \textcolor{coeffectColor}{\square_{ \SYSTEMnt{r} } }  \SYSTEMnt{A}  $
formed from promotion of $\Pi'$
such that $ \textcolor{LGcolor}{\llparenthesis}  \Pi  \textcolor{LGcolor}{\rrparenthesis}  =   \Lambda   \beta   .   \lambda  \SYSTEMmv{f}  .  \SYSTEMmv{f}   \,   \textcolor{LGcolor}{\llparenthesis}  \Pi'  \textcolor{LGcolor}{\rrparenthesis}  $
then $ \beta  \not\in \mathsf{fv}(  \textcolor{LGcolor}{\llparenthesis}  \Pi  \textcolor{LGcolor}{\rrparenthesis}  )$.
\end{restatable}

\begin{proof}
By case analysis on $\Pi'$; there is no part of the translation that
introduces a free type variable.
\end{proof}

Operational correspondence then follows by the following proof:

\begin{proof}
\begin{itemize}
\item (betaBox)
\[
\SYSTEMdruleSemLinbetaBox{}
\]
with typing:
\[
\inferrule*[right=\SYSTEMRenameRuleLinlet{}]
{  \textcolor{coeffectColor}{ \SYSTEMnt{r}  \cdot}  \Gamma_{{\mathrm{1}}}   \vdash_{\textsc{l} }   \textcolor{coeffectColor}{[}  \SYSTEMnt{t_{{\mathrm{1}}}}  \textcolor{coeffectColor}{]}   :   \textcolor{coeffectColor}{\square_{ \SYSTEMnt{r} } }  \SYSTEMnt{A}   \quad
   \Gamma_{{\mathrm{2}}} ,   \SYSTEMmv{x}  : \textcolor{coeffectColor}{[}  \SYSTEMnt{A} {\textcolor{coeffectColor}{]_{ \SYSTEMnt{r} } } }    \vdash_{\textsc{l} }  \SYSTEMnt{t_{{\mathrm{2}}}}  :  \SYSTEMnt{B} }
{  \textcolor{coeffectColor}{ \SYSTEMnt{r}  \cdot}  \Gamma_{{\mathrm{1}}}   \SYSTEMsym{+}  \Gamma_{{\mathrm{2}}}  \vdash_{\textsc{l} }   \mathsf{let} \, \textcolor{coeffectColor}{[}  \SYSTEMmv{x}  \textcolor{coeffectColor}{]} =   \textcolor{coeffectColor}{[}  \SYSTEMnt{t_{{\mathrm{1}}}}  \textcolor{coeffectColor}{]}   \, \mathsf{in} \,  \SYSTEMnt{t_{{\mathrm{2}}}}   :  \SYSTEMnt{B} }
\]
such that $\Pi_{{\mathrm{1}}} :  \Gamma_{{\mathrm{1}}}  \vdash_{\textsc{l} }  \SYSTEMnt{t_{{\mathrm{1}}}}  :  \SYSTEMnt{A} $
and $\Pi_{{\mathrm{2}}} :   \Gamma_{{\mathrm{2}}} ,   \SYSTEMmv{x}  : \textcolor{coeffectColor}{[}  \SYSTEMnt{A} {\textcolor{coeffectColor}{]_{ \SYSTEMnt{r} } } }    \vdash_{\textsc{l} }  \SYSTEMnt{t_{{\mathrm{2}}}}  :  \SYSTEMnt{B} $
in the premises of this derivation
and
with $\Pi' :   \textcolor{coeffectColor}{ \SYSTEMnt{r}  \cdot}  \Gamma_{{\mathrm{1}}}   \SYSTEMsym{+}  \Gamma_{{\mathrm{2}}}  \vdash_{\textsc{l} }   [  \SYSTEMnt{t_{{\mathrm{1}}}}  /  \SYSTEMmv{x}  ]  \SYSTEMnt{t_{{\mathrm{2}}}}   :  \SYSTEMnt{B} $.

The interpretation of the reducing term's proof is:
\[
 \textcolor{LGcolor}{\llparenthesis}  \Pi  \textcolor{LGcolor}{\rrparenthesis}  \equiv
 \SYSTEMsym{(}   \SYSTEMsym{(}    \Lambda   \beta   .   \lambda  \SYSTEMmv{f}  .  \SYSTEMmv{f}   \,   \textcolor{LGcolor}{\llparenthesis}  \Pi_{{\mathrm{1}}}  \textcolor{LGcolor}{\rrparenthesis}    \SYSTEMsym{)}  \text{@}   \textcolor{LGcolor}{\llparenthesis} \smidge  \SYSTEMnt{B}  \smidge \textcolor{LGcolor}{\rrparenthesis}    \SYSTEMsym{)} \,  \SYSTEMsym{(}   \lambda  \SYSTEMmv{x}  .   \textcolor{LGcolor}{\llparenthesis}  \Pi_{{\mathrm{2}}}  \textcolor{LGcolor}{\rrparenthesis}    \SYSTEMsym{)} 
\]
Then we have the following reduction sequence in Graded Base:
\begin{align*}
\inferrule*[right=\SYSTEMRenameRuleSemGrdPolytyBeta{}]
{\quad}
{  \SYSTEMsym{(}   \SYSTEMsym{(}    \Lambda   \beta   .   \lambda  \SYSTEMmv{f}  .  \SYSTEMmv{f}   \,   \textcolor{LGcolor}{\llparenthesis}  \Pi_{{\mathrm{1}}}  \textcolor{LGcolor}{\rrparenthesis}    \SYSTEMsym{)}  \text{@}   \textcolor{LGcolor}{\llparenthesis} \smidge  \SYSTEMnt{B}  \smidge \textcolor{LGcolor}{\rrparenthesis}    \SYSTEMsym{)} \,  \SYSTEMsym{(}   \lambda  \SYSTEMmv{x}  .   \textcolor{LGcolor}{\llparenthesis}  \Pi_{{\mathrm{2}}}  \textcolor{LGcolor}{\rrparenthesis}    \SYSTEMsym{)}   \rightsquigarrow_{\textsc{g} }   \SYSTEMsym{(}   [   \textcolor{LGcolor}{\llparenthesis} \smidge  \SYSTEMnt{B}  \smidge \textcolor{LGcolor}{\rrparenthesis}   /   \beta   ]  \SYSTEMsym{(}    \lambda  \SYSTEMmv{f}  .  \SYSTEMmv{f}  \,   \textcolor{LGcolor}{\llparenthesis}  \Pi_{{\mathrm{1}}}  \textcolor{LGcolor}{\rrparenthesis}    \SYSTEMsym{)}   \SYSTEMsym{)} \,  \SYSTEMsym{(}   \lambda  \SYSTEMmv{x}  .   \textcolor{LGcolor}{\llparenthesis}  \Pi_{{\mathrm{2}}}  \textcolor{LGcolor}{\rrparenthesis}    \SYSTEMsym{)}  }
\end{align*}
\begin{align*}
\inferrule*[right=\SYSTEMRenameRuleSemGrdbeta{}]
{\quad}
{   \SYSTEMsym{(}   [   \textcolor{LGcolor}{\llparenthesis} \smidge  \SYSTEMnt{B}  \smidge \textcolor{LGcolor}{\rrparenthesis}   /   \beta   ]  \SYSTEMsym{(}    \lambda  \SYSTEMmv{f}  .  \SYSTEMmv{f}  \,   \textcolor{LGcolor}{\llparenthesis}  \Pi_{{\mathrm{1}}}  \textcolor{LGcolor}{\rrparenthesis}    \SYSTEMsym{)}   \SYSTEMsym{)} \,  \SYSTEMsym{(}   \lambda  \SYSTEMmv{x}  .   \textcolor{LGcolor}{\llparenthesis}  \Pi_{{\mathrm{2}}}  \textcolor{LGcolor}{\rrparenthesis}    \SYSTEMsym{)}   \rightsquigarrow_{\textsc{g} }   [   \textcolor{LGcolor}{\llparenthesis} \smidge  \SYSTEMnt{B}  \smidge \textcolor{LGcolor}{\rrparenthesis}   /   \beta   ]  \SYSTEMsym{(}   \SYSTEMsym{(}   \lambda  \SYSTEMmv{x}  .   \textcolor{LGcolor}{\llparenthesis}  \Pi_{{\mathrm{2}}}  \textcolor{LGcolor}{\rrparenthesis}    \SYSTEMsym{)} \,   \textcolor{LGcolor}{\llparenthesis}  \Pi_{{\mathrm{1}}}  \textcolor{LGcolor}{\rrparenthesis}    \SYSTEMsym{)}  
}
\end{align*}
\begin{align*}
\inferrule*[right=\SYSTEMRenameRuleSemGrdbeta{}]
{\quad}
{   [   \textcolor{LGcolor}{\llparenthesis} \smidge  \SYSTEMnt{B}  \smidge \textcolor{LGcolor}{\rrparenthesis}   /   \beta   ]  \SYSTEMsym{(}   \SYSTEMsym{(}   \lambda  \SYSTEMmv{x}  .   \textcolor{LGcolor}{\llparenthesis}  \Pi_{{\mathrm{2}}}  \textcolor{LGcolor}{\rrparenthesis}    \SYSTEMsym{)} \,   \textcolor{LGcolor}{\llparenthesis}  \Pi_{{\mathrm{1}}}  \textcolor{LGcolor}{\rrparenthesis}    \SYSTEMsym{)}   \rightsquigarrow_{\textsc{g} }   [   \textcolor{LGcolor}{\llparenthesis} \smidge  \SYSTEMnt{B}  \smidge \textcolor{LGcolor}{\rrparenthesis}   /   \beta   ]   [   \textcolor{LGcolor}{\llparenthesis}  \Pi_{{\mathrm{1}}}  \textcolor{LGcolor}{\rrparenthesis}   /  \SYSTEMmv{x}  ]   \textcolor{LGcolor}{\llparenthesis}  \Pi_{{\mathrm{2}}}  \textcolor{LGcolor}{\rrparenthesis}    
}
\end{align*}
thus $\SYSTEMnt{t''} =  [   \textcolor{LGcolor}{\llparenthesis} \smidge  \SYSTEMnt{B}  \smidge \textcolor{LGcolor}{\rrparenthesis}   /   \beta   ]   [   \textcolor{LGcolor}{\llparenthesis}  \Pi_{{\mathrm{1}}}  \textcolor{LGcolor}{\rrparenthesis}   /  \SYSTEMmv{x}  ]   \textcolor{LGcolor}{\llparenthesis}  \Pi_{{\mathrm{2}}}  \textcolor{LGcolor}{\rrparenthesis}   $.

By Lemma~\ref{lemma:substIrrelevant}, type variables are irrelevant in the translation
and thus $ [   \textcolor{LGcolor}{\llparenthesis} \smidge  \SYSTEMnt{B}  \smidge \textcolor{LGcolor}{\rrparenthesis}   /   \beta   ]   [   \textcolor{LGcolor}{\llparenthesis}  \Pi_{{\mathrm{1}}}  \textcolor{LGcolor}{\rrparenthesis}   /  \SYSTEMmv{x}  ]   \textcolor{LGcolor}{\llparenthesis}  \Pi_{{\mathrm{2}}}  \textcolor{LGcolor}{\rrparenthesis}   
=  [   \textcolor{LGcolor}{\llparenthesis}  \Pi_{{\mathrm{1}}}  \textcolor{LGcolor}{\rrparenthesis}   /  \SYSTEMmv{x}  ]   \textcolor{LGcolor}{\llparenthesis}  \Pi_{{\mathrm{2}}}  \textcolor{LGcolor}{\rrparenthesis}  $.

The goal is now that $\Pi' =  [   \textcolor{LGcolor}{\llparenthesis}  \Pi_{{\mathrm{1}}}  \textcolor{LGcolor}{\rrparenthesis}   /  \SYSTEMmv{x}  ]   \textcolor{LGcolor}{\llparenthesis}  \Pi_{{\mathrm{2}}}  \textcolor{LGcolor}{\rrparenthesis}  $
which holds by Lemma~\ref{lemma:substGradPoly}.

\item (congLetL)
\[
\SYSTEMdruleSemLincongLetL{}
\]
with typing:
\[
\SYSTEMdruleLinlet{}
\]
such that $\Pi_{{\mathrm{1}}} :  \Gamma_{{\mathrm{1}}}  \vdash_{\textsc{l} }  \SYSTEMnt{t_{{\mathrm{1}}}}  :   \textcolor{coeffectColor}{\square_{ \SYSTEMnt{r} } }  \SYSTEMnt{A}  $
and $\Pi_{{\mathrm{2}}} :   \Gamma_{{\mathrm{2}}} ,   \SYSTEMmv{x}  : \textcolor{coeffectColor}{[}  \SYSTEMnt{A} {\textcolor{coeffectColor}{]_{ \SYSTEMnt{r} } } }    \vdash_{\textsc{l} }  \SYSTEMnt{t_{{\mathrm{2}}}}  :  \SYSTEMnt{B} $
and
$\Pi' :  \Gamma_{{\mathrm{1}}}  \SYSTEMsym{+}  \Gamma_{{\mathrm{2}}}  \vdash_{\textsc{l} }   \mathsf{let} \, \textcolor{coeffectColor}{[}  \SYSTEMmv{x}  \textcolor{coeffectColor}{]} =  \SYSTEMnt{t'_{{\mathrm{1}}}}  \, \mathsf{in} \,  \SYSTEMnt{t_{{\mathrm{2}}}}   :  \SYSTEMnt{B} $.

The interpretation of the reducing term and its target's proofs are:
\begin{align*}
 \textcolor{LGcolor}{\llparenthesis}  \Pi  \textcolor{LGcolor}{\rrparenthesis}  & \equiv
 \SYSTEMsym{(}    \textcolor{LGcolor}{\llparenthesis}  \Pi_{{\mathrm{1}}}  \textcolor{LGcolor}{\rrparenthesis}   \text{@}   \textcolor{LGcolor}{\llparenthesis} \smidge  \SYSTEMnt{B}  \smidge \textcolor{LGcolor}{\rrparenthesis}    \SYSTEMsym{)} \,  \SYSTEMsym{(}   \lambda  \SYSTEMmv{x}  .   \textcolor{LGcolor}{\llparenthesis}  \Pi_{{\mathrm{2}}}  \textcolor{LGcolor}{\rrparenthesis}    \SYSTEMsym{)}  \\
 \textcolor{LGcolor}{\llparenthesis}  \Pi'  \textcolor{LGcolor}{\rrparenthesis}  & \equiv
 \SYSTEMsym{(}    \textcolor{LGcolor}{\llparenthesis}  \Pi'_{{\mathrm{1}}}  \textcolor{LGcolor}{\rrparenthesis}   \text{@}   \textcolor{LGcolor}{\llparenthesis} \smidge  \SYSTEMnt{B}  \smidge \textcolor{LGcolor}{\rrparenthesis}    \SYSTEMsym{)} \,  \SYSTEMsym{(}   \lambda  \SYSTEMmv{x}  .   \textcolor{LGcolor}{\llparenthesis}  \Pi_{{\mathrm{2}}}  \textcolor{LGcolor}{\rrparenthesis}    \SYSTEMsym{)} 
\end{align*}
since by type preservation with $ \SYSTEMnt{t_{{\mathrm{1}}}}  \rightsquigarrow_{\textsc{l} }  \SYSTEMnt{t'_{{\mathrm{1}}}} $ then
$\Pi'_{{\mathrm{1}}} :  \Gamma_{{\mathrm{1}}}  \vdash_{\textsc{l} }  \SYSTEMnt{t'_{{\mathrm{1}}}}  :   \textcolor{coeffectColor}{\square_{ \SYSTEMnt{r} } }  \SYSTEMnt{A}  $.

By induction on $\Pi_{{\mathrm{1}}} :  \Gamma_{{\mathrm{1}}}  \vdash_{\textsc{l} }  \SYSTEMnt{t_{{\mathrm{1}}}}  :   \textcolor{coeffectColor}{\square_{ \SYSTEMnt{r} } }  \SYSTEMnt{A}  $ with $ \SYSTEMnt{t_{{\mathrm{1}}}}  \rightsquigarrow_{\textsc{l} }  \SYSTEMnt{t'_{{\mathrm{1}}}} $
and $\Pi'_{{\mathrm{1}}} :  \Gamma_{{\mathrm{1}}}  \vdash_{\textsc{l} }  \SYSTEMnt{t'_{{\mathrm{1}}}}  :   \textcolor{coeffectColor}{\square_{ \SYSTEMnt{r} } }  \SYSTEMnt{A}  $, we then have
$  \textcolor{LGcolor}{\llparenthesis}  \Pi_{{\mathrm{1}}}  \textcolor{LGcolor}{\rrparenthesis}   \rightsquigarrow_{\textsc{g} }   \textcolor{LGcolor}{\llparenthesis}  \Pi'_{{\mathrm{1}}}  \textcolor{LGcolor}{\rrparenthesis}  $.

Therefore, we construct
the reduction in Graded Base:
\[
\inferrule*[right=\SYSTEMRenameRuleSemGrdModcongLetL{}]
{  \textcolor{LGcolor}{\llparenthesis}  \Pi_{{\mathrm{1}}}  \textcolor{LGcolor}{\rrparenthesis}   \rightsquigarrow_{\textsc{g} }   \textcolor{LGcolor}{\llparenthesis}  \Pi'_{{\mathrm{1}}}  \textcolor{LGcolor}{\rrparenthesis}  }
{  \SYSTEMsym{(}    \textcolor{LGcolor}{\llparenthesis}  \Pi_{{\mathrm{1}}}  \textcolor{LGcolor}{\rrparenthesis}   \text{@}   \textcolor{LGcolor}{\llparenthesis} \smidge  \SYSTEMnt{B}  \smidge \textcolor{LGcolor}{\rrparenthesis}    \SYSTEMsym{)} \,  \SYSTEMsym{(}   \lambda  \SYSTEMmv{x}  .   \textcolor{LGcolor}{\llparenthesis}  \Pi_{{\mathrm{2}}}  \textcolor{LGcolor}{\rrparenthesis}    \SYSTEMsym{)}   \rightsquigarrow_{\textsc{g} }   \SYSTEMsym{(}    \textcolor{LGcolor}{\llparenthesis}  \Pi'_{{\mathrm{1}}}  \textcolor{LGcolor}{\rrparenthesis}   \text{@}   \textcolor{LGcolor}{\llparenthesis} \smidge  \SYSTEMnt{B}  \smidge \textcolor{LGcolor}{\rrparenthesis}    \SYSTEMsym{)} \,  \SYSTEMsym{(}   \lambda  \SYSTEMmv{x}  .   \textcolor{LGcolor}{\llparenthesis}  \Pi_{{\mathrm{2}}}  \textcolor{LGcolor}{\rrparenthesis}    \SYSTEMsym{)}  }
\]
satisfying the goal.

\end{itemize}

\end{proof}

\subsubsection{Equational correspondence}
\label{app:proofs-lin-to-grad-poly-eqs}

Note, $\eta$-equality for $\Box$ is not preserved as described in the theorem statement.
\begin{proof}
  \begin{itemize}
      \item \[\SYSTEMdruleLinEqbetaBox{}\]
        Follows trivially from preservation of the operational semantics (Theorem~\ref{thrm:linToGradTranslationPoly}) since reduction is a subset of the equational theory.

      \item \[\SYSTEMdruleLinEqletCommBox{}\]
      \begin{align*}
      \begin{array}{rll}
        &  \textcolor{LGcolor}{\llparenthesis} \smidge   \mathsf{let} \, \textcolor{coeffectColor}{[}  \SYSTEMmv{x}  \textcolor{coeffectColor}{]} =   \textcolor{coeffectColor}{[}  \SYSTEMnt{t_{{\mathrm{1}}}}  \textcolor{coeffectColor}{]}   \, \mathsf{in} \,   \textcolor{coeffectColor}{[}  \SYSTEMnt{t_{{\mathrm{2}}}}  \textcolor{coeffectColor}{]}    \smidge \textcolor{LGcolor}{\rrparenthesis}  &  \\
  \textit{\{defn. translation}\} & =  \SYSTEMsym{(}    \textcolor{LGcolor}{\llparenthesis} \smidge   \textcolor{coeffectColor}{[}  \SYSTEMnt{t_{{\mathrm{1}}}}  \textcolor{coeffectColor}{]}   \smidge \textcolor{LGcolor}{\rrparenthesis}   \text{@}   \textcolor{LGcolor}{\llparenthesis} \smidge  \SYSTEMnt{B}  \smidge \textcolor{LGcolor}{\rrparenthesis}    \SYSTEMsym{)} \,  \SYSTEMsym{(}   \lambda  \SYSTEMmv{x}  .   \textcolor{LGcolor}{\llparenthesis} \smidge   \textcolor{coeffectColor}{[}  \SYSTEMnt{t_{{\mathrm{2}}}}  \textcolor{coeffectColor}{]}   \smidge \textcolor{LGcolor}{\rrparenthesis}    \SYSTEMsym{)}  \\
  \textit{\{defn. translation\}} & =  \SYSTEMsym{(}   \SYSTEMsym{(}    \Lambda   \beta'   .   \lambda  \SYSTEMmv{f}  .  \SYSTEMmv{f}   \,   \textcolor{LGcolor}{\llparenthesis} \smidge  \SYSTEMnt{t_{{\mathrm{1}}}}  \smidge \textcolor{LGcolor}{\rrparenthesis}    \SYSTEMsym{)}  \text{@}   \textcolor{LGcolor}{\llparenthesis} \smidge  \SYSTEMnt{B}  \smidge \textcolor{LGcolor}{\rrparenthesis}    \SYSTEMsym{)} \,  \SYSTEMsym{(}    \lambda  \SYSTEMmv{x}  .   \Lambda   \beta   .   \lambda  \SYSTEMmv{f}  .  \SYSTEMmv{f}    \,   \textcolor{LGcolor}{\llparenthesis} \smidge  \SYSTEMnt{t_{{\mathrm{2}}}}  \smidge \textcolor{LGcolor}{\rrparenthesis}    \SYSTEMsym{)}  \\
  \textit{\{\SYSTEMRenameRuleGradPolyEqbetaTy{}}\} & =  \SYSTEMsym{(}   \lambda  \SYSTEMmv{f}  .   [   \textcolor{LGcolor}{\llparenthesis} \smidge  \SYSTEMnt{B}  \smidge \textcolor{LGcolor}{\rrparenthesis}   /   \beta'   ]  \SYSTEMsym{(}   \SYSTEMmv{f} \,   \textcolor{LGcolor}{\llparenthesis} \smidge  \SYSTEMnt{t_{{\mathrm{1}}}}  \smidge \textcolor{LGcolor}{\rrparenthesis}    \SYSTEMsym{)}    \SYSTEMsym{)} \,  \SYSTEMsym{(}    \lambda  \SYSTEMmv{x}  .   \Lambda   \beta   .   \lambda  \SYSTEMmv{f}  .  \SYSTEMmv{f}    \,   \textcolor{LGcolor}{\llparenthesis} \smidge  \SYSTEMnt{t_{{\mathrm{2}}}}  \smidge \textcolor{LGcolor}{\rrparenthesis}    \SYSTEMsym{)}  \\
  \textit{\{\SYSTEMRenameRuleGradEqbeta{}}\} & =  [   \textcolor{LGcolor}{\llparenthesis} \smidge  \SYSTEMnt{B}  \smidge \textcolor{LGcolor}{\rrparenthesis}   /   \beta'   ]  \SYSTEMsym{(}   \SYSTEMsym{(}    \lambda  \SYSTEMmv{x}  .   \Lambda   \beta   .   \lambda  \SYSTEMmv{f}  .  \SYSTEMmv{f}    \,   \textcolor{LGcolor}{\llparenthesis} \smidge  \SYSTEMnt{t_{{\mathrm{2}}}}  \smidge \textcolor{LGcolor}{\rrparenthesis}    \SYSTEMsym{)} \,   \textcolor{LGcolor}{\llparenthesis} \smidge  \SYSTEMnt{t_{{\mathrm{1}}}}  \smidge \textcolor{LGcolor}{\rrparenthesis}    \SYSTEMsym{)}  \\
  \textit{\{\SYSTEMRenameRuleGradEqbeta{}}\} & =  [   \textcolor{LGcolor}{\llparenthesis} \smidge  \SYSTEMnt{B}  \smidge \textcolor{LGcolor}{\rrparenthesis}   /   \beta'   ]  \SYSTEMsym{(}   \Lambda   \beta   .   \lambda  \SYSTEMmv{f}  .   [   \textcolor{LGcolor}{\llparenthesis} \smidge  \SYSTEMnt{t_{{\mathrm{1}}}}  \smidge \textcolor{LGcolor}{\rrparenthesis}   /  \SYSTEMmv{x}  ]  \SYSTEMsym{(}   \SYSTEMmv{f} \,   \textcolor{LGcolor}{\llparenthesis} \smidge  \SYSTEMnt{t_{{\mathrm{2}}}}  \smidge \textcolor{LGcolor}{\rrparenthesis}    \SYSTEMsym{)}     \SYSTEMsym{)}  \\
  \textit{\{defn. subst}\} & =  [   \textcolor{LGcolor}{\llparenthesis} \smidge  \SYSTEMnt{B}  \smidge \textcolor{LGcolor}{\rrparenthesis}   /   \beta'   ]  \SYSTEMsym{(}    \Lambda   \beta   .   \lambda  \SYSTEMmv{f}  .  \SYSTEMmv{f}   \,  \SYSTEMsym{(}   [   \textcolor{LGcolor}{\llparenthesis} \smidge  \SYSTEMnt{t_{{\mathrm{1}}}}  \smidge \textcolor{LGcolor}{\rrparenthesis}   /  \SYSTEMmv{x}  ]   \textcolor{LGcolor}{\llparenthesis} \smidge  \SYSTEMnt{t_{{\mathrm{2}}}}  \smidge \textcolor{LGcolor}{\rrparenthesis}    \SYSTEMsym{)}   \SYSTEMsym{)}  \\
  \textit{\{defn. subst}\} & =   \Lambda   \beta   .   \lambda  \SYSTEMmv{f}  .  \SYSTEMmv{f}   \,  \SYSTEMsym{(}   [   \textcolor{LGcolor}{\llparenthesis} \smidge  \SYSTEMnt{B}  \smidge \textcolor{LGcolor}{\rrparenthesis}   /   \beta'   ]  \SYSTEMsym{(}   [   \textcolor{LGcolor}{\llparenthesis} \smidge  \SYSTEMnt{t_{{\mathrm{1}}}}  \smidge \textcolor{LGcolor}{\rrparenthesis}   /  \SYSTEMmv{x}  ]   \textcolor{LGcolor}{\llparenthesis} \smidge  \SYSTEMnt{t_{{\mathrm{2}}}}  \smidge \textcolor{LGcolor}{\rrparenthesis}    \SYSTEMsym{)}   \SYSTEMsym{)}  \\
  \\
&  \textcolor{LGcolor}{\llparenthesis} \smidge   \textcolor{coeffectColor}{[}   \mathsf{let} \, \textcolor{coeffectColor}{[}  \SYSTEMmv{x}  \textcolor{coeffectColor}{]} =   \textcolor{coeffectColor}{[}  \SYSTEMnt{t_{{\mathrm{1}}}}  \textcolor{coeffectColor}{]}   \, \mathsf{in} \,  \SYSTEMnt{t_{{\mathrm{2}}}}   \textcolor{coeffectColor}{]}   \smidge \textcolor{LGcolor}{\rrparenthesis}  \\
\textit{\{defn. translation\}} & =   \Lambda   \beta   .   \lambda  \SYSTEMmv{f}  .  \SYSTEMmv{f}   \,   \textcolor{LGcolor}{\llparenthesis} \smidge   \mathsf{let} \, \textcolor{coeffectColor}{[}  \SYSTEMmv{x}  \textcolor{coeffectColor}{]} =   \textcolor{coeffectColor}{[}  \SYSTEMnt{t_{{\mathrm{1}}}}  \textcolor{coeffectColor}{]}   \, \mathsf{in} \,  \SYSTEMnt{t_{{\mathrm{2}}}}   \smidge \textcolor{LGcolor}{\rrparenthesis}   \\
\textit{\{defn. translation\}} & =   \Lambda   \beta   .   \lambda  \SYSTEMmv{f}  .  \SYSTEMmv{f}   \,  \SYSTEMsym{(}   \SYSTEMsym{(}   \SYSTEMsym{(}    \Lambda   \beta'   .   \lambda  \SYSTEMmv{f}  .  \SYSTEMmv{f}   \,   \textcolor{LGcolor}{\llparenthesis} \smidge  \SYSTEMnt{t_{{\mathrm{1}}}}  \smidge \textcolor{LGcolor}{\rrparenthesis}    \SYSTEMsym{)}  \text{@}   \textcolor{LGcolor}{\llparenthesis} \smidge  \SYSTEMnt{B}  \smidge \textcolor{LGcolor}{\rrparenthesis}    \SYSTEMsym{)} \,  \SYSTEMsym{(}   \lambda  \SYSTEMmv{x}  .   \textcolor{LGcolor}{\llparenthesis} \smidge  \SYSTEMnt{t_{{\mathrm{2}}}}  \smidge \textcolor{LGcolor}{\rrparenthesis}    \SYSTEMsym{)}   \SYSTEMsym{)}  \\ 
\textit{\{\SYSTEMRenameRuleGradPolyEqbetaTy{}\}} & =   \Lambda   \beta   .   \lambda  \SYSTEMmv{f}  .  \SYSTEMmv{f}   \,  \SYSTEMsym{(}   \SYSTEMsym{(}   \lambda  \SYSTEMmv{f}  .   [   \textcolor{LGcolor}{\llparenthesis} \smidge  \SYSTEMnt{B}  \smidge \textcolor{LGcolor}{\rrparenthesis}   /   \beta'   ]  \SYSTEMsym{(}   \SYSTEMmv{f} \,   \textcolor{LGcolor}{\llparenthesis} \smidge  \SYSTEMnt{t_{{\mathrm{1}}}}  \smidge \textcolor{LGcolor}{\rrparenthesis}    \SYSTEMsym{)}    \SYSTEMsym{)} \,  \SYSTEMsym{(}   \lambda  \SYSTEMmv{x}  .   \textcolor{LGcolor}{\llparenthesis} \smidge  \SYSTEMnt{t_{{\mathrm{2}}}}  \smidge \textcolor{LGcolor}{\rrparenthesis}    \SYSTEMsym{)}   \SYSTEMsym{)}  \\ 
\textit{\{\SYSTEMRenameRuleGradEqbeta{}\}} & =   \Lambda   \beta   .   \lambda  \SYSTEMmv{f}  .  \SYSTEMmv{f}   \,  \SYSTEMsym{(}   [   \textcolor{LGcolor}{\llparenthesis} \smidge  \SYSTEMnt{B}  \smidge \textcolor{LGcolor}{\rrparenthesis}   /   \beta'   ]  \SYSTEMsym{(}   \SYSTEMsym{(}   \lambda  \SYSTEMmv{x}  .   \textcolor{LGcolor}{\llparenthesis} \smidge  \SYSTEMnt{t_{{\mathrm{2}}}}  \smidge \textcolor{LGcolor}{\rrparenthesis}    \SYSTEMsym{)} \,   \textcolor{LGcolor}{\llparenthesis} \smidge  \SYSTEMnt{t_{{\mathrm{1}}}}  \smidge \textcolor{LGcolor}{\rrparenthesis}    \SYSTEMsym{)}   \SYSTEMsym{)}  \\ 
\textit{\{\SYSTEMRenameRuleGradEqbeta{}\}} & =   \Lambda   \beta   .   \lambda  \SYSTEMmv{f}  .  \SYSTEMmv{f}   \,  \SYSTEMsym{(}   [   \textcolor{LGcolor}{\llparenthesis} \smidge  \SYSTEMnt{B}  \smidge \textcolor{LGcolor}{\rrparenthesis}   /   \beta'   ]  \SYSTEMsym{(}   [   \textcolor{LGcolor}{\llparenthesis} \smidge  \SYSTEMnt{t_{{\mathrm{1}}}}  \smidge \textcolor{LGcolor}{\rrparenthesis}   /  \SYSTEMmv{x}  ]   \textcolor{LGcolor}{\llparenthesis} \smidge  \SYSTEMnt{t_{{\mathrm{2}}}}  \smidge \textcolor{LGcolor}{\rrparenthesis}    \SYSTEMsym{)}   \SYSTEMsym{)}  \\ 
\end{array}
      \end{align*}
      i.e., both terms are equal to the same common term under $\beta$-equalities and the
      definitions of the translation and substitution.

      \item \[\SYSTEMdruleLinEqletCommOne{}\]

            \begin{align*}
      \begin{array}{rll}
        &  \textcolor{LGcolor}{\llparenthesis} \smidge   \SYSTEMnt{t_{{\mathrm{0}}}} \,  \SYSTEMsym{(}   \mathsf{let} \, \textcolor{coeffectColor}{[}  \SYSTEMmv{x}  \textcolor{coeffectColor}{]} =  \SYSTEMnt{t_{{\mathrm{1}}}}  \, \mathsf{in} \,  \SYSTEMnt{t_{{\mathrm{2}}}}   \SYSTEMsym{)}   \smidge \textcolor{LGcolor}{\rrparenthesis}  &  \\
        \textit{\{defn. translation\}} & =   \textcolor{LGcolor}{\llparenthesis} \smidge  \SYSTEMnt{t_{{\mathrm{0}}}}  \smidge \textcolor{LGcolor}{\rrparenthesis}  \,  \SYSTEMsym{(}   \SYSTEMsym{(}    \textcolor{LGcolor}{\llparenthesis} \smidge  \SYSTEMnt{t_{{\mathrm{1}}}}  \smidge \textcolor{LGcolor}{\rrparenthesis}   \text{@}   \textcolor{LGcolor}{\llparenthesis} \smidge  \SYSTEMnt{B}  \smidge \textcolor{LGcolor}{\rrparenthesis}    \SYSTEMsym{)} \,  \SYSTEMsym{(}   \lambda  \SYSTEMmv{x}  .   \textcolor{LGcolor}{\llparenthesis} \smidge  \SYSTEMnt{t_{{\mathrm{2}}}}  \smidge \textcolor{LGcolor}{\rrparenthesis}    \SYSTEMsym{)}   \SYSTEMsym{)}  \\
        \textit{$\eta$-reasoning} & = \SYSTEMsym{(}   \SYSTEMsym{(}    \textcolor{LGcolor}{\llparenthesis} \smidge  \SYSTEMnt{t_{{\mathrm{1}}}}  \smidge \textcolor{LGcolor}{\rrparenthesis}   \text{@}   \textcolor{LGcolor}{\llparenthesis} \smidge  \SYSTEMnt{B}  \smidge \textcolor{LGcolor}{\rrparenthesis}    \SYSTEMsym{)} \,  \SYSTEMsym{(}    \lambda  \SYSTEMmv{x}  .   \textcolor{LGcolor}{\llparenthesis} \smidge  \SYSTEMnt{t_{{\mathrm{0}}}}  \smidge \textcolor{LGcolor}{\rrparenthesis}   \,   \textcolor{LGcolor}{\llparenthesis} \smidge  \SYSTEMnt{t_{{\mathrm{2}}}}  \smidge \textcolor{LGcolor}{\rrparenthesis}    \SYSTEMsym{)}   \SYSTEMsym{)} \\
        \textit{\{defn. translation\}} & =  \textcolor{LGcolor}{\llparenthesis} \smidge    \mathsf{let} \, \textcolor{coeffectColor}{[}  \SYSTEMmv{x}  \textcolor{coeffectColor}{]} =  \SYSTEMnt{t_{{\mathrm{1}}}}  \, \mathsf{in} \,  \SYSTEMnt{t_{{\mathrm{0}}}}  \,  \SYSTEMnt{t_{{\mathrm{2}}}}   \smidge \textcolor{LGcolor}{\rrparenthesis} 
      \end{array}
            \end{align*}

      \item \[\SYSTEMdruleLinEqletCommTwo{}\]

            \begin{align*}
      \begin{array}{rll}
        &  \textcolor{LGcolor}{\llparenthesis} \smidge   \SYSTEMsym{(}   \mathsf{let} \, \textcolor{coeffectColor}{[}  \SYSTEMmv{x}  \textcolor{coeffectColor}{]} =  \SYSTEMnt{t_{{\mathrm{1}}}}  \, \mathsf{in} \,  \SYSTEMnt{t_{{\mathrm{2}}}}   \SYSTEMsym{)} \,  \SYSTEMnt{t_{{\mathrm{0}}}}   \smidge \textcolor{LGcolor}{\rrparenthesis}  &  \\
        \textit{\{defn. translation\}} & =  \SYSTEMsym{(}   \SYSTEMsym{(}    \textcolor{LGcolor}{\llparenthesis} \smidge  \SYSTEMnt{t_{{\mathrm{1}}}}  \smidge \textcolor{LGcolor}{\rrparenthesis}   \text{@}   \textcolor{LGcolor}{\llparenthesis} \smidge  \SYSTEMnt{B}  \smidge \textcolor{LGcolor}{\rrparenthesis}    \SYSTEMsym{)} \,  \SYSTEMsym{(}   \lambda  \SYSTEMmv{x}  .   \textcolor{LGcolor}{\llparenthesis} \smidge  \SYSTEMnt{t_{{\mathrm{2}}}}  \smidge \textcolor{LGcolor}{\rrparenthesis}    \SYSTEMsym{)}   \SYSTEMsym{)} \,   \textcolor{LGcolor}{\llparenthesis} \smidge  \SYSTEMnt{t_{{\mathrm{0}}}}  \smidge \textcolor{LGcolor}{\rrparenthesis}   \\
        \textit{$\eta$-reasoning} & = \SYSTEMsym{(}   \SYSTEMsym{(}    \textcolor{LGcolor}{\llparenthesis} \smidge  \SYSTEMnt{t_{{\mathrm{1}}}}  \smidge \textcolor{LGcolor}{\rrparenthesis}   \text{@}   \textcolor{LGcolor}{\llparenthesis} \smidge  \SYSTEMnt{B}  \smidge \textcolor{LGcolor}{\rrparenthesis}    \SYSTEMsym{)} \,  \SYSTEMsym{(}    \lambda  \SYSTEMmv{x}  .   \textcolor{LGcolor}{\llparenthesis} \smidge  \SYSTEMnt{t_{{\mathrm{2}}}}  \smidge \textcolor{LGcolor}{\rrparenthesis}   \,   \textcolor{LGcolor}{\llparenthesis} \smidge  \SYSTEMnt{t_{{\mathrm{0}}}}  \smidge \textcolor{LGcolor}{\rrparenthesis}    \SYSTEMsym{)}   \SYSTEMsym{)} \\
        \textit{\{defn. translation\}} & =  \textcolor{LGcolor}{\llparenthesis} \smidge    \mathsf{let} \, \textcolor{coeffectColor}{[}  \SYSTEMmv{x}  \textcolor{coeffectColor}{]} =  \SYSTEMnt{t_{{\mathrm{1}}}}  \, \mathsf{in} \,  \SYSTEMnt{t_{{\mathrm{2}}}}  \,  \SYSTEMnt{t_{{\mathrm{0}}}}   \smidge \textcolor{LGcolor}{\rrparenthesis} 
      \end{array}
            \end{align*}

    \item The congruences follow by induction and congruences in the target.
  \end{itemize}
    \end{proof}

\subsection{Proof of Soundness for Graded Modal Core to Linear Push Core}
\label{app:proofs-grad-core-to-lin-core}

\ifextended\gradToLinTranslationExt*\fi

\subsubsection{Type preservation}
\label{app:proofs-grad-core-to-line-core-typ}

\begin{proof}
  The proof subsumes the earlier proofs; we add the additional cases here due to the
  extension of the language from Graded Modal Base to Graded Modal Core.
  \begin{itemize}
\item (prod$_i$)
$$
\SYSTEMdruleGradprodi{}
$$
$$
\SYSTEMdruleLinprodi{}
$$
By induction on both premises we have
$  \textcolor{GLcolor}{\llbracket}  \Delta_{{\mathrm{1}}}  \textcolor{GLcolor}{\rrbracket}   \vdash_{\textsc{l} }   \textcolor{GLcolor}{\llbracket}  \SYSTEMnt{t_{{\mathrm{1}}}}  \textcolor{GLcolor}{\rrbracket}   :   \textcolor{GLcolor}{\llbracket}  \SYSTEMnt{A}  \textcolor{GLcolor}{\rrbracket}  $
and
$  \textcolor{GLcolor}{\llbracket}  \Delta_{{\mathrm{2}}}  \textcolor{GLcolor}{\rrbracket}   \vdash_{\textsc{l} }   \textcolor{GLcolor}{\llbracket}  \SYSTEMnt{t_{{\mathrm{2}}}}  \textcolor{GLcolor}{\rrbracket}   :   \textcolor{GLcolor}{\llbracket}  \SYSTEMnt{B}  \textcolor{GLcolor}{\rrbracket}  $.

Therefore we can construct:
$$
\inferrule*[Right=\SYSTEMRenameRuleLinprodi{}]
{  \textcolor{GLcolor}{\llbracket}  \Delta_{{\mathrm{1}}}  \textcolor{GLcolor}{\rrbracket}   \vdash_{\textsc{l} }   \textcolor{GLcolor}{\llbracket}  \SYSTEMnt{t_{{\mathrm{1}}}}  \textcolor{GLcolor}{\rrbracket}   :   \textcolor{GLcolor}{\llbracket}  \SYSTEMnt{A}  \textcolor{GLcolor}{\rrbracket}   \\
   \textcolor{GLcolor}{\llbracket}  \Delta_{{\mathrm{2}}}  \textcolor{GLcolor}{\rrbracket}   \vdash_{\textsc{l} }   \textcolor{GLcolor}{\llbracket}  \SYSTEMnt{t_{{\mathrm{2}}}}  \textcolor{GLcolor}{\rrbracket}   :   \textcolor{GLcolor}{\llbracket}  \SYSTEMnt{B}  \textcolor{GLcolor}{\rrbracket}  }
{  \textcolor{GLcolor}{\llbracket}  \Delta_{{\mathrm{1}}}  \textcolor{GLcolor}{\rrbracket}   \SYSTEMsym{+}   \textcolor{GLcolor}{\llbracket}  \Delta_{{\mathrm{2}}}  \textcolor{GLcolor}{\rrbracket}   \vdash_{\textsc{l} }   \langle   \textcolor{GLcolor}{\llbracket}  \SYSTEMnt{t_{{\mathrm{1}}}}  \textcolor{GLcolor}{\rrbracket}  ,   \textcolor{GLcolor}{\llbracket}  \SYSTEMnt{t_{{\mathrm{2}}}}  \textcolor{GLcolor}{\rrbracket}   \rangle   :    \textcolor{GLcolor}{\llbracket}  \SYSTEMnt{A}  \textcolor{GLcolor}{\rrbracket}   \otimes   \textcolor{GLcolor}{\llbracket}  \SYSTEMnt{B}  \textcolor{GLcolor}{\rrbracket}    }
$$

\item (prod$_e$)
$$
\SYSTEMdruleGradprode{}
$$
% $$
% \SYSTEMdruleLinprode{}
% $$
By induction on both premises we have
$  \textcolor{GLcolor}{\llbracket}  \Delta_{{\mathrm{1}}}  \textcolor{GLcolor}{\rrbracket}   \vdash_{\textsc{l} }   \textcolor{GLcolor}{\llbracket}  \SYSTEMnt{t_{{\mathrm{1}}}}  \textcolor{GLcolor}{\rrbracket}   :    \textcolor{GLcolor}{\llbracket}  \SYSTEMnt{A}  \textcolor{GLcolor}{\rrbracket}   \otimes   \textcolor{GLcolor}{\llbracket}  \SYSTEMnt{B}  \textcolor{GLcolor}{\rrbracket}   $
and
$    \textcolor{GLcolor}{\llbracket}  \Delta_{{\mathrm{2}}}  \textcolor{GLcolor}{\rrbracket}  ,   \SYSTEMmv{x}  : \textcolor{coeffectColor}{[}   \textcolor{GLcolor}{\llbracket}  \SYSTEMnt{A}  \textcolor{GLcolor}{\rrbracket}  {\textcolor{coeffectColor}{]_{ \SYSTEMsym{1} } } }   ,   \SYSTEMmv{y}  : \textcolor{coeffectColor}{[}   \textcolor{GLcolor}{\llbracket}  \SYSTEMnt{B}  \textcolor{GLcolor}{\rrbracket}  {\textcolor{coeffectColor}{]_{ \SYSTEMsym{1} } } }    \vdash_{\textsc{l} }   \textcolor{GLcolor}{\llbracket}  \SYSTEMnt{t_{{\mathrm{2}}}}  \textcolor{GLcolor}{\rrbracket}   :   \textcolor{GLcolor}{\llbracket}  \SYSTEMnt{C}  \textcolor{GLcolor}{\rrbracket}  $.

Therefore we can construct with $ \SYSTEMmv{x'} ,   \SYSTEMmv{y'}  \,\#\,  \SYSTEMnt{t_{{\mathrm{2}}}}  $:

\begin{gather*}
    \inferrule*[Right=\SYSTEMRenameRuleLinprode{}]
      {
        \inferrule*[Right=\SYSTEMRenameRuleLinpushprod{}, vdots=6em, rightskip=15em]
          {
            \inferrule*[Right=\SYSTEMRenameRuleLinpr{}]
              {
                \inferrule*
                {\textit{ih}}
                {  \textcolor{GLcolor}{\llbracket}  \Delta_{{\mathrm{1}}}  \textcolor{GLcolor}{\rrbracket}   \vdash_{\textsc{l} }   \textcolor{GLcolor}{\llbracket}  \SYSTEMnt{t_{{\mathrm{1}}}}  \textcolor{GLcolor}{\rrbracket}   :    \textcolor{GLcolor}{\llbracket}  \SYSTEMnt{A}  \textcolor{GLcolor}{\rrbracket}   \otimes   \textcolor{GLcolor}{\llbracket}  \SYSTEMnt{B}  \textcolor{GLcolor}{\rrbracket}   }
              }
              {   \textcolor{coeffectColor}{ \SYSTEMnt{r}  \cdot}   \textcolor{GLcolor}{\llbracket}  \Delta_{{\mathrm{1}}}  \textcolor{GLcolor}{\rrbracket}    \vdash_{\textsc{l} }   \textcolor{coeffectColor}{[}   \textcolor{GLcolor}{\llbracket}  \SYSTEMnt{t_{{\mathrm{1}}}}  \textcolor{GLcolor}{\rrbracket}   \textcolor{coeffectColor}{]}   :   \textcolor{coeffectColor}{\square_{ \SYSTEMnt{r} } }   (    \textcolor{GLcolor}{\llbracket}  \SYSTEMnt{A}  \textcolor{GLcolor}{\rrbracket}   \otimes   \textcolor{GLcolor}{\llbracket}  \SYSTEMnt{B}  \textcolor{GLcolor}{\rrbracket}    )   }
          }
          {  \textcolor{coeffectColor}{ \SYSTEMnt{r}  \cdot}   \textcolor{GLcolor}{\llbracket}  \Delta_{{\mathrm{1}}}  \textcolor{GLcolor}{\rrbracket}    \vdash_{\textsc{l} }   \textsf{push}_\otimes   \textcolor{coeffectColor}{[}   \textcolor{GLcolor}{\llbracket}  \SYSTEMnt{t_{{\mathrm{1}}}}  \textcolor{GLcolor}{\rrbracket}   \textcolor{coeffectColor}{]}    :    \textcolor{coeffectColor}{\square_{ \SYSTEMnt{r} } }   \textcolor{GLcolor}{\llbracket}  \SYSTEMnt{A}  \textcolor{GLcolor}{\rrbracket}    \otimes   \textcolor{coeffectColor}{\square_{ \SYSTEMnt{r} } }   \textcolor{GLcolor}{\llbracket}  \SYSTEMnt{B}  \textcolor{GLcolor}{\rrbracket}    }
          \inferrule*[Right=\SYSTEMRenameRuleLinlet{}, width=300pt]
            {
              \inferrule*[Right=\SYSTEMRenameRuleLinvar{}]{ }{   \SYSTEMmv{x'}  :   \textcolor{coeffectColor}{\square_{ \SYSTEMnt{r} } }   \textcolor{GLcolor}{\llbracket}  \SYSTEMnt{A}  \textcolor{GLcolor}{\rrbracket}      \vdash_{\textsc{l} }  \SYSTEMmv{x'}  :   \textcolor{coeffectColor}{\square_{ \SYSTEMnt{r} } }   \textcolor{GLcolor}{\llbracket}  \SYSTEMnt{A}  \textcolor{GLcolor}{\rrbracket}   }
              \hspace{4em}
              \inferrule*[Right=\SYSTEMRenameRuleLinlet{}]
                {
                  \inferrule*[Right=\SYSTEMRenameRuleLinvar{}]{ }{   \SYSTEMmv{y}  :   \textcolor{coeffectColor}{\square_{ \SYSTEMnt{r} } }   \textcolor{GLcolor}{\llbracket}  \SYSTEMnt{B}  \textcolor{GLcolor}{\rrbracket}      \vdash_{\textsc{l} }  \SYSTEMmv{y}  :   \textcolor{coeffectColor}{\square_{ \SYSTEMnt{r} } }   \textcolor{GLcolor}{\llbracket}  \SYSTEMnt{B}  \textcolor{GLcolor}{\rrbracket}   }
                  \hspace{4em}
                  \inferrule*
                    { \textit{ih} }
                    {    \textcolor{GLcolor}{\llbracket}  \Delta_{{\mathrm{2}}}  \textcolor{GLcolor}{\rrbracket}  ,   \SYSTEMmv{x}  : \textcolor{coeffectColor}{[}   \textcolor{GLcolor}{\llbracket}  \SYSTEMnt{A}  \textcolor{GLcolor}{\rrbracket}  {\textcolor{coeffectColor}{]_{ \SYSTEMnt{r} } } }   ,   \SYSTEMmv{y}  : \textcolor{coeffectColor}{[}   \textcolor{GLcolor}{\llbracket}  \SYSTEMnt{B}  \textcolor{GLcolor}{\rrbracket}  {\textcolor{coeffectColor}{]_{ \SYSTEMnt{r} } } }    \vdash_{\textsc{l} }   \textcolor{GLcolor}{\llbracket}  \SYSTEMnt{t_{{\mathrm{2}}}}  \textcolor{GLcolor}{\rrbracket}   :   \textcolor{GLcolor}{\llbracket}  \SYSTEMnt{C}  \textcolor{GLcolor}{\rrbracket}  }
                }
                {   \textcolor{GLcolor}{\llbracket}  \Delta_{{\mathrm{2}}}  \textcolor{GLcolor}{\rrbracket}  ,   \SYSTEMmv{x}  : \textcolor{coeffectColor}{[}   \textcolor{GLcolor}{\llbracket}  \SYSTEMnt{A}  \textcolor{GLcolor}{\rrbracket}  {\textcolor{coeffectColor}{]_{ \SYSTEMnt{r} } } }    \vdash_{\textsc{l} }   \mathsf{let} \, \textcolor{coeffectColor}{[}  \SYSTEMmv{y}  \textcolor{coeffectColor}{]} =  \SYSTEMmv{y'}  \, \mathsf{in} \,   \textcolor{GLcolor}{\llbracket}  \SYSTEMnt{t_{{\mathrm{2}}}}  \textcolor{GLcolor}{\rrbracket}    :   \textcolor{GLcolor}{\llbracket}  \SYSTEMnt{C}  \textcolor{GLcolor}{\rrbracket}  }
            }
            {     \textcolor{GLcolor}{\llbracket}  \Delta_{{\mathrm{2}}}  \textcolor{GLcolor}{\rrbracket}  ,   \SYSTEMmv{x'}  :   \textcolor{coeffectColor}{\square_{ \SYSTEMnt{r} } }   \textcolor{GLcolor}{\llbracket}  \SYSTEMnt{A}  \textcolor{GLcolor}{\rrbracket}     ,   \SYSTEMmv{y}  :   \textcolor{coeffectColor}{\square_{ \SYSTEMnt{r} } }   \textcolor{GLcolor}{\llbracket}  \SYSTEMnt{B}  \textcolor{GLcolor}{\rrbracket}      \vdash_{\textsc{l} }   \mathsf{let} \, \textcolor{coeffectColor}{[}  \SYSTEMmv{x}  \textcolor{coeffectColor}{]} =  \SYSTEMmv{x'}  \, \mathsf{in} \,   \mathsf{let} \, \textcolor{coeffectColor}{[}  \SYSTEMmv{y}  \textcolor{coeffectColor}{]} =  \SYSTEMmv{y'}  \, \mathsf{in} \,   \textcolor{GLcolor}{\llbracket}  \SYSTEMnt{t_{{\mathrm{2}}}}  \textcolor{GLcolor}{\rrbracket}     :   \textcolor{GLcolor}{\llbracket}  \SYSTEMnt{C}  \textcolor{GLcolor}{\rrbracket}   }
      }
      {  \textcolor{coeffectColor}{ \SYSTEMnt{r}  \cdot}   \textcolor{GLcolor}{\llbracket}  \Delta_{{\mathrm{1}}}  \textcolor{GLcolor}{\rrbracket}    \SYSTEMsym{+}   \textcolor{GLcolor}{\llbracket}  \Delta_{{\mathrm{2}}}  \textcolor{GLcolor}{\rrbracket}   \vdash_{\textsc{l} }   \mathsf{let} \, \langle  \SYSTEMmv{x'} ,  \SYSTEMmv{y'}  \rangle =   \textsf{push}_\otimes   \textcolor{coeffectColor}{[}   \textcolor{GLcolor}{\llbracket}  \SYSTEMnt{t_{{\mathrm{1}}}}  \textcolor{GLcolor}{\rrbracket}   \textcolor{coeffectColor}{]}    \, \mathsf{in} \,   \mathsf{let} \, \textcolor{coeffectColor}{[}  \SYSTEMmv{x}  \textcolor{coeffectColor}{]} =  \SYSTEMmv{x'}  \, \mathsf{in} \,   \mathsf{let} \, \textcolor{coeffectColor}{[}  \SYSTEMmv{y}  \textcolor{coeffectColor}{]} =  \SYSTEMmv{y'}  \, \mathsf{in} \,   \textcolor{GLcolor}{\llbracket}  \SYSTEMnt{t_{{\mathrm{2}}}}  \textcolor{GLcolor}{\rrbracket}      :   \textcolor{GLcolor}{\llbracket}  \SYSTEMnt{C}  \textcolor{GLcolor}{\rrbracket}   }
\end{gather*}

\item (unit$_i$)
$$
\SYSTEMdruleGraduniti{}
$$

Therefore we construct the goal typing:
$$
\SYSTEMdruleLinuniti{}
$$

\item (unit$_e$)
$$
\SYSTEMdruleGradunite{}
$$
$$
\SYSTEMdruleLinunite{}
$$
By induction on both premises we have
$  \textcolor{GLcolor}{\llbracket}  \Delta_{{\mathrm{1}}}  \textcolor{GLcolor}{\rrbracket}   \vdash_{\textsc{l} }   \textcolor{GLcolor}{\llbracket}  \SYSTEMnt{t_{{\mathrm{1}}}}  \textcolor{GLcolor}{\rrbracket}   :   \textcolor{GLcolor}{\llbracket}   \mathrm{unit}   \textcolor{GLcolor}{\rrbracket}  $
and
$  \textcolor{GLcolor}{\llbracket}  \Delta_{{\mathrm{2}}}  \textcolor{GLcolor}{\rrbracket}   \vdash_{\textsc{l} }   \textcolor{GLcolor}{\llbracket}  \SYSTEMnt{t_{{\mathrm{2}}}}  \textcolor{GLcolor}{\rrbracket}   :   \textcolor{GLcolor}{\llbracket}  \SYSTEMnt{A}  \textcolor{GLcolor}{\rrbracket}  $.

Therefore we can construct:
$$
\inferrule*[Right=\SYSTEMRenameRuleLinprode{}]
{  \textcolor{GLcolor}{\llbracket}  \Delta_{{\mathrm{1}}}  \textcolor{GLcolor}{\rrbracket}   \vdash_{\textsc{l} }   \textcolor{GLcolor}{\llbracket}  \SYSTEMnt{t_{{\mathrm{1}}}}  \textcolor{GLcolor}{\rrbracket}   :   \textcolor{GLcolor}{\llbracket}   \mathrm{unit}   \textcolor{GLcolor}{\rrbracket}  \\  \textcolor{GLcolor}{\llbracket}  \Delta_{{\mathrm{2}}}  \textcolor{GLcolor}{\rrbracket}   \vdash_{\textsc{l} }   \textcolor{GLcolor}{\llbracket}  \SYSTEMnt{t_{{\mathrm{2}}}}  \textcolor{GLcolor}{\rrbracket}   :   \textcolor{GLcolor}{\llbracket}  \SYSTEMnt{A}  \textcolor{GLcolor}{\rrbracket}  }
{  \textcolor{GLcolor}{\llbracket}  \Delta_{{\mathrm{1}}}  \textcolor{GLcolor}{\rrbracket}   \SYSTEMsym{+}   \textcolor{GLcolor}{\llbracket}  \Delta_{{\mathrm{2}}}  \textcolor{GLcolor}{\rrbracket}   \vdash_{\textsc{l} }   \mathsf{let} \, \langle \rangle =   \textcolor{GLcolor}{\llbracket}  \SYSTEMnt{t_{{\mathrm{1}}}}  \textcolor{GLcolor}{\rrbracket}   \, \mathsf{in} \,   \textcolor{GLcolor}{\llbracket}  \SYSTEMnt{t_{{\mathrm{2}}}}  \textcolor{GLcolor}{\rrbracket}    :   \textcolor{GLcolor}{\llbracket}  \SYSTEMnt{A}  \textcolor{GLcolor}{\rrbracket}   }
$$

\item (sum$_{i1}$)
$$
\SYSTEMdruleGradsumiOne{}
$$
% $$
% \SYSTEMdruleLinsumiOne{}
% $$
By induction on the premise we have
$  \textcolor{GLcolor}{\llbracket}  \Delta  \textcolor{GLcolor}{\rrbracket}   \vdash_{\textsc{l} }   \textcolor{GLcolor}{\llbracket}  \SYSTEMnt{t}  \textcolor{GLcolor}{\rrbracket}   :   \textcolor{GLcolor}{\llbracket}  \SYSTEMnt{A}  \textcolor{GLcolor}{\rrbracket}  $.

Therefore we can construct:
$$
\inferrule*[Right=\SYSTEMRenameRuleLinsumiOne{}]
{  \textcolor{GLcolor}{\llbracket}  \Delta  \textcolor{GLcolor}{\rrbracket}   \vdash_{\textsc{l} }   \textcolor{GLcolor}{\llbracket}  \SYSTEMnt{t}  \textcolor{GLcolor}{\rrbracket}   :   \textcolor{GLcolor}{\llbracket}  \SYSTEMnt{A}  \textcolor{GLcolor}{\rrbracket}  }
{  \textcolor{GLcolor}{\llbracket}  \Delta  \textcolor{GLcolor}{\rrbracket}   \vdash_{\textsc{l} }   \mathsf{inj}_1 \,   \textcolor{GLcolor}{\llbracket}  \SYSTEMnt{t}  \textcolor{GLcolor}{\rrbracket}    :    \textcolor{GLcolor}{\llbracket}  \SYSTEMnt{A}  \textcolor{GLcolor}{\rrbracket}   +   \textcolor{GLcolor}{\llbracket}  \SYSTEMnt{B}  \textcolor{GLcolor}{\rrbracket}    }
$$

\item (sum$_{i2}$)
$$
\SYSTEMdruleGradsumiTwo{}
$$
% $$
% \SYSTEMdruleLinsumiTwo{}
% $$
By induction on the premise we have
$  \textcolor{GLcolor}{\llbracket}  \Delta  \textcolor{GLcolor}{\rrbracket}   \vdash_{\textsc{l} }   \textcolor{GLcolor}{\llbracket}  \SYSTEMnt{t}  \textcolor{GLcolor}{\rrbracket}   :   \textcolor{GLcolor}{\llbracket}  \SYSTEMnt{B}  \textcolor{GLcolor}{\rrbracket}  $.

Therefore we can construct:
$$
\inferrule*[Right=\SYSTEMRenameRuleLinsumiTwo{}]
{  \textcolor{GLcolor}{\llbracket}  \Delta  \textcolor{GLcolor}{\rrbracket}   \vdash_{\textsc{l} }   \textcolor{GLcolor}{\llbracket}  \SYSTEMnt{t}  \textcolor{GLcolor}{\rrbracket}   :   \textcolor{GLcolor}{\llbracket}  \SYSTEMnt{B}  \textcolor{GLcolor}{\rrbracket}  }
{  \textcolor{GLcolor}{\llbracket}  \Delta  \textcolor{GLcolor}{\rrbracket}   \vdash_{\textsc{l} }   \mathsf{inj}_2 \,   \textcolor{GLcolor}{\llbracket}  \SYSTEMnt{t}  \textcolor{GLcolor}{\rrbracket}    :    \textcolor{GLcolor}{\llbracket}  \SYSTEMnt{A}  \textcolor{GLcolor}{\rrbracket}   +   \textcolor{GLcolor}{\llbracket}  \SYSTEMnt{B}  \textcolor{GLcolor}{\rrbracket}    }
$$

\item (sum$_e$)
$$
\SYSTEMdruleGradsume{}
$$
% $$
% \SYSTEMdruleLinsume{}
% $$
By induction on the premises we have
$  \textcolor{GLcolor}{\llbracket}  \Delta_{{\mathrm{1}}}  \textcolor{GLcolor}{\rrbracket}   \vdash_{\textsc{l} }   \textcolor{GLcolor}{\llbracket}  \SYSTEMnt{t}  \textcolor{GLcolor}{\rrbracket}   :    \textcolor{GLcolor}{\llbracket}  \SYSTEMnt{A}  \textcolor{GLcolor}{\rrbracket}   \oplus   \textcolor{GLcolor}{\llbracket}  \SYSTEMnt{B}  \textcolor{GLcolor}{\rrbracket}   $,
$   \textcolor{GLcolor}{\llbracket}  \Delta_{{\mathrm{2}}}  \textcolor{GLcolor}{\rrbracket}  ,   \SYSTEMmv{x}  : \textcolor{coeffectColor}{[}   \textcolor{GLcolor}{\llbracket}  \SYSTEMnt{A}  \textcolor{GLcolor}{\rrbracket}  {\textcolor{coeffectColor}{]_{ \SYSTEMnt{r} } } }    \vdash_{\textsc{l} }   \textcolor{GLcolor}{\llbracket}  \SYSTEMnt{t_{{\mathrm{1}}}}  \textcolor{GLcolor}{\rrbracket}   :   \textcolor{GLcolor}{\llbracket}  \SYSTEMnt{C}  \textcolor{GLcolor}{\rrbracket}  $
and
$   \textcolor{GLcolor}{\llbracket}  \Delta_{{\mathrm{2}}}  \textcolor{GLcolor}{\rrbracket}  ,   \SYSTEMmv{y}  : \textcolor{coeffectColor}{[}   \textcolor{GLcolor}{\llbracket}  \SYSTEMnt{B}  \textcolor{GLcolor}{\rrbracket}  {\textcolor{coeffectColor}{]_{ \SYSTEMnt{r} } } }    \vdash_{\textsc{l} }   \textcolor{GLcolor}{\llbracket}  \SYSTEMnt{t_{{\mathrm{2}}}}  \textcolor{GLcolor}{\rrbracket}   :   \textcolor{GLcolor}{\llbracket}  \SYSTEMnt{C}  \textcolor{GLcolor}{\rrbracket}  $.

Therefore we can construct with $ \SYSTEMmv{x'}  \,\#\,  \SYSTEMnt{t_{{\mathrm{1}}}} $ and $ \SYSTEMmv{y'}  \,\#\,  \SYSTEMnt{t_{{\mathrm{2}}}} $:
\begin{gather*}
\inferrule*[Right=\SYSTEMRenameRuleLinsume{}]
{
  \inferrule*[Right=\SYSTEMRenameRuleLinpushprod{}, vdots=5em, rightskip=15em]
    {
      \inferrule*[Right=\SYSTEMRenameRuleLinpr{}]
        {
          \inferrule*
          {\textit{ih}}
          {  \textcolor{GLcolor}{\llbracket}  \Delta_{{\mathrm{1}}}  \textcolor{GLcolor}{\rrbracket}   \vdash_{\textsc{l} }   \textcolor{GLcolor}{\llbracket}  \SYSTEMnt{t}  \textcolor{GLcolor}{\rrbracket}   :    \textcolor{GLcolor}{\llbracket}  \SYSTEMnt{A}  \textcolor{GLcolor}{\rrbracket}   \oplus   \textcolor{GLcolor}{\llbracket}  \SYSTEMnt{B}  \textcolor{GLcolor}{\rrbracket}   }
        }
        {   \textcolor{coeffectColor}{ \SYSTEMnt{r}  \cdot}   \textcolor{GLcolor}{\llbracket}  \Delta_{{\mathrm{1}}}  \textcolor{GLcolor}{\rrbracket}    \vdash_{\textsc{l} }   \textcolor{coeffectColor}{[}   \textcolor{GLcolor}{\llbracket}  \SYSTEMnt{t}  \textcolor{GLcolor}{\rrbracket}   \textcolor{coeffectColor}{]}   :   \textcolor{coeffectColor}{\square_{ \SYSTEMnt{r} } }   (    \textcolor{GLcolor}{\llbracket}  \SYSTEMnt{A}  \textcolor{GLcolor}{\rrbracket}   \oplus   \textcolor{GLcolor}{\llbracket}  \SYSTEMnt{B}  \textcolor{GLcolor}{\rrbracket}    )   }
    }
    {  \textcolor{coeffectColor}{ \SYSTEMnt{r}  \cdot}   \textcolor{GLcolor}{\llbracket}  \Delta_{{\mathrm{1}}}  \textcolor{GLcolor}{\rrbracket}    \vdash_{\textsc{l} }   \textsf{push}_\oplus   \textcolor{coeffectColor}{[}   \textcolor{GLcolor}{\llbracket}  \SYSTEMnt{t}  \textcolor{GLcolor}{\rrbracket}   \textcolor{coeffectColor}{]}    :    \textcolor{coeffectColor}{\square_{ \SYSTEMnt{r} } }   \textcolor{GLcolor}{\llbracket}  \SYSTEMnt{A}  \textcolor{GLcolor}{\rrbracket}    \oplus   \textcolor{coeffectColor}{\square_{ \SYSTEMnt{r} } }   \textcolor{GLcolor}{\llbracket}  \SYSTEMnt{B}  \textcolor{GLcolor}{\rrbracket}    }
    \inferrule*[Right=\SYSTEMRenameRuleLinlet{}]
      { \inferrule*[Right=\SYSTEMRenameRuleLinvar{}]{ }{   \SYSTEMmv{x'}  :   \textcolor{coeffectColor}{\square_{ \SYSTEMnt{r} } }   \textcolor{GLcolor}{\llbracket}  \SYSTEMnt{A}  \textcolor{GLcolor}{\rrbracket}      \vdash_{\textsc{l} }  \SYSTEMmv{x'}  :   \textcolor{coeffectColor}{\square_{ \SYSTEMnt{r} } }   \textcolor{GLcolor}{\llbracket}  \SYSTEMnt{A}  \textcolor{GLcolor}{\rrbracket}   }
        \hspace{4em}
        \inferrule*
          {
            \textit{ih}
          }
          {   \textcolor{GLcolor}{\llbracket}  \Delta_{{\mathrm{2}}}  \textcolor{GLcolor}{\rrbracket}  ,   \SYSTEMmv{x}  : \textcolor{coeffectColor}{[}   \textcolor{GLcolor}{\llbracket}  \SYSTEMnt{A}  \textcolor{GLcolor}{\rrbracket}  {\textcolor{coeffectColor}{]_{ \SYSTEMnt{r} } } }    \vdash_{\textsc{l} }   \textcolor{GLcolor}{\llbracket}  \SYSTEMnt{t_{{\mathrm{1}}}}  \textcolor{GLcolor}{\rrbracket}   :   \textcolor{GLcolor}{\llbracket}  \SYSTEMnt{C}  \textcolor{GLcolor}{\rrbracket}  }
      }
      {    \textcolor{GLcolor}{\llbracket}  \Delta_{{\mathrm{2}}}  \textcolor{GLcolor}{\rrbracket}  ,   \SYSTEMmv{x'}  :   \textcolor{coeffectColor}{\square_{ \SYSTEMnt{r} } }   \textcolor{GLcolor}{\llbracket}  \SYSTEMnt{A}  \textcolor{GLcolor}{\rrbracket}      \vdash_{\textsc{l} }   \mathsf{let} \, \textcolor{coeffectColor}{[}  \SYSTEMmv{x}  \textcolor{coeffectColor}{]} =  \SYSTEMmv{x'}  \, \mathsf{in} \,   \textcolor{GLcolor}{\llbracket}  \SYSTEMnt{t_{{\mathrm{1}}}}  \textcolor{GLcolor}{\rrbracket}    :   \textcolor{GLcolor}{\llbracket}  \SYSTEMnt{C}  \textcolor{GLcolor}{\rrbracket}  
      }
      \hspace{4em}
      \inferrule*[Right=\SYSTEMRenameRuleLinlet{}, vdots=6em, leftskip=13em]
        { \inferrule*[Right=\SYSTEMRenameRuleLinvar{}]
            { }
            {    \SYSTEMmv{y'}  :   \textcolor{coeffectColor}{\square_{ \SYSTEMnt{r} } }   \textcolor{GLcolor}{\llbracket}  \SYSTEMnt{B}  \textcolor{GLcolor}{\rrbracket}      \vdash_{\textsc{l} }  \SYSTEMmv{y'}  :   \textcolor{coeffectColor}{\square_{ \SYSTEMnt{r} } }   \textcolor{GLcolor}{\llbracket}  \SYSTEMnt{B}  \textcolor{GLcolor}{\rrbracket}    }
          \hspace{4em}
          \inferrule*
            {
              \textit{ih}
            }
            {   \textcolor{GLcolor}{\llbracket}  \Delta_{{\mathrm{2}}}  \textcolor{GLcolor}{\rrbracket}  ,   \SYSTEMmv{y}  : \textcolor{coeffectColor}{[}   \textcolor{GLcolor}{\llbracket}  \SYSTEMnt{B}  \textcolor{GLcolor}{\rrbracket}  {\textcolor{coeffectColor}{]_{ \SYSTEMnt{r} } } }    \vdash_{\textsc{l} }   \textcolor{GLcolor}{\llbracket}  \SYSTEMnt{t_{{\mathrm{2}}}}  \textcolor{GLcolor}{\rrbracket}   :   \textcolor{GLcolor}{\llbracket}  \SYSTEMnt{C}  \textcolor{GLcolor}{\rrbracket}  }
        }
        {    \textcolor{GLcolor}{\llbracket}  \Delta_{{\mathrm{2}}}  \textcolor{GLcolor}{\rrbracket}  ,   \SYSTEMmv{y'}  :   \textcolor{coeffectColor}{\square_{ \SYSTEMnt{r} } }   \textcolor{GLcolor}{\llbracket}  \SYSTEMnt{B}  \textcolor{GLcolor}{\rrbracket}      \vdash_{\textsc{l} }   \mathsf{let} \, \textcolor{coeffectColor}{[}  \SYSTEMmv{y}  \textcolor{coeffectColor}{]} =  \SYSTEMmv{y'}  \, \mathsf{in} \,   \textcolor{GLcolor}{\llbracket}  \SYSTEMnt{t_{{\mathrm{2}}}}  \textcolor{GLcolor}{\rrbracket}    :   \textcolor{GLcolor}{\llbracket}  \SYSTEMnt{C}  \textcolor{GLcolor}{\rrbracket}   }
}
{  \textcolor{coeffectColor}{ \SYSTEMnt{r}  \cdot}   \textcolor{GLcolor}{\llbracket}  \Delta_{{\mathrm{1}}}  \textcolor{GLcolor}{\rrbracket}    \SYSTEMsym{+}   \textcolor{GLcolor}{\llbracket}  \Delta_{{\mathrm{2}}}  \textcolor{GLcolor}{\rrbracket}   \vdash_{\textsc{l} }   \mathsf{case} \,   \textsf{push}_\oplus   \textcolor{coeffectColor}{[}   \textcolor{GLcolor}{\llbracket}  \SYSTEMnt{t}  \textcolor{GLcolor}{\rrbracket}   \textcolor{coeffectColor}{]}    \, \mathsf{of} \, \{ \mathsf{inj1} \,  \SYSTEMmv{x'}  \rightarrow   \mathsf{let} \, \textcolor{coeffectColor}{[}  \SYSTEMmv{x}  \textcolor{coeffectColor}{]} =  \SYSTEMmv{x'}  \, \mathsf{in} \,   \textcolor{GLcolor}{\llbracket}  \SYSTEMnt{t_{{\mathrm{1}}}}  \textcolor{GLcolor}{\rrbracket}    ; \, \mathsf{inj2} \,  \SYSTEMmv{y'}  \rightarrow   \mathsf{let} \, \textcolor{coeffectColor}{[}  \SYSTEMmv{y}  \textcolor{coeffectColor}{]} =  \SYSTEMmv{y'}  \, \mathsf{in} \,   \textcolor{GLcolor}{\llbracket}  \SYSTEMnt{t_{{\mathrm{2}}}}  \textcolor{GLcolor}{\rrbracket}    \}   :   \textcolor{GLcolor}{\llbracket}  \SYSTEMnt{C}  \textcolor{GLcolor}{\rrbracket}   }
\end{gather*}

  \end{itemize}
\end{proof}

\subsubsection{Operational correspondence}
\label{app:proofs-grad-core-to-line-core-ops}

\begin{lemma}[Interpretation preserves substitution]
\label{lemma:interp-grd-core-to-lin-core-preserves-subst}
For all Graded Base terms $\SYSTEMnt{t}, \SYSTEMnt{t'}$ then
$ \textcolor{GLcolor}{\llbracket}   [  \SYSTEMnt{t}  /  \SYSTEMmv{x}  ]  \SYSTEMnt{t'}   \textcolor{GLcolor}{\rrbracket}  \equiv  [   \textcolor{GLcolor}{\llbracket}  \SYSTEMnt{t}  \textcolor{GLcolor}{\rrbracket}   /  \SYSTEMmv{x}  ]   \textcolor{GLcolor}{\llbracket}  \SYSTEMnt{t'}  \textcolor{GLcolor}{\rrbracket}  $.
\end{lemma}

\begin{proof}
  Extending the previous proof of  Lemma~\ref{lemma:interp-grd-modal-to-lin-preserves-subst}:
  \begin{itemize}
\item ($\times_i$) $\SYSTEMnt{t'} \equiv  \langle  \SYSTEMnt{t_{{\mathrm{1}}}} ,  \SYSTEMnt{t_{{\mathrm{2}}}}  \rangle $;
\begin{gather*}
\begin{align*}
\begin{array}{rcrlll}
\textit{(goal)} & &  \textcolor{GLcolor}{\llbracket}   [  \SYSTEMnt{t}  /  \SYSTEMmv{x}  ]   \langle  \SYSTEMnt{t_{{\mathrm{1}}}} ,  \SYSTEMnt{t_{{\mathrm{2}}}}  \rangle    \textcolor{GLcolor}{\rrbracket}  & \eqStep &  [   \textcolor{GLcolor}{\llbracket}  \SYSTEMnt{t}  \textcolor{GLcolor}{\rrbracket}   /  \SYSTEMmv{x}  ]   \textcolor{GLcolor}{\llbracket}   \langle  \SYSTEMnt{t_{{\mathrm{1}}}} ,  \SYSTEMnt{t_{{\mathrm{2}}}}  \rangle   \textcolor{GLcolor}{\rrbracket}   \\
\textit{(defn. subst [lhs])} & \Rightarrow &  \textcolor{GLcolor}{\llbracket}   \langle   [  \SYSTEMnt{t}  /  \SYSTEMmv{x}  ]  \SYSTEMnt{t_{{\mathrm{1}}}}  ,   [  \SYSTEMnt{t}  /  \SYSTEMmv{x}  ]  \SYSTEMnt{t_{{\mathrm{2}}}}   \rangle   \textcolor{GLcolor}{\rrbracket}  & \eqStep &  [   \textcolor{GLcolor}{\llbracket}  \SYSTEMnt{t}  \textcolor{GLcolor}{\rrbracket}   /  \SYSTEMmv{x}  ]   \textcolor{GLcolor}{\llbracket}   \langle  \SYSTEMnt{t_{{\mathrm{1}}}} ,  \SYSTEMnt{t_{{\mathrm{2}}}}  \rangle   \textcolor{GLcolor}{\rrbracket}   \\
\textit{(defn. interp [lhs])} & \Rightarrow &  \langle   \textcolor{GLcolor}{\llbracket}   [  \SYSTEMnt{t}  /  \SYSTEMmv{x}  ]  \SYSTEMnt{t_{{\mathrm{1}}}}   \textcolor{GLcolor}{\rrbracket}  ,   \textcolor{GLcolor}{\llbracket}   [  \SYSTEMnt{t}  /  \SYSTEMmv{x}  ]  \SYSTEMnt{t_{{\mathrm{2}}}}   \textcolor{GLcolor}{\rrbracket}   \rangle  & \eqStep &  [   \textcolor{GLcolor}{\llbracket}  \SYSTEMnt{t}  \textcolor{GLcolor}{\rrbracket}   /  \SYSTEMmv{x}  ]   \textcolor{GLcolor}{\llbracket}   \langle  \SYSTEMnt{t_{{\mathrm{1}}}} ,  \SYSTEMnt{t_{{\mathrm{2}}}}  \rangle   \textcolor{GLcolor}{\rrbracket}   \\
\end{array}
\end{align*}
\end{gather*}
\item ($\times_e$) $\SYSTEMnt{t'} \equiv  \mathsf{let} \, \langle  \SYSTEMmv{y} ,  \SYSTEMmv{z}  \rangle =  \SYSTEMnt{t_{{\mathrm{1}}}}  \, \mathsf{in} \,  \SYSTEMnt{t_{{\mathrm{2}}}} $ (with $ \SYSTEMmv{y} ,   \SYSTEMmv{y'} ,   \SYSTEMmv{z} ,   \SYSTEMmv{z'}  \,\#\,  \SYSTEMnt{t}    $);
\begin{gather*}
\begin{align*}
\begin{array}{rcl}
\textit{(goal)} & &  \textcolor{GLcolor}{\llbracket}   [  \SYSTEMnt{t}  /  \SYSTEMmv{x}  ]   \mathsf{let} \, \langle  \SYSTEMmv{y} ,  \SYSTEMmv{z}  \rangle =  \SYSTEMnt{t_{{\mathrm{1}}}}  \, \mathsf{in} \,  \SYSTEMnt{t_{{\mathrm{2}}}}    \textcolor{GLcolor}{\rrbracket}  \\
                              &             & \quad \equiv  [   \textcolor{GLcolor}{\llbracket}  \SYSTEMnt{t}  \textcolor{GLcolor}{\rrbracket}   /  \SYSTEMmv{x}  ]   \textcolor{GLcolor}{\llbracket}   \mathsf{let} \, \langle  \SYSTEMmv{y} ,  \SYSTEMmv{z}  \rangle =  \SYSTEMnt{t_{{\mathrm{1}}}}  \, \mathsf{in} \,  \SYSTEMnt{t_{{\mathrm{2}}}}   \textcolor{GLcolor}{\rrbracket}   \\
\textit{(defn. subst [lhs])}  & \Rightarrow &  \textcolor{GLcolor}{\llbracket}   \mathsf{let} \, \langle  \SYSTEMmv{y} ,  \SYSTEMmv{z}  \rangle =   [  \SYSTEMnt{t}  /  \SYSTEMmv{x}  ]  \SYSTEMnt{t_{{\mathrm{1}}}}   \, \mathsf{in} \,   [  \SYSTEMnt{t}  /  \SYSTEMmv{x}  ]  \SYSTEMnt{t_{{\mathrm{2}}}}    \textcolor{GLcolor}{\rrbracket}  \\
                              &             & \quad \equiv  [   \textcolor{GLcolor}{\llbracket}  \SYSTEMnt{t}  \textcolor{GLcolor}{\rrbracket}   /  \SYSTEMmv{x}  ]   \textcolor{GLcolor}{\llbracket}   \mathsf{let} \, \langle  \SYSTEMmv{y} ,  \SYSTEMmv{z}  \rangle =  \SYSTEMnt{t_{{\mathrm{1}}}}  \, \mathsf{in} \,  \SYSTEMnt{t_{{\mathrm{2}}}}   \textcolor{GLcolor}{\rrbracket}   \\
\end{array}
\end{align*}
\end{gather*}
\item ($ \mathrm{unit} _i$) $\SYSTEMnt{t'} \equiv  \langle \rangle $;
\begin{align*}
\begin{array}{rcrlll}
\textit{(goal)} & &  \textcolor{GLcolor}{\llbracket}   [  \SYSTEMnt{t}  /  \SYSTEMmv{x}  ]   \langle \rangle    \textcolor{GLcolor}{\rrbracket}  & \eqStep &  [   \textcolor{GLcolor}{\llbracket}  \SYSTEMnt{t}  \textcolor{GLcolor}{\rrbracket}   /  \SYSTEMmv{x}  ]   \textcolor{GLcolor}{\llbracket}   \langle \rangle   \textcolor{GLcolor}{\rrbracket}   \\
\textit{(defn. subst [lhs])} & \Rightarrow &  \textcolor{GLcolor}{\llbracket}   \langle \rangle   \textcolor{GLcolor}{\rrbracket}  & \eqStep &  [   \textcolor{GLcolor}{\llbracket}  \SYSTEMnt{t}  \textcolor{GLcolor}{\rrbracket}   /  \SYSTEMmv{x}  ]   \textcolor{GLcolor}{\llbracket}   \langle \rangle   \textcolor{GLcolor}{\rrbracket}   \\
\textit{(defn. subst [rhs])} & \Rightarrow &  \textcolor{GLcolor}{\llbracket}   \langle \rangle   \textcolor{GLcolor}{\rrbracket}  & \eqStep &  \textcolor{GLcolor}{\llbracket}   \langle \rangle   \textcolor{GLcolor}{\rrbracket}  \\
\textit{(reflexivity)} & \Rightarrow & \top
\end{array}
\end{align*}
\item ($ \mathrm{unit} _e$) $\SYSTEMnt{t'} \equiv  \mathsf{let} \, \langle \rangle =  \SYSTEMnt{t_{{\mathrm{1}}}}  \, \mathsf{in} \,  \SYSTEMnt{t_{{\mathrm{2}}}} $;
\begin{gather*}
\begin{align*}
\begin{array}{rcrlll}
  \textit{(goal)} & &  \textcolor{GLcolor}{\llbracket}   [  \SYSTEMnt{t}  /  \SYSTEMmv{x}  ]   \mathsf{let} \, \langle \rangle =  \SYSTEMnt{t_{{\mathrm{1}}}}  \, \mathsf{in} \,  \SYSTEMnt{t_{{\mathrm{2}}}}    \textcolor{GLcolor}{\rrbracket}  & \eqStep &  [   \textcolor{GLcolor}{\llbracket}  \SYSTEMnt{t}  \textcolor{GLcolor}{\rrbracket}   /  \SYSTEMmv{x}  ]   \textcolor{GLcolor}{\llbracket}   \mathsf{let} \, \langle \rangle =  \SYSTEMnt{t_{{\mathrm{1}}}}  \, \mathsf{in} \,  \SYSTEMnt{t_{{\mathrm{2}}}}   \textcolor{GLcolor}{\rrbracket}   \\
  \textit{(defn. subst [lhs])} & \Rightarrow &  \textcolor{GLcolor}{\llbracket}   \mathsf{let} \, \langle \rangle =   [  \SYSTEMnt{t}  /  \SYSTEMmv{x}  ]  \SYSTEMnt{t_{{\mathrm{1}}}}   \, \mathsf{in} \,   [  \SYSTEMnt{t}  /  \SYSTEMmv{x}  ]  \SYSTEMnt{t_{{\mathrm{2}}}}    \textcolor{GLcolor}{\rrbracket}  & \eqStep &  [   \textcolor{GLcolor}{\llbracket}  \SYSTEMnt{t}  \textcolor{GLcolor}{\rrbracket}   /  \SYSTEMmv{x}  ]   \textcolor{GLcolor}{\llbracket}   \mathsf{let} \, \langle \rangle =  \SYSTEMnt{t_{{\mathrm{1}}}}  \, \mathsf{in} \,  \SYSTEMnt{t_{{\mathrm{2}}}}   \textcolor{GLcolor}{\rrbracket}   \\
  \textit{(defn. interp [lhs])} & \Rightarrow &  \mathsf{let} \, \langle \rangle =  \SYSTEMsym{(}   \textsf{push}_{\mathrm{unit} }   \textcolor{coeffectColor}{[}   \textcolor{GLcolor}{\llbracket}   [  \SYSTEMnt{t}  /  \SYSTEMmv{x}  ]  \SYSTEMnt{t_{{\mathrm{1}}}}   \textcolor{GLcolor}{\rrbracket}   \textcolor{coeffectColor}{]}    \SYSTEMsym{)}  \, \mathsf{in} \,   \textcolor{GLcolor}{\llbracket}   [  \SYSTEMnt{t}  /  \SYSTEMmv{x}  ]  \SYSTEMnt{t_{{\mathrm{2}}}}   \textcolor{GLcolor}{\rrbracket}   & \eqStep &  [   \textcolor{GLcolor}{\llbracket}  \SYSTEMnt{t}  \textcolor{GLcolor}{\rrbracket}   /  \SYSTEMmv{x}  ]   \textcolor{GLcolor}{\llbracket}   \mathsf{let} \, \langle \rangle =  \SYSTEMnt{t_{{\mathrm{1}}}}  \, \mathsf{in} \,  \SYSTEMnt{t_{{\mathrm{2}}}}   \textcolor{GLcolor}{\rrbracket}   \\
\end{array}
\end{align*}
\end{gather*}
\item ($+_{i1}$) $\SYSTEMnt{t'} \equiv  \mathsf{inj}_1 \,  \SYSTEMnt{t_{{\mathrm{1}}}} $;
\begin{align*}
\begin{array}{rcrlll}
\textit{(goal)}               &             &  \textcolor{GLcolor}{\llbracket}   [  \SYSTEMnt{t}  /  \SYSTEMmv{x}  ]  \SYSTEMsym{(}   \mathsf{inj}_1 \,  \SYSTEMnt{t_{{\mathrm{1}}}}   \SYSTEMsym{)}   \textcolor{GLcolor}{\rrbracket}  & \eqStep &  [   \textcolor{GLcolor}{\llbracket}  \SYSTEMnt{t}  \textcolor{GLcolor}{\rrbracket}   /  \SYSTEMmv{x}  ]   \textcolor{GLcolor}{\llbracket}   \mathsf{inj}_1 \,  \SYSTEMnt{t_{{\mathrm{1}}}}   \textcolor{GLcolor}{\rrbracket}   \\
\textit{(defn. subst [lhs])}  & \Rightarrow &  \textcolor{GLcolor}{\llbracket}   \mathsf{inj}_1 \,  \SYSTEMsym{(}   [  \SYSTEMnt{t}  /  \SYSTEMmv{x}  ]  \SYSTEMnt{t_{{\mathrm{1}}}}   \SYSTEMsym{)}   \textcolor{GLcolor}{\rrbracket}  & \eqStep &  [   \textcolor{GLcolor}{\llbracket}  \SYSTEMnt{t}  \textcolor{GLcolor}{\rrbracket}   /  \SYSTEMmv{x}  ]   \textcolor{GLcolor}{\llbracket}   \mathsf{inj}_1 \,  \SYSTEMnt{t_{{\mathrm{1}}}}   \textcolor{GLcolor}{\rrbracket}   \\
\textit{(defn. interp [lhs])} & \Rightarrow &  \mathsf{inj}_1 \,   \textcolor{GLcolor}{\llbracket}   [  \SYSTEMnt{t}  /  \SYSTEMmv{x}  ]  \SYSTEMnt{t_{{\mathrm{1}}}}   \textcolor{GLcolor}{\rrbracket}     & \eqStep &  [   \textcolor{GLcolor}{\llbracket}  \SYSTEMnt{t}  \textcolor{GLcolor}{\rrbracket}   /  \SYSTEMmv{x}  ]   \textcolor{GLcolor}{\llbracket}   \mathsf{inj}_1 \,  \SYSTEMnt{t_{{\mathrm{1}}}}   \textcolor{GLcolor}{\rrbracket}   \\
\textit{(defn. interp [rhs])} & \Rightarrow &  \mathsf{inj}_1 \,   \textcolor{GLcolor}{\llbracket}   [  \SYSTEMnt{t}  /  \SYSTEMmv{x}  ]  \SYSTEMnt{t_{{\mathrm{1}}}}   \textcolor{GLcolor}{\rrbracket}     & \eqStep &  [   \textcolor{GLcolor}{\llbracket}  \SYSTEMnt{t}  \textcolor{GLcolor}{\rrbracket}   /  \SYSTEMmv{x}  ]  \SYSTEMsym{(}   \mathsf{inj}_1 \,   \textcolor{GLcolor}{\llbracket}  \SYSTEMnt{t_{{\mathrm{1}}}}  \textcolor{GLcolor}{\rrbracket}    \SYSTEMsym{)}  \\
\end{array}
\end{align*}
\item ($+_{i2}$) $\SYSTEMnt{t'} \equiv  \mathsf{inj}_2 \,  \SYSTEMnt{t_{{\mathrm{2}}}} $;
\begin{align*}
\begin{array}{rcrlll}
\textit{(goal)}               &             &  \textcolor{GLcolor}{\llbracket}   [  \SYSTEMnt{t}  /  \SYSTEMmv{x}  ]  \SYSTEMsym{(}   \mathsf{inj}_2 \,  \SYSTEMnt{t_{{\mathrm{2}}}}   \SYSTEMsym{)}   \textcolor{GLcolor}{\rrbracket}  & \eqStep &  [   \textcolor{GLcolor}{\llbracket}  \SYSTEMnt{t}  \textcolor{GLcolor}{\rrbracket}   /  \SYSTEMmv{x}  ]   \textcolor{GLcolor}{\llbracket}   \mathsf{inj}_2 \,  \SYSTEMnt{t_{{\mathrm{2}}}}   \textcolor{GLcolor}{\rrbracket}   \\
\textit{(defn. subst [lhs])}  & \Rightarrow &  \textcolor{GLcolor}{\llbracket}   \mathsf{inj}_2 \,  \SYSTEMsym{(}   [  \SYSTEMnt{t}  /  \SYSTEMmv{x}  ]  \SYSTEMnt{t_{{\mathrm{2}}}}   \SYSTEMsym{)}   \textcolor{GLcolor}{\rrbracket}  & \eqStep &  [   \textcolor{GLcolor}{\llbracket}  \SYSTEMnt{t}  \textcolor{GLcolor}{\rrbracket}   /  \SYSTEMmv{x}  ]   \textcolor{GLcolor}{\llbracket}   \mathsf{inj}_2 \,  \SYSTEMnt{t_{{\mathrm{2}}}}   \textcolor{GLcolor}{\rrbracket}   \\
\textit{(defn. interp [lhs])} & \Rightarrow &  \mathsf{inj}_2 \,   \textcolor{GLcolor}{\llbracket}   [  \SYSTEMnt{t}  /  \SYSTEMmv{x}  ]  \SYSTEMnt{t_{{\mathrm{2}}}}   \textcolor{GLcolor}{\rrbracket}     & \eqStep &  [   \textcolor{GLcolor}{\llbracket}  \SYSTEMnt{t}  \textcolor{GLcolor}{\rrbracket}   /  \SYSTEMmv{x}  ]   \textcolor{GLcolor}{\llbracket}   \mathsf{inj}_2 \,  \SYSTEMnt{t_{{\mathrm{2}}}}   \textcolor{GLcolor}{\rrbracket}   \\
\textit{(defn. interp [rhs])} & \Rightarrow &  \mathsf{inj}_2 \,   \textcolor{GLcolor}{\llbracket}   [  \SYSTEMnt{t}  /  \SYSTEMmv{x}  ]  \SYSTEMnt{t_{{\mathrm{2}}}}   \textcolor{GLcolor}{\rrbracket}     & \eqStep &  [   \textcolor{GLcolor}{\llbracket}  \SYSTEMnt{t}  \textcolor{GLcolor}{\rrbracket}   /  \SYSTEMmv{x}  ]  \SYSTEMsym{(}   \mathsf{inj}_2 \,   \textcolor{GLcolor}{\llbracket}  \SYSTEMnt{t_{{\mathrm{2}}}}  \textcolor{GLcolor}{\rrbracket}    \SYSTEMsym{)}  \\
\end{array}
\end{align*}
\item ($+_e$) $\SYSTEMnt{t'} \equiv  \mathsf{case} \,  \SYSTEMnt{t_{{\mathrm{1}}}}  \, \mathsf{of} \, \{ \mathsf{inj1} \,  \SYSTEMmv{y}  \rightarrow  \SYSTEMnt{t_{{\mathrm{2}}}}  ; \, \mathsf{inj2} \,  \SYSTEMmv{z}  \rightarrow  \SYSTEMnt{t_{{\mathrm{3}}}}  \} $ (with $ \SYSTEMmv{y} ,   \SYSTEMmv{z} ,   \SYSTEMmv{y'} ,   \SYSTEMmv{z'}  \,\#\,  \SYSTEMnt{t}    $);
\begin{gather*}
  \begin{align*}
  \begin{array}{rcl}
  \textit{(goal)}               &             &  \textcolor{GLcolor}{\llbracket}   [  \SYSTEMnt{t}  /  \SYSTEMmv{x}  ]   \mathsf{case} \,  \SYSTEMnt{t_{{\mathrm{1}}}}  \, \mathsf{of} \, \{ \mathsf{inj1} \,  \SYSTEMmv{y}  \rightarrow  \SYSTEMnt{t_{{\mathrm{2}}}}  ; \, \mathsf{inj2} \,  \SYSTEMmv{z}  \rightarrow  \SYSTEMnt{t_{{\mathrm{3}}}}  \}    \textcolor{GLcolor}{\rrbracket}  \\
                                &             & \quad \equiv  [   \textcolor{GLcolor}{\llbracket}  \SYSTEMnt{t}  \textcolor{GLcolor}{\rrbracket}   /  \SYSTEMmv{x}  ]   \textcolor{GLcolor}{\llbracket}   \mathsf{case} \,  \SYSTEMnt{t_{{\mathrm{1}}}}  \, \mathsf{of} \, \{ \mathsf{inj1} \,  \SYSTEMmv{y}  \rightarrow  \SYSTEMnt{t_{{\mathrm{2}}}}  ; \, \mathsf{inj2} \,  \SYSTEMmv{z}  \rightarrow  \SYSTEMnt{t_{{\mathrm{3}}}}  \}   \textcolor{GLcolor}{\rrbracket}   \\
  \textit{(defn. subst [lhs])}  & \Rightarrow &  \textcolor{GLcolor}{\llbracket}   \mathsf{case} \,   [  \SYSTEMnt{t}  /  \SYSTEMmv{x}  ]  \SYSTEMnt{t_{{\mathrm{1}}}}   \, \mathsf{of} \, \{ \mathsf{inj1} \,  \SYSTEMmv{y}  \rightarrow   [  \SYSTEMnt{t}  /  \SYSTEMmv{x}  ]  \SYSTEMnt{t_{{\mathrm{2}}}}   ; \, \mathsf{inj2} \,  \SYSTEMmv{z}  \rightarrow   [  \SYSTEMnt{t}  /  \SYSTEMmv{x}  ]  \SYSTEMnt{t_{{\mathrm{3}}}}   \}   \textcolor{GLcolor}{\rrbracket}  \\
                                &             & \quad \equiv  [   \textcolor{GLcolor}{\llbracket}  \SYSTEMnt{t}  \textcolor{GLcolor}{\rrbracket}   /  \SYSTEMmv{x}  ]   \textcolor{GLcolor}{\llbracket}   \mathsf{case} \,  \SYSTEMnt{t_{{\mathrm{1}}}}  \, \mathsf{of} \, \{ \mathsf{inj1} \,  \SYSTEMmv{y}  \rightarrow  \SYSTEMnt{t_{{\mathrm{2}}}}  ; \, \mathsf{inj2} \,  \SYSTEMmv{z}  \rightarrow  \SYSTEMnt{t_{{\mathrm{3}}}}  \}   \textcolor{GLcolor}{\rrbracket}   \\
\end{array}
  \end{align*}
\end{gather*}
\end{itemize}
\end{proof}

The operational correspondence then follows via the following proof:

\begin{proof}
  The proof subsumes the earlier proofs; we add the additional cases here due to the
  extension of the language from Graded Modal Base to Graded Modal Core.

  For convenience, we expand the statement to $ \SYSTEMnt{t}  \rightsquigarrow_{\textsc{g} }  \SYSTEMnt{t'} $
  then $\exists t'' .   \textcolor{GLcolor}{\llbracket}  \SYSTEMnt{t}  \textcolor{GLcolor}{\rrbracket}   \rightsquigarrow_{\textsc{l} }^\ast  \SYSTEMnt{t''} $
   $\, \wedge \,  \textcolor{GLcolor}{\llbracket}  \SYSTEMnt{t'}  \textcolor{GLcolor}{\rrbracket}  \equiv \SYSTEMnt{t''}$.
  \begin{itemize}
\item (prodCong)
\[
\SYSTEMdruleSemGrdprodCong{}
\]
By induction, analogously to the case for (congAppL).

\item (prodBeta)
\[
\SYSTEMdruleSemGrdprodBeta{}
\]
The interpretation of the reducing term is:
\[
  \textcolor{GLcolor}{\llbracket}   \mathsf{let} \, \langle  \SYSTEMmv{x} ,  \SYSTEMmv{y}  \rangle =   \langle  \SYSTEMnt{t_{{\mathrm{1}}}} ,  \SYSTEMnt{t_{{\mathrm{2}}}}  \rangle   \, \mathsf{in} \,  \SYSTEMnt{t_{{\mathrm{3}}}}   \textcolor{GLcolor}{\rrbracket}   \equiv   \mathsf{let} \, \langle  \SYSTEMmv{x} ,  \SYSTEMmv{y}  \rangle =   \langle   \textcolor{GLcolor}{\llbracket}  \SYSTEMnt{t_{{\mathrm{1}}}}  \textcolor{GLcolor}{\rrbracket}  ,   \textcolor{GLcolor}{\llbracket}  \SYSTEMnt{t_{{\mathrm{2}}}}  \textcolor{GLcolor}{\rrbracket}   \rangle   \, \mathsf{in} \,   \textcolor{GLcolor}{\llbracket}  \SYSTEMnt{t_{{\mathrm{3}}}}  \textcolor{GLcolor}{\rrbracket}   
\]
where $ \SYSTEMmv{x} ,   \SYSTEMmv{y}  \,\#\,  \SYSTEMnt{t}  $.
Then we construct the reduction in Linear Base:
\begin{align*}
\inferrule*[right=\SYSTEMRenameRuleSemLinprodBeta{}]
{ }
{  \mathsf{let} \, \langle  \SYSTEMmv{x} ,  \SYSTEMmv{y}  \rangle =   \langle   \textcolor{GLcolor}{\llbracket}  \SYSTEMnt{t_{{\mathrm{1}}}}  \textcolor{GLcolor}{\rrbracket}  ,   \textcolor{GLcolor}{\llbracket}  \SYSTEMnt{t_{{\mathrm{2}}}}  \textcolor{GLcolor}{\rrbracket}   \rangle   \, \mathsf{in} \,   \textcolor{GLcolor}{\llbracket}  \SYSTEMnt{t_{{\mathrm{3}}}}  \textcolor{GLcolor}{\rrbracket}    \rightsquigarrow_{\textsc{l} }   [   \textcolor{GLcolor}{\llbracket}  \SYSTEMnt{t_{{\mathrm{1}}}}  \textcolor{GLcolor}{\rrbracket}   /  \SYSTEMmv{x}  ]   [   \textcolor{GLcolor}{\llbracket}  \SYSTEMnt{t_{{\mathrm{2}}}}  \textcolor{GLcolor}{\rrbracket}   /  \SYSTEMmv{y}  ]   \textcolor{GLcolor}{\llbracket}  \SYSTEMnt{t_{{\mathrm{3}}}}  \textcolor{GLcolor}{\rrbracket}    }
\end{align*}
By Lemma~\ref{lemma:interp-grd-to-lin-preserves-subst} then interpreting
the linear-base reduced term is equal to the graded-base reduced term:
\begin{align*}
 \textcolor{GLcolor}{\llbracket}   [  \SYSTEMnt{t_{{\mathrm{1}}}}  /  \SYSTEMmv{x}  ]   [  \SYSTEMnt{t_{{\mathrm{2}}}}  /  \SYSTEMmv{y}  ]  \SYSTEMnt{t_{{\mathrm{3}}}}    \textcolor{GLcolor}{\rrbracket}  & \equiv  [   \textcolor{GLcolor}{\llbracket}  \SYSTEMnt{t_{{\mathrm{1}}}}  \textcolor{GLcolor}{\rrbracket}   /  \SYSTEMmv{x}  ]   \textcolor{GLcolor}{\llbracket}   [  \SYSTEMnt{t_{{\mathrm{2}}}}  /  \SYSTEMmv{y}  ]  \SYSTEMnt{t_{{\mathrm{3}}}}   \textcolor{GLcolor}{\rrbracket}   \\
                                     & \equiv  [   \textcolor{GLcolor}{\llbracket}  \SYSTEMnt{t_{{\mathrm{1}}}}  \textcolor{GLcolor}{\rrbracket}   /  \SYSTEMmv{x}  ]   [   \textcolor{GLcolor}{\llbracket}  \SYSTEMnt{t_{{\mathrm{2}}}}  \textcolor{GLcolor}{\rrbracket}   /  \SYSTEMmv{y}  ]   \textcolor{GLcolor}{\llbracket}  \SYSTEMnt{t_{{\mathrm{3}}}}  \textcolor{GLcolor}{\rrbracket}   
\end{align*}

\item (unitCong)
\[
\SYSTEMdruleSemGrdunitCong{}
\]
By induction, analogously to the case for (congAppL).

\item (unitBeta)
\[
\SYSTEMdruleSemGrdunitBeta{}
\]
The interpretation of the reducing term is:
\[
  \textcolor{GLcolor}{\llbracket}   \mathsf{let} \, \langle \rangle =   \langle \rangle   \, \mathsf{in} \,  \SYSTEMnt{t_{{\mathrm{2}}}}   \textcolor{GLcolor}{\rrbracket}   \equiv   \mathsf{let} \, \langle \rangle =   \textsf{push}_{\mathrm{unit} }   \textcolor{coeffectColor}{[}   \langle \rangle   \textcolor{coeffectColor}{]}    \, \mathsf{in} \,   \textcolor{GLcolor}{\llbracket}  \SYSTEMnt{t_{{\mathrm{2}}}}  \textcolor{GLcolor}{\rrbracket}   
\]
Then we construct the following reduction sequence (two reductions) in Linear Base:
\begin{align*}
\inferrule*[right=\SYSTEMRenameRuleSemLinpushUnit{}]
{ }
{  \mathsf{let} \, \langle \rangle =   \textsf{push}_{\mathrm{unit} }   \textcolor{coeffectColor}{[}   \langle \rangle   \textcolor{coeffectColor}{]}    \, \mathsf{in} \,   \textcolor{GLcolor}{\llbracket}  \SYSTEMnt{t_{{\mathrm{2}}}}  \textcolor{GLcolor}{\rrbracket}    \rightsquigarrow_{\textsc{l} }   \mathsf{let} \, \langle \rangle =   \langle \rangle   \, \mathsf{in} \,   \textcolor{GLcolor}{\llbracket}  \SYSTEMnt{t_{{\mathrm{2}}}}  \textcolor{GLcolor}{\rrbracket}   }
\quad
  \inferrule*[right=\SYSTEMRenameRuleSemLinbetaBox{}]
{ }
{  \mathsf{let} \, \langle \rangle =   \langle \rangle   \, \mathsf{in} \,   \textcolor{GLcolor}{\llbracket}  \SYSTEMnt{t_{{\mathrm{2}}}}  \textcolor{GLcolor}{\rrbracket}    \rightsquigarrow_{\textsc{l} }   \textcolor{GLcolor}{\llbracket}  \SYSTEMnt{t_{{\mathrm{2}}}}  \textcolor{GLcolor}{\rrbracket}  }
\end{align*}
By reflexivity then interpreting the linear-base reduced term is equal to the
graded-base reduced term.

\item (congCase)
\[
\SYSTEMdruleSemGrdcongCase{}
\]
By induction, analogously to the case for (congAppL).

\item (caseInj1)
\[
\SYSTEMdruleSemGrdcaseInjOne{}
\]
The interpretation of the reducing term is:
\[
  \textcolor{GLcolor}{\llbracket}   \mathsf{case} \,   \mathsf{inj}_1 \,  \SYSTEMnt{t}   \, \mathsf{of} \, \{ \mathsf{inj1} \,  \SYSTEMmv{x}  \rightarrow  \SYSTEMnt{t_{{\mathrm{1}}}}  ; \, \mathsf{inj2} \,  \SYSTEMmv{y}  \rightarrow  \SYSTEMnt{t_{{\mathrm{2}}}}  \}   \textcolor{GLcolor}{\rrbracket}   \equiv   \mathsf{case} \,   \mathsf{inj}_1 \,   \textcolor{GLcolor}{\llbracket}  \SYSTEMnt{t}  \textcolor{GLcolor}{\rrbracket}    \, \mathsf{of} \, \{ \mathsf{inj1} \,  \SYSTEMmv{x}  \rightarrow   \textcolor{GLcolor}{\llbracket}  \SYSTEMnt{t_{{\mathrm{1}}}}  \textcolor{GLcolor}{\rrbracket}   ; \, \mathsf{inj2} \,  \SYSTEMmv{y}  \rightarrow   \textcolor{GLcolor}{\llbracket}  \SYSTEMnt{t_{{\mathrm{2}}}}  \textcolor{GLcolor}{\rrbracket}   \}  
\]
Then we construct the reduction in Linear Base:
\begin{align*}
\inferrule*[right=\SYSTEMRenameRuleSemLincaseInjOne{}]
{ }
{  \mathsf{case} \,   \mathsf{inj}_1 \,   \textcolor{GLcolor}{\llbracket}  \SYSTEMnt{t}  \textcolor{GLcolor}{\rrbracket}    \, \mathsf{of} \, \{ \mathsf{inj1} \,  \SYSTEMmv{x}  \rightarrow   \textcolor{GLcolor}{\llbracket}  \SYSTEMnt{t_{{\mathrm{1}}}}  \textcolor{GLcolor}{\rrbracket}   ; \, \mathsf{inj2} \,  \SYSTEMmv{y}  \rightarrow   \textcolor{GLcolor}{\llbracket}  \SYSTEMnt{t_{{\mathrm{2}}}}  \textcolor{GLcolor}{\rrbracket}   \}   \rightsquigarrow_{\textsc{l} }   [   \textcolor{GLcolor}{\llbracket}  \SYSTEMnt{t}  \textcolor{GLcolor}{\rrbracket}   /  \SYSTEMmv{x}  ]   \textcolor{GLcolor}{\llbracket}  \SYSTEMnt{t_{{\mathrm{1}}}}  \textcolor{GLcolor}{\rrbracket}   }
\end{align*}
By Lemma~\ref{lemma:interp-grd-to-lin-preserves-subst} then interpreting
the linear-base reduced term is equal to the graded-base
reduced term: $  \textcolor{GLcolor}{\llbracket}   [  \SYSTEMnt{t}  /  \SYSTEMmv{x}  ]  \SYSTEMnt{t_{{\mathrm{1}}}}   \textcolor{GLcolor}{\rrbracket}   \equiv   [   \textcolor{GLcolor}{\llbracket}  \SYSTEMnt{t}  \textcolor{GLcolor}{\rrbracket}   /  \SYSTEMmv{x}  ]   \textcolor{GLcolor}{\llbracket}  \SYSTEMnt{t_{{\mathrm{1}}}}  \textcolor{GLcolor}{\rrbracket}   $.

\item (caseInj2)
\[
\SYSTEMdruleSemGrdcaseInjTwo{}
\]
Analogously to the case for (caseInj1).
  \end{itemize}
\end{proof}

\subsubsection{Equation correspondence}
\label{app:proofs-grad-core-to-line-core-eqs}

\begin{proof}
  The proof subsumes the earlier proofs; we add the additional cases here due to the
  extension of the language from Graded Modal Base to Graded Modal Core.
  \begin{itemize}
  \item \[\SYSTEMdruleGradEqbetaUnit{}\]
     \begin{align*}
      \begin{array}{rl}
        &  \textcolor{GLcolor}{\llbracket}   \mathsf{let} \, \langle \rangle =   \langle \rangle   \, \mathsf{in} \,  \SYSTEMnt{t}   \textcolor{GLcolor}{\rrbracket}  \\
        \textit{\{defn. translation\}} & =  \mathsf{let} \, \langle \rangle =   \textsf{push}_{\mathrm{unit} }   \textcolor{coeffectColor}{[}   \langle \rangle   \textcolor{coeffectColor}{]}    \, \mathsf{in} \,   \textcolor{GLcolor}{\llbracket}  \SYSTEMnt{t}  \textcolor{GLcolor}{\rrbracket}   \\
        \textit{\{\SYSTEMRenameRuleLinEqpushUnitBeta{}\}} &  \equiv_{\textsc{l} }   \mathsf{let} \, \langle \rangle =   \langle \rangle   \, \mathsf{in} \,   \textcolor{GLcolor}{\llbracket}  \SYSTEMnt{t}  \textcolor{GLcolor}{\rrbracket}   \\
        \textit{\{\SYSTEMRenameRuleLinEqbetaUnit{}\}} &  \equiv_{\textsc{l} }   \textcolor{GLcolor}{\llbracket}  \SYSTEMnt{t}  \textcolor{GLcolor}{\rrbracket} 
      \end{array}
    \end{align*}

  % Failed eta preservation; keeping as notes
  % \item \[\SYSTEMdruleGradEqetaUnit{}\]

  %   \begin{align*}
  %     \begin{array}{rl}
  %       &  \textcolor{GLcolor}{\llbracket}   \mathsf{let} \, \langle \rangle =  \SYSTEMnt{t}  \, \mathsf{in} \,   \langle \rangle    \textcolor{GLcolor}{\rrbracket}  \\
  %       \textit{\{defn. translation\}} & =  \mathsf{let} \, \langle \rangle =   \textcolor{GLcolor}{\llbracket}  \SYSTEMnt{t}  \textcolor{GLcolor}{\rrbracket}   \, \mathsf{in} \,   \langle \rangle   \\
  %       \textit{\{\SYSTEMRenameRuleLinEqetaUnit{}\}} &  \equiv_{\textsc{l} }   \textcolor{GLcolor}{\llbracket}  \SYSTEMnt{t}  \textcolor{GLcolor}{\rrbracket} 
  %     \end{array}
  %   \end{align*}

  \item \[ \SYSTEMdruleGradEqbetaProd{}  \]
    \begin{align*}
      \begin{array}{rl}
        &  \textcolor{GLcolor}{\llbracket}   \mathsf{let} \, \langle  \SYSTEMmv{x} ,  \SYSTEMmv{y}  \rangle =   \langle  \SYSTEMnt{t_{{\mathrm{1}}}} ,  \SYSTEMnt{t_{{\mathrm{2}}}}  \rangle   \, \mathsf{in} \,  \SYSTEMnt{t}   \textcolor{GLcolor}{\rrbracket}  \\
        \textit{\{defn. translation\}} & =  \mathsf{let} \, \langle  \SYSTEMmv{x'} ,  \SYSTEMmv{y'}  \rangle =   \textsf{push}_\otimes   \textcolor{coeffectColor}{[}   \langle   \textcolor{GLcolor}{\llbracket}  \SYSTEMnt{t_{{\mathrm{1}}}}  \textcolor{GLcolor}{\rrbracket}  ,   \textcolor{GLcolor}{\llbracket}  \SYSTEMnt{t_{{\mathrm{2}}}}  \textcolor{GLcolor}{\rrbracket}   \rangle   \textcolor{coeffectColor}{]}    \, \mathsf{in} \,   \mathsf{let} \, \textcolor{coeffectColor}{[}  \SYSTEMmv{x}  \textcolor{coeffectColor}{]} =  \SYSTEMmv{x'}  \, \mathsf{in} \,   \mathsf{let} \, \textcolor{coeffectColor}{[}  \SYSTEMmv{y}  \textcolor{coeffectColor}{]} =  \SYSTEMmv{y'}  \, \mathsf{in} \,   \textcolor{GLcolor}{\llbracket}  \SYSTEMnt{t}  \textcolor{GLcolor}{\rrbracket}     \\
        \textit{\{\SYSTEMRenameRuleLinEqpushProdBeta{}\}} &  \equiv_{\textsc{l} }   \mathsf{let} \, \langle  \SYSTEMmv{x'} ,  \SYSTEMmv{y'}  \rangle =   \langle   \textcolor{coeffectColor}{[}   \textcolor{GLcolor}{\llbracket}  \SYSTEMnt{t_{{\mathrm{1}}}}  \textcolor{GLcolor}{\rrbracket}   \textcolor{coeffectColor}{]}  ,   \textcolor{coeffectColor}{[}   \textcolor{GLcolor}{\llbracket}  \SYSTEMnt{t_{{\mathrm{2}}}}  \textcolor{GLcolor}{\rrbracket}   \textcolor{coeffectColor}{]}   \rangle   \, \mathsf{in} \,   \mathsf{let} \, \textcolor{coeffectColor}{[}  \SYSTEMmv{x}  \textcolor{coeffectColor}{]} =  \SYSTEMmv{x'}  \, \mathsf{in} \,   \mathsf{let} \, \textcolor{coeffectColor}{[}  \SYSTEMmv{y}  \textcolor{coeffectColor}{]} =  \SYSTEMmv{y'}  \, \mathsf{in} \,   \textcolor{GLcolor}{\llbracket}  \SYSTEMnt{t}  \textcolor{GLcolor}{\rrbracket}     \\
        \textit{\{\SYSTEMRenameRuleLinEqbetaProd{}\}} &  \equiv_{\textsc{l} }   \mathsf{let} \, \textcolor{coeffectColor}{[}  \SYSTEMmv{x}  \textcolor{coeffectColor}{]} =   \textcolor{coeffectColor}{[}   \textcolor{GLcolor}{\llbracket}  \SYSTEMnt{t_{{\mathrm{1}}}}  \textcolor{GLcolor}{\rrbracket}   \textcolor{coeffectColor}{]}   \, \mathsf{in} \,   \mathsf{let} \, \textcolor{coeffectColor}{[}  \SYSTEMmv{y}  \textcolor{coeffectColor}{]} =   \textcolor{coeffectColor}{[}   \textcolor{GLcolor}{\llbracket}  \SYSTEMnt{t_{{\mathrm{2}}}}  \textcolor{GLcolor}{\rrbracket}   \textcolor{coeffectColor}{]}   \, \mathsf{in} \,   \textcolor{GLcolor}{\llbracket}  \SYSTEMnt{t}  \textcolor{GLcolor}{\rrbracket}    \\
        \textit{\{\SYSTEMRenameRuleLinEqbetaBox{}\}} &  \equiv_{\textsc{l} }   \mathsf{let} \, \textcolor{coeffectColor}{[}  \SYSTEMmv{y}  \textcolor{coeffectColor}{]} =   \textcolor{coeffectColor}{[}   \textcolor{GLcolor}{\llbracket}  \SYSTEMnt{t_{{\mathrm{2}}}}  \textcolor{GLcolor}{\rrbracket}   \textcolor{coeffectColor}{]}   \, \mathsf{in} \,   [   \textcolor{GLcolor}{\llbracket}  \SYSTEMnt{t_{{\mathrm{1}}}}  \textcolor{GLcolor}{\rrbracket}   /  \SYSTEMmv{x}  ]   \textcolor{GLcolor}{\llbracket}  \SYSTEMnt{t}  \textcolor{GLcolor}{\rrbracket}    \\
        \textit{\{\SYSTEMRenameRuleLinEqbetaBox{}\}} &  \equiv_{\textsc{l} }   [   \textcolor{GLcolor}{\llbracket}  \SYSTEMnt{t_{{\mathrm{1}}}}  \textcolor{GLcolor}{\rrbracket}   /  \SYSTEMmv{x}  ]   [   \textcolor{GLcolor}{\llbracket}  \SYSTEMnt{t_{{\mathrm{2}}}}  \textcolor{GLcolor}{\rrbracket}   /  \SYSTEMmv{y}  ]   \textcolor{GLcolor}{\llbracket}  \SYSTEMnt{t}  \textcolor{GLcolor}{\rrbracket}    \\
      \end{array}
      \end{align*}

  \item \[\SYSTEMdruleGradEqbetaSumOne{}\]
        \begin{align*}
      \begin{array}{rl}
        &  \textcolor{GLcolor}{\llbracket}   \mathsf{case} \,   \mathsf{inj}_1 \,  \SYSTEMnt{t}   \, \mathsf{of} \, \{ \mathsf{inj1} \,  \SYSTEMmv{x}  \rightarrow  \SYSTEMnt{t_{{\mathrm{1}}}}  ; \, \mathsf{inj2} \,  \SYSTEMmv{y}  \rightarrow  \SYSTEMnt{t_{{\mathrm{2}}}}  \}   \textcolor{GLcolor}{\rrbracket}  \\
        \textit{\{defn. translation\}} & =  \mathsf{case} \,   \textsf{push}_\oplus   \textcolor{coeffectColor}{[}   \mathsf{inj}_1 \,   \textcolor{GLcolor}{\llbracket}  \SYSTEMnt{t}  \textcolor{GLcolor}{\rrbracket}    \textcolor{coeffectColor}{]}    \, \mathsf{of} \, \{ \mathsf{inj1} \,  \SYSTEMmv{x'}  \rightarrow   \mathsf{let} \, \textcolor{coeffectColor}{[}  \SYSTEMmv{x}  \textcolor{coeffectColor}{]} =  \SYSTEMmv{x'}  \, \mathsf{in} \,   \textcolor{GLcolor}{\llbracket}  \SYSTEMnt{t_{{\mathrm{1}}}}  \textcolor{GLcolor}{\rrbracket}    ; \, \mathsf{inj2} \,  \SYSTEMmv{y'}  \rightarrow   \mathsf{let} \, \textcolor{coeffectColor}{[}  \SYSTEMmv{y}  \textcolor{coeffectColor}{]} =  \SYSTEMmv{y'}  \, \mathsf{in} \,   \textcolor{GLcolor}{\llbracket}  \SYSTEMnt{t_{{\mathrm{2}}}}  \textcolor{GLcolor}{\rrbracket}    \}  \\
        \textit{\{\SYSTEMRenameRuleLinEqpushProdBeta{}\}} & =  \mathsf{case} \,  \SYSTEMsym{(}   \mathsf{inj}_1 \,   \textcolor{coeffectColor}{[}   \textcolor{GLcolor}{\llbracket}  \SYSTEMnt{t}  \textcolor{GLcolor}{\rrbracket}   \textcolor{coeffectColor}{]}    \SYSTEMsym{)}  \, \mathsf{of} \, \{ \mathsf{inj1} \,  \SYSTEMmv{x'}  \rightarrow   \mathsf{let} \, \textcolor{coeffectColor}{[}  \SYSTEMmv{x}  \textcolor{coeffectColor}{]} =  \SYSTEMmv{x'}  \, \mathsf{in} \,   \textcolor{GLcolor}{\llbracket}  \SYSTEMnt{t_{{\mathrm{1}}}}  \textcolor{GLcolor}{\rrbracket}    ; \, \mathsf{inj2} \,  \SYSTEMmv{y'}  \rightarrow   \mathsf{let} \, \textcolor{coeffectColor}{[}  \SYSTEMmv{y}  \textcolor{coeffectColor}{]} =  \SYSTEMmv{y'}  \, \mathsf{in} \,   \textcolor{GLcolor}{\llbracket}  \SYSTEMnt{t_{{\mathrm{2}}}}  \textcolor{GLcolor}{\rrbracket}    \}  \\
        \textit{\{\SYSTEMRenameRuleLinEqbetaSumOne{}\}} & =  \mathsf{let} \, \textcolor{coeffectColor}{[}  \SYSTEMmv{x}  \textcolor{coeffectColor}{]} =   \textcolor{coeffectColor}{[}   \textcolor{GLcolor}{\llbracket}  \SYSTEMnt{t}  \textcolor{GLcolor}{\rrbracket}   \textcolor{coeffectColor}{]}   \, \mathsf{in} \,   \textcolor{GLcolor}{\llbracket}  \SYSTEMnt{t_{{\mathrm{1}}}}  \textcolor{GLcolor}{\rrbracket}   \\
        \textit{\{\SYSTEMRenameRuleLinEqbetaBox{}\}} & =  [   \textcolor{GLcolor}{\llbracket}  \SYSTEMnt{t}  \textcolor{GLcolor}{\rrbracket}   /  \SYSTEMmv{x}  ]   \textcolor{GLcolor}{\llbracket}  \SYSTEMnt{t_{{\mathrm{1}}}}  \textcolor{GLcolor}{\rrbracket}  
      \end{array}
      \end{align*}

      \item  \[\SYSTEMdruleGradEqbetaSumTwo{}\]
        As above but symmetrically for the other beta rule.

  \item Preservation of congruence equations follows simply by induction and congruence
    in the target calculus.

  \end{itemize}
\end{proof}

\subsection{Proof of Soundness for Linear Core to Graded Modal Core}
\label{app:proofs-lin-core-to-grad-core}

\ifextended\linToGradTranslationExt*\fi

\subsubsection{Type preservation}
\label{app:proofs-lin-core-to-grad-core-typ}

\begin{proof}
  The proof subsumes the earlier proofs; we add the additional cases here due to the
  extension of the language from Linear Base to Linear Core.
  \begin{itemize}
\item (prod$_i$)
$$
\SYSTEMdruleLinprodi{}
$$
% $$
% \SYSTEMdruleGradprodi{}
% $$
By induction on both premises we have
$  \textcolor{LGcolor}{\llparenthesis}  \Gamma_{{\mathrm{1}}}  \textcolor{LGcolor}{\rrparenthesis}   \vdash_{\textsc{g} }   \textcolor{LGcolor}{\llparenthesis} \smidge  \SYSTEMnt{t_{{\mathrm{1}}}}  \smidge \textcolor{LGcolor}{\rrparenthesis}   :   \textcolor{LGcolor}{\llparenthesis} \smidge  \SYSTEMnt{A}  \smidge \textcolor{LGcolor}{\rrparenthesis}  $
and
$  \textcolor{LGcolor}{\llparenthesis}  \Gamma_{{\mathrm{2}}}  \textcolor{LGcolor}{\rrparenthesis}   \vdash_{\textsc{g} }   \textcolor{LGcolor}{\llparenthesis} \smidge  \SYSTEMnt{t_{{\mathrm{2}}}}  \smidge \textcolor{LGcolor}{\rrparenthesis}   :   \textcolor{LGcolor}{\llparenthesis} \smidge  \SYSTEMnt{B}  \smidge \textcolor{LGcolor}{\rrparenthesis}  $.

Therefore we can construct:
$$
\inferrule*[Right=\SYSTEMRenameRuleGradprodi{}]
{  \textcolor{LGcolor}{\llparenthesis}  \Gamma_{{\mathrm{1}}}  \textcolor{LGcolor}{\rrparenthesis}   \vdash_{\textsc{g} }   \textcolor{LGcolor}{\llparenthesis} \smidge  \SYSTEMnt{t_{{\mathrm{1}}}}  \smidge \textcolor{LGcolor}{\rrparenthesis}   :   \textcolor{LGcolor}{\llparenthesis} \smidge  \SYSTEMnt{A}  \smidge \textcolor{LGcolor}{\rrparenthesis}   \\
   \textcolor{LGcolor}{\llparenthesis}  \Gamma_{{\mathrm{2}}}  \textcolor{LGcolor}{\rrparenthesis}   \vdash_{\textsc{g} }   \textcolor{LGcolor}{\llparenthesis} \smidge  \SYSTEMnt{t_{{\mathrm{2}}}}  \smidge \textcolor{LGcolor}{\rrparenthesis}   :   \textcolor{LGcolor}{\llparenthesis} \smidge  \SYSTEMnt{B}  \smidge \textcolor{LGcolor}{\rrparenthesis}  }
{  \textcolor{LGcolor}{\llparenthesis}  \Gamma_{{\mathrm{1}}}  \textcolor{LGcolor}{\rrparenthesis}   \SYSTEMsym{+}   \textcolor{LGcolor}{\llparenthesis}  \Gamma_{{\mathrm{2}}}  \textcolor{LGcolor}{\rrparenthesis}   \vdash_{\textsc{g} }   \langle   \textcolor{LGcolor}{\llparenthesis} \smidge  \SYSTEMnt{t_{{\mathrm{1}}}}  \smidge \textcolor{LGcolor}{\rrparenthesis}  ,   \textcolor{LGcolor}{\llparenthesis} \smidge  \SYSTEMnt{t_{{\mathrm{2}}}}  \smidge \textcolor{LGcolor}{\rrparenthesis}   \rangle   :    \textcolor{LGcolor}{\llparenthesis} \smidge  \SYSTEMnt{A}  \smidge \textcolor{LGcolor}{\rrparenthesis}   \times   \textcolor{LGcolor}{\llparenthesis} \smidge  \SYSTEMnt{B}  \smidge \textcolor{LGcolor}{\rrparenthesis}    }
$$

\item (prod$_e$)
$$
\SYSTEMdruleLinprode{}
$$
% $$
% \SYSTEMdruleGradprode{}
% $$
By induction on both premises we have
$  \textcolor{LGcolor}{\llparenthesis}  \Gamma_{{\mathrm{1}}}  \textcolor{LGcolor}{\rrparenthesis}   \vdash_{\textsc{g} }   \textcolor{LGcolor}{\llparenthesis} \smidge  \SYSTEMnt{t_{{\mathrm{1}}}}  \smidge \textcolor{LGcolor}{\rrparenthesis}   :    \textcolor{LGcolor}{\llparenthesis} \smidge  \SYSTEMnt{A}  \smidge \textcolor{LGcolor}{\rrparenthesis}   \times   \textcolor{LGcolor}{\llparenthesis} \smidge  \SYSTEMnt{B}  \smidge \textcolor{LGcolor}{\rrparenthesis}   $
and
$    \textcolor{LGcolor}{\llparenthesis}  \Gamma_{{\mathrm{2}}}  \textcolor{LGcolor}{\rrparenthesis}  ,   \SYSTEMmv{x}  :_{\textcolor{coeffectColor}{ \SYSTEMsym{1} } }   \textcolor{LGcolor}{\llparenthesis} \smidge  \SYSTEMnt{A}  \smidge \textcolor{LGcolor}{\rrparenthesis}    ,   \SYSTEMmv{y}  :_{\textcolor{coeffectColor}{ \SYSTEMsym{1} } }   \textcolor{LGcolor}{\llparenthesis} \smidge  \SYSTEMnt{B}  \smidge \textcolor{LGcolor}{\rrparenthesis}     \vdash_{\textsc{g} }   \textcolor{LGcolor}{\llparenthesis} \smidge  \SYSTEMnt{t_{{\mathrm{2}}}}  \smidge \textcolor{LGcolor}{\rrparenthesis}   :   \textcolor{LGcolor}{\llparenthesis} \smidge  \SYSTEMnt{C}  \smidge \textcolor{LGcolor}{\rrparenthesis}  $.

Therefore we can construct:
$$
\inferrule*[Right=\SYSTEMRenameRuleGradprode{}]
{  \textcolor{LGcolor}{\llparenthesis}  \Gamma_{{\mathrm{1}}}  \textcolor{LGcolor}{\rrparenthesis}   \vdash_{\textsc{g} }   \textcolor{LGcolor}{\llparenthesis} \smidge  \SYSTEMnt{t_{{\mathrm{1}}}}  \smidge \textcolor{LGcolor}{\rrparenthesis}   :    \textcolor{LGcolor}{\llparenthesis} \smidge  \SYSTEMnt{A}  \smidge \textcolor{LGcolor}{\rrparenthesis}   \times   \textcolor{LGcolor}{\llparenthesis} \smidge  \SYSTEMnt{B}  \smidge \textcolor{LGcolor}{\rrparenthesis}    \\
     \textcolor{LGcolor}{\llparenthesis}  \Gamma_{{\mathrm{2}}}  \textcolor{LGcolor}{\rrparenthesis}  ,   \SYSTEMmv{x}  :_{\textcolor{coeffectColor}{ \SYSTEMsym{1} } }   \textcolor{LGcolor}{\llparenthesis} \smidge  \SYSTEMnt{A}  \smidge \textcolor{LGcolor}{\rrparenthesis}    ,   \SYSTEMmv{y}  :_{\textcolor{coeffectColor}{ \SYSTEMsym{1} } }   \textcolor{LGcolor}{\llparenthesis} \smidge  \SYSTEMnt{B}  \smidge \textcolor{LGcolor}{\rrparenthesis}     \vdash_{\textsc{g} }   \textcolor{LGcolor}{\llparenthesis} \smidge  \SYSTEMnt{t_{{\mathrm{2}}}}  \smidge \textcolor{LGcolor}{\rrparenthesis}   :   \textcolor{LGcolor}{\llparenthesis} \smidge  \SYSTEMnt{C}  \smidge \textcolor{LGcolor}{\rrparenthesis}  }
{  \textcolor{coeffectColor}{ \SYSTEMsym{1}  \cdot}   \textcolor{LGcolor}{\llparenthesis}  \Gamma_{{\mathrm{1}}}  \textcolor{LGcolor}{\rrparenthesis}    \SYSTEMsym{+}   \textcolor{LGcolor}{\llparenthesis}  \Gamma_{{\mathrm{2}}}  \textcolor{LGcolor}{\rrparenthesis}   \vdash_{\textsc{g} }   \mathsf{let} \, \langle  \SYSTEMmv{x} ,  \SYSTEMmv{y}  \rangle =   \textcolor{LGcolor}{\llparenthesis} \smidge  \SYSTEMnt{t_{{\mathrm{1}}}}  \smidge \textcolor{LGcolor}{\rrparenthesis}   \, \mathsf{in} \,   \textcolor{LGcolor}{\llparenthesis} \smidge  \SYSTEMnt{t_{{\mathrm{2}}}}  \smidge \textcolor{LGcolor}{\rrparenthesis}    :   \textcolor{LGcolor}{\llparenthesis} \smidge  \SYSTEMnt{C}  \smidge \textcolor{LGcolor}{\rrparenthesis}   }
$$

\item (unit$_i$)
$$
\SYSTEMdruleLinuniti{}
$$

Therefore we construct the goal typing:
$$
\SYSTEMdruleGraduniti{}
$$

\item (unit$_e$)
$$
\SYSTEMdruleLinunite{}
$$
% $$
% \SYSTEMdruleGradunite{}
% $$
By induction on both premises we have
$  \textcolor{LGcolor}{\llparenthesis}  \Gamma_{{\mathrm{1}}}  \textcolor{LGcolor}{\rrparenthesis}   \vdash_{\textsc{g} }   \textcolor{LGcolor}{\llparenthesis} \smidge  \SYSTEMnt{t_{{\mathrm{1}}}}  \smidge \textcolor{LGcolor}{\rrparenthesis}   :   \textcolor{LGcolor}{\llparenthesis} \smidge   \mathrm{unit}   \smidge \textcolor{LGcolor}{\rrparenthesis}  $
and
$  \textcolor{LGcolor}{\llparenthesis}  \Gamma_{{\mathrm{2}}}  \textcolor{LGcolor}{\rrparenthesis}   \vdash_{\textsc{g} }   \textcolor{LGcolor}{\llparenthesis} \smidge  \SYSTEMnt{t_{{\mathrm{2}}}}  \smidge \textcolor{LGcolor}{\rrparenthesis}   :   \textcolor{LGcolor}{\llparenthesis} \smidge  \SYSTEMnt{A}  \smidge \textcolor{LGcolor}{\rrparenthesis}  $.

Therefore we can construct:
$$
\inferrule*[Right=\SYSTEMRenameRuleGradprode{}]
{  \textcolor{LGcolor}{\llparenthesis}  \Gamma_{{\mathrm{1}}}  \textcolor{LGcolor}{\rrparenthesis}   \vdash_{\textsc{g} }   \textcolor{LGcolor}{\llparenthesis} \smidge  \SYSTEMnt{t_{{\mathrm{1}}}}  \smidge \textcolor{LGcolor}{\rrparenthesis}   :   \textcolor{LGcolor}{\llparenthesis} \smidge   \mathrm{unit}   \smidge \textcolor{LGcolor}{\rrparenthesis}   \\
   \textcolor{LGcolor}{\llparenthesis}  \Gamma_{{\mathrm{2}}}  \textcolor{LGcolor}{\rrparenthesis}   \vdash_{\textsc{g} }   \textcolor{LGcolor}{\llparenthesis} \smidge  \SYSTEMnt{t_{{\mathrm{2}}}}  \smidge \textcolor{LGcolor}{\rrparenthesis}   :   \textcolor{LGcolor}{\llparenthesis} \smidge  \SYSTEMnt{A}  \smidge \textcolor{LGcolor}{\rrparenthesis}  }
{  \textcolor{LGcolor}{\llparenthesis}  \Gamma_{{\mathrm{1}}}  \textcolor{LGcolor}{\rrparenthesis}   \SYSTEMsym{+}   \textcolor{LGcolor}{\llparenthesis}  \Gamma_{{\mathrm{2}}}  \textcolor{LGcolor}{\rrparenthesis}   \vdash_{\textsc{g} }   \mathsf{let} \, \langle \rangle =   \textcolor{LGcolor}{\llparenthesis} \smidge  \SYSTEMnt{t_{{\mathrm{1}}}}  \smidge \textcolor{LGcolor}{\rrparenthesis}   \, \mathsf{in} \,   \textcolor{LGcolor}{\llparenthesis} \smidge  \SYSTEMnt{t_{{\mathrm{2}}}}  \smidge \textcolor{LGcolor}{\rrparenthesis}    :   \textcolor{LGcolor}{\llparenthesis} \smidge  \SYSTEMnt{A}  \smidge \textcolor{LGcolor}{\rrparenthesis}   }
$$

\item (sum$_{i1}$)
$$
\SYSTEMdruleLinsumiOne{}
$$
% $$
% \SYSTEMdruleGradsumiOne{}
% $$
By induction on the premise we have
$  \textcolor{LGcolor}{\llparenthesis}  \Gamma  \textcolor{LGcolor}{\rrparenthesis}   \vdash_{\textsc{g} }   \textcolor{LGcolor}{\llparenthesis} \smidge  \SYSTEMnt{t}  \smidge \textcolor{LGcolor}{\rrparenthesis}   :   \textcolor{LGcolor}{\llparenthesis} \smidge  \SYSTEMnt{A}  \smidge \textcolor{LGcolor}{\rrparenthesis}  $.

Therefore we can construct:
$$
\inferrule*[Right=\SYSTEMRenameRuleGradsumiOne{}]
{  \textcolor{LGcolor}{\llparenthesis}  \Gamma  \textcolor{LGcolor}{\rrparenthesis}   \vdash_{\textsc{g} }   \textcolor{LGcolor}{\llparenthesis} \smidge  \SYSTEMnt{t}  \smidge \textcolor{LGcolor}{\rrparenthesis}   :   \textcolor{LGcolor}{\llparenthesis} \smidge  \SYSTEMnt{A}  \smidge \textcolor{LGcolor}{\rrparenthesis}  }
{  \textcolor{LGcolor}{\llparenthesis}  \Gamma  \textcolor{LGcolor}{\rrparenthesis}   \vdash_{\textsc{g} }   \mathsf{inj}_1 \,   \textcolor{LGcolor}{\llparenthesis} \smidge  \SYSTEMnt{t}  \smidge \textcolor{LGcolor}{\rrparenthesis}    :    \textcolor{LGcolor}{\llparenthesis} \smidge  \SYSTEMnt{A}  \smidge \textcolor{LGcolor}{\rrparenthesis}   +   \textcolor{LGcolor}{\llparenthesis} \smidge  \SYSTEMnt{B}  \smidge \textcolor{LGcolor}{\rrparenthesis}    }
$$

\item (sum$_{i2}$)
$$
\SYSTEMdruleLinsumiTwo{}
$$
% $$
% \SYSTEMdruleGradsumiTwo{}
% $$
By induction on the premise we have
$  \textcolor{LGcolor}{\llparenthesis}  \Gamma  \textcolor{LGcolor}{\rrparenthesis}   \vdash_{\textsc{g} }   \textcolor{LGcolor}{\llparenthesis} \smidge  \SYSTEMnt{t}  \smidge \textcolor{LGcolor}{\rrparenthesis}   :   \textcolor{LGcolor}{\llparenthesis} \smidge  \SYSTEMnt{B}  \smidge \textcolor{LGcolor}{\rrparenthesis}  $.

Therefore we can construct:
$$
\inferrule*[Right=\SYSTEMRenameRuleGradsumiTwo{}]
{  \textcolor{LGcolor}{\llparenthesis}  \Gamma  \textcolor{LGcolor}{\rrparenthesis}   \vdash_{\textsc{g} }   \textcolor{LGcolor}{\llparenthesis} \smidge  \SYSTEMnt{t}  \smidge \textcolor{LGcolor}{\rrparenthesis}   :   \textcolor{LGcolor}{\llparenthesis} \smidge  \SYSTEMnt{B}  \smidge \textcolor{LGcolor}{\rrparenthesis}  }
{  \textcolor{LGcolor}{\llparenthesis}  \Gamma  \textcolor{LGcolor}{\rrparenthesis}   \vdash_{\textsc{g} }   \mathsf{inj}_2 \,   \textcolor{LGcolor}{\llparenthesis} \smidge  \SYSTEMnt{t}  \smidge \textcolor{LGcolor}{\rrparenthesis}    :    \textcolor{LGcolor}{\llparenthesis} \smidge  \SYSTEMnt{A}  \smidge \textcolor{LGcolor}{\rrparenthesis}   +   \textcolor{LGcolor}{\llparenthesis} \smidge  \SYSTEMnt{B}  \smidge \textcolor{LGcolor}{\rrparenthesis}    }
$$

\item (sum$_e$)
$$
\SYSTEMdruleLinsume{}
$$
% $$
% \SYSTEMdruleGradsume{}
% $$
By induction on the premises we have
$  \textcolor{LGcolor}{\llparenthesis}  \Gamma_{{\mathrm{1}}}  \textcolor{LGcolor}{\rrparenthesis}   \vdash_{\textsc{g} }   \textcolor{LGcolor}{\llparenthesis} \smidge  \SYSTEMnt{t}  \smidge \textcolor{LGcolor}{\rrparenthesis}   :    \textcolor{LGcolor}{\llparenthesis} \smidge  \SYSTEMnt{A}  \smidge \textcolor{LGcolor}{\rrparenthesis}   +   \textcolor{LGcolor}{\llparenthesis} \smidge  \SYSTEMnt{B}  \smidge \textcolor{LGcolor}{\rrparenthesis}   $,
$   \textcolor{LGcolor}{\llparenthesis}  \Gamma_{{\mathrm{2}}}  \textcolor{LGcolor}{\rrparenthesis}  ,   \SYSTEMmv{x}  :_{\textcolor{coeffectColor}{ \SYSTEMsym{1} } }   \textcolor{LGcolor}{\llparenthesis} \smidge  \SYSTEMnt{A}  \smidge \textcolor{LGcolor}{\rrparenthesis}     \vdash_{\textsc{g} }   \textcolor{LGcolor}{\llparenthesis} \smidge  \SYSTEMnt{t_{{\mathrm{1}}}}  \smidge \textcolor{LGcolor}{\rrparenthesis}   :   \textcolor{LGcolor}{\llparenthesis} \smidge  \SYSTEMnt{C}  \smidge \textcolor{LGcolor}{\rrparenthesis}  $
and
$   \textcolor{LGcolor}{\llparenthesis}  \Gamma_{{\mathrm{2}}}  \textcolor{LGcolor}{\rrparenthesis}  ,   \SYSTEMmv{y}  :_{\textcolor{coeffectColor}{ \SYSTEMsym{1} } }   \textcolor{LGcolor}{\llparenthesis} \smidge  \SYSTEMnt{B}  \smidge \textcolor{LGcolor}{\rrparenthesis}     \vdash_{\textsc{g} }   \textcolor{LGcolor}{\llparenthesis} \smidge  \SYSTEMnt{t_{{\mathrm{2}}}}  \smidge \textcolor{LGcolor}{\rrparenthesis}   :   \textcolor{LGcolor}{\llparenthesis} \smidge  \SYSTEMnt{C}  \smidge \textcolor{LGcolor}{\rrparenthesis}  $.

Using the unit property on graded contexts we can construct:
$$
\inferrule*[Right=\SYSTEMRenameRuleGradsume{}]
{ \SYSTEMsym{1}  \, \textcolor{coeffectColor}{\sqsubseteq} \,  \SYSTEMsym{1}  \qquad   \textcolor{LGcolor}{\llparenthesis}  \Gamma_{{\mathrm{1}}}  \textcolor{LGcolor}{\rrparenthesis}   \vdash_{\textsc{g} }   \textcolor{LGcolor}{\llparenthesis} \smidge  \SYSTEMnt{t}  \smidge \textcolor{LGcolor}{\rrparenthesis}   :    \textcolor{LGcolor}{\llparenthesis} \smidge  \SYSTEMnt{A}  \smidge \textcolor{LGcolor}{\rrparenthesis}   +   \textcolor{LGcolor}{\llparenthesis} \smidge  \SYSTEMnt{B}  \smidge \textcolor{LGcolor}{\rrparenthesis}   \\
    \textcolor{LGcolor}{\llparenthesis}  \Gamma_{{\mathrm{2}}}  \textcolor{LGcolor}{\rrparenthesis}  ,   \SYSTEMmv{x}  :_{\textcolor{coeffectColor}{ \SYSTEMsym{1} } }   \textcolor{LGcolor}{\llparenthesis} \smidge  \SYSTEMnt{A}  \smidge \textcolor{LGcolor}{\rrparenthesis}     \vdash_{\textsc{g} }   \textcolor{LGcolor}{\llparenthesis} \smidge  \SYSTEMnt{t_{{\mathrm{1}}}}  \smidge \textcolor{LGcolor}{\rrparenthesis}   :   \textcolor{LGcolor}{\llparenthesis} \smidge  \SYSTEMnt{C}  \smidge \textcolor{LGcolor}{\rrparenthesis}  \\
    \textcolor{LGcolor}{\llparenthesis}  \Gamma_{{\mathrm{2}}}  \textcolor{LGcolor}{\rrparenthesis}  ,   \SYSTEMmv{y}  :_{\textcolor{coeffectColor}{ \SYSTEMsym{1} } }   \textcolor{LGcolor}{\llparenthesis} \smidge  \SYSTEMnt{B}  \smidge \textcolor{LGcolor}{\rrparenthesis}     \vdash_{\textsc{g} }   \textcolor{LGcolor}{\llparenthesis} \smidge  \SYSTEMnt{t_{{\mathrm{2}}}}  \smidge \textcolor{LGcolor}{\rrparenthesis}   :   \textcolor{LGcolor}{\llparenthesis} \smidge  \SYSTEMnt{C}  \smidge \textcolor{LGcolor}{\rrparenthesis}  }
{  \textcolor{coeffectColor}{ \SYSTEMsym{1}  \cdot}   \textcolor{LGcolor}{\llparenthesis}  \Gamma_{{\mathrm{1}}}  \textcolor{LGcolor}{\rrparenthesis}    \SYSTEMsym{+}   \textcolor{LGcolor}{\llparenthesis}  \Gamma_{{\mathrm{2}}}  \textcolor{LGcolor}{\rrparenthesis}   \vdash_{\textsc{g} }   \mathsf{case} \,   \textcolor{LGcolor}{\llparenthesis} \smidge  \SYSTEMnt{t}  \smidge \textcolor{LGcolor}{\rrparenthesis}   \, \mathsf{of} \, \{ \mathsf{inj1} \,  \SYSTEMmv{x}  \rightarrow   \textcolor{LGcolor}{\llparenthesis} \smidge  \SYSTEMnt{t_{{\mathrm{1}}}}  \smidge \textcolor{LGcolor}{\rrparenthesis}   ; \, \mathsf{inj2} \,  \SYSTEMmv{y}  \rightarrow   \textcolor{LGcolor}{\llparenthesis} \smidge  \SYSTEMnt{t_{{\mathrm{2}}}}  \smidge \textcolor{LGcolor}{\rrparenthesis}   \}   :   \textcolor{LGcolor}{\llparenthesis} \smidge  \SYSTEMnt{C}  \smidge \textcolor{LGcolor}{\rrparenthesis}   }
$$

\item (push$_\otimes$)
  $$
  \SYSTEMdruleLinpushprod{}
  $$

  By induction on the premise we have (ih):
  $$  \textcolor{LGcolor}{\llparenthesis}  \Gamma  \textcolor{LGcolor}{\rrparenthesis}   \vdash_{\textsc{g} }   \textcolor{LGcolor}{\llparenthesis} \smidge  \SYSTEMnt{t}  \smidge \textcolor{LGcolor}{\rrparenthesis}   :   \textcolor{coeffectColor}{\square_{ \SYSTEMnt{r} } }   (    \textcolor{LGcolor}{\llparenthesis} \smidge  \SYSTEMnt{A}  \smidge \textcolor{LGcolor}{\rrparenthesis}   \times   \textcolor{LGcolor}{\llparenthesis} \smidge  \SYSTEMnt{B}  \smidge \textcolor{LGcolor}{\rrparenthesis}    )   $$

  Then we can construct:
  \begin{gather*}
  \inferrule*[Right=\SYSTEMRenameRuleGradBoxlet{}]
  {
    (ih) \quad \inferrule*[Right=\SYSTEMRenameRuleGradprode{}]
    {
    \inferrule*[Right=\SYSTEMRenameRuleGradvar{}]
         {\quad}
         {   \SYSTEMmv{x}  :_{\textcolor{coeffectColor}{ \SYSTEMsym{1} } }    \textcolor{LGcolor}{\llparenthesis} \smidge  \SYSTEMnt{A}  \smidge \textcolor{LGcolor}{\rrparenthesis}   \times   \textcolor{LGcolor}{\llparenthesis} \smidge  \SYSTEMnt{B}  \smidge \textcolor{LGcolor}{\rrparenthesis}      \vdash_{\textsc{g} }  \SYSTEMmv{x}  :    \textcolor{LGcolor}{\llparenthesis} \smidge  \SYSTEMnt{A}  \smidge \textcolor{LGcolor}{\rrparenthesis}   \times   \textcolor{LGcolor}{\llparenthesis} \smidge  \SYSTEMnt{B}  \smidge \textcolor{LGcolor}{\rrparenthesis}   }
       \qquad
       \inferrule*[Right=\SYSTEMRenameRuleGradprodi{}]
        {\inferrule*[Right=\SYSTEMRenameRuleGradBoxpr{}]
         {\inferrule*[Right=\SYSTEMRenameRuleGradvar{}]
          {\quad}
          {   \SYSTEMmv{y}  :_{\textcolor{coeffectColor}{ \SYSTEMsym{1} } }   \textcolor{LGcolor}{\llparenthesis} \smidge  \SYSTEMnt{A}  \smidge \textcolor{LGcolor}{\rrparenthesis}     \vdash_{\textsc{g} }  \SYSTEMmv{y}  :   \textcolor{LGcolor}{\llparenthesis} \smidge  \SYSTEMnt{A}  \smidge \textcolor{LGcolor}{\rrparenthesis}  }
         }
         {   \SYSTEMmv{y}  :_{\textcolor{coeffectColor}{ \SYSTEMnt{r} } }   \textcolor{LGcolor}{\llparenthesis} \smidge  \SYSTEMnt{A}  \smidge \textcolor{LGcolor}{\rrparenthesis}     \vdash_{\textsc{g} }   \textcolor{coeffectColor}{[}  \SYSTEMmv{y}  \textcolor{coeffectColor}{]}   :   \textcolor{coeffectColor}{\square_{ \SYSTEMnt{r} } }   \textcolor{LGcolor}{\llparenthesis} \smidge  \SYSTEMnt{A}  \smidge \textcolor{LGcolor}{\rrparenthesis}   }
         \qquad
         \inferrule*[Right=\SYSTEMRenameRuleGradBoxpr{}]
         {\inferrule*[Right=\SYSTEMRenameRuleGradvar{}]
          {\quad}
          {   \SYSTEMmv{z}  :_{\textcolor{coeffectColor}{ \SYSTEMsym{1} } }   \textcolor{LGcolor}{\llparenthesis} \smidge  \SYSTEMnt{B}  \smidge \textcolor{LGcolor}{\rrparenthesis}     \vdash_{\textsc{g} }  \SYSTEMmv{z}  :   \textcolor{LGcolor}{\llparenthesis} \smidge  \SYSTEMnt{B}  \smidge \textcolor{LGcolor}{\rrparenthesis}  }
         }
         {   \SYSTEMmv{z}  :_{\textcolor{coeffectColor}{ \SYSTEMnt{r} } }   \textcolor{LGcolor}{\llparenthesis} \smidge  \SYSTEMnt{B}  \smidge \textcolor{LGcolor}{\rrparenthesis}     \vdash_{\textsc{g} }   \textcolor{coeffectColor}{[}  \SYSTEMmv{z}  \textcolor{coeffectColor}{]}   :   \textcolor{coeffectColor}{\square_{ \SYSTEMnt{r} } }   \textcolor{LGcolor}{\llparenthesis} \smidge  \SYSTEMnt{B}  \smidge \textcolor{LGcolor}{\rrparenthesis}   }
       }
       {    \SYSTEMmv{y}  :_{\textcolor{coeffectColor}{ \SYSTEMnt{r} } }   \textcolor{LGcolor}{\llparenthesis} \smidge  \SYSTEMnt{A}  \smidge \textcolor{LGcolor}{\rrparenthesis}    ,   \SYSTEMmv{z}  :_{\textcolor{coeffectColor}{ \SYSTEMnt{r} } }   \textcolor{LGcolor}{\llparenthesis} \smidge  \SYSTEMnt{B}  \smidge \textcolor{LGcolor}{\rrparenthesis}     \vdash_{\textsc{g} }   \langle   \textcolor{coeffectColor}{[}  \SYSTEMmv{y}  \textcolor{coeffectColor}{]}  ,   \textcolor{coeffectColor}{[}  \SYSTEMmv{z}  \textcolor{coeffectColor}{]}   \rangle   :    \textcolor{coeffectColor}{\square_{ \SYSTEMnt{r} } }   \textcolor{LGcolor}{\llparenthesis} \smidge  \SYSTEMnt{A}  \smidge \textcolor{LGcolor}{\rrparenthesis}    \times   \textcolor{coeffectColor}{\square_{ \SYSTEMnt{r} } }   \textcolor{LGcolor}{\llparenthesis} \smidge  \SYSTEMnt{B}  \smidge \textcolor{LGcolor}{\rrparenthesis}    }
    }
    {   \textcolor{LGcolor}{\llparenthesis}  \Gamma  \textcolor{LGcolor}{\rrparenthesis}  ,   \SYSTEMmv{x}  :_{\textcolor{coeffectColor}{ \SYSTEMnt{r} } }   \textcolor{LGcolor}{\llparenthesis} \smidge  \SYSTEMnt{A}  \smidge \textcolor{LGcolor}{\rrparenthesis}     \vdash_{\textsc{g} }   \mathsf{let} \, \langle  \SYSTEMmv{y} ,  \SYSTEMmv{z}  \rangle =  \SYSTEMmv{x}  \, \mathsf{in} \,   \langle   \textcolor{coeffectColor}{[}  \SYSTEMmv{y}  \textcolor{coeffectColor}{]}  ,   \textcolor{coeffectColor}{[}  \SYSTEMmv{z}  \textcolor{coeffectColor}{]}   \rangle    :    \textcolor{coeffectColor}{\square_{ \SYSTEMnt{r} } }   \textcolor{LGcolor}{\llparenthesis} \smidge  \SYSTEMnt{A}  \smidge \textcolor{LGcolor}{\rrparenthesis}    \times   \textcolor{coeffectColor}{\square_{ \SYSTEMnt{r} } }   \textcolor{LGcolor}{\llparenthesis} \smidge  \SYSTEMnt{B}  \smidge \textcolor{LGcolor}{\rrparenthesis}    }
  }
  {  \textcolor{LGcolor}{\llparenthesis}  \Gamma  \textcolor{LGcolor}{\rrparenthesis}   \vdash_{\textsc{g} }   \mathsf{let} \, \textcolor{coeffectColor}{[}  \SYSTEMmv{x}  \textcolor{coeffectColor}{]} =   \textcolor{LGcolor}{\llparenthesis} \smidge  \SYSTEMnt{t}  \smidge \textcolor{LGcolor}{\rrparenthesis}   \, \mathsf{in} \,   \mathsf{let} \, \langle  \SYSTEMmv{y} ,  \SYSTEMmv{z}  \rangle =  \SYSTEMmv{x}  \, \mathsf{in} \,   \langle   \textcolor{coeffectColor}{[}  \SYSTEMmv{y}  \textcolor{coeffectColor}{]}  ,   \textcolor{coeffectColor}{[}  \SYSTEMmv{z}  \textcolor{coeffectColor}{]}   \rangle     :    \textcolor{coeffectColor}{\square_{ \SYSTEMnt{r} } }   \textcolor{LGcolor}{\llparenthesis} \smidge  \SYSTEMnt{A}  \smidge \textcolor{LGcolor}{\rrparenthesis}    \times   \textcolor{coeffectColor}{\square_{ \SYSTEMnt{r} } }   \textcolor{LGcolor}{\llparenthesis} \smidge  \SYSTEMnt{B}  \smidge \textcolor{LGcolor}{\rrparenthesis}    }
  \end{gather*}
  satisfying the goal.

\item (push$_\oplus$)
  $$
  \SYSTEMdruleLinpushsum{}
  $$
  By induction on the premise we have (ih):
  $$  \textcolor{LGcolor}{\llparenthesis}  \Gamma  \textcolor{LGcolor}{\rrparenthesis}   \vdash_{\textsc{g} }   \textcolor{LGcolor}{\llparenthesis} \smidge  \SYSTEMnt{t}  \smidge \textcolor{LGcolor}{\rrparenthesis}   :   \textcolor{coeffectColor}{\square_{ \SYSTEMnt{r} } }   (    \textcolor{LGcolor}{\llparenthesis} \smidge  \SYSTEMnt{A}  \smidge \textcolor{LGcolor}{\rrparenthesis}   +   \textcolor{LGcolor}{\llparenthesis} \smidge  \SYSTEMnt{B}  \smidge \textcolor{LGcolor}{\rrparenthesis}    )   $$

  We then construct:
  \begin{gather*}
    \inferrule*[Right=\SYSTEMRenameRuleGradBoxlet{}]
    {
      (ih) \quad \inferrule*[Right=\SYSTEMRenameRuleGradsume{}]
            {
         \inferrule*[Right=\SYSTEMRenameRuleGradvar{}]
           {\quad}
           {   \SYSTEMmv{x}  :_{\textcolor{coeffectColor}{ \SYSTEMsym{1} } }    \textcolor{LGcolor}{\llparenthesis} \smidge  \SYSTEMnt{A}  \smidge \textcolor{LGcolor}{\rrparenthesis}   +   \textcolor{LGcolor}{\llparenthesis} \smidge  \SYSTEMnt{B}  \smidge \textcolor{LGcolor}{\rrparenthesis}      \vdash_{\textsc{g} }  \SYSTEMmv{x}  :    \textcolor{LGcolor}{\llparenthesis} \smidge  \SYSTEMnt{A}  \smidge \textcolor{LGcolor}{\rrparenthesis}   +   \textcolor{LGcolor}{\llparenthesis} \smidge  \SYSTEMnt{B}  \smidge \textcolor{LGcolor}{\rrparenthesis}   }
         \qquad
          {\inferrule*[Right=\SYSTEMRenameRuleGradsumiOne{}]
           {\inferrule*[Right=\SYSTEMRenameRuleGradBoxpr{}]
           {\inferrule*[Right=\SYSTEMRenameRuleGradvar{}]
            {\quad}
            {   \SYSTEMmv{y}  :_{\textcolor{coeffectColor}{ \SYSTEMsym{1} } }   \textcolor{LGcolor}{\llparenthesis} \smidge  \SYSTEMnt{A}  \smidge \textcolor{LGcolor}{\rrparenthesis}     \vdash_{\textsc{g} }  \SYSTEMmv{y}  :   \textcolor{LGcolor}{\llparenthesis} \smidge  \SYSTEMnt{A}  \smidge \textcolor{LGcolor}{\rrparenthesis}  }
           }
           {   \SYSTEMmv{y}  :_{\textcolor{coeffectColor}{ \SYSTEMnt{r} } }   \textcolor{LGcolor}{\llparenthesis} \smidge  \SYSTEMnt{A}  \smidge \textcolor{LGcolor}{\rrparenthesis}     \vdash_{\textsc{g} }   \textcolor{coeffectColor}{[}  \SYSTEMmv{y}  \textcolor{coeffectColor}{]}   :   \textcolor{coeffectColor}{\square_{ \SYSTEMnt{r} } }   \textcolor{LGcolor}{\llparenthesis} \smidge  \SYSTEMnt{A}  \smidge \textcolor{LGcolor}{\rrparenthesis}   }
           }
           {   \SYSTEMmv{y}  :_{\textcolor{coeffectColor}{ \SYSTEMnt{r} } }   \textcolor{LGcolor}{\llparenthesis} \smidge  \SYSTEMnt{A}  \smidge \textcolor{LGcolor}{\rrparenthesis}     \vdash_{\textsc{g} }   \mathsf{inj}_1 \,   \textcolor{coeffectColor}{[}  \SYSTEMmv{y}  \textcolor{coeffectColor}{]}    :    \textcolor{coeffectColor}{\square_{ \SYSTEMnt{r} } }   \textcolor{LGcolor}{\llparenthesis} \smidge  \SYSTEMnt{A}  \smidge \textcolor{LGcolor}{\rrparenthesis}    +   \textcolor{coeffectColor}{\square_{ \SYSTEMnt{r} } }   \textcolor{LGcolor}{\llparenthesis} \smidge  \SYSTEMnt{B}  \smidge \textcolor{LGcolor}{\rrparenthesis}    }
          }
          \qquad
           \inferrule*[Right=\SYSTEMRenameRuleGradsumiTwo{}]
           {\inferrule*[Right=\SYSTEMRenameRuleGradBoxpr{}]
           {\inferrule*[Right=\SYSTEMRenameRuleGradvar{}]
            {\quad}
            {   \SYSTEMmv{z}  :_{\textcolor{coeffectColor}{ \SYSTEMsym{1} } }   \textcolor{LGcolor}{\llparenthesis} \smidge  \SYSTEMnt{B}  \smidge \textcolor{LGcolor}{\rrparenthesis}     \vdash_{\textsc{g} }  \SYSTEMmv{z}  :   \textcolor{LGcolor}{\llparenthesis} \smidge  \SYSTEMnt{B}  \smidge \textcolor{LGcolor}{\rrparenthesis}  }
           }
           {   \SYSTEMmv{z}  :_{\textcolor{coeffectColor}{ \SYSTEMnt{r} } }   \textcolor{LGcolor}{\llparenthesis} \smidge  \SYSTEMnt{B}  \smidge \textcolor{LGcolor}{\rrparenthesis}     \vdash_{\textsc{g} }   \textcolor{coeffectColor}{[}  \SYSTEMmv{z}  \textcolor{coeffectColor}{]}   :   \textcolor{coeffectColor}{\square_{ \SYSTEMnt{r} } }   \textcolor{LGcolor}{\llparenthesis} \smidge  \SYSTEMnt{B}  \smidge \textcolor{LGcolor}{\rrparenthesis}   }
           }
           {   \SYSTEMmv{z}  :_{\textcolor{coeffectColor}{ \SYSTEMnt{r} } }   \textcolor{LGcolor}{\llparenthesis} \smidge  \SYSTEMnt{B}  \smidge \textcolor{LGcolor}{\rrparenthesis}     \vdash_{\textsc{g} }   \mathsf{inj}_2 \,   \textcolor{coeffectColor}{[}  \SYSTEMmv{z}  \textcolor{coeffectColor}{]}    :    \textcolor{coeffectColor}{\square_{ \SYSTEMnt{r} } }   \textcolor{LGcolor}{\llparenthesis} \smidge  \SYSTEMnt{A}  \smidge \textcolor{LGcolor}{\rrparenthesis}    +   \textcolor{coeffectColor}{\square_{ \SYSTEMnt{r} } }   \textcolor{LGcolor}{\llparenthesis} \smidge  \SYSTEMnt{B}  \smidge \textcolor{LGcolor}{\rrparenthesis}    }
         }
      {   \SYSTEMmv{x}  :_{\textcolor{coeffectColor}{ \SYSTEMnt{r} } }   (    \textcolor{LGcolor}{\llparenthesis} \smidge  \SYSTEMnt{A}  \smidge \textcolor{LGcolor}{\rrparenthesis}   +   \textcolor{LGcolor}{\llparenthesis} \smidge  \SYSTEMnt{B}  \smidge \textcolor{LGcolor}{\rrparenthesis}    )     \vdash_{\textsc{g} }   \mathsf{case} \,  \SYSTEMmv{x}  \, \mathsf{of} \, \{ \mathsf{inj1} \,  \SYSTEMmv{y}  \rightarrow   \mathsf{inj}_1 \,   \textcolor{coeffectColor}{[}  \SYSTEMmv{y}  \textcolor{coeffectColor}{]}    ; \, \mathsf{inj2} \,  \SYSTEMmv{z}  \rightarrow   \mathsf{inj}_2 \,   \textcolor{coeffectColor}{[}  \SYSTEMmv{z}  \textcolor{coeffectColor}{]}    \}   :    \textcolor{coeffectColor}{\square_{ \SYSTEMnt{r} } }   \textcolor{LGcolor}{\llparenthesis} \smidge  \SYSTEMnt{A}  \smidge \textcolor{LGcolor}{\rrparenthesis}    +   \textcolor{coeffectColor}{\square_{ \SYSTEMnt{r} } }   \textcolor{LGcolor}{\llparenthesis} \smidge  \SYSTEMnt{B}  \smidge \textcolor{LGcolor}{\rrparenthesis}    }
    }
    {  \textcolor{LGcolor}{\llparenthesis}  \Gamma  \textcolor{LGcolor}{\rrparenthesis}   \vdash_{\textsc{g} }   \mathsf{let} \, \textcolor{coeffectColor}{[}  \SYSTEMmv{x}  \textcolor{coeffectColor}{]} =   \textcolor{LGcolor}{\llparenthesis} \smidge  \SYSTEMnt{t}  \smidge \textcolor{LGcolor}{\rrparenthesis}   \, \mathsf{in} \,   \mathsf{case} \,  \SYSTEMmv{x}  \, \mathsf{of} \, \{ \mathsf{inj1} \,  \SYSTEMmv{y}  \rightarrow   \mathsf{inj}_1 \,   \textcolor{coeffectColor}{[}  \SYSTEMmv{y}  \textcolor{coeffectColor}{]}    ; \, \mathsf{inj2} \,  \SYSTEMmv{z}  \rightarrow   \mathsf{inj}_2 \,   \textcolor{coeffectColor}{[}  \SYSTEMmv{z}  \textcolor{coeffectColor}{]}    \}    :    \textcolor{coeffectColor}{\square_{ \SYSTEMnt{r} } }   \textcolor{LGcolor}{\llparenthesis} \smidge  \SYSTEMnt{A}  \smidge \textcolor{LGcolor}{\rrparenthesis}    +   \textcolor{coeffectColor}{\square_{ \SYSTEMnt{r} } }   \textcolor{LGcolor}{\llparenthesis} \smidge  \SYSTEMnt{B}  \smidge \textcolor{LGcolor}{\rrparenthesis}    }
    \end{gather*}

\item ($ \textsf{push}_{\mathrm{unit} }     $)
  $$
  \SYSTEMdruleLinpushunit{}
  $$
  By induction on the premise we have (ih):
  $$  \textcolor{LGcolor}{\llparenthesis}  \Gamma  \textcolor{LGcolor}{\rrparenthesis}   \vdash_{\textsc{g} }   \textcolor{LGcolor}{\llparenthesis} \smidge  \SYSTEMnt{t}  \smidge \textcolor{LGcolor}{\rrparenthesis}   :   \textcolor{coeffectColor}{\square_{ \SYSTEMnt{r} } }   \mathrm{unit}   $$

  Then we construct the typing:
\begin{gather*}
\inferrule*[Right=\SYSTEMRenameRuleGradBoxlet{}]
{
  (ih) :   \textcolor{LGcolor}{\llparenthesis}  \Gamma  \textcolor{LGcolor}{\rrparenthesis}   \vdash_{\textsc{g} }   \textcolor{LGcolor}{\llparenthesis} \smidge  \SYSTEMnt{t}  \smidge \textcolor{LGcolor}{\rrparenthesis}   :   \textcolor{coeffectColor}{\square_{ \SYSTEMnt{r} } }   \mathrm{unit}   
  \quad
  \inferrule*[Right=\SYSTEMRenameRuleGradunite{}]
    {
      \inferrule*[Right=\SYSTEMRenameRuleGradvar{}]
        {\quad}
        {   \SYSTEMmv{x}  :_{\textcolor{coeffectColor}{ \SYSTEMsym{1} } }   \mathrm{unit}     \vdash_{\textsc{g} }  \SYSTEMmv{x}  :   \mathrm{unit}  }
      \quad\quad
      \inferrule*[Right=\SYSTEMRenameRuleGraduniti{}]
        {\quad}
        { \emptyset  \vdash_{\textsc{g}}  \langle \rangle  :  \mathrm{unit} }
    }
    {   \SYSTEMmv{x}  :_{\textcolor{coeffectColor}{ \SYSTEMnt{r} } }   \mathrm{unit}     \vdash_{\textsc{g} }   \mathsf{let} \, \langle \rangle =  \SYSTEMmv{x}  \, \mathsf{in} \,   \langle \rangle    :   \mathrm{unit}  }
  }
{  \textcolor{LGcolor}{\llparenthesis}  \Gamma  \textcolor{LGcolor}{\rrparenthesis}   \vdash_{\textsc{g} }   \mathsf{let} \, \textcolor{coeffectColor}{[}  \SYSTEMmv{x}  \textcolor{coeffectColor}{]} =   \textcolor{LGcolor}{\llparenthesis} \smidge  \SYSTEMnt{t}  \smidge \textcolor{LGcolor}{\rrparenthesis}   \, \mathsf{in} \,   \mathsf{let} \, \langle \rangle =  \SYSTEMmv{x}  \, \mathsf{in} \,   \langle \rangle     :   \mathrm{unit}  }
\end{gather*}

  \end{itemize}
\end{proof}

\subsubsection{Operational correspondence}
\label{app:proofs-lin-core-to-grad-core-ops}

\begin{lemma}[Interpretation preserves substitution]
\label{lemma:interp-lin-to-grd-core-preserves-subst}
For all Linear Base terms $\SYSTEMnt{t}, \SYSTEMnt{t'}$ then
$ \textcolor{LGcolor}{\llparenthesis} \smidge   [  \SYSTEMnt{t}  /  \SYSTEMmv{x}  ]  \SYSTEMnt{t'}   \smidge \textcolor{LGcolor}{\rrparenthesis}  \equiv  [   \textcolor{LGcolor}{\llparenthesis} \smidge  \SYSTEMnt{t}  \smidge \textcolor{LGcolor}{\rrparenthesis}   /  \SYSTEMmv{x}  ]   \textcolor{LGcolor}{\llparenthesis} \smidge  \SYSTEMnt{t'}  \smidge \textcolor{LGcolor}{\rrparenthesis}  $.
\end{lemma}

\begin{proof}
  We extend the proof of Lemma~\ref{lemma:interp-lin-to-grd-mod-preserves-subst} with the new cases:
  \begin{itemize}
    \item ($\otimes_i$) $\SYSTEMnt{t'} \equiv  \langle  \SYSTEMnt{t_{{\mathrm{1}}}} ,  \SYSTEMnt{t_{{\mathrm{2}}}}  \rangle $;
\begin{gather*}
\begin{align*}
\begin{array}{rcrlll}
\textit{(goal)} & &  \textcolor{LGcolor}{\llparenthesis} \smidge   [  \SYSTEMnt{t}  /  \SYSTEMmv{x}  ]   \langle  \SYSTEMnt{t_{{\mathrm{1}}}} ,  \SYSTEMnt{t_{{\mathrm{2}}}}  \rangle    \smidge \textcolor{LGcolor}{\rrparenthesis}  & \eqStep &  [   \textcolor{LGcolor}{\llparenthesis} \smidge  \SYSTEMnt{t}  \smidge \textcolor{LGcolor}{\rrparenthesis}   /  \SYSTEMmv{x}  ]   \textcolor{LGcolor}{\llparenthesis} \smidge   \langle  \SYSTEMnt{t_{{\mathrm{1}}}} ,  \SYSTEMnt{t_{{\mathrm{2}}}}  \rangle   \smidge \textcolor{LGcolor}{\rrparenthesis}   \\
\textit{(defn. subst [lhs])} & \Rightarrow &  \textcolor{LGcolor}{\llparenthesis} \smidge   \langle   [  \SYSTEMnt{t}  /  \SYSTEMmv{x}  ]  \SYSTEMnt{t_{{\mathrm{1}}}}  ,   [  \SYSTEMnt{t}  /  \SYSTEMmv{x}  ]  \SYSTEMnt{t_{{\mathrm{2}}}}   \rangle   \smidge \textcolor{LGcolor}{\rrparenthesis}  & \eqStep &  [   \textcolor{LGcolor}{\llparenthesis} \smidge  \SYSTEMnt{t}  \smidge \textcolor{LGcolor}{\rrparenthesis}   /  \SYSTEMmv{x}  ]   \textcolor{LGcolor}{\llparenthesis} \smidge   \langle  \SYSTEMnt{t_{{\mathrm{1}}}} ,  \SYSTEMnt{t_{{\mathrm{2}}}}  \rangle   \smidge \textcolor{LGcolor}{\rrparenthesis}   \\
\textit{(defn. interp [lhs])} & \Rightarrow &  \langle   \textcolor{LGcolor}{\llparenthesis} \smidge   [  \SYSTEMnt{t}  /  \SYSTEMmv{x}  ]  \SYSTEMnt{t_{{\mathrm{1}}}}   \smidge \textcolor{LGcolor}{\rrparenthesis}  ,   \textcolor{LGcolor}{\llparenthesis} \smidge   [  \SYSTEMnt{t}  /  \SYSTEMmv{x}  ]  \SYSTEMnt{t_{{\mathrm{2}}}}   \smidge \textcolor{LGcolor}{\rrparenthesis}   \rangle  & \eqStep &  [   \textcolor{LGcolor}{\llparenthesis} \smidge  \SYSTEMnt{t}  \smidge \textcolor{LGcolor}{\rrparenthesis}   /  \SYSTEMmv{x}  ]   \textcolor{LGcolor}{\llparenthesis} \smidge   \langle  \SYSTEMnt{t_{{\mathrm{1}}}} ,  \SYSTEMnt{t_{{\mathrm{2}}}}  \rangle   \smidge \textcolor{LGcolor}{\rrparenthesis}   \\
\end{array}
\end{align*}
\end{gather*}
\item ($\otimes_e$) $\SYSTEMnt{t'} \equiv  \mathsf{let} \, \langle  \SYSTEMmv{y} ,  \SYSTEMmv{z}  \rangle =  \SYSTEMnt{t_{{\mathrm{1}}}}  \, \mathsf{in} \,  \SYSTEMnt{t_{{\mathrm{2}}}} $ (with $ \SYSTEMmv{y} ,   \SYSTEMmv{z}  \,\#\,  \SYSTEMnt{t}  $);
\begin{gather*}
\begin{align*}
\begin{array}{rcrlll}
\textit{(goal)} & &  \textcolor{LGcolor}{\llparenthesis} \smidge   [  \SYSTEMnt{t}  /  \SYSTEMmv{x}  ]   \mathsf{let} \, \langle  \SYSTEMmv{y} ,  \SYSTEMmv{z}  \rangle =  \SYSTEMnt{t_{{\mathrm{1}}}}  \, \mathsf{in} \,  \SYSTEMnt{t_{{\mathrm{2}}}}    \smidge \textcolor{LGcolor}{\rrparenthesis}  & \eqStep &  [   \textcolor{LGcolor}{\llparenthesis} \smidge  \SYSTEMnt{t}  \smidge \textcolor{LGcolor}{\rrparenthesis}   /  \SYSTEMmv{x}  ]   \textcolor{LGcolor}{\llparenthesis} \smidge   \mathsf{let} \, \langle  \SYSTEMmv{y} ,  \SYSTEMmv{z}  \rangle =  \SYSTEMnt{t_{{\mathrm{1}}}}  \, \mathsf{in} \,  \SYSTEMnt{t_{{\mathrm{2}}}}   \smidge \textcolor{LGcolor}{\rrparenthesis}   \\
\textit{(defn. subst [lhs])} & \Rightarrow &  \textcolor{LGcolor}{\llparenthesis} \smidge   \mathsf{let} \, \langle  \SYSTEMmv{y} ,  \SYSTEMmv{z}  \rangle =   [  \SYSTEMnt{t}  /  \SYSTEMmv{x}  ]  \SYSTEMnt{t_{{\mathrm{1}}}}   \, \mathsf{in} \,   [  \SYSTEMnt{t}  /  \SYSTEMmv{x}  ]  \SYSTEMnt{t_{{\mathrm{2}}}}    \smidge \textcolor{LGcolor}{\rrparenthesis}  & \eqStep &  [   \textcolor{LGcolor}{\llparenthesis} \smidge  \SYSTEMnt{t}  \smidge \textcolor{LGcolor}{\rrparenthesis}   /  \SYSTEMmv{x}  ]   \textcolor{LGcolor}{\llparenthesis} \smidge   \mathsf{let} \, \langle  \SYSTEMmv{y} ,  \SYSTEMmv{z}  \rangle =  \SYSTEMnt{t_{{\mathrm{1}}}}  \, \mathsf{in} \,  \SYSTEMnt{t_{{\mathrm{2}}}}   \smidge \textcolor{LGcolor}{\rrparenthesis}   \\
\end{array}
\end{align*}
\end{gather*}
\item ($ \mathrm{unit} _i$) $\SYSTEMnt{t'} \equiv  \langle \rangle $;
\begin{align*}
\begin{array}{rcrlll}
\textit{(goal)} & &  \textcolor{LGcolor}{\llparenthesis} \smidge   [  \SYSTEMnt{t}  /  \SYSTEMmv{x}  ]   \langle \rangle    \smidge \textcolor{LGcolor}{\rrparenthesis}  & \eqStep &  [   \textcolor{LGcolor}{\llparenthesis} \smidge  \SYSTEMnt{t}  \smidge \textcolor{LGcolor}{\rrparenthesis}   /  \SYSTEMmv{x}  ]   \textcolor{LGcolor}{\llparenthesis} \smidge   \langle \rangle   \smidge \textcolor{LGcolor}{\rrparenthesis}   \\
\textit{(defn. subst [lhs])} & \Rightarrow &  \textcolor{LGcolor}{\llparenthesis} \smidge   \langle \rangle   \smidge \textcolor{LGcolor}{\rrparenthesis}  & \eqStep &  [   \textcolor{LGcolor}{\llparenthesis} \smidge  \SYSTEMnt{t}  \smidge \textcolor{LGcolor}{\rrparenthesis}   /  \SYSTEMmv{x}  ]   \textcolor{LGcolor}{\llparenthesis} \smidge   \langle \rangle   \smidge \textcolor{LGcolor}{\rrparenthesis}   \\
\textit{(defn. subst [rhs])} & \Rightarrow &  \textcolor{LGcolor}{\llparenthesis} \smidge   \langle \rangle   \smidge \textcolor{LGcolor}{\rrparenthesis}  & \eqStep &  \textcolor{LGcolor}{\llparenthesis} \smidge   \langle \rangle   \smidge \textcolor{LGcolor}{\rrparenthesis}  \\
\textit{(reflexivity)} & \Rightarrow & \top
\end{array}
\end{align*}
\item ($ \mathrm{unit} _e$) $\SYSTEMnt{t'} \equiv  \mathsf{let} \, \langle \rangle =  \SYSTEMnt{t_{{\mathrm{1}}}}  \, \mathsf{in} \,  \SYSTEMnt{t_{{\mathrm{2}}}} $;
\begin{gather*}
  \begin{align*}
  \begin{array}{rcrlll}
  \textit{(goal)} & &  \textcolor{LGcolor}{\llparenthesis} \smidge   [  \SYSTEMnt{t}  /  \SYSTEMmv{x}  ]   \mathsf{let} \, \langle \rangle =  \SYSTEMnt{t_{{\mathrm{1}}}}  \, \mathsf{in} \,  \SYSTEMnt{t_{{\mathrm{2}}}}    \smidge \textcolor{LGcolor}{\rrparenthesis}  & \eqStep &  [   \textcolor{LGcolor}{\llparenthesis} \smidge  \SYSTEMnt{t}  \smidge \textcolor{LGcolor}{\rrparenthesis}   /  \SYSTEMmv{x}  ]   \textcolor{LGcolor}{\llparenthesis} \smidge   \mathsf{let} \, \langle \rangle =  \SYSTEMnt{t_{{\mathrm{1}}}}  \, \mathsf{in} \,  \SYSTEMnt{t_{{\mathrm{2}}}}   \smidge \textcolor{LGcolor}{\rrparenthesis}   \\
  \textit{(defn. subst [lhs])} & \Rightarrow &  \textcolor{LGcolor}{\llparenthesis} \smidge   \mathsf{let} \, \langle \rangle =   [  \SYSTEMnt{t}  /  \SYSTEMmv{x}  ]  \SYSTEMnt{t_{{\mathrm{1}}}}   \, \mathsf{in} \,   [  \SYSTEMnt{t}  /  \SYSTEMmv{x}  ]  \SYSTEMnt{t_{{\mathrm{2}}}}    \smidge \textcolor{LGcolor}{\rrparenthesis}  & \eqStep &  [   \textcolor{LGcolor}{\llparenthesis} \smidge  \SYSTEMnt{t}  \smidge \textcolor{LGcolor}{\rrparenthesis}   /  \SYSTEMmv{x}  ]   \textcolor{LGcolor}{\llparenthesis} \smidge   \mathsf{let} \, \langle \rangle =  \SYSTEMnt{t_{{\mathrm{1}}}}  \, \mathsf{in} \,  \SYSTEMnt{t_{{\mathrm{2}}}}   \smidge \textcolor{LGcolor}{\rrparenthesis}   \\
\end{array}
  \end{align*}
\end{gather*}
\item ($\oplus_{i1}$) $\SYSTEMnt{t'} \equiv  \mathsf{inj}_1 \,  \SYSTEMnt{t_{{\mathrm{1}}}} $;
\begin{align*}
\begin{array}{rcrlll}
\textit{(goal)}               &             &  \textcolor{LGcolor}{\llparenthesis} \smidge   [  \SYSTEMnt{t}  /  \SYSTEMmv{x}  ]  \SYSTEMsym{(}   \mathsf{inj}_1 \,  \SYSTEMnt{t_{{\mathrm{1}}}}   \SYSTEMsym{)}   \smidge \textcolor{LGcolor}{\rrparenthesis}  & \eqStep &  [   \textcolor{LGcolor}{\llparenthesis} \smidge  \SYSTEMnt{t}  \smidge \textcolor{LGcolor}{\rrparenthesis}   /  \SYSTEMmv{x}  ]   \textcolor{LGcolor}{\llparenthesis} \smidge   \mathsf{inj}_1 \,  \SYSTEMnt{t_{{\mathrm{1}}}}   \smidge \textcolor{LGcolor}{\rrparenthesis}   \\
\textit{(defn. subst [lhs])}  & \Rightarrow &  \textcolor{LGcolor}{\llparenthesis} \smidge   \mathsf{inj}_1 \,  \SYSTEMsym{(}   [  \SYSTEMnt{t}  /  \SYSTEMmv{x}  ]  \SYSTEMnt{t_{{\mathrm{1}}}}   \SYSTEMsym{)}   \smidge \textcolor{LGcolor}{\rrparenthesis}  & \eqStep &  [   \textcolor{LGcolor}{\llparenthesis} \smidge  \SYSTEMnt{t}  \smidge \textcolor{LGcolor}{\rrparenthesis}   /  \SYSTEMmv{x}  ]   \textcolor{LGcolor}{\llparenthesis} \smidge   \mathsf{inj}_1 \,  \SYSTEMnt{t_{{\mathrm{1}}}}   \smidge \textcolor{LGcolor}{\rrparenthesis}   \\
\textit{(defn. interp [lhs])} & \Rightarrow &  \mathsf{inj}_1 \,   \textcolor{LGcolor}{\llparenthesis} \smidge   [  \SYSTEMnt{t}  /  \SYSTEMmv{x}  ]  \SYSTEMnt{t_{{\mathrm{1}}}}   \smidge \textcolor{LGcolor}{\rrparenthesis}     & \eqStep &  [   \textcolor{LGcolor}{\llparenthesis} \smidge  \SYSTEMnt{t}  \smidge \textcolor{LGcolor}{\rrparenthesis}   /  \SYSTEMmv{x}  ]   \textcolor{LGcolor}{\llparenthesis} \smidge   \mathsf{inj}_1 \,  \SYSTEMnt{t_{{\mathrm{1}}}}   \smidge \textcolor{LGcolor}{\rrparenthesis}   \\
\end{array}
\end{align*}
\item ($\oplus_{i2}$) $\SYSTEMnt{t'} \equiv  \mathsf{inj}_2 \,  \SYSTEMnt{t_{{\mathrm{2}}}} $;
\begin{align*}
\begin{array}{rcrlll}
\textit{(goal)}               &             &  \textcolor{LGcolor}{\llparenthesis} \smidge   [  \SYSTEMnt{t}  /  \SYSTEMmv{x}  ]  \SYSTEMsym{(}   \mathsf{inj}_2 \,  \SYSTEMnt{t_{{\mathrm{2}}}}   \SYSTEMsym{)}   \smidge \textcolor{LGcolor}{\rrparenthesis}  & \eqStep &  [   \textcolor{LGcolor}{\llparenthesis} \smidge  \SYSTEMnt{t}  \smidge \textcolor{LGcolor}{\rrparenthesis}   /  \SYSTEMmv{x}  ]   \textcolor{LGcolor}{\llparenthesis} \smidge   \mathsf{inj}_2 \,  \SYSTEMnt{t_{{\mathrm{2}}}}   \smidge \textcolor{LGcolor}{\rrparenthesis}   \\
\textit{(defn. subst [lhs])}  & \Rightarrow &  \textcolor{LGcolor}{\llparenthesis} \smidge   \mathsf{inj}_2 \,  \SYSTEMsym{(}   [  \SYSTEMnt{t}  /  \SYSTEMmv{x}  ]  \SYSTEMnt{t_{{\mathrm{2}}}}   \SYSTEMsym{)}   \smidge \textcolor{LGcolor}{\rrparenthesis}  & \eqStep &  [   \textcolor{LGcolor}{\llparenthesis} \smidge  \SYSTEMnt{t}  \smidge \textcolor{LGcolor}{\rrparenthesis}   /  \SYSTEMmv{x}  ]   \textcolor{LGcolor}{\llparenthesis} \smidge   \mathsf{inj}_2 \,  \SYSTEMnt{t_{{\mathrm{2}}}}   \smidge \textcolor{LGcolor}{\rrparenthesis}   \\
\textit{(defn. interp [lhs])} & \Rightarrow &  \mathsf{inj}_2 \,   \textcolor{LGcolor}{\llparenthesis} \smidge   [  \SYSTEMnt{t}  /  \SYSTEMmv{x}  ]  \SYSTEMnt{t_{{\mathrm{2}}}}   \smidge \textcolor{LGcolor}{\rrparenthesis}     & \eqStep &  [   \textcolor{LGcolor}{\llparenthesis} \smidge  \SYSTEMnt{t}  \smidge \textcolor{LGcolor}{\rrparenthesis}   /  \SYSTEMmv{x}  ]   \textcolor{LGcolor}{\llparenthesis} \smidge   \mathsf{inj}_2 \,  \SYSTEMnt{t_{{\mathrm{2}}}}   \smidge \textcolor{LGcolor}{\rrparenthesis}   \\
\end{array}
\end{align*}
\item ($\oplus_e$) $\SYSTEMnt{t'} \equiv  \mathsf{case} \,  \SYSTEMnt{t_{{\mathrm{1}}}}  \, \mathsf{of} \, \{ \mathsf{inj1} \,  \SYSTEMmv{y}  \rightarrow  \SYSTEMnt{t_{{\mathrm{2}}}}  ; \, \mathsf{inj2} \,  \SYSTEMmv{z}  \rightarrow  \SYSTEMnt{t_{{\mathrm{3}}}}  \} $ (with $ \SYSTEMmv{y} ,   \SYSTEMmv{z}  \,\#\,  \SYSTEMnt{t}  $);
\begin{gather*}
  \begin{align*}
  \begin{array}{rcl}
  \textit{(goal)}               &             &  \textcolor{LGcolor}{\llparenthesis} \smidge   [  \SYSTEMnt{t}  /  \SYSTEMmv{x}  ]   \mathsf{case} \,  \SYSTEMnt{t_{{\mathrm{1}}}}  \, \mathsf{of} \, \{ \mathsf{inj1} \,  \SYSTEMmv{y}  \rightarrow  \SYSTEMnt{t_{{\mathrm{2}}}}  ; \, \mathsf{inj2} \,  \SYSTEMmv{z}  \rightarrow  \SYSTEMnt{t_{{\mathrm{3}}}}  \}    \smidge \textcolor{LGcolor}{\rrparenthesis}                                  \\
                                &             & \quad \equiv  [   \textcolor{LGcolor}{\llparenthesis} \smidge  \SYSTEMnt{t}  \smidge \textcolor{LGcolor}{\rrparenthesis}   /  \SYSTEMmv{x}  ]   \textcolor{LGcolor}{\llparenthesis} \smidge   \mathsf{case} \,  \SYSTEMnt{t_{{\mathrm{1}}}}  \, \mathsf{of} \, \{ \mathsf{inj1} \,  \SYSTEMmv{y}  \rightarrow  \SYSTEMnt{t_{{\mathrm{2}}}}  ; \, \mathsf{inj2} \,  \SYSTEMmv{z}  \rightarrow  \SYSTEMnt{t_{{\mathrm{3}}}}  \}   \smidge \textcolor{LGcolor}{\rrparenthesis}   \\
  \textit{(defn. subst [lhs])}  & \Rightarrow &  \textcolor{LGcolor}{\llparenthesis} \smidge   \mathsf{case} \,   [  \SYSTEMnt{t}  /  \SYSTEMmv{x}  ]  \SYSTEMnt{t_{{\mathrm{1}}}}   \, \mathsf{of} \, \{ \mathsf{inj1} \,  \SYSTEMmv{y}  \rightarrow   [  \SYSTEMnt{t}  /  \SYSTEMmv{x}  ]  \SYSTEMnt{t_{{\mathrm{2}}}}   ; \, \mathsf{inj2} \,  \SYSTEMmv{z}  \rightarrow   [  \SYSTEMnt{t}  /  \SYSTEMmv{x}  ]  \SYSTEMnt{t_{{\mathrm{3}}}}   \}   \smidge \textcolor{LGcolor}{\rrparenthesis}             \\
                                &             & \quad \equiv  [   \textcolor{LGcolor}{\llparenthesis} \smidge  \SYSTEMnt{t}  \smidge \textcolor{LGcolor}{\rrparenthesis}   /  \SYSTEMmv{x}  ]   \textcolor{LGcolor}{\llparenthesis} \smidge   \mathsf{case} \,  \SYSTEMnt{t_{{\mathrm{1}}}}  \, \mathsf{of} \, \{ \mathsf{inj1} \,  \SYSTEMmv{y}  \rightarrow  \SYSTEMnt{t_{{\mathrm{2}}}}  ; \, \mathsf{inj2} \,  \SYSTEMmv{z}  \rightarrow  \SYSTEMnt{t_{{\mathrm{3}}}}  \}   \smidge \textcolor{LGcolor}{\rrparenthesis}   \\
\end{array}
  \end{align*}
\end{gather*}
\item ($ \textsf{push}_\otimes     $) $\SYSTEMnt{t'} \equiv  \textsf{push}_\otimes  \SYSTEMnt{t_{{\mathrm{1}}}} $ (with $ \SYSTEMmv{y}  \,\#\,  \SYSTEMnt{t}  \wedge  \SYSTEMmv{z}  \,\#\,  \SYSTEMnt{t} $);
\begin{gather*}
\begin{align*}
\begin{array}{rcrlll}
\textit{(goal)} & &  \textcolor{LGcolor}{\llparenthesis} \smidge   [  \SYSTEMnt{t}  /  \SYSTEMmv{x}  ]   \textsf{push}_\otimes  \SYSTEMnt{t_{{\mathrm{1}}}}    \smidge \textcolor{LGcolor}{\rrparenthesis}  & \eqStep &  [   \textcolor{LGcolor}{\llparenthesis} \smidge  \SYSTEMnt{t}  \smidge \textcolor{LGcolor}{\rrparenthesis}   /  \SYSTEMmv{x}  ]   \textcolor{LGcolor}{\llparenthesis} \smidge   \textsf{push}_\otimes  \SYSTEMnt{t_{{\mathrm{1}}}}   \smidge \textcolor{LGcolor}{\rrparenthesis}   \\
\textit{(defn. subst [lhs])} & \Rightarrow &  \textcolor{LGcolor}{\llparenthesis} \smidge   \textsf{push}_\otimes  \SYSTEMsym{(}   [  \SYSTEMnt{t}  /  \SYSTEMmv{x}  ]  \SYSTEMnt{t_{{\mathrm{1}}}}   \SYSTEMsym{)}   \smidge \textcolor{LGcolor}{\rrparenthesis}  & \eqStep &  [   \textcolor{LGcolor}{\llparenthesis} \smidge  \SYSTEMnt{t}  \smidge \textcolor{LGcolor}{\rrparenthesis}   /  \SYSTEMmv{x}  ]   \textcolor{LGcolor}{\llparenthesis} \smidge   \textsf{push}_\otimes  \SYSTEMnt{t_{{\mathrm{1}}}}   \smidge \textcolor{LGcolor}{\rrparenthesis}   \\
\textit{(defn. interp [lhs])} & \Rightarrow &  \mathsf{let} \, \textcolor{coeffectColor}{[}  \SYSTEMmv{x}  \textcolor{coeffectColor}{]} =   \textcolor{LGcolor}{\llparenthesis} \smidge   [  \SYSTEMnt{t}  /  \SYSTEMmv{x}  ]  \SYSTEMnt{t_{{\mathrm{1}}}}   \smidge \textcolor{LGcolor}{\rrparenthesis}   \, \mathsf{in} \,   \mathsf{let} \, \langle  \SYSTEMmv{y} ,  \SYSTEMmv{z}  \rangle =  \SYSTEMmv{x}  \, \mathsf{in} \,   \langle   \textcolor{coeffectColor}{[}  \SYSTEMmv{y}  \textcolor{coeffectColor}{]}  ,   \textcolor{coeffectColor}{[}  \SYSTEMmv{z}  \textcolor{coeffectColor}{]}   \rangle    & \eqStep &  [   \textcolor{LGcolor}{\llparenthesis} \smidge  \SYSTEMnt{t}  \smidge \textcolor{LGcolor}{\rrparenthesis}   /  \SYSTEMmv{x}  ]   \textcolor{LGcolor}{\llparenthesis} \smidge   \textsf{push}_\otimes  \SYSTEMnt{t_{{\mathrm{1}}}}   \smidge \textcolor{LGcolor}{\rrparenthesis}   \\
\end{array}
\end{align*}
\end{gather*}
\item ($ \textsf{push}_\oplus     $) $\SYSTEMnt{t'} \equiv  \textsf{push}_\oplus  \SYSTEMnt{t_{{\mathrm{1}}}} $ (with $ \SYSTEMmv{y}  \,\#\,  \SYSTEMnt{t}  \wedge  \SYSTEMmv{z}  \,\#\,  \SYSTEMnt{t} $);
\begin{gather*}
\begin{align*}
\begin{array}{rcrlll}
\textit{(goal)} & &  \textcolor{LGcolor}{\llparenthesis} \smidge   [  \SYSTEMnt{t}  /  \SYSTEMmv{x}  ]   \textsf{push}_\oplus  \SYSTEMnt{t_{{\mathrm{1}}}}    \smidge \textcolor{LGcolor}{\rrparenthesis}  & \eqStep &  [   \textcolor{LGcolor}{\llparenthesis} \smidge  \SYSTEMnt{t}  \smidge \textcolor{LGcolor}{\rrparenthesis}   /  \SYSTEMmv{x}  ]   \textcolor{LGcolor}{\llparenthesis} \smidge   \textsf{push}_\oplus  \SYSTEMnt{t_{{\mathrm{1}}}}   \smidge \textcolor{LGcolor}{\rrparenthesis}   \\
\textit{(defn. subst [lhs])} & \Rightarrow &  \textcolor{LGcolor}{\llparenthesis} \smidge   \textsf{push}_\oplus  \SYSTEMsym{(}   [  \SYSTEMnt{t}  /  \SYSTEMmv{x}  ]  \SYSTEMnt{t_{{\mathrm{1}}}}   \SYSTEMsym{)}   \smidge \textcolor{LGcolor}{\rrparenthesis}  & \eqStep &  [   \textcolor{LGcolor}{\llparenthesis} \smidge  \SYSTEMnt{t}  \smidge \textcolor{LGcolor}{\rrparenthesis}   /  \SYSTEMmv{x}  ]   \textcolor{LGcolor}{\llparenthesis} \smidge   \textsf{push}_\oplus  \SYSTEMnt{t_{{\mathrm{1}}}}   \smidge \textcolor{LGcolor}{\rrparenthesis}   \\
\end{array}
\end{align*}
\end{gather*}
\item ($ \textsf{push}_{\mathrm{unit} }     $) $\SYSTEMnt{t'} \equiv  \textsf{push}_{\mathrm{unit} }  \SYSTEMnt{t_{{\mathrm{1}}}} $ (with $ \SYSTEMmv{y}  \,\#\,  \SYSTEMnt{t} $);
\begin{gather*}
\begin{align*}
\begin{array}{rcrlll}
\textit{(goal)} & &  \textcolor{LGcolor}{\llparenthesis} \smidge   [  \SYSTEMnt{t}  /  \SYSTEMmv{x}  ]   \textsf{push}_{\mathrm{unit} }  \SYSTEMnt{t_{{\mathrm{1}}}}    \smidge \textcolor{LGcolor}{\rrparenthesis}  & \eqStep &  [   \textcolor{LGcolor}{\llparenthesis} \smidge  \SYSTEMnt{t}  \smidge \textcolor{LGcolor}{\rrparenthesis}   /  \SYSTEMmv{x}  ]   \textcolor{LGcolor}{\llparenthesis} \smidge   \textsf{push}_{\mathrm{unit} }  \SYSTEMnt{t_{{\mathrm{1}}}}   \smidge \textcolor{LGcolor}{\rrparenthesis}   \\
\textit{(defn. subst [lhs])} & \Rightarrow &  \textcolor{LGcolor}{\llparenthesis} \smidge   \textsf{push}_{\mathrm{unit} }  \SYSTEMsym{(}   [  \SYSTEMnt{t}  /  \SYSTEMmv{x}  ]  \SYSTEMnt{t_{{\mathrm{1}}}}   \SYSTEMsym{)}   \smidge \textcolor{LGcolor}{\rrparenthesis}  & \eqStep &  [   \textcolor{LGcolor}{\llparenthesis} \smidge  \SYSTEMnt{t}  \smidge \textcolor{LGcolor}{\rrparenthesis}   /  \SYSTEMmv{x}  ]   \textcolor{LGcolor}{\llparenthesis} \smidge   \textsf{push}_{\mathrm{unit} }  \SYSTEMnt{t_{{\mathrm{1}}}}   \smidge \textcolor{LGcolor}{\rrparenthesis}   \\
\textit{(defn. interp [lhs])} & \Rightarrow &  \mathsf{let} \, \textcolor{coeffectColor}{[}  \SYSTEMmv{y}  \textcolor{coeffectColor}{]} =   \textcolor{LGcolor}{\llparenthesis} \smidge   [  \SYSTEMnt{t}  /  \SYSTEMmv{x}  ]  \SYSTEMnt{t_{{\mathrm{1}}}}   \smidge \textcolor{LGcolor}{\rrparenthesis}   \, \mathsf{in} \,   \mathsf{let} \, \langle \rangle =  \SYSTEMmv{y}  \, \mathsf{in} \,   \langle \rangle    & \eqStep &  [   \textcolor{LGcolor}{\llparenthesis} \smidge  \SYSTEMnt{t}  \smidge \textcolor{LGcolor}{\rrparenthesis}   /  \SYSTEMmv{x}  ]   \textcolor{LGcolor}{\llparenthesis} \smidge   \textsf{push}_{\mathrm{unit} }  \SYSTEMnt{t_{{\mathrm{1}}}}   \smidge \textcolor{LGcolor}{\rrparenthesis}   \\
\end{array}
\end{align*}
\end{gather*}
\end{itemize}
\end{proof}

The operational correspondence then follows according to the following proof:

\begin{proof}
  The proof subsumes the earlier proofs; we add the additional cases here due to the
  extension of the language from Linear Base to Linear Core.

  We consider the theorem in an expanded form: for every $ \SYSTEMnt{t}  \rightsquigarrow_{\textsc{l} }  \SYSTEMnt{t'} $
  then $\exists t'' .   \textcolor{LGcolor}{\llparenthesis} \smidge  \SYSTEMnt{t}  \smidge \textcolor{LGcolor}{\rrparenthesis}   \rightsquigarrow_{\textsc{g} }^\ast  \SYSTEMnt{t''}  \wedge  \textcolor{LGcolor}{\llparenthesis} \smidge  \SYSTEMnt{t'}  \smidge \textcolor{LGcolor}{\rrparenthesis}  \equiv \SYSTEMnt{t''}$.
  \begin{itemize}
\item (prodCong)
\[
\SYSTEMdruleSemLinprodCong{}
\]
By induction, analogously to the case for (congAppL).

\item (prodBeta)
\[
\SYSTEMdruleSemLinprodBeta{}
\]
The interpretation of the reducing term is:
\[
  \textcolor{LGcolor}{\llparenthesis} \smidge   \mathsf{let} \, \langle  \SYSTEMmv{x} ,  \SYSTEMmv{y}  \rangle =   \langle  \SYSTEMnt{t_{{\mathrm{1}}}} ,  \SYSTEMnt{t_{{\mathrm{2}}}}  \rangle   \, \mathsf{in} \,  \SYSTEMnt{t_{{\mathrm{3}}}}   \smidge \textcolor{LGcolor}{\rrparenthesis}   \equiv   \mathsf{let} \, \langle  \SYSTEMmv{x} ,  \SYSTEMmv{y}  \rangle =   \langle   \textcolor{LGcolor}{\llparenthesis} \smidge  \SYSTEMnt{t_{{\mathrm{1}}}}  \smidge \textcolor{LGcolor}{\rrparenthesis}  ,   \textcolor{LGcolor}{\llparenthesis} \smidge  \SYSTEMnt{t_{{\mathrm{2}}}}  \smidge \textcolor{LGcolor}{\rrparenthesis}   \rangle   \, \mathsf{in} \,   \textcolor{LGcolor}{\llparenthesis} \smidge  \SYSTEMnt{t_{{\mathrm{3}}}}  \smidge \textcolor{LGcolor}{\rrparenthesis}   
\]
where $ \SYSTEMmv{x} ,   \SYSTEMmv{y}  \,\#\,  \SYSTEMnt{t}  $.
Then we construct the reduction in Graded Base:
\begin{align*}
\inferrule*[right=\SYSTEMRenameRuleSemGrdprodBeta{}]
{ }
{  \mathsf{let} \, \langle  \SYSTEMmv{x} ,  \SYSTEMmv{y}  \rangle =   \langle   \textcolor{LGcolor}{\llparenthesis} \smidge  \SYSTEMnt{t_{{\mathrm{1}}}}  \smidge \textcolor{LGcolor}{\rrparenthesis}  ,   \textcolor{LGcolor}{\llparenthesis} \smidge  \SYSTEMnt{t_{{\mathrm{2}}}}  \smidge \textcolor{LGcolor}{\rrparenthesis}   \rangle   \, \mathsf{in} \,   \textcolor{LGcolor}{\llparenthesis} \smidge  \SYSTEMnt{t_{{\mathrm{3}}}}  \smidge \textcolor{LGcolor}{\rrparenthesis}    \rightsquigarrow_{\textsc{l} }   [   \textcolor{LGcolor}{\llparenthesis} \smidge  \SYSTEMnt{t_{{\mathrm{1}}}}  \smidge \textcolor{LGcolor}{\rrparenthesis}   /  \SYSTEMmv{x}  ]   [   \textcolor{LGcolor}{\llparenthesis} \smidge  \SYSTEMnt{t_{{\mathrm{2}}}}  \smidge \textcolor{LGcolor}{\rrparenthesis}   /  \SYSTEMmv{y}  ]   \textcolor{LGcolor}{\llparenthesis} \smidge  \SYSTEMnt{t_{{\mathrm{3}}}}  \smidge \textcolor{LGcolor}{\rrparenthesis}    }
\end{align*}
By Lemma~\ref{lemma:interp-lin-to-grd-mod-preserves-subst} then interpreting
the graded-base reduced term is equal to the linear-base reduced term:
\begin{align*}
 \textcolor{LGcolor}{\llparenthesis} \smidge   [  \SYSTEMnt{t_{{\mathrm{1}}}}  /  \SYSTEMmv{x}  ]   [  \SYSTEMnt{t_{{\mathrm{2}}}}  /  \SYSTEMmv{y}  ]  \SYSTEMnt{t_{{\mathrm{3}}}}    \smidge \textcolor{LGcolor}{\rrparenthesis}  & \equiv  [   \textcolor{LGcolor}{\llparenthesis} \smidge  \SYSTEMnt{t_{{\mathrm{1}}}}  \smidge \textcolor{LGcolor}{\rrparenthesis}   /  \SYSTEMmv{x}  ]   \textcolor{LGcolor}{\llparenthesis} \smidge   [  \SYSTEMnt{t_{{\mathrm{2}}}}  /  \SYSTEMmv{y}  ]  \SYSTEMnt{t_{{\mathrm{3}}}}   \smidge \textcolor{LGcolor}{\rrparenthesis}   \\
                                     & \equiv  [   \textcolor{LGcolor}{\llparenthesis} \smidge  \SYSTEMnt{t_{{\mathrm{1}}}}  \smidge \textcolor{LGcolor}{\rrparenthesis}   /  \SYSTEMmv{x}  ]   [   \textcolor{LGcolor}{\llparenthesis} \smidge  \SYSTEMnt{t_{{\mathrm{2}}}}  \smidge \textcolor{LGcolor}{\rrparenthesis}   /  \SYSTEMmv{y}  ]   \textcolor{LGcolor}{\llparenthesis} \smidge  \SYSTEMnt{t_{{\mathrm{3}}}}  \smidge \textcolor{LGcolor}{\rrparenthesis}   
\end{align*}

\item (unitCong)
\[
\SYSTEMdruleSemLinunitCong{}
\]
By induction, analogously to the case for (congAppL).

\item (unitBeta)
\[
\SYSTEMdruleSemLinunitBeta{}
\]
The interpretation of the reducing term is:
\[
  \textcolor{LGcolor}{\llparenthesis} \smidge   \mathsf{let} \, \langle \rangle =   \langle \rangle   \, \mathsf{in} \,  \SYSTEMnt{t_{{\mathrm{2}}}}   \smidge \textcolor{LGcolor}{\rrparenthesis}   \equiv   \mathsf{let} \, \langle \rangle =   \langle \rangle   \, \mathsf{in} \,   \textcolor{LGcolor}{\llparenthesis} \smidge  \SYSTEMnt{t_{{\mathrm{2}}}}  \smidge \textcolor{LGcolor}{\rrparenthesis}   
\]
Then we construct the reduction in Graded Base:
\begin{align*}
\inferrule*[right=\SYSTEMRenameRuleSemGrdModbetaBox{}]
{ }
{  \mathsf{let} \, \langle \rangle =   \langle \rangle   \, \mathsf{in} \,   \textcolor{LGcolor}{\llparenthesis} \smidge  \SYSTEMnt{t_{{\mathrm{2}}}}  \smidge \textcolor{LGcolor}{\rrparenthesis}    \rightsquigarrow_{\textsc{l} }   \textcolor{LGcolor}{\llparenthesis} \smidge  \SYSTEMnt{t_{{\mathrm{2}}}}  \smidge \textcolor{LGcolor}{\rrparenthesis}  }
\end{align*}
By reflexivity then interpreting the graded-base reduced term is equal to the
linear-base reduced term.

\item (congCase)
\[
\SYSTEMdruleSemLincongCase{}
\]
By induction, analogously to the case for (congAppL).

\item (caseInj1)
\[
\SYSTEMdruleSemLincaseInjOne{}
\]
The interpretation of the reducing term is:
\[
  \textcolor{LGcolor}{\llparenthesis} \smidge   \mathsf{case} \,   \mathsf{inj}_1 \,  \SYSTEMnt{t}   \, \mathsf{of} \, \{ \mathsf{inj1} \,  \SYSTEMmv{x}  \rightarrow  \SYSTEMnt{t_{{\mathrm{1}}}}  ; \, \mathsf{inj2} \,  \SYSTEMmv{y}  \rightarrow  \SYSTEMnt{t_{{\mathrm{2}}}}  \}   \smidge \textcolor{LGcolor}{\rrparenthesis}   \equiv   \mathsf{case} \,   \mathsf{inj}_1 \,   \textcolor{LGcolor}{\llparenthesis} \smidge  \SYSTEMnt{t}  \smidge \textcolor{LGcolor}{\rrparenthesis}    \, \mathsf{of} \, \{ \mathsf{inj1} \,  \SYSTEMmv{x}  \rightarrow   \textcolor{LGcolor}{\llparenthesis} \smidge  \SYSTEMnt{t_{{\mathrm{1}}}}  \smidge \textcolor{LGcolor}{\rrparenthesis}   ; \, \mathsf{inj2} \,  \SYSTEMmv{y}  \rightarrow   \textcolor{LGcolor}{\llparenthesis} \smidge  \SYSTEMnt{t_{{\mathrm{2}}}}  \smidge \textcolor{LGcolor}{\rrparenthesis}   \}  
\]
Then we construct the reduction in Graded Base:
\begin{align*}
\inferrule*[right=\SYSTEMRenameRuleSemGrdcaseInjOne{}]
{ }
{  \mathsf{case} \,   \mathsf{inj}_1 \,   \textcolor{LGcolor}{\llparenthesis} \smidge  \SYSTEMnt{t}  \smidge \textcolor{LGcolor}{\rrparenthesis}    \, \mathsf{of} \, \{ \mathsf{inj1} \,  \SYSTEMmv{x}  \rightarrow   \textcolor{LGcolor}{\llparenthesis} \smidge  \SYSTEMnt{t_{{\mathrm{1}}}}  \smidge \textcolor{LGcolor}{\rrparenthesis}   ; \, \mathsf{inj2} \,  \SYSTEMmv{y}  \rightarrow   \textcolor{LGcolor}{\llparenthesis} \smidge  \SYSTEMnt{t_{{\mathrm{2}}}}  \smidge \textcolor{LGcolor}{\rrparenthesis}   \}   \rightsquigarrow_{\textsc{l} }   [   \textcolor{LGcolor}{\llparenthesis} \smidge  \SYSTEMnt{t}  \smidge \textcolor{LGcolor}{\rrparenthesis}   /  \SYSTEMmv{x}  ]   \textcolor{LGcolor}{\llparenthesis} \smidge  \SYSTEMnt{t_{{\mathrm{1}}}}  \smidge \textcolor{LGcolor}{\rrparenthesis}   }
\end{align*}
By Lemma~\ref{lemma:interp-lin-to-grd-mod-preserves-subst} then interpreting
the graded-base reduced term is equal to the linear-base
reduced term: $  \textcolor{LGcolor}{\llparenthesis} \smidge   [  \SYSTEMnt{t}  /  \SYSTEMmv{x}  ]  \SYSTEMnt{t_{{\mathrm{1}}}}   \smidge \textcolor{LGcolor}{\rrparenthesis}   \equiv   [   \textcolor{LGcolor}{\llparenthesis} \smidge  \SYSTEMnt{t}  \smidge \textcolor{LGcolor}{\rrparenthesis}   /  \SYSTEMmv{x}  ]   \textcolor{LGcolor}{\llparenthesis} \smidge  \SYSTEMnt{t_{{\mathrm{1}}}}  \smidge \textcolor{LGcolor}{\rrparenthesis}   $.

\item (caseInj2)
\[
\SYSTEMdruleSemLincaseInjTwo{}
\]
Analogously to the case for (caseInj1).

\item (pushProdCong)
\[
\SYSTEMdruleSemLinpushProdCong{}
\]
By induction, analogously to the case for (congAppL).

\item (pushProdBoxCong)
\[
\SYSTEMdruleSemLinpushProdBoxCong{}
\]
By induction, analogously to the case for (congAppL).

\item (pushProd)
\[
\SYSTEMdruleSemLinpushProd{}
\]
The interpretation of the reducing term is (with $ \begin{array}{cc}    \SYSTEMmv{z}  \,\#\,  \SYSTEMnt{t_{{\mathrm{1}}}}    \; & \;    \SYSTEMmv{z}  \,\#\,  \SYSTEMnt{t_{{\mathrm{2}}}}    \end{array} $):
\[
  \textcolor{LGcolor}{\llparenthesis} \smidge   \textsf{push}_\otimes   \textcolor{coeffectColor}{[}   \langle  \SYSTEMnt{t_{{\mathrm{1}}}} ,  \SYSTEMnt{t_{{\mathrm{2}}}}  \rangle   \textcolor{coeffectColor}{]}    \smidge \textcolor{LGcolor}{\rrparenthesis}   \equiv   \mathsf{let} \, \textcolor{coeffectColor}{[}  \SYSTEMmv{z}  \textcolor{coeffectColor}{]} =   \textcolor{coeffectColor}{[}   \langle   \textcolor{LGcolor}{\llparenthesis} \smidge  \SYSTEMnt{t_{{\mathrm{1}}}}  \smidge \textcolor{LGcolor}{\rrparenthesis}  ,   \textcolor{LGcolor}{\llparenthesis} \smidge  \SYSTEMnt{t_{{\mathrm{2}}}}  \smidge \textcolor{LGcolor}{\rrparenthesis}   \rangle   \textcolor{coeffectColor}{]}   \, \mathsf{in} \,   \mathsf{let} \, \langle  \SYSTEMmv{x} ,  \SYSTEMmv{y}  \rangle =  \SYSTEMmv{z}  \, \mathsf{in} \,   \langle   \textcolor{coeffectColor}{[}  \SYSTEMmv{x}  \textcolor{coeffectColor}{]}  ,   \textcolor{coeffectColor}{[}  \SYSTEMmv{y}  \textcolor{coeffectColor}{]}   \rangle    
\]
Then we construct the following reduction sequence in Graded Base:
\begin{align*}
& \inferrule*[right=\SYSTEMRenameRuleSemGrdModbetaBox{}]
{ }
{  \mathsf{let} \, \textcolor{coeffectColor}{[}  \SYSTEMmv{z}  \textcolor{coeffectColor}{]} =   \textcolor{coeffectColor}{[}   \langle   \textcolor{LGcolor}{\llparenthesis} \smidge  \SYSTEMnt{t_{{\mathrm{1}}}}  \smidge \textcolor{LGcolor}{\rrparenthesis}  ,   \textcolor{LGcolor}{\llparenthesis} \smidge  \SYSTEMnt{t_{{\mathrm{2}}}}  \smidge \textcolor{LGcolor}{\rrparenthesis}   \rangle   \textcolor{coeffectColor}{]}   \, \mathsf{in} \,   \mathsf{let} \, \langle  \SYSTEMmv{x} ,  \SYSTEMmv{y}  \rangle =  \SYSTEMmv{z}  \, \mathsf{in} \,   \langle   \textcolor{coeffectColor}{[}  \SYSTEMmv{x}  \textcolor{coeffectColor}{]}  ,   \textcolor{coeffectColor}{[}  \SYSTEMmv{y}  \textcolor{coeffectColor}{]}   \rangle     \rightsquigarrow_{\textsc{l} }   [   \langle   \textcolor{LGcolor}{\llparenthesis} \smidge  \SYSTEMnt{t_{{\mathrm{1}}}}  \smidge \textcolor{LGcolor}{\rrparenthesis}  ,   \textcolor{LGcolor}{\llparenthesis} \smidge  \SYSTEMnt{t_{{\mathrm{2}}}}  \smidge \textcolor{LGcolor}{\rrparenthesis}   \rangle   /  \SYSTEMmv{z}  ]   \mathsf{let} \, \langle  \SYSTEMmv{x} ,  \SYSTEMmv{y}  \rangle =  \SYSTEMmv{z}  \, \mathsf{in} \,   \langle   \textcolor{coeffectColor}{[}  \SYSTEMmv{x}  \textcolor{coeffectColor}{]}  ,   \textcolor{coeffectColor}{[}  \SYSTEMmv{y}  \textcolor{coeffectColor}{]}   \rangle    } \\
& \inferrule*[right=\SYSTEMRenameRuleSemGrdprodBeta{}]
{ }
{  \mathsf{let} \, \langle  \SYSTEMmv{x} ,  \SYSTEMmv{y}  \rangle =   \textcolor{coeffectColor}{[}   \langle   \textcolor{LGcolor}{\llparenthesis} \smidge  \SYSTEMnt{t_{{\mathrm{1}}}}  \smidge \textcolor{LGcolor}{\rrparenthesis}  ,   \textcolor{LGcolor}{\llparenthesis} \smidge  \SYSTEMnt{t_{{\mathrm{2}}}}  \smidge \textcolor{LGcolor}{\rrparenthesis}   \rangle   \textcolor{coeffectColor}{]}   \, \mathsf{in} \,   \langle   \textcolor{coeffectColor}{[}  \SYSTEMmv{x}  \textcolor{coeffectColor}{]}  ,   \textcolor{coeffectColor}{[}  \SYSTEMmv{y}  \textcolor{coeffectColor}{]}   \rangle    \rightsquigarrow_{\textsc{l} }   \langle   \textcolor{coeffectColor}{[}   \textcolor{LGcolor}{\llparenthesis} \smidge  \SYSTEMnt{t_{{\mathrm{1}}}}  \smidge \textcolor{LGcolor}{\rrparenthesis}   \textcolor{coeffectColor}{]}  ,   \textcolor{coeffectColor}{[}   \textcolor{LGcolor}{\llparenthesis} \smidge  \SYSTEMnt{t_{{\mathrm{2}}}}  \smidge \textcolor{LGcolor}{\rrparenthesis}   \textcolor{coeffectColor}{]}   \rangle  }
\end{align*}
By the definition of interpretation, $  \textcolor{LGcolor}{\llparenthesis} \smidge   \langle  \SYSTEMnt{t_{{\mathrm{1}}}} ,  \SYSTEMnt{t_{{\mathrm{2}}}}  \rangle   \smidge \textcolor{LGcolor}{\rrparenthesis}   \equiv   \langle   \textcolor{coeffectColor}{[}   \textcolor{LGcolor}{\llparenthesis} \smidge  \SYSTEMnt{t_{{\mathrm{1}}}}  \smidge \textcolor{LGcolor}{\rrparenthesis}   \textcolor{coeffectColor}{]}  ,   \textcolor{coeffectColor}{[}   \textcolor{LGcolor}{\llparenthesis} \smidge  \SYSTEMnt{t_{{\mathrm{2}}}}  \smidge \textcolor{LGcolor}{\rrparenthesis}   \textcolor{coeffectColor}{]}   \rangle  $.

\item (pushUnitCong)
\[
\SYSTEMdruleSemLinpushUnitCong{}
\]
By induction, analogously to the case for (congAppL).

\item (pushUnitBoxCong)
\[
\SYSTEMdruleSemLinpushUnitBoxCong{}
\]
By induction, analogously to the case for (congAppL).

\item (pushUnit)
\[
\SYSTEMdruleSemLinpushUnit{}
\]
The interpretation of the reducing term is:
\[
  \textcolor{LGcolor}{\llparenthesis} \smidge   \textsf{push}_\otimes   \textcolor{coeffectColor}{[}   \langle \rangle   \textcolor{coeffectColor}{]}    \smidge \textcolor{LGcolor}{\rrparenthesis}   \equiv   \mathsf{let} \, \textcolor{coeffectColor}{[}  \SYSTEMmv{x}  \textcolor{coeffectColor}{]} =   \textcolor{coeffectColor}{[}   \langle \rangle   \textcolor{coeffectColor}{]}   \, \mathsf{in} \,   \mathsf{let} \, \langle \rangle =  \SYSTEMmv{x}  \, \mathsf{in} \,   \langle \rangle    
\]
Then we construct the following reduction sequence in Graded Base:
\begin{align*}
& \inferrule*[right=\SYSTEMRenameRuleSemGrdModbetaBox{}]
{ }
{  \mathsf{let} \, \textcolor{coeffectColor}{[}  \SYSTEMmv{x}  \textcolor{coeffectColor}{]} =   \textcolor{coeffectColor}{[}   \langle \rangle   \textcolor{coeffectColor}{]}   \, \mathsf{in} \,   \mathsf{let} \, \langle \rangle =  \SYSTEMmv{x}  \, \mathsf{in} \,   \langle \rangle     \rightsquigarrow_{\textsc{l} }   [   \langle \rangle   /  \SYSTEMmv{x}  ]   \mathsf{let} \, \langle \rangle =  \SYSTEMmv{x}  \, \mathsf{in} \,   \langle \rangle    } \\
& \inferrule*[right=\SYSTEMRenameRuleSemGrdunitBeta{}]
{ }
{  \mathsf{let} \, \langle \rangle =   \langle \rangle   \, \mathsf{in} \,   \langle \rangle    \rightsquigarrow_{\textsc{l} }   \langle \rangle  }
\end{align*}
By the definition of interpretation, $  \textcolor{LGcolor}{\llparenthesis} \smidge   \langle \rangle   \smidge \textcolor{LGcolor}{\rrparenthesis}   \equiv   \langle \rangle  $.

\item (pushSumCong)
\[
\SYSTEMdruleSemLinpushSumCong{}
\]
By induction, analogously to the case for (congAppL).

\item (pushSumBoxCong)
\[
\SYSTEMdruleSemLinpushSumCong{}
\]
By induction, analogously to the case for (congAppL).

\item (pushSumInj1)
\[
\SYSTEMdruleSemLinpushSumInjOne{}
\]
The interpretation of the reducing term is ($ \SYSTEMmv{z}  \,\#\,  \SYSTEMnt{t_{{\mathrm{1}}}} $):
\[
  \textcolor{LGcolor}{\llparenthesis} \smidge   \textsf{push}_\oplus   \textcolor{coeffectColor}{[}   \mathsf{inj}_1 \,  \SYSTEMnt{t_{{\mathrm{1}}}}   \textcolor{coeffectColor}{]}    \smidge \textcolor{LGcolor}{\rrparenthesis}   \equiv   \mathsf{let} \, \textcolor{coeffectColor}{[}  \SYSTEMmv{z}  \textcolor{coeffectColor}{]} =   \textcolor{coeffectColor}{[}   \mathsf{inj}_1 \,   \textcolor{LGcolor}{\llparenthesis} \smidge  \SYSTEMnt{t_{{\mathrm{1}}}}  \smidge \textcolor{LGcolor}{\rrparenthesis}    \textcolor{coeffectColor}{]}   \, \mathsf{in} \,   \mathsf{case} \,  \SYSTEMmv{z}  \, \mathsf{of} \, \{ \mathsf{inj1} \,  \SYSTEMmv{x}  \rightarrow   \textcolor{coeffectColor}{[}  \SYSTEMmv{x}  \textcolor{coeffectColor}{]}   ; \, \mathsf{inj2} \,  \SYSTEMmv{y}  \rightarrow   \textcolor{coeffectColor}{[}  \SYSTEMmv{y}  \textcolor{coeffectColor}{]}   \}   
\]
Then we construct the following reduction sequence in Graded Base:
\begin{gather*}
    \begin{align*}
    &\inferrule*[right=\SYSTEMRenameRuleSemGrdModbetaBox{}]
    { }
    {  \mathsf{let} \, \textcolor{coeffectColor}{[}  \SYSTEMmv{z}  \textcolor{coeffectColor}{]} =   \textcolor{coeffectColor}{[}   \mathsf{inj}_1 \,   \textcolor{LGcolor}{\llparenthesis} \smidge  \SYSTEMnt{t_{{\mathrm{1}}}}  \smidge \textcolor{LGcolor}{\rrparenthesis}    \textcolor{coeffectColor}{]}   \, \mathsf{in} \,   \mathsf{case} \,  \SYSTEMmv{z}  \, \mathsf{of} \, \{ \mathsf{inj1} \,  \SYSTEMmv{x}  \rightarrow   \textcolor{coeffectColor}{[}  \SYSTEMmv{x}  \textcolor{coeffectColor}{]}   ; \, \mathsf{inj2} \,  \SYSTEMmv{y}  \rightarrow   \textcolor{coeffectColor}{[}  \SYSTEMmv{y}  \textcolor{coeffectColor}{]}   \}    \rightsquigarrow_{\textsc{l} }   \mathsf{case} \,   \mathsf{inj}_1 \,   \textcolor{LGcolor}{\llparenthesis} \smidge  \SYSTEMnt{t_{{\mathrm{1}}}}  \smidge \textcolor{LGcolor}{\rrparenthesis}    \, \mathsf{of} \, \{ \mathsf{inj1} \,  \SYSTEMmv{x}  \rightarrow   \textcolor{coeffectColor}{[}  \SYSTEMmv{x}  \textcolor{coeffectColor}{]}   ; \, \mathsf{inj2} \,  \SYSTEMmv{y}  \rightarrow   \textcolor{coeffectColor}{[}  \SYSTEMmv{y}  \textcolor{coeffectColor}{]}   \}  } \\
    &\inferrule*[right=\SYSTEMRenameRuleSemGrdcaseInjOne{}]
    { }
    {  \mathsf{case} \,   \mathsf{inj}_1 \,   \textcolor{LGcolor}{\llparenthesis} \smidge  \SYSTEMnt{t_{{\mathrm{1}}}}  \smidge \textcolor{LGcolor}{\rrparenthesis}    \, \mathsf{of} \, \{ \mathsf{inj1} \,  \SYSTEMmv{x}  \rightarrow   \textcolor{coeffectColor}{[}  \SYSTEMmv{x}  \textcolor{coeffectColor}{]}   ; \, \mathsf{inj2} \,  \SYSTEMmv{y}  \rightarrow   \textcolor{coeffectColor}{[}  \SYSTEMmv{y}  \textcolor{coeffectColor}{]}   \}   \rightsquigarrow_{\textsc{l} }   \mathsf{inj}_1 \,   \textcolor{coeffectColor}{[}   \textcolor{LGcolor}{\llparenthesis} \smidge  \SYSTEMnt{t_{{\mathrm{1}}}}  \smidge \textcolor{LGcolor}{\rrparenthesis}   \textcolor{coeffectColor}{]}   }
    \end{align*}
\end{gather*}
By the definition of interpretation, $  \textcolor{LGcolor}{\llparenthesis} \smidge   \mathsf{inj}_1 \,   \textcolor{coeffectColor}{[}  \SYSTEMnt{t_{{\mathrm{1}}}}  \textcolor{coeffectColor}{]}    \smidge \textcolor{LGcolor}{\rrparenthesis}   \equiv   \mathsf{inj}_1 \,   \textcolor{coeffectColor}{[}   \textcolor{LGcolor}{\llparenthesis} \smidge  \SYSTEMnt{t_{{\mathrm{1}}}}  \smidge \textcolor{LGcolor}{\rrparenthesis}   \textcolor{coeffectColor}{]}   $.

\item (pushSumInj2)
\[
\SYSTEMdruleSemLinpushSumInjTwo{}
\]
Analogously to the case for (pushSumInj1).
  \end{itemize}
\end{proof}

\subsubsection{Equation correspondence}
\label{app:proofs-lin-core-to-grad-core-eqs}

\begin{proof}
  The proof subsumes the earlier proofs; we add the additional cases here due to the
  extension of the language from Linear Base to Linear Core.
  \begin{itemize}
  \item \[\SYSTEMdruleLinEqbetaUnit{}\]
     \begin{align*}
      \begin{array}{rl}
        &  \textcolor{LGcolor}{\llparenthesis} \smidge   \mathsf{let} \, \langle \rangle =   \langle \rangle   \, \mathsf{in} \,  \SYSTEMnt{t}   \smidge \textcolor{LGcolor}{\rrparenthesis}  \\
        \textit{\{defn. translation\}} & =  \mathsf{let} \, \langle \rangle =   \langle \rangle   \, \mathsf{in} \,   \textcolor{LGcolor}{\llparenthesis} \smidge  \SYSTEMnt{t}  \smidge \textcolor{LGcolor}{\rrparenthesis}   \\
        \textit{\{\SYSTEMRenameRuleGradEqbetaUnit{}\}} &  \equiv_{\textsc{g} }   \textcolor{LGcolor}{\llparenthesis} \smidge  \SYSTEMnt{t}  \smidge \textcolor{LGcolor}{\rrparenthesis} 
      \end{array}
    \end{align*}

  \item \[\SYSTEMdruleLinEqetaUnit{}\]

    \begin{align*}
      \begin{array}{rl}
        &  \textcolor{LGcolor}{\llparenthesis} \smidge   \mathsf{let} \, \langle \rangle =  \SYSTEMnt{t}  \, \mathsf{in} \,   \langle \rangle    \smidge \textcolor{LGcolor}{\rrparenthesis}  \\
        \textit{\{defn. translation\}} & =  \mathsf{let} \, \langle \rangle =   \textcolor{LGcolor}{\llparenthesis} \smidge  \SYSTEMnt{t}  \smidge \textcolor{LGcolor}{\rrparenthesis}   \, \mathsf{in} \,   \langle \rangle   \\
        \textit{\{\SYSTEMRenameRuleGradEqetaUnit{}\}} &  \equiv_{\textsc{g} }   \textcolor{LGcolor}{\llparenthesis} \smidge  \SYSTEMnt{t}  \smidge \textcolor{LGcolor}{\rrparenthesis} 
      \end{array}
    \end{align*}

  \item \[ \SYSTEMdruleLinEqbetaProd{}  \]
    \begin{align*}
      \begin{array}{rl}
        &  \textcolor{LGcolor}{\llparenthesis} \smidge   \mathsf{let} \, \langle  \SYSTEMmv{x} ,  \SYSTEMmv{y}  \rangle =   \langle  \SYSTEMnt{t_{{\mathrm{1}}}} ,  \SYSTEMnt{t_{{\mathrm{2}}}}  \rangle   \, \mathsf{in} \,  \SYSTEMnt{t}   \smidge \textcolor{LGcolor}{\rrparenthesis}  \\
        \textit{\{defn. translation\}} & =  \mathsf{let} \, \langle  \SYSTEMmv{x'} ,  \SYSTEMmv{y'}  \rangle =   \langle   \textcolor{LGcolor}{\llparenthesis} \smidge  \SYSTEMnt{t_{{\mathrm{1}}}}  \smidge \textcolor{LGcolor}{\rrparenthesis}  ,   \textcolor{LGcolor}{\llparenthesis} \smidge  \SYSTEMnt{t_{{\mathrm{2}}}}  \smidge \textcolor{LGcolor}{\rrparenthesis}   \rangle   \, \mathsf{in} \,   \textcolor{LGcolor}{\llparenthesis} \smidge  \SYSTEMnt{t}  \smidge \textcolor{LGcolor}{\rrparenthesis}   \\
        \textit{\{\SYSTEMRenameRuleGradEqbetaProd{}\}} &  \equiv_{\textsc{g} }   [   \textcolor{LGcolor}{\llparenthesis} \smidge  \SYSTEMnt{t_{{\mathrm{1}}}}  \smidge \textcolor{LGcolor}{\rrparenthesis}   /  \SYSTEMmv{x}  ]   [   \textcolor{LGcolor}{\llparenthesis} \smidge  \SYSTEMnt{t_{{\mathrm{2}}}}  \smidge \textcolor{LGcolor}{\rrparenthesis}   /  \SYSTEMmv{y}  ]   \textcolor{LGcolor}{\llparenthesis} \smidge  \SYSTEMnt{t}  \smidge \textcolor{LGcolor}{\rrparenthesis}    \\
      \end{array}
      \end{align*}

  \item \[\SYSTEMdruleLinEqetaProd{}\]

    \begin{align*}
      \begin{array}{rl}
        &  \textcolor{LGcolor}{\llparenthesis} \smidge   \mathsf{let} \, \langle  \SYSTEMmv{x} ,  \SYSTEMmv{y}  \rangle =  \SYSTEMnt{t}  \, \mathsf{in} \,   \langle  \SYSTEMmv{x} ,  \SYSTEMmv{y}  \rangle    \smidge \textcolor{LGcolor}{\rrparenthesis}  \\
        \textit{\{defn. translation\}} & =  \mathsf{let} \, \langle  \SYSTEMmv{x} ,  \SYSTEMmv{y}  \rangle =   \textcolor{LGcolor}{\llparenthesis} \smidge  \SYSTEMnt{t}  \smidge \textcolor{LGcolor}{\rrparenthesis}   \, \mathsf{in} \,   \langle  \SYSTEMmv{x} ,  \SYSTEMmv{y}  \rangle   \\
        \textit{\SYSTEMRenameRuleGradEqetaProd{}} &  \equiv_{\textsc{g} }   \textcolor{LGcolor}{\llparenthesis} \smidge  \SYSTEMnt{t}  \smidge \textcolor{LGcolor}{\rrparenthesis} 
      \end{array}
      \end{align*}

  \item \[\SYSTEMdruleLinEqbetaSumOne{}\]
        \begin{align*}
      \begin{array}{rl}
        &  \textcolor{LGcolor}{\llparenthesis} \smidge   \mathsf{case} \,   \mathsf{inj}_1 \,  \SYSTEMnt{t}   \, \mathsf{of} \, \{ \mathsf{inj1} \,  \SYSTEMmv{x}  \rightarrow  \SYSTEMnt{t_{{\mathrm{1}}}}  ; \, \mathsf{inj2} \,  \SYSTEMmv{y}  \rightarrow  \SYSTEMnt{t_{{\mathrm{2}}}}  \}   \smidge \textcolor{LGcolor}{\rrparenthesis}  \\
        \textit{\{defn. translation\}} & =  \mathsf{case} \,   \mathsf{inj}_1 \,   \textcolor{LGcolor}{\llparenthesis} \smidge  \SYSTEMnt{t}  \smidge \textcolor{LGcolor}{\rrparenthesis}    \, \mathsf{of} \, \{ \mathsf{inj1} \,  \SYSTEMmv{x}  \rightarrow   \textcolor{LGcolor}{\llparenthesis} \smidge  \SYSTEMnt{t_{{\mathrm{1}}}}  \smidge \textcolor{LGcolor}{\rrparenthesis}   ; \, \mathsf{inj2} \,  \SYSTEMmv{y}  \rightarrow   \textcolor{LGcolor}{\llparenthesis} \smidge  \SYSTEMnt{t_{{\mathrm{2}}}}  \smidge \textcolor{LGcolor}{\rrparenthesis}   \}  \\
        \textit{\{\SYSTEMRenameRuleGradEqbetaSumOne{}\}} &  \equiv_{\textsc{g} }   [   \textcolor{LGcolor}{\llparenthesis} \smidge  \SYSTEMnt{t}  \smidge \textcolor{LGcolor}{\rrparenthesis}   /  \SYSTEMmv{x}  ]   \textcolor{LGcolor}{\llparenthesis} \smidge  \SYSTEMnt{t_{{\mathrm{1}}}}  \smidge \textcolor{LGcolor}{\rrparenthesis}  
      \end{array}
      \end{align*}

      \item  \[\SYSTEMdruleLinEqbetaSumTwo{}\]
        As above but symmetrically for the other beta rule.

      \item \[\SYSTEMdruleLinEqetaSum{}\]

            \begin{align*}
      \begin{array}{rl}
        &  \textcolor{LGcolor}{\llparenthesis} \smidge   \mathsf{case} \,  \SYSTEMnt{t}  \, \mathsf{of} \, \{ \mathsf{inj1} \,  \SYSTEMmv{x}  \rightarrow   \mathsf{inj}_1 \,  \SYSTEMmv{x}   ; \, \mathsf{inj2} \,  \SYSTEMmv{y}  \rightarrow   \mathsf{inj}_2 \,  \SYSTEMmv{y}   \}   \smidge \textcolor{LGcolor}{\rrparenthesis}  \\
        \textit{\{defn. translation\}} & =  \mathsf{case} \,   \textcolor{LGcolor}{\llparenthesis} \smidge  \SYSTEMnt{t}  \smidge \textcolor{LGcolor}{\rrparenthesis}   \, \mathsf{of} \, \{ \mathsf{inj1} \,  \SYSTEMmv{x}  \rightarrow   \mathsf{inj}_1 \,  \SYSTEMmv{x}   ; \, \mathsf{inj2} \,  \SYSTEMmv{y}  \rightarrow   \mathsf{inj}_2 \,  \SYSTEMmv{y}   \}  \\
        \textit{\{\SYSTEMRenameRuleGradEqetaSum{}\}} &  \equiv_{\textsc{g} }   \textcolor{LGcolor}{\llparenthesis} \smidge  \SYSTEMnt{t}  \smidge \textcolor{LGcolor}{\rrparenthesis} 
      \end{array}
            \end{align*}

  \item \[\SYSTEMdruleLinEqpushUnitBeta{}\]

    \begin{align*}
      \begin{array}{rl}
        &  \textcolor{LGcolor}{\llparenthesis} \smidge   \textsf{push}_{\mathrm{unit} }   \textcolor{coeffectColor}{[}   \langle \rangle   \textcolor{coeffectColor}{]}    \smidge \textcolor{LGcolor}{\rrparenthesis}  \\
        \textit{\{defn. translation\}} & =  \mathsf{let} \, \textcolor{coeffectColor}{[}  \SYSTEMmv{x}  \textcolor{coeffectColor}{]} =   \textcolor{coeffectColor}{[}   \langle \rangle   \textcolor{coeffectColor}{]}   \, \mathsf{in} \,   \mathsf{let} \, \langle \rangle =  \SYSTEMmv{x}  \, \mathsf{in} \,   \langle \rangle    \\
        \textit{\{\SYSTEMRenameRuleGradMEqbetaBox \}} &  \equiv_{\textsc{g} }   \mathsf{let} \, \langle \rangle =   \langle \rangle   \, \mathsf{in} \,   \langle \rangle   \\
        \textit{\{\SYSTEMRenameRuleGradEqbetaUnit{}\}} &  \equiv_{\textsc{g} }   \langle \rangle  \\
        \textit{\{defn. translation\}} & =  \textcolor{LGcolor}{\llparenthesis} \smidge   \langle \rangle   \smidge \textcolor{LGcolor}{\rrparenthesis} 
      \end{array}
    \end{align*}

  \item \[\SYSTEMdruleLinEqpushProdBeta{}\]

    \begin{align*}
      \begin{array}{rl}
        &  \textcolor{LGcolor}{\llparenthesis} \smidge   \textsf{push}_\otimes   \textcolor{coeffectColor}{[}   \langle  \SYSTEMnt{t_{{\mathrm{1}}}} ,  \SYSTEMnt{t_{{\mathrm{2}}}}  \rangle   \textcolor{coeffectColor}{]}    \smidge \textcolor{LGcolor}{\rrparenthesis}  \\
        \textit{\{defn. translation\}} & =  \mathsf{let} \, \textcolor{coeffectColor}{[}  \SYSTEMmv{x}  \textcolor{coeffectColor}{]} =   \textcolor{coeffectColor}{[}   \langle   \textcolor{LGcolor}{\llparenthesis} \smidge  \SYSTEMnt{t_{{\mathrm{1}}}}  \smidge \textcolor{LGcolor}{\rrparenthesis}  ,   \textcolor{LGcolor}{\llparenthesis} \smidge  \SYSTEMnt{t_{{\mathrm{2}}}}  \smidge \textcolor{LGcolor}{\rrparenthesis}   \rangle   \textcolor{coeffectColor}{]}   \, \mathsf{in} \,   \mathsf{let} \, \langle  \SYSTEMmv{y} ,  \SYSTEMmv{z}  \rangle =  \SYSTEMmv{x}  \, \mathsf{in} \,   \langle   \textcolor{coeffectColor}{[}  \SYSTEMmv{y}  \textcolor{coeffectColor}{]}  ,   \textcolor{coeffectColor}{[}  \SYSTEMmv{z}  \textcolor{coeffectColor}{]}   \rangle    \\
        \textit{\{\SYSTEMRenameRuleGradMEqbetaBox \}} &  \equiv_{\textsc{g} }   \mathsf{let} \, \langle  \SYSTEMmv{y} ,  \SYSTEMmv{z}  \rangle =   \langle   \textcolor{LGcolor}{\llparenthesis} \smidge  \SYSTEMnt{t_{{\mathrm{1}}}}  \smidge \textcolor{LGcolor}{\rrparenthesis}  ,   \textcolor{LGcolor}{\llparenthesis} \smidge  \SYSTEMnt{t_{{\mathrm{2}}}}  \smidge \textcolor{LGcolor}{\rrparenthesis}   \rangle   \, \mathsf{in} \,   \langle   \textcolor{coeffectColor}{[}  \SYSTEMmv{y}  \textcolor{coeffectColor}{]}  ,   \textcolor{coeffectColor}{[}  \SYSTEMmv{z}  \textcolor{coeffectColor}{]}   \rangle   \\
        \textit{\{\SYSTEMRenameRuleGradEqbetaProd{}\}} &  \equiv_{\textsc{g} }   \langle   \textcolor{coeffectColor}{[}   \textcolor{LGcolor}{\llparenthesis} \smidge  \SYSTEMnt{t_{{\mathrm{1}}}}  \smidge \textcolor{LGcolor}{\rrparenthesis}   \textcolor{coeffectColor}{]}  ,   \textcolor{coeffectColor}{[}   \textcolor{LGcolor}{\llparenthesis} \smidge  \SYSTEMnt{t_{{\mathrm{2}}}}  \smidge \textcolor{LGcolor}{\rrparenthesis}   \textcolor{coeffectColor}{]}   \rangle  \\
        \textit{\{defn. translation\}} & =  \textcolor{LGcolor}{\llparenthesis} \smidge   \langle   \textcolor{coeffectColor}{[}  \SYSTEMnt{t_{{\mathrm{1}}}}  \textcolor{coeffectColor}{]}  ,   \textcolor{coeffectColor}{[}  \SYSTEMnt{t_{{\mathrm{2}}}}  \textcolor{coeffectColor}{]}   \rangle   \smidge \textcolor{LGcolor}{\rrparenthesis} 
      \end{array}
    \end{align*}

  \item \[\SYSTEMdruleLinEqpushSumBetaOne{}\]

    \begin{align*}
      \begin{array}{rl}
        &  \textcolor{LGcolor}{\llparenthesis} \smidge   \textsf{push}_\oplus   \textcolor{coeffectColor}{[}   \mathsf{inj}_1 \,  \SYSTEMnt{t}   \textcolor{coeffectColor}{]}    \smidge \textcolor{LGcolor}{\rrparenthesis}  \\
        \textit{\{defn. translation\}} & =  \mathsf{let} \, \textcolor{coeffectColor}{[}  \SYSTEMmv{x}  \textcolor{coeffectColor}{]} =   \textcolor{coeffectColor}{[}   \mathsf{inj}_1 \,   \textcolor{LGcolor}{\llparenthesis} \smidge  \SYSTEMnt{t}  \smidge \textcolor{LGcolor}{\rrparenthesis}    \textcolor{coeffectColor}{]}   \, \mathsf{in} \,   \mathsf{case} \,  \SYSTEMmv{x}  \, \mathsf{of} \, \{ \mathsf{inj1} \,  \SYSTEMmv{y}  \rightarrow   \mathsf{inj}_1 \,   \textcolor{coeffectColor}{[}  \SYSTEMmv{y}  \textcolor{coeffectColor}{]}    ; \, \mathsf{inj2} \,  \SYSTEMmv{z}  \rightarrow   \mathsf{inj}_2 \,   \textcolor{coeffectColor}{[}  \SYSTEMmv{z}  \textcolor{coeffectColor}{]}    \}   \\
        \textit{\{\SYSTEMRenameRuleGradMEqbetaBox \}} &  \equiv_{\textsc{g} }   \mathsf{case} \,   \mathsf{inj}_1 \,   \textcolor{LGcolor}{\llparenthesis} \smidge  \SYSTEMnt{t}  \smidge \textcolor{LGcolor}{\rrparenthesis}    \, \mathsf{of} \, \{ \mathsf{inj1} \,  \SYSTEMmv{y}  \rightarrow   \mathsf{inj}_1 \,   \textcolor{coeffectColor}{[}  \SYSTEMmv{y}  \textcolor{coeffectColor}{]}    ; \, \mathsf{inj2} \,  \SYSTEMmv{z}  \rightarrow   \mathsf{inj}_2 \,   \textcolor{coeffectColor}{[}  \SYSTEMmv{z}  \textcolor{coeffectColor}{]}    \}  \\
        \textit{\{\SYSTEMRenameRuleGradEqbetaSumOne{}\}} &  \equiv_{\textsc{g} }   \mathsf{inj}_1 \,   \textcolor{coeffectColor}{[}   \textcolor{LGcolor}{\llparenthesis} \smidge  \SYSTEMnt{t}  \smidge \textcolor{LGcolor}{\rrparenthesis}   \textcolor{coeffectColor}{]}   \\
        \textit{\{defn. translation\}} & =  \textcolor{LGcolor}{\llparenthesis} \smidge   \mathsf{inj}_1 \,   \textcolor{coeffectColor}{[}  \SYSTEMnt{t}  \textcolor{coeffectColor}{]}    \smidge \textcolor{LGcolor}{\rrparenthesis} 
      \end{array}
    \end{align*}

  \item \[\SYSTEMdruleLinEqpushSumBetaTwo{}\]

    \begin{align*}
      \begin{array}{rl}
        &  \textcolor{LGcolor}{\llparenthesis} \smidge   \textsf{push}_\oplus   \textcolor{coeffectColor}{[}   \mathsf{inj}_2 \,  \SYSTEMnt{t}   \textcolor{coeffectColor}{]}    \smidge \textcolor{LGcolor}{\rrparenthesis}  \\
        \textit{\{defn. translation\}} & =  \mathsf{let} \, \textcolor{coeffectColor}{[}  \SYSTEMmv{x}  \textcolor{coeffectColor}{]} =   \textcolor{coeffectColor}{[}   \mathsf{inj}_2 \,   \textcolor{LGcolor}{\llparenthesis} \smidge  \SYSTEMnt{t}  \smidge \textcolor{LGcolor}{\rrparenthesis}    \textcolor{coeffectColor}{]}   \, \mathsf{in} \,   \mathsf{case} \,  \SYSTEMmv{x}  \, \mathsf{of} \, \{ \mathsf{inj1} \,  \SYSTEMmv{y}  \rightarrow   \mathsf{inj}_1 \,   \textcolor{coeffectColor}{[}  \SYSTEMmv{y}  \textcolor{coeffectColor}{]}    ; \, \mathsf{inj2} \,  \SYSTEMmv{z}  \rightarrow   \mathsf{inj}_2 \,   \textcolor{coeffectColor}{[}  \SYSTEMmv{z}  \textcolor{coeffectColor}{]}    \}   \\
        \textit{\{\SYSTEMRenameRuleGradMEqbetaBox \}} &  \equiv_{\textsc{g} }   \mathsf{case} \,   \mathsf{inj}_2 \,   \textcolor{LGcolor}{\llparenthesis} \smidge  \SYSTEMnt{t}  \smidge \textcolor{LGcolor}{\rrparenthesis}    \, \mathsf{of} \, \{ \mathsf{inj1} \,  \SYSTEMmv{y}  \rightarrow   \mathsf{inj}_1 \,   \textcolor{coeffectColor}{[}  \SYSTEMmv{y}  \textcolor{coeffectColor}{]}    ; \, \mathsf{inj2} \,  \SYSTEMmv{z}  \rightarrow   \mathsf{inj}_2 \,   \textcolor{coeffectColor}{[}  \SYSTEMmv{z}  \textcolor{coeffectColor}{]}    \}  \\
        \textit{\{\SYSTEMRenameRuleGradEqbetaSumTwo{}\}} &  \equiv_{\textsc{g} }   \mathsf{inj}_2 \,   \textcolor{coeffectColor}{[}   \textcolor{LGcolor}{\llparenthesis} \smidge  \SYSTEMnt{t}  \smidge \textcolor{LGcolor}{\rrparenthesis}   \textcolor{coeffectColor}{]}   \\
        \textit{\{defn. translation\}} & =  \textcolor{LGcolor}{\llparenthesis} \smidge   \mathsf{inj}_2 \,   \textcolor{coeffectColor}{[}  \SYSTEMnt{t}  \textcolor{coeffectColor}{]}    \smidge \textcolor{LGcolor}{\rrparenthesis} 
      \end{array}
    \end{align*}

  \item Preservation of congruence equations follows simply by induction and congruence
    in the target calculus.

  \end{itemize}
\end{proof}

\fi

\end{document}